\documentclass[b5paper,twoside,english]{memoir}
\setsecnumdepth{subsection}
\maxtocdepth{subsection}

\usepackage[b5paper,top=2.5cm,bottom=2.5cm,left=2.5cm,right=2.0cm]{geometry}
\usepackage{graphicx}
\usepackage{epstopdf}
\usepackage{epsfig}
\usepackage{hyperref}
\usepackage{cite,multirow}
\usepackage{url}
\usepackage{xcolor}
\usepackage{cancel}
\usepackage{amssymb}
\usepackage{amsmath}
\usepackage{amsfonts}
\usepackage{bm}
\usepackage{tikz}
\usepackage{feynarts}
\usetikzlibrary{shapes,arrows}
\hypersetup{
    colorlinks,
    urlcolor=blue,
    anchorcolor=blue,
    citecolor=blue,
    filecolor=blue,
    linkcolor=blue,
    menucolor=blue,
    pagecolor=blue,
    linktocpage=true,
    pdfproducer=medialab,
    pdfa=true
}
\makeindex

\begin{document}

\frontmatter
\pagestyle{empty}

\begin{center} 
  \begin{figure}[ht!]
    \centering
    \includegraphics[height=2.8cm]{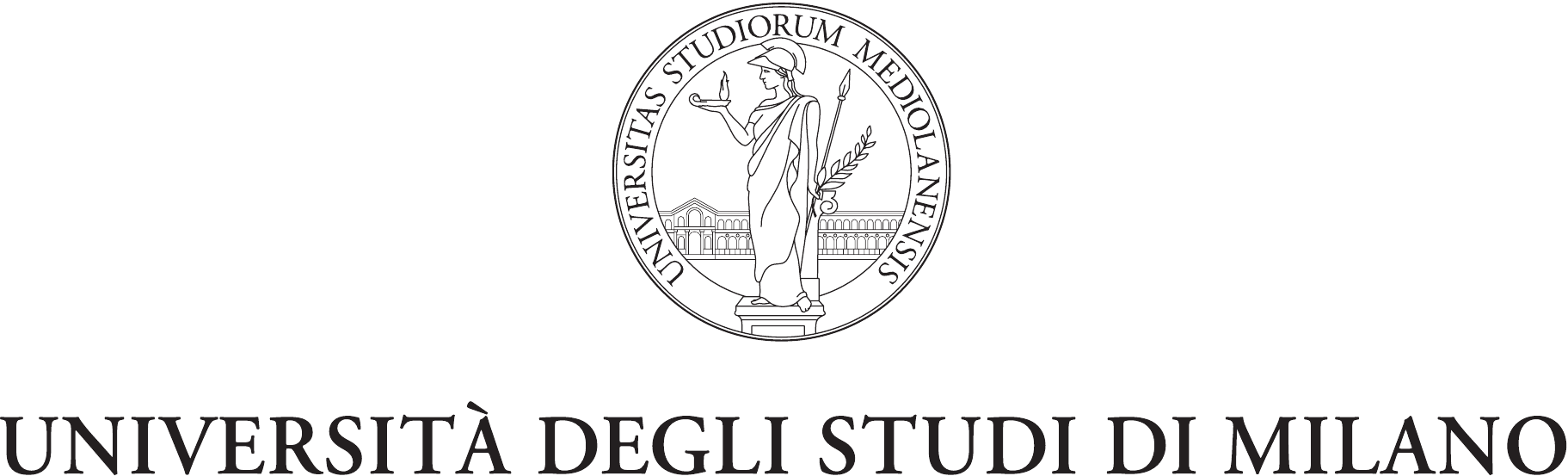}
  \end{figure}
  Scuola di Dottorato in Fisica, Astrofisica e Fisica Applicata

  \medskip
  Dipartimento di Fisica  

  \medskip
  \medskip
  \medskip
  Corso di Dottorato in Fisica, Astrofisica e Fisica Applicata

  \medskip
  Ciclo XXVII
  \bigskip\vfill
  \bigskip
  \bigskip
  {\vfil
    \begin{Huge}
      \begin{bfseries}
        {\renewcommand{\\}{\cr} \halign{\hbox
            to\textwidth{\strut\hfil#\hfil}\cr Parton distribution
            functions\\ with QED corrections\cr}}
      \end{bfseries}
    \end{Huge}}
  \bigskip
  \bigskip
  \bigskip
  \vfill
  Settore Scientifico Disciplinare FIS/02
  
  \bigskip
  \bigskip
  \bigskip
  \bigskip
  \bigskip
  \bigskip
  \bigskip
  \bigskip
  \begin{tabular*}{\textwidth}{@{}l@{\extracolsep{\fill}}l@{}}
    &Tesi di Dottorato di:\vspace{0.2cm}\\
    &%
    \vtop{\halign{\strut\hfil#\hfil\cr
        Stefano~\uppercase{CARRAZZA}\cr}}
    \vspace{1.0cm}\\
    Supervisore: 
    \vtop{\halign{\strut\hfil#\hfil\cr
        Prof. Stefano \uppercase{FORTE}\cr}} &
    \vspace{0.5cm}\\
    Coordinatore: 
    \vtop{\halign{\strut\hfil#\hfil\cr
        Prof. Marco \uppercase{BERSANELLI}\cr}}
  \end{tabular*}
\end{center}
\begin{center}
  \vspace{3cm}
  Anno Accademico 2014-2015\\
  \bigskip
  \bigskip
\end{center}
\clearpage

\pagestyle{empty}
\noindent{\textbf{Committee of the final examination:}}\vspace{0.4cm}

\noindent{External Referee:}

\noindent{Prof. Jos\'e~I.~\uppercase{LATORRE}}\vspace{0.2cm}

\noindent{External Members:}

\noindent{Dr. Michelangelo~L.~\uppercase{MANGANO}}

\noindent{Prof. Paolo~\uppercase{GAMBINO}}\vspace{0.2cm}

\noindent{Internal Members:}

\noindent{Prof. Alessandro~\uppercase{VICINI}}

\vspace{0.6cm}

\noindent{\textbf{Final examination:}}\vspace{0.4cm}

\noindent{July 6, 2015}

\noindent{Universit\`{a} degli Studi di Milano, Dipartimento di
  Fisica, Milano, Italy}\vspace{0.6cm}

\noindent{\textbf{MIUR subjects:}} \vspace{0.2cm}

\noindent{FIS/02 - Fisica Teorica, Modelli e Metodi Matematici}\vspace{0.6cm}

\noindent{{\textbf{PACS:}}} \vspace{0.2cm}

\noindent{02.70.Uu, 07.05.Mh, 12.38.-t, 13.60.Hb, 89.20.Ft, 12.15.Lk} \vspace{0.6cm}

\noindent{\textbf{Keywords:}} \vspace{0.2cm}

\noindent{Parton Distribution Functions (PDFs), Electroweak
  corrections to PDFs, Photon PDF}

\vspace{0.6cm}

\noindent{\textbf{Official Collaboration logos:}}\vspace{0.2cm}

\begin{figure}[h]
  \includegraphics[scale=0.3]{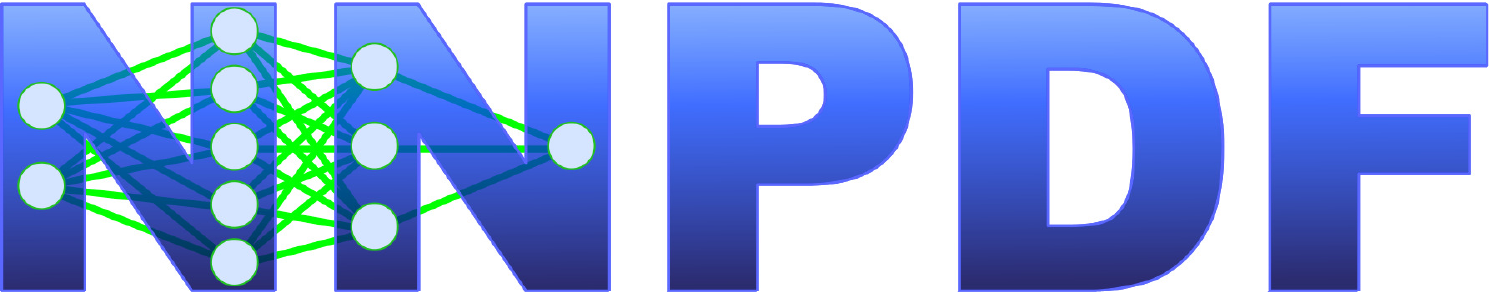}
\end{figure}

\begin{figure}[h]
  \includegraphics[scale=0.45]{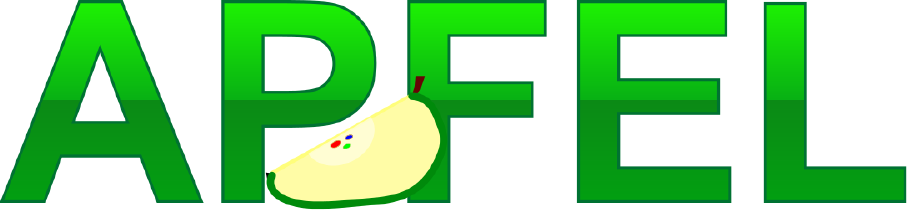}
\end{figure}

\vspace{0.2cm}

\vfill
\noindent{\textbf{Cover illustration:}}\vspace{0.2cm}

\noindent{S. Carrazza, \textit{PDF covariance}, generated with
  1000 replicas of NNPDF3.0 NLO.}
\vspace{0.6cm}

\noindent{\textbf{Internal illustrations:}} \vspace{0.2cm}

\noindent{All plots have been produced by Stefano~Carrazza}


\noindent{Typeset by Stefano Carrazza using \LaTeX}
\clearpage

\vspace{5cm}
{\hfill\textit{To Vera Lucia, Dante e Roxana}}
\cleardoublepage

\rmfamily
\normalfont

\chapter{Abstract}
\thispagestyle{plain}

We present the first unbiased determination of parton distribution
functions (PDFs) with electroweak corrections. 
The aim of this thesis is to provide an exhaustive description of the
theoretical framework and the technical implementation which leads to
the determination of a set of PDFs which includes the photon PDF and
quantum electrodynamics (QED) contributions to parton evolution.
First, we introduce and motivate the need of including electroweak
corrections to PDFs, providing phenomenological examples and
presenting an overview of the current state of the art in PDF fits.
The theoretical implications of such corrections are then described
through the implementation of the combined QCD$\otimes$QED evolution
in \texttt{APFEL}, a public code for the solution of the PDF evolution
developed particularly for this thesis.
We proceed by presenting the new structure of the Neural-Network PDF
(NNPDF) methodology used for the extraction of this set of PDFs with
QED corrections. We then provide a first determination of the full set
of PDFs based on deep-inelastic scattering data and LHC data for $W$
and $Z/\gamma^*$ Drell-Yan production, using leading-order QED and NLO
or NNLO QCD: the so-called NNPDF2.3QED set of PDFs.
We perform a preliminary investigation of the phenomenological
implications of NNPDF2.3QED set, in particular, focusing on the
photon-induced corrections to direct photon production at HERA,
high-mass dilepton and $W$ pair production at the LHC and finally,
providing a first determination of lepton PDFs through the \texttt{APFEL}
evolution.
We conclude with a summary of the technological upgrades required for
the improvement of future PDF determinations with electroweak
corrections.

\cleardoublepage

\chapter{List of Publications}
\label{sec:listpub}
\textbf{Refereed publications}

\begin{itemize}
\item V.~Bertone, S.~Carrazza and E.~R.~Nocera, \textit{Reference
    results for time-like evolution up to
    $\mathcal{O}(\alpha_s^3)$}, JHEP \textbf{1503} (2015) 46
  \newline
  \href{http://arxiv.org/abs/1501.00494}{arXiv:1501.00494},
  \href{10.1007/JHEP03(2015)046}{DOI:10.1007/JHEP03(2015)046}
  
\item R.~D.~Ball \textit{et al.}, \textit{Parton distributions for the
    LHC Run II}, JHEP (2015) 
  \newline
  \href{http://arxiv.org/abs/1410.8849}{arXiv:1410:8849},
  \href{10.1007/JHEP04(2015)040}{DOI:10.1007/JHEP04(2015)040}

\item S.~Carrazza \textit{et al.}, \textit{APFEL Web: a web-based
    application for the graphical visualization of parton distribution
    functions}, J. Phys. G: Nucl. Part. Phys. \textbf{45} 057001 
  \newline
  \href{http://arxiv.org/abs/1410.5456}{arXiv:1410.5456},
  \href{10.1088/0954-3899/42/5/057001}{DOI:10.1088/0954-3899/42/5/057001}

\item S.~Carrazza and J.~Pires, \textit{Perturbative QCD description
    of jet data from LHC Run-I and Tevatron Run-II}, JHEP
  \textbf{1410} (2014) 145 \newline
  \href{http://arxiv.org/abs/1407.7031}{arXiv:1407.7031},
  \href{http://dx.doi.org/10.1007/JHEP10(2014)145}{DOI:
    10.1007/JHEP10(2014)145}

\item P.~Skands, S.~Carrazza and J.~Rojo, \textit{Tuning PYTHIA 8.1:
    the Monash 2013 Tune}, Eur. Phys. J. C \textbf{74} (2014) 8, 3024  
  \newline
  \href{http://arxiv.org/abs/1404.5630 }{arXiv:1404.5630},
  \href{http://dx.doi.org/10.1140/epjc/s10052-014-3024-y}{DOI: 10.1140/epjc/s10052-014-3024-y}

\item V.~Bertone, S.~Carrazza and J.~Rojo, \textit{APFEL: A PDF
    Evolution Library with QED corrections}, Comput. Phys. Commun.
  \textbf{185} (2014) 1647 \newline
  \href{http://arxiv.org/abs/1310.1394}{arXiv:1310.1394},
  \href{http://dx.doi.org/10.1016/j.cpc.2014.03.007}{DOI: 10.1016/j.cpc.2014.03.007}

\item R.~D.~Ball \textit{et al.}, \textit{Parton distributions with
    QED corrections}, Nucl. Phys. B \textbf{877} (2013) 290
  \newline
  \href{http://arxiv.org/abs/1308.0598}{arXiv:1308.0598},
  \href{http://dx.doi.org/10.1016/j.nuclphysb.2013.10.010}{DOI: 10.1016/j.nuclphysb.2013.10.010}

\item R.~D.~Ball \textit{et al.}, \textit{Parton Distribution
    Benchmarking with LHC Data}, JHEP \textbf{1304} (2013) 125
  \newline 
  \href{http://arxiv.org/abs/1211.5142}{arXiv:1211.5142},
  \href{http://dx.doi.org/10.1007/JHEP04(2013)125}{DOI: 10.1007/JHEP04(2013)125}

\item R.~D.~Ball \textit{et al.}, \textit{Parton distributions with
    LHC data}, Nucl. Phys. B \textbf{867} (2013) 244
  \newline
  \href{http://arxiv.org/abs/1207.1303}{arXiv:1207.1303},
  \href{http://dx.doi.org/10.1016/j.nuclphysb.2012.10.003}{DOI: 10.1016/j.nuclphysb.2012.10.003}

\end{itemize}
\newpage
\noindent
\textbf{Publications in reports}
\begin{itemize}

\item J.~M.~Campbell \textit{et. al.}, \textit{Working Group Report:
    Quantum Chromodynamics}, Snowmass community summer study 2013
  \newline
  \href{http://arxiv.org/abs/1310.5189}{arXiv:1310.5189}

\end{itemize}
\textbf{Publications in conference proceedings}

\begin{itemize}

\item S.~Carrazza, \textit{Disentangling electroweak effects in
    Z-boson production}, Nuovo Cimento, La Thuile 2014
  \newline
  \href{http://arxiv.org/abs/1405.1728}{arXiv:1405.1728}

\item S.~Carrazza, S.~Forte and J.~Rojo, \textit{Parton Distributions
    and Event Generators}, ISMD 2013
  \newline
  \href{http://arxiv.org/abs/1311.5887}{arXiv:1311.5887}

\item S.~Carrazza, \textit{Towards the determination of the photon
    parton distribution function constrained by LHC data}, PoS DIS
  \textbf{2013} (2013) 279 \newline
  \href{http://arxiv.org/abs/1307.1131}{arXiv:1307.1131}

\item S.~Carrazza, \textit{Towards an unbiased determination of parton
    distributions with QED corrections}, Les Rencontres de Moriond, p.357-360
  2013 \newline \href{http://arxiv.org/abs/1305.4179}{arXiv:1305.4179}

\end{itemize}

\chapter{Acknowledgements}
\pagestyle{headings}

First of all, I am infinitely grateful to my parents who have always
taken part of my life and supported my activities. They have provided
me an excellent cultural environment during my existence and they have
always helped me to fulfill my dreams. I dedicate this thesis to them
and to my sister.\vspace{0.1cm}

\noindent I also would like to thank my supervisor, Prof.~Stefano
Forte, who guided me during my doctorate with passion, determination
and a lot patience. I am immensely thankful for his encouragement in
becoming part of the scientific community, in particular by motivating
and supporting my participation at several doctoral schools and
international conferences.\vspace{0.1cm}

\noindent I thank all my collaborators for their support and
motivation too. In particular, a special thanks to Juan Rojo, Valerio
Bertone, Nathan Hartland and Emanuele Nocera. Thanks to these
interactions I have learned and developed a solid scientific
approach.\vspace{0.1cm}

\noindent I would like to thank Prof.~Jos\'e~I.~Latorre for accepting
to be the referee of this work. I thank as well the external and
internal members of the final examination committee:
Dr.~Michelangelo~L.~Mangano, which provided me an exciting experience
as visitor at CERN, Prof. Paolo Gambino and Prof. Alessandro
Vicini.\vspace{0.1cm}

\noindent Finally, I would like to thank all people from the Milan
Department of Physics with whom I have interacted during these last
wonderful years, in particular the members of the Milan Theoretical
Physics Group: Giancarlo Ferrera, Jo\~{a}o Pires. I take this
opportunity also to thank Alfio Ferrara, Silvia Salini and Massimo
Florio in the context of the EIBURS project.  \vspace{0.1cm}

\begin{flushright}
Stefano~Carrazza
\end{flushright}
\cleardoublepage

\pagestyle{headings}
\tableofcontents

\mainmatter
\chapter*{Introduction}
\addcontentsline{toc}{chapter}{Introduction}
\label{chap:intro}

During the past years we have witnessed several discoveries predicted
by the Standard Model (SM) of particle physics, for example, the
discovery of the tau neutrino~\cite{Decamp:1989tu}, the top
quark~\cite{Abe:1995hr}, and recently in 2012, the Higgs
boson~\cite{Aad:2012tfa,Chatrchyan:2012ufa} and its subsequent
phenomenological characterization, thanks to the measurements
performed by the Large Hadron Collider (LHC) at CERN.

The great success of the SM is practically due to the two theories
which it is based: the Quantum Chromodynamics (QCD) and the
Electroweak theory. The QCD is the theory of strong interactions
between quarks, antiquarks and gluons (\emph{partons}), meanwhile
Quantum Electrodynamics (QED) and Weak interactions are described by
the unified Electroweak theory. Both theories are in a continuous
development of calculation techniques which improves the accuracy of
theoretical predictions since the latter half of the 20th century. It
is interesting to remark that even if the gauge groups of such
theories are factored ($SU(3) \times SU(2) \times U(1)$) the theory is
unique. First of all, we notice that these interactions are connected
to each other through the mediation of common fundamental particles,
moreover, from a technical point of view, the CKM matrix mixes the
strong and weak interactions and the commutation of the generators of
the strong and weak interactions constraints their form. This remark
is essential in the context of this thesis where we present explicitly
the combination of the QCD and QED theories.

Parton distribution functions (PDFs) are one of the most important
ingredients for a realistic computation of any particle physics
observable thanks to the collinear factorization property of QCD
states. This formalism expresses any cross-section, $\sigma$, as the
convolution product
\begin{equation}
  \sigma = \hat{\sigma} \otimes f,
\end{equation}
where the elementary hard cross-section $\hat{\sigma}$ is convoluted
with $f$ the PDF. The hard cross-section is computed in QCD and it
depends just on the physical process, meanwhile PDFs cannot be
computed using perturbative QCD because of the confinement property of
QCD. PDFs carry the probability that a nucleon contains a parton with
a certain momentum fraction, this information is process independent
and thus are extracted from experimental data.

Motivated by the need for greater precision phenomenology at the LHC,
the inclusion of electroweak corrections, in particular QED, to hadron
collider processes is essential. From the technological point of view,
this goal requires the development of computational tools which
include such corrections in the hard cross-section
calculations~\cite{CarloniCalame:2007cd,Alwall:2014hca,Li:2012wna} and
on the other hand, it also requires a precise determination of sets of
PDFs with QED corrections.

In these last three years I have been working on topics which cover an
extensive and detailed set of arguments from collider phenomenology
with emphasis on PDFs. In this thesis I focus the discussion on the
inclusion of QED corrections to PDFs, taking the opportunity to
present several studies performed during my PhD. 

We start from the theoretical aspects of this implementation, such as
the upgrade of parton evolution equations, the inclusion of
photon-related contributions in the computation of deep inelastic
scattering (DIS) and Drell-Yan processes. This discussion is then
followed by a technical description of the framework required for the
determination of a set of PDFs with QED corrections.

Here we include QED corrections up to leading order (LO) in
$\mathcal{O}(\alpha)$, to next-to-leading (NLO,
\textit{i.e.}~$\mathcal{O}(\alpha_{s}^{2})$) and
next-to-next-to-leading (NNLO) order QCD computations. This choice can
be motivated by the na\"{i}ve comparison of the similar magnitude of
the coupling constants $\alpha_{s}^{2}(M_{Z}^{2})$ and
$\alpha(M_{Z}^{2})$, which suggests that LO QED corrections and NLO
QCD corrections are of a similar size, \textit{e.g.}
\begin{equation}
  \frac{\mathcal{O}(\alpha_{s}^2)}{\mathcal{O}(\alpha)} \rightarrow
  \frac{\alpha_{s}^{2}(M_{Z}^{2})}{\alpha(M_{Z}^{2})} = \frac{0.118^2}{1/127} \sim 1.78.
\end{equation}
and thus non-negligible QED effects are expected when computing
predictions beyond the NLO QCD. This observation also suggests that
measurements from the LHC contain useful information for an accurate
determination of sets of PDFs with electroweak corrections.

The inclusion of QED corrections to PDFs assumes the presence of the
photon particle as an additional parton of the nucleon which interacts
with other partons. This assumption is translated by the definition of
a photon parton distribution function which could be obtained from a
fit to experimental data.

In this work we determine a set of PDFs with QED corrections which
includes a photon PDF and its uncertainties extracted from DIS and LHC
hadronic data using the Monte Carlo approach adopted by the Neural
Network PDF (NNPDF) methodology. In fact, a precise determination of
the photon PDF is needed for reliable computations of high mass
searches, $W$ mass determination, $WW$ production and for several new
physics signals, such as the cross-section for $Z'$ and $W'$
production. It is important to highlight that the methodology for PDF
determination is a complex topic subjected to studies and discussions,
so in this thesis we present some of these aspects, such as the most
remarkable methodological choices adopted by the most active groups of
PDFs. The reader is invited to check the results presented in the
works listed in (\textit{vii}) for a complete overview of the
technical aspects of such issues.

Finally, the phenomenological implications of this set of QED
corrected PDFs and the impact of the photon PDF are presented at the
end of the discussion. We conclude with a short summary of potential
improvements which are required from the point of view of new data
measurements and theory developments which are expected to improve the
accuracy of this set of PDFs.

\begin{figure}
  \renewcommand\thefigure{A} 
  \begin{center}
  \includegraphics[scale=0.35]{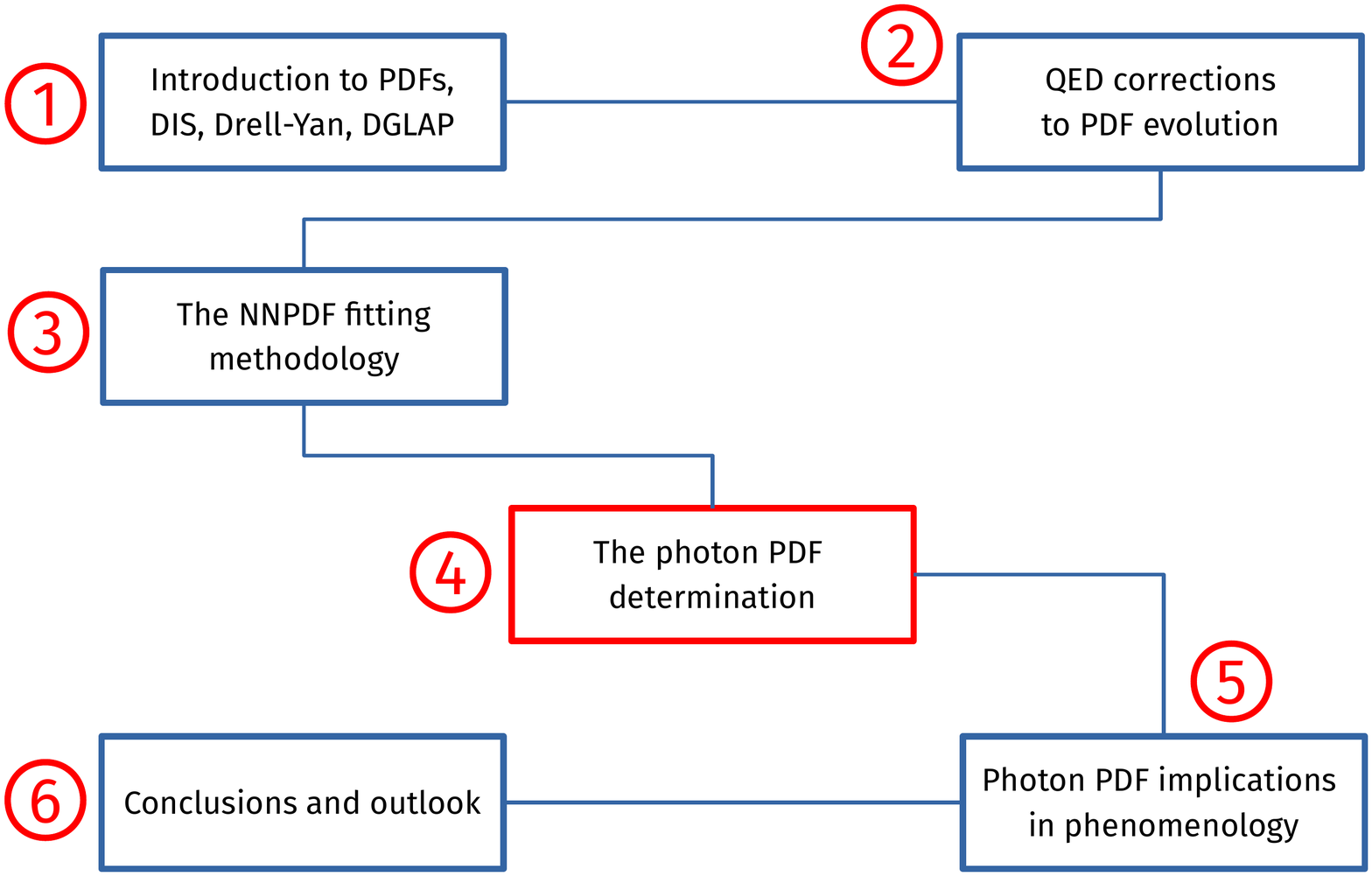}
  \end{center}
  \caption{\label{fig:thesis_scheme} Schematic summary of the topics
    discussed in this thesis.}
\end{figure}

We organize the discussion of this thesis following the scheme
presented in Figure~\ref{fig:thesis_scheme}: 
\begin{description}
\item {\textbf{Chapter~\ref{sec:chap1}: Parton distribution
      functions.}}  We review the theoretical formalism of the parton
  model, providing a brief description of DIS and Drell-Yan processes,
  together with the DGLAP formulation. These concepts are then used in
  the determination of a set of PDFs with QED corrections. The
  formalism is then followed by an overview of the general features of
  modern PDF determination, following the layout of the benchmarking
  exercise with LHC data performed in Ref.~\cite{Ball:2012wy}. 

\item {\textbf{Chapter~\ref{sec:chap2}: QED corrections to PDF
      evolution.}}  The combined QCD$\otimes$QED DGLAP evolution
  equations are presented together with the numerical implementation
  in \texttt{APFEL}, a PDF evolution
  library~\cite{Bertone:2013vaa}. We describe the features and the
  upgrades that \texttt{APFEL} received since its initial publication,
  such as the \emph{combined} and \emph{unified} evolution
  solutions. We validate the results by comparing the \texttt{APFEL}
  evolution with other public codes. Finally, we present \texttt{APFEL
    Web}~\cite{Carrazza:2014gfa}, a web-based application for the
  graphical visualization of parton distribution functions that
  regroups in a centralized system tools for the manipulation of PDFs.

\item {\textbf{Chapter~\ref{sec:chap3}: The NNPDF methodology.}}  We
  review the methodology used for the determination of PDFs from a
  global fit to experimental data. The discussion starts from the
  presentation of the NNPDF
  methodology~\cite{Ball:2012cx,Ball:2014uwa} which is complemented by
  a description of the new code structure in \texttt{C++}. This
  code was developed in order to improve the performance and simplify
  the determination of modern sets of PDFs with LHC data. We conclude
  this chapter with the description of the NNPDF2.3 set of
  PDFs, which was employed as the baseline technology for the
  determination of the set of PDFs with QED corrections.

\item {\textbf{Chapter~\ref{sec:chap4}: The photon PDF
      determination.}}  We present the details of the first
  determination of a set of PDFs with QED corrections and the
  respective photon PDF based on the NNPDF methodology: the so-called
  NNPDF2.3QED set. The results presented in this chapter are
  partially based on the published
  Refs.~\cite{Ball:2013lla,Carrazza:2013wua,Carrazza:2013bra}. The
  photon PDF is first extracted from a fit to DIS data and then,
  consequently, reweighted by LHC $\gamma^*/Z$ high mass and low mass
  measurements and $W,Z$ rapidity distributions. We show PDF
  comparison plots for the photon PDF and we measure the impact of QED
  corrections to sets of PDFs without QED corrections.

\item {\textbf{Chapter~\ref{sec:chap5}: Phenomenological implications
      of the photon PDF.}} We investigate the impact of the
  NNPDF2.3QED set of PDFs, with emphasis on the photon PDF, looking at
  several observables, such as direct photon production at HERA,
  searches for new massive electroweak gauge boson, $W$ pair
  production at the LHC presented and high and low mass Drell-Yan in
  Ref.~\cite{Ball:2013lla,Carrazza:2014cpa}. We conclude the
  discussion with a preliminary guess for the lepton PDFs obtained
  through the \texttt{APFEL} evolution and fully documented in
  Ref.~\cite{Bertone:2015lqa}.

\item {\textbf{Chapter~\ref{sec:conclusions}. Conclusions and
      outlook.}}  We conclude with a summary of the most relevant
  results presented in this thesis. Furthermore, we provide an outlook
  about future technical developments, in terms of experimental data
  and theory developments required to constraint and reduce the
  uncertainties of the photon PDF and improve the accuracy of sets of
  PDFs with QED corrections.

\end{description}

\chapter{Parton distribution functions}
\label{sec:chap1}

In the first part of this chapter we review the basic concepts of
deep-inelastic scattering (DIS) process and the definition of parton
distribution functions. Then we present the Drell-Yan process in
hadron collisions~\cite{Ellis:1991qj,opac-b1131978} and the DGLAP
evolution equations which are essential in PDF determination and
particularly important when including QED corrections to PDFs.

On the second half of this chapter we discuss about the general
features of modern parton distributions, presenting the current state
of the art in PDF determination through the results of the
benchmarking exercise of Ref.~\cite{Ball:2012wy} performed between the
most active PDF groups.

\section{Deep-inelastic scattering}

Deep-inelastic scattering is a fundamental process which have been
used for testing the validity of perturbative QCD. This process played
an important role in the historical development of the theory but it
still has a relevant role in PDF determination from experimental data,
for example measurements at HERA (H1 and
ZEUS~\cite{h1fl,ZEUS:2012bx}), SLAC~\cite{Whitlow:1991uw} and
BCDMS~\cite{bcdms2}.

\subsection{DIS kinematics and the parton model}

\begin{figure}
  \begin{centering}
    \includegraphics[scale=0.5]{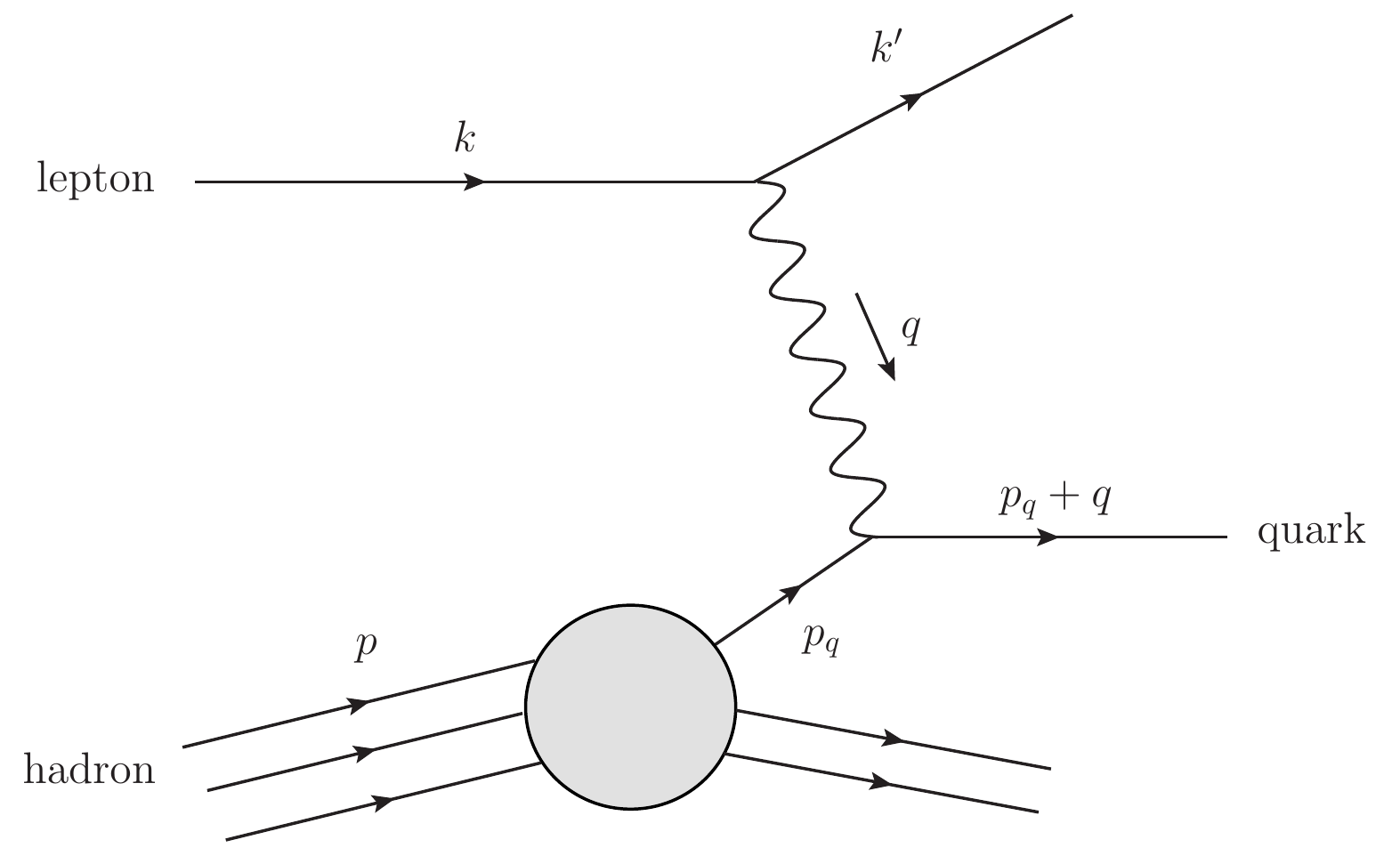}
    \par\end{centering}

  \caption{\label{fig:Example-of-DIS} Example of deep-inelastic
    scattering in QCD.}
\end{figure}

We consider the scattering of a charged lepton $l(k^\mu)$, with
four-momenta $k^\mu$, off a hadron target $h(p^\mu)$, such as
\begin{equation}
  l(k^\mu) + h(p^\mu) \rightarrow l'(k'^{\mu}) + X,
\end{equation}
where $l'(k')$ is the scattered lepton and $X$ is the hadronic final
state, see Figure~\ref{fig:Example-of-DIS} for a graphical
representation of this process. We define the space-like lepton
momentum transfer $q=k-k'$ in terms of differences between the
incoming and outgoing leptons four-momenta. Then, the standard
variables used in DIS are
\begin{eqnarray}
\begin{aligned}
  Q^{2}=-q^2,\\
  p^2 = M^2,\\
  W^{2}=\left(p+q\right)^{2},\\
  s = \left(p+k\right)^{2},
\end{aligned}
\end{eqnarray}
where $Q^2$ is the virtuality of electroweak vector boson exchange,
$M$ the hadron mass, $W$ the invariant mass of the hadronic final
state and $s$ the square of the lepton-hadron center of mass
energy. 

If we consider the picture of a hadron composed by pointlike massless
partons, then it is natural to introduce the \emph{longitudinal
  momentum fraction} $\xi$, where $0 \leq \xi \leq 1$, of the hadron's
total momentum $p$. This suggests that for a given hadron composed by
$n_f$ partons, there exists a probability distribution function
$f_i(\xi)$ which translates the probability that the hadron contains a
parton $i$ carrying a longitudinal fraction $\xi$. This concept is the
basis of the parton model which was proposed by Feynman in
1969~\cite{Feynman:1989dd} even before the formulation of QCD.

In this framework, we note $p_q=\xi p$ the momentum of the scattered
parton. Satisfying the mass-shell constraint for the outgoing parton,
we define the so-called Bjorken variable $x$ as the momentum fraction
$\xi$ of a parton inside the hadron
\begin{equation}
  m_q^2 = (p_q+q)^2 \simeq 2 \xi p \cdot q - Q^2 = \frac{Q^2}{x}\xi -
  Q^2, \quad \xi = \left( 1 + \frac{m_q^2}{Q^2}\right)x \simeq x.
\end{equation}

This picture, also noted as the Bjorken limit, defined when $Q^2,p
\cdot q \rightarrow \infty$ with $x$ fixed, probes the structure of
the incoming hadron at short distances. The Bjorken variable $x$ and
the energy fraction transferred by the scattered lepton are defined as
\begin{equation}
  x=\frac{Q^{2}}{2p \cdot q},\quad y=\frac{p \cdot q}{p \cdot
    k}=\frac{Q^{2}}{x \cdot s},
\end{equation}
where $0\leq x \leq1$ and if $x=1$ the scattering is totally elastic.

As we have already mentioned, the determination of parton distribution
functions from DIS data is possible through a fitting procedure. Since
we have introduced the parton model concept for PDFs, if we consider
the proton as the target hadron, there are the so-called sum rules
which implies some constraints to PDF fits. The proton consists of
three valence quarks $uud$, this yields to the following rules
\begin{eqnarray}
  \int_0^1 dx \, [ f_u(x)-f_{\bar{u}}(x) ] = 2, \quad \int_0^1 dx \, [
  f_d(x)-f_{\bar{d}}(x) ] = 1, \label{eq:valencesum} \\
  \textrm{MSR} = \int_0^1 dx \, x\, \left( \sum_{i=q} \left[f_i(x)+f_{\bar{i}}(x)\right] + f_g(x) \right) = 1.\label{eq:msrint}
\end{eqnarray}
Equations~\ref{eq:valencesum} and~\ref{eq:msrint} are respectively known
as the valence and momentum sum rules. It is important to highlight
that thanks to the isospin symmetry in QCD, the proton PDFs are also
expressed in terms of neutron PDFs: $f^n_u = f_d$, $f^n_d = f_u$ and
$f^n_{\bar{u}} = f_{\bar{d}}$. The validity of these equalities is
limited to the framework of pure QCD processes. Indeed, when
considering QED corrections to QCD the isospin symmetry breaking
introduces the electric charge of quarks and hence such simplification
is no more possible.

\subsection{DIS in perturbative QCD}

Given the basic idea behind the na\"{i}ve parton model, it is possible to
formulate the DIS process through the quantum field theory
formalism. For simplicity let us consider the neutral current
electron-proton scattering process with a virtual photon $\gamma^*$
exchange. At the first order in perturbation theory the matrix element
of this process is
\begin{equation}
  T = \frac{ie^2}{q^2}[\bar{u}(k')\gamma^\mu u(k)] \langle X | j_\mu(0) |
  p \rangle.
\end{equation}
where $j_\mu$ is the electromagnetic current. From the last expression
we observe that the amplitude squared is factored into the leptonic
and hadronic tensors
\begin{equation}
  |T|^2 \propto L_{\mu\nu} W^{\mu\nu}.
\end{equation}
The leptonic tensor expression is trivially extracted from a simple
QED computation, meanwhile the hadronic tensor cannot be completely
determined, \textit{i.e.}
\begin{eqnarray}
  L_{\mu\nu} = e^2 \textrm{Tr}[\cancel{k}' \gamma_\mu \cancel{k}
  \gamma_\nu] = 4 e^2 (k_\mu k'_\nu + k_\nu k'_\mu - g_{\mu\nu} k \cdot
  k'),\\
  W_{\mu\nu}= \frac{1}{4\pi} \int d^4x\,e^{iq \cdot x}
  \langle p | \left[j_\mu(x),j_\nu(0)\right] | p \rangle.
\end{eqnarray}

However, requiring the current conservation, $q \cdot W = 0$, one may
parametrize the hadronic tensor in terms of two real scalar structure
functions $F_1$ and $F_2$
\begin{equation}
  W^{\mu\nu}=\left(g^{\mu\nu}-\frac{q^{\mu}q^{\nu}}{q^{2}}\right)F_{1}(x,\, Q^{2})+\left(p^{\mu}+\frac{q^{\mu}}{2x}\right)\left(p^{\nu}+\frac{q^{\nu}}{2x}\right)\frac{1}{p\cdot q}F_{2}(x,\, Q^{2}).
\end{equation}

Both functions, $F_1(x,Q^2)$ and $F_2(x,Q^2)$, parametrizes the
structure of the target hadron in terms of $x,Q^2$ which are
correlated to $p,q$. If we compute explicitly the leptonic tensor at
leading order (LO), negleting the proton mass, one obtains
\begin{equation}
  \frac{d^2\sigma}{dxdQ^2} = \frac{4\pi\alpha^2}{Q^4} \left[
    [1+(1-y)^2] F_1(x,Q^2) + \frac{(1-y)}{x}F_L(x,Q^2)\right]
  \label{eq:sigmadis}
\end{equation}
where $F_L=F_2 - 2xF_1$ is the \emph{longitudinal structure
  function}. Furthermore, $F_L=0 \Leftrightarrow F_2=2xF_1$ is the
Callan-Gross relation, a consequence of quarks having spin 1/2. In
this equation we clearly see that the dynamics of strong interactions
are represented by the structure functions of the incoming hadron.

When considering the parton model with no QCD corrections the
structure functions are simply given by
\begin{equation}
  F_2(x,Q^2) = 2xF_1(x,Q^2) = \sum_{q,\bar{q}} \int_0^1 \frac{dy}{y} f_q(y)
  x e_q^2 \delta\left(1-\frac{x}{y}\right) = \sum_{q,\bar{q}} e_q^2 x
  f_q(x).
  \label{eq:f2lo}
\end{equation}

This result anticipates the factorization theorem~\cite{Ellis:1991qj}
which generalizes Eq.~(\ref{eq:f2lo}) to all order in QCD. The
factorization theorem states that any structure function $F$ is
factorized by weighting the parton structure functions with PDFs
\begin{eqnarray}
\begin{split}
    F(x,Q^2) & = \sum_{i=q,\bar{q},g} \int_{0}^{1} dy \int_{0}^{1} dz \,
    C_i(z,Q^2) \, f_i(y) \, \delta(x-yz) \\
    & = \sum_{i=q,\bar{q},g} \int_{x}^{1} \frac{dy}{y} \,
    C_i\left(\frac{x}{y}, Q^2\right) \, f_i(y) \\
    & = \sum_{i=q,\bar{q},g} C_i(x,Q^2) \otimes f_i(x),
    \label{eq:strucfunc}
\end{split}
\end{eqnarray}
where the $C_i(x,Q^2)$ are the so-called coefficient functions or
Wilson coefficients. We have also introduced the Mellin convolution
product $\otimes$ which is defined as
\begin{equation}
  \label{eq:convolution}
  f(x) \otimes g(x) \equiv \int_0^1 dy \int_0^1 dz
  f(y)g(z)\delta(x-yz) = \int_x^1 \frac{dy}{y} f\left( \frac{x}{y}
  \right) g(y).
\end{equation}

In Eq.~(\ref{eq:strucfunc}) the Wilson coefficients carries the
information from high-energy contributions and so their exact
formulation is process depend and it is calculable in perturbation
theory. On the other hand the functions $f_i$, the PDFs, enclose the
low-energy contributions and thus are non-perturbative and universal
quantities which characterizes the intrinsic components of the hadron.

The calculation of the coefficient functions beyond the LO shows an
\emph{ultraviolet} (UV) and \emph{infrared} (IR) divergences. The
complete calculation of such divergences which are fully documented in
Refs.~\cite{Ellis:1991qj,opac-b1131978} is beyond the scope of this
short review, however we summarize in the next paragraphs the most
important results.

The UV divergences arising from the loop contribution are typically
treated using the dimensional regularization and renormalization
techniques. Concerning the IR divergences, we observe the cancellation
of soft and final state collinear singularities thanks to the
completely inclusive final state, which is IR safe.

\begin{figure}
  \begin{centering}
    \includegraphics[scale=0.5]{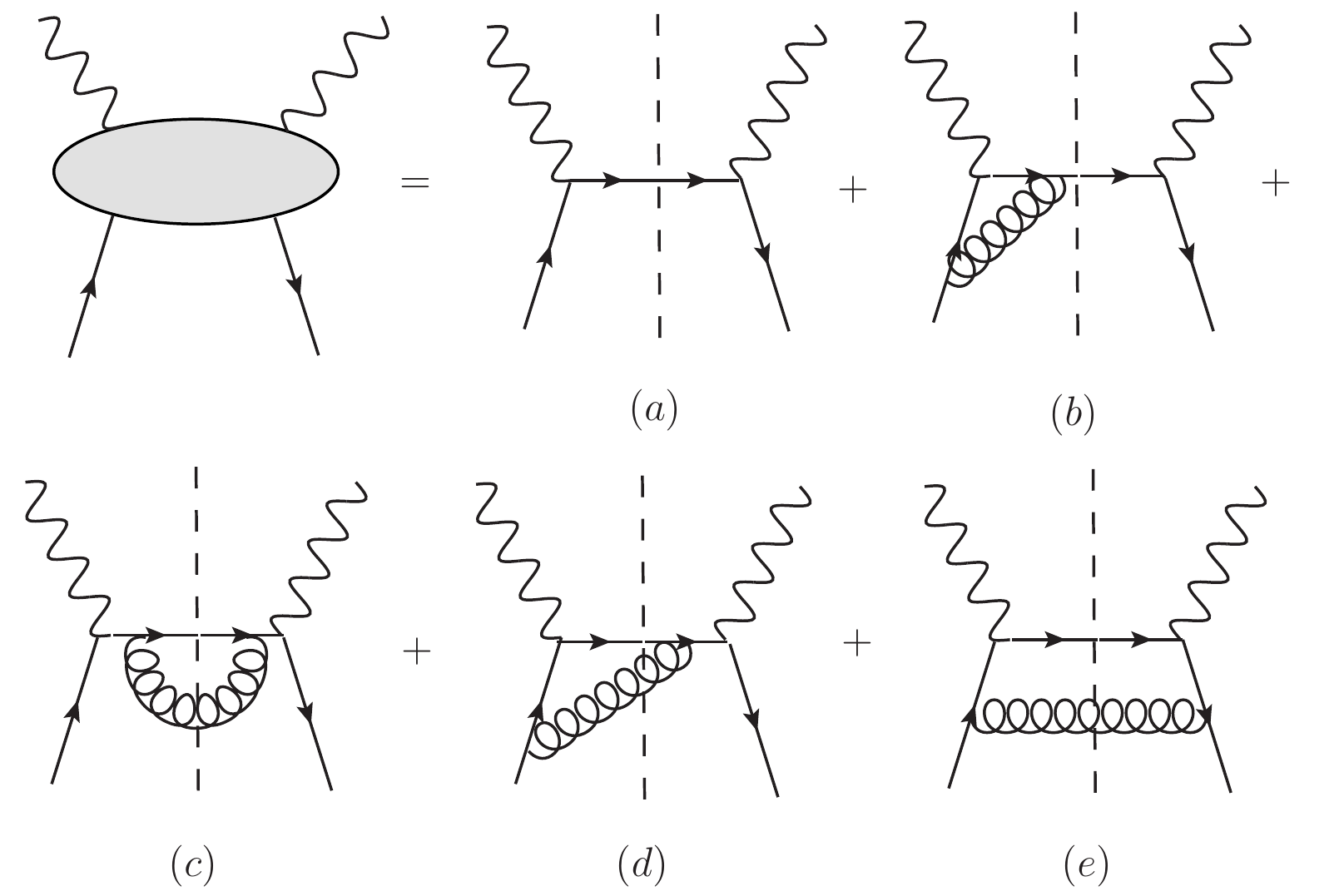}
    \par\end{centering}
  
  \caption{\label{fig:QCDcorr} Diagrams contributing to the
    $\mathcal{O}(\alpha_s)$ QCD corrections with initial state quarks
    and anti-quarks.}
\end{figure}

In order to provide an example of the removal of the uncancelled
initial state collinear singularities, lets consider the $\gamma^*q
\rightarrow gq $ processes. In Figure~\ref{fig:QCDcorr} we present the
diagrams which contribute to the lowest-order corrections to the
partonic cross-section $\mathcal{O}(\alpha_s)$. The usual procedure
consists in introducing the infrared cutoff $\mu^2$ which can be
chosen arbitrarily small and a bare distribution $f_q^{(0)}$ of a
quark in a proton, thus we define $f_q(x, \mu^2_F)$ as the
renormalized and measurable distribution
\begin{equation}
  f_q(x, \mu^2_F) = f_q^{(0)}(x) + \frac{\alpha_s}{2\pi} \int_x^1
  \frac{d\xi}{\xi} f_q^{(0)}(\xi) \left[ P_{qq}\left( \frac{x}{\xi}\right)
    \ln \frac{Q^2}{\mu^2} + \kappa\left( \frac{x}{\xi}\right) \right]
  + \cdots,
  \label{eq:pdfrenorm}
\end{equation}
where $\mu_F^{2} \geq \mu^2$ is the mass factorization scale at which
the quark distribution is measured, $\kappa(x)$ is a calculable
function and $P_{qq}(x)$ is known as the $q \rightarrow q$ splitting
function.

Using the expression of Eq.~(\ref{eq:pdfrenorm}) into the QCD corrected
structure function we obtain
\begin{equation}
  F_2(x,Q^2) = x \sum_{q,\bar{q}} e_q^2 \int_x^1 \frac{d\xi}{\xi}
  f_q(\xi, \mu_F^2) \left[ \delta \left( 1 - \frac{x}{\xi}\right) +
    \frac{\alpha_s}{2\pi} P_{qq} \left( \frac{x}{\xi} \right) \ln
    \frac{Q^2}{\mu_F^2} + \cdots \right],
\end{equation}
which is independent of the infrared cutoff $\mu^2$ and when setting
$\mu_F^2 = Q^2$ as usual in DIS computations, the $f_q(x,Q^2)$ can be
determined from structure function data at any scale.

This discussion is easily generalized also to the initial state
gluons, and to other renormalization schemes, for example the
$\overline{\textrm{MS}}$ scheme.

\section{Hard processes in hadron collisions}
\label{sec:dy}

Another important result of the factorization theorem is the study
of processes and observables at hadron colliders such as the LHC. The
high energy collision of hadrons induce soft interactions of the
constituent partons, and therefore such interactions cannot be treated
with perturbative QCD, but as in DIS the lowest-order QCD prediction
should accurately describe the process.

The parton model cross-section for hadron-hadron colliders is defined as
\begin{equation}
  \sigma_{AB} = \sum_{i,j\in\{q,\bar{q},g\}} \int dx_1 dx_2 f_i(x_1) f_j(x_2)
  \hat{\sigma}_{ij \rightarrow X},
  \label{eq:hadroncoll}
\end{equation}
where two partons enter into a hard collision from which a final state
$X$ emerges. In this equation, the subprocess cross-section
$\hat{\sigma}$ is weighted by the PDFs extracted respectively from the
beam $A$ and target $B$. The formal domain of validity of this
definition is the asymptotic scaling limit $M^2,\hat{s} \rightarrow
\infty$, with $\tau = M^2/\hat{s}$ fixed, which is the analogous of
the Bjorken limit in DIS.

\begin{figure}
  \begin{centering}
    \includegraphics[scale=0.6]{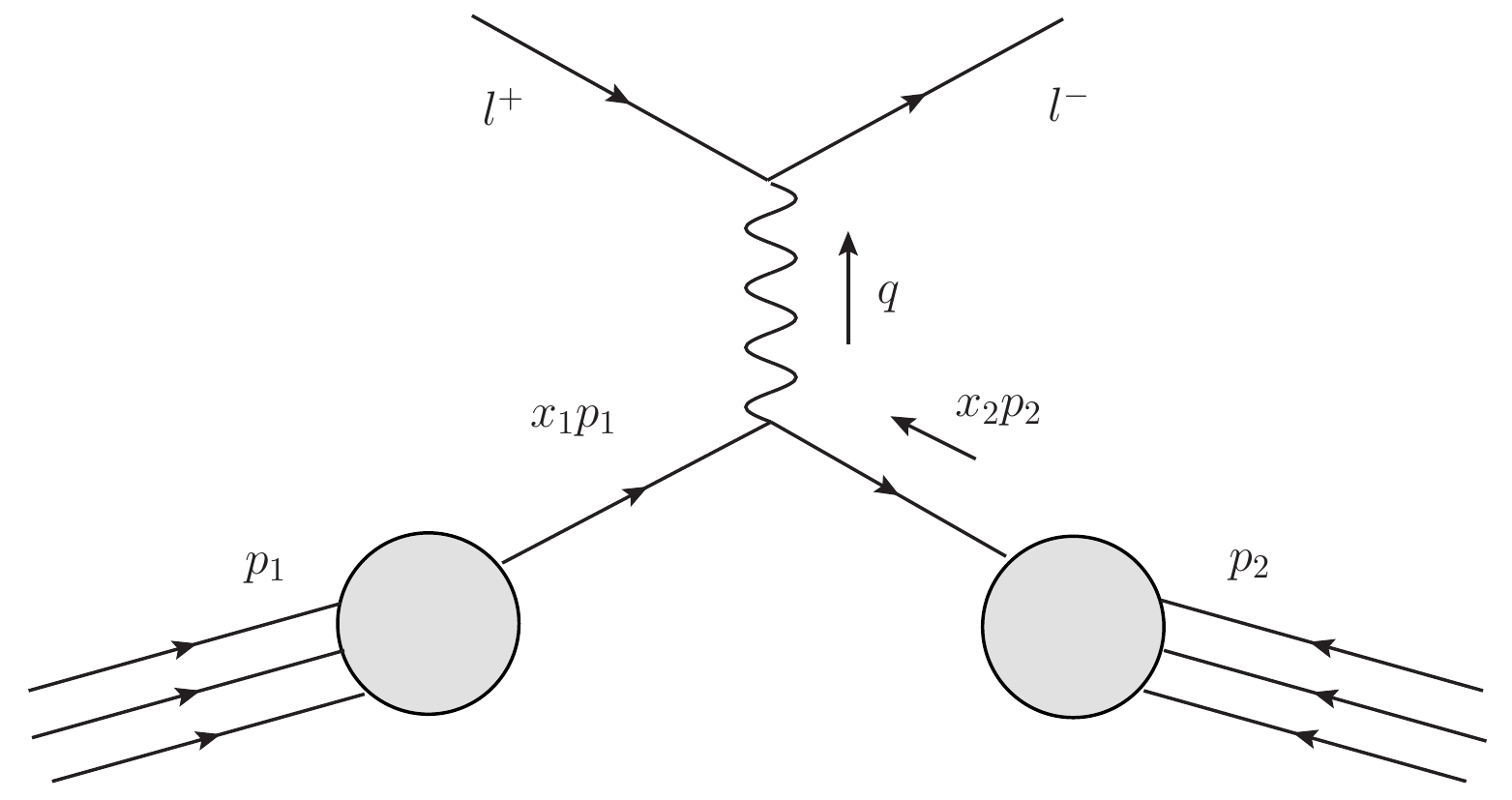}
    \par\end{centering}
  
  \caption{\label{fig:dy} A pictorial representation of the Drell-Yan
    process.}
\end{figure}

One of the most relevant process in hadron-hadron collision is the
production of lepton pair $l^+l^-$ with large invariant mass-squared,
$M^2 = (p_{l^+}+p_{l^-})^2 \gg 1\,\textrm{GeV}^2$, through
quark-antiquark annihilation, the so-called Drell-Yan (DY) process
represented in Figure~\ref{fig:dy}. Such process is extremely
important to describe $Z/\gamma^*$ and $W$ production in high-energy
collisions. It is possible to proof that the inclusion of QCD
corrections to this process generates the same IR behavior observed in
DIS, where PDFs have been defined as renormalized scale dependent
objects as in Eq.~(\ref{eq:pdfrenorm}). Thus, this is also the case
for hard scattering process in hadron collisions. In this particular
setup then Eq.~(\ref{eq:hadroncoll}) becomes
\begin{equation}
  \sigma_{\textrm{DY}} = \sum_{q} \int dx_1 dx_2 f_q(x_1,
  M^2) f_{\bar{q}}(x_2, M^2)
  \hat{\sigma}_{q\bar{q} \rightarrow l^+ l^-}.
\end{equation}
where the PDFs are called at the $M^2$ scale. In this framework we
identify the square of the invariance mass as
\begin{equation}
  M^2 = x_1 x_2 \hat{s},
\end{equation}
where the variables $x_1$ and $x_2$ are defined as
\begin{equation}
  x_1 = \frac{M}{\sqrt{\hat{s}}} e^y, \quad x_2 =
  \frac{M}{\sqrt{\hat{s}}} e^{-y},
\end{equation}
where $y$ is the rapidity of the virtual photon.

Important measurements from the LHC have been performed during the
last years which are relevant in PDF determination, \textit{e.g.}~the
ATLAS measurements of the $Z/\gamma^*$ high-mass~\cite{Aad:2013iua}
and $W,Z$ rapidity distributions ~\cite{Aad:2011dm} and LHCb low-mass
measurements~\cite{LHCb-CONF-2012-013}. In Chapter~\ref{sec:chap4}
data from these experiments are used in order to provide a reliable
constraint on the photon PDF uncertainties.

\section{DGLAP evolution equations}
\label{sec:dglapintro}

The definition of the renormalized PDFs presented in the previous
sections shows the need of evolution equations which describes the
variation of $f_q(x,\mu_F^2)$ with $\mu_F^2$. By differentiating
Eq.~(\ref{eq:pdfrenorm}) with respect to $\ln \mu_F^2$ we obtain the
renormalization group equation for the quark distribution:
\begin{equation}
  \mu_F^2 \frac{\partial}{\partial \mu_F^2} f_q(x,\mu_F^2) =
  \frac{\alpha_s(\mu_F^2)}{2\pi} \int_x^1 \frac{d\xi}{\xi} P_{qq}\left(
    \frac{x}{\xi}, \alpha_s(\mu_F^2) \right) f_q(\xi, \mu_F^2).
  \label{eq:dglapbasis}
\end{equation}

This is the so-called Dokshitzer-Gribov-Lipatov-Altarelli-Parisi
(DGLAP) equation. With DGLAP evolution equations we compute PDFs
distributions at any given value of $\mu_F^2$, by solving the system
of integro-differential equation which requires just the initial
condition of the PDFs.

The most important ingredients of DGLAP equations are the splitting
functions. The splitting functions depend on the type of the parton
splitting, and they have a perturbative expansion in the running
coupling $\alpha_s(\mu_F^2)$. Currently, in the QCD framework
splitting functions have been computed up to
$\mathcal{O}(\alpha_s^3)$~\cite{Vogt:2004mw,Moch:2004pa}. In
Chapter~\ref{sec:chap2} we discuss in detail the solution of this
system of equation in the framework of combined QCD$\otimes$QED
evolution, meanwhile in the next lines we present some basic concepts
about the solution of the DGLAP equations.

At leading-order the splitting functions contributions are
\begin{eqnarray}
  P^{(0)}_{qq}(x) & = & C_F \left[ \frac{1+x^2}{(1-x)_+} + \frac{3}{2}
    \delta(1-x) \right],\\
  P^{(0)}_{qg}(x) & = & T_R \left[ x^2 + (1-x)^2 \right],\,T_R=\frac{1}{2},\\
  P^{(0)}_{gq}(x) & = & C_F \left[ \frac{1+(1-x)^2}{x} \right],\\
  P^{(0)}_{gg}(x) & = & 2C_A \left[ \frac{x}{(1-x)_+} + \frac{1-x}{x} +
    x(1-x) \right] \\
  &  &  + \left( \frac{11}{6} C_A - \frac{4}{6} n_f T_R \right) \delta(1-x),
\end{eqnarray}
where $C_F=4/3$ and $C_A=3$ are the QCD color factors, and the
\emph{plus} refers to the prescription
\begin{equation}
  \int_0^1 dx \frac{f(x)}{(1-x)_+} = \int_0^1 dx \frac{f(x)-f(1)}{1-x}.
\end{equation}

In order to solve the DGLAP evolution equations in an efficient way,
we split the system of equations into two subsystems: the singlet and
non-singlet sectors. Given a system with $n_f=6$ flavors, where
$f_i=u,d,s,c,b,t$, we introduce a new PDF basis, known as evolution
basis, by first defining
\begin{equation}
  f_i^\pm = f_i \pm \bar{f}_i. 
  \label{eq:fpm}
\end{equation}
The non-singlet sector evolves accordingly to
Eq.~(\ref{eq:dglapbasis}), and it is composed by valences and triplets
\begin{equation}
  \textrm{Valences:} \quad V_i \equiv f_i^-, \quad \textrm{Triplets:}
  \quad \begin{cases}
    T_{3}\equiv u^+ - d^+\\
    T_{8}\equiv u^+ + d^+ - 2 s^+\\
    T_{15}\equiv u^+ + d^+ + s^+ - 3 c^+\\
    T_{24}\equiv  u^+ + d^+ + s^+ + c^+ - 4 b^+\\
    T_{35}\equiv u^+ + d^+ + s^+ + c^+ + b^+ - 5 t^+\\
    \end{cases}.
    \label{eq:nonsingletpdf}
\end{equation}
On the other hand, the non singlet sector couples all quarks to the
gluon PDF, so we define the singlet PDF as
\begin{equation}
  \Sigma(x,\mu_F^2) = \sum_i f_i^+ \equiv \sum_i f_i(x,\mu_F^2) +
  f_{\bar{i}}(x,\mu_F^2).
  \label{eq:singletpdf}
\end{equation}

Then, the coupled singlet system reads
\begin{equation}
  \mu_F^2 \frac{\partial}{\partial \mu_F^2} \left(
  \begin{array}{c}
    \Sigma (x, \mu_F^2)\\
    g (x, \mu_F^2)\\
  \end{array}
  \right) = \frac{\alpha_s(\mu_F^2)}{2\pi} \int_x^1
  \frac{d\xi}{\xi} 
\left(
  \begin{array}{cc}
    P_{qq} \left( \frac{x}{\xi},\alpha_s \right) & 2n_fP_{qg}\left( \frac{x}{\xi},\alpha_s \right) \\
    P_{qg} \left( \frac{x}{\xi},\alpha_s \right)& P_{gg}\left( \frac{x}{\xi},\alpha_s \right)
  \end{array}
\right)
\left(
  \begin{array}{c}
    \Sigma (\xi, \mu_F^2)\\
    g (\xi, \mu_F^2)\\
  \end{array}
  \right).
\end{equation}

\begin{figure}
  \begin{centering}
    \includegraphics[scale=0.5]{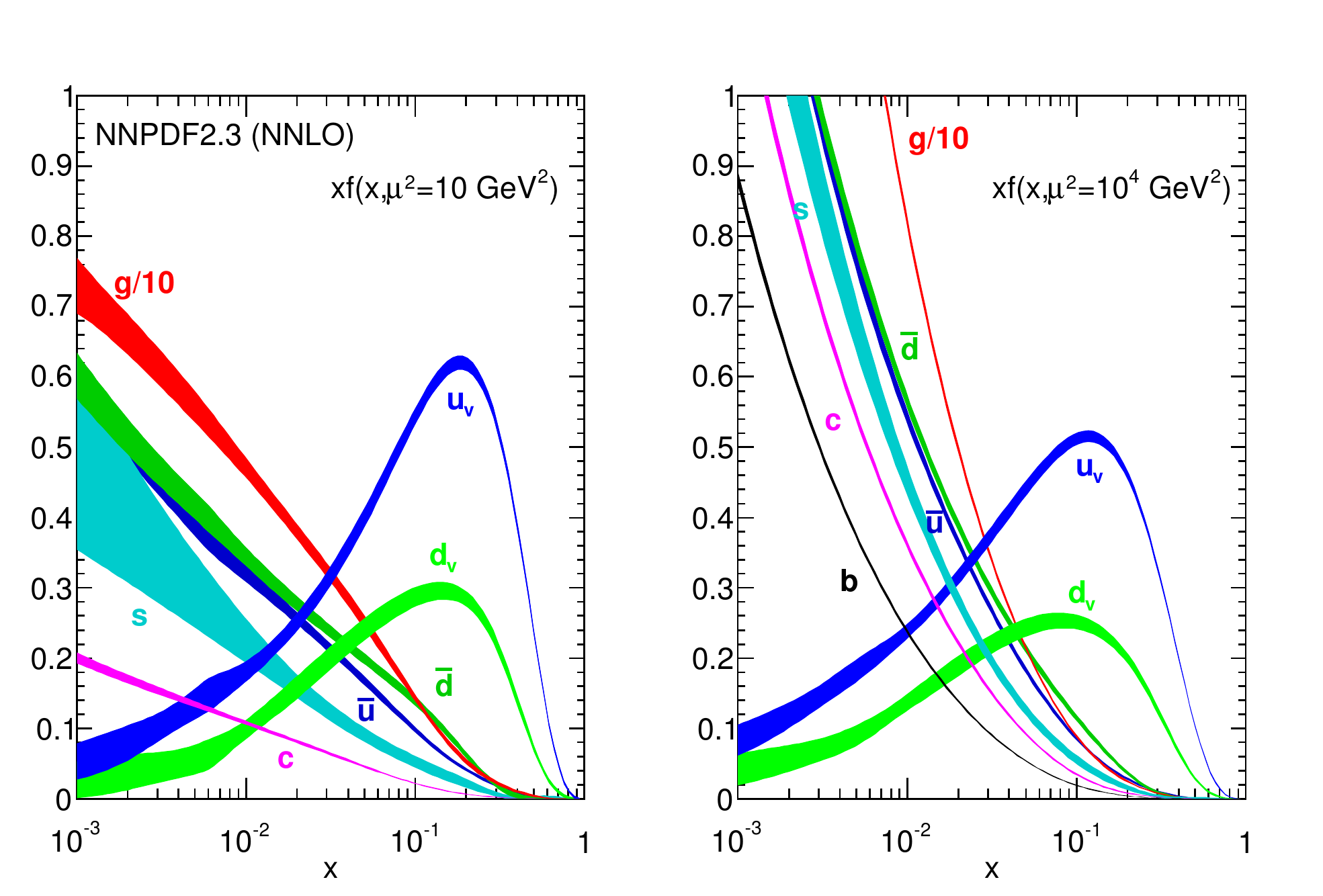}
    \par\end{centering}
  
  \caption{\label{fig:pdgplot} Example of PDFs evolution obtained in
    NNLO NNPDF2.3 global analysis~\cite{Ball:2012cx} at scales
    $\mu_F^2=10\,\textrm{GeV}^2$ (left plot) and
    $\mu_F^2=10^4\,\textrm{GeV}^2$ (right plot), with
    $\alpha_s(M_Z^2)=0.118$. This plot was produced for the PDG 2013
    edition.}
\end{figure}

%
In the last paragraphs we have presented the DGLAP equations in
$x$-space, however by looking at Eq.~(\ref{eq:dglapbasis}) we identify
the Mellin convolution and so we are able to translate the same set of
equations in the Mellin $N$-space where the DGLAP has an analytic
solution. For example, for the non-singlet we obtain
\begin{eqnarray}
  \mu_F^2 \frac{\partial}{\partial \mu_F^2} f_{\textrm{NS}}(N,\mu_F^2)
  & = & \frac{\alpha_s(\mu_F^2)}{2\pi} \gamma_{qq} (N, \alpha_s(\mu_F^2))
  f_{\textrm{NS}} (N,\mu_F^2),
\end{eqnarray}
where we applied the Mellin transform to the PDFs and splitting
functions, defining the so-called anomalous dimension
\begin{equation}
  f_i(N,\mu_F^2) = \int_0^1 dx\,x^{N-1} f_i(x,\mu_F^2), \quad
  \gamma_{ij} (N, \mu_F^2) = \int_0^1 dx\, x^{N-1} P_{ij} (x, \alpha_s(\mu_F^2)).
\end{equation}

The anomalous dimension at LO are given by
\begin{eqnarray}
  \gamma^{(0)}_{qq}(N) & = & C_F \left[ -\frac{1}{2} + \frac{1}{N(N+1)} - 2
    \sum_{k=2}^N \frac{1}{k} \right],\\
  \gamma^{(0)}_{qg}(N) & = & T_R \left[ \frac{2+N+N^2}{N(N+1)(N+2)} \right],\\
  \gamma^{(0)}_{gq}(N) & = & C_F \left[ \frac{2+N+N^2}{N(N^2-1)} \right],\\
  \gamma^{(0)}_{gg}(N) & = & 2C_A \left[
    -\frac{1}{12}+\frac{1}{N^2-N}+\frac{1}{(N+1)(N+2)} - \sum_{k=2}^N
  \frac{1}{k} \right] - \frac{2n_f T_R}{3}.
\end{eqnarray}

In both spaces, the solution of the DGLAP evolution is possible to
derive by solving the respective integro-differential systems of
equations. The $N$-space solution is trivial to obtain through the
simple analytic solution for both sectors when using the basis
presented in Eqs.~\ref{eq:nonsingletpdf} and~\ref{eq:singletpdf}. On
the other hand, the solution in the $x$-space representation is highly
non-trivial, so a numerical approach solution, based \textit{e.g.}~on
the Runge-Kutta method is preferable. Technical details of this
solution will be presented in Chapter~\ref{sec:chap2}.

The approach used here is easily generalized by the Wilson Operator
Product Expansion (OPE) which provides a powerful computational tool
for the determination of the anomalous dimensions, and it provides a
more abstract determination of the DGLAP equation from the the
renormalization group equations~\cite{Georgi:1951sr,Gross:1974cs}.

Finally, in Figure~\ref{fig:pdgplot} we show an example of PDF
evolution in function of $x$, using the physical basis where $f_v =
f^-$. In this case, the evolution is performed from the initial scale
$\mu_0^2= 2$ GeV$^2$ to $\mu_F^2= 10$ GeV$^2$ (left plot) and
$\mu_F^2= 10^4$ GeV$^2$ (right plot). Thanks to the DGLAP evolution,
PDF determination from data with different energy scales is much
simpler because we are able to select a initial scale $\mu_F^2 =
Q_0^2$ where the PDF is parametrized when comparing predictions to the
data evolve the PDF to the experiment energy value.

A final remark concerns the notation, in the next sections and
chapters the factorization scale will be noted in terms of the energy
of the processes: $\mu_F^2=Q^2$.

\section{Characterization of modern PDFs}
\label{sec:generalfeatures}

\begin{figure}
    \begin{center}
      \includegraphics[width=0.6\textwidth]{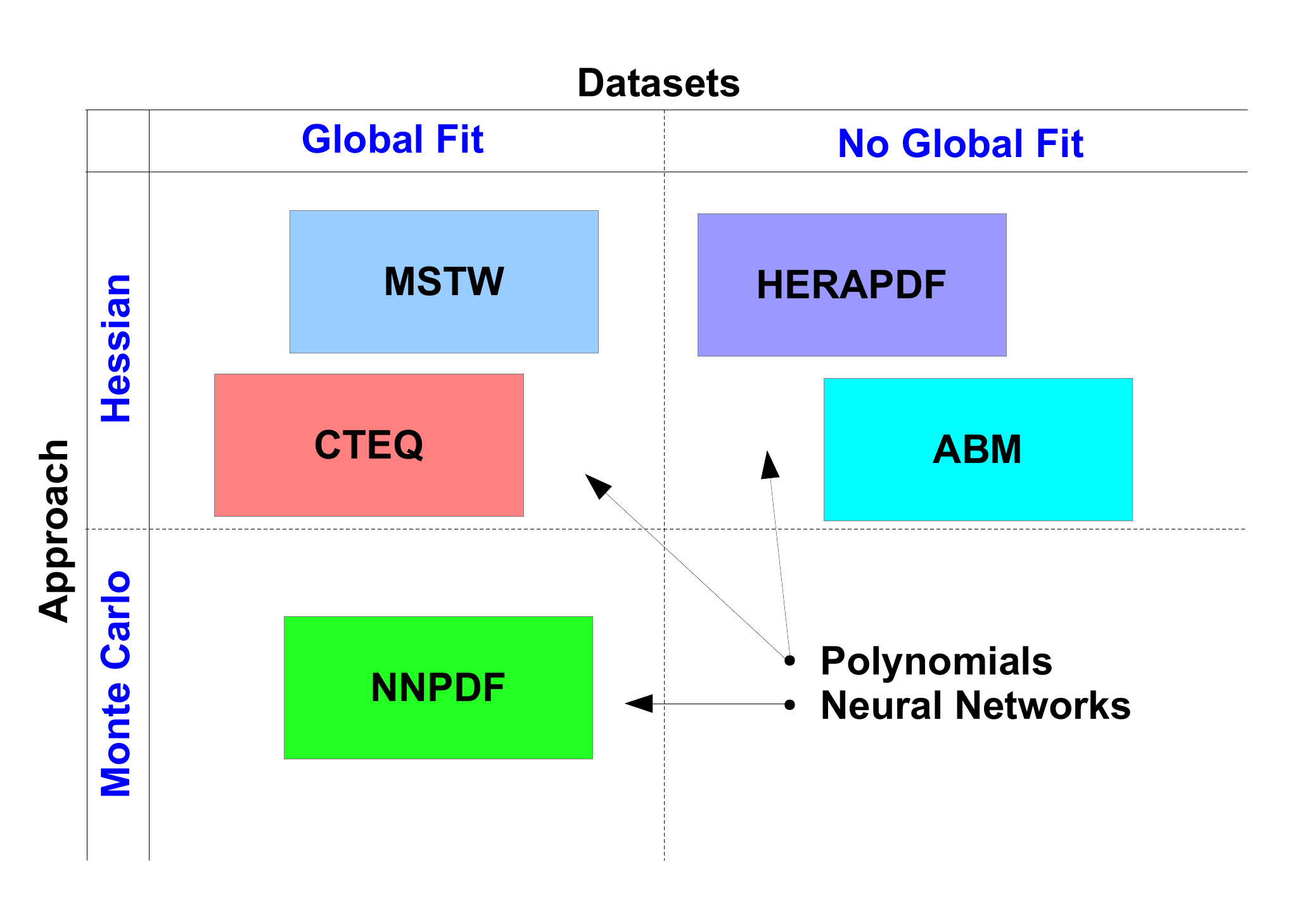}\quad
    \end{center}
    \caption{Pictorial representation of the PDF groups discussed in
      this section.}
      \label{fig:pdfgroups}
\end{figure}

After introducing the origin and definition of PDFs, we conclude this
chapter by showing some general features of modern PDF determination
from experimental data. Nowadays this topic is studied by several
groups and each group provides its own sets of PDFs. The main
differences between these sets are due to the technical choices of
each group, \textit{i.e.}~the experimental data included in the fit,
the theoretical choices for the computation of predictions, the PDF
functional form parametrization and finally, the fitting algorithm. In
the next paragraph we present an overview of the most active groups of
PDFs.
\begin{itemize}

\item The ABM collaboration provides sets of PDFs based on DIS and
  Drell-Yan data at NLO and NNLO. The ABM PDFs are parametrized by 6
  independent PDFs using polynomials (25 free parameters). The
  minimization algorithm is based in the Hessian method, where the PDF
  uncertainties are given by symmetric eigenvectors. This
  collaboration has released ABM11~\cite{Alekhin:2012ig} which uses
  the combined HERA-I data, $\overline{\mbox{MS}}$ running heavy quark
  masses for DIS structure functions~\cite{Alekhin:2010sv}, and
  provides PDF sets for a range of values of $\alpha_s$ in a fixed
  flavor number scheme (FFNS) with $n_f=5$.

\item The CT collaboration extracts PDFs from a global dataset that
  includes DIS, Drell-Yan, $W,Z$ production and jet data using the
  Hessian approach at LO, NLO and NNLO. The PDFs are also parametrized
  by 6 polynomials (26 free parameters) and the uncertainties are
  delivered through eigenvectors. The CT collaboration has released
  the CT10 set of PDFs~\cite{Nadolsky:2012ia,Lai:2010vv} using the
  NNLO implementation of the S-ACOT-$\chi$ variable flavor number
  scheme (VFNS) for heavy quark structure
  functions~\cite{Guzzi:2011ew}.

\item The HERAPDF collaboration provides PDF sets based on HERA-only
  DIS data at NLO and NNLO. The approach is the Hessian one, in
  combination with 5 polynomial independent PDFs (14 free
  parameters). The recent HERAPDF1.5
  set~\cite{Radescu:2010zz,CooperSarkar:2011aa} contains the combined
  HERA-I dataset and the inclusive HERA-II data from
  H1~\cite{Aaron:2012qi} and ZEUS~\cite{ZEUS:2012bx}. This is the only
  set of PDFs where uncertainties are provided in terms of variations
  of fit parameters and experimental uncertainties.

\item The MSTW collaboration releases PDF sets using a global dataset
  at LO, NLO and NNLO. The fit is performed by 7 independent PDFs,
  parametrized by polynomials (20 free parameters). The MSTW PDFs are
  based on the Hessian approach. Here we use the MSTW08
  PDFs~\cite{Martin:2009iq} which was available together with the
  other sets considered in this section, although the
  MMHT2014~\cite{Harland-Lang:2014zoa} set of PDFs has been released
  recently.

\item The NNPDF collaboration determines PDFs at LO, NLO and NNLO from
  a global dataset like CT and MSTW collaborations. The NNPDF approach
  uses the Monte Carlo sampling method for the determination of PDF
  uncertainties. The parametrization consists in 7 PDFs based on
  artificial neural networks (ANN), for a total of 259 free parameters
  trained by a genetic algorithm (GA). A complete description of the
  NNPDF methodology is presented in Chapter~\ref{sec:chap3}. In this
  thesis we focus the discussion on the NNPDF2.3~\cite{Ball:2012cx}
  set even if this set has been recently superseded by the
  NNPDF3.0~\cite{Ball:2014uwa,Carrazza:2015hva,Carrazza:2015aoa,Bonvini:2015ira}. The
  NNPDF2.3 set implements the FONLL VFNS at NNLO~\cite{Forte:2010ta},
  and it also includes relevant LHC data for which the experimental
  correlation matrix is available.

\end{itemize}

\begin{table}
  \centering
  \small
  \begin{tabular}{c||c|c|c|c|c}
    \hline
    PDF set  & Ref. & $\alpha_s^{(0)}$(NLO) & $\alpha_s$ range (NLO)
    & $\alpha_s^{(0)}$(NNLO) & $\alpha_s$ range (NNLO) \\
    \hline
    \hline
    ABM11      & \cite{Alekhin:2012ig}  & 0.1181 & $[0.110,0.130]$ & 0.1134   & $[ 0.104,0.120]$    \\
    CT10       & \cite{Nadolsky:2012ia} & 0.118  & $[0.112,0.127]$  & 0.118   & 
    $[ 0.112,0.127 ]$  \\
    HERAPDF1.5 & \cite{Radescu:2010zz,CooperSarkar:2011aa}  & 0.1176 & $[0.114,0.122]$  & 0.1176   & $[ 0.114,0.122 ]$   \\
    MSTW08     &  \cite{Martin:2009iq} & 0.1202 & $[ 0.110,0.130 ]$  & 0.1171   & $[ 0.107,0.127 ]$   \\
    NNPDF2.3   & \cite{Ball:2012cx} & all  & $[ 0.114,0.124 ]$  & all   &   $[ 0.114,0.124 ]$ \\
    \hline
  \end{tabular}
  \caption{PDF sets described in this
    section. The table contains information about the available
    $\alpha_s$ range at NLO and NNLO for the PDF central value together with
    $\alpha_s^{(0)}$ for which PDF uncertainties are provided. 
    For ABM11 the $\alpha_s$ varying PDF 
    sets are only available for the $n_f=5$ set. NNPDF always provides 
    uncertanties for every $\alpha_s$ in the range. \label{tab:as}}
\end{table}

In Figure~\ref{fig:pdfgroups} we show a pictorial representation of
the PDF groups listed above. In Table~\ref{tab:as} we summarize the
PDF sets that will be compared with the common value of
$\alpha_s(M_Z^2) = 0.118$. We will show results for PDFs, parton
luminosities and physical cross-sections. We do not include in this
comparison the JR09 PDF set~\cite{JimenezDelgado:2009tv} because it is
available only for a single value of $\alpha_s(M_Z^2)$.

All the above groups provide versions of the respective PDF sets both
at NLO and at NNLO, however here we will show only the NNLO
PDFs. Results at NLO and for a wider range of $\alpha_s$ values is
available from an online catalog of plots at \texttt{HepForge}:
\begin{center} \textbf{
  \url{http://nnpdf.hepforge.org/html/pdfbench/catalog}}.
\end{center}

\subsection{Parton distributions and parton luminosities}

In this section we compare the PDFs of the groups presented in
Section~\ref{sec:generalfeatures} and then parton luminosities at NNLO
for $\alpha_s=0.118$. Some of the sets provide PDF errors exclusively
for some default value of $\alpha_s$. For those sets we take the
central replica for the PDFs values at $\alpha_s=0.118$ but we use the
uncertainties of the PDF set at the default value of $\alpha_s$.

\subsubsection*{PDF comparison}

We compare PDFs at $Q^2=$ 25 GeV$^2$, which is above the $b$-quark
threshold knowing that the ABM11 set provides multiple values of
$\alpha_s$ only when $n_f=5$. The comparisons are organized in the
following:
\begin{itemize}

\item For each PDFs flavor and combination we compare two sub-groups
  of sets: NNPDF2.3, CT10 and MSTW08 and then NNPDF2.3, ABM11 and
  HERAPDF1.5. The first sub-group considers sets determined from fits
  to a global dataset, meanwhile in the second group we still use
  NNPDF2.3 as reference which is compared to PDFs obtained from
  reduced datasets.

\item In all plots, PDF uncertainties do not contain the $\alpha_s$
  uncertainty, except for the ABM11 PDFs, where the $\alpha_s$
  uncertainty is treated on a equal footing to the PDF parameters in
  the covariance matrix. The ABM11 and HERAPDF results also include an
  uncertainty on quark masses, while other groups provide sets with a
  variety of masses.
\end{itemize}

\begin{figure}
    \begin{center}
      \includegraphics[width=0.48\textwidth]{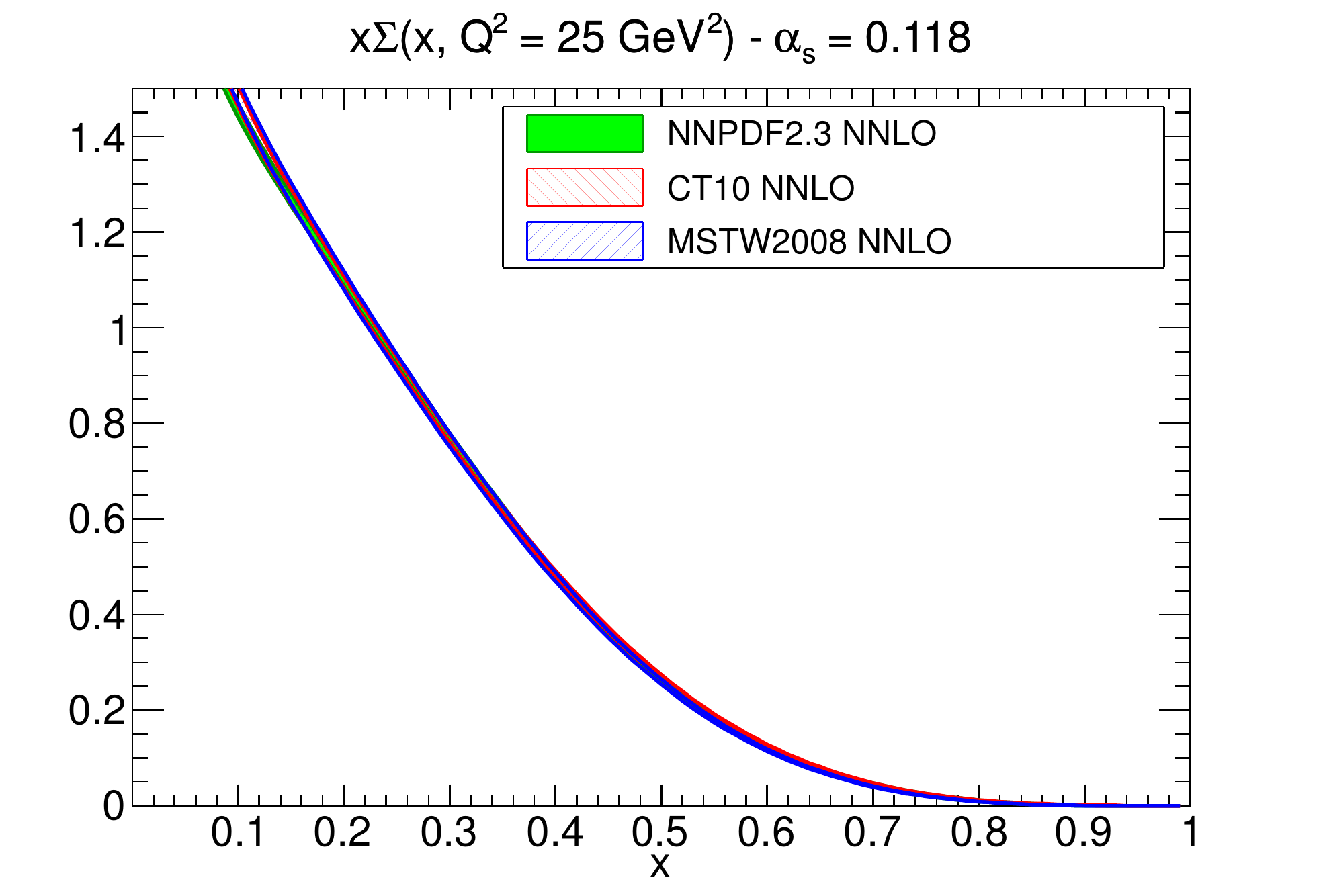}\quad
      \includegraphics[width=0.48\textwidth]{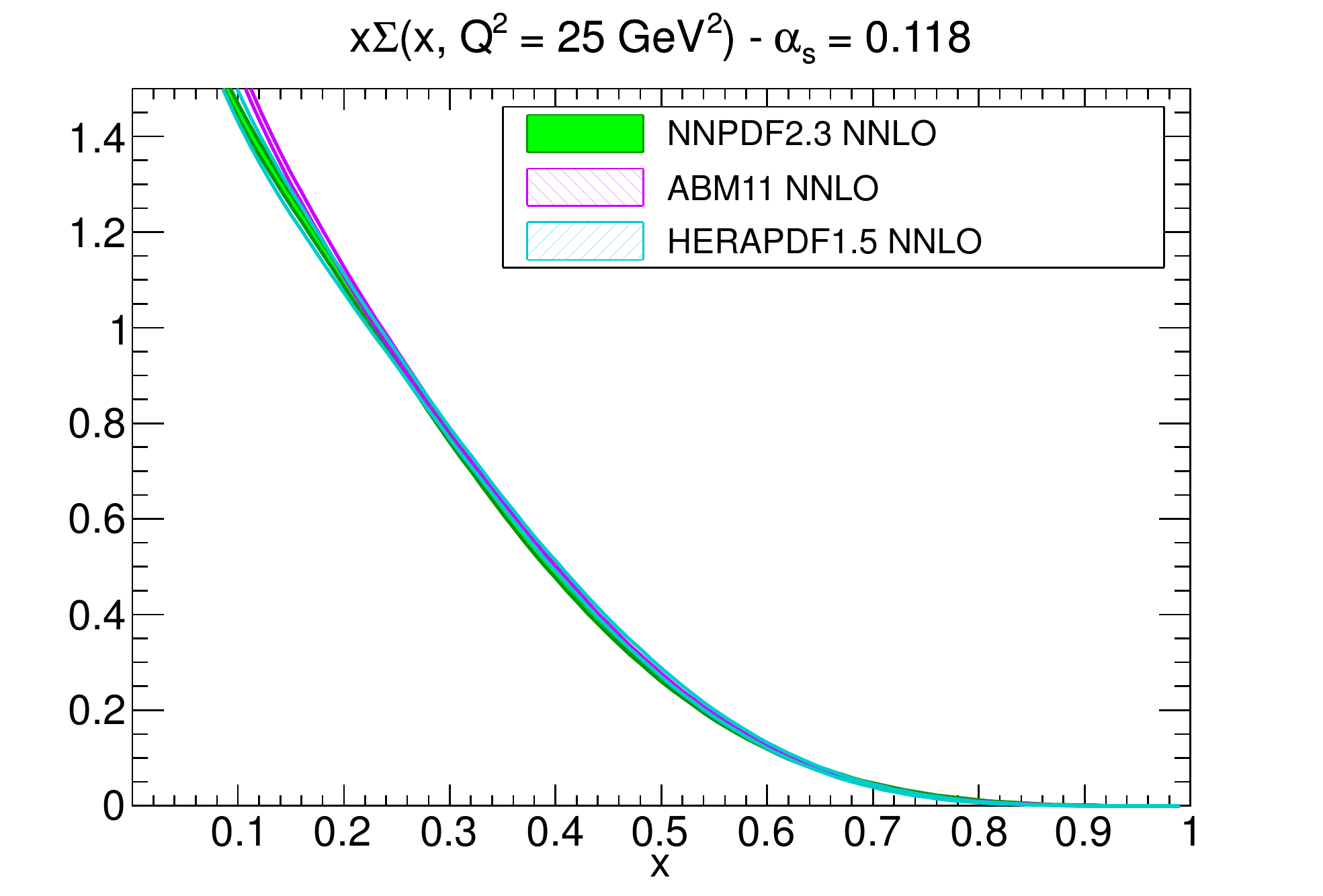}\\
      \includegraphics[width=0.48\textwidth]{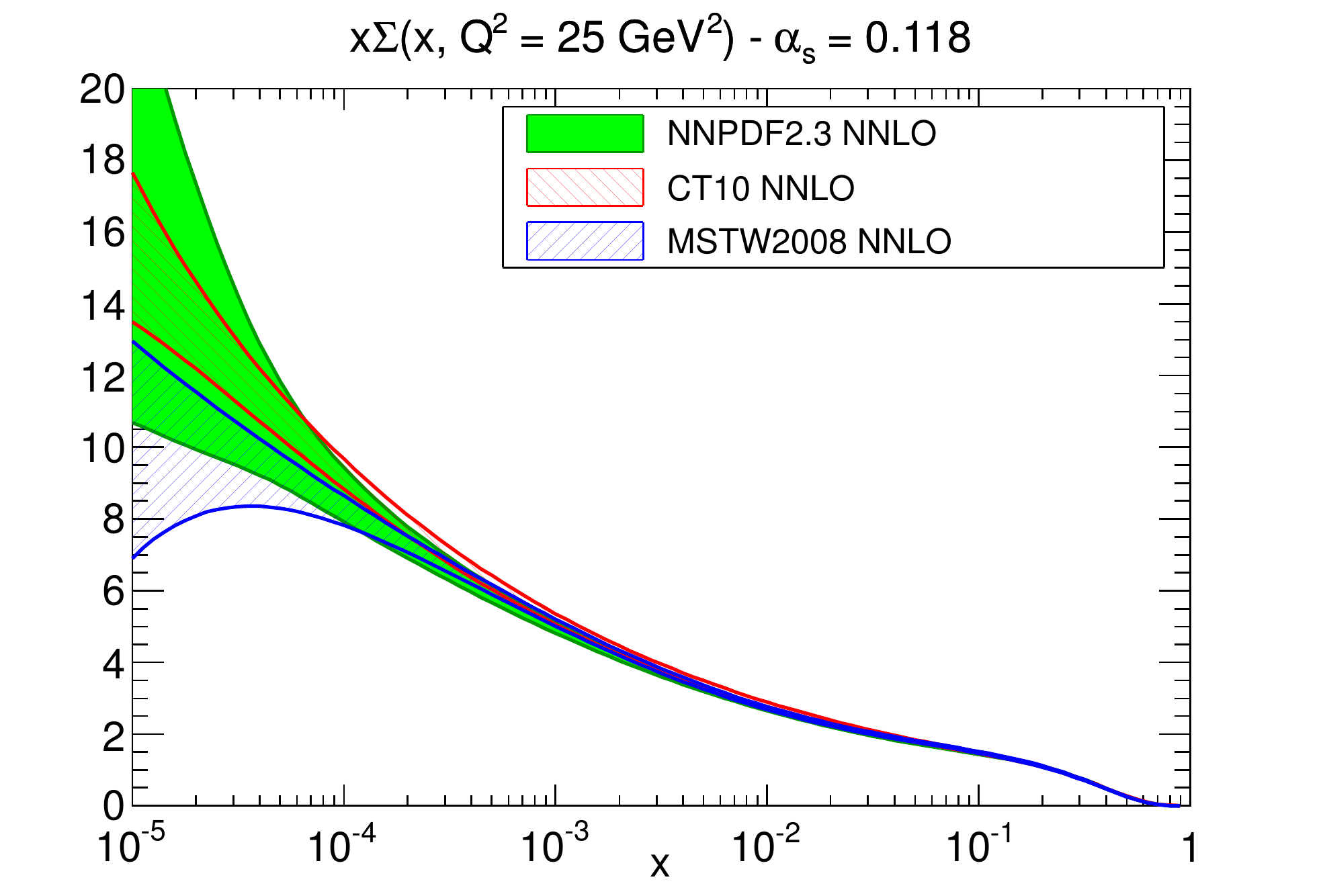}\quad
      \includegraphics[width=0.48\textwidth]{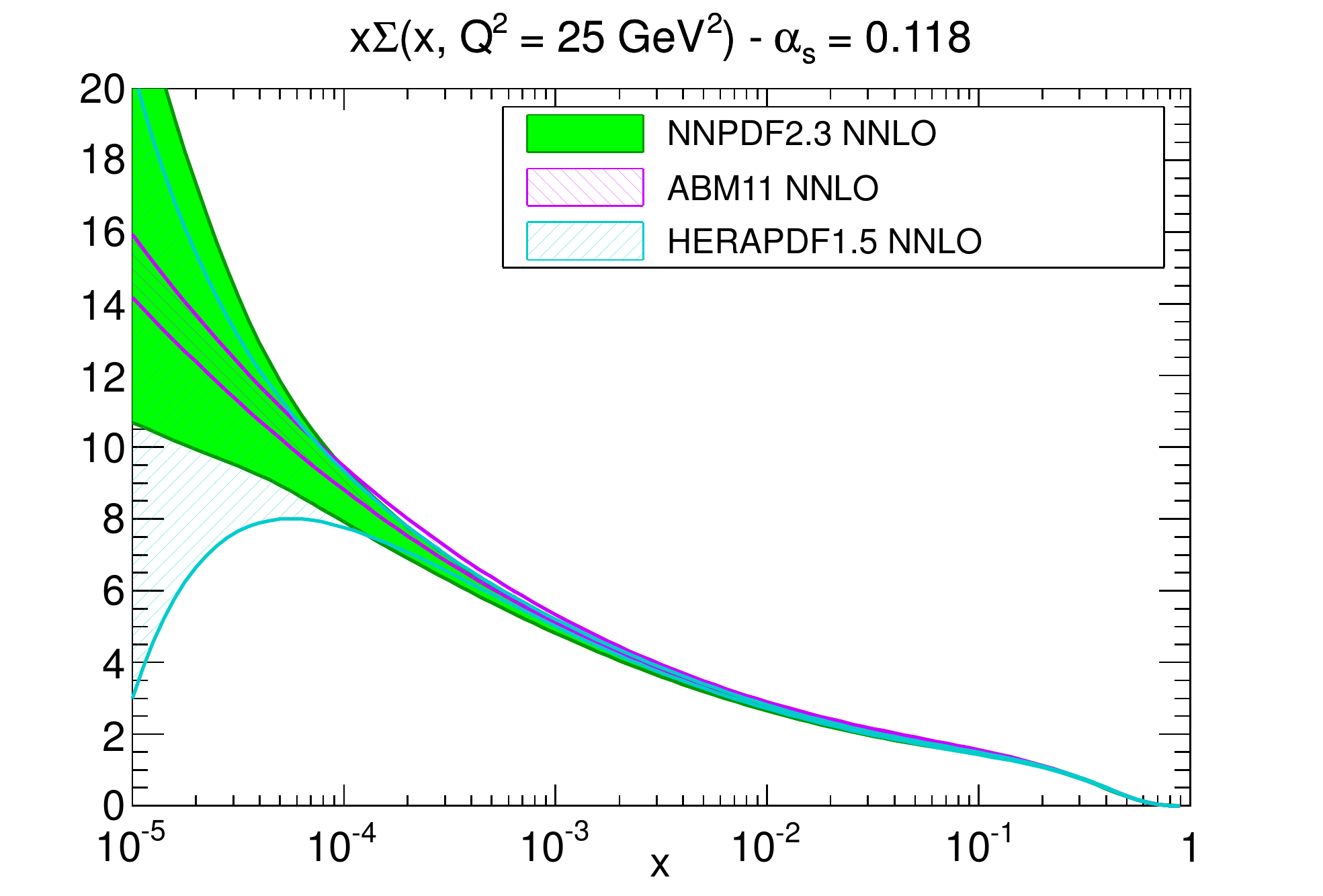}
    \end{center}
      \caption{ The singlet PDFs comparison at $Q^2 = 25$ GeV$^2$
        between the NNLO PDF sets with $\alpha_s=0.118$, on a linear
        scale (upper plots) and on a logarithmic scale (lower
        plots). The plots on the left show the comparison between
        NNPDF2.3, CT10 and MSTW08, while the plots on the right
        compare NNPDF2.3, HERAPDF1.5 and ABM11.}
      \label{fig:PDFcomp-initscale-singlet}
\end{figure}

\begin{figure}
    \begin{center}
      \includegraphics[width=0.48\textwidth]{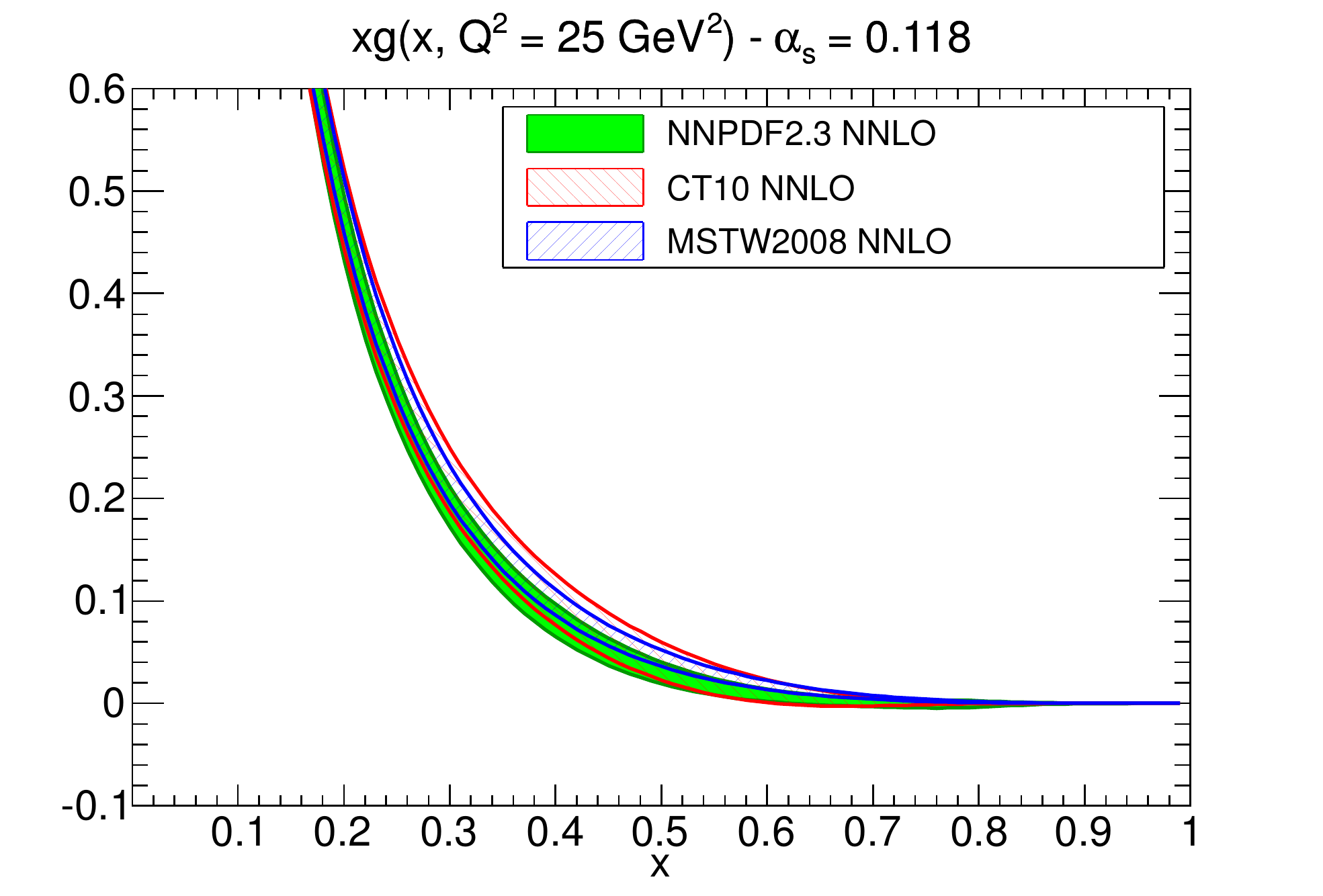}\quad
      \includegraphics[width=0.48\textwidth]{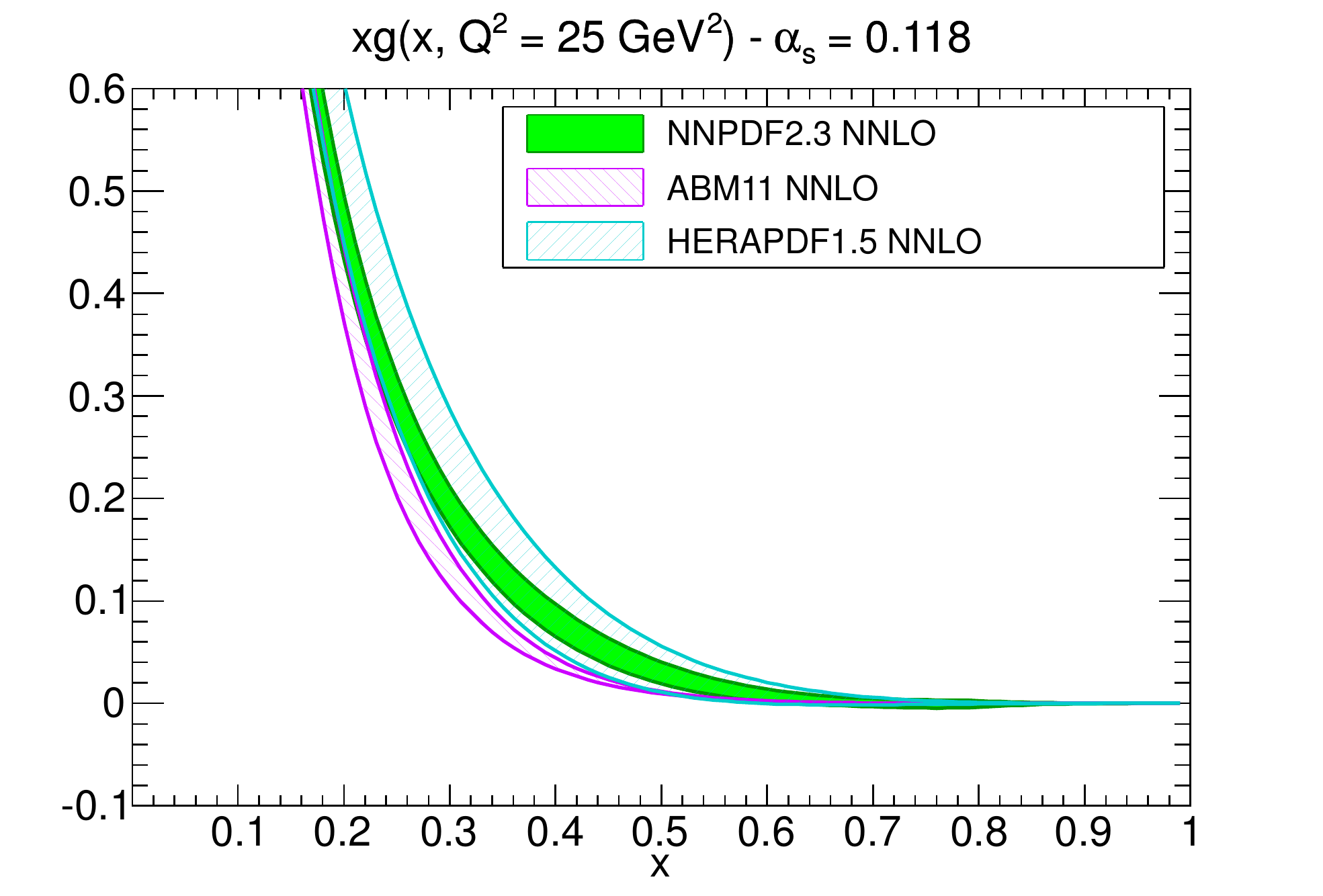}\\
      \includegraphics[width=0.48\textwidth]{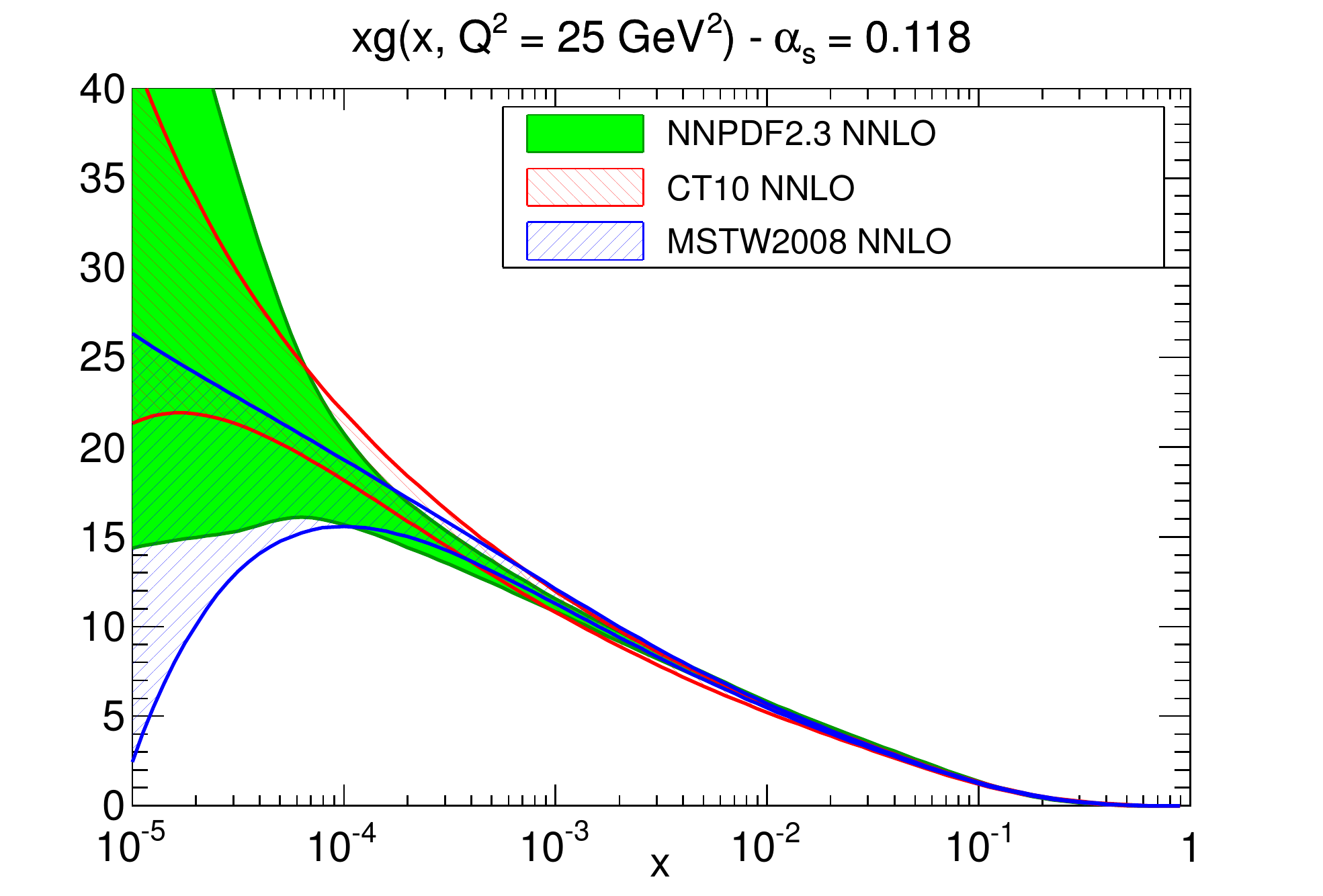}\quad
      \includegraphics[width=0.48\textwidth]{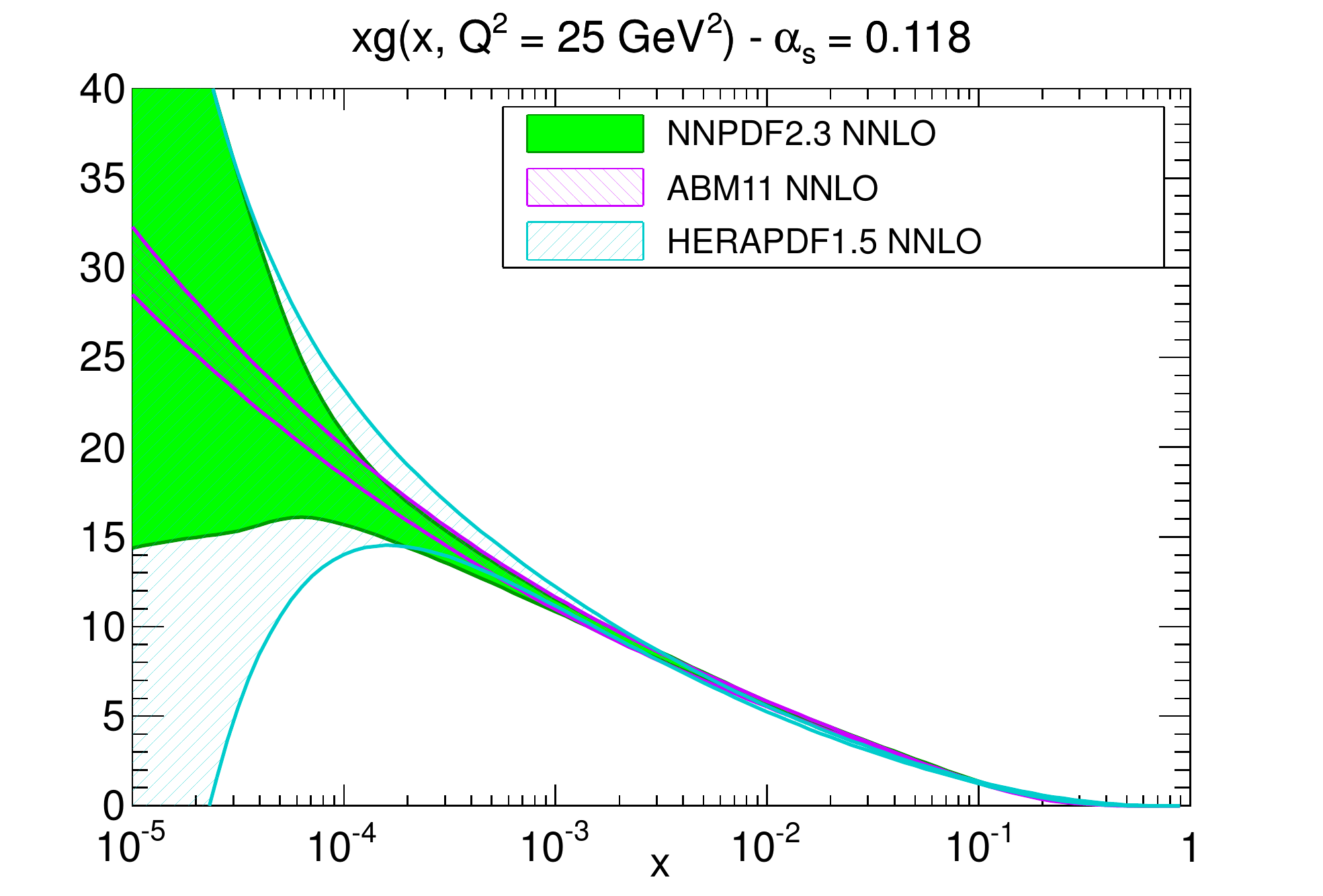}
      \end{center}
     \caption{
    \label{fig:PDFcomp-initscale-g} 
    Same as Figure~\ref{fig:PDFcomp-initscale-singlet}, but for the
    gluon PDF.}
\end{figure}

In Figure~\ref{fig:PDFcomp-initscale-singlet} we show the singlet PDF,
as defined in Eq.~(\ref{eq:singletpdf}), both on linear (upper plots)
and on logarithmic (lower plots) scales, while in
Figure~\ref{fig:PDFcomp-initscale-g} we show the equivalent comparison
for the gluon PDFs.

The agreement is good between all the sets for the singlet, though the
uncertainty band at small $x$ is rather wider for NNPDF and
HERAPDF. There is also reasonable agreement for the gluon between
CT10, MSTW and NNPDF sets, where the PDF 1-$\sigma$ uncertainty bands
overlap for all the range of $x$. Differences are larger for ABM11, in
particular, at small $x$ the ABM11 gluon has smaller uncertainties
than other groups, even for $x$ values where there is little
constraint from the data, due perhaps to the more restrictive
underlying PDF parametrization. The ABM11 gluon at high $x$ is smaller
than that of CT, MSTW and NNPDF, meanwhile the uncertainty band
overlaps that of HERAPDF in most places.  The HERAPDF1.5 gluon at
large $x$ has larger uncertainties due to the lack of collider data,
while at small $x$ it is close to the other PDF sets as expected,
since in this region it is only the precise HERA-I data that provides
constraints to the gluon.

\begin{figure}
    \begin{center}
\includegraphics[width=0.48\textwidth]{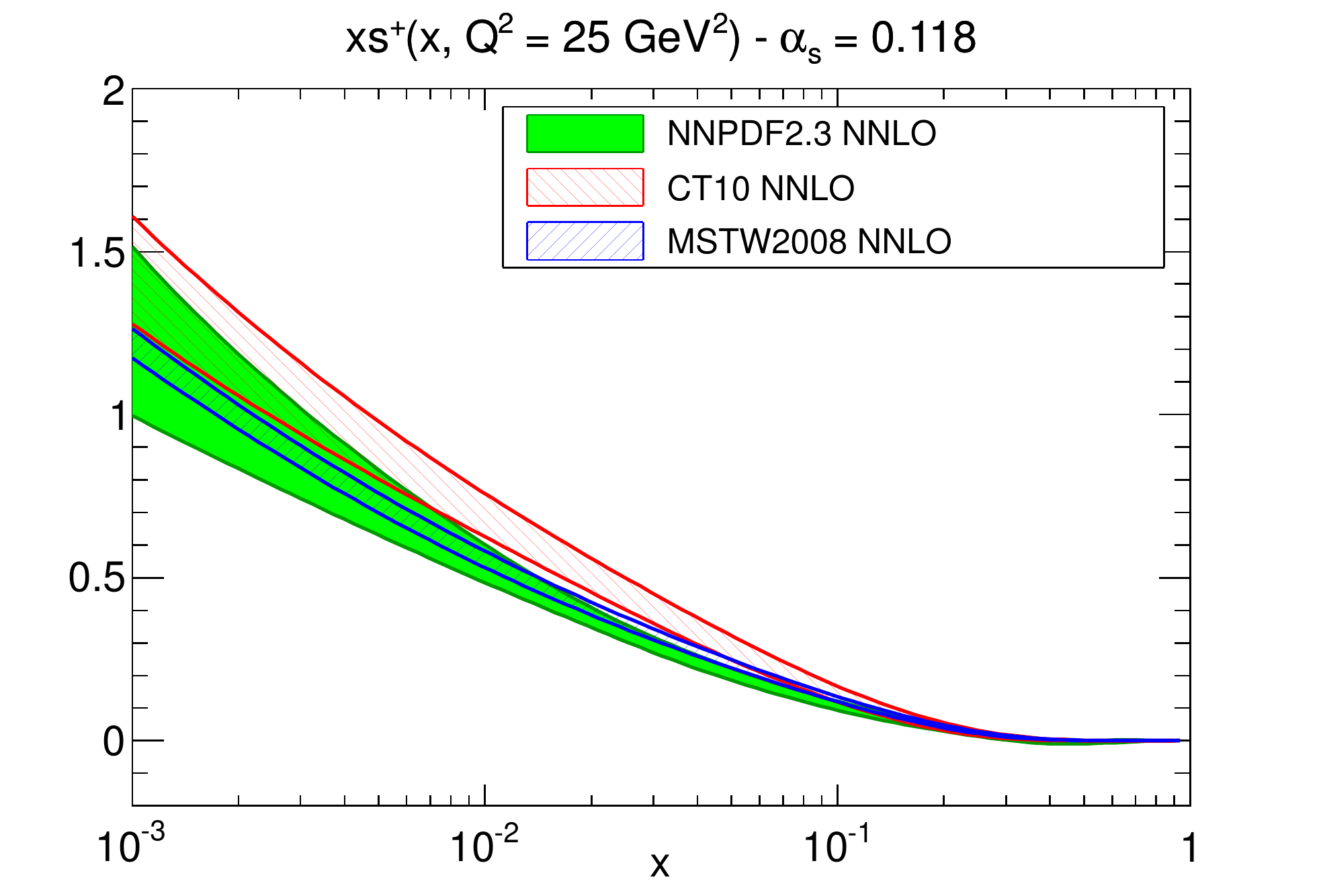}\quad
\includegraphics[width=0.48\textwidth]{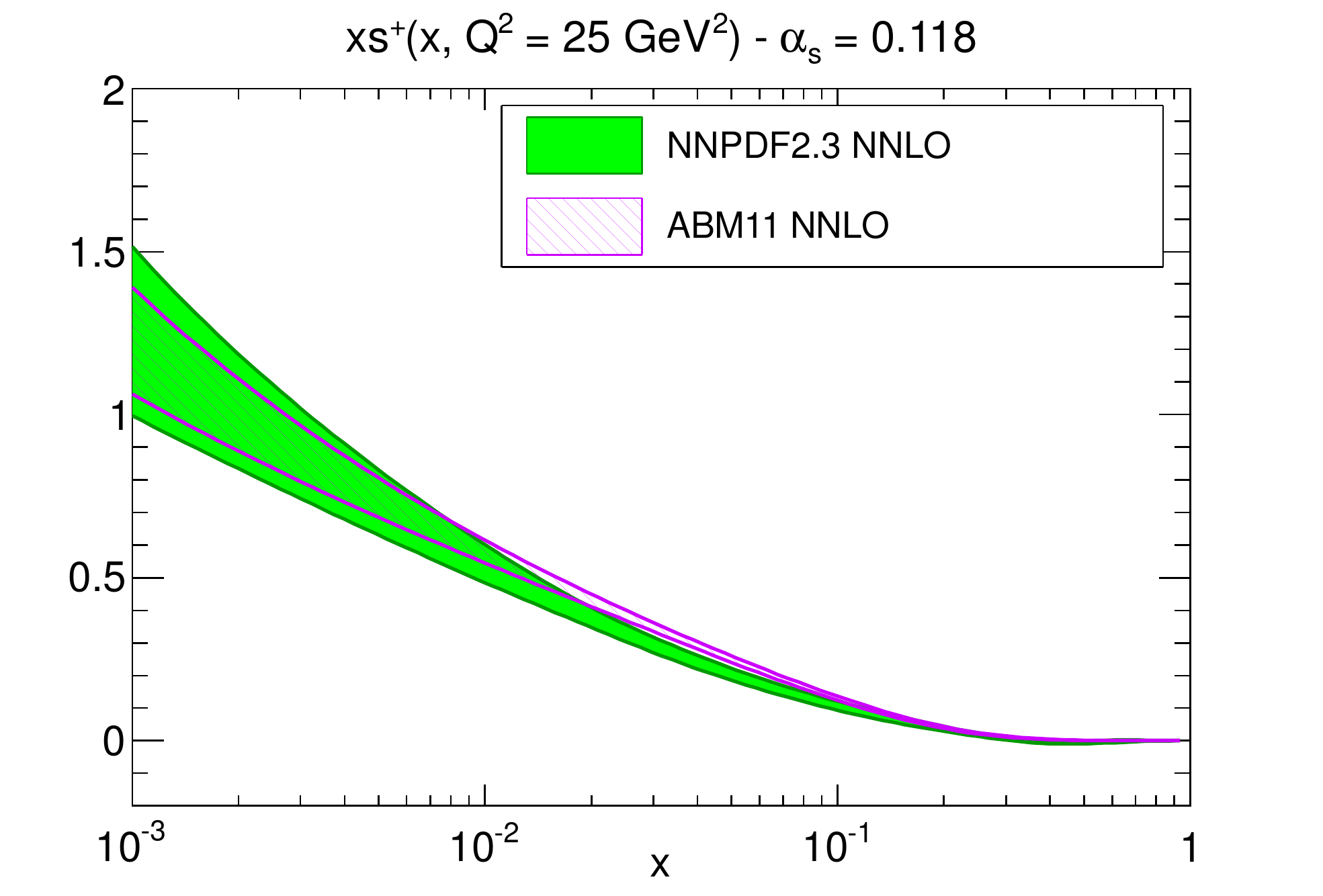}
      \end{center}
     \caption{Comparison of the total strange PDFs
at $Q^2 = 25$ GeV$^2$ between different NNLO PDF sets on a 
 logarithmic scale. On the left plot we show the comparison between
 NNPDF2.3, CT10 and MSTW08, while in the plot on the right we compare only
 NNPDF2.3 and ABM11. HERAPDF1.5 is not show because it does not have
 an independent parametrization of strangeness.
 \label{fig:PDFcomp-initscale-str} }
\end{figure}

In Figure~\ref{fig:PDFcomp-initscale-str} we show the total
strangeness $s^+(x,Q^2)$, see Eq.~(\ref{eq:fpm}), on a logarithmic
scale. NNPDF2.3, MSTW08, ABM11 agree at the 1-$\sigma$ level, however
CT10 is slightly higher, which is justified by the different
treatments of heavy-quark mass effects near threshold in charged
current structure functions and implementation of NuTeV data. The
CT10, MSTW, and NNPDF groups use a general-mass variable flavor number
(GM-VFNS) scheme, which in the case of MSTW and NNPDF turns out to be
close to the fixed-flavor number scheme (FFNS) in neutrino charm
production in the region relevant to
data~\cite{Ball:2011mu,Ball:2011uy}. The ABM11 uses FFNS for neutrino
charm production, while HERAPDF1.5 does not use the dimuon data and
fixes strangeness to be a fraction of the total quark sea.

Studies from the ATLAS collaboration have shown that the inclusive
$W,Z$ production with the 36 pb$^{-1}$ data prefers a larger strange
PDF~\cite{Aad:2012sb} with large uncertainties than the one typically
extracted from the neutrino dimuon data. In the NNPDF2.3
analysis~\cite{Ball:2012cx} this behavior is confirmed, ATLAS data
prefers a larger strangeness, but the uncertainties are still sizable
so the global fit still prefers the softer strange PDF favored by the
NuTeV dimuon data. This issue should be clarified in future when
including more data from the LHC, from more inclusive electroweak
vector boson production data and the exclusive $W+c$ data.

\begin{figure}
    \begin{center}
      \includegraphics[width=0.48\textwidth]{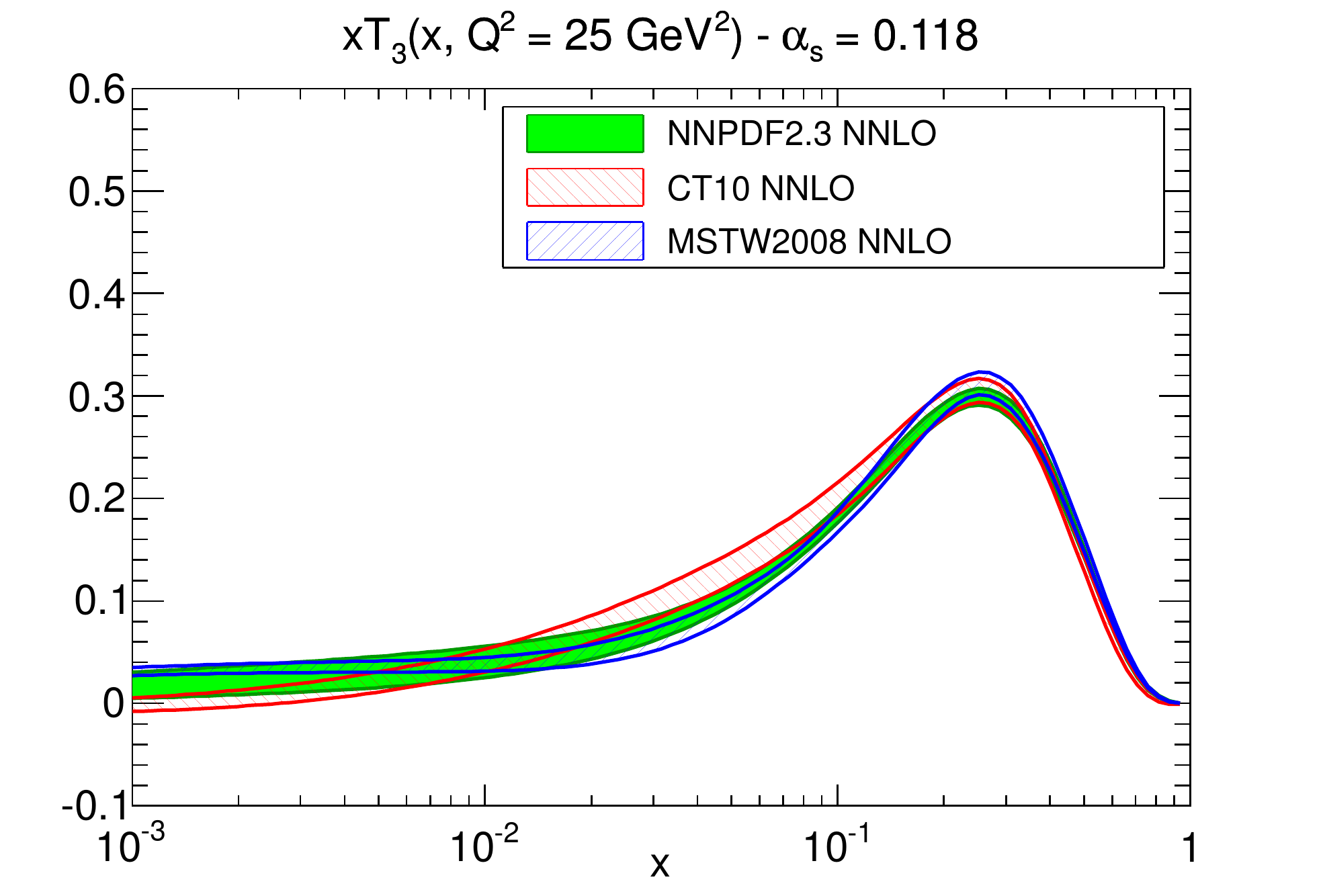}\quad
\includegraphics[width=0.48\textwidth]{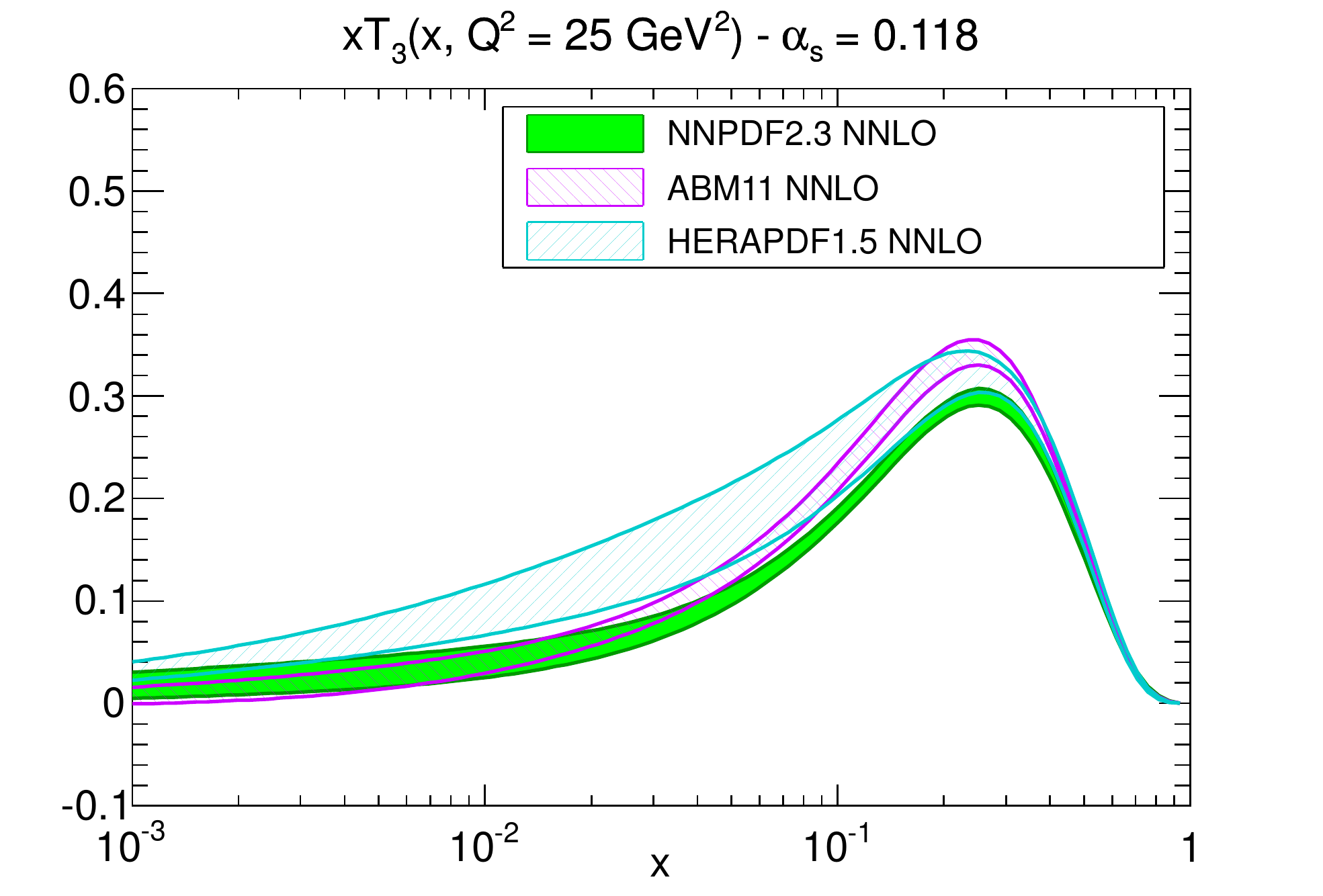}\\
\includegraphics[width=0.48\textwidth]{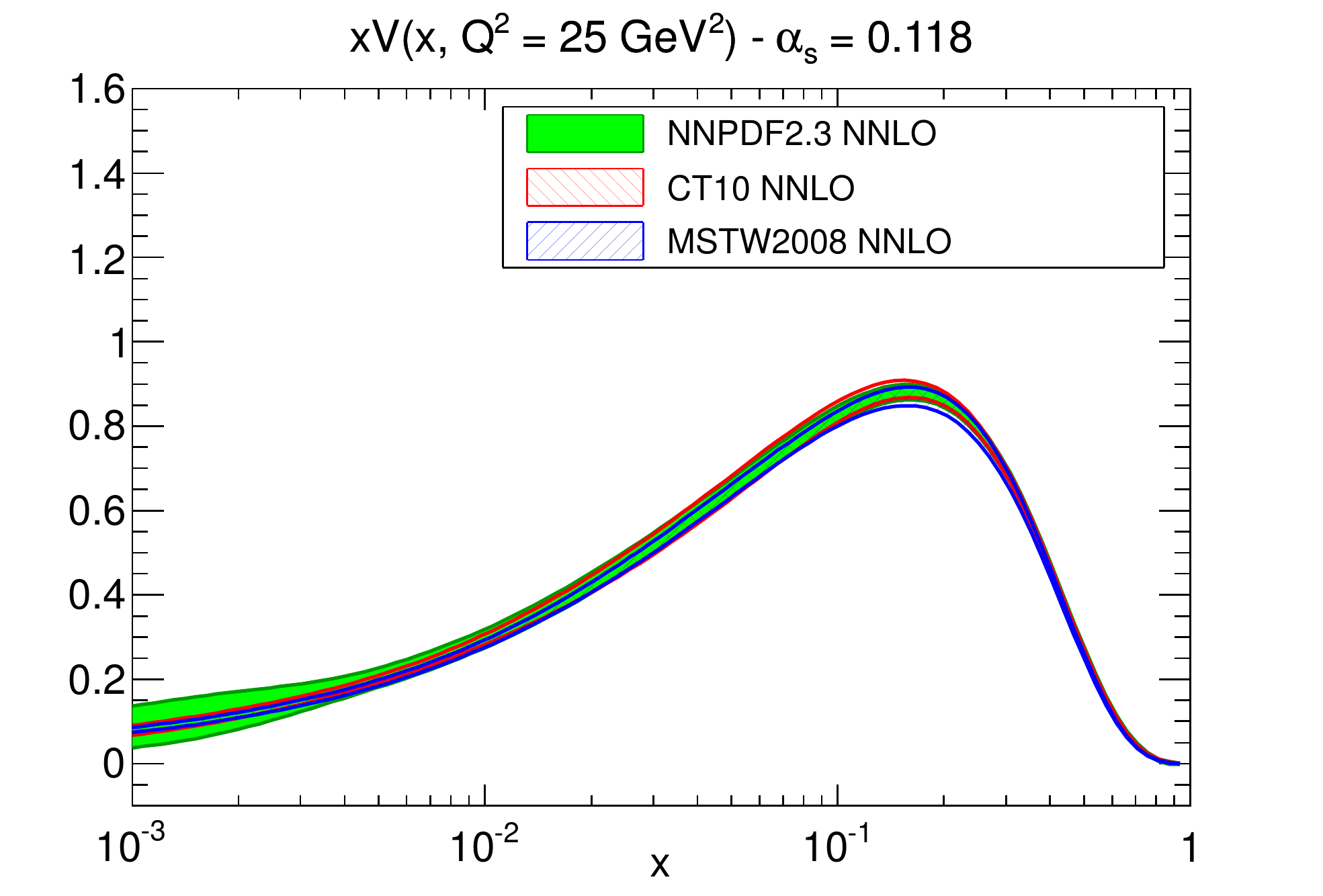}\quad
\includegraphics[width=0.48\textwidth]{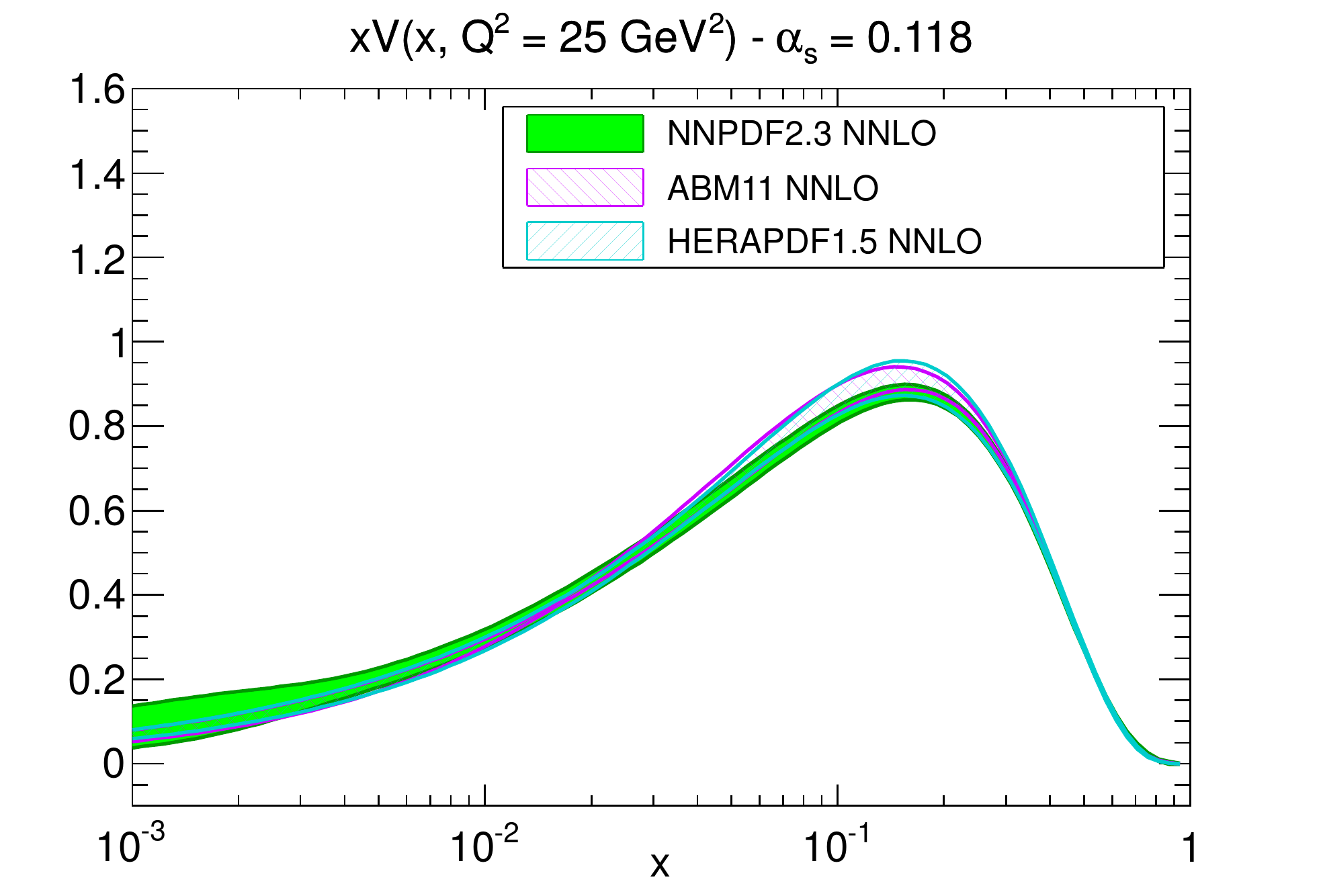}
      \end{center}
      \caption{Same as Figure~\ref{fig:PDFcomp-initscale-singlet} for
        the non-singlet triplet $T_3(x)$ and the total valence $V(x)$
        PDFs.
        \label{fig:PDFcomp-initscale-t3} }
\end{figure}

\begin{figure}
  \begin{center}
    \includegraphics[width=0.48\textwidth]{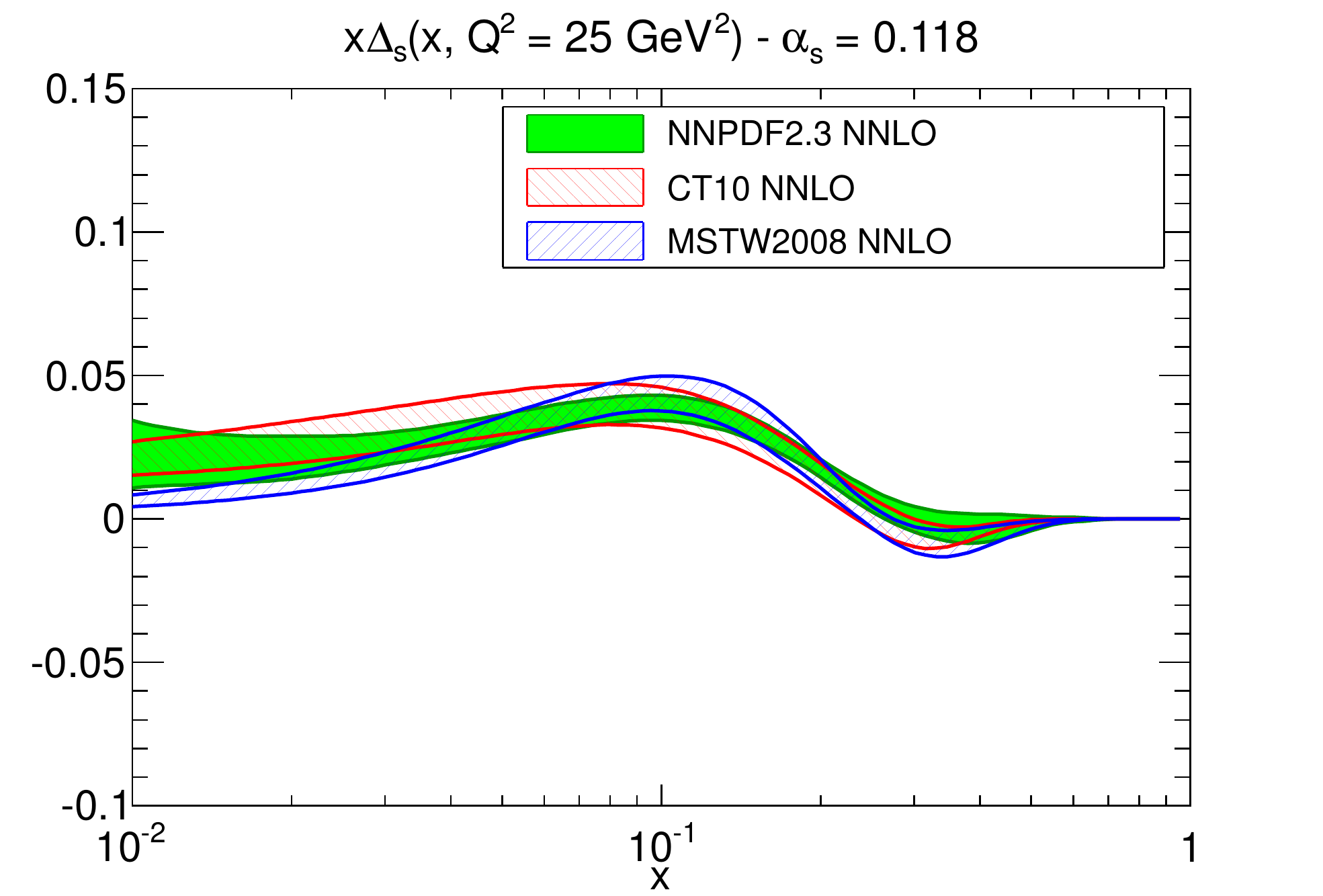}\quad
    \includegraphics[width=0.48\textwidth]{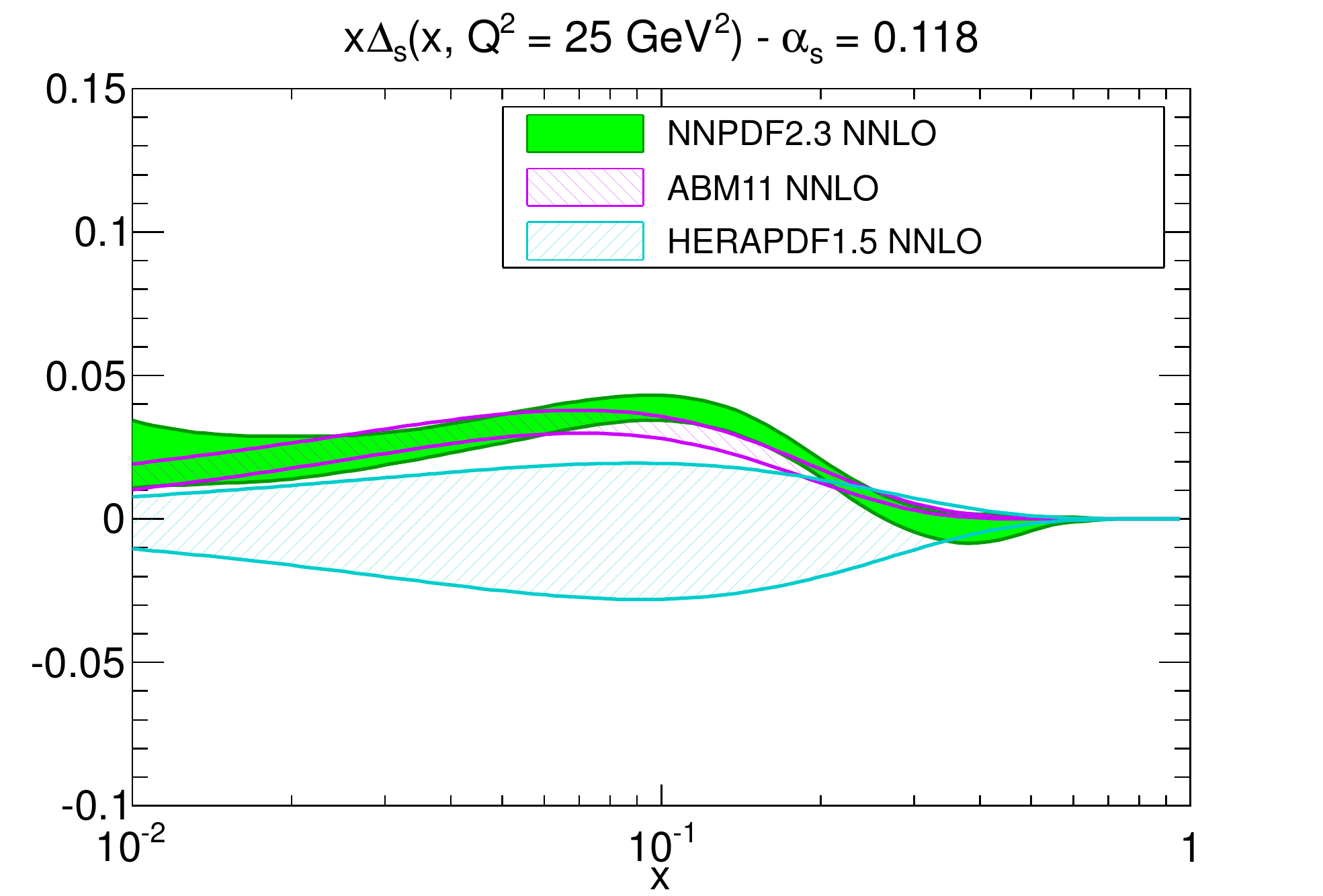}\\
    \includegraphics[width=0.48\textwidth]{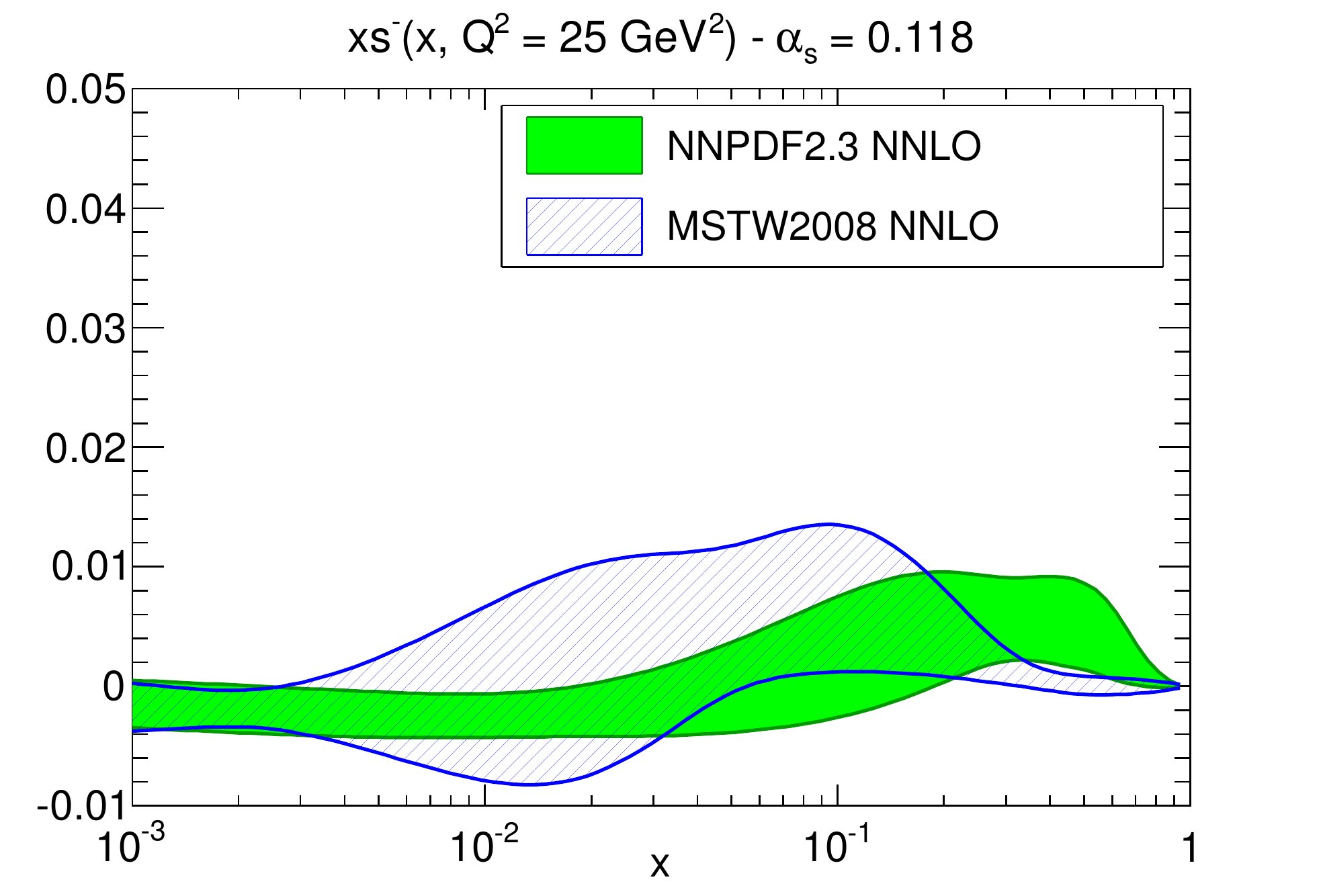}
  \end{center}
  \caption{Same as Figure~\ref{fig:PDFcomp-initscale-t3} for the the
    sea asymmetry $\Delta_S$ and the strange asymmetry $s^-$ PDFs. In
    the latter case we show only the results for MSTW08 and NNPDF2.3,
    the only PDF sets that introduce an independent parametrization of
    the strangeness asymmetry.}
  \label{fig:PDFcomp-initscale-deltas}
\end{figure}

We conclude this comparison analysis section with other flavor
combinations:
\begin{itemize}
\item The non-singlet distributions $T_3$ and $V$ PDFs, defined in
  Eq.~(\ref{eq:nonsingletpdf}) in Figure~\ref{fig:PDFcomp-initscale-t3}.
  
\item The quark sea asymmetry $\Delta_S = \bar{d}-\bar{u}$ and the
  strangeness asymmetry $s^- = s-\bar{s}$ in
  Figure~\ref{fig:PDFcomp-initscale-deltas}.
\end{itemize}

We observe a reasonable agreement for $T_3$ and $V$, except for ABM11,
where $T_3$ is higher at large $x$ due to a larger $u$
distribution. The HERAPDF1.5 PDF uncertainties in $T_3$ are rather
larger, reflecting the fact that HERA data does not provide much
information on quark flavor separation. Concerning the quark sea
asymmetry all sets are in a agreement apart from the HERAPDF1.5, which
does not include the Drell-Yan and electroweak boson production data
and cannot separate $\bar u$ and $\bar d$ flavors. Finally, the only
sets that provide an independent parametrization of the strangeness
asymmetry PDF are MSTW08 and NNPDF2.3, showing a reasonable agreement
within uncertainties.

\subsubsection*{PDF luminosities}

At a hadron collider the factorized observables for production of a
heavy final state with mass $M_X$ depend on parton distributions
through a parton luminosity, which, following our introduction in
Sect.~\ref{sec:dy} and Ref.~\cite{Campbell:2006wx}, is defined as
\begin{equation}
  \Phi_{ij}\left( M_X^2\right) = \frac{1}{s}\int_{\tau}^1
  \frac{dx_1}{x_1} f_i\left( x_1,M_X^2\right) f_j\left( \tau/x_1,M_X^2\right) \ ,
\label{eq:lumdef}
\end{equation}
where $f_i(x,M^2)$ is a PDF at a scale $M^2$, and $\tau \equiv
M_X^2/s$.  Following the criteria applied to the PDF comparison, also
here all parton luminosities are compared at $\alpha_s=0.118$. The
NNPDF2.3 set is used as reference for the parton luminosities ratios,
and we assume a center-of-mass energy of $\sqrt{s} = 8$ TeV which is
close to the energy achieved by the LHC Run-I collisions.

\begin{figure}
    \begin{center}
      \includegraphics[width=0.48\textwidth]{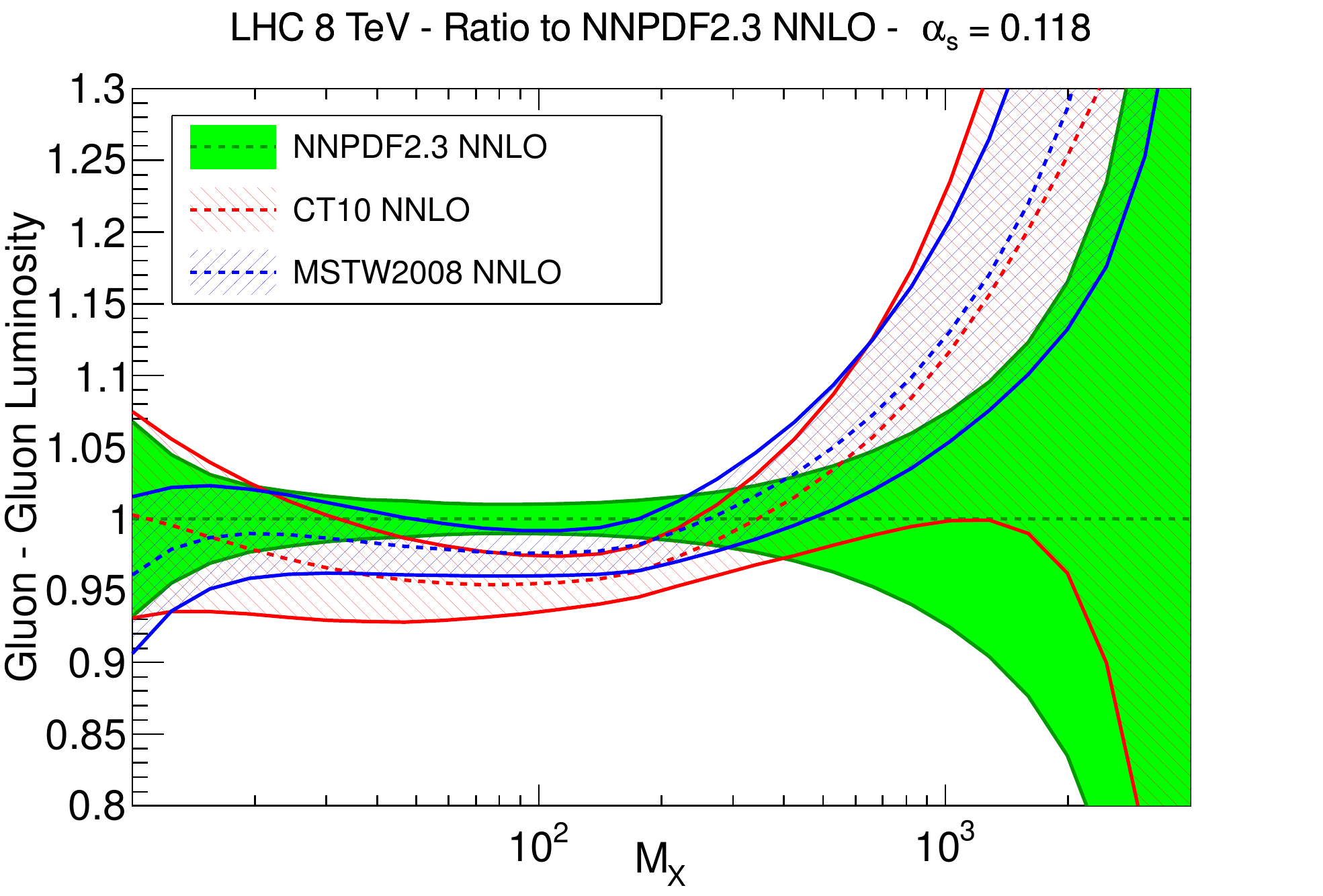}\quad
      \includegraphics[width=0.48\textwidth]{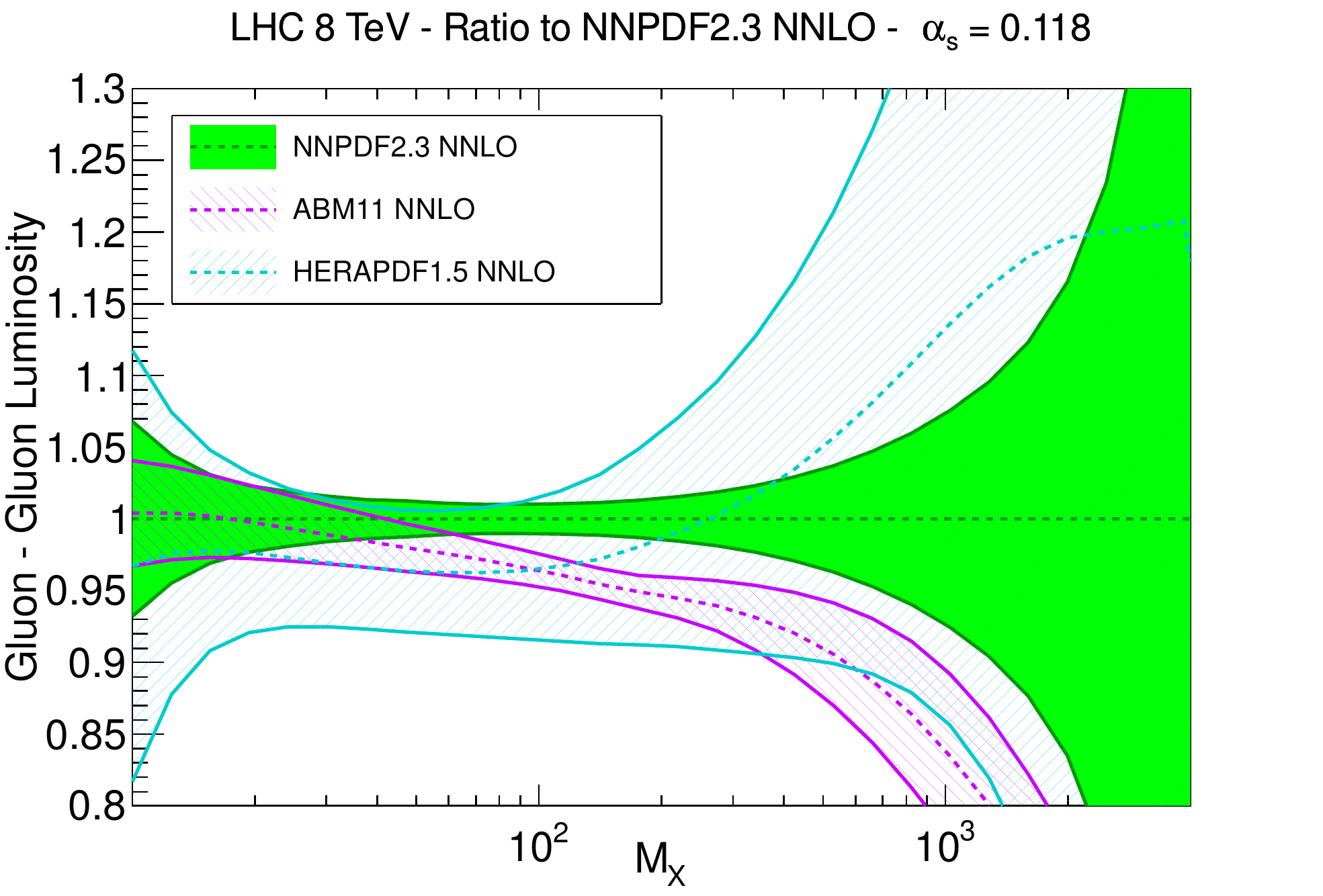}\\
      \includegraphics[width=0.48\textwidth]{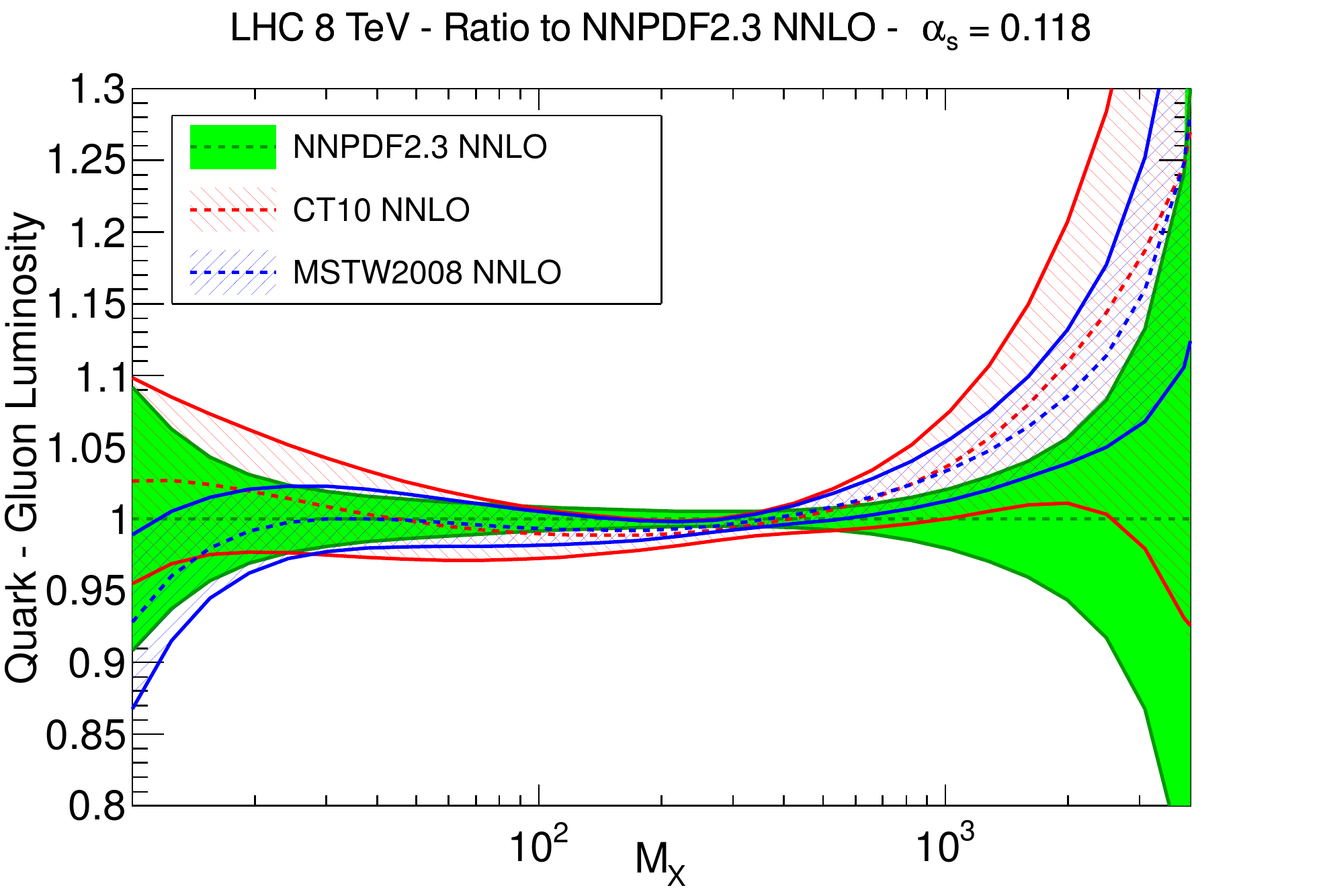}\quad
      \includegraphics[width=0.48\textwidth]{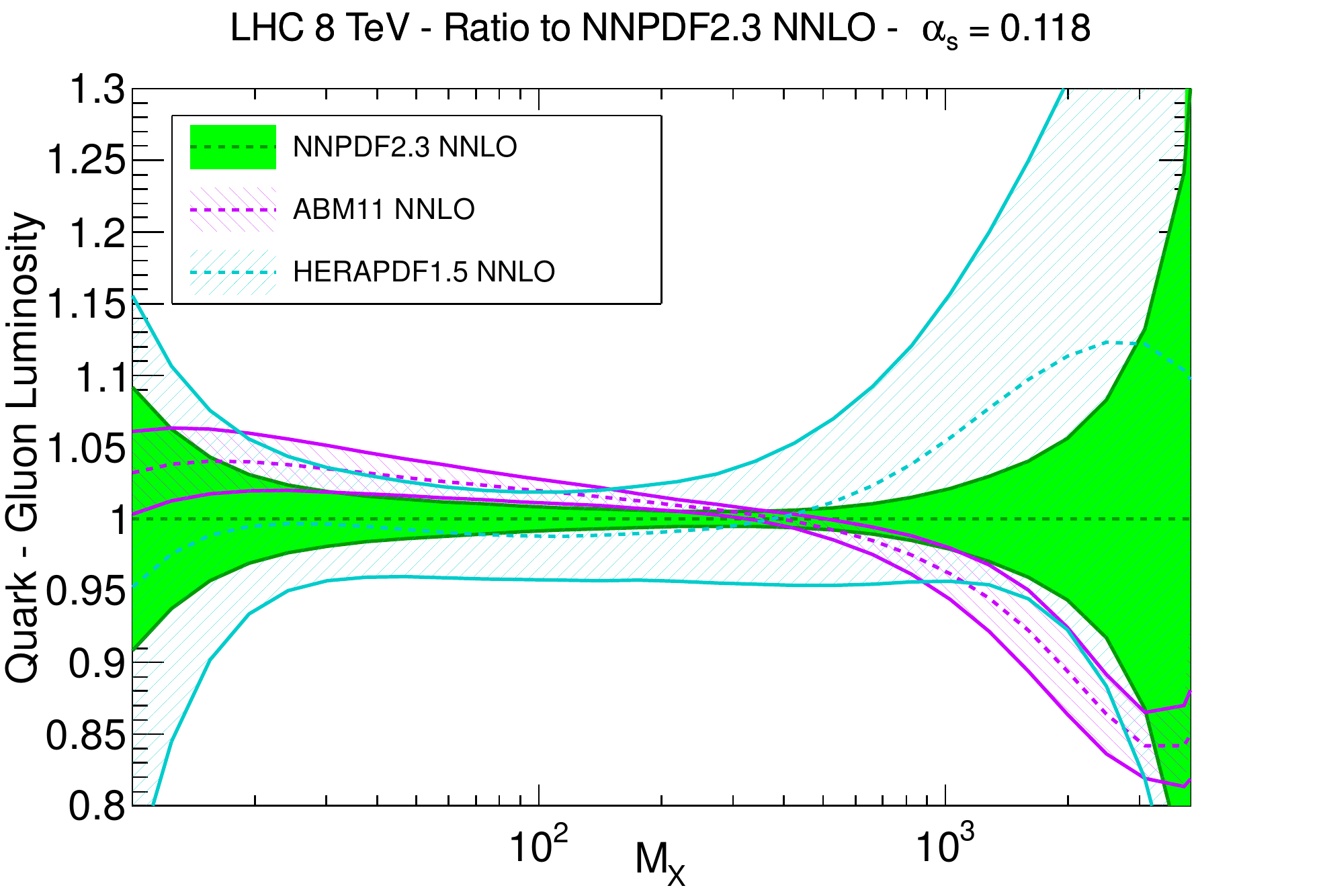}
    \end{center}
    
    \caption{The gluon-gluon (upper plots) and quark-gluon (lower
      plots) luminosities, Eq.~(\ref{eq:lumdef}), with $\alpha_s=0.118$,
      at LHC $\sqrt{s} = 8$ TeV. The NNPDF2.3 set is used as reference for both comparison groups.
      \label{fig:PDFlumi-gg} }
\end{figure}

\begin{figure}
    \begin{center}
      \includegraphics[width=0.48\textwidth]{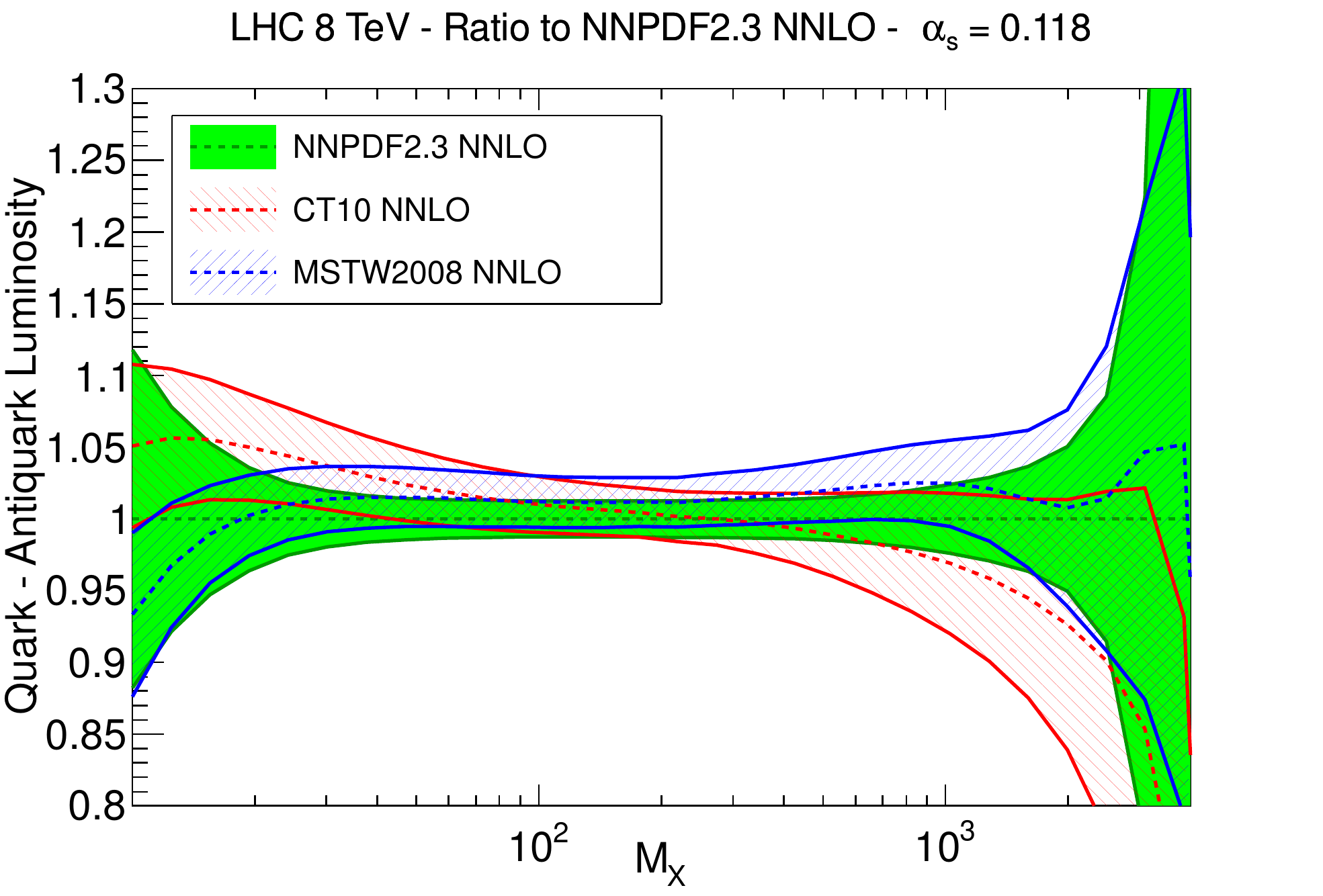}\quad
\includegraphics[width=0.48\textwidth]{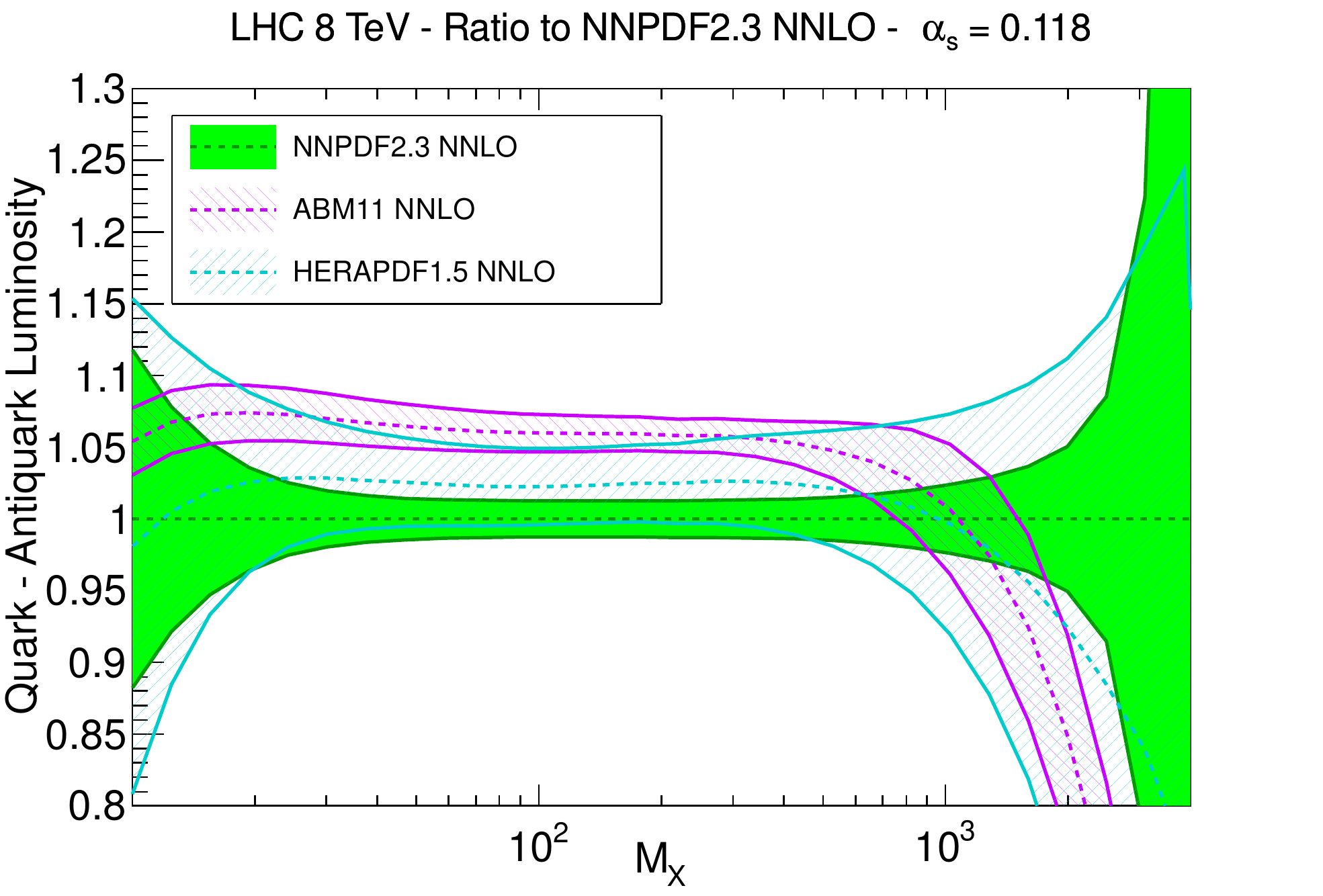}\\
  \includegraphics[width=0.48\textwidth]{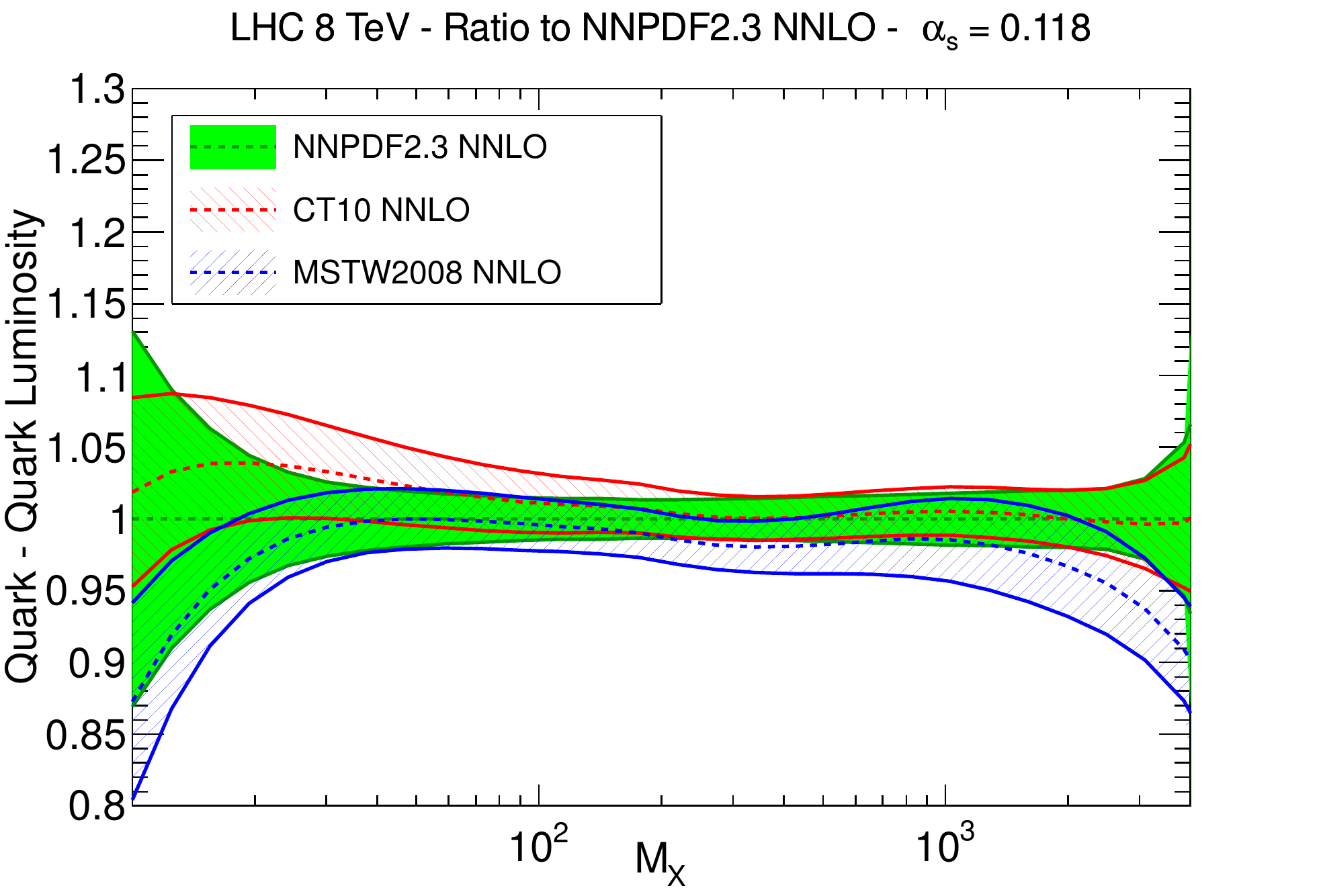}\quad
\includegraphics[width=0.48\textwidth]{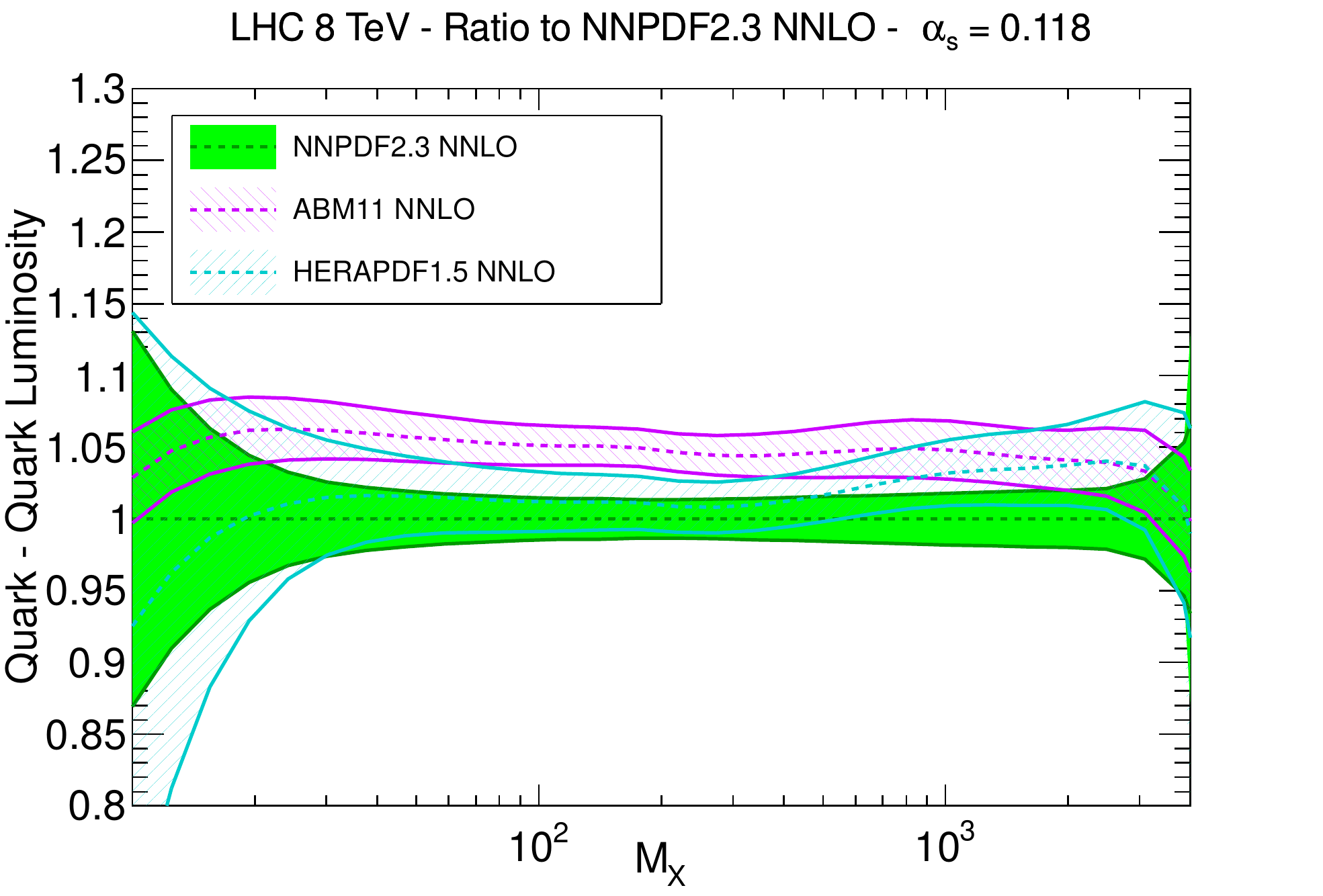}
      \end{center}
     \caption{
Same as  Figure~\ref{fig:PDFlumi-gg} 
for the quark-antiquark (upper plots)
and quark-quark (lower plots) luminosities.
\label{fig:PDFlumi-qq}}
\end{figure}
 
The gluon-gluon and quark-gluon luminosities are shown in
Figure~\ref{fig:PDFlumi-gg}, and the quark-quark and quark-antiquark
luminosities are shown in Figure~\ref{fig:PDFlumi-qq}. A reasonably
good agreement is observed between the NNPDF2.3, MSTW08 and CT10 PDF
sets for the full range of invariant masses. The PDF uncertainties
increase dramatically at $M_X > 1$~TeV. Future data from the LHC such
as the high-mass Drell-Yan process should be able to provide
constraints in this important region. For HERAPDF1.5, there is
generally an agreement in central values, but the uncertainty is
rather larger in some $x$ ranges, particularly for the gluon
luminosity, but also to some extent for the quark-antiquark one. For
ABM11 instead, the quark-quark and quark-antiquark luminosity are
systematically higher by over 5\% below 1 TeV, and above this the
quark-antiquark luminosity becomes much softer than either NNPDF2.3 or
MSTW08. The gluon-gluon luminosity becomes smaller than all the other
PDFs at high invariant masses, overlapping only with the very large
HERAPDF1.5 uncertainty.

\begin{figure}
    \begin{center}
      \includegraphics[width=0.48\textwidth]{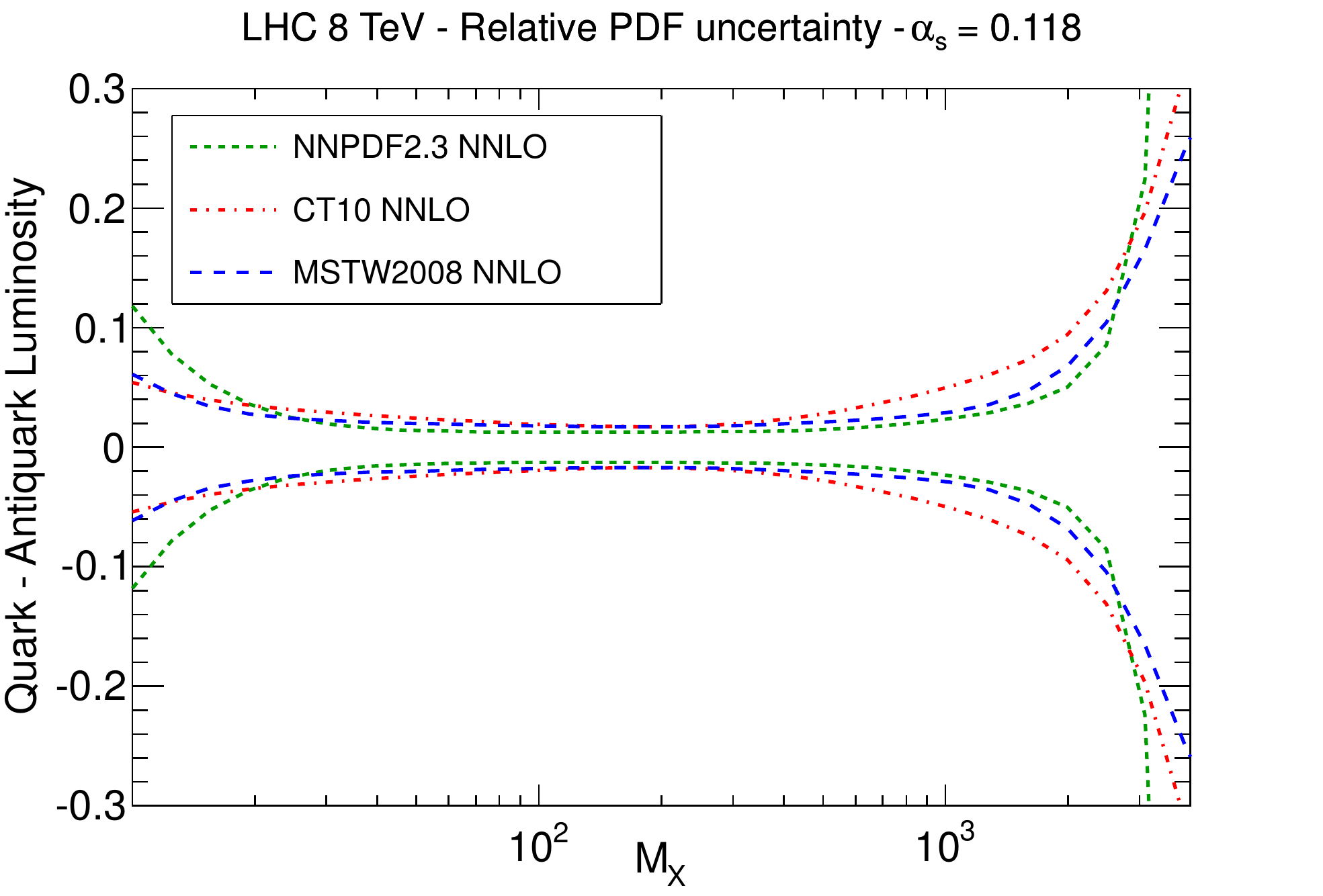}\quad
      \includegraphics[width=0.48\textwidth]{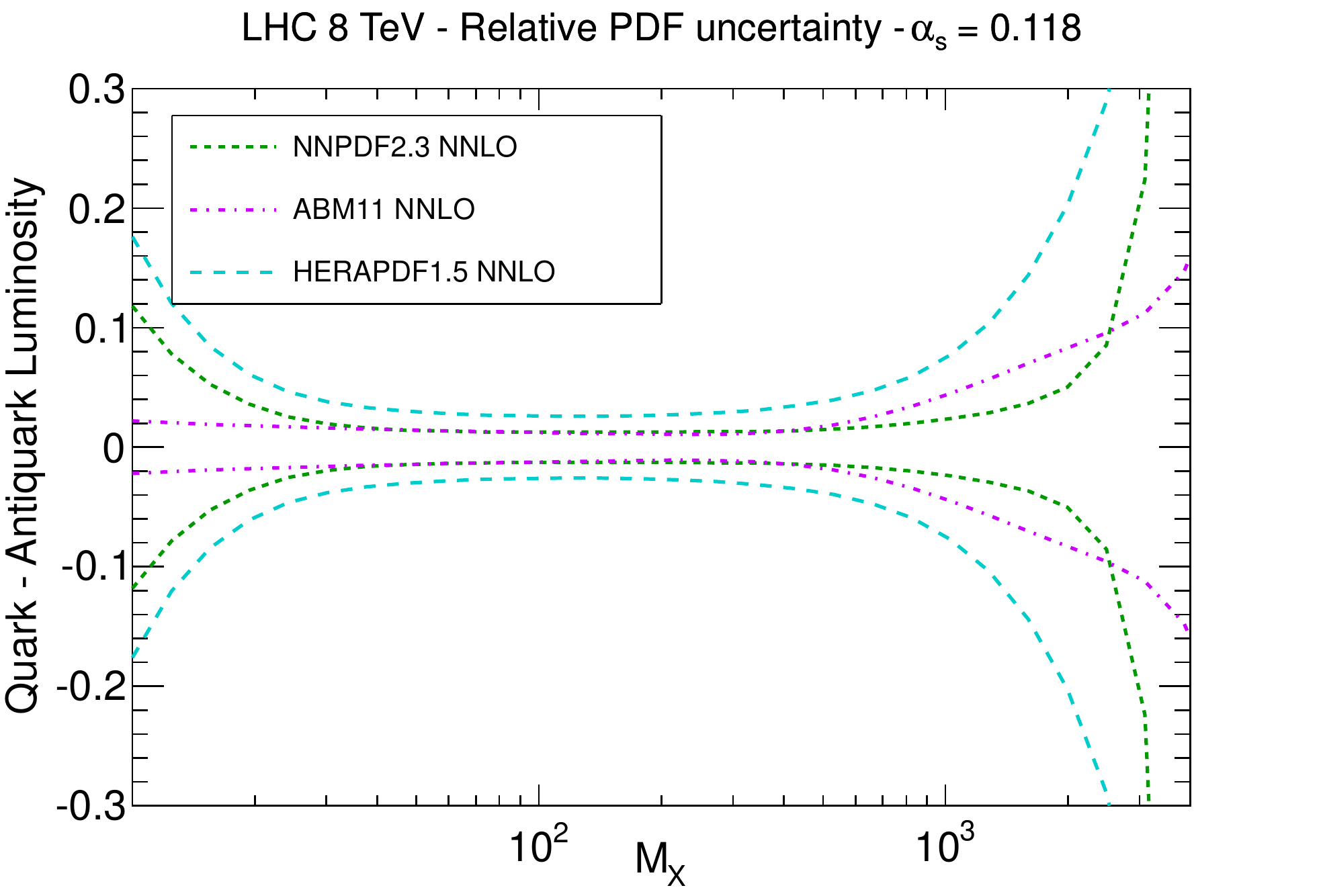}\\
      \includegraphics[width=0.48\textwidth]{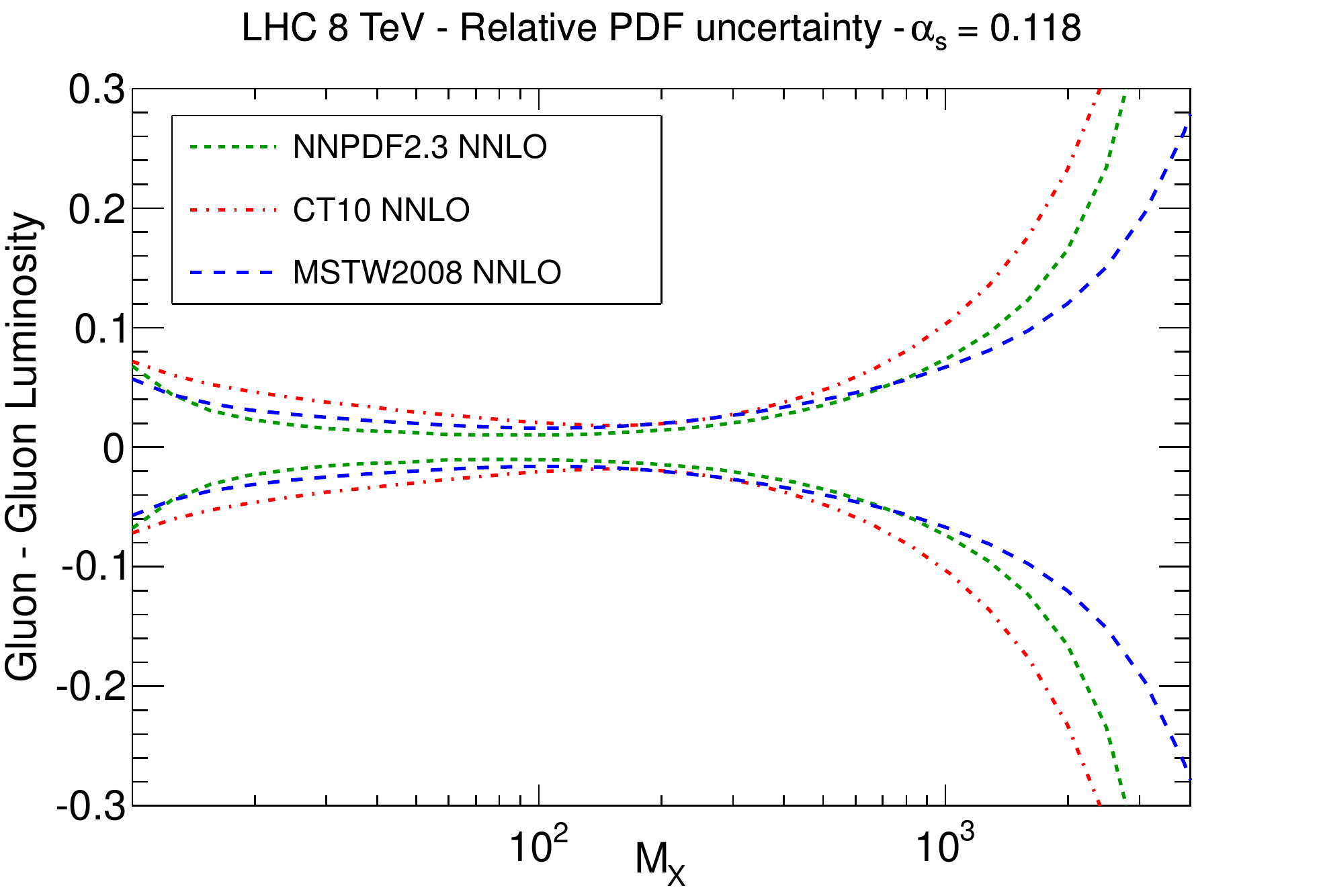}\quad
      \includegraphics[width=0.48\textwidth]{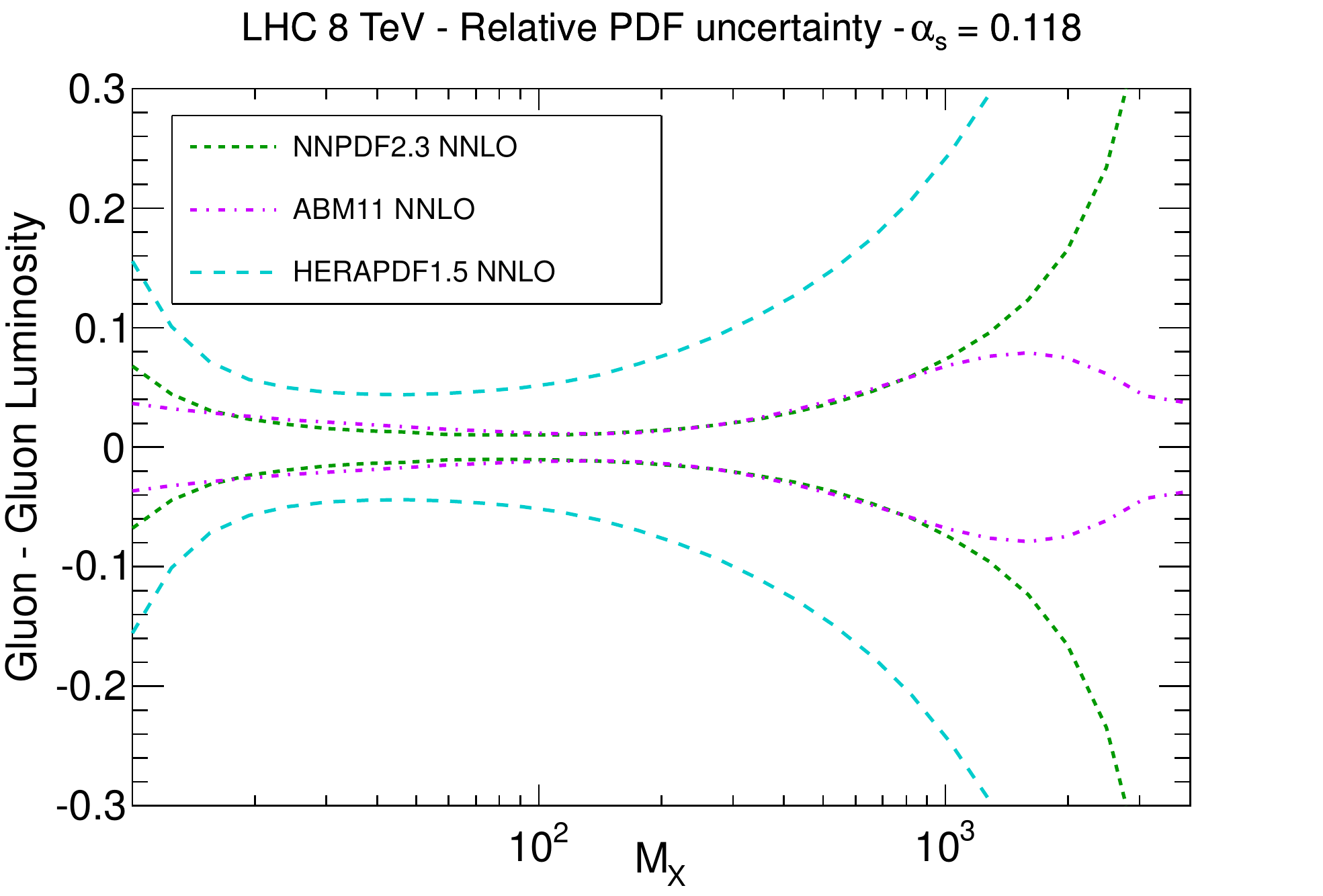}
    \end{center}
    \caption{The relative PDF uncertainties in the
      quark-antiquark luminosity (upper plots) and in the gluon-gluon
      luminosity (lower plots), with $\alpha_s= 0.118$ at the LHC $\sqrt{s} = 8$ TeV.
      \label{fig:PDFlumi-rel} }
\end{figure}

It is also useful to compare the relative PDF uncertainties in the
parton luminosities. In Figure~\ref{fig:PDFlumi-rel} we show this
relative PDF uncertainty for the quark-antiquark and gluon-gluon
luminosities. We see clearly the much larger HERAPDF1.5 uncertainty,
and that at high invariant mass, the uncertainty in the ABM11
gluon-gluon luminosity becomes smaller.

The larger quark-antiquark luminosity from ABM11 as compared to the
other PDF sets could be inferred from the PDF comparison plots at
lower $Q^2$, the ABM gluon is a little larger than the central value
of the other groups below about $x=0.05$, and this drives more quark
and antiquark evolution at small $x$ values.  It has been recently
suggested~\cite{Thorne:2012az} from the results of a NLO fit to DIS
data only, that some of these features could receive a contribution
from the different ABM treatment of heavy-quark masses (see
also~\cite{CooperSarkar:2007ny}).  While CT, MSTW and NNPDF use
different versions of the variable flavour number
scheme~\cite{Forte:2010ta,Thorne:2006qt,Guzzi:2011ew}, which are
broadly equivalent to one another up to small subleading terms, ABM11
uses a fixed flavour number scheme for heavy-quark PDFs.  This may
explain the increase in the medium-$x$ and small-$x$ light quarks and
gluons, and the corresponding softer large-$x$ gluon required by the
momentum sum rule, found in the ABM fits~\cite{Thorne:2012az}, though
more studies are required to confirm this point.

\begin{table}
\centering
\begin{tabular}{|c|c|c|c|}
  \hline
  & $Q_0^2$ (GeV$^2$)& $Q^2_{\textrm{min}}$ (GeV$^2$) & $W^2_{\textrm{min}}$ (GeV$^2$)\\
  \hline
  \hline
  ABM11 & 9  & 2.5 & 3.24 \\
  CT10 & 1.69 & 4.0 & 12.25 \\
  HERAPDF1.5 & 1.9 & 3.5 & - \\
  MSTW08 & 1  & 2.0 & 15.0\\
  NNPDF2.3 & 2.0 & 3.0 & 12.5 \\
  \hline
\end{tabular}
\caption{The values of the initial evolution scale where the
  PDFs are parametrized, $Q^2_0$, and the kinematical cuts in $Q^2$
  and $W^2=Q^2\ \left(1/x-1\right)$ 
  applied to the fitted DIS dataset, $Q^2_{\textrm{min}}$
  and $W^2_{\textrm{min}}$, in the present work and in other recent
  PDF determinations.}
\label{tab:kincuts}
\end{table}

An alternative interpretation proposed to explain these differences
between ABM11 and the other groups resides on the treatment of the
kinematical cuts of the DIS data. These cuts control the impact of
higher twists contributions. All groups undertake measures to minimize
the impact of higher twists, in particular the CT10, MSTW08 and
NNPDF2.3 fits suppress this contribution with a minimal cut on
$W^2=Q^2 \left(1/x-1\right)$. 

In Table~\ref{tab:kincuts} we show a summary of the values of the
initial evolution scale $Q^2_0$ where the PDFs are parametrized,
together with the lower kinematical cuts $Q^2_{\textrm{min}}$ and
$W^2_{\textrm{min}}$ applied to the fitted DIS data sets for each PDF
group. The ABM11 fit also imposes an upper cut
$Q^2_{\textrm{max}}=10^3$ GeV$^2$ on the HERA data. It is well known
that, the larger the dataset, the more robust are the PDFs with
respect to variations in these cuts. For instance, stability under
variation of the default MSTW08 kinematical cuts was studied in
Ref.~\cite{Thorne:2011kq}. The inclusion of higher twists in MRST fits
has previously been shown to lead to only a small effect on high-$Q^2$
PDFs~\cite{Martin:2003sk}, and an ongoing extension of the study
in~\cite{Thorne:2012az} suggests this is qualitatively the same with
more up-to-date PDFs. This conclusion has been confirmed in similar
studies by NNPDF~\cite{Ball:2013gsa}.

\subsection{LHC inclusive cross-sections}
\label{sec:LHCincl}

We conclude this chapter by describing the behavior of the
cross-sections predictions at 8 TeV for various benchmark processes
and compare the results for all NNLO PDF sets used in the previous
section. Also here we consider only PDF uncertainties, negleting a
careful assessment of all relevant theoretical uncertainties into
consideration for each of the studied processes.

\begin{figure}
\centering
\includegraphics[width=0.47\textwidth]{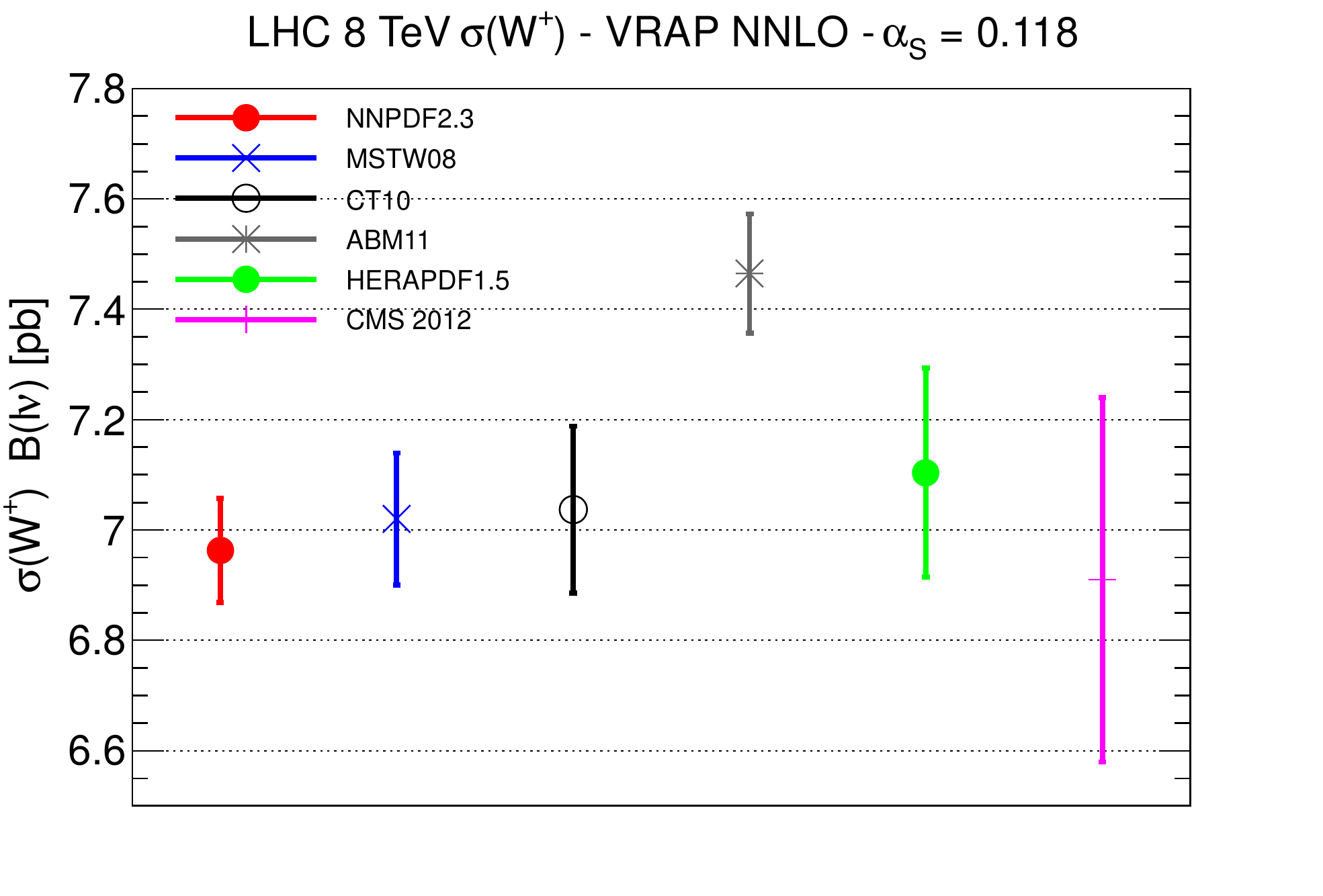}\quad
\includegraphics[width=0.47\textwidth]{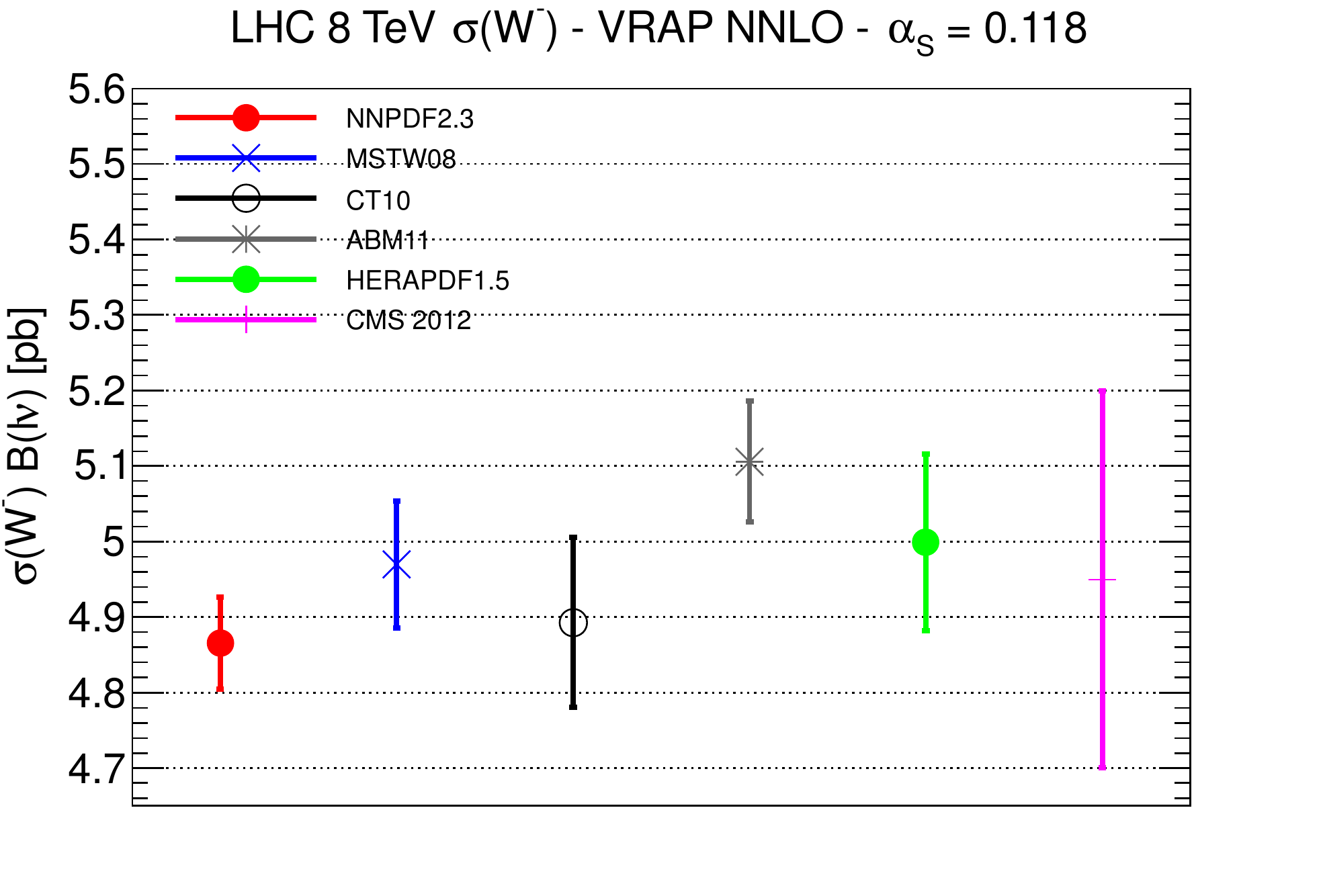}
\includegraphics[width=0.47\textwidth]{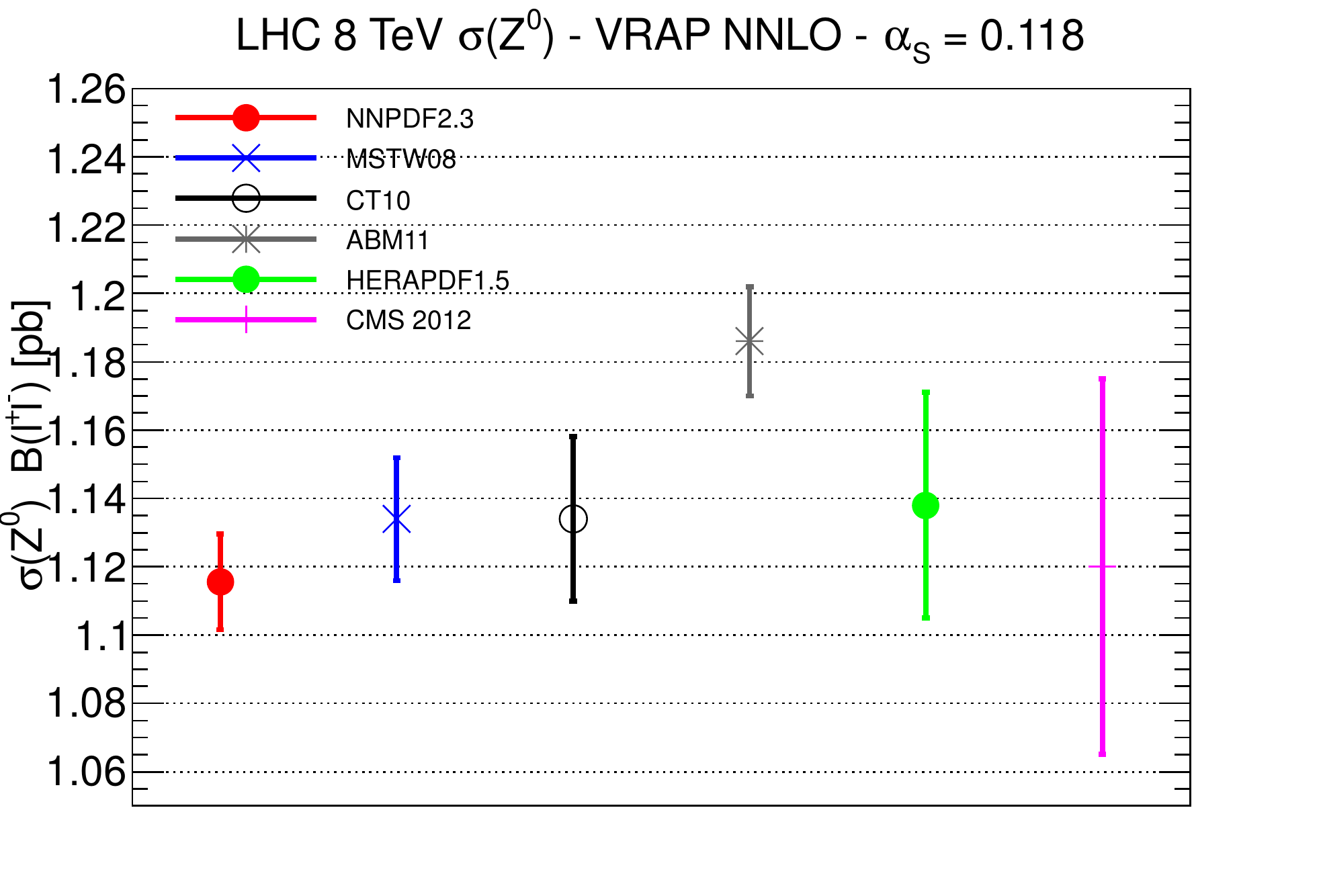}
\caption{Comparison of the 
predictions for inclusive cross-sections for
electroweak gauge boson production
between different PDF sets at LHC 8 TeV. In all cases
the branching ratios to leptons have already been
taken into account. From top to
bottom and from left to right we show the 
$W^+$, $W^-$, and $Z$ inclusive cross-sections. 
All cross-sections are compared
at a common value of $\alpha_s(M_Z)=0.118$. We also
show the recent CMS 8 TeV measurements.
\label{fig:8tev-ewk}}
\end{figure}

\begin{figure}
  \centering
  \includegraphics[width=0.47\textwidth]{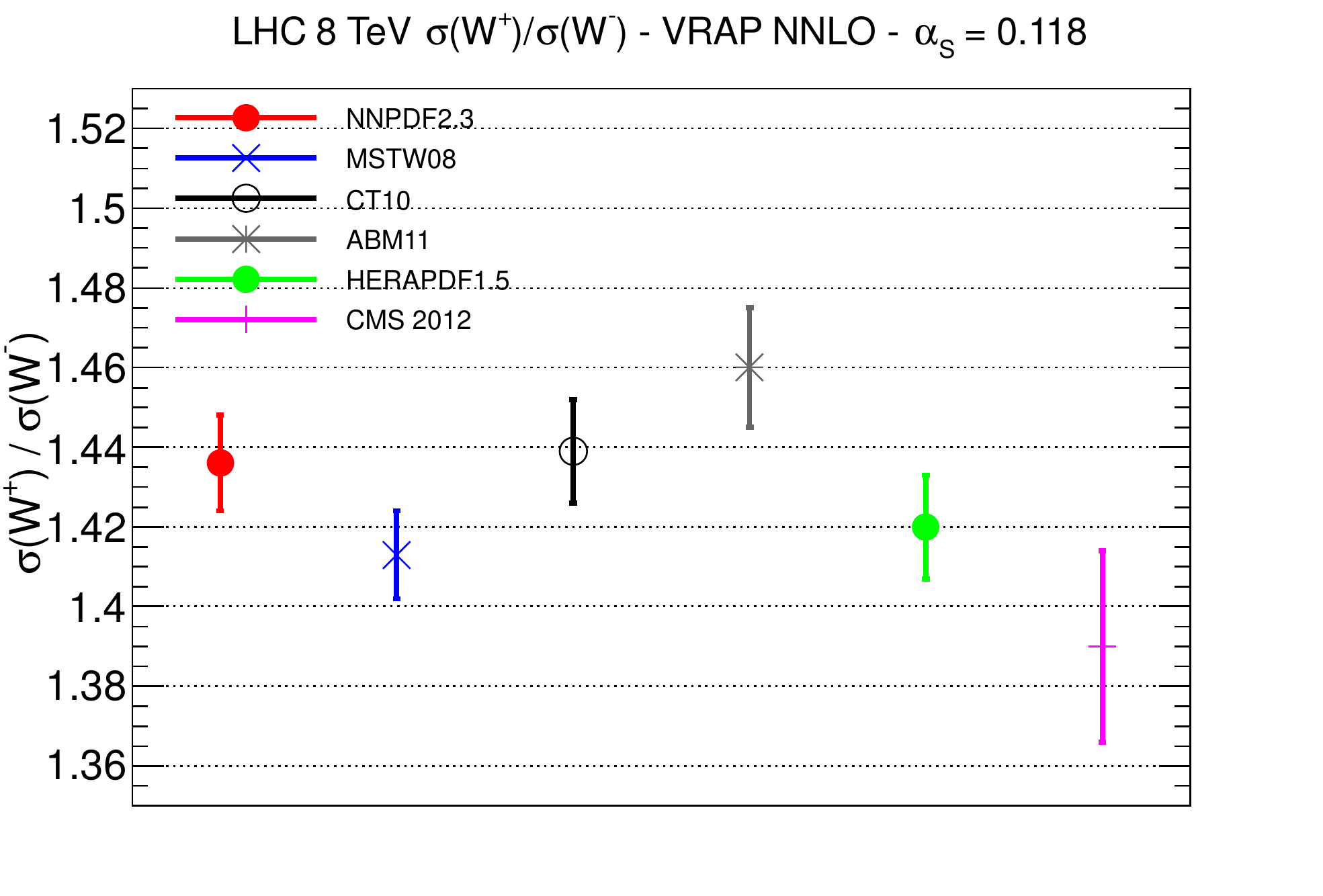}\quad
  \includegraphics[width=0.47\textwidth]{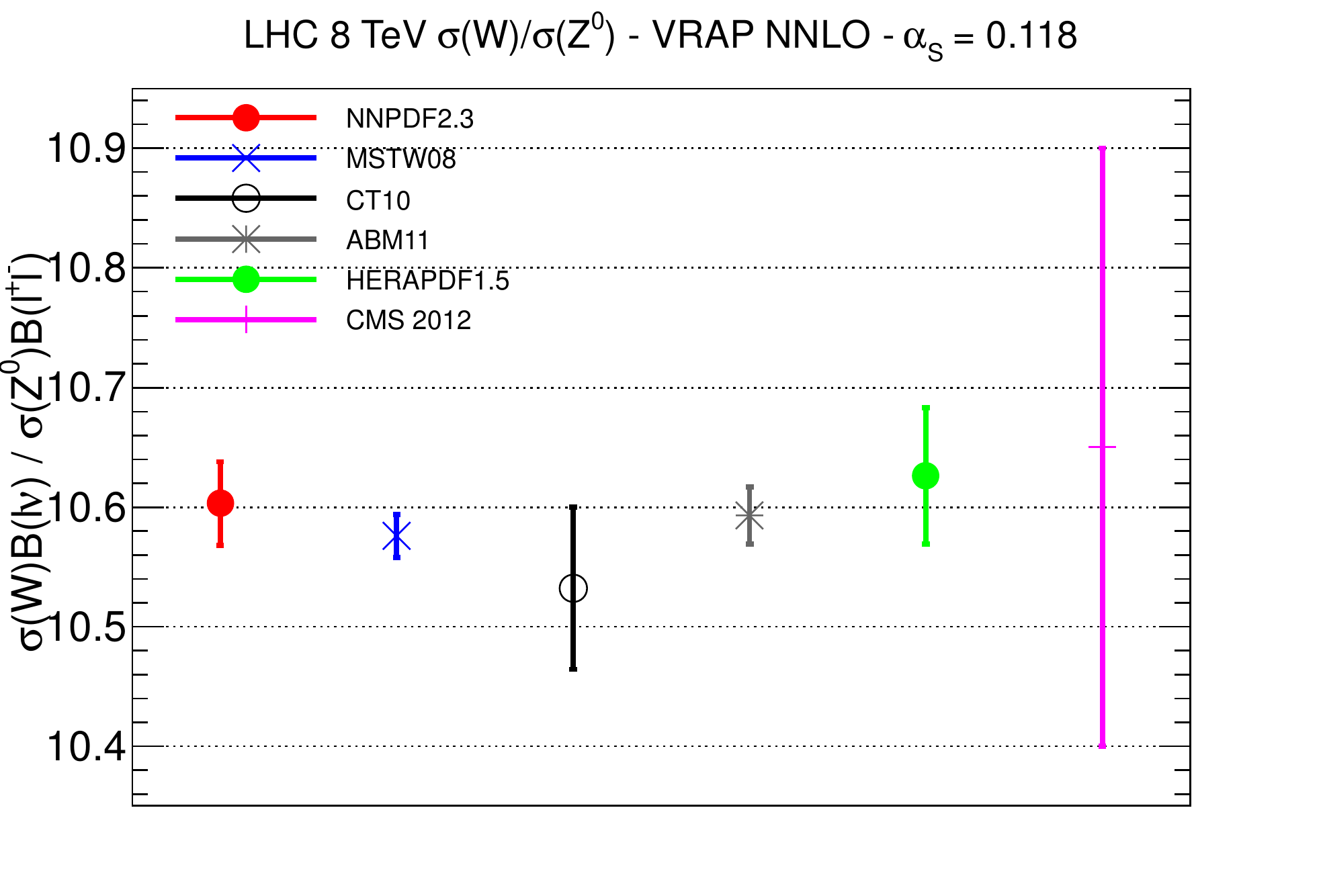}
  \caption{Same as Figure~\ref{fig:8tev-ewk} for the $W^+/W^-$ and 
  $W/Z$ ratios.}
\label{fig:8tev-ewk2}
\end{figure}

In Figure~\ref{fig:8tev-ewk} we show the inclusive cross-sections for
electroweak gauge boson production, $W^+,W^-$ and $Z$, at 8 TeV with
$\alpha_s(M_Z^2)=0.118$, meanwhile in Figure~\ref{fig:8tev-ewk2} we
present the $W^+/W^-$ and $W/Z$ cross-section ratios. In both cases,
the predictions have been computed at NNLO using the \texttt{Vrap}
code~\cite{Anastasiou:2003ds} with the central scale choice
$\mu_R=\mu_F=M_V$. The CMS measurements~\cite{cmsewk} are plotted
together with the theoretical predictions showing a good agreement
between NNPDF2.3, CT10, MSTW08 and HERAPDF1.5, as already observed in
the quark-antiquark luminosity in Figure~\ref{fig:PDFlumi-qq}. The
comparison with ABM11 leads to systematically higher cross-sections
which is also consistent with the larger luminosities.

\begin{figure}
\centering
\includegraphics[width=0.47\textwidth]{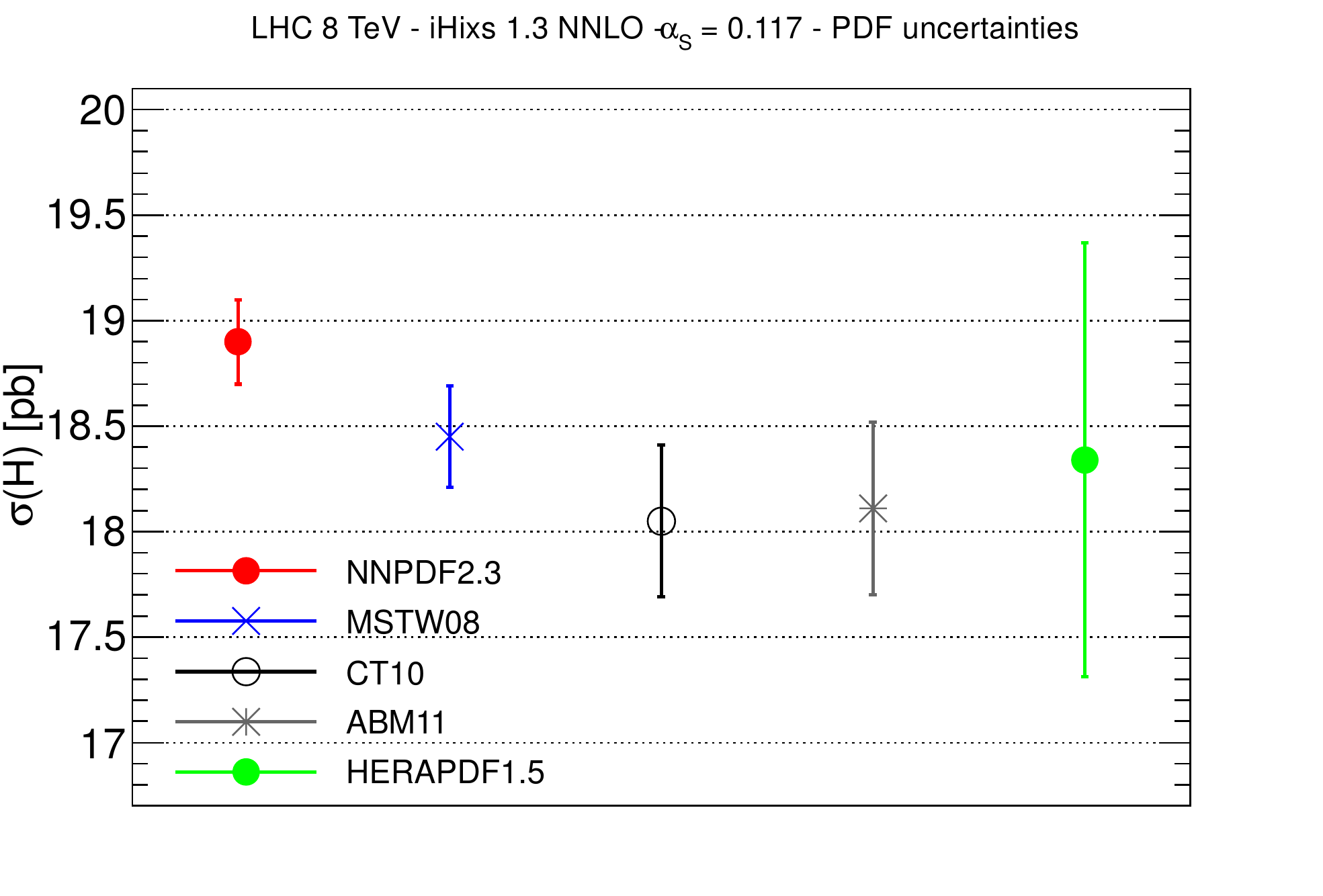}\quad
\includegraphics[width=0.47\textwidth]{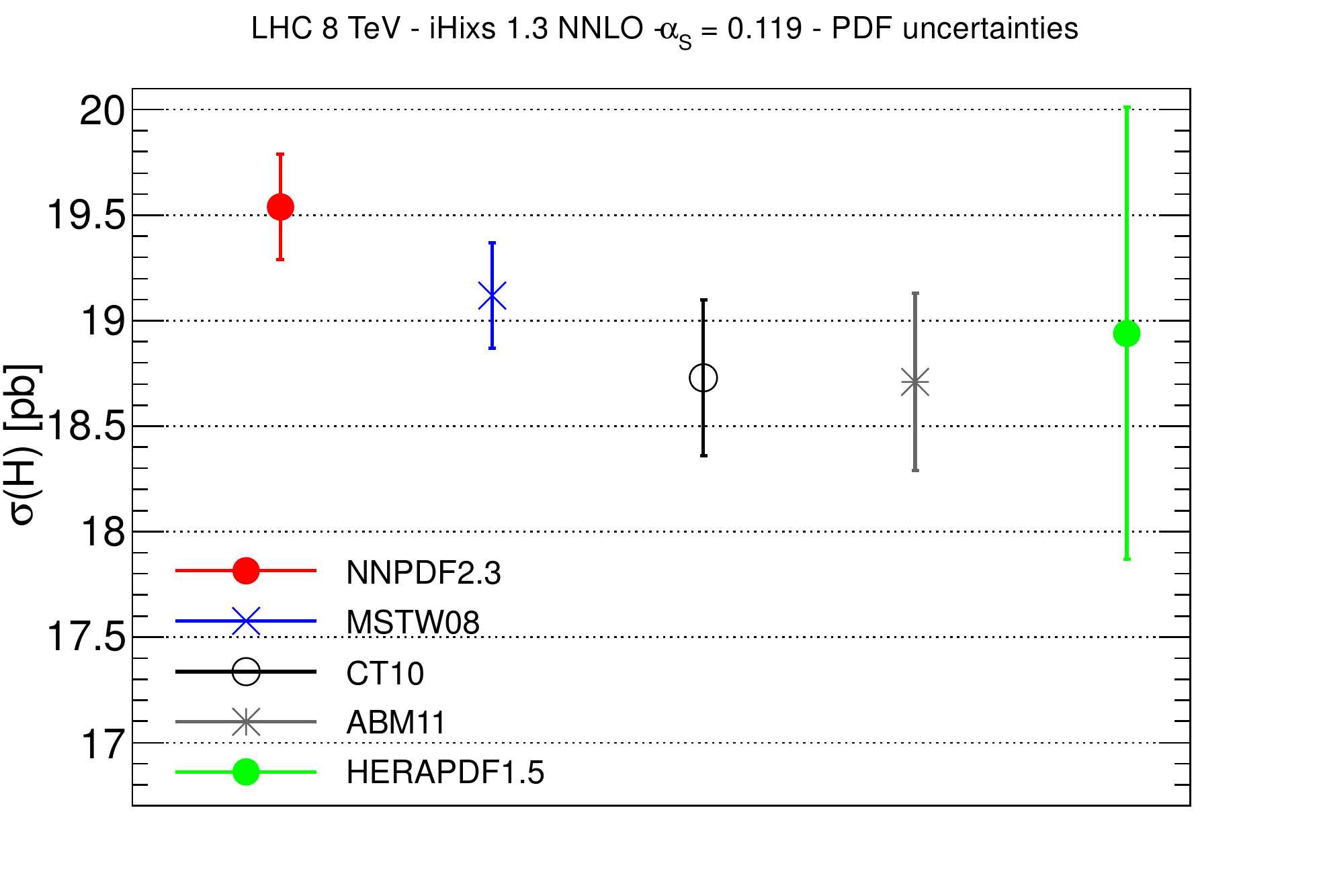}
\includegraphics[width=0.47\textwidth]{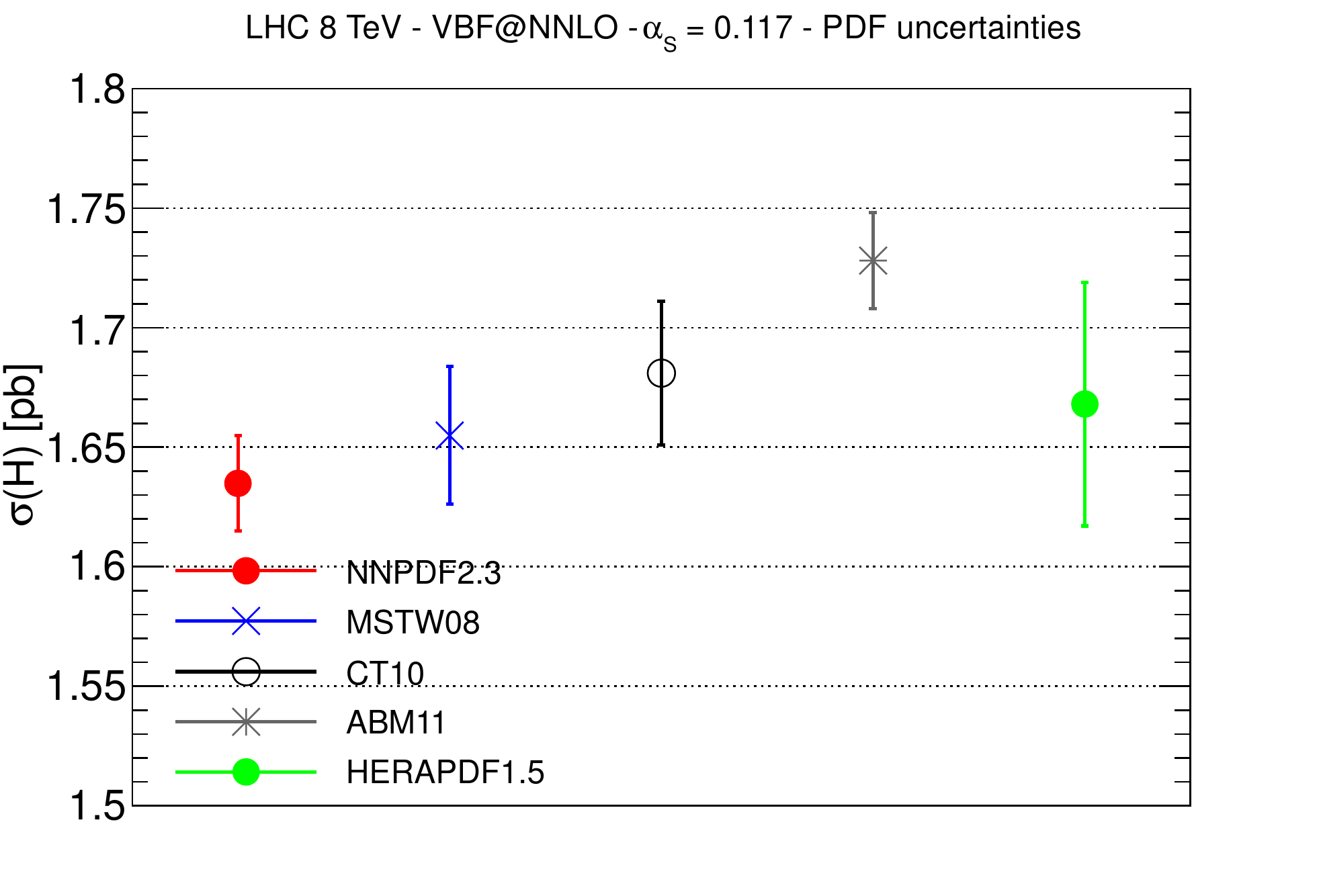}\quad
\includegraphics[width=0.47\textwidth]{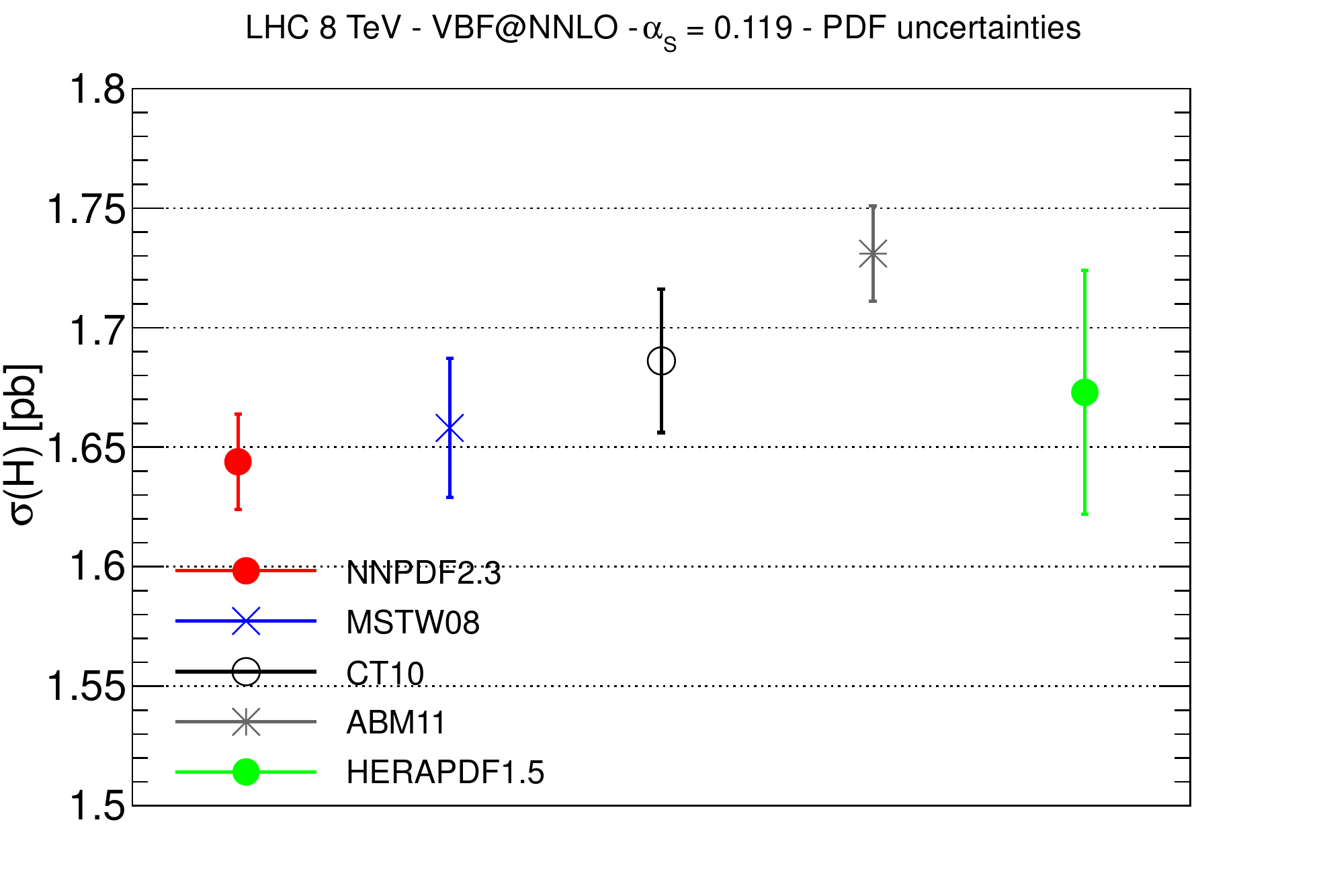}
\includegraphics[width=0.47\textwidth]{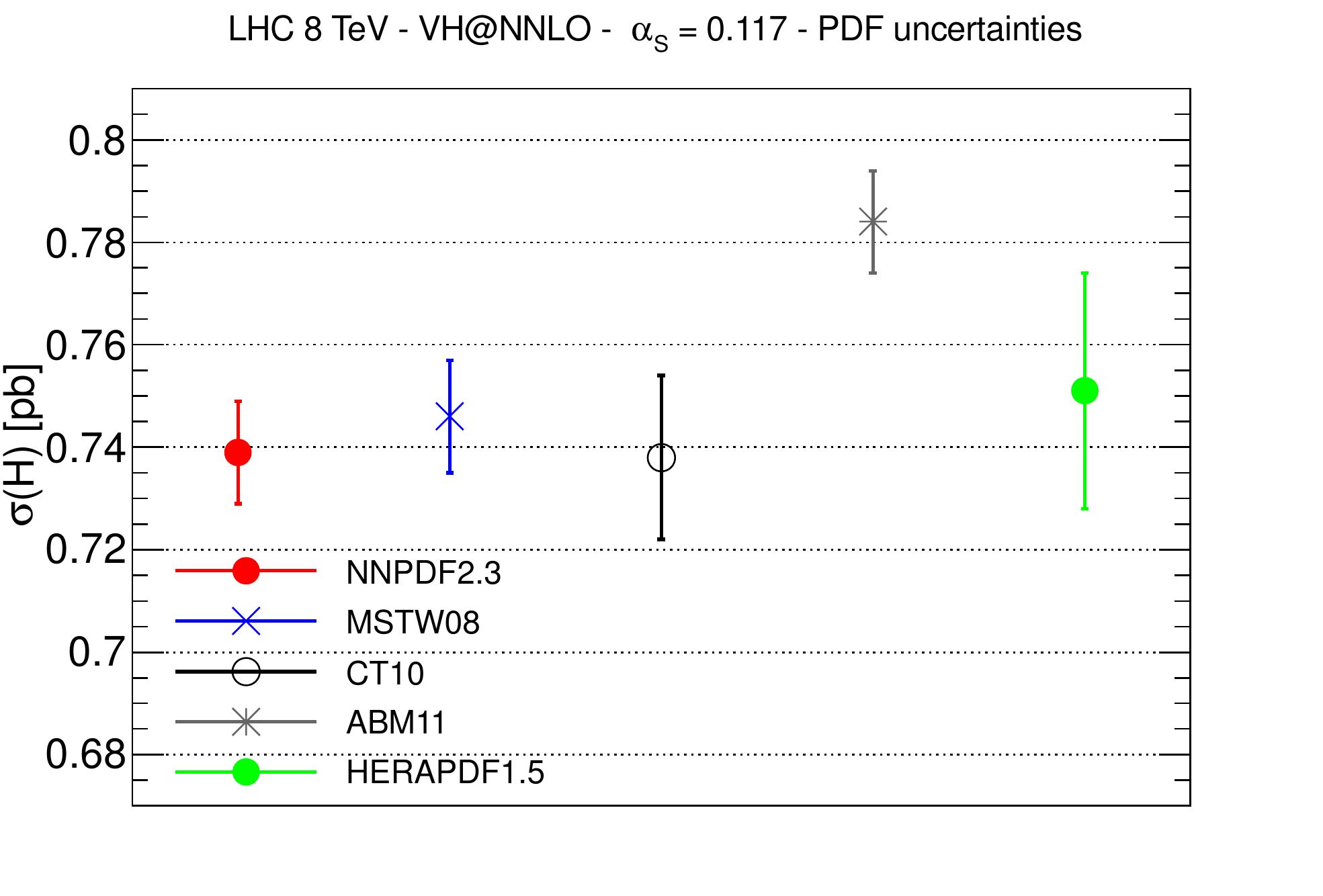}\quad
\includegraphics[width=0.47\textwidth]{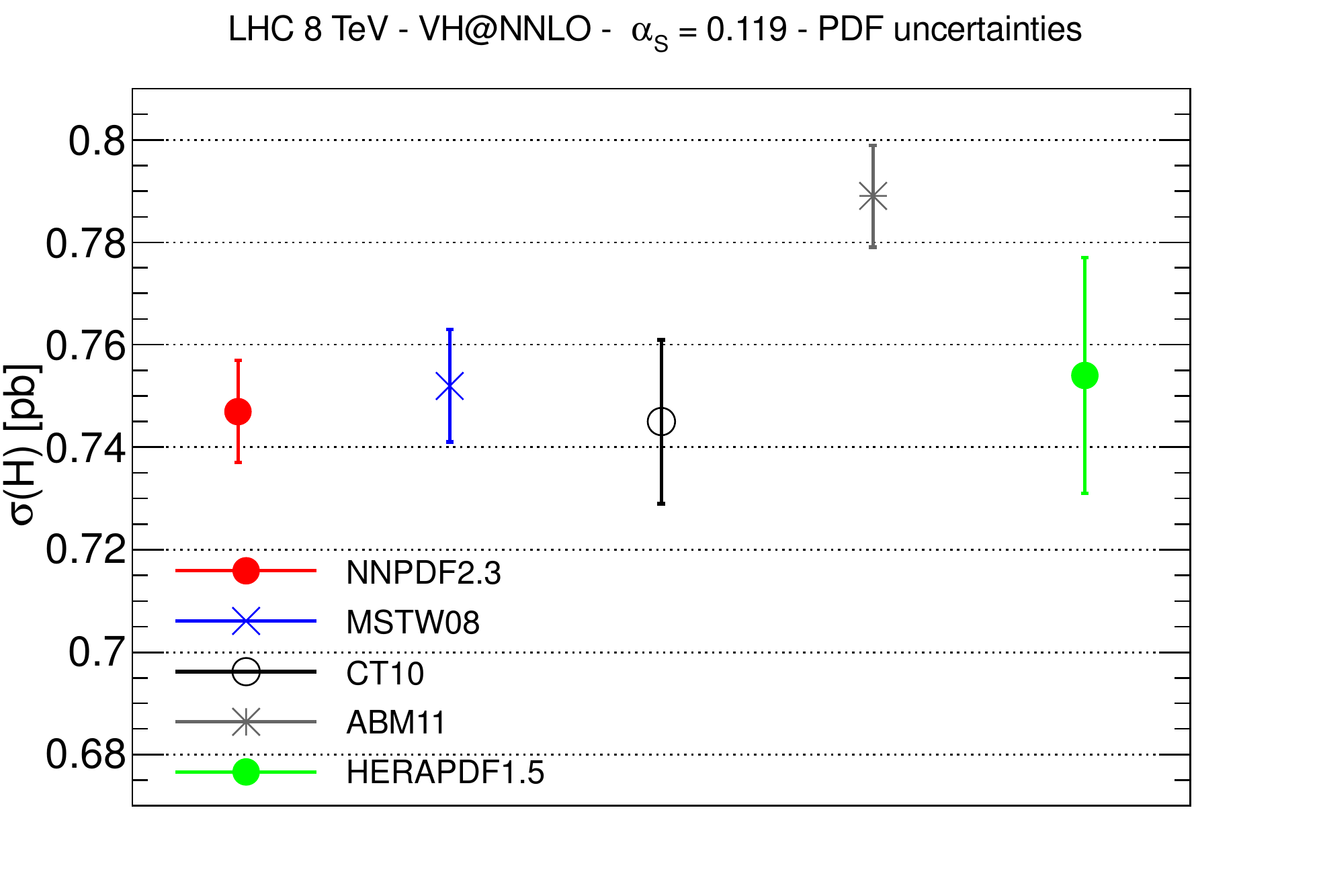}
\includegraphics[width=0.47\textwidth]{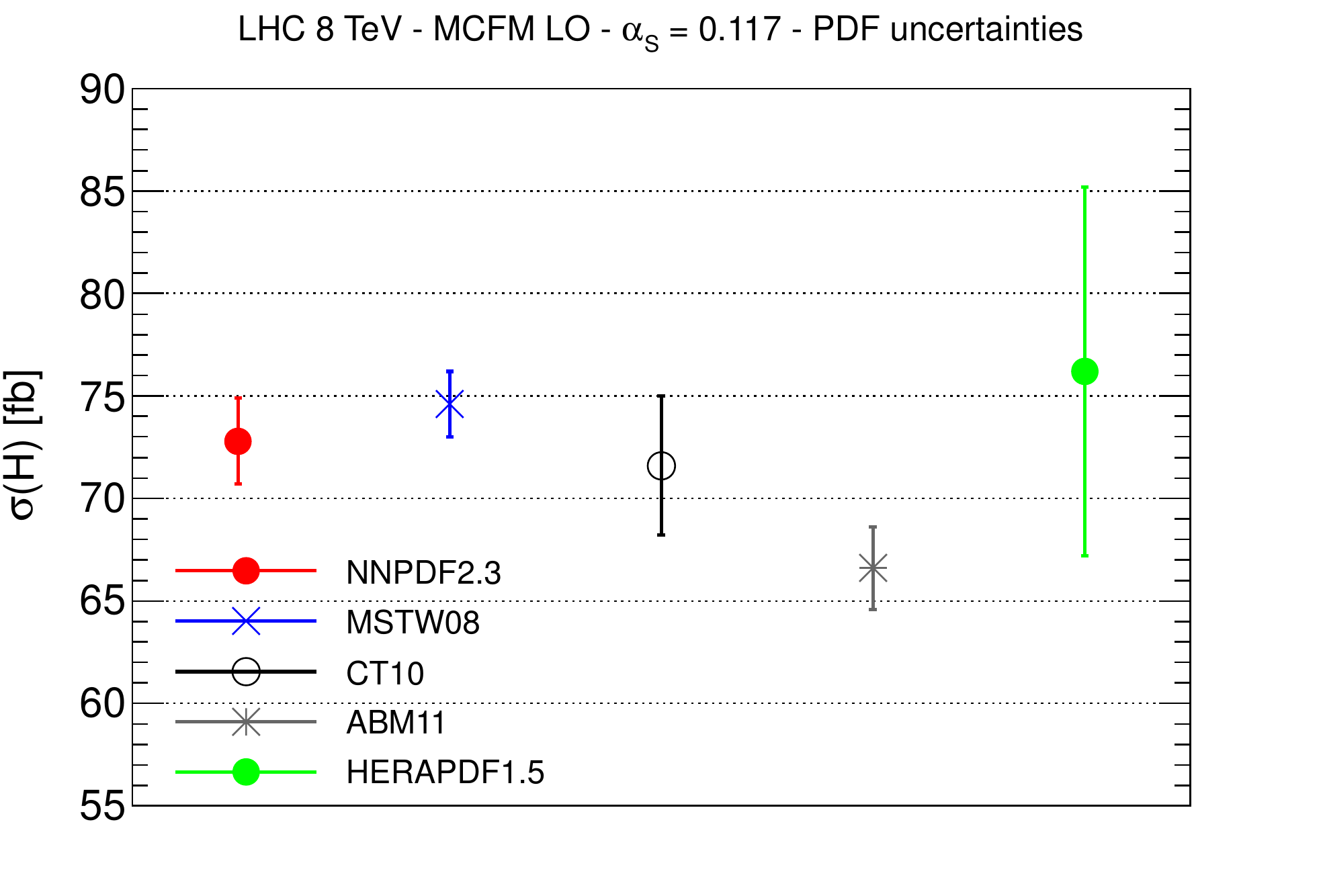}\quad
\includegraphics[width=0.47\textwidth]{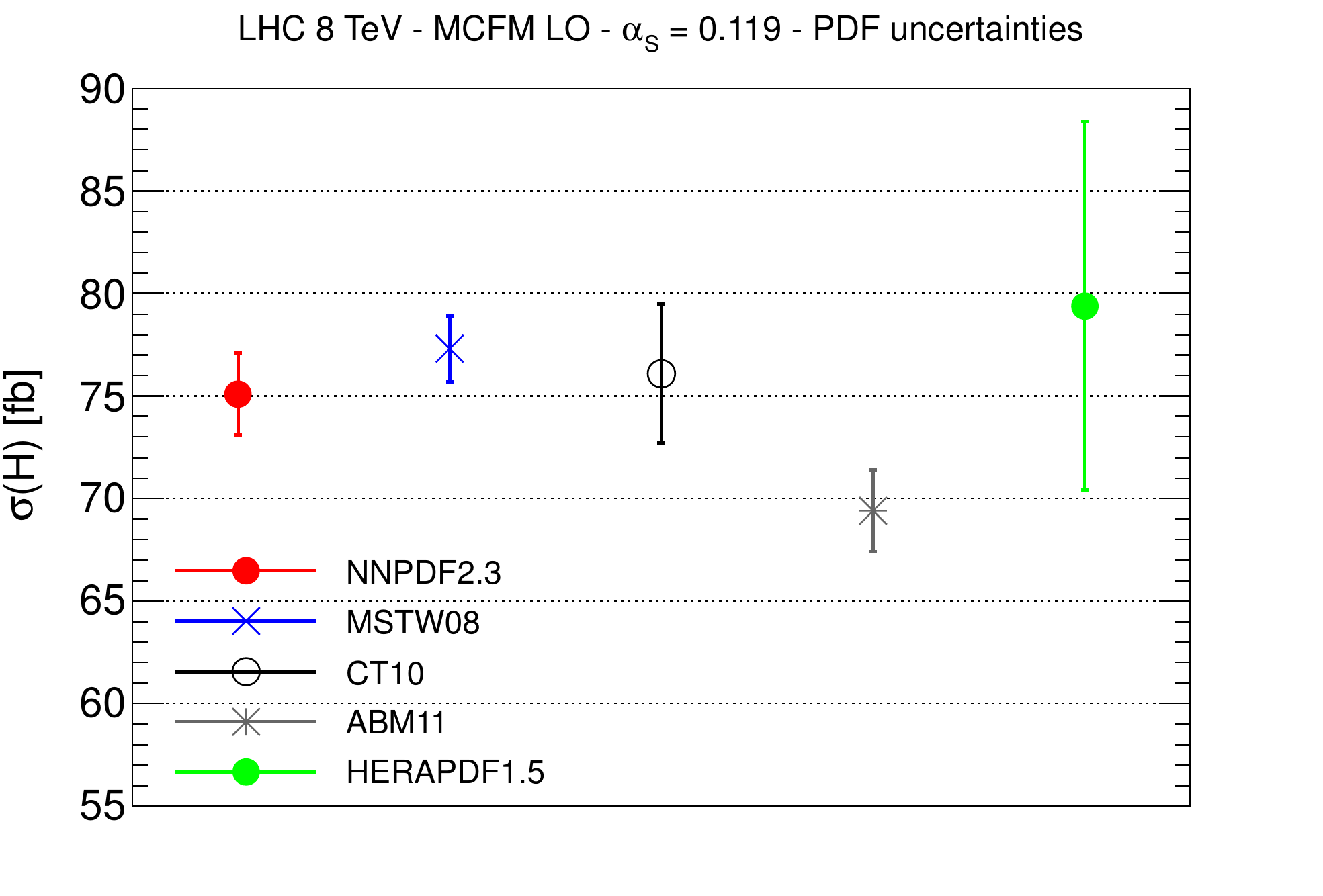}
\caption{Comparison of the predictions for the LHC Standard Model
  Higgs boson cross-sections at 8 TeV between various NNLO PDF
  sets. From top to bottom we show gluon fusion, vector boson fusion
  (VBF), associated production (with $W$), and associated production
  with a $t\bar{t}$ pair. The left hand plots show results for
  $\alpha_S(M_Z)=0.117$, while on the right we have
  $\alpha_S(M_Z)=0.119$.}
\label{fig:8tev-higgs}
\end{figure}

The Higgs boson production cross-section is another important process
for LHC phenomenology. In Figure~\ref{fig:8tev-higgs} we compare
several predictions for the LHC Standard Model Higgs boson
cross-section at 8 TeV between the NNLO PDF sets. The left hand plots
show results for $\alpha_s(M_Z^2) = 0.117$, while on the right
$\alpha_s(M_Z^2) = 0.119$. This choice is made in order to quantify
the impact of the $\alpha_s$ variation. In all cases the same value of
$\alpha_S$ is used consistently in both the PDFs and in the matrix
element calculation and we take $m_H=125$ GeV. The codes and setups
used for the formulation of these plots are listed below:

\begin{itemize}

\item The Higgs boson production cross-sections in the gluon fusion
  channel have been computed with the \texttt{iHixs}
  code~\cite{Anastasiou:2011pi} where the central scale has been taken
  to be $\mu_F=\mu_R=m_H$, consistent with the Higgs cross-section
  working group recommendations~\cite{Dittmaier:2011ti}.
  
\item The Higgs production in the Vector Boson Fusion (VBF) channel
  has been computed at NNLO with the \texttt{VBF@NNLO}
  code~\cite{Bolzoni:2010xr}, with $\mu_F=\mu_R=m_H$.
  
\item The Higgs production in association with $W$ and $Z$ bosons has
  been computed at NNLO with the \texttt{VH@NNLO}
  program~\cite{Brein:2003wg,Brein:2012ne}. Also here the scale choice
  is $\mu_F=\mu_R=m_H$.
  
\item The Higgs production in association with a top quark pair,
  $t\bar{t}H$, has been computed at LO with the \texttt{MCFM}
  program~\cite{Campbell:2002tg}, with the scale choice $\mu_F=\mu_R=2m_t+m_H$.

\end{itemize}

In summary we observe that the comparison between PDF sets is
unaffected by changes in $\alpha_s$. The variation of $\alpha_s$
produces shifts of the absolute value of the cross-section. ABM11 and
HERAPDF1.5 for the gluon fusion fall within the envelope composed by
the NNPDF2.3, CT10 and MSTW08 PDFs.  However, the HERAPDF1.5
uncertainty is bigger than this envelope. For VBF, $WH$ and
$t\bar{t}H$ production, there is a reasonable agreement between CT10,
MSTW08 and NNPDF2.3 both in central values and in the size of PDF
uncertainties. ABM11, on the other hand, leads to rather different
results, despite the fact that a common value of $\alpha_s$ is being
used. For quark-initiated processes, like VBF and $WH$, the ABM11
cross-section is higher than that of the other sets, specially for
$WH$ production.  For $t\bar{t}H$, which has its largest contribution
from gluon-initiated diagrams, the ABM11 cross-section is smaller.
The HERAPDF1.5 PDF uncertainties are distinctly larger compared to
three global fits, especially for $ggH$ and $t \bar tH$.  This can be
attributed to the poorly constrained large-$x$ gluon in the HERA-only
fits and, in the case of $t\bar tH$, less constrained sea quarks.

Finally, we conclude the comparisons with the inclusive top quark pair
production cross-section, which has been computed at
NNLO$_{\textrm{approx}}$+NNLL with the \texttt{top++}
code~\cite{Czakon:2011xx,Baernreuther:2012ws,Czakon:2012pz,Czakon:2012zr,Aliev:2010zk,Moch:2012mk}
as implemented in \texttt{v1.3}, which includes the complete NNLO
corrections to the $q\bar{q}\rightarrow t \bar{t}$, with the central
scale $\mu_F=\mu_R=m_t$.  The settings of the theoretical calculations
are the default ones in Ref.~\cite{Cacciari:2011hy}. In all
calculations we use $m_t=173.2$ GeV.

\begin{figure}
\centering
\includegraphics[width=0.47\textwidth]{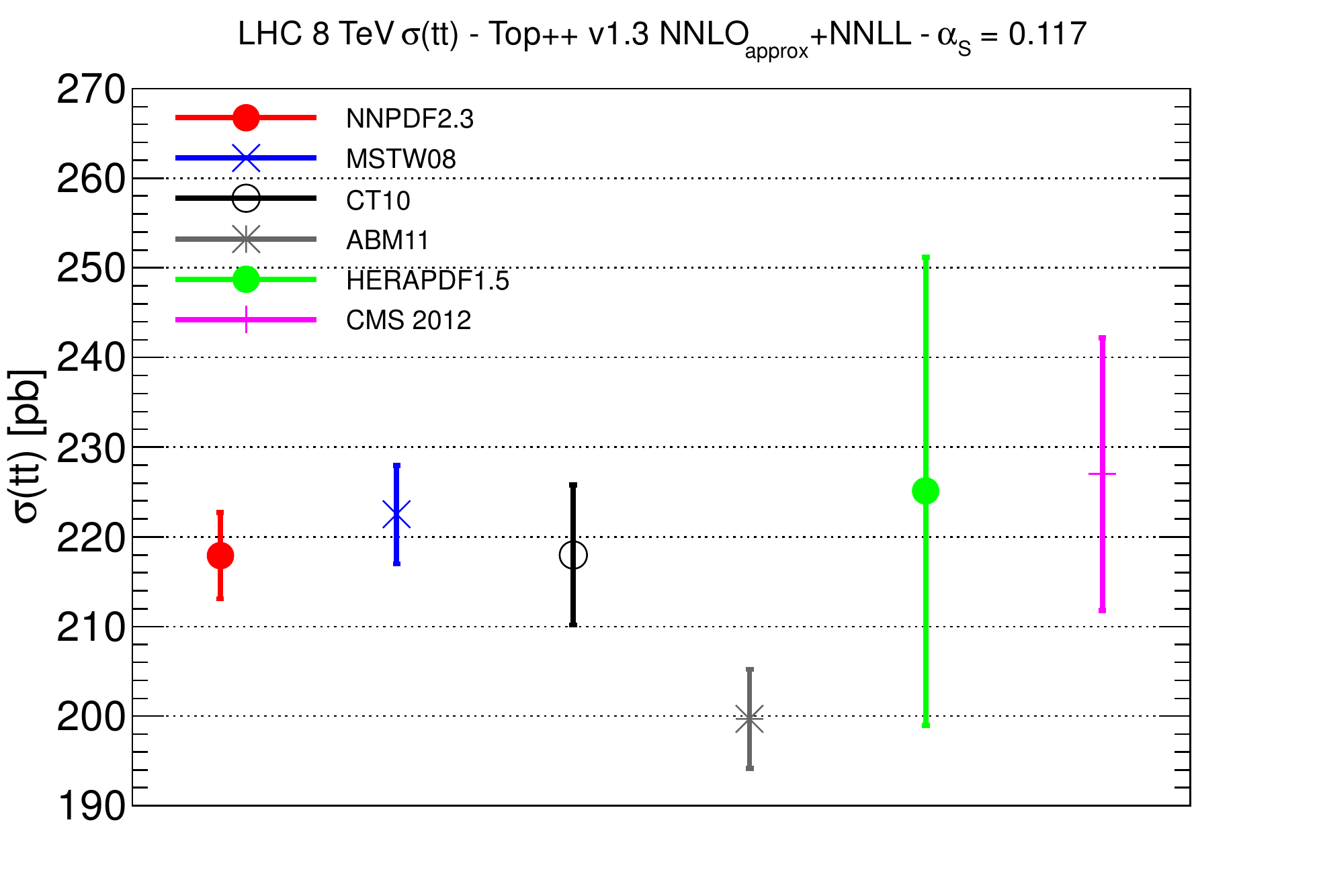}\quad
\includegraphics[width=0.47\textwidth]{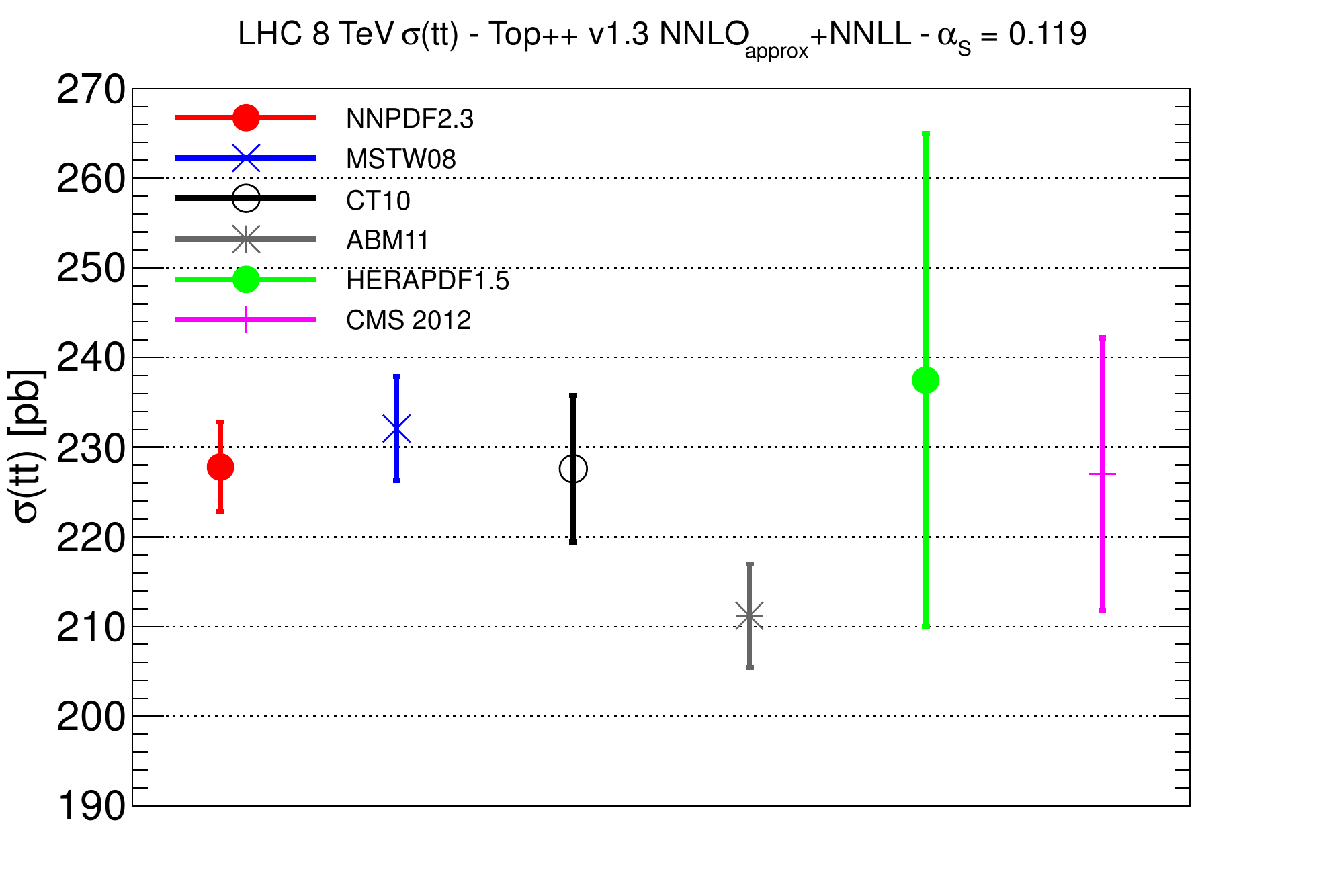}
\caption{Comparison of the predictions for the top quark pair
  production at LHC 8 TeV between various NNLO PDF sets. Left plot:
  results for $\alpha_S(M_Z)=0.117$.  Right plot: results for
  $\alpha_S(M_Z)=0.119$.  In both cases we also show the CMS 8 TeV
  measurement.}
\label{fig:8tev-ttbar}
\end{figure}

In Figure~\ref{fig:8tev-ttbar} we show the approximate NNLO top quark
pair production cross-section at 8 TeV for different NNLO PDF sets
with $\alpha_s(M_Z^2)=0.117$ and $\alpha_s(M_Z^2)=0.119$. Also in this
case theoretical predictions are compared to the recent CMS
measurements~\cite{cmstop} in terms of the average of the cross-section in the dilepton and lepton+jets final states. The $t\bar{t}$
total cross-section has some sensitivity to the value of $\alpha_s$.
This sensitivity has been recently used by CMS to provide the first
ever determination of $\alpha_s$ from top cross-sections~\cite{cmstopas}.  For the $t\bar{t}$ cross-section, we see a
reasonable agreement between NNPDF2.3, CT10 and MSTW08, while ABM11 is
somewhat lower. The HERAPDF1.5 central value is in good agreement with
the global fits but, as usual, the PDF uncertainties are larger.
\chapter{QED corrections to PDF evolution}
\label{sec:chap2}

In this chapter we introduce the theoretical framework and the
numerical implementation of QED corrections to parton evolution
equations. We organize this chapter as follows: in
Sect.~\ref{sec:apfelintro} we motivate the inclusion of QED
corrections to DGLAP and we introduce the \texttt{APFEL} library which
was developed specifically for this project. Then, in
Sect.~\ref{sec:apfelevolution}, the combined DGLAP equations are
presented explicitly and the solution strategy is discussed in
detail. The numerical techniques used in \texttt{APFEL} are summarized
in Sect.~\ref{sec:apfelnumeric}, validation and benchmarking results
against other public codes are presented in
Sect.~\ref{sec:apfelvalidation}. Finally, in Sect.~\ref{sec:apfelweb}
we conclude the discussion with the description of \texttt{APFEL Web},
a spin-off of the \texttt{APFEL} library. The DGLAP solution developed
in this chapter have been applied to the photon PDF determination in
Chapter~\ref{sec:chap4}.

\section{Introduction to APFEL}
\label{sec:apfelintro}

Following the discussion started in the introduction of this thesis,
we recall that during the last years a great effort has been made for
the achievement of PDFs determined using NLO and NNLO QCD
theory~\cite{Forte:2013wc,Ball:2012wy,DeRoeck:2011na}. However, at
present, the level of accuracy in theoretical predictions and
experimental uncertainties is such that QED and electroweak (EW)
corrections are required for the precision physics at the LHC.

There are several examples of predictions for hadron collider
processes with QED and EW corrections, which have been computed in the
last years. A full review of such processes is presented in
Ref.~\cite{Mishra:2013una}, from which we can mention:
\begin{itemize}
\item the inclusive $W$ and $Z$
  production~\cite{Baur:1998kt,Zykunov:2001mn,Dittmaier:2001ay,Baur:2001ze,Baur:2004ig,Arbuzov:2007db,Arbuzov:2005dd,Brensing:2007qm,Balossini:2009sa,CarloniCalame:2007cd,Dittmaier:2009cr}
  
\item the $W$ and $Z$ boson production in association with
  jets~\cite{Denner:2009gj,Denner:2011vu,Denner:2012ts}, diboson
  production~\cite{Baglio:2013toa,Bierweiler:2012kw,Luszczak:2013ata},
  dijet production~\cite{Moretti:2005ut,Dittmaier:2012kx} and top
  quark pair
  production~\cite{Bernreuther:2005is,Kuhn:2005it,Hollik:2007sw,Hollik:2011ps,Kuhn:2013zoa}
\end{itemize}

The combination of QCD and QED calculations at hadron colliders
requires PDFs with QCD$\otimes$QED DGLAP evolution
equations~\cite{DeRujula:1979jj,Kripfganz:1988bd,Blumlein:1989gk}. Many
studies have been performed during the last 20 years about the
numerical solution and optimization of the QCD DGLAP evolution
equations, many of which have become public
tools~\cite{Salam:2008qg,Cafarella:2008du,Botje:2010ay,Ratcliffe:2000kp,Schoeffel:1998tz,Pascaud:2001bi,pegasus,Kosower:1997hg}. On
the other hand, much less effort has been invested to the solutions of
QCD$\otimes$QED
DGLAP~\cite{Spiesberger:1994dm,Roth:2004ti,Martin:2004dh}, in
particular, to the best of our knowledge, before the release of this
work, the only public codes which offered the possibility to obtain an
estimation of such corrections were
\begin{itemize}

\item the \texttt{partonevolution}~\cite{Weinzierl:2002mv,Roth:2004ti}
  library, which is limited to NLO QCD corrections and it does not
  contain a modern interface the
  \texttt{LHAPDF}~\cite{Buckley:2014ana} library (used for accessing
  all published PDF sets), and in addition it does not allow to
  explore different possibilities for the combination of the QCD and
  QED evolution equations.

\item the MRST2004QED set of PDFs~\cite{Martin:2004dh}, which until
  the set of PDFs presented in this thesis, was the only set which
  included QED corrections, where the photon PDF is based on model
  assumptions. However, this set of PDFs delivers precomputed
  evolution encoded in the grid, which denies the possibility to
  perform systematic studies of the DGLAP equation for different
  initial conditions.
\end{itemize}

Therefore, in this work we present
\texttt{APFEL}~\cite{Bertone:2013vaa}, which stands for \textit{A
  Parton distribution Function Evolution Library}. \texttt{APFEL}'s
goal is to fill the need of a public tool, accurate and flexible that
can be used to perform PDF evolution up to NNLO in QCD and LO in QED,
both in the fixed-flavor-number (FFN) and in the
variable-flavor-number (VFN) schemes, and using either pole or
$\overline{\textrm{MS}}$ heavy quark masses. \texttt{APFEL} is
designed to meet the needs of PDF fits, providing large control of
evolution parameters like the heavy quark thresholds, the coupling
running solution, and many others.

\texttt{APFEL} is implemented in \texttt{Fortran77} with wrappers in
\texttt{C++} and \texttt{Python}. It is publicly available from the
\texttt{HepForge}
website\footnote{\url{http://apfel.hepforge.org/}}. \texttt{APFEL} is
part of the family of codes which solves the DGLAP equations using
$x$-space methods, which typically use a representation of the PDFs on
a grid in $x$ and $\mu_F^2$ together with higher-order interpolation
techniques for the solution of the intergro-differential
equations~\cite{Salam:2008qg,Botje:2010ay,Schoeffel:1998tz,Pascaud:2001bi,Ratcliffe:2000kp}.

This methodology is widely used by other pure QCD evolution libraries
such as: \texttt{HOPPET}~\cite{Salam:2008qg} and
\texttt{QCDNUM}~\cite{Botje:2010ay}. Other tools, like the best-know
\texttt{PEGASUS}~\cite{pegasus}, solve the DGLAP equations in
$N$-space, by transforming the evolution equations into Mellin space
(see Sect.~\ref{sec:dglapintro}) which are then analytically solved
and inverted back to $x$-space using complex-variable
methods~\cite{pegasus,Cafarella:2008du,Weinzierl:2002mv,Roth:2004ti,Kosower:1997hg}.
The main drawback of the $N$-space methods, however, is the fact that
they require the analytical Mellin transform of the initial PDFs which
is possible only for some very specific functional forms, which is
unlikely the case for the PDF sets in \texttt{LHAPDF} which are
delivered in function of the $x$ variable.
A third approach is provided by the hybrid method adopted in the
\texttt{FastKernel} methodology, the internal code used in the NNPDF
fits~\cite{DelDebbio:2007ee,Ball:2008by}, where DGLAP equations are
solved in Mellin space and then used to determine the $x$-space
evolution operators, which are convoluted with the $x$-space PDFs to
perform the evolution.

\section{DGLAP evolution with QED corrections}
\label{sec:apfelevolution}

In this section we present the strategy that \texttt{APFEL} adopts in
order to perform the DGLAP evolution of PDFs when QCD and QED effects
are taken into account.

First, we present the QED evolution equations, and then we show how to
define an evolution basis which solves the system. In this work we
suggest two different approaches for the solution of the combined
system: the \emph{unified} solution and the so-called \emph{coupled}
solution where QCD and QED equations are solved separately and then
combined. We show that the coupled approach provides a good
approximation of the unified solution.

In next sections we limit the discussion to the QED sector, however
more details about QCD corrections to DGLAP evolution equations up to
NNLO are available in
Refs.~\cite{ap,gl,dok,gNLOa,gNLOb,gNLOc,gNLOe,gNLOf,gNNLOa,gNNLOb,Buza:1995ie},
and the structure of their solutions has also been discussed in great
detail in the literature, see for example
Refs.~\cite{Salam:2008qg,pegasus,Forte:2013wc}.

\subsection{Solving the QED evolution equations}
\label{SolvingQEDEquation}

The implementation of the QED corrections to the DGLAP evolution
equations leads to the inclusion of additional terms which contain QED
splitting
functions~\cite{DeRujula:1979jj,Kripfganz:1988bd,Blumlein:1989gk},
proportional to the QED coupling $\alpha$, convoluted with the PDFs.
There are several possibilities to solve the combined QCD$\otimes$QED
DGLAP evolution equations, and, as opposed to previous works,
\texttt{APFEL} adopts: 
\begin{itemize}

\item the \emph{coupled} solution: a fully factorized approach where
  the QCD and the QED factorization procedures can be regarded as two
  independent steps that lead to two independent factorization scales
  on which all PDFs depend.

\item the \emph{unified} solution: the procedure where the QCD and QED
  sectors are solved by an unique system of equations.

\end{itemize}

In the next paragraphs and in Sect.~\ref{sec:coupledsol} we describe
the coupled approach, meanwhile we devote Sect.~\ref{sec:exactsol} for
the unified method. For simplicity, in both discussions we assume that
no heavy quark threshold is crossed during the DGLAP evolution
\textit{i.e.}~it is valid only when PDF evolution is performed in the
FFN scheme, however the generalization to the VFN scheme is in
\texttt{APFEL}, the documented in Sect. 2.3 of
Ref.~\cite{Bertone:2013vaa}.

In the case where QED corrections are included up to
$\mathcal{O}(\alpha)$ and the mixed subleading terms
$\mathcal{O}(\alpha\alpha_s)$ are neglected, the QCD evolution with
respect to $\mu$ and the QED evolution with respect to $\nu$ will be
given by two fully decoupled equations:
\begin{equation}\label{DGLAPequationsQCD_QED}
\begin{array}{rcl}
\displaystyle \mu^{2}\frac{\partial}{\partial \mu^{2}}{\mathbf q}(x,\mu,\nu) &=&
\displaystyle {\mathbf P}^\textrm{QCD} (x,\alpha_s(\mu))\otimes {\mathbf
q}(x,\mu,\nu)\,,\\
\\
\displaystyle \nu^{2}\frac{\partial}{\partial \nu^{2}}{\mathbf
 q}(x,\mu,\nu) &=&
\displaystyle {\mathbf P}^\textrm{QED} (x,\alpha(\nu))\otimes {\mathbf
q}(x,\mu,\nu)\,,
\end{array}
\end{equation}
where ${\mathbf P}^\textrm{QCD}$ and ${\mathbf P}^\textrm{QED}$ are respectively 
the QCD and
QED matrices of splitting functions and 
${\mathbf q}(x,\mu,\nu)$ is a vector
containing all the parton distribution functions.
Let us recall that in the presence of QED corrections, the photon PDF
$\gamma(x,\mu,\nu)$ should also be included in ${\mathbf
  q}(x,\mu,\nu)$. The independent solutions of the differential
equations in Eq.~(\ref{DGLAPequationsQCD_QED}), irrespective of the
numerical technique used, will give as a result two different
evolution operators: ${\bm \Gamma}^\textrm{QCD}$, that evolves the
array $\mathbf q$ in $\mu$ while keeping $\nu$ constant, and ${\bm
  \Gamma}^\textrm{QED}$, that evolves $\mathbf q$ in $\nu$ while
keeping $\mu$ constant.
If the QCD evolution takes place between $\mu_0$ and $\mu_1$ and the
QED evolution between $\nu_0$ and $\nu_1$, we will have that:
\begin{equation}\label{DecoupledDGLAPequations}
\begin{array}{l}
  \displaystyle {\mathbf q}(x,\mu_1,\nu)={\bm \Gamma}^\textrm{QCD}(x|\mu_1,\mu_0)\otimes{\mathbf q}(x,\mu_0,\nu)\,,\\
  \\
  \displaystyle {\mathbf q}(x,\mu,\nu_1)={\bm \Gamma}^\textrm{QED}(x|\nu_1,\nu_0)\otimes{\mathbf q}(x,\mu,\nu_0)\,.\\
\end{array}
\end{equation}
Once the QCD and QED evolution operators in Eq.~(\ref{DecoupledDGLAPequations}) have been
calculated, one can combine them to obtain a coupled
evolution operator ${\bm \Gamma}^{\textrm{QCD} \otimes \textrm{QED}}$ that evolves PDFs
both in the QCD and in the QED scales, that is:
\begin{equation} {\mathbf q}(x,\mu_1,\nu_1)={\bm \Gamma}^{\textrm{QCD}
    \otimes \textrm{QED}}(x|\mu_1,\mu_0;\nu_1,\nu_0)\otimes{\mathbf
    q}(x,\mu_0,\nu_0)\,.
\end{equation}

Before discussing the derivation of the combined evolution operator
${\bm \Gamma}^{\textrm{QCD} \otimes \textrm{QED}}$, we present the
strategy used in \texttt{APFEL} to solve the QED DGLAP equations in
Eq.~(\ref{DGLAPequationsQCD_QED}).
At leading order, the QED equations for the evolution of the quark and photon
PDFs, dropping for simplicity the dependence on the QCD factorization scale $\mu$, read:
\begin{equation}\label{QED_DGLAP}
\begin{array}{rcl}
\displaystyle \nu^{2}\frac{\partial}{\partial \nu^{2}}\gamma(x,\nu)
&=& \displaystyle \frac{\alpha(\nu)}{4\pi} \left[\left(\sum_{i}N_c e_{i}^{2}\right)
P_{\gamma\gamma}^{(0)}(x)\otimes\gamma(x,\nu)+\sum_ie_{i}^{2}P_{\gamma
  q}^{(0)}(x)\otimes (q_{i}+\bar{q}_{i})(x,\nu)\right]\,,\\
\\
\displaystyle \nu^{2}\frac{\partial}{\partial
   \nu^{2}} q_{i}(x,\nu)&=&\displaystyle \frac{\alpha(\nu)}{4\pi}
 \left[ N_c e_{i}^{2} P_{q\gamma}^{(0)}(x)\otimes \gamma(x,\nu)+e_{i}^{2}
 P_{qq}^{(0)}(x)\otimes q_i(x,\nu)\right]\,,\\
\\
\displaystyle \nu^{2}\frac{\partial}{\partial
   \nu^{2}} \bar{q}_{i}(x,\nu)&=&\displaystyle \frac{\alpha(\nu)}{4\pi}
 \left[ N_c e_{i}^{2} P_{q\gamma}^{(0)}(x)\otimes \gamma(x,\nu)+e_{i}^{2}
 P_{qq}^{(0)}(x)\otimes \bar{q}_i(x,\nu)\right]\,, 
\end{array}
\end{equation}
where $\gamma(x,\nu)$, $q_i(x,\nu)$ and $\bar{q}_i(x,\nu)$ are
respectively the PDFs of the photon, the $i$-th quark and the $i$-th
antiquark, $e_i$ the quark electric charge, $N_c=3$ the number of
colors and $\alpha(\nu)$ the running fine structure constant. In this
work we neglect the impact of lepton PDFs.
Note that at this order  the gluon PDF does not enter the
QED evolution equations.
The leading-order QED splitting functions $P^{(0)}_{ij}(x)$ are given by:
\begin{equation}\label{QEDLOsplittingFunctions}
\begin{array}{rcl}
\displaystyle P_{q\gamma}^{(0)}(x) &=& \displaystyle 2\left[x^2+(1-x)^2\right],\\
\\
\displaystyle P_{\gamma q}^{(0)}(x) &=& \displaystyle
2\left[\frac{1+(1-x)^2}{x}\right],\\
\\
\displaystyle P_{\gamma\gamma}^{(0)}(x) &=& \displaystyle
-\frac{4}{3}\delta(1-x),\\
\\
\displaystyle P_{qq}^{(0)}(x) &=& \displaystyle
2\frac{1+x^2}{(1-x)_+}+3\delta(1-x)\,.
\end{array}
\end{equation}
The index $i$ in Eq.~(\ref{QED_DGLAP}) runs over the active quark flavors
at a given scale $\nu$. 

It should be noted that, in the presence of QED effects, the usual
momentum sum rule is modified to take into account the contribution
coming from the photon PDF. Therefore, provided that
the input PDFs respect the momentum sum rule, the QED evolution should
satisfy the equality:
 \begin{equation}
 \int_{0}^{1}dx~x\left\{ \underset{i}{\sum}(q_{i}+\bar{q}_{i})(x,\mu,\nu)+g(x,\mu,\nu)+\gamma(x,\mu,\nu)\right\} =1 \ ,
 \label{eq:momsr}
 \end{equation}
for any value of the scales $\mu$ and $\nu$. An important
test of the numerical implementation of DGLAP evolution in the
presence of QED effects is to check that Eq.~(\ref{eq:momsr}) indeed
holds at all scales. 

As in the case of QCD, an important practical issue 
that needs to be addressed
when solving the QED DGLAP evolution
equations is the choice of the PDF basis. 
The use of the flavor basis ${\mathbf q} =\{\gamma,u,\bar{u},d,\bar{d},...\}$
requires the solution of a system of thirteen coupled equations which in turns 
leads to a cumbersome numerical implementation.  
This problem can be overcome by choosing a suitable PDF
basis, the evolution basis, that maximally diagonalizes the
QED splitting function matrix.
Note that this optimized basis will be different from that used in
QCD, due to the presence of the electric charges $e_i$ in
Eq.~(\ref{QED_DGLAP}) that are different between up- and down-type
quarks. This difference between up- and down-type quarks, in the
presence of QED effects, is also responsible for the dynamical
generation of isospin symmetry breaking between proton and neutron
PDFs.

\subsection{Basis for the coupled QCD$\otimes$QED solution}
\label{sec:coupledsol}
For the coupled approach we adopt a PDF basis for the QED evolution
which was originally suggested in Ref.~\cite{Roth:2004ti}, defined by
the following singlet and non-singlet PDF combinations:
\begin{equation}\label{QEDEvolutionBasis}
\begin{array}{rcl}
\text{Singlet}: &\quad& \mathbf{q}^\textrm{SG}=
\begin{pmatrix}
\gamma\\
\Sigma\equiv u^{+}+c^{+}+t^{+}+d^{+}+s^{+}+b^{+}\\
\Delta_{\Sigma} \equiv u^{+}+c^{+}+t^{+}-d^{+}-s^{+}-b^{+}
\end{pmatrix}\,,\\
\\
\text{ Non-Singlet}: &\quad& q_i^\textrm{NS}=\left\{\begin{array}{c}
\Delta_{uc}\equiv u^{+}-c^{+},\\
\Delta_{ds} \equiv  d^{+}-s^{+},\\
\Delta_{sb} \equiv s^{+}-b^{+},\\
\Delta_{ct} \equiv c^{+}-t^{+},\\
u^{-},\\
d^{-},\\
s^{-},\\
c^{-},\\
b^{-},\\
t^{-}
\end{array}\right\}\,,\quad i = 1,\dots,10\,,
\end{array}
\end{equation}
where $q^{\pm} \equiv q \pm \overline{q}$.
Similarly to the QCD notation introduced in
Sect.~\ref{sec:dglapintro}, the singlet distributions are those that
couple to the photon PDF $\gamma(x,\nu)$, while the non-singlet
distributions evolve multiplicatively and do not couple to the photon.

With the choice of basis of Eq.~(\ref{QEDEvolutionBasis}), the
original thirteen-by-thirteen system of coupled equations in the
flavor basis reduce to a three-by-three system of coupled equations
and ten additional decoupled differential equations.
Expressing the QED DGLAP equations given in Eq.~(\ref{QED_DGLAP}) in
terms of this evolution basis, we find that the singlet PDFs evolve as
follows:
\begin{equation}\label{SingletEvolution}
  \nu^{2}\frac{\partial}{\partial \nu^{2}}
  \begin{pmatrix}
    \gamma\\
    \Sigma\\
    \Delta_{\Sigma}
  \end{pmatrix}=\frac{\alpha(\nu)}{4\pi}
  \begin{pmatrix}
    e_{\Sigma}^{2}P_{\gamma\gamma}^{(0)} & \eta^{+}P_{\gamma q}^{(0)} & \eta^{-}P_{\gamma q}^{(0)}\\
    \theta^{-}P_{q\gamma}^{(0)} & \eta^{+}P_{qq}^{(0)} & \eta^{-}P_{qq}^{(0)}\\
    \theta^{+}P_{q\gamma}^{(0)} & \eta^{-}P_{qq}^{(0)} & \eta^{+}P_{qq}^{(0)}
  \end{pmatrix}\otimes
  \begin{pmatrix}
    \gamma\\
    \Sigma\\
    \Delta_{\Sigma}
  \end{pmatrix}\,,
\end{equation}
where, using the fact that $e_{u}^{2}=e_{c}^{2}=e_{t}^{2}$ and
$e_{d}^{2}=e_{s}^{2}=e_{b}^{2}$, we have defined:
\begin{equation}
  \begin{array}{rcl}
    e_{\Sigma}^{2}& \equiv &\displaystyle
    N_c(n_{u}e_{u}^{2}+n_{d}e_{d}^{2})\,,\\
    \\
    \eta^{\pm} & \equiv & \displaystyle \frac{1}{2}\left(e_{u}^{2}\pm
      e_{d}^{2}\right)\,,\\
    \\
    \theta^{\pm} & \equiv & \displaystyle
    2N_{c}n_{f}\left[\left(\frac{n_{u}-n_{d}}{n_{f}}\right)\eta^{\pm}+\eta^{\mp}\right]\,,
  \end{array}
\end{equation}
where $n_{u}$ and $n_{d}$ are the number of up-
and down-type active quark flavors, respectively, and
$n_{f}=n_{u}+n_{d}$.
The non-singlet PDFs, instead, obey the multiplicative evolution equation:
\begin{equation}\label{NonSingletEvolution}
  \nu^{2}\frac{\partial}{\partial \nu^{2}} q_i^\textrm{NS}(x,\nu) = e_i^2 P_{qq}^{(0)}(x)
  \otimes q_i^\textrm{NS}(x,\nu)\,,
\end{equation}
where the electric charge $e_i^2 = e_u^2$ for the up-type
distributions $q_i^\textrm{NS} = \Delta_{uc}, \Delta_{ct},
u^-,c^-,t^-$ while $e_i^2 = e_d^2$ for the down-type distributions
$q_i^\textrm{NS} = \Delta_{ds}, \Delta_{sb}, d^-,s^-,b^-$.
Let us mention that strictly speaking Eq.~(\ref{NonSingletEvolution})
is valid only if all the quark flavors are present in the evolution,
that is for $n_f=6$.
For $3 \le n_f \le 5$, some non-singlet PDF ($\Delta_{uc}$,
$\Delta_{sb}$ and $\Delta_{ct}$) will not evolve independently, since
they can be written as a linear combination of singlet PDFs.
For instance, below the charm threshold,
$\Delta_{uc}=u^+=(\Sigma+\Delta_{\Sigma})/2$.

The solution of Eqs.~(\ref{SingletEvolution})
and~(\ref{NonSingletEvolution}) determines the QED evolution operators
that evolve the singlet and non-singlet PDFs from the initial scale
$\nu_0$ to some final scale $\nu$ according to the equations:
\begin{equation}\label{SolutionQED}
\begin{array}{rcl}
{\mathbf q}^\textrm{SG}(x,\nu) &=& {\bm \Gamma}^\textrm{SG}_\textrm{QED}(x|\nu,\nu_0)
\otimes {\mathbf q}^\textrm{SG}(x,\nu_0)\,,\\
\\
q_i^\textrm{NS}(x,\nu) &=& \Gamma_{\textrm{QED},i}^\textrm{NS}(x|\nu,\nu_0) \otimes q_i^\textrm{NS}(x,\nu_0)\,,
\end{array}
\end{equation}
where the singlet evolution operator ${\bm
  \Gamma}^\textrm{SG}_\textrm{QED}$ is a three-by-three matrix while
the non-singlet evolution operators
$\Gamma_{\textrm{QED},i}^\textrm{NS}$ form an scalar array.
In Sect.~\ref{sec:apfelnumeric} we will show how to compute
numerically these evolution operators solving the corresponding
integro-differential equations by means of higher-order interpolation
techniques.

Once the QED evolution operators in Eq.~(\ref{SolutionQED}) have been
computed by means of some suitable numerical method, one needs to
combine them with the corresponding QCD evolution operators.
In order to perform the combination, we can write
Eq.~(\ref{SolutionQED}) in a matrix form introducing in the PDF basis
also the gluon PDF $g(x,\nu,\mu)$.
Taking into account the fact that at leading order in QED the gluon
PDF does not evolve, reintroducing the dependence on the QCD
factorization scales $\mu$ and dropping for simplicity the dependence
on $x$, we can write Eq.~(\ref{SolutionQED}) as follows:
\begin{equation}\label{SolutionQEDMatr}
\underbrace{\begin{pmatrix}
g(\mu,\nu)\\
{\mathbf q}^\textrm{SG}(\mu,\nu)\\
q_1^\textrm{NS}(\mu,\nu)\\
\vdots \\
q_{10}^\textrm{NS}(\mu,\nu)\\
\end{pmatrix}}_{{\mathbf q}^{}(\mu,\nu)} =
\underbrace{\begin{pmatrix}
1 & 0 & 0 & 0 & 0\\
0 & {\bm \Gamma}^\textrm{SG}_\textrm{QED} & {0} & \dots & {0} \\
0 & {0}  & \Gamma_{\textrm{QED},1}^\textrm{NS} & \dots & 0 \\
\vdots &  \vdots  & \vdots & \ddots & \vdots \\
0 & {0}  & 0 & \dots   & \Gamma_{\textrm{QED},10}^\textrm{NS}
\end{pmatrix}}_{{\bm \Gamma}^\textrm{QED}(\nu,\nu_0)}
\otimes
\underbrace{\begin{pmatrix}
g(\mu,\nu_0)\\
{\mathbf q}^\textrm{SG}(\mu,\nu_0)\\
q_1^\textrm{NS}(\mu,\nu_0)\\
\vdots \\
q_{10}^\textrm{NS}(\mu,\nu_0)\\
\end{pmatrix}}_{{\mathbf q}^\textrm{}(\mu,\nu_0)}\,.
\end{equation}
In the above expression, we have denoted by ${\mathbf
  q}^\textrm{}(\mu,\nu)$ the fourteen-dimensional vector that contains
all PDF combinations in the QED evolution basis of
Eq.~(\ref{QEDEvolutionBasis}) plus the gluon PDF.
Of course, a similar expression as that of Eq.~(\ref{SolutionQED})
will hold for the solution of the QCD DGLAP evolution equations:
\begin{equation}\label{SolutionQCDMatr1}
\widetilde{\mathbf q}^\textrm{}(\mu,\nu) = \widetilde{\bm \Gamma}^\textrm{QCD}(\mu,\mu_0) \otimes \widetilde{\mathbf q}^\textrm{}(\mu_0,\nu)\,,
\end{equation}
where in this case the vector ${ \widetilde{\mathbf q} }$ is given in
the QCD evolution basis, which is a different linear combination of
the quark, anti-quark, gluon and photon PDFs as compared to the
corresponding QED evolution basis.
The two basis are related by an invertible fourteen-by-fourteen
rotation matrix $\mathbf T$ that transforms the vector
$\widetilde{\mathbf q}^\textrm{}$ into the vector ${\mathbf
  q}^\textrm{}$:
\begin{equation}\label{TransformationQCDxQED}
{\mathbf q}^\textrm{} = {\mathbf T}\cdot \widetilde{\mathbf q}^\textrm{} \quad
\Longrightarrow \quad \widetilde{\mathbf q}^\textrm{} = {\mathbf T}^{-1}\cdot {\mathbf q}^\textrm{}\,.
\end{equation}
Using Eq.~(\ref{TransformationQCDxQED}) and the condition ${\mathbf
  T}\cdot{\mathbf T}^{-1} = {\bm 1}$, the solution of the QED
evolution equations Eq.~(\ref{SolutionQEDMatr}) can be rotated as
follows:
\begin{equation}
  \label{SolutionQEDMatr1}
\widetilde{\mathbf q}^\textrm{}(\mu,\nu) = \underbrace{\left[{\mathbf
  T}^{-1}\cdot {\bm \Gamma}^\textrm{QED}(\nu,\nu_0) \cdot {\mathbf T}\right]}_{\widetilde{{\bm \Gamma}}^\textrm{QED}(\nu,\nu_0)}
\otimes ~\widetilde{\mathbf q}^\textrm{}(\mu,\nu_0)\,.
\end{equation}
where $\widetilde{{\bm \Gamma}}^\textrm{QED}(\nu,\nu_0)$ is now the
QED evolution operator expressed in the QCD evolution basis.
Eqs.~(\ref{SolutionQCDMatr1}) and~(\ref{SolutionQEDMatr1}) determine
the QCD and the QED evolution, respectively, of PDFs in the QCD
evolution basis and can therefore be consistently used to construct a
combined QCD$\otimes$QED evolution operator.
In the following, we drop all the tildes since it is understood that
PDFs and evolution operators are always expressed in the QCD evolution
basis.

Now, when combining QCD and QED evolution operators we are faced with
an inherent ambiguity.
Given that QCD and QED evolutions take place by means of the matrix
evolution operators ${\bm \Gamma}^\textrm{QCD}$ and ${{\bm
    \Gamma}}^\textrm{QED}$ that do not commute,
\begin{equation}\label{commutator}
[{\bm \Gamma}^\textrm{QCD},{{\bm \Gamma}}^\textrm{QED}] \neq 0\,,
\end{equation}
this implies  that performing first the QCD evolution followed by the
QED evolution leads to a different result if the opposite order is
assumed. 
We can then define the two possible cases:
\begin{eqnarray}
  \label{eq:QCED}
  {\bm \Gamma}^{\textrm{QCED}}(\mu,\mu_0;\nu,\nu_0) \equiv {\bm
    \Gamma}^{\textrm{QED}}(\nu,\nu_0)\otimes{\bm \Gamma}^{\textrm{QCD}}(\mu,\mu_0)\,,\\
\label{eq:QECD}
{\bm \Gamma}^\textrm{QECD}(\mu,\mu_0;\nu,\nu_0) \equiv {\bm \Gamma}^\textrm{QCD}(\mu,\mu_0)\otimes {{\bm \Gamma}}^\textrm{QED}(\nu,\nu_0)\,,
\end{eqnarray}
and the condition in Eq.~(\ref{commutator}) implies that:
\begin{equation}
\label{eq:diff}
{\bm \Gamma}^\textrm{QCED}(\mu,\mu_0;\nu,\nu_0) \otimes {\mathbf q}(\mu_0,\nu_0) \neq {\bm \Gamma}^\textrm{QECD}(\mu,\mu_0;\nu,\nu_0) \otimes {\mathbf q}(\mu_0,\nu_0)\,.
\end{equation}
However, using the analytical solution of the QCD and QED DGLAP
equations in Mellin space and the Baker-Campbell-Hausdorff formula, it
is possible to show that:
\begin{equation}\label{commutator1}
[{\bm \Gamma}^\textrm{QCD},{{\bm \Gamma}}^\textrm{QED}] = \mathcal{O}(\alpha\alpha_s)\,,
\end{equation}

A careful analysis of the expansions of the two combined evolution
operators in Eqs.~(\ref{eq:QCED}) and~(\ref{eq:QECD}) shows that they
have a similar perturbative structure:
\begin{eqnarray}\label{ExpansionQCED}
{\bm \Gamma}^\textrm{QCED} = \sum_{n=0}^{\infty} (\alpha{\bm A} +
\alpha_s {\bm B})^n + \alpha \alpha_s {\bm C} + \mathcal{O}(\alpha^2)\,,\\
\label{ExpansionQECD}
{\bm \Gamma}^\textrm{QECD} = \sum_{n=0}^{\infty} (\alpha {\bm A} +
\alpha_s {\bm B})^n - \alpha\alpha_s{\bm C} + \mathcal{O}(\alpha^2)\,.
\end{eqnarray}
These expansions suggest a third possibility for the combined
evolution operator given by the average of the ${\bm
  \Gamma}^\textrm{QCED}$ and ${\bm \Gamma}^\textrm{QECD}$ operators:
\begin{equation}\label{AveragedSolution}
{\bm \Gamma}^\textrm{QavD} \equiv \frac{{\bm \Gamma}^\textrm{QCED} + {\bm \Gamma}^\textrm{QECD}}{2}\,,
\end{equation}
so that the subleading terms $\mathcal{O}(\alpha\alpha_s)$ cancel and
the perturbative remainder is $\mathcal{O}(\alpha^2)$.

A possible objection to this approach is that, in the case in which
$\mu$ and $\nu$ are very different from each other, this procedure
might lead to the presence of numerically large, unresummed
logarithms. So, in order to suppress the impact of these potentially
large (subleading) logarithms, we have implemented in \texttt{APFEL}
the combination of QCD and QED evolutions not over the whole (possibly
large) $\left[ Q_0, Q\right]$ range, but rather dividing it in small
intervals $\left[ Q_0,Q_1\right]$, $\left[ Q_1,Q_2\right]$, $\ldots$,
$\left[ Q_N,Q\right]$, and performing the combination on each
interval. This procedure ensures that no artificially large logarithm
of two widely different scales appears in the solution.

In Sect.~\ref{sec:apfelvalidation} we will show that the
\texttt{QavD}, \texttt{QCED} and \texttt{QECD} solutions implemented
with this strategy turn out to be good approximations to the unified
approach, which is equivalent to the MRST2004QED~\cite{Martin:2004dh}
and \texttt{partonevolution}~\cite{Weinzierl:2002mv,Roth:2004ti}
implementations, all of them different by $\mathcal{O}(\alpha^2)$
terms only.

\subsection{Basis for the unified QCD$\otimes$QED solution}
\label{sec:exactsol}

Another common choice for the solution of the combined DGLAP consists
in solving the unified QCD$\otimes$QED system with a specific basis
which satisfies both evolution equations. In order to diagonlize as
much as possible the evolution matrix in the presence of QED
corrections avoiding unnecessary couplings between parton
distributions, we propose the following evolution basis
\begin{equation}
\begin{array}{ll}
\mbox{\texttt{1} : }g & \\
\mbox{\texttt{2} : }\gamma & \\
\mbox{\texttt{3} : }\displaystyle \Sigma = \Sigma_u + \Sigma_d & \quad
\mbox{\texttt{ 9} : }\displaystyle V =V_u +  V_d\\
\mbox{\texttt{4} : }\displaystyle \Delta_\Sigma = \Sigma_u - \Sigma_d
& \quad \mbox{\texttt{10} : }\displaystyle \Delta_V = V_u - V_d\\
\mbox{\texttt{5} : }T_1^u = u^+ - c^+ &\quad \mbox{\texttt{11} : }V_1^u = u^- - c^- \\
\mbox{\texttt{6} : }T_2^u = u^+ + c^+ - 2t^+ &\quad \mbox{\texttt{12} : }V_2^u = u^- + c^- - 2t^-\\
\mbox{\texttt{7} : }T_1^d = d^+ - s^+ &\quad \mbox{\texttt{13} : }V_1^d = d^- - s^- \\
\mbox{\texttt{8} : }T_2^d = d^+ + s^+ - 2b^+ &\quad \mbox{\texttt{14} : }V_2^d = d^- + s^- - 2b^-\,,
\end{array}
\end{equation}
where we have introduced the singlet, triplets and valences combinations
\begin{eqnarray}
  \Sigma_u = \sum_{k=1}^{n_u} u^+_k\,,\quad \Sigma_d =
  \sum_{k=1}^{n_d} d^+_k\,,\\
  V_u = \sum_{k=1}^{n_u} u^-_k\,,\quad V_d = \sum_{k=1}^{n_d} d^-_k\,.
\end{eqnarray}

When considering leading-order QED corrections to DGLAP, the equations
for this basis are divided into three sub-systems: the singlet, the
valence-sector and the non-singlet. For the singlet sector we have
\begin{equation}\label{APFELsys}
\begin{array}{rcl}
\displaystyle\mu^2\frac{\partial}{\partial \mu^2}
\begin{pmatrix}
g\\
\gamma\\
\Sigma\\
\Delta_\Sigma
\end{pmatrix} &=& \displaystyle \left[
\begin{pmatrix}
\widetilde{P}_{gg} & 0 & \widetilde{P}_{gq} & 0 \\
0 & 0 & 0 & 0 \\
2n_f\widetilde{P}_{qg} & 0 & \widetilde{P}_{qq} & 0 \\
\frac{n_u-n_d}{n_f} 2n_f\widetilde{P}_{qg} & 0 & \frac{n_u-n_d}{n_f}(\widetilde{P}_{qq}-\widetilde{P}^+) & \widetilde{P}^+
\end{pmatrix}\right.
\\
\\
&+&\left.\begin{pmatrix}
0 & 0 & 0 & 0 \\
0 & e_\Sigma^2 P^{(0)}_{\gamma\gamma} & \eta^+P^{(0)}_{\gamma q} &\eta^-P^{(0)}_{\gamma q} \\
0 & \theta^- P^{(0)}_{q\gamma} & \eta^+P^{(0)}_{qq} & \eta^-P^{(0)}_{qq}\\
0 & \theta^+ P^{(0)}_{q\gamma}
&\eta^-P^{(0)}_{qq}
&\eta^+P^{(0)}_{qq}
\end{pmatrix}
\right]
\begin{pmatrix}
g\\
\gamma\\
\Sigma\\
\Delta_\Sigma
\end{pmatrix},
\end{array}
\end{equation}
where we separate the splitting matrix into two elements: the first
matrix contains the QCD splitting functions $\widetilde{P}_{ij}$
meanwhile the second contains LO QED splittings $P^{(0)}_{ij}$. Note
that the QED splitting matrix is identical to
Eq.~(\ref{SingletEvolution}). For the QCD sector we have introduce the
usual
notation~\cite{ap,gl,dok,gNLOa,gNLOb,gNLOc,gNLOe,gNLOf,gNNLOa,gNNLOb,Buza:1995ie}
in terms of flavor singlet ($S$) and non-singlet $(V)$ quantities:
\begin{equation}\label{decomposition}
\begin{array}{l}
\widetilde{P}_{q_iq_j} = \widetilde{P}_{\overline{q}_i\overline{q}_j} = \delta_{ij} \widetilde{P}_{qq}^V+\widetilde{P}_{qq}^S\\
\widetilde{P}_{\overline{q}_iq_j} = \widetilde{P}_{q_i\overline{q}_j}  = \delta_{ij} \widetilde{P}_{\overline{q}q}^V+\widetilde{P}_{\overline{q}q}^S\\
\widetilde{P}_{q_ig}  =\widetilde{P}_{\overline{q}_ig} = \widetilde{P}_{qg} \\
\widetilde{P}_{gq_i}  =\widetilde{P}_{g\overline{q}_i} = \widetilde{P}_{gq} \,.
\end{array}
\end{equation}

It follows the definition of $\widetilde{P}^\pm$, $\widetilde{P}_{qq}$
and $\widetilde{P}^V$ as
\begin{equation}
\begin{array}{l}
  \widetilde{P}^\pm \equiv \widetilde{P}_{qq}^V \pm \widetilde{P}_{q\overline{q}}^V \\
  \\
  \widetilde{P}_{qq} \equiv \widetilde{P}^+ + n_f (\widetilde{P}_{qq}^S + \widetilde{P}_{q\overline{q}}^S)\\
  \\
  \widetilde{P}^V \equiv \widetilde{P}^- + n_f (\widetilde{P}_{qq}^S - \widetilde{P}_{q\overline{q}}^S)
\end{array}\,,
\end{equation}

The second system to solve is the valence-sector defined as
\begin{equation}
\displaystyle\mu^2\frac{\partial}{\partial \mu^2}
\begin{pmatrix}
V\\
\Delta_V
\end{pmatrix} = 
\left[
\begin{pmatrix}
\widetilde{P}^V & 0 \\
\frac{n_u-n_d}{n_f}(\widetilde{P}^V-\widetilde{P}^-)  & \widetilde{P}^- 
\end{pmatrix}
+
\begin{pmatrix}
  \eta^+P^{(0)}_{qq} & \eta^-P^{(0)}_{qq} \\
  \eta^-P^{(0)}_{qq} & \eta^+P^{(0)}_{qq}
\end{pmatrix}
\right]
\begin{pmatrix}
V\\
\Delta_V
\end{pmatrix}.
\end{equation}

Finally, we have the non-singlet equations for the remaining evolution
flavors:
\begin{equation}
\begin{array}{l}
\begin{array}{rcl}
\\
\displaystyle \mu^2\frac{\partial T^u_{1,2}}{\partial \mu^2} &=&
\displaystyle (\widetilde{P}^+ + e_u^2P^{(0)}_{qq}) T^u_{1,2}\,,\\
\\
\displaystyle \mu^2\frac{\partial T^d_{1,2}}{\partial \mu^2} &=&
\displaystyle (\widetilde{P}^+ + e_d^2P^{(0)}_{qq}) T^d_{1,2}\,,
\end{array}
\\
\\
\begin{array}{rcl}
\displaystyle \mu^2\frac{\partial V^u_{1,2}}{\partial \mu^2} &=&
\displaystyle (\widetilde{P}^- + e_u^2P^{(0)}_{qq}) V^u_{1,2}\,,\\
\\
\displaystyle \mu^2\frac{\partial V^d_{1,2}}{\partial \mu^2} &=&
\displaystyle (\widetilde{P}^- + e_d^2P^{(0)}_{qq}) V^d_{1,2}\,.
\end{array}
\end{array}
\end{equation}

The basis presented here is just an example of possible choice for the
unified solution, which is implemented in \texttt{APFEL} as
\texttt{QUniD} solution, however many other choices are possible.

\section{Numerical techniques}
\label{sec:apfelnumeric}

In this section we will present the numerical techniques that
\texttt{APFEL} uses to solve the DGLAP evolution equations. The same
numerical techniques presented here are applied to both QCD and QED
DGLAP evolution equations thanks to the same formal structure.
In order to show the general strategy, we will see how \texttt{APFEL}
solves the QCD evolution equations but keeping in mind that the same
procedure applies to the QED ones as well.

The DGLAP evolution equations can be written as:
\begin{equation}\label{dglap}
\mu^{2}\frac{\partial q_{i}(x,\mu)}{\partial \mu^2}=\int^{1}_{x}\frac{dy}y P_{ij}\left(\frac{x}{y},\alpha_{s}(\mu)\right)q_{j}(y,\mu)\,, 
\end{equation}
where $ P_{ij}\left(x,\alpha_{s}(\mu)\right)$ are the usual splitting
functions up to some perturbative order in $\alpha_s$.  If we make the
following definitions:
\begin{equation}
\begin{array}{rcl}
t &\equiv&\ln(\mu^{2})\,,\\
\tilde{q}(x,t)&\equiv&xq(x,\mu)\,,\\
\tilde{P}_{ij}(x,t)&\equiv&xP_{ij}(x,\alpha_{s}(\mu))\,,
\end{array}
\end{equation}
Eq.~(\ref{dglap}) becomes:
\begin{equation}\label{dglap2}
\frac{\partial \tilde{q}_{i}(x,t)}{\partial t}=\int^{1}_{x}\frac{dy}y \tilde{P}_{ij}\left(\frac{x}{y},t\right)\tilde{q}_{j}(y,t) \ .
\end{equation}
In order to numerically solve the above equation, we choose to express PDFs 
in terms of an interpolation basis over an $x$ grid with $N_x+1$ points. 
This way we can write:
\begin{equation}
\tilde{q}(y,t)=\sum^{N_{x}}_{\alpha=0}w_{\alpha}^{(k)}(y)\tilde{q}(x_{\alpha},t)\,,
\end{equation}
where $\{w_{\alpha}^{(k)}(y)\}$  is a set of interpolation functions
of degree $k$. In \texttt{APFEL} we have chosen to use the Lagrange
interpolation method and therefore the interpolation functions read:
\begin{equation}\label{LagrangeFormula}
w_{\alpha}^{(k)}(x) = \sum_{j=0,j \leq \alpha}^{k}\theta(x-x_{\alpha-j})\theta(x_{\alpha-j+1}-x)\prod^{k}_{\delta=0,\delta\ne j}\left[\frac{x-x_{\alpha-j+\delta}}{x_{\alpha}-x_{\alpha-j+\delta}}\right]\,.
\end{equation}
Notice that Eq. (\ref{LagrangeFormula}) implies that:
\begin{equation}\label{nonzero}
w_{\alpha}^{(k)}(x) \neq 0 \quad\mbox{for}\quad x_{\alpha-k} < x < x_{\alpha+1} \,.
\end{equation}
Now we can rewrite Eq.~(\ref{dglap2}) as follows:
\begin{equation}\label{dglap3}
\frac{\partial \tilde{q}_{i}(x,t)}{\partial t}=\sum_{\alpha}\left[\int^{1}_{x}\frac{dy}y \tilde{P}_{ij}\left(\frac{x}{y},t\right)w_{\alpha}^{(k)}(y)\right]\tilde{q}_{j}(x_{\alpha},t) \, .
\end{equation}
In the particular case in which the $x$ variable in Eq. (\ref{dglap3}) coincides with one of the $x$-grid nodes, say
$x_\beta$, the evolution equations take the following discretized form:
\begin{equation}\label{dglap4}
\frac{\partial \tilde{q}_{i}(x_\beta,t)}{\partial t}=\sum_{\alpha}\underbrace{\left[\int^{1}_{x_\beta}\frac{dy}y \tilde{P}_{ij}\left(\frac{x_\beta}{y},t\right)w_{\alpha}^{(k)}(y)\right]}_{\Pi_{ij,\beta\alpha}(t)}\tilde{q}_{j}(x_{\alpha},t)\,.
\end{equation}
From Eq.~(\ref{nonzero}) follows the condition:
\begin{equation}\label{nonzero2}
\Pi_{ij,\beta\alpha}(t) \neq 0 \quad\mbox{for}\quad \beta \leq \alpha\,.
\end{equation}
In addition, the computation $\Pi_{ij,\beta\alpha}$ in  Eq.~(\ref{dglap4}) can be simplified to:
\begin{equation}\label{optimization}
\Pi_{ij,\beta\alpha}(t) = \int^{b}_{a}\frac{dy}y \tilde{P}_{ij}\left(\frac{x_\beta}{y},t\right)w_{\alpha}^{(k)}(y)\,,
\end{equation}
where the integration bounds are given by:
\begin{equation}
a \equiv \mbox{max}(x_\beta,x_{\alpha-k})\quad\mbox{and}\quad b \equiv \mbox{min}(1,x_{\alpha+1})\,.
\end{equation}
Alternatively, by means of a change of variable, the integral in
Eq. (\ref{optimization}) can be rearranged as follows:
\begin{equation}\label{optimization2}
\Pi_{ij,\beta\alpha}(t) = \int^{d}_{c}\frac{dy}y \tilde{P}_{ij}(y,t)w_{\alpha}\left(\frac{x_\beta}{y}\right)\,,
\end{equation}
where the new integration bounds are defined as:
\begin{equation}\label{bounds2}
c \equiv \mbox{max}(x_\beta,x_\beta/x_{\alpha+1}) \quad\mbox{and}\quad d \equiv \mbox{min}(1,x_\beta/x_{\alpha-k}) \,.
\end{equation}

One central aspect of the numerical methods used in
\texttt{APFEL} is the use of an interpolation over a
logarithmically distributed $x$ grid.
In this case, the interpolation
coefficients in Eq.~(\ref{LagrangeFormula}) can be expressed as
\begin{equation}\label{LagrangeFormulaLog}
w_{\alpha}^{(k)}(x) = \sum_{j=0,j \leq \alpha}^{k}\theta(x-x_{\alpha-j})\theta(x_{\alpha-j+1}-x)\prod^{k}_{\delta=0,\delta\ne j}\left[\frac{\ln(x)-\ln(x_{\alpha-j+\delta})}{\ln(x_{\alpha})-\ln(x_{\alpha-j+\delta})}\right]\,.
\end{equation}
If in addition the $x$ grid is logarithmically distributed,
\textit{i.e.} such that
$\ln(x_{\beta})-\ln(x_{\alpha})=(\beta-\alpha)\Delta$, where the step
$\Delta$ is a constant, one has that the interpolating functions read:
\begin{equation}\label{LagrangeFormulaLog2}
w_{\alpha}^{(k)}(x) = \sum_{j=0,\,j \leq \alpha}^{k}\theta(x-x_{\alpha-j})\theta(x_{\alpha-j+1}-x)\prod^{k}_{\delta=0,\delta\ne j}\left[\frac{1}{\Delta} \ln\left(\frac{x}{x_\alpha}\right)\frac{1}{j-\delta}+1\right]\,,
\end{equation}
so that the dependence on $x$ of the interpolating function $w_\alpha^{(k)}(x)$ is
through the function $ \ln(x/x_\alpha)$ only. Therefore,
it can be shown that in Eq.~(\ref{optimization2})
$w_{\alpha}^{(k)}\left(x_\beta/y\right)$ depends only on the
combination $\left[ \left( \beta-\alpha \right) \Delta-\ln y\right]$ and thus
$\Pi_{ij,\beta\alpha}$ depends only on the difference $(\beta-\alpha)$. 

One can use this information, together with the condition in
Eq.~(\ref{nonzero2}), to represent $\Pi_{ij,\beta\alpha}(t)$ as a
matrix, where $\beta$ is the row index and $\alpha$ the column index.
Such a representation of $\Pi_{ij,\beta\alpha}(t)$ reads:
\begin{equation}\label{MatrixRep}
\displaystyle \Pi_{ij,\beta\alpha}(t) = 
\begin{pmatrix}
a_0 &  a_1 & a_2 & \cdots & a_{N_x} \\
 0  & a_0 & a_1 & \cdots & a_{N_x-1} \\
 0  & 0   &  a_0 & \cdots & a_{N_x-2} \\
\vdots & \vdots & \vdots & \ddots & \vdots \\
 0  &   0  &   0 & \cdots & a_0 
\end{pmatrix}\,.
\end{equation}

The knowledge of the first row of the matrix $\Pi_{ij,\beta\alpha}(t)$
is enough to determine all the other entries.  This feature, which is
based on the particular choice of the interpolation procedure, leads
to a more efficient computation of the evolution operators since it
reduces by a factor $N_x$ the number of integrals to be computed.

After the presentation of the interpolation method, we turn to discuss the 
actual computation of the evolution operators. 
Any splitting function,
be it QED or QCD at any given perturbative order, has the following general structure:
\begin{equation}\label{SlittingFunctionDecomposition}
\tilde{P}_{ij}(x,t) = xP_{ij}^{R}(x,t) + \frac{xP_{ij}^{S}(x,t)}{(1-x)_+} + P_{ij}^{L}(t)x\delta(1-x)\,,
\end{equation}
where $P_{ij}^{R}(x,t)$ is the regular term,  $P_{ij}^{S}(x,t)$ is the
coefficient of the plus-distribution term, and $P_{ij}^{L}(t)$ is the
coefficient of the local term proportional to the delta functions.
It is useful to recall here that the general definition of
plus-distribution in the presence of arbitrary integration bounds is given by:
\begin{equation}\label{PlusPrescriptionDef}
\int_c^d dy \frac{f(y)}{(1-y)_+} = \int_c^d dy \frac{f(y) - f(1)\theta(d-1)}{1-y} + f(1)\ln(1-c)\theta(d-1)\,.
\end{equation}
Moreover, each of the functions $P_{ij}$ appearing in
Eq. (\ref{SlittingFunctionDecomposition}) has the usual perturbative
expansion that at N$^k$LO reads:
\begin{equation}
P_{ij}^{J}(x,t) = \sum_{n=0}^{k}a_s^{n+1}(t)P_{ij}^{J,(n)}(x),\quad\mbox{with}\quad J=R,S,L \,,
\label{eq:expansion}
\end{equation}
where we have defined $a_s\equiv \alpha_s/4\pi$.

Taking the above considerations into account and using the fact that
$w_{\alpha}^{(k)}(x_\beta)=\delta_{\beta\alpha}$, we can write the evolution operators
in terms of the various parts of the splitting functions as follows:
\begin{equation} \label{pertExp}
\begin{array}{c}
\displaystyle \Pi_{ij,\beta\alpha}(t) = \\
\\
\displaystyle \sum_{n=0}^{k} a_s^{n+1}(t) \bigg\{\int^{d}_{c}dy\left[{P}_{ij}^{R,(n)}(y)w_{\alpha}\left(\frac{x_\beta}{y}\right)+\frac{{P}_{ij}^{S,(n)}(y)}{1-y}\left(w_{\alpha}\left(\frac{x_\beta}{y}\right)-\delta_{\beta\alpha}\theta(d-1)\right)\right]\\
\\
\displaystyle +\left[{P}_{ij}^{S,(n)}(1)\ln(1-c)\theta(d-1)+{P}_{ij}^{L,(n)}\right]\delta_{\beta\alpha}\bigg\}\equiv \sum_{n=0}^{k} a_s^{n+1}(t) \Pi_{ij,\beta\alpha}^{(n)}\,,
\end{array}
\end{equation}
where the coefficients $\Pi_{ij,\beta\alpha}^{(n)}$ are independent of
the energy scale $t$,  and need to be evaluated a single time once the
$x$ interpolation grid and the evolution parameters have been defined.

Now we will show that Eq.~(\ref{pertExp}) respects the symmetry
conditions of Eq.~(\ref{MatrixRep}). We can distinguish two cases: 1)
$d < 1$ and  2) $d = 1$. In the case 1), due to the presence of the
Heaviside functions $\theta(d-1)$, Eq.~(\ref{pertExp}) reduces to:
\begin{equation}\label{pertExp2}
\Pi_{ij,\beta\alpha}^{(n)} = \int^{d}_{c}dy\left[{P}_{ij}^{R,(n)}(y)+\frac{{P}_{ij}^{S,(n)}(y)}{1-y}\right]w_{\alpha}\left(\frac{x_\beta}{y}\right) + {P}_{ij}^{L,(n)}\delta_{\beta\alpha}\,,
\end{equation}
which clearly follows Eq.~(\ref{MatrixRep}). In the case 2), instead, we have:
\begin{equation}\label{pertExp3}
\begin{array}{c}
\displaystyle \Pi_{ij,\beta\alpha}^{(n)} = \int^{1}_{c}dy\left[{P}_{ij}^{R,(n)}(y)w_{\alpha}\left(\frac{x_\beta}{y}\right)+\frac{{P}_{ij}^{S,(n)}(y)}{1-y}\left(w_{\alpha}\left(\frac{x_\beta}{y}\right)-\delta_{\beta\alpha}\right)\right]\\
\\
\displaystyle +\left[{P}_{ij}^{S,(n)}(1)\ln(1-c)+{P}_{ij}^{L,(n)}\right]\delta_{\beta\alpha}\,,
\end{array}
\end{equation}
and apparently, if $\alpha=\beta$, the term proportional to $\ln(1-c)$
could break the symmetry. However, from Eq.~(\ref{bounds2}), we know
that in this case:
\begin{equation}
c = \mbox{max}(x_\beta,x_\beta/x_{\beta+1}) = \frac{x_\beta}{x_{\beta+1}} \,,
\end{equation}
because $x_{\beta+1} \leq 1$. 
In addition, on a logarithmically distributed grid we have that
 $x_{\beta+1}=x_{\beta}\exp(\Delta)$. 
Therefore, it turns out that:
\begin{equation}
\ln(1-c) = \ln\left(1-\frac{x_{\beta}}{x_{\beta+1}}\right) = \ln[1 - \exp(-\Delta)]\,,
\end{equation}
which is a constant which does not depend on the indices $\alpha$ and $\beta$ and therefore
satisfies Eq.~(\ref{MatrixRep}).

At this point, the DGLAP equations imply that the discretized PDFs evolve between two scales $t$ and $t_0$
according to the following matrix equation:
\begin{equation}\label{EvolutionOperatorsDef}
\tilde{q}_{i}(x_\beta,t) = \sum_{\gamma,k} \Gamma_{ik,\beta\gamma}(t,t_0)\tilde{q}_{k}(x_\gamma,t_0) \ ,
\end{equation}
where it follows from Eq.~(\ref{dglap4}) that the evolution operators
are given by the solution of the system:
\begin{equation}\label{tosolve}
\left\{\begin{array}{l}
\displaystyle \frac{\partial  \Gamma_{ij,\alpha\beta}(t,t_0)}{\partial t}=\sum_{\gamma,k} \Pi_{ik,\alpha\gamma}(t)\Gamma_{kj,\gamma\beta}(t,t_0)\\
\\
\displaystyle \Gamma_{ij,\alpha\beta}(t_0,t_0)=\delta_{ij}\delta_{\alpha\beta}
\end{array}\right.
\end{equation}
Eq.~(\ref{tosolve}) is a set of coupled first order ordinary linear differential equations for the
evolution operators $\Gamma_{ij,\alpha\beta}(t,t_0)$.
In \texttt{APFEL} Eq.~(\ref{tosolve}) is solved  using a fourth-order
adaptive step-size control Runge-Kutta (RK) algorithm.
Note that no interpolation in $t$ is involved, the solution of the
differential equations in $t$ is only limited by the precision of the RK method.
Once the evolved PDFs at the grid values $\tilde{q}_{i}(x_\beta,t)$ 
have been determined by means of the evolution operators in Eq.~(\ref{EvolutionOperatorsDef}), 
the value of these same PDFs for arbitrary values of $x$ will be computed
using again higher-order interpolation.

A final consideration concerning the choice of interpolating grid in
$x$ is needed.
As is well known, an accurate solution of the  DGLAP equations
requires a denser grid at large $x$, where PDFs have more structure than at
small-$x$.
In \texttt{APFEL} it is not possible to use an $x$-grid with variable
spacing that allows to have a denser grid at large $x$ and at the same time to maintain
the symmetry that allows to substantially reduce the number of integrals to be
evaluated, see Eq.~(\ref{MatrixRep}).
In fact, a logarithmically distributed $x$ grid necessarily
leads to a looser grid in the large-$x$ region, thus potentially
degrading the evolution accuracy there.
To overcome this problem, \texttt{APFEL} implements the possibility of using
different interpolation grids according to the value of $x$ in which
PDFs need to be evaluated.

The basic idea is the following. The evolution of a given set of PDFs
from the initial condition at the scale $\mu_0$ up to some other scale
$\mu$ is determined by the convolution between the evolution operators
and the boundary conditions, which implies performing and integral
between $x$ and one.
This convolution, when discretized on an interpolation $x$ grid,
corresponds to Eq.~(\ref{EvolutionOperatorsDef}).
It is clear that such operation will use only those $x_{\beta}$ nodes of
the interpolation grid that fall in the range between $x$ and one. 

Therefore, the computation of the PDF evolution in the large-$x$
region using a logarithmically spaced interpolation grid with a small value of $x_\textrm{min}$ will
be certainly inefficient, since the convolution would use only a small
number of points
in the large-$x$ region such that $x_{\beta}\le x \le 1$, discarding those with  $x< x_{\beta}$.
In order to avoid this problem and simultaneously achieve a good
accuracy and performance over the whole range in $x$, 
\texttt{APFEL} gives the possibility to use different interpolating
grids, each with a different value of $x_\textrm{min}$, interpolation degree and number of points.
Then, to compute the evolution of the PDFs for the point $x$, the program
will automatically select the grid with the largest value of $x_{\textrm{min}}$ compatible with the condition $x_\textrm{min}\le x$. 

The use of $n\ge 2$ subgrids increases slightly the time taken by initialization phase,
since more evolution operators need to be precomputed, and also the actual
evolution is somewhat slower than in the case with a single grid ($n=1$), 
with the important trade-off of a much
more accurate result in the large-$x$ region. 
As default settings, \texttt{APFEL} uses $n=3$ interpolation
grids, with interpolation order $3,5$ and $5$, number of points $N_x=80,50$ and
$40$ and $x_\textrm{min}=10^{-5},0.1$ and $0.8$ respectively.

\section{Validation and benchmarking}

\label{sec:apfelvalidation}

In this section we first perform a detailed benchmarking of
\texttt{APFEL} against \texttt{HOPPET} finding good agreement for the
QCD evolution up to NNLO, both with pole and $\overline{\textrm{MS}}$
heavy quark masses.
Then we turn to the validation of the combined QCD$\otimes$QED
evolution. We verify the consistency of the different methods for the
solution of the combined QCD$\otimes$QED evolution equations, showing
that the coupled solution is numerically equivalent to the unified
solution when constructed iteratively in small steps in $Q$.
Finally, we compare the predictions of \texttt{APFEL} with: the
\texttt{partonevolution} code, the internal MRST2004QED evolution and
the \texttt{QCDNUM} library~\cite{Botje:2010ay,Sadykov:2014aua}.

\subsection{QCD evolution}
\label{APFELvsHOPPET}

We validate the QCD evolution in \texttt{APFEL} by comparing it with
the results from the \texttt{HOPPET} program, version 1.1.5, up to
NNLO, and using both pole and $\overline{\textrm{MS}}$ heavy quark
masses.
The settings are the same as in the original Les Houches PDF evolution
benchmark~\cite{Dittmar:2005ed}.
In the case of $\overline{\textrm{MS}}$ masses, we take the
$\overline{\textrm{MS}}$ Renormalization-Group-Invariant charm mass
$m_c(m_c)$ to have the same numerical values as the pole masses.
In all the comparisons in this section, the interpolation settings in
\texttt{APFEL} are the default ones discussed in
Sect.~\ref{sec:apfelnumeric}.

\begin{figure}
  \centering
  \includegraphics[scale=0.34]{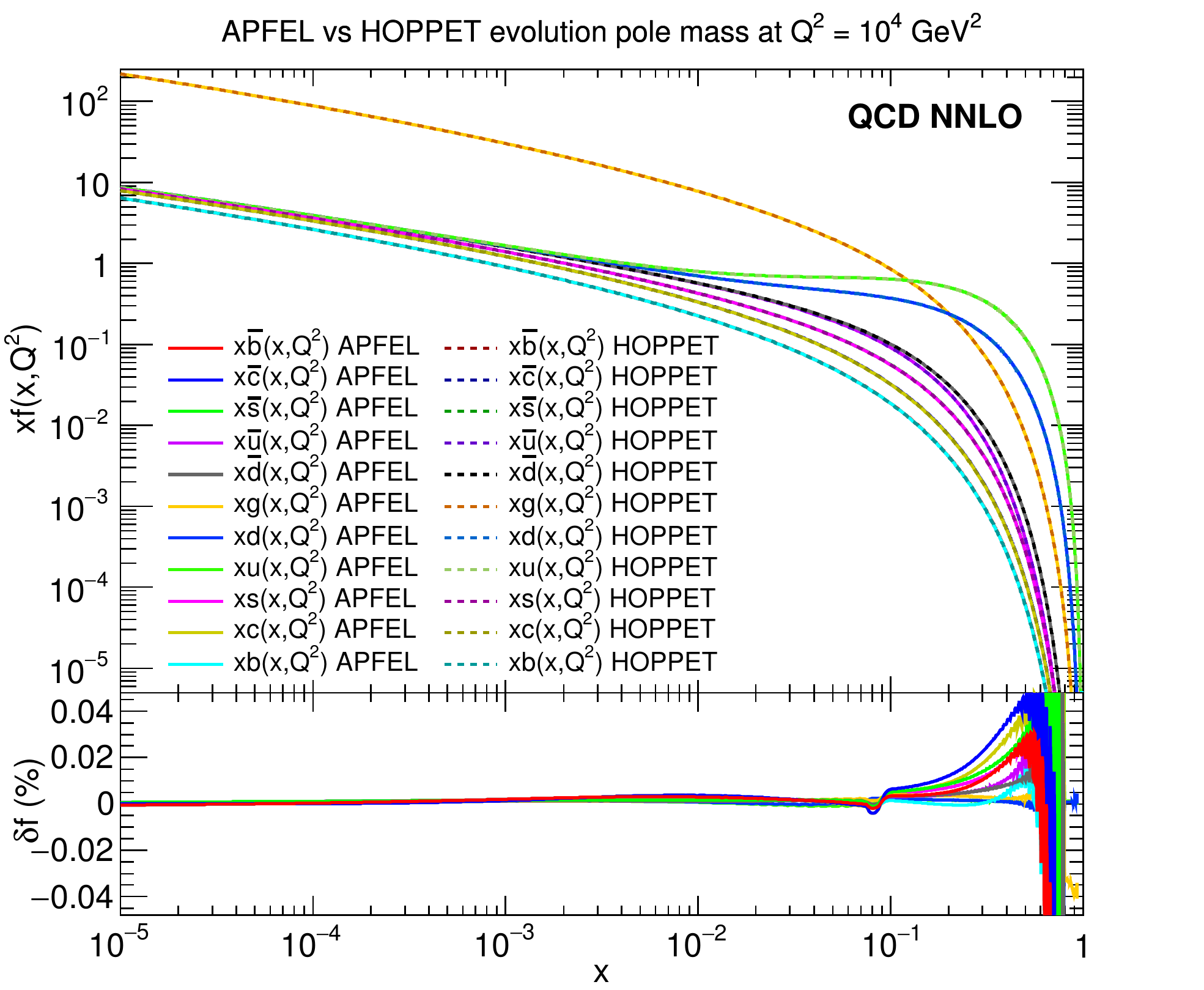}\includegraphics[scale=0.34]{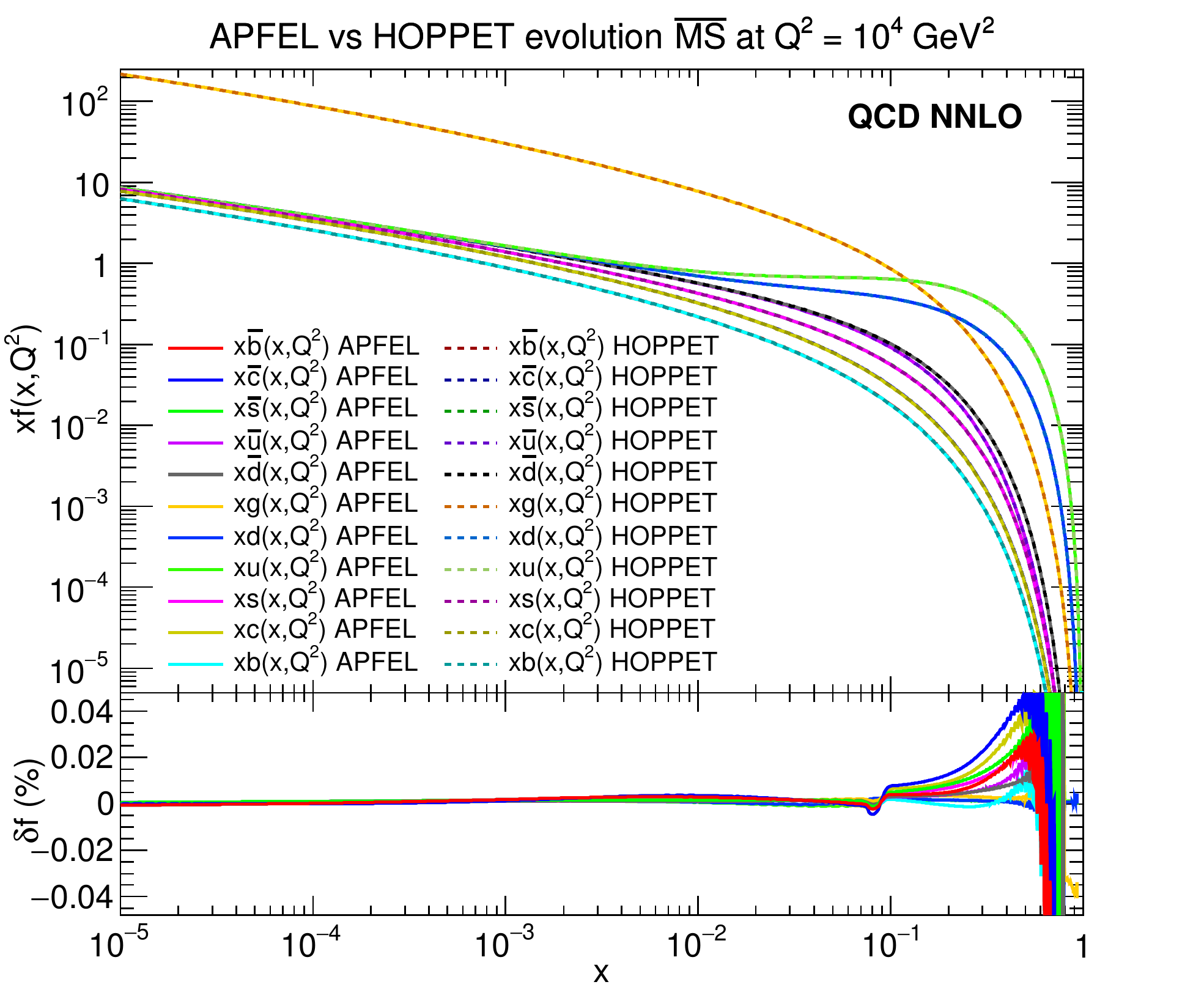}
  \caption{Comparison between PDFs evolved at NNLO in QCD
    using \texttt{APFEL}
    and \texttt{HOPPET}, from $Q_0^2=$2 GeV$^2$ up to
    $Q^2=$10$^4$ GeV$^2$, using the Les Houches  PDF benchmark settings.
    The comparison is performed 
    in the pole mass scheme (left) and  in the $\overline{\textrm{MS}}$ scheme (right).
    The lower plots show the percent differences between the two
    codes.  }
\label{fig:APFELvsHOPPET-one}
\end{figure}

Results for the evolved PDFs at $Q^2=10^{4}$ GeV$^2$ for both
\texttt{HOPPET} and \texttt{APFEL} are shown in
Fig.~\ref{fig:APFELvsHOPPET-one}.
The left plot shows the results using pole masses, while the right
plot corresponds to the case of $\overline{\textrm{MS}}$
masses. Fig.~\ref{fig:APFELvsHOPPET-one} also shows the percent
difference between both predictions, to show the excellent agreement
obtained for the whole range in $x$, being at most $\sim 0.04\%$ at
large-$x$, where PDFs have more structure.

\subsection{QCD$\otimes$QED evolution}

\subsubsection{Consistency of the coupled solution}

\begin{figure}
  \centering
  \includegraphics[scale=0.34]{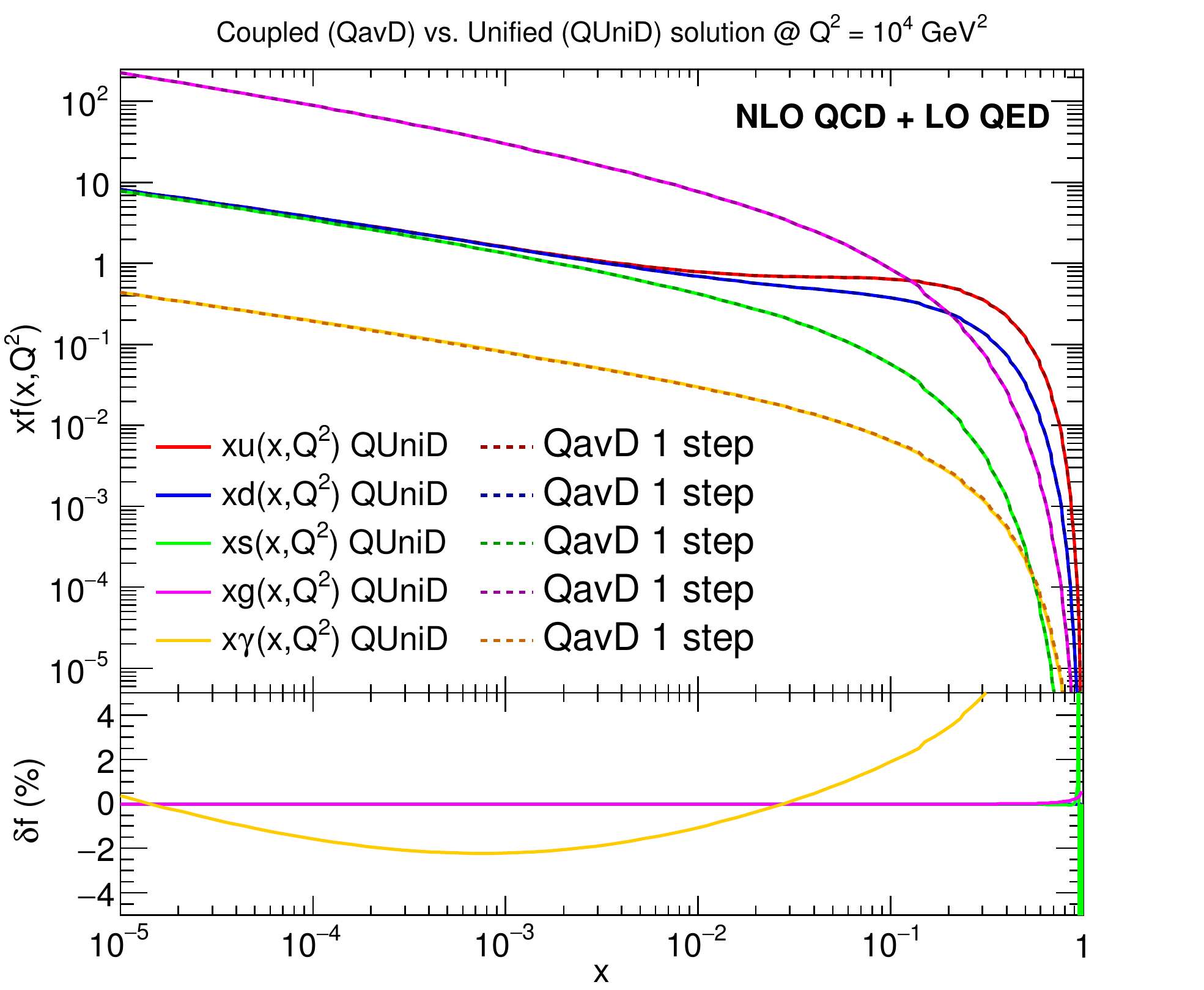}\includegraphics[scale=0.34]{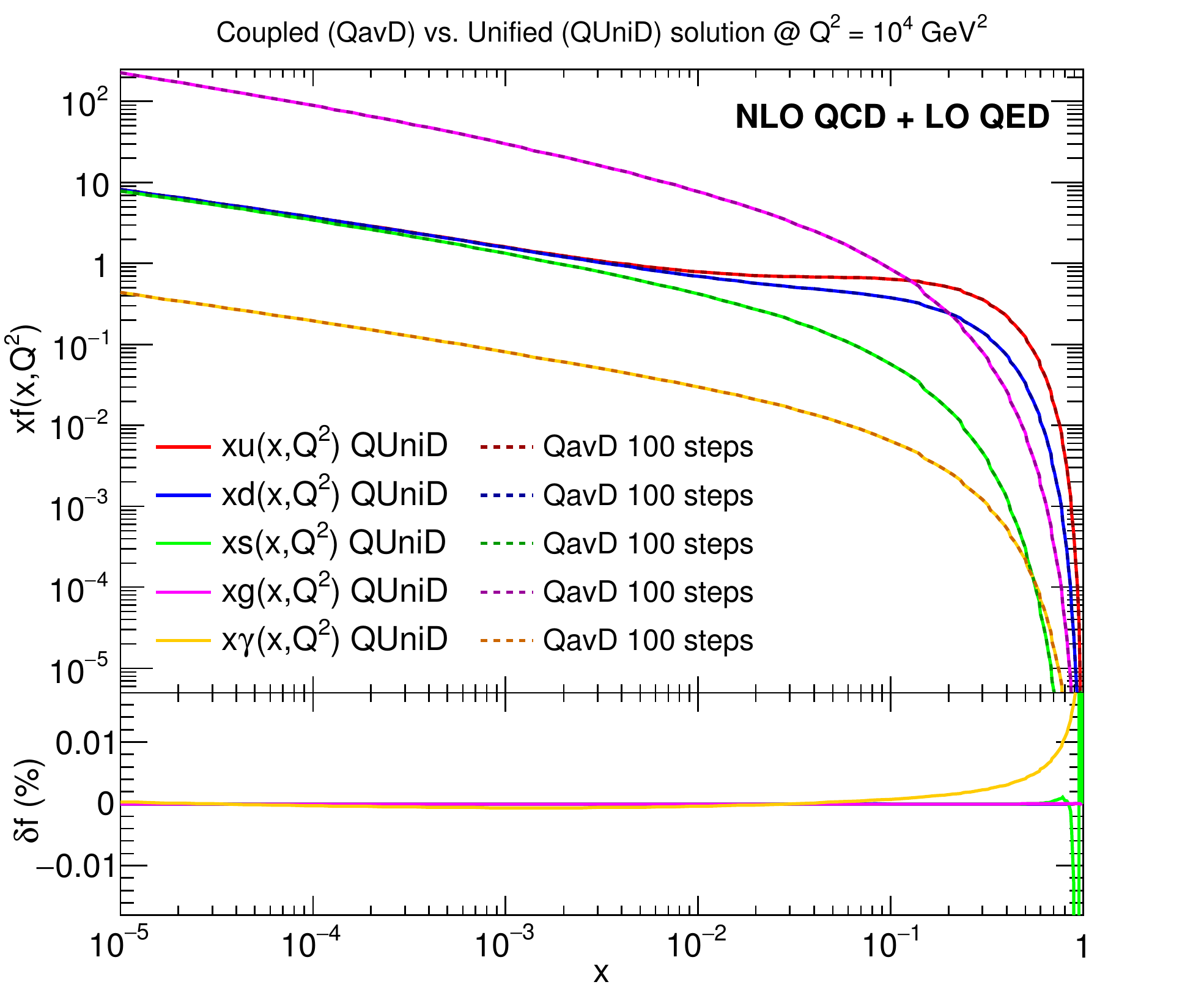}\\
  \includegraphics[scale=0.34]{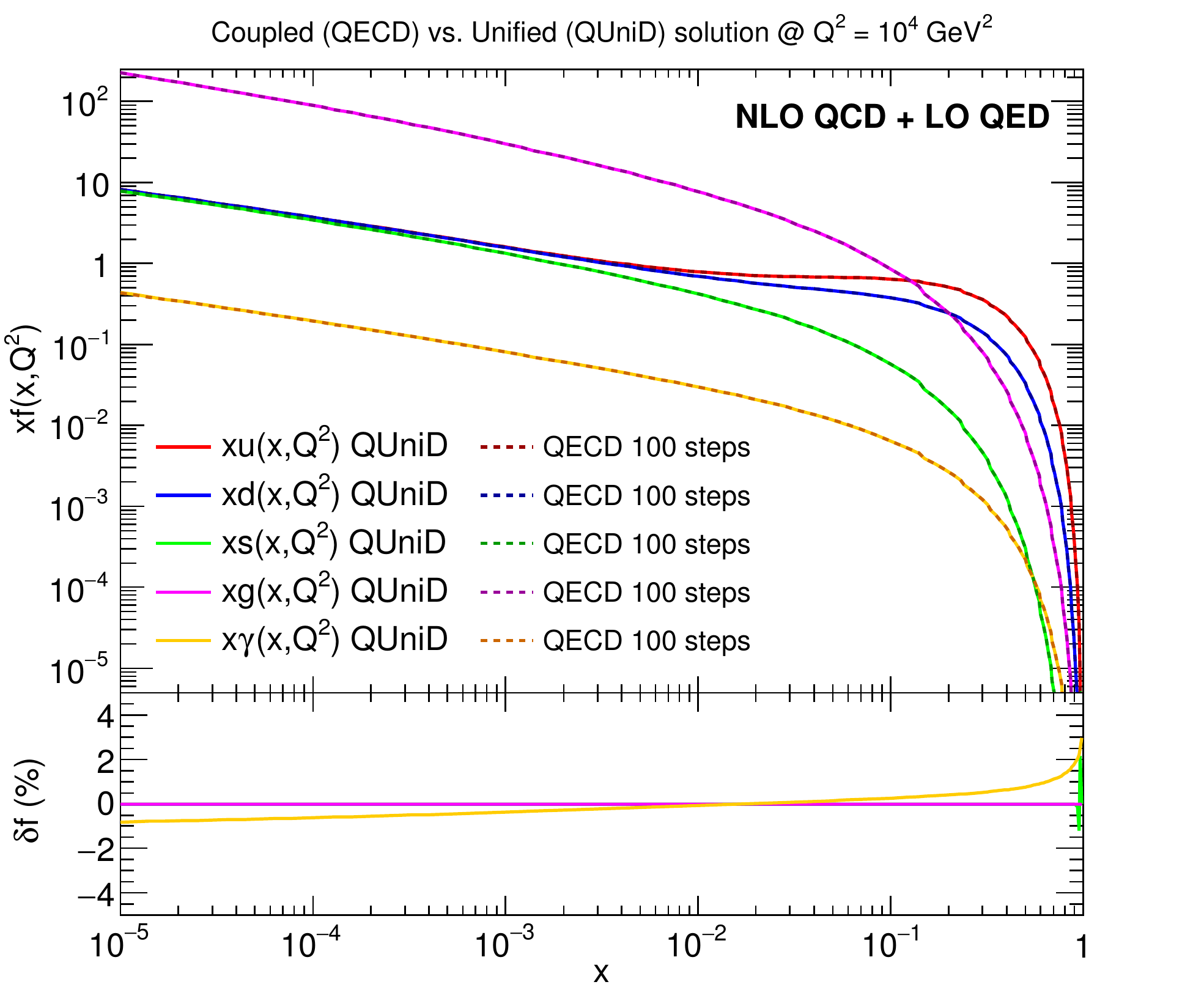}\includegraphics[scale=0.34]{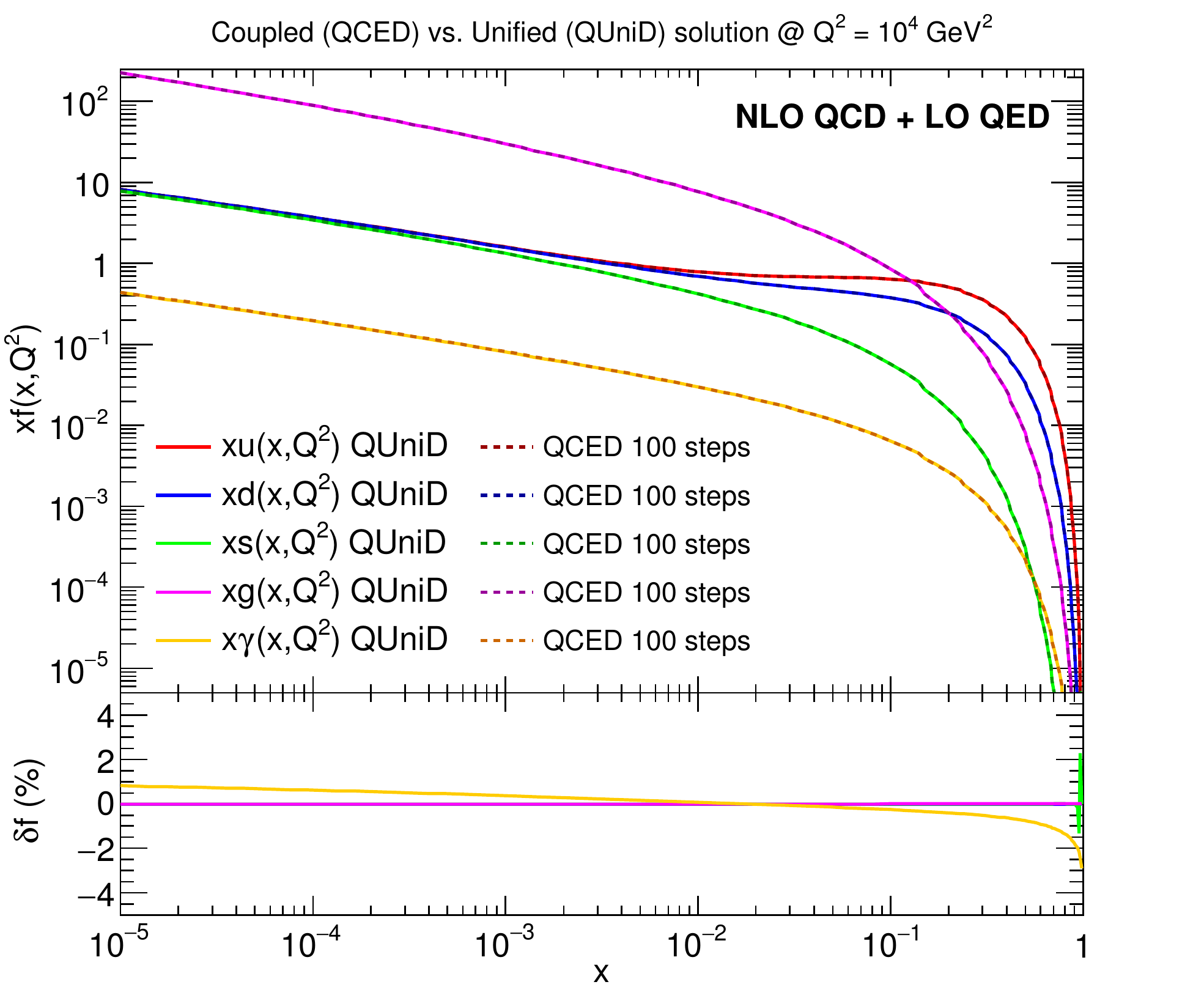}
  \caption{Comparison between PDFs evolved with APFEL with the
    combined QCD$\otimes$QED DGLAP. We show in the plots of the top
    the comparison between the \texttt{QUniD} and the \texttt{QavD}
    solutions using 1 step (left plot) and 100 steps (right plot). The
    bottom plots show the comparison of \texttt{QUniD} and
    \texttt{QECD} (plot on the left) and \texttt{QCED} (plot on the
    right) both using 100 steps. The evolution is performed between
    $Q_0^2 = 2$ GeV$^2$ and $Q^2=10^{4}$ GeV$^2$ in the VFN scheme at
    NLO in QCD and LO in QED using the Les Houches PDF
    setup~\cite{Dittmar:2005ed}, supplemented by the ansatz
    $\gamma(x,Q_0)=0$.}
\label{fig:consistencyapfel}
\end{figure}

Before comparing \texttt{APFEL} to other libraries we first analyze
the numerical impact of the coupled approach in function of the
operators introduced in Sect.~\ref{sec:coupledsol}.

Using the same settings of the Les Houches PDF evolution
benchmark~\cite{Dittmar:2005ed}, supplemented by the ansatz
$\gamma(x,Q_0)=0$ we evolve the PDFs at NLO in QCD and LO in QED
between $Q_0^2=2$ GeV$^2$ and $Q^2=10^{4}$ GeV$^2$ using the VFN
scheme. In the top left plot of Figure~\ref{fig:consistencyapfel} we
show the comparison between the unified solution \texttt{QUniD} and
the average solution \texttt{QavD}, performed with a single step
between $Q_0^2$ and $Q^2$. In the bottom panel of each plot we show
the percentage difference between the two results: a good agreement is
found for all flavors except for the photon PDF, where at $x \sim
10^{-3}$ a peak of -2\% difference is observed, followed by more
important discrepancies at large $x$. The right plot of the same
figure shows the comparison where the full range $[Q_0^2,Q^2]$ has
been divided into 100 logarithmically spaced intervals. In this
condition we obtain a good agreement between both solutions for all
flavors. This result confirms that the average solution \texttt{QavD}
is free of numerically large scale logarithms when introducing a
moderate number of $Q^2$ intervals.

At this point we turn to consider the \texttt{QECD} and the
\texttt{QCED} solutions. The result for 100 steps evolution is shown
in the bottom plots of Fig.~\ref{fig:consistencyapfel}, where the
\texttt{QECD} (left plot) and the \texttt{QCED} (right plot) solutions
are compared to the unified solution \texttt{QUniD}. We observe
evident discrepancies for the photon PDF: the \texttt{QECD} solution
underestimates the photon evolution at small $x$ meanwhile the
\texttt{QCED} solution overestimates in the same region. It is
important to highlight that both solution are not able to reproduce
the same level of accuracy of the average solution, even if we require
the same number of steps. This suggests that these solutions introduce
artificially large logarithms and that an effective way to cancel them
is to perform the evolution in smaller steps combining sequentially the
results. In this regime the \texttt{QECD} and \texttt{QCED} solutions
coincide to a good approximation with that of the \texttt{QavD}
solution, so that all three strategies lead to the same numerical
accuracy.

In the next paragraphs, in order to simplify the analysis, we will use
exclusively the \texttt{QUniD} solution when comparing the combined
QCD$\otimes$QED evolution to other codes.

\subsubsection{Comparison with \texttt{partonevolution}}

We start the benchmarking exercise by comparing the results of the
combined QCD$\otimes$QED DGLAP evolution in \texttt{APFEL} with those
of the public \texttt{partonevolution}
code~\cite{Weinzierl:2002mv,Roth:2004ti}, version 1.1.3.
%
\begin{figure}
\centering
\includegraphics[scale=0.34]{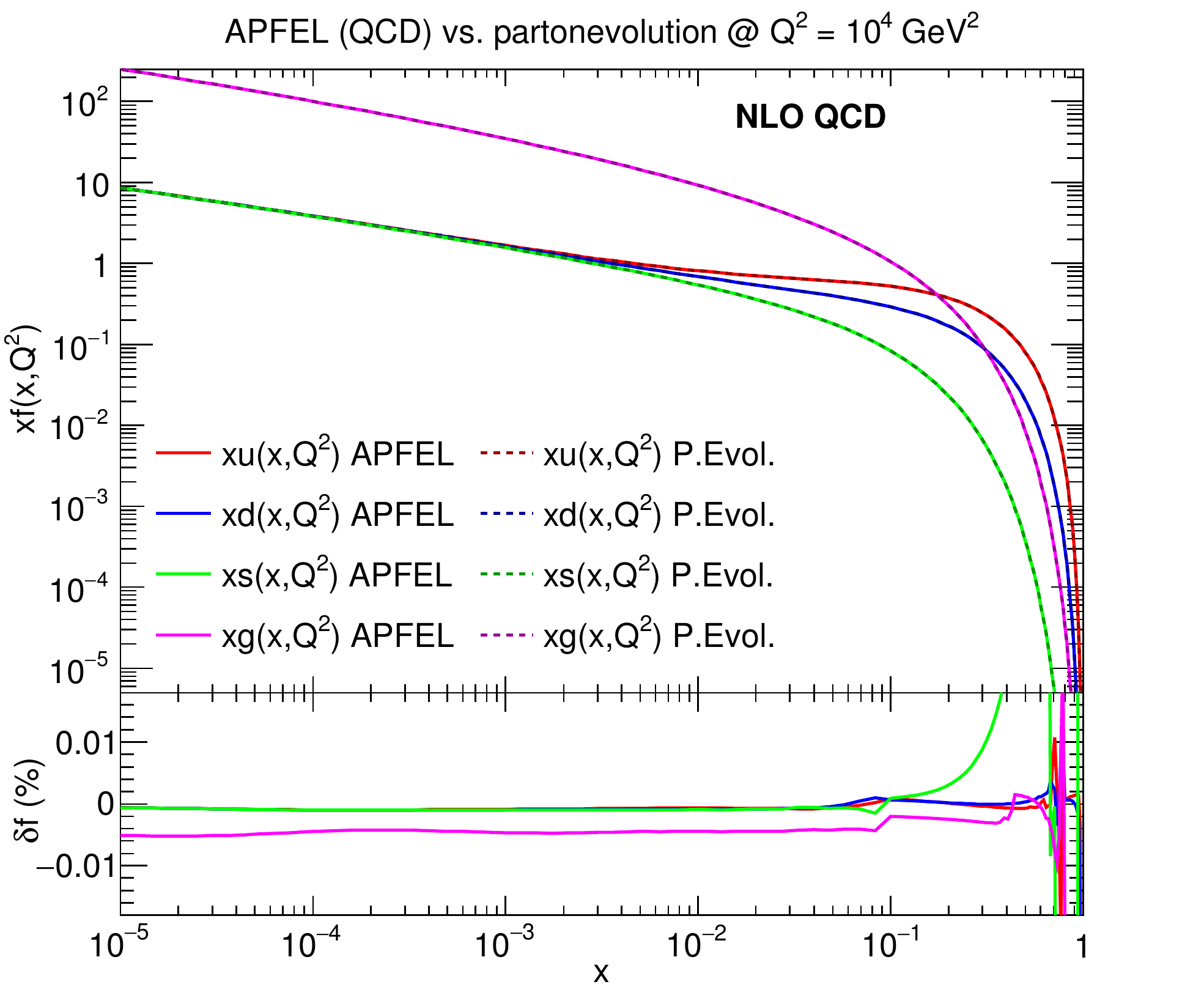}\includegraphics[scale=0.34]{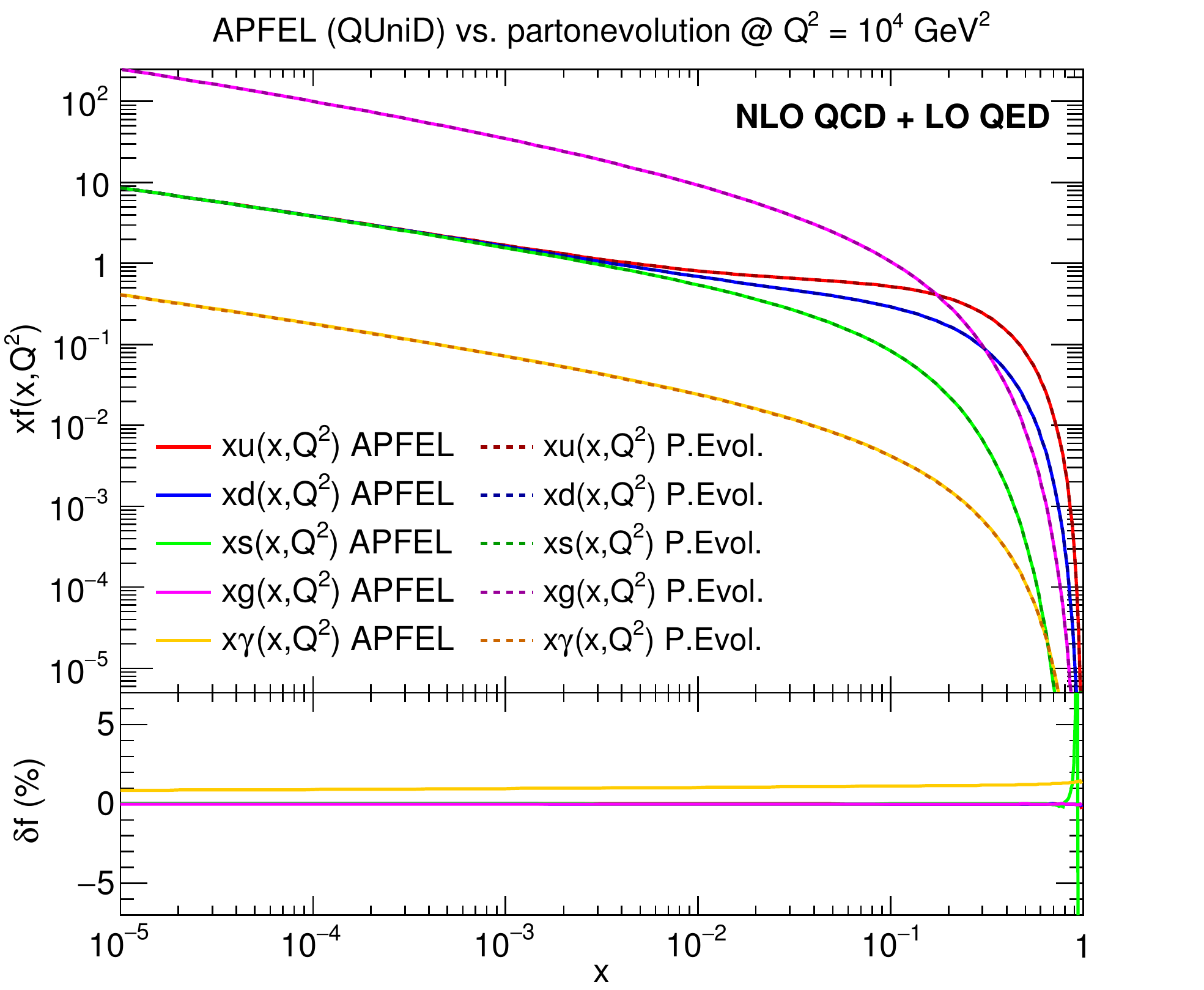}
\caption{Comparison between PDFs evolved using \texttt{APFEL}
  and \texttt{partonevolution}, from $Q_0^2=$4 GeV$^2$ up 
$Q_0^2=$10$^4$ GeV$^2$. The same settings of the PDF benchmark study of
Ref.~\cite{Blumlein:1996rp} have been used.
On the left plot we show evolution at NLO in QCD (without QED
corrections), meanwhile on the right plot we consider the
QCD$\otimes$QED evolution.
For each comparison, we also show the percent differences with respect
to the \texttt{partonevolution} results.}
\label{fig:APFELvsPARTONEVOLUTION}
\end{figure}

To perform the benchmark, we use \texttt{APFEL} with the same settings
used in the original publication~\cite{Roth:2004ti} to present the
numerical results of \texttt{partonevolution}, \textit{i.e.}  we take
the input PDFs from the toy model used in the benchmarking exercise of
Ref.~\cite{Blumlein:1996rp}, given by:
\begin{eqnarray}
xu_v(x)=A_ux^{0.5}(1-x)^3\, , \quad &\,& xd_v(x)=A_dx^{0.5}(1-x)^4 \, ,\nonumber \\
xS(x)=A_Sx^{-0.2}(1-x)^7 \, ,\quad &\,& xg(x)=A_gx^{-0.2}(1-x)^5 \, , \nonumber \\
xc(x)=0 \, , \quad &\,& x\bar{c}(x)=0 \, , 
\end{eqnarray}
at the initial scale $Q_0^2=4$~GeV$^2$, with a SU(3) symmetric sea
that carries 15\% of the proton's momentum at $Q_0^2$, and only four
active quarks are considered even above the bottom threshold.
This toy model should not be confused with that used in the Les
Houches PDF benchmark study, used elsewhere in this paper.
In addition, the photon PDF is set to zero at the initial scale, that
is $\gamma(x,Q_0^2)=0$.

In order to set up the baseline, we ran the two codes at NLO QCD only,
switching off the QED corrections, in the FFN scheme with $n_f=4$. As
can be seen from the left plot of
Fig.~\ref{fig:APFELvsPARTONEVOLUTION}, a good agreement is achieved.
The results about the combined QCD$\otimes$QED evolution are
summarized on the right plot of Fig.~\ref{fig:APFELvsPARTONEVOLUTION},
where we compare the evolution of quark, gluon and photon PDFs given
by the two codes, using the \texttt{QUniD} solution implemented in
\texttt{APFEL}.
With these settings the evolution of quarks and gluon is essentially
identical, with differences at most being $\mathcal{O}( 0.01\%)$,
while differences in the evolution of $\gamma(x,Q^2)$ are below the
few percent level except at the largest values of $x$.
More substantial differences appear for the photon PDF, in this case
the solutions differ by up to 1\%, both at small and large-$x$,
however this level of agreement is still acceptable in view of the
technical differences between both codes.
As we will see in the next paragraphs the quality of the comparison is
much better when using codes with a modern implementations of the
QCD$\otimes$QED combined evolution.

\subsubsection{Comparison with MRST2004QED}

At this point we compare \texttt{APFEL} to the QED evolution used in
the determination of the MRST2004QED parton
distributions~\cite{Martin:2004dh}.
Though the original evolution code is not publicly available, the
evolution which was used can be indirectly accessed via the public
\texttt{LHAPDF} grids.
In this case, it is not possible to use the Les Houches benchmark
settings, and we are instead forced to use the same boundary
conditions for the PDFs at $Q_0$ as those used in the MRST2004QED fit
as well as the same values of the heavy quark masses and reference
coupling constants.
The available MRST2004QED fit was obtained at NLO in QCD in the VFN
scheme, therefore it is possible to perform a meaningful comparison
with the results of their evolution by using \texttt{APFEL} at NLO
with the same settings.

\begin{figure}
\centering
\includegraphics[scale=0.34]{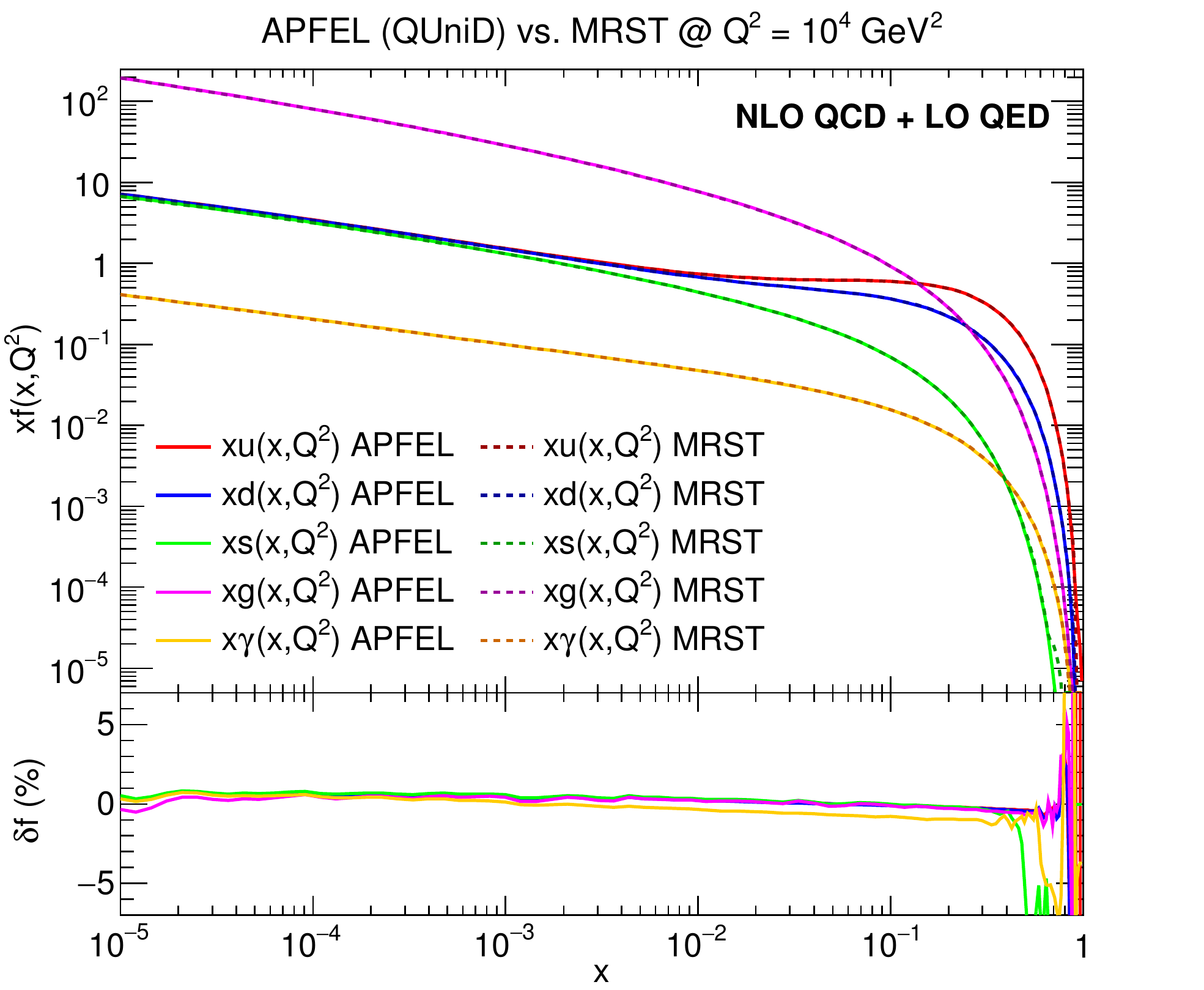}
\caption{Comparison between PDFs evolved using \texttt{APFEL} and the
  internal MRST2004QED parton evolution, from $Q_0^2=$1.25 GeV$^2$ up
  to $Q^2=$10$^4$ GeV$^2$ at NLO in QCD and LO in QED using the VFN scheme.
The boundary conditions for the PDFs are the same as those of the MRST2004QED fit.}
\label{fig:APFELvsMRST-one}
\end{figure}

The comparison between the \texttt{APFEL} predictions and the
MRST2004QED evolution is shown in Fig.~\ref{fig:APFELvsMRST-one},
where PDFs have been evolved using \texttt{APFEL} and the internal
MRST evolution from $Q_0^2=$1.25 GeV$^2$ up to $Q^2=$10$^4$ GeV$^2$.
An excellent agreement is found for all flavors. We observe
differences of -1\% at most for the $\gamma(x,Q^2)$ PDF at large-$x$,
meanwhile for quark and gluon PDFs the discrepancies are smaller.

\subsubsection{Comparison with QCDNUM}

We conclude this section by performing the comparison with the recent
implementation of the combined QCD$\otimes$QED evolution in the
\texttt{QCDNUM} library~\cite{Botje:2010ay,Sadykov:2014aua}.

\begin{figure}
\centering
\includegraphics[scale=0.34]{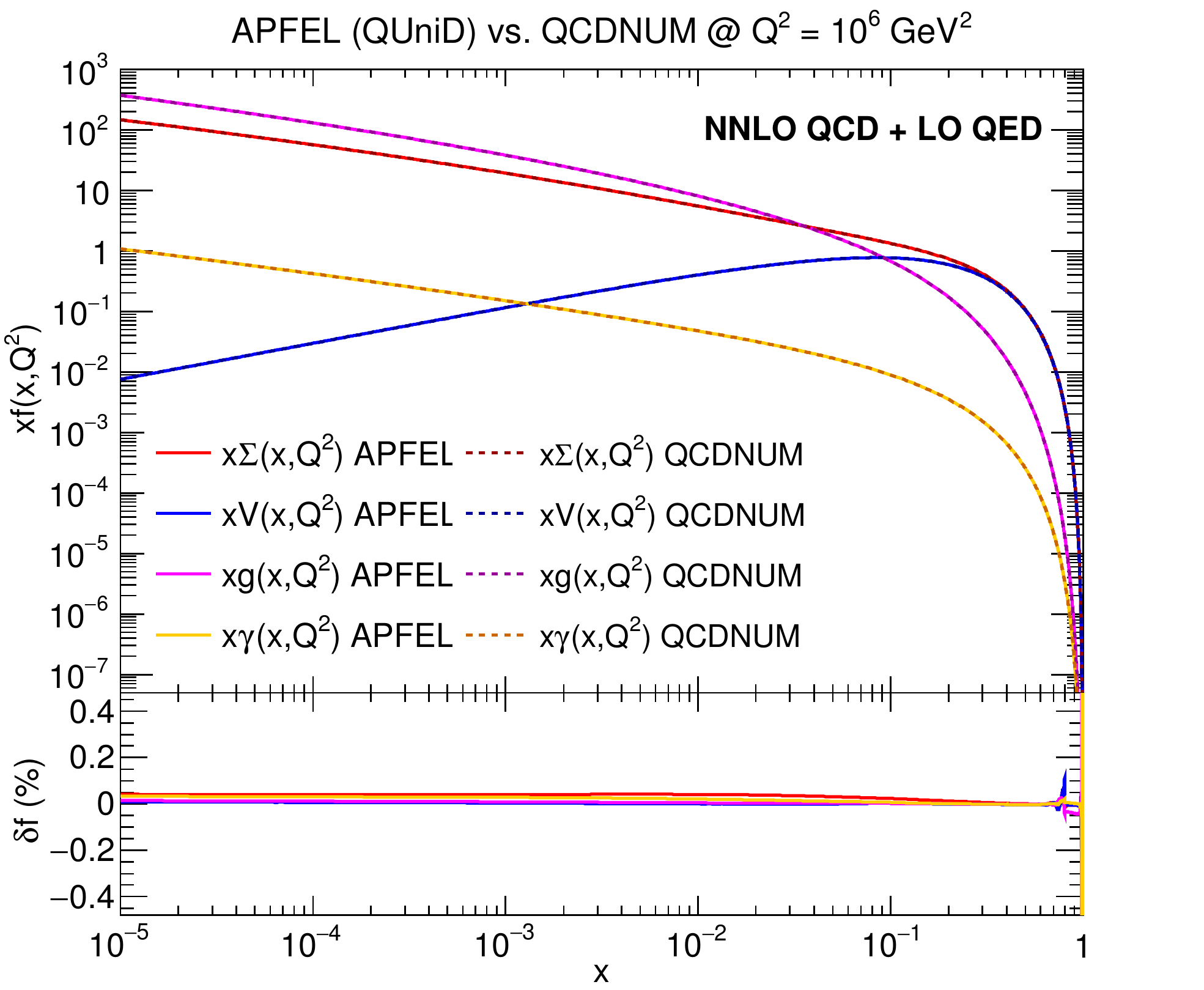}
\caption{Comparison between PDFs evolved using \texttt{APFEL} and
  \texttt{QCDNUM} evolution, from $Q_0^2=$2 GeV$^2$ up to $Q^2=$10$^6$
  GeV$^2$ at NNLO in QCD and LO in QED using the VFN scheme. The same settings of the PDF benchmark study of
Ref.~\cite{Blumlein:1996rp} have been used. We show PDFs in the
evolution basis presented in Sect.~\ref{sec:exactsol}.
}
\label{fig:APFELvsQCDNUM}
\end{figure}

In Figure~\ref{fig:APFELvsQCDNUM} we show the comparison of both codes
from $Q_0^2 = 2$ GeV$^2$ up to $Q^2 = 10^{6}$ GeV$^{2}$ at NNLO in QCD
and LO in QED, using the VFN scheme. The boundary condition for the
input PDFs are the same of the PDF benchmark study of
Ref.~\cite{Blumlein:1996rp}. In this case, instead of plotting the
single quark PDFs we have plotted the singlet and valence PDFs defined
in Sect.~\ref{sec:exactsol}. The level of agreement between
\texttt{APFEL} and \texttt{QCDNUM} is extremely good for all flavors,
we observe differences of 0.02\% at most in all cases.

In conclusion, the found a good level of agreement for all comparison
performed in this section. This guarantees that \texttt{APFEL}
implements correctly the QCD and QCD$\otimes$QED evolutions, therefore
it can be used in PDF fits.

\section{APFEL Web}
\label{sec:apfelweb}

We conclude this chapter by presenting \texttt{APFEL Web}, a spin-off
of the \texttt{APFEL} library, which has been ported to an online
centralized server system. This service is designed with the objective
to provide a fast and complete set of tools for PDF comparison,
luminosities, DIS observables and theoretical prediction computed
through the \texttt{APPLgrid} interface~\cite{Carli:2010rw} with an
user-friendly Web-application interface. The advantage of this system
resides on the possibility to setup PDF evolution in real time, and
perform quick comparison of the effects due to different
configurations. In this respect, \texttt{APFEL Web} provides also a
timely replacement to the \texttt{HepData} online PDF
plotter\footnote{\url{http://hepdata.cedar.ac.uk/pdf/pdf3.html}}.

\texttt{APFEL Web} is a Web-based application attached to a computer
cluster, available online at:
\begin{center}
  \textbf{\url{http://apfel.mi.infn.it/}~}
\end{center}
It contains PDF grids from \texttt{LHAPDF5} and \texttt{LHAPDF6}
libraries and it allows users to evolve PDFs using custom
configurations provided by the \texttt{APFEL} library. Computational
results are presented in the format of plots which are produced by the
\texttt{ROOT} framework.

This article is organized as follows. In Sect.~\ref{sec:design} we
document the application design and we explain the model scheme
developed for this project. In Sect.~\ref{sec:results} we discuss how
to use the Web-application and obtain results. Finally, in
Sect.~\ref{sec:conclusion} we present our conclusion and directions
for future work.

\subsection{Application design}
\label{sec:design}

The \texttt{APFEL Web} project is divided into two parts: the server-side
and the cluster-side. The separation is a real requirement because the
service needs to interact with multiple users and computational jobs
at the same time. In the following we start from the description of
the Web framework developed for the server-side and then we show how
the combination is performed.

\subsubsection{The Web framework and interface}

\begin{figure}
  \begin{centering}
  \includegraphics[scale=0.5]{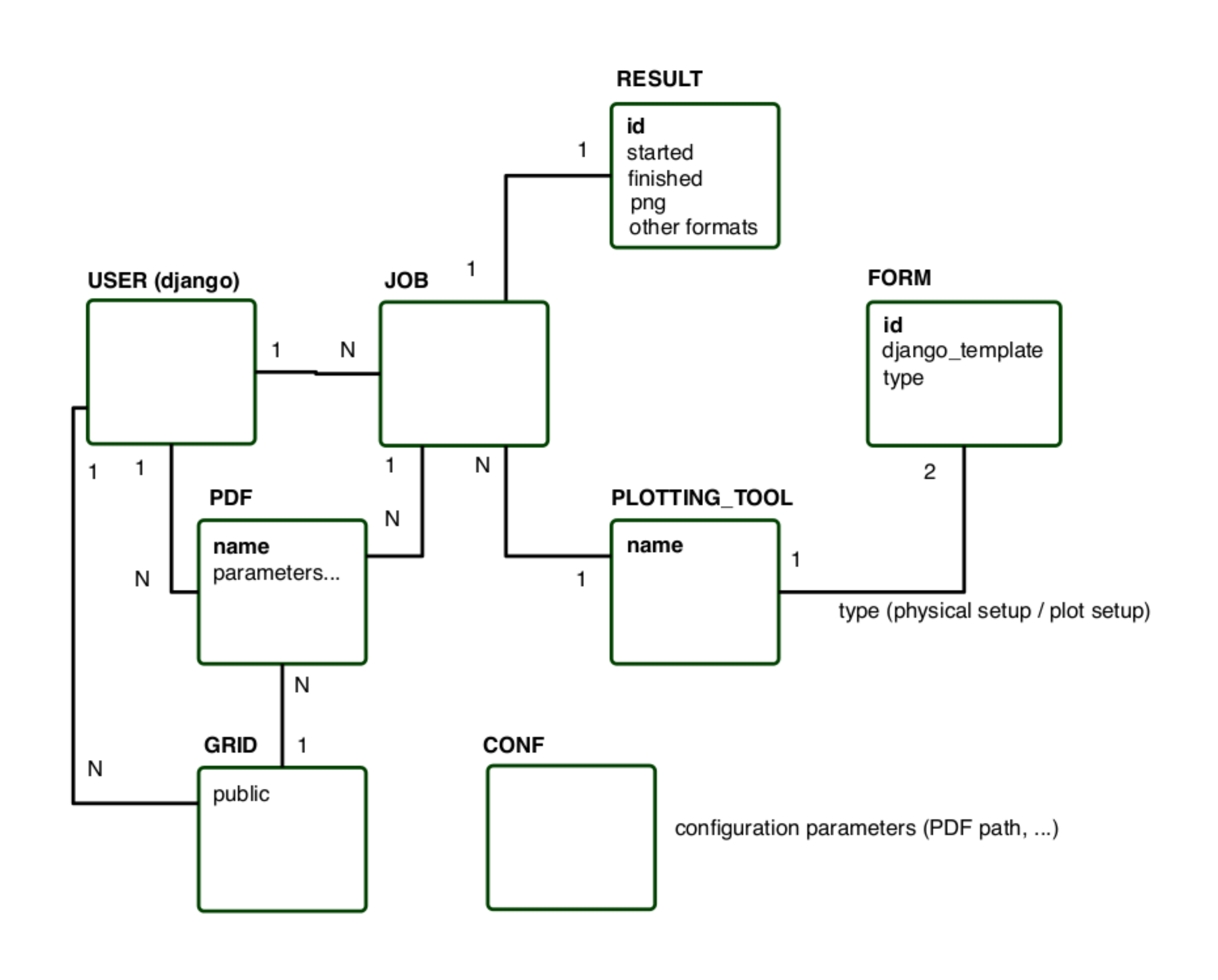}
  \par\end{centering}
\caption{A static design scheme of the \texttt{APFEL Web} application
  model. The boxes represent a simplified view of the main components
  of this Web-application. Solid lines with 1/N labels represent the
  one/many relationships for each component of the application. Labels
  inside the boxes are examples of the database entry keys associated
  to the model.}
\label{fig:model}       
\end{figure}

For the development of the Web interface we have used the \texttt{Django}
Web framework\footnote{\url{https://www.djangoproject.com/}}. \texttt{
  Django} is a high-level \texttt{Python} Web framework which provides a
high-performing solution for custom and flexible
Web-applications. Moreover the choice of \texttt{Python} as programming
language instead of \texttt{PHP} or \texttt{Java}, is motivated by the need
of a simple interface to interact with the server system, by
simplifying the implementation of the communication between server and
cluster sides.

Following the \texttt{Django} data model we have chosen to stored data in
a \texttt{PostgreSQL}\footnote{\url{http://www.postgresql.org}} database
which should provide a good performance for our query requirements. We
use the authentication system provided by the \texttt{Django} framework
in order to create a personal user Web-space, so users can save
privately personal configurations and start long jobs without need to
be connected over the whole calculation time.

In Figure~\ref{fig:model} we show a schematic view of the
Web-application model used in \texttt{APFEL Web}. Starting from the
top-left element, users have access to \texttt{PDF} objects which store
in the database the information about the PDF: e.g. the set name, the
PDF uncertainty treatment and the library for the treatment of PDF
evolution. Users have the option to choose PDF sets from the \texttt{
  LHAPDF} library or, if preferred, upload their own private grid
using the \texttt{LHAPDF5 LHgrid} and \texttt{LHAPDF6} formats. Users are
able to run jobs after setting up the PDF grid objects: there are seven
job types which are classified in the image as plotting tools and will
be described in detail in Sect.~\ref{sec:results}. For each plotting
tool there are customized input Web-forms, implemented with the \texttt{Django} \texttt{models} framework, which collect information and store
it in the database before the job submission.  When a job finalizes,
it stores images and \texttt{ROOT} files to the server disk, which are
then downloaded by the user. General configuration information such as
the path of the PDF grids and libraries are stored directly into the
\texttt{Django} settings. 

Concerning Web-security in \texttt{APFEL Web}, the user's account and its
information are protected by the \texttt{Django Middleware}
framework. Undesirable users, such as spambots, are filtered by a
security question during the registration form. Finally, all users
have a limited disk quota which disable job submissions when exceeded.

\subsubsection{Computation engine and server deployment}

\begin{figure}
  \begin{centering}
  \includegraphics[scale=0.4]{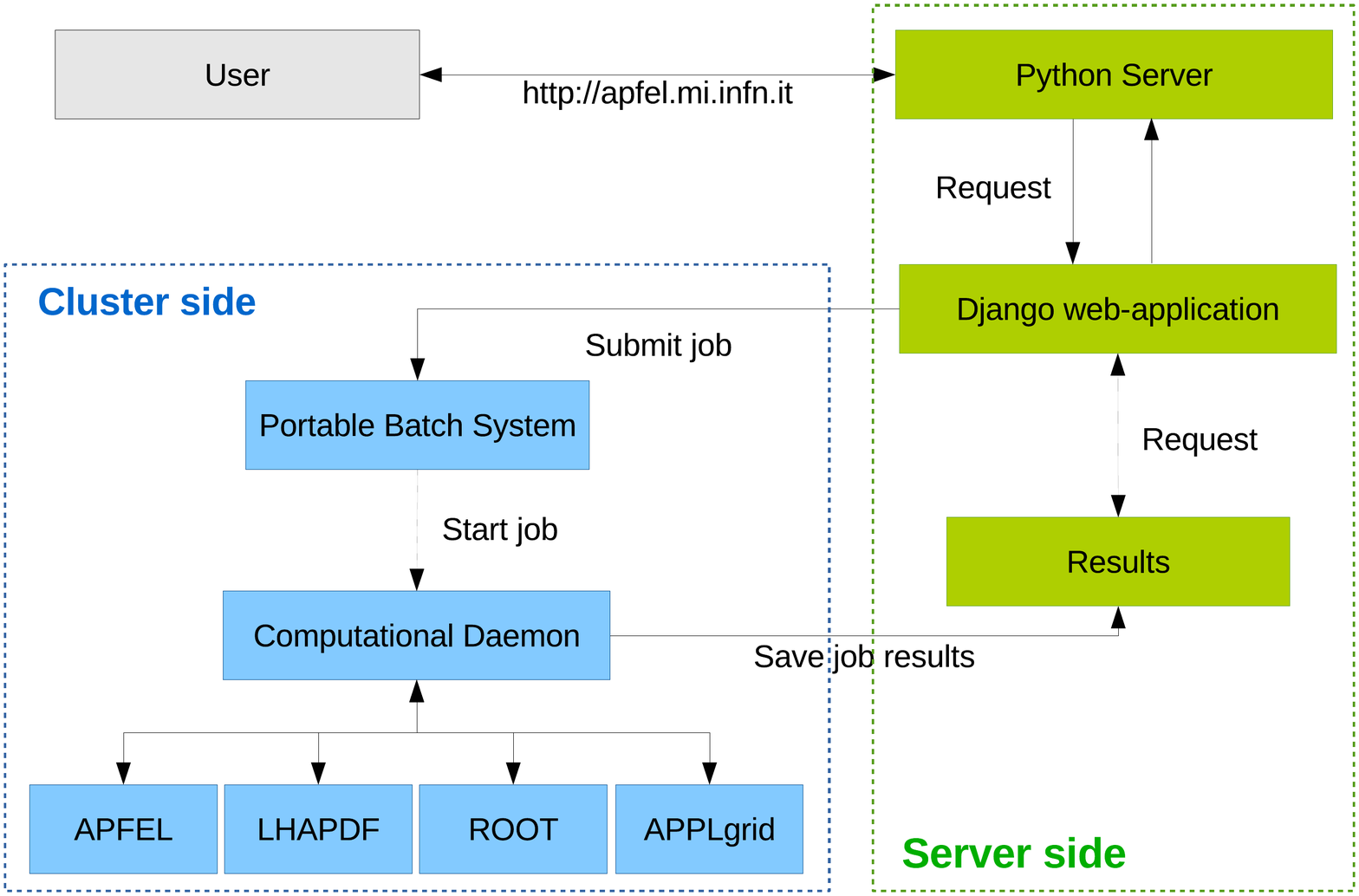}
  \par\end{centering}
\caption{Deployment layout of \texttt{APFEL Web}.}
\label{fig:system}       
\end{figure}

In parallel to the Web development, the most important component of
\texttt{APFEL Web} is the computational engine that we called \texttt{
  apfeldaemon}. The program is a generalization of the open source
\texttt{APFEL GUI} code in \texttt{C++} with the inclusion of the database
I/O procedures. The job configuration and the PDF grids are read from
the database, and the computation is performed upon request by the
user. In order to solve the problem correlated with the usage of two
different interfaces to PDF grids, i.e. \texttt{LHAPDF5} and \texttt{
  LHAPDF6}, the \texttt{apfeldaemon} is composed by two binaries which
are linked to the respective libraries: the Web-application checks the
PDF grid version and it starts the computation procedure with the
correct daemon.

In Figure~\ref{fig:system} we show the scheme of the Web-application
structure. Users from Web browsers send requests to a \texttt{Python}
server which in our case is implemented by \texttt{
  gunicorn} and \texttt{nginx}\footnote{\url{http://gunicorn.org} and \url{http://nginx.org/}}. The \texttt{Python} server
performs the request using the \texttt{Django} framework, at this level
PDF objects and jobs are prepared and saved in the database,
additionally eventual job results are collected in a dedicated
view. From the computational point of view the layout is very simple
and clearly illustrated by the left side of Fig.~\ref{fig:system}. We
have set up a Portable Batch System (PBS)\footnote{An example of PBS
  open source implementation: \url{http://www.adaptivecomputing.com/}}
for the multi-core server which receives job submissions and is able
to automatically handle the job queue, avoiding the unpleasant
situation of server overloading. Jobs are submitted by the \texttt{
  Django} application which passes the job identification number, this
value is read by the \texttt{apfeldaemon} and it performs a query at the
corresponding database entry, then it collects the relevant
information to start the correct job. When the job finalizes the \texttt{
  apfeldaemon} modifies the job status in the database, so the Web
interface notifies the user of the job status.

The \texttt{apfeldaemon} program was designed and compiled with
performance as priority, in fact there are relevant computational
speed improvements when comparing to the previous \texttt{APFEL GUI}
program almost due to the clear separation between the \texttt{GUI} and
the calculation engine. In order to provide to the reader an idea of
the typical processing time per job, we estimate that one job requires
two seconds to process a single PDF set when producing a PDF
comparison plots, meanwhile for luminosity and observables jobs, the
system takes up to one minute per PDF set when including the
uncertainty treatment.

\subsection{Plotting tools}
\label{sec:results}

\begin{figure}
  \begin{centering}
    \includegraphics[scale=0.3]{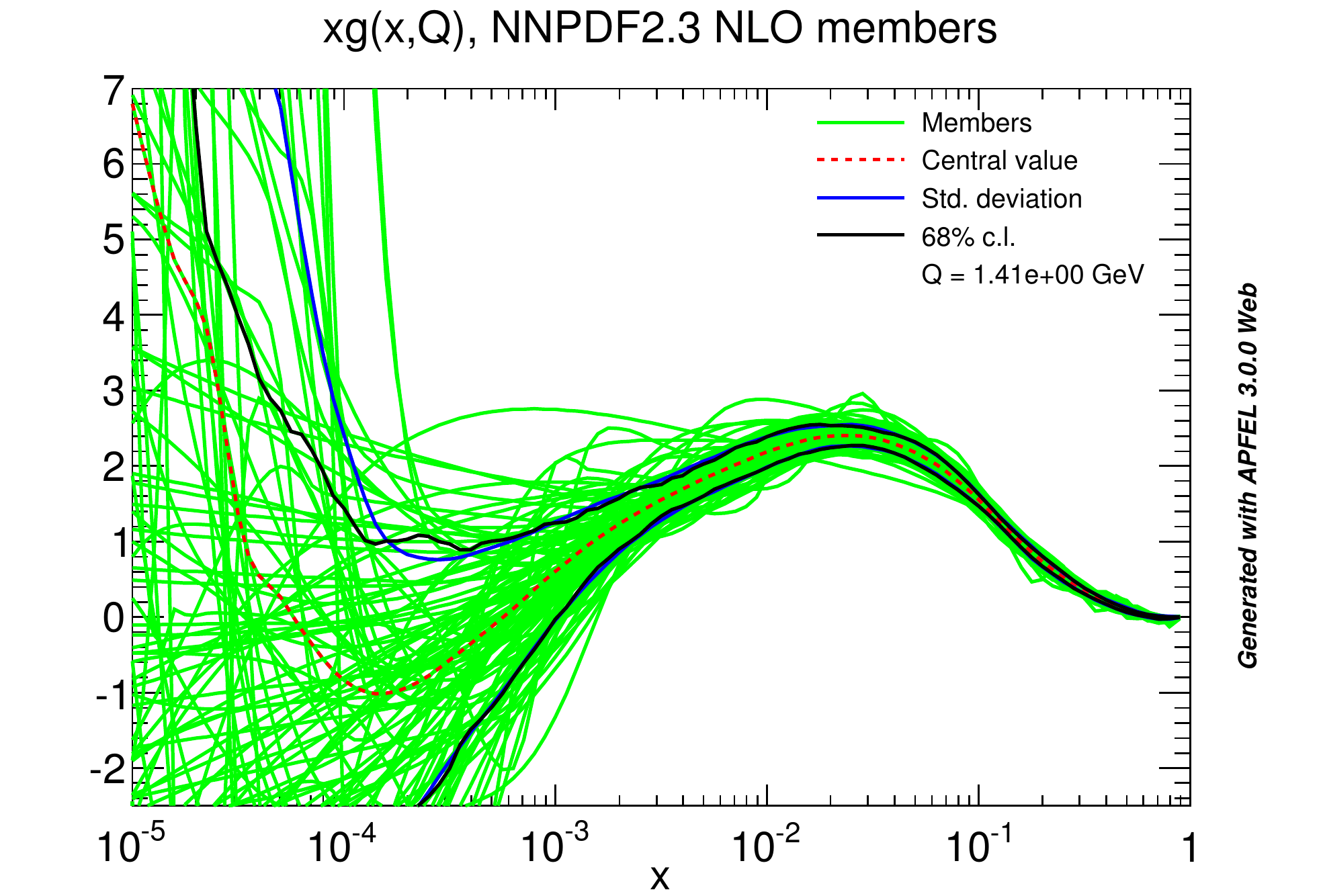}\includegraphics[scale=0.3]{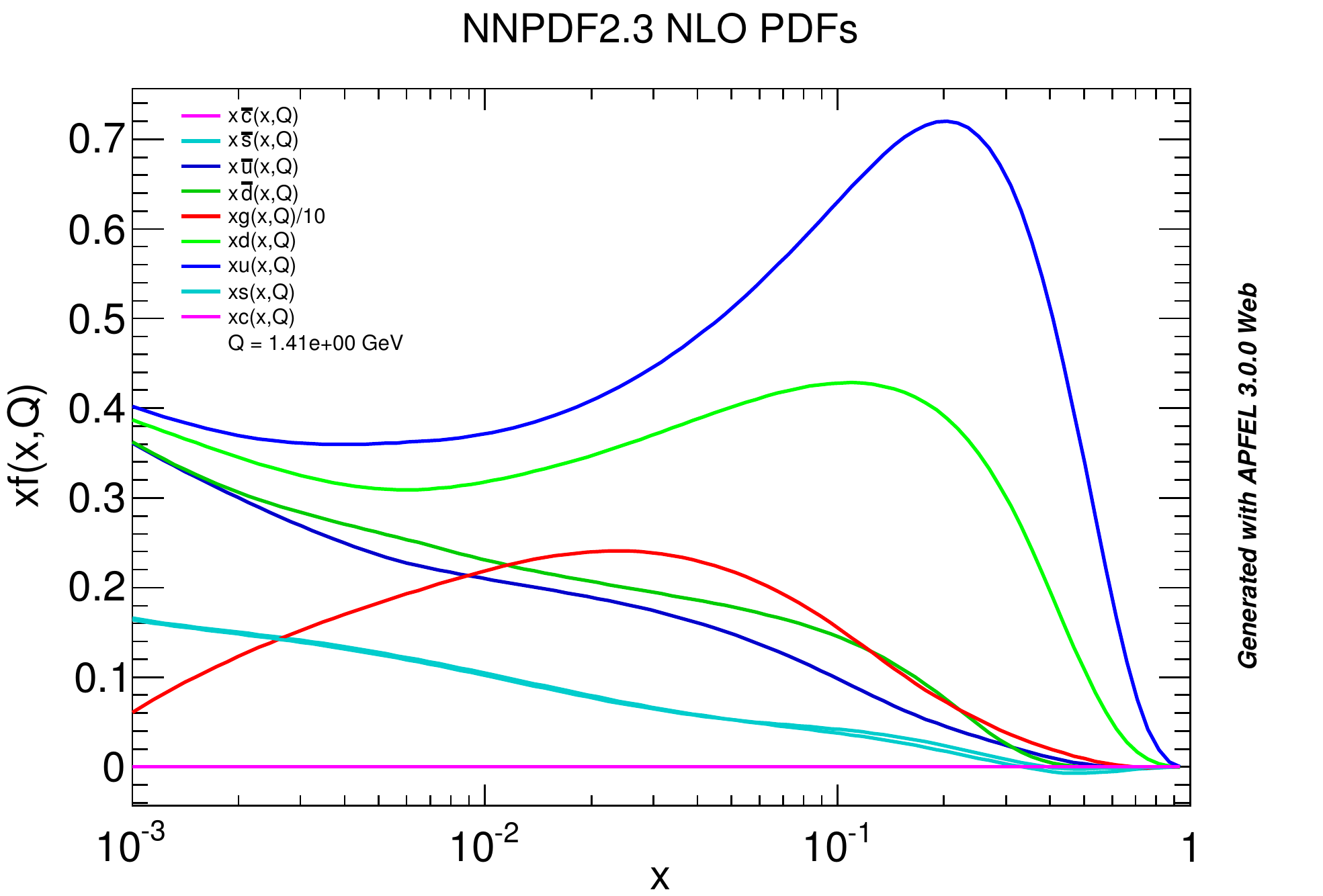}
    \includegraphics[scale=0.3]{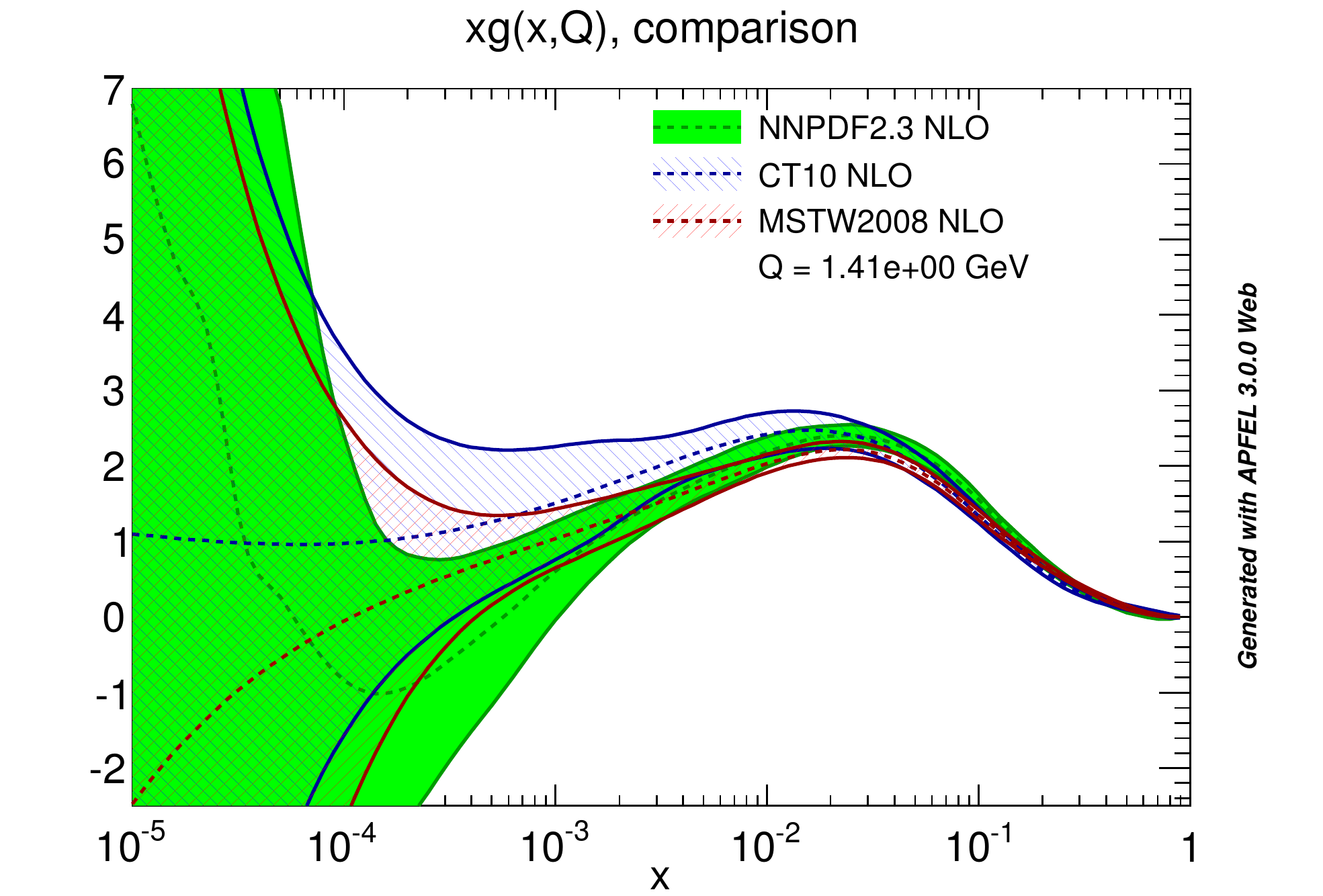}\includegraphics[scale=0.3]{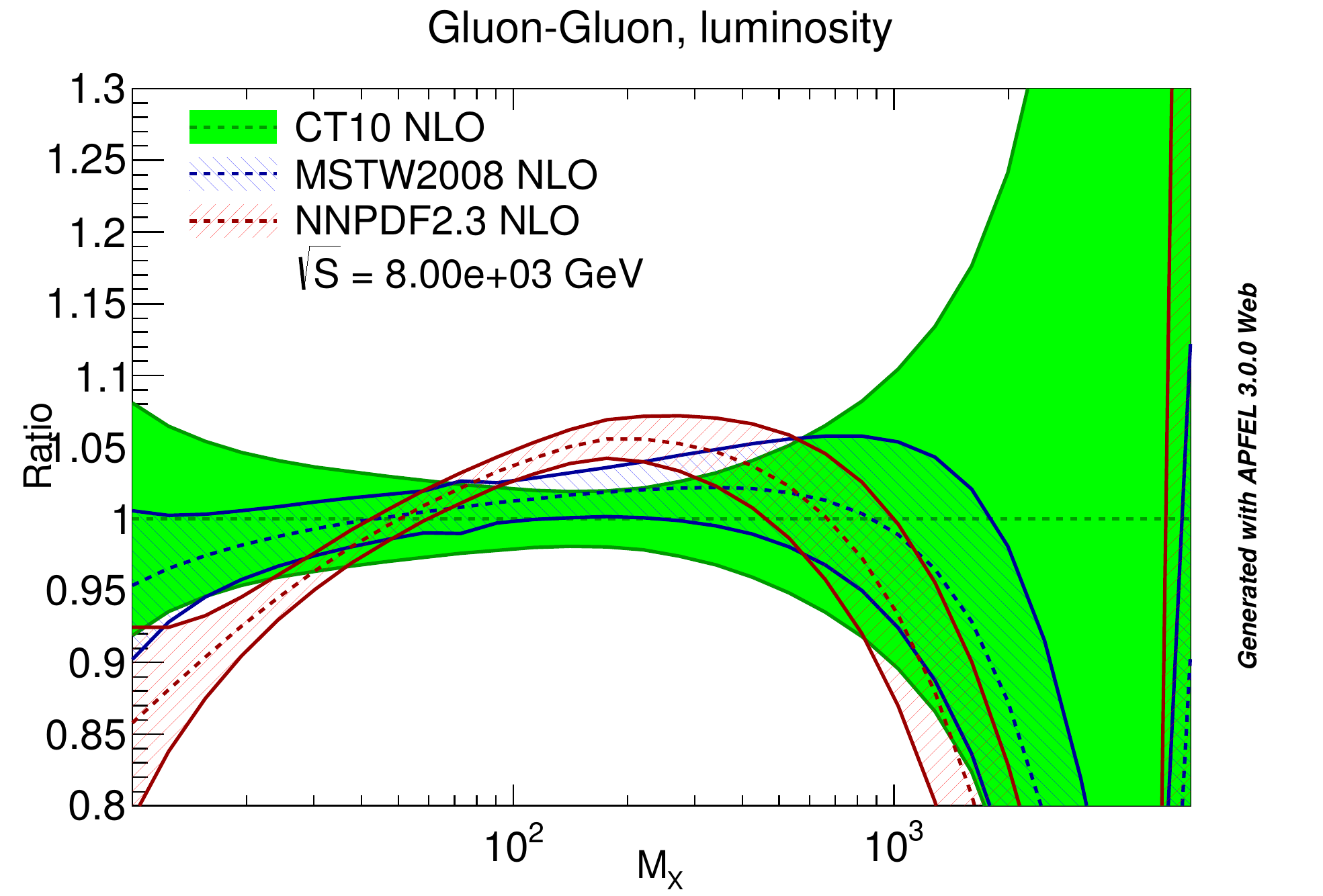}
    \includegraphics[scale=0.3]{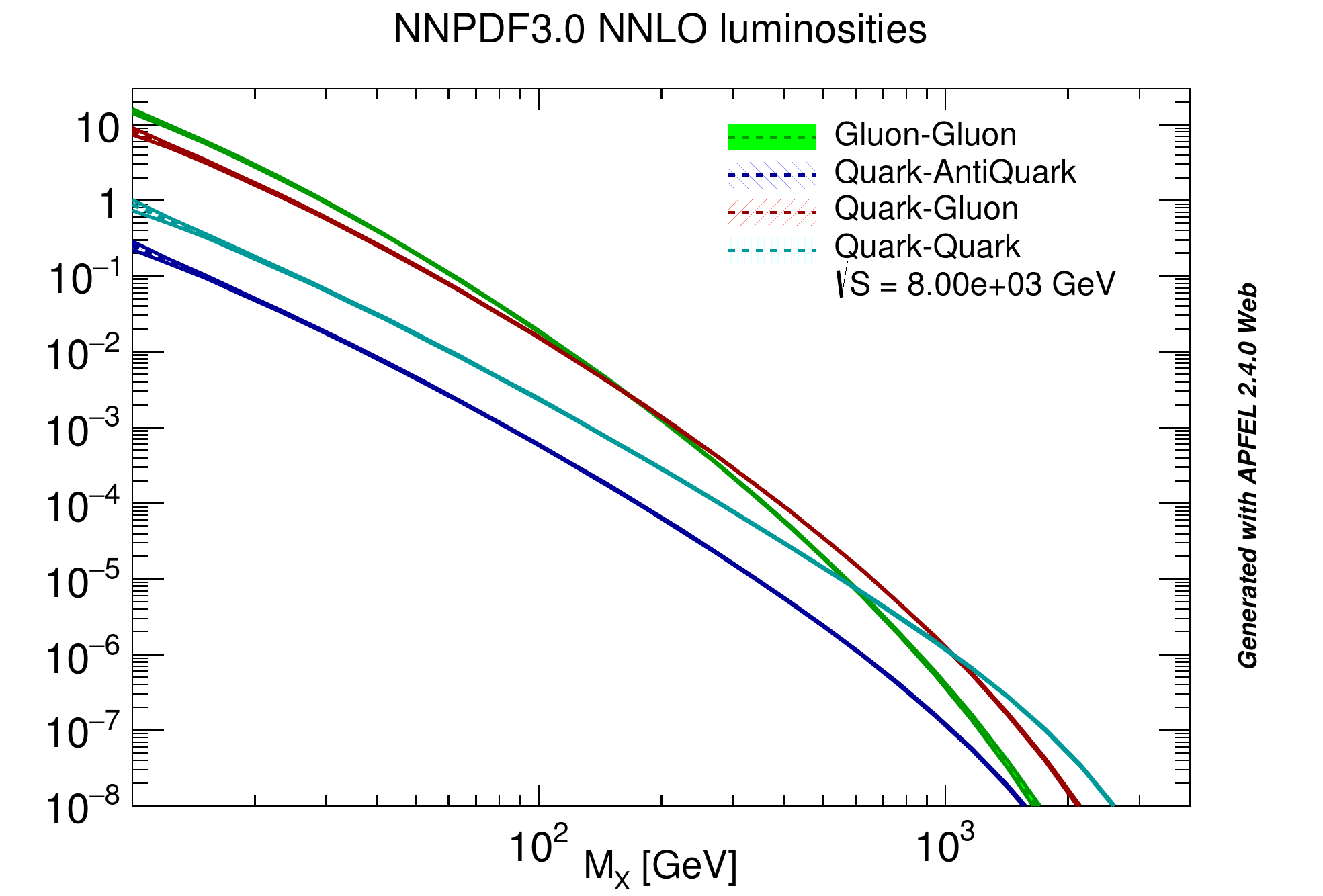}\includegraphics[scale=0.3]{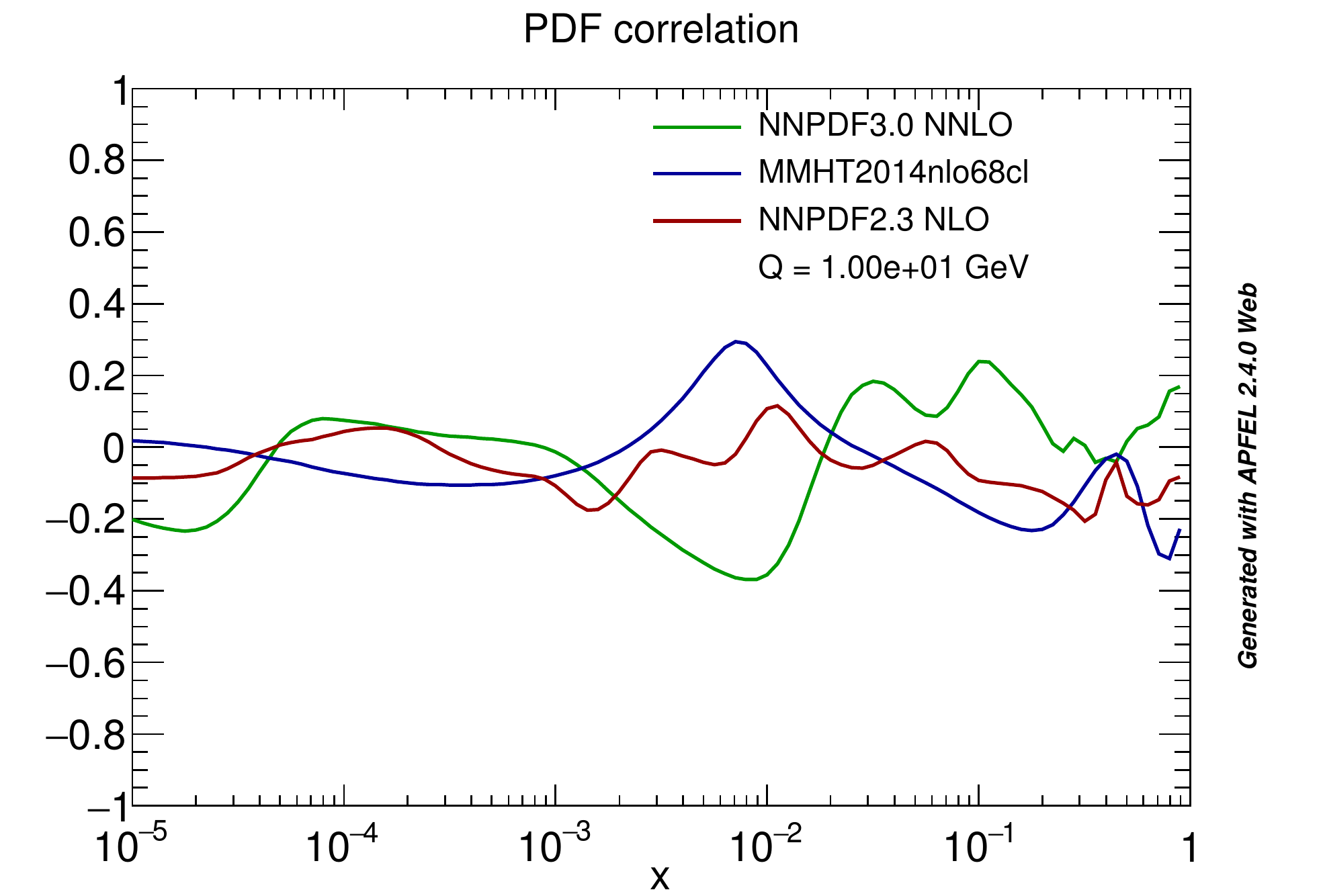}
    \includegraphics[scale=0.25]{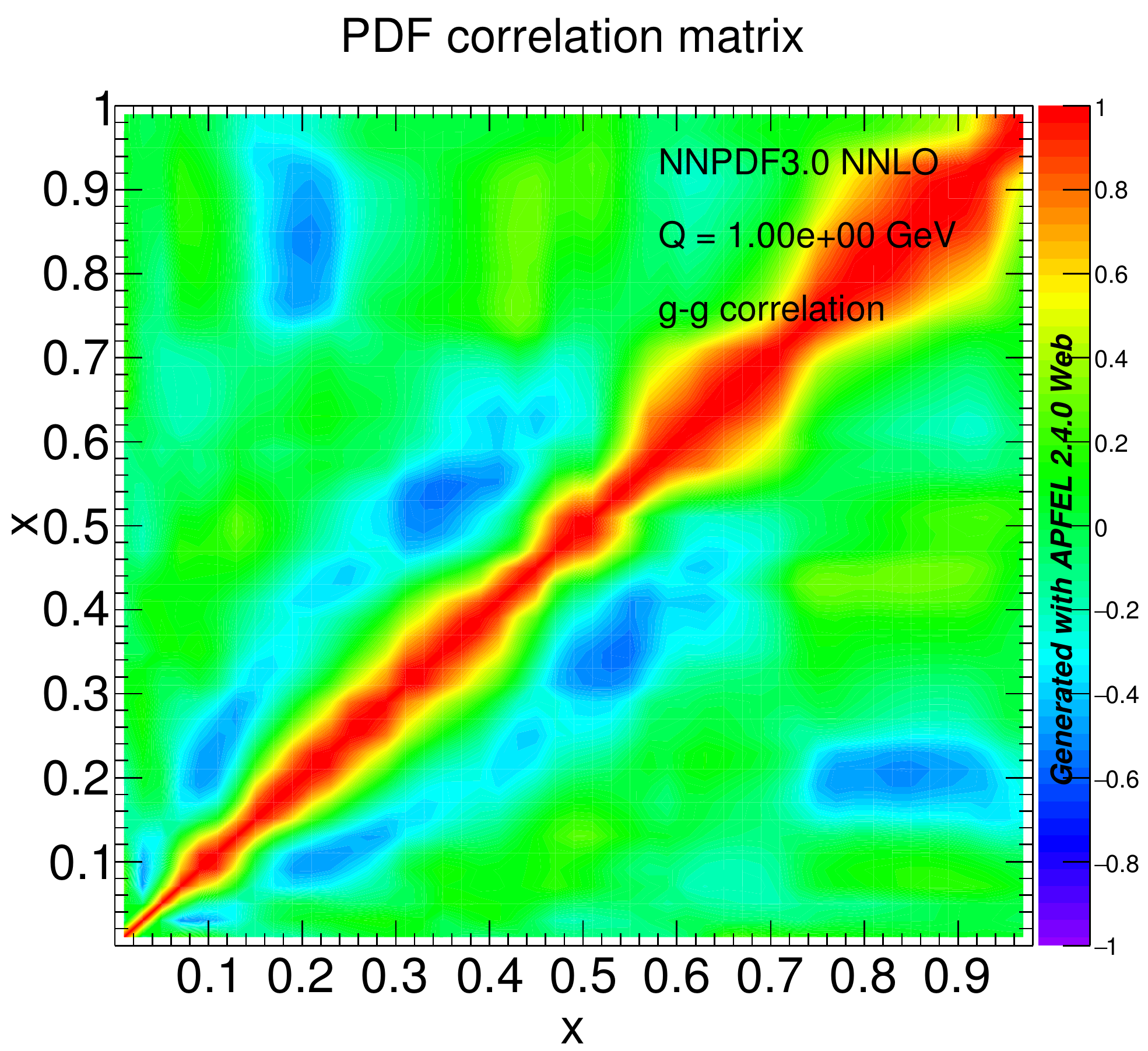}
    \par\end{centering}
  \caption{Examples of output generated with \texttt{APFEL Web}. Plots
    are presented in the following order, clockwise from top-left:
    PDF members, multiple PDF flavors, PDF comparison in $x$, $gg$-channel
    luminosity, all luminosities, PDF correlations and correlation matrix.}
\label{fig:results}       
\end{figure}

While the use of the Web-interface should be self-explanatory, here we
describe and show examples of job results that a user is able to
obtain from \texttt{APFEL Web}.

The first step consists in the creation of custom ``PDF objects'' in
the user's workspace. The following points explain how to create such
objects:
\begin{enumerate}

\item select a PDF grid from the \texttt{LHAPDF5} and/or \texttt{LHAPDF6}
  libraries and determine the treatment of the PDF uncertainty among:
  no error, Monte Carlo approach, Hessian eigenvectors (68 and 90\%
  c.l.)  and symmetric eigenvectors. When selecting a PDF set the
  system proposes automatically an uncertainty type based on the PDF
  collaboration name.

\item import a new \texttt{LHAPDF} grid file, with the only requirement
  that it is provided either in the \texttt{LHAPDF5 LHgrid} or in the
  \texttt{LHAPDF6} format. The main target for this feature are the
  members of the PDF collaborations which can perform comparisons with
  preliminary sets of PDFs before the publication in \texttt{LHAPDF}.

\item set the evolution library by choosing between the \texttt{LHAPDF}
  interpolation routines or the \texttt{APFEL} custom evolution.

\end{enumerate}

\begin{figure}
  \begin{centering}
    \includegraphics[scale=0.3]{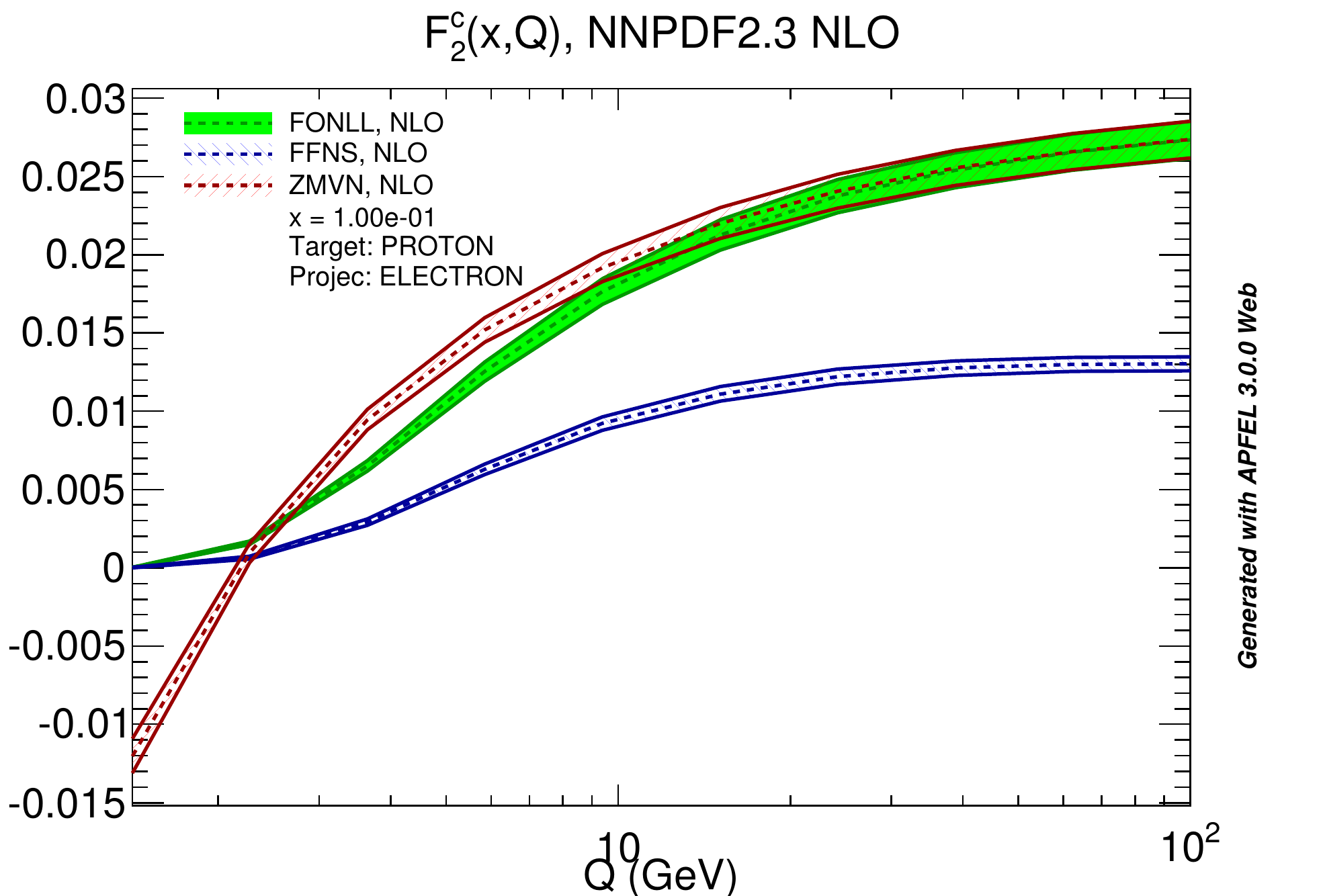}\includegraphics[scale=0.3]{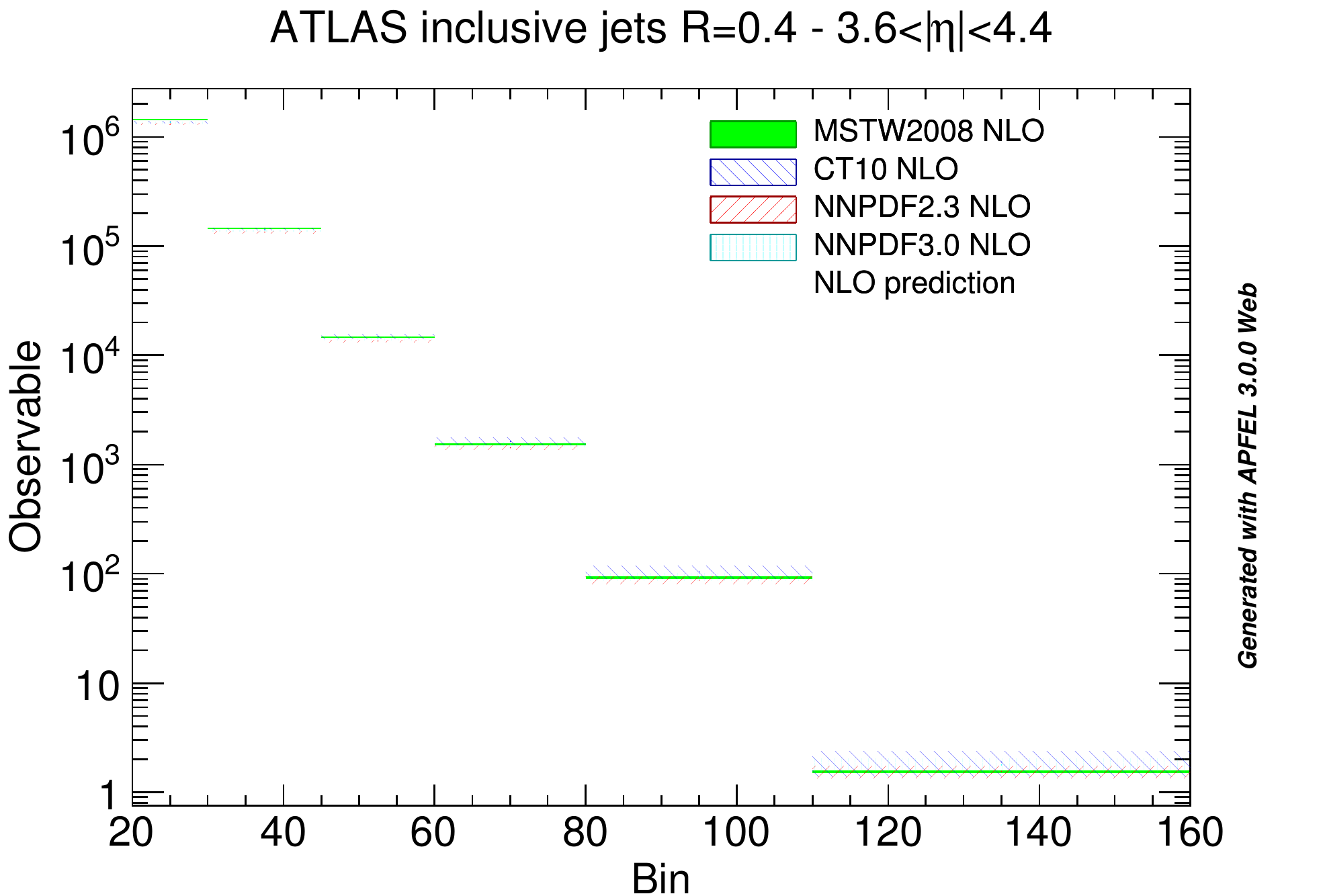}
  \par\end{centering}
\caption{On the left, an example of DIS observable computed by
  \texttt{APFEL Web}: $F^{c}_{2}(x,Q)$. On the right, another example
  about the \texttt{APPLgrid} observables tool used for the
  computation of predictions for ATLAS 2010 inclusive jets $R=0.4$ at
  $\sqrt{s}=7$ TeV~\cite{Aad:2011fc}.}
\label{fig:results2}       
\end{figure}

We provide the following computational functions, which are
illustrated in Figures~\ref{fig:results} and~\ref{fig:results2}:
\begin{itemize}

\item ``\texttt{Plot PDF Members}'': it plots for projections in $x$ all
  the members of a PDF set for a single parton flavor at a given
  energy scale $Q$. See the top-left image in Fig.~\ref{fig:results}
  where we show the replicas of \texttt{NNPDF2.3 NLO}~\cite{Ball:2012cx}
  together with its central value and Monte Carlo uncertainty band,
  these last features are options which can be disabled by the
  user. This plotting tool accepts only a single PDF set at each time
  in order to avoid too many information in a single plot. We provide
  the possibility to choose between the usual parton flavors,
  i.e. $b,t,c,s,d,u,g,\gamma,q_i^{\pm}=q_i\pm\bar{q_i}$ with $q_i=u,d,s,c,b,t$, and the combination of them:
  ($\Sigma,V,V_{3},V_{15},V_{24},V_{35},T_{3},T_{15},T_{24},T_{35},\Delta_{s}$)~\cite{Ball:2011uy},
  the so called evolution basis.
  
\item ``\texttt{Plot multiple PDF flavors}'': each PDF flavor is plotted
  together in the same canvas at a fixed energy scale. We also provide
  the possibility to scale PDF flavors by a predetermined numeric
  factor in order to produce plots similar to the
  PDG~\cite{Agashe:2014kda}. An example of PDF flavor plot is
  presented in the top-right of Fig.~\ref{fig:results} where the gluon
  PDF is scaled by a factor 10.
 
\item ``\texttt{Compare PDFs in $x$}'': this tool compares the same
  flavor of multiple PDF sets and the respective uncertainties at a
  given energy scale for projections in $x$. We provide the
  possibility to compute the absolute value or the just the ratio
  respect to a reference PDF set. The second row left image of
  Fig.~\ref{fig:results} shows the comparison between \texttt{NNPDF2.3
    NLO}~\cite{Ball:2012cx}, \texttt{CT10 NLO}~\cite{Lai:2010vv} and
  \texttt{ MSTW2008 NLO}~\cite{Martin:2009iq} sets at $Q=1$ GeV.

\item ``\texttt{Compare PDFs in $Q$}'': this tool compares the same
  flavor of multiple PDF sets and the respective uncertainties at a
  fixed $x$-value as a function of the energy scale $Q$.

\item ``\texttt{Compare PDF Luminosity}'': it performs the computation
  of parton luminosities~\cite{Ball:2010de} normalized to a reference
  PDF set at a given center of mass energy. There are several channels
  available: $gg$, $q\bar{q}$, $qg$, $cg$, $bg$, $qq$, $c\bar{c}$,
  $b\bar{b}$, $\gamma \gamma$, $\gamma g$, etc. In the second row
  right plot of Fig.~\ref{fig:results} we show an example of
  $gg$-luminosity at $\sqrt{s}=8$ TeV using the PDF sets presented
  above with \texttt{CT10 NLO} as reference PDF set.

\item ``\texttt{All PDF Luminosities}'': for a given set of PDFs this
  tool compares the $gg$, $q\bar{q}$, $qg$ and $qq$ luminosities in a
  single plot. The third row left image of Fig.~\ref{fig:results}
  shows an example of the output for \texttt{NNPDF3.0 NLO} at
  $\sqrt{s}=8$ TeV.

\item ``\texttt{Compare PDF Correlations}'': it performs the
  comparison of PDF correlations between pairs of PDFs flavors for
  multiple sets of PDFs. The correlation coefficients are obtained
  through the \texttt{LHAPDF6} interface. The third row right image of
  Fig.~\ref{fig:results} shows an example of the output for this
  plotting tool.

\item ``\texttt{PDF Correlations Matrix}'': for a given set of PDFs
  this tool computes the correlation matrix for pairs of PDF flavors
  in a grid of $x$-points. The correlation coefficients are computed
  automatically through the \texttt{LHAPDF6} interface. An example of
  such tool is shown in the bottom image of Fig.~\ref{fig:results}.

\item ``\texttt{DIS in $x$/DIS in $Q$}'': it computes DIS observables as
  functions of $x$ or $Q$ for different heavy quark schemes and
  perturbative orders, including the Fixed Flavor Number scheme
  (FFNS), the Zero Mass Variable Number scheme (ZMVN), and the FONLL
  scheme~\cite{Forte:2010ta} where the choice of a NLO prediction
  implies using the FONLL-A scheme, while choosing NNLO leads to using
  the FONLL-C scheme. A detailed explanation of all possible
  configurations is presented in Sect.~4.3 of
  Ref.~\cite{Bertone:2013vaa}. An example of such tool is presented in
  the left plot of Fig.~\ref{fig:results2}.

\item ``\texttt{APPLgrid observables}'': this tool provides a simple a
  fast interface to theoretical predictions through the \texttt{APPLgrid}
  library~\cite{Carli:2010rw}. The system already provides several
  grids that are available from the \texttt{APPLgrid}
  website\footnote{\url{http://applgrid.hepforge.org/}} but also from
  the NNPDF collaboration~\cite{Ball:2014uwa} and \texttt{
    aMCfast}~\cite{Bertone:2014zva}.  This function allows users to
  compute the central value and the respective uncertainties for
  multiple PDF sets. On the right plot of Fig.~\ref{fig:results2} we
  show the output of this tool for the predictions of ATLAS 2010
  inclusive jets $R=0.4$ at $\sqrt{s}=7$ TeV~\cite{Aad:2011fc}.

\end{itemize}

For all the tools presented above, the Web interface provides options
for customizing the graphics, like setting the plot title, axis
ranges, axis titles and curve colors. \texttt{APFEL Web} also provides
the possibility to save plots and the associated data in multiple
formats, including: PNG, EPS, PDF,~.C (\texttt{ROOT}) and~.root (\texttt{ROOT}).

Finally, it is important to highlight that the results produced by
\texttt{APFEL Web} for PDF comparison and parton luminosities from
different PDF sets have been verified against the PDF benchmarking
exercise of Ref.~\cite{Ball:2012wy}.

\subsection{Usage statistics}
\label{sec:conclusion}

\begin{figure}
  \begin{centering}
    \includegraphics[scale=0.3]{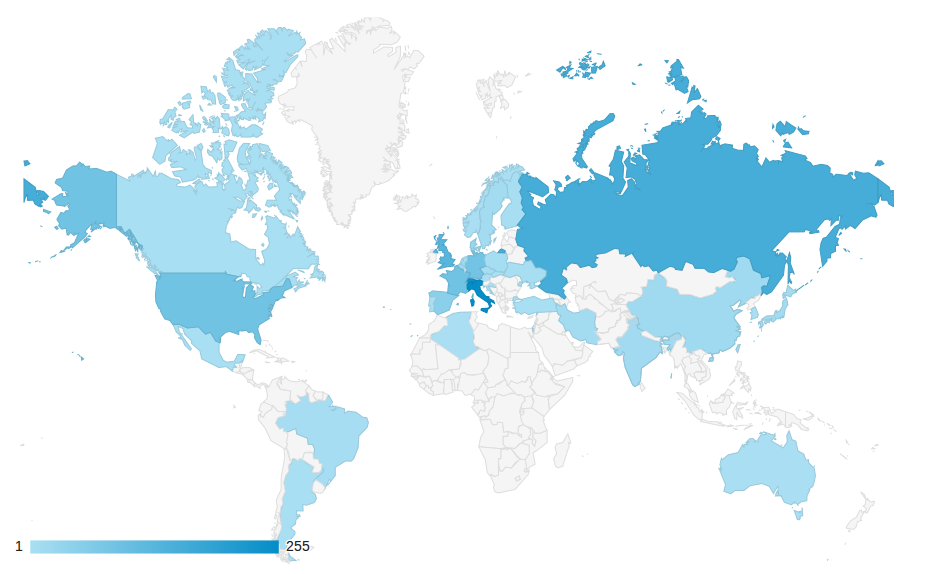}
    \par\end{centering}
  \caption{Unique sessions by country from October 2014 to March
    2015 (1293 visits).}
  \label{fig:visits}
\end{figure}

The \texttt{APFEL Web} application was released on October 7,
2014. Five months after the release we already have 131 registered
users from 20 countries, and an average of 258 visits each month. In
Figure~\ref{fig:visits} we show a pictorial representation of the
total unique visits by country during the period between the release
date to March 2015.

\begin{figure}
  \begin{centering}
    \begin{tabular}{c|c}
    \includegraphics[scale=0.36]{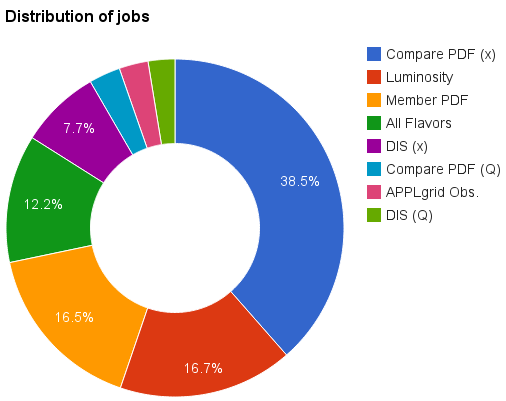}
    &
    \includegraphics[scale=0.35]{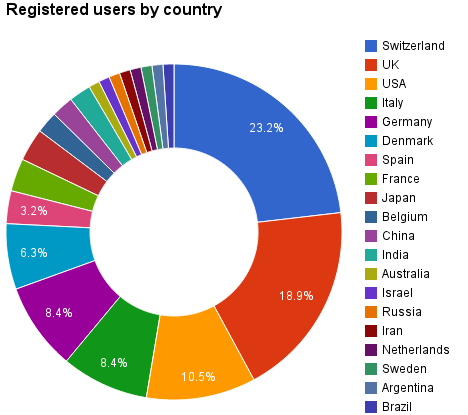}
  \end{tabular}
    \par\end{centering}
  \caption{On the left plot, the distribution of the plotting tools
    selected by the users when sending jobs. On the right plot, the
    fraction of the registered users organized by country. In both
    cases, the legend elements are organized in descending order. The
    results refer for the period from October 2014 to March 2015.}
  \label{fig:userjobhisto}
\end{figure}

At the current date, the server has successfully completed more than
3500 jobs. In the left plot of Figure~\ref{fig:userjobhisto} the
distribution of jobs selected by the users is shown in
percentages. The PDF comparison and luminosity are the most popular
tools, followed by PDF members and all PDF flavors plots. The right
plot of Figure~\ref{fig:userjobhisto} presents a pie chart with the
country affiliation of users registered in the \texttt{APFEL}
website. Top users are from Switzerland (mainly from CERN), UK, USA
followed by users spread across all continents. These results,
obtained in a relative short time period, are rewarding showing that
there is an international community of physicist interested in the
features provided by \texttt{APFEL Web}.

\begin{figure}
  \begin{centering}
    \includegraphics[scale=0.36]{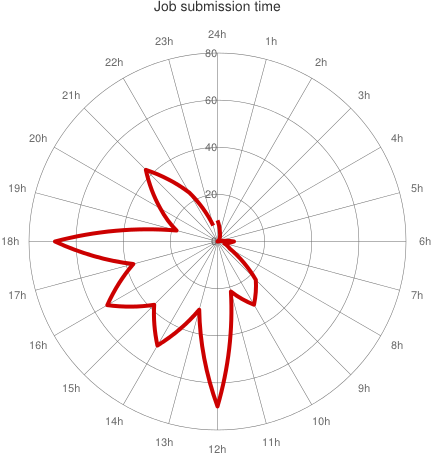}
    \par\end{centering}
  \caption{Number of jobs submitted in function of the local time of day.}
  \label{fig:timejob}
\end{figure}

Finally, in Figure~\ref{fig:timejob} we show the local time of the day
preferred by users for the submission of jobs. The polar axis shows
the time of day, meanwhile the radius the total number of job
submissions. There are two peaks of activity, the first at 12am and
the second at 6pm. Furthermore we observe a continuous operation cycle
from 9am to 10pm. Possibly, these results can be interpreted as
another advantage of having an online server interface, accessible
from any device connected to internet at any time.

Thanks to its flexibility and user-friendliness, we believe that in
the coming months and years \texttt{APFEL Web} has the potential to
become a widely used tool worldwide.

\chapter{The NNPDF methodology}
\label{sec:chap3}

In this chapter we present the Neural Network Parton Distribution
Function (NNPDF) methodology. We provide an overview of the NNPDF
approach to PDFs, which is then used in Chapter~\ref{sec:chap4} for
the determination of a set of PDFs with QED corrections. In
Sect.~\ref{sec:nnmethodology} we begin with the description of the
NNPDF methodology which is then followed in
Sect.~\ref{sec:methodology} by a technical presentation of the new
modern framework developed specifically for this project and for the
next generation of NNPDF fits. In Sect.~\ref{sec:nnpdf23}, we conclude
the chapter with the characterization of the NNPDF2.3 set of PDFs,
which is the baseline configuration used in the fit with QED
corrections.

\section{Introduction to NNPDF}
\label{sec:nnmethodology}

The NNPDF Collaboration is the only group which implements the Monte
Carlo approach to a global fit of PDFs instead of the usual Hessian
method. The goal of this strategy is to provide an unbiased
determination of PDFs with reliable uncertainty. The approach
implemented in NNPDF is based on advanced computational techniques,
such as:
\begin{itemize}
\item The Monte Carlo treatment of experimental data.
\item The parametrization of PDFs with artificial neural networks.
\item The minimization strategy based on Genetic Algorithm.
\end{itemize}

In an initial step, the original experimental data is transformed into
a Monte Carlo ensemble of replicas. In this procedure, the ensemble of
artificial data replicas follows a multi-Gaussian distribution
centered around the central value of each data point and with the
variance based on the statistical, systematic and normalization
uncertainties, encoded in the experimental covariance matrix. The
total number of replicas is selected in such a way that it is large
enough to produce the statistical properties of the original data to
the desired accuracy.

Each of the Monte Carlo data replica is then fitted by PDFs
parametrized with artificial neural networks (ANN). The use of ANNs
instead of selecting a specific functional forms, \textit{e.g.}~based
on polynomial, guarantees no bias due to the parametrization. In fact,
neural networks with large architectures are able to imitate the
behavior of any functional form.

The last stage of the NNPDF methodology is the fitting strategy. As in
any other fitting procedure, we define a figure of merit which
compares the theoretical predictions of physical observables, obtained
through the convolution of PDFs, to the respective data replica. In
this case, as we are dealing with a large number of parameters, the
ANNs are trained by a Genetic Algorithm (GA) which shows a good
performance in comparison to algorithms based on Newton's methods.

The Monte Carlo representation of the underlying probability density
associated to a given set of PDFs has several advantages as compared
with the traditional Hessian approach. The most important advantage of
the MC method is that it does not require the selection of a fixed
functional form. This feature lets discard any bias associated to the
PDF parametrization. Moreover, it also does not assume that the
underlying PDF uncertainties are Gaussian, as the Hessian method does,
and so, it does not rely on the linear approximation to propagate
uncertainties from the original data to the PDFs.
Technical details about each of the previous points will be addressed
in the next section.

\section{A modern implemention of the NNPDF framework}
\label{sec:methodology}

We show the details of the NNPDF methodology from the point of view of
the implementation of a new code framework.
The main motivation for updating the NNPDF code resides on the need
for flexibility and performance. There are several advantages in
reformulating the methodology in a modern object-oriented
approach. First of all, the possibility to have more expressiveness,
which allows the inheritance of data structures, introducing layers of
abstraction between several components of the code.
From the NNPDF practical point of view, this strategy is translated by
a huge simplification of the framework, where data, theory and fitting
are completely independent elements, which can be easily extended and
optimized.
These technical advantages reflect an easy a fast development of
specific projects, for example the QED corrections to PDFs, as
presented here, the determination of Nuclear PDFs and Fragmentation
Functions~\cite{Bertone:2015cwa}.

On the other hand, with the current inclusion of a substantial number
of LHC datasets in a global PDF determination, we face performance
issues due to the complexity of adding new hadronic observables into
the fitting framework.
These issues reflect an increasingly computational cost of running
fits. This trend is supposed to grow in the next years, due to future
new LHC measurements.
The main cause of these performance issues resides on the NNPDF
computationally intensive Genetic Algorithm minimization.
So, in order to deal with such problems, we have developed a modern
fitting code based on two object-oriented languages: \texttt{C++} and
\texttt{Python}. This choice, as already mentioned before, allows the
inclusion of new datasets achieving a highly efficient implementation
of the minimization algorithms which is not possible to achieve in the
previous \texttt{Fortran77} implementation.

In what follows we describe the technical choices and code structure
of the new code framework through the description of the NNPDF
methodology. In this thesis, we focus on the NNPDF2.3 setup because
the QED corrections have been applied to this fitting configuration
using a preliminary version of the updated framework.
However, note that the NNPDF Collaboration have recently presented a
new set of PDFs, the NNPDF3.0~\cite{Ball:2014uwa}, where the new
framework is used by default.
 
\subsection{Data treatment}
\label{sec:nnpdfdata}
\begin{figure}
  \centering
  \includegraphics[scale=0.4]{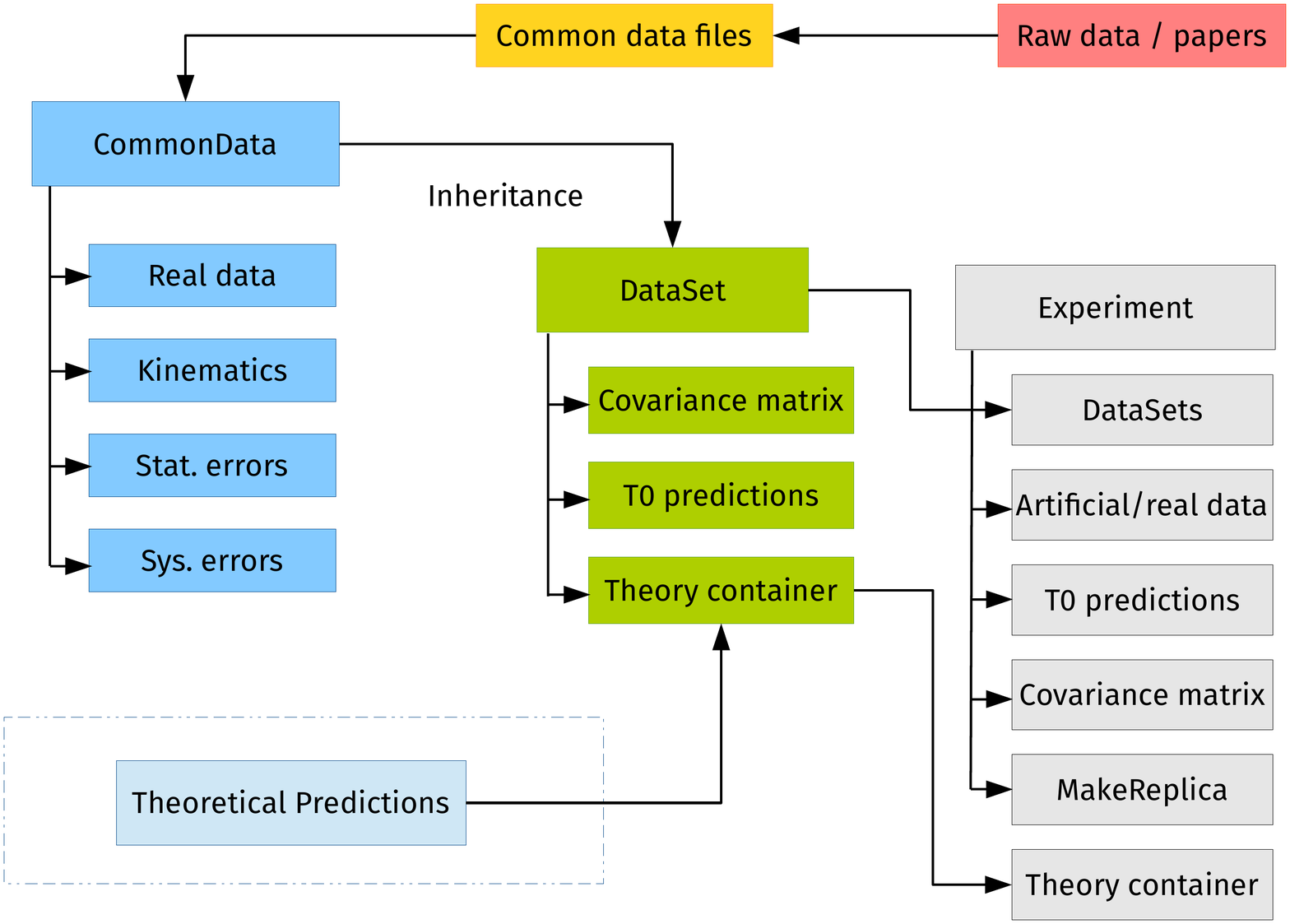}
  \caption{Pictorial representation of the NNPDF data management
    layout. The dashed blue box indicates a simplification in the
    diagram.}
  \label{fig:datascheme}
\end{figure}

The implementation of the Monte Carlo artificial data generation
starts from the construction of the experimental covariance
matrix. 
For each experiment, the current framework first groups together the
respective datasets, in order to take into account eventual
cross-correlations, and then creates the final covariance matrix.
For a given experiment let us consider the measurement of two
observables $O_{I}$ and $O_{J}$, so, the experimental covariance
matrix reads
\begin{equation}
  \textrm{cov}_{ij} =  O_{I,i} O_{J,j} \left( \sum_{l=1}^{N_c} \sigma_{i,l} \sigma_{j,l} + \sum_{n=1}^{N_a} \sigma_{i,n} \sigma_{j,n} + \sum_{n=1}^{N_r}
    \sigma_{i,n} \sigma_{j,n} +  \delta_{ij} \sigma^{2}_{i,s}
  \right)\,,
  \label{eq:covmat}
\end{equation}
where $i$ and $j$ run over the experimental points, and the various
uncertainties given as relative values, are: 
\begin{itemize}
  \item $\sigma_{i,l}$, the $N_c$ correlated systematic uncertainties,
  \item $\sigma_{i,n}$ the $N_a$ absolute and $N_r$ relative
    normalization uncertainties,
  \item $\sigma_{i,s}$, the statistical uncertainty.
\end{itemize}
Before defining the artificial replica generation we introduce the
total uncertainty for the $i-$th point, in terms of
\begin{equation}
  \sigma_{i,\textrm{tot}} = \sqrt{\sigma_{i,s}^2 + \sigma_{i,c}^2 + \sigma_{i,N}^2}\,,
\end{equation}
where $\sigma_{i,c}^2$ and $\sigma_{i,N}^2$ are respectively the total
correlated and the total normalization uncertainties defined as
\begin{equation}
  \sigma_{i,c}^2 = \sum_{l=1}^{N_c} \sigma_{i,l}^2\,,
\end{equation}
and
\begin{equation}
  \sigma_{i,N}^2 = \sum_{n=1}^{N_a} \sigma_{i,n}^2 + \sum_{n=1}^{N_r} \left(
  \frac{1}{2} \sigma_{i,n} \right)^2\,.
\end{equation}

Note that in Eq.~(\ref{eq:covmat}) we have introduced the definition
of the experimental covariance matrix, however in a real fit, such
matrix is replaced by the so called $t_0$ covariance matrix where the
observables $O_I$ are extracted from predictions obtained with a prior
set of PDFs, rater than the original data, avoiding the known bias
presented in Ref.~\cite{dagos}.

At this stage, we generate $k=1,\ldots,N_{\textrm{rep}}$ artificial
replicas of the original data points by shifting with a multi-Gaussian
distribution defined as
\begin{equation}
  O_{I,i}^{(\textrm{art})(k)} = O_{I,i} \left( 1+\sum_{l=1}^{N_c}
    r_{i,l}^{(k)} \sigma_{i,l} + r_{i}^{(k)} \sigma_{i,s} \right)
  \prod_{n=1}^{N_a} \left( 1+r_{i,n}^{(k)}\sigma_{i,n} \right)
  \prod_{n=1}^{N_r} \sqrt{1+r_{i,n}^{(k)}\sigma_{i,n}} \,,
  \label{eq:artdata}
\end{equation}
where the univariate Gaussian random numbers,
$r_{i,l}^{(k)},r_{i}^{(k)},r_{i,n}^{(k)}$, generate fluctuations of
the artificial data around the central value given by the
experiments. For each replica $k$, if two experimental data points
have correlated systematics or normalization uncertainties, then the
fluctuations associated to such uncertainty are taken the same for
both points.

In Figure~\ref{fig:datascheme} we show a simplified picture of the
code structure used for the manipulation of data and the generation of
MC artificial replicas.
Experimental data is stored in files with a common layout, which
contains the process type information, the experimental kinematics for
each data point, the experimental central values, the full breakdown
of experimental systematic uncertainties and the choice of
additive/multiplicative treatment of systematic uncertainties. These
files are obtained from the conversion of raw data information
extracted directly from publications of experimental collaborations.
From a programatically point of view, this information is read from
the common data files when the \texttt{CommonData} container is
initialized and allocated in memory.
From \texttt{CommonData} we have created the inherited
\texttt{DataSet} class which implements the covariance matrix using
both definitions: experimental and $t_0$.
This class also loads in memory the associated theoretical prediction
model which will be discussed in details in
Sect.~\ref{sec:thpredictions}.
Note that the information contained in \texttt{DataSet} is not used
directly in the fit.
The final element of the data layout is the \texttt{Experiment} class,
which groups together datasets from the same experiment, constructs
the covariance matrix taking into account eventual cross-correlations,
and provide the algorithm for generating the MC artificial replicas
given by Eq.~(\ref{eq:artdata}).
This class is used directly in the fit of PDFs, and it is easily
generalized for any kind of experimental data.

\subsection{PDF parametrization}

\begin{figure}
  \centering
  \includegraphics[scale=0.4]{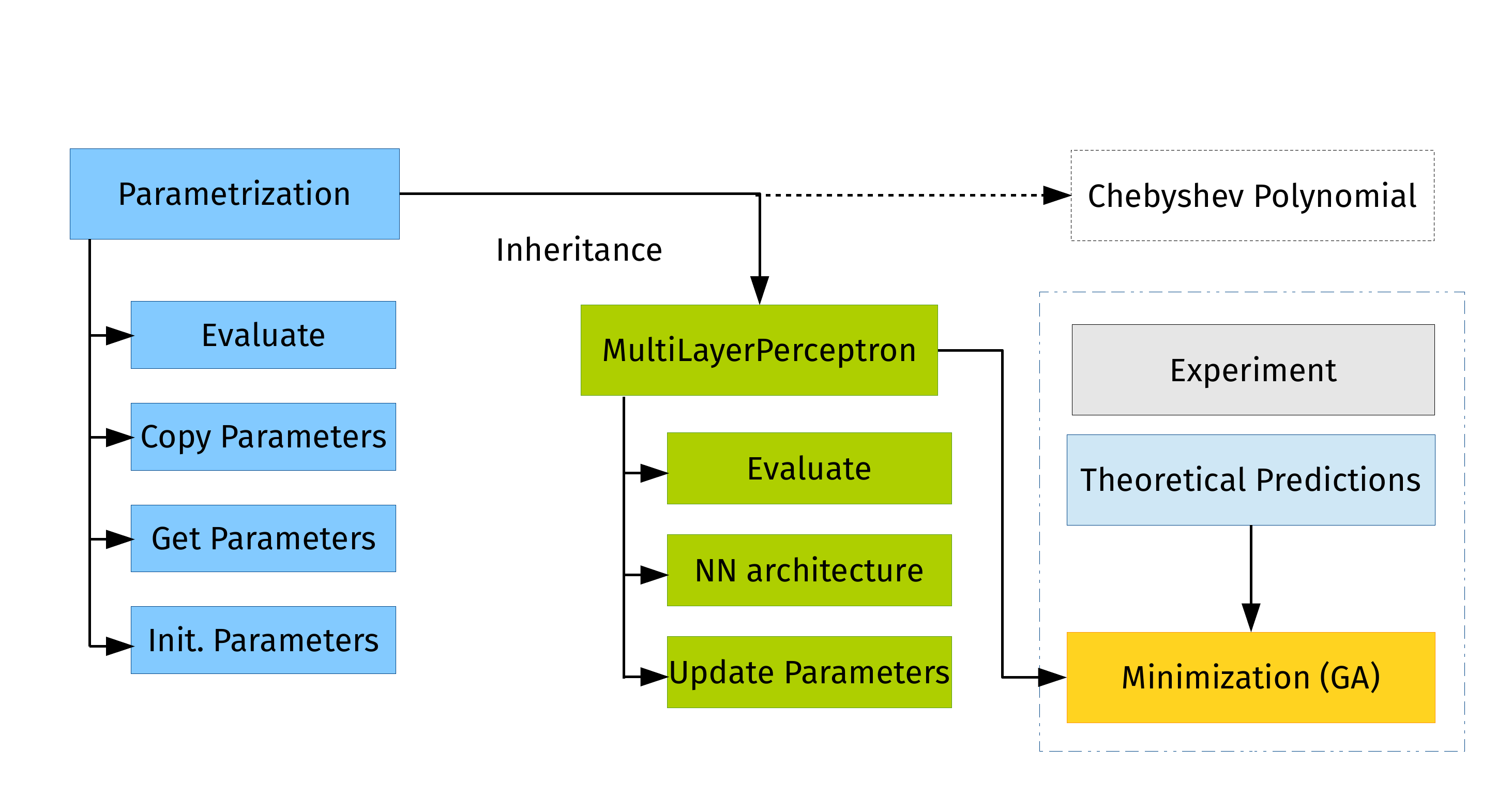}
  \caption{Pictorial representation of the NNPDF parametrization
    layout.}
  \label{fig:pdfparam}
\end{figure}

Concerning the PDF parametrization, the artificial neural networks
used in NNPDF fits consist of connected nodes organized in layers. In
order to evaluate the network, the nodes in the input layer are set
with the required $x$ and $\log x$ values and then the activation of
nodes in successive layers are calculated according to
\begin{eqnarray}
  \xi_i^{(l)} &= &g \left(\sum\limits_j w_{ij}^{(l)} \xi_j^{(l-1)} + \theta_i^{l}\right) \\
  \label{eq:gfun}
  g(a) &= &\frac{1}{1+e^{-a}}
\end{eqnarray}
where $\xi_i^{(l)}$ is the activation of the $i$-th node in the $l$-th
layer of the network, $w_{ij}^{(l)}$ are the weights from that node to
the nodes in the previous layer and $\theta_i^{l}$ is the threshold
for that node. The weights and the thresholds are the parameters in
the fit which are changed during the Genetic Algorithm minimization.
This implementation is known as a multi-layer feed-forward neural
network model (MLP).
There is an exception to Eq.~(\ref{eq:gfun}) in the last layer, where
in order to allow for an unbounded output a linear activation function
$g(a)=a$ is used instead. The flexibility of the fitting code allows
us to easily explore other choices, for instance a quadratic output of
the last layer, $g(a)=a^2$, has been used in studies of the PDF
positivity in leading order fits, including special configurations
where only a single PDF flavors is positive defined,~\textit{e.g.}~the
photon PDF (cfr. Chap.~\ref{sec:chap4}).

In Figure~\ref{fig:pdfparam} we present the parametrization layout
implemented in the new framework. An abstract container, called
\texttt{Parametrization} implements \emph{virtual} methods for the
evaluation and manipulation of parameters for a generic input
function. From this class we can inherit different functions, in
particular, for the NNPDF methodology we have implemented the neural
networks of Eq.~(\ref{eq:gfun}) in the \texttt{MultiLayerPerceptron}
container. In the diagram we show a dashed line with another example
of parametrization, the Chebyshev polynomial. This container provides
methods for the evaluation, and modification of weights and thresholds
of a given ANN architecture by the minimization algorithm.
This new framework also provides several features such as the
possibility to choose the input scale, the parametrization basis,
preprocessing, and the implementation of PDF positivity.

\subsubsection{Parametrization basis}
\label{sec:pdfparam}

In the NNPDF fits, PDFs are parametrized at a reference scale $Q_0^2$.
The choice of $Q_0^{2}$ has no effect whatsoever on the results of the
fit because the DGLAP evolution evolves the input parametrization from
the initial scale to the energy of the experimental data point.
PDFs are expressed in terms of a set of basis functions for quark,
antiquark and gluon PDFs already introduced in Chap.~\ref{sec:chap1}.
For the NNPDF2.3 we define the following basis:
\begin{eqnarray}
\Sigma(x,Q_0^2) &=&  \left( u+ \bar{u} + d +\bar{d}+ s + \bar{s}
\right)(x,Q_0^2) 
\nonumber \\ 
T_3(x,Q_0^2) &=& \left( u+ \bar{u} - d -\bar{d} \right)(x,Q_0^2) \nonumber \\ 
V(x,Q_0^2) &=& \left( u- \bar{u} + d -\bar{d} + s - \bar{s} 
\right)(x,Q_0^2)  \nonumber\\ 
\Delta_S(x,Q_0^2) &=& \left(  \bar{d} - \bar{u}  \right)(x,Q_0^2) \label{eq:nnpdf23basis} \\ 
s^+(x,Q_0^2) &=& \left( s+ \bar{s} \right)(x,Q_0^2)  \nonumber\\ 
s^-(x,Q_0^2) &=& \left( s- \bar{s} \right)(x,Q_0^2)  \nonumber\\
g(x,Q_0^2)\,. \nonumber
\label{eq:nn23basis}
\end{eqnarray}
In the PDF basis above we do not introduce an independent
parametrization for the charm and anticharm PDFs (intrinsic charm),
however the new framework provides the possibility to easily activate
any combination or flavor parametrization.

This basis was chosen in NNPDF2.3 because it directly relates physical
observables to PDFs, by making the leading order expression of some
physical observables in terms of the basis functions particularly
simple: for example, $T_3$ is directly related to the difference in
proton and deuteron deep-inelastic structure functions $F_2^p-F_2^d$,
and $\Delta_S$ is simply expressed in terms of Drell-Yan production in
proton-proton and proton-deuteron collisions, for which there is data
for example from the E866 experiment.
On the other hand, with the current code we can show that several
other basis choices does not affect the results: our results are
independent of the basis change, as recently presented in details in
the NNPDF3.0 paper~\cite{Ball:2014uwa}.

Each PDF is then parametrized by the ANN of Eq.~(\ref{eq:gfun}) with
architecture 2-5-3-1 at the reference scale $Q_0^{2}$ times a
preprocessing factor:
\begin{equation}
\label{eq:preproc}
f_i(x,Q_0) = A_i \hat f_i(x,Q^2_0);\quad   \hat f_i(x,Q^2_0)=\,x^{-\alpha_i} (1-x)^{\beta_i} \,\textrm{NN}_i(x)
\end{equation}
where $A_i$ is an overall normalization constant, and $f_i$ and $\hat
f_i$ denote the normalized and un-normalized PDF respectively.  The
preprocessing term $x^{-\alpha_i} (1-x)^{\beta_i}$ is simply there to
speed up the minimization, without biasing the fit. In the case of the
$s^-$ parametrization we introduce an auxiliary term such as
\begin{equation}
  s^{-}(x,Q_0^2) = A_{s^{-}} \hat s^{-}(x,Q_0^2) - s_{\textrm{aux}}(x,Q_0^2)\,,
\end{equation}
where $s_{\textrm{aux}}(x,Q_0^2) = A_{s^-} x^{-\gamma_{s^-}}
(1-x)^{\delta_{s^-}}$, with exponents chosen in such a way that
$s_{\textrm{aux}}(x,Q_0^2)$ peaks in the valence region, not
interfering with the small-$x$ and large-$x$ behavior of
$s^{-}(x,Q_0^2)$.

Out of the seven normalization constants, $A_i$ in
Eq.~(\ref{eq:preproc}), three can be constrained by the valence sum
rules, sea asymmetry and the momentum sum rule.
Which particular combinations depends of course of the choice of
basis.
With the basis, Eq.~(\ref{eq:nn23basis}), these constraints lead to
\begin{equation}
  A_g = \frac{1-\int_0^1 dx  x \Sigma(x,Q_0^2)}{\int_0^1 dx \; x\,\hat{g}(x,Q_0^2)};\quad
  A_{V} = \frac{3}{\int_0^1 dx \, \hat{V}(x,Q_0^2)}; \quad \label{eq:sumrules}
A_{\Delta_S} = \frac{1-\int_0^1 dx \, \hat{T}_3(x,Q_0^2)}{\int_0^1 dx \, 2 \hat{\Delta}_S(x,Q_0^2)}\,.
\end{equation} 
The other normalization constants can be set arbitrarily to unity,
that is $A_{\Sigma}=A_{T_3}=A_{s^-}=A_{s^+}=1$: the overall size of
these PDFs is then determined by the size of the fitted network.
The finiteness of sum rule integrals Eq.~(\ref{eq:sumrules}) is
enforced by discarding during the Genetic Algorithm minimization (see
Sect.~\ref{subsec:ga} below) any mutation for which the integrals
would diverge.
This condition, in particular, takes care of those NN configurations
that lead to a too singular behavior at small-$x$.

\subsubsection{Effective preprocessing exponents}
\label{sec:preproc}

We have introduced in Eq.~(\ref{eq:preproc}) the preprocessing concept
which absorbs in a prefactor the bulk of the fitted behavior so that
ANN only has to fit deviations from it. This choice is motivated by a
performance improvement during the fit. However, it is important to
implement an automatic mechanism that performs the choice of these
coefficients without biasing the result.
As in previous NNPDF fits, this is done by randomizing the
preprocessing exponents, choosing a different value for each replica
within a suitable range.
We first define the effective asymptotic exponents as follows:
\begin{equation}
  \label{effalpha}
  \alpha_{\textrm{eff},i}(x) = \frac{\ln f_i(x)}{\ln 1/x}\,,\quad
  \beta_{\textrm{eff},i}(x) = \frac{\ln f_i(x)}{\ln (1-x)}\,.
\end{equation}

Then, we perform a fit where the algorithm chooses a random set of
coefficients between a wide starting range for the preprocessing
exponents for each PDF. The effective exponents Eq.~(\ref{effalpha})
are then computed for all replicas at $x=10^{-6}$ and $10^{-3}$ for
the low-$x$ exponent $\alpha_i$ and at $x=0.95$ and $0.65$ for the
large-$x$ exponent $\beta_i$, for all PDFs (except for the gluon and
singlet small-$x$ exponent, $\alpha_i$, which is computed at
$x=10^{-6}$).
The fit is then repeated by taking as new range for each exponent the
envelope of twice the 68\% confidence interval for each $x$ value. The
process is then iterated until convergence, with a tolerance of few
percent.
From a practical point of view, the convergence i typically fast,
even in the cases where the fitted dataset is varied significantly or
for example when the minimization algorithm is modified.

This procedure ensures that the final effective exponents are well
within the range of variation both in the region of the smallest and
largest $x$ data points, and in the asymptotic region (these two
regions coincide for the gluon and singlet at small $x$), thereby
ensuring that the allowed range of effective exponents is not
artificially reduced by the preprocessing, either asymptotically or
sub-asymptotically.

\subsection{Theoretical predictions}
\label{sec:thpredictions}

As we have anticipated at the beginning of this section, the most
computationally intensive task for the PDF fitting technology is the
computation of theoretical predictions.
Indeed, any PDF determination involves an iterative procedure where
all the data points included in the fit need to be recomputed a very
large number of times for different functional forms of the input
PDFs.
The computation of physical observables in the NNPDF framework is
based upon the \texttt{FastKernel} method introduced in
Refs.~\cite{Ball:2010de,Ball:2012cx}.
Here we recall the basic concepts necessary to explain the structure
of the new code.

\subsubsection{The FastKernel methodology}
\label{sec:fk}

Let us consider a grid of points in $x$, where each PDF flavor at a
given scale $Q^2$ is represented in terms of $f_i(x_\alpha, Q^2)$ with
$\alpha=1,\ldots,N_x$ where the index $i$ identifies the parton
flavor, and the index $\alpha$ enumerates the points on the grid.
DIS observables, which are linear in the PDFs, can be computed using a
precomputed kernel $\hat{\sigma}^{I,J}_{\alpha j}$:
\begin{equation}
  \label{eq:disFK}
  O_I(x_J,Q_J^2) = \sum_{j=1}^{N_\mathrm{pdf}}
  \sum_{\alpha=1}^{N_x} \hat{\sigma}^{I,J}_{\alpha j}
  f_j(x_\alpha,Q_0^2)\, ,
\end{equation}
where the index $I$ labels the physical observable, $x_J$ and $Q_J$
are the corresponding kinematical variables for each specific
experimental data point $J$, $j$ runs over the parton flavors and
$\alpha$ runs over the $x$-grid points.
The kernel $\hat{\sigma}$ just introduced is referred to as an
\texttt{FKTable}. A similar expression is available for the hadronic
observables, which are written as a convolution of two PDFs, and
computed in terms of an (hadronic) \texttt{FKTable}
$\hat{W}^{I,J}_{kl\gamma\delta}$:
\begin{equation}
  \label{eq:colFK}
  O_I(x_J,Q_J^2) = \sum_{k,l=1}^{N_\textrm{pdf}} \sum_{\gamma,\delta=1}^{N_x}
  \hat{W}^{I,J}_{kl\gamma\delta} f_k(x_\gamma,Q_0^2)
  f_l(x_\delta,Q_0^2)\, ,
\end{equation}
where the indices $k,l$ run over the parton flavors, and the indices
$\gamma,\delta$ count the points on the interpolating grids.

\begin{figure}
  \centering
  \includegraphics[scale=0.4]{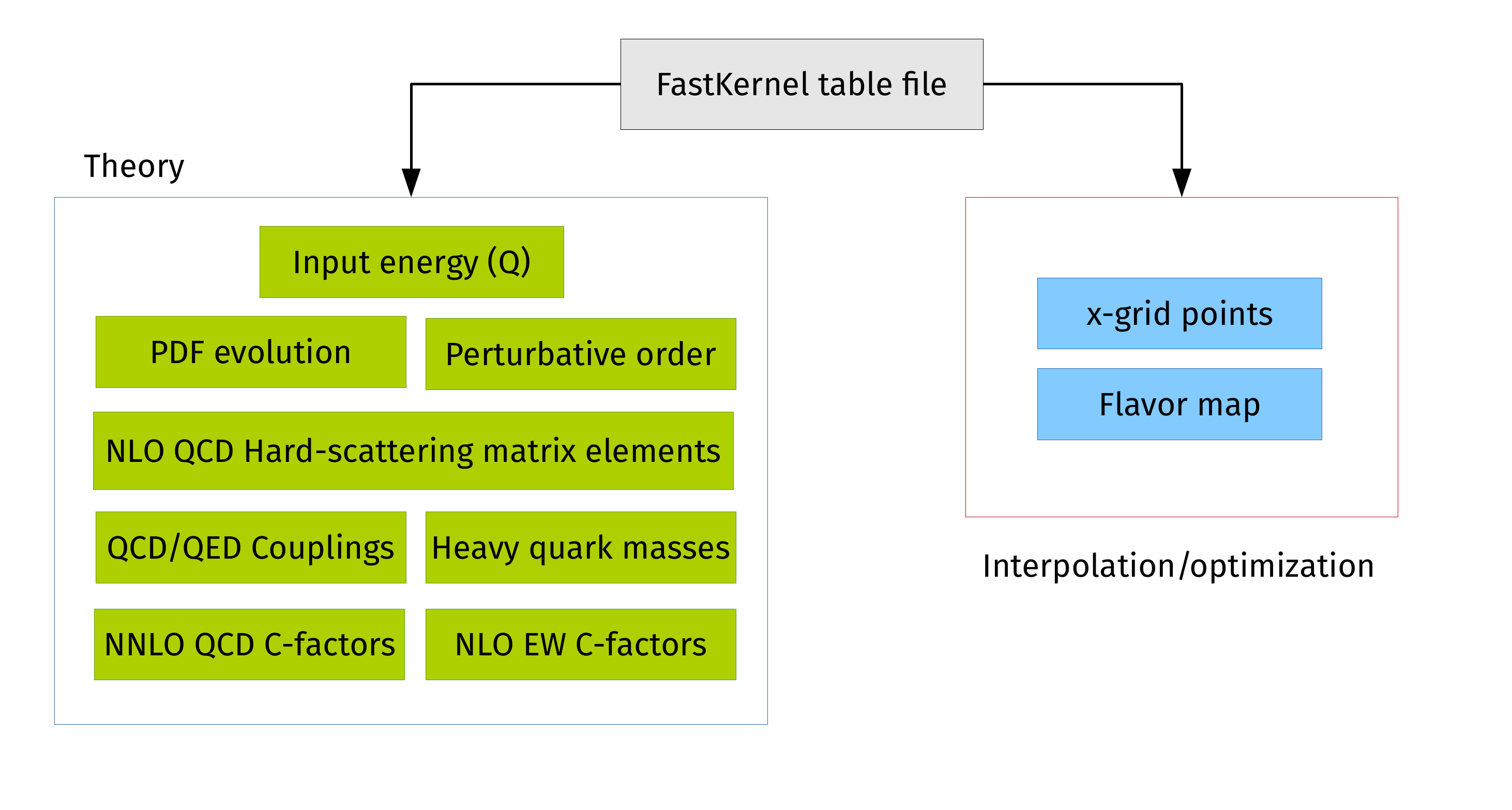}
  \caption{Graphical summary of the \texttt{FKTable} layout.}
  \label{fig:fktable}
\end{figure}

In the fitting code, for each experimental dataset $I$ we have a
separate \texttt{FKTable} that encodes all the theory information. In
Figure~\ref{fig:fktable} we show the components encoded in a
\texttt{FKTable} file.
These tables encode all the information about the theoretical
description of the observables such as: the perturbative order, the
value of the strong coupling, the choice of scales, the QCD and
electroweak perturbative corrections ($C$-factors), or the prescription
for the evolution.
The modification of any of the of theoretical description of a given
observable is reflected in a new \texttt{FKtable}. The convolutions of
the \texttt{FastKernel} tables with the PDFs at the initial scale are
generic, and do not require any knowledge about the theoretical
framework.
On the other hand, the tables also contain information about the grid
of points in $x$ used for the interpolation and the so-called flavor
map which optimizes the grids size by indicating all the available
non-zero flavor channels.
Notice that this layout implements a clean separation of the
theoretical assumption from the fitting procedure. In particular,
during the fitting procedure the tables are always kept fixed and
treated as an external input. The only shared information between
these tables and fit is the initial scale $Q^2$.

One important remark about the differences between the
\texttt{FastKernel} approach in comparison to fast NLO calculators
such as \texttt{FastNLO}~\cite{Kluge:2006xs},
\texttt{APPLgrid}~\cite{Carli:2010rw} and
\texttt{aMCfast}~\cite{Bertone:2014zva}, is that it includes PDF
evolution into the precomputed tables, while the other approaches
require as input the PDFs evolved at the scales where experimental
data is provided.
The inclusion of PDF evolution is essential to reduce drastically the
computational cost of running PDF fits.
Note also that the generic structure of the \texttt{FastKernel}
methodology holds for any fast NLO calculator as well as for any PDF
evolution code. For example, in NNPDF2.3 and later we use our own
internal Mellin-space \texttt{FKgenerator} code for PDF evolution and
DIS observables. A future version of this combination, planned for the
next NNPDF release, combines the $x$-space evolution from
\texttt{APFEL} with the usual \texttt{FastKernel} combination
algorithm (the so-called \texttt{APFELcomb} project).
This shows how flexible the code is: the \texttt{FastKernel} tables
are independent elements from the NNPDF framework, which can be
computed with external tools, specialized in the computation of
theoretical predictions.

The main advantage of the \texttt{FastKernel} methodology in
comparison to \textit{e.g.}~\texttt{APPLgrid} or \texttt{FastNLO} is
that PDF evolution is precomputed and stored in the \texttt{FKTable}
itself.
This point is particularly relevant when performing a fit to data
distributed along a large range of $Q^2$ values, \textit{e.g.}~the
inclusive jet production, where an equivalent large number of PDF
evolutions are needed.
In these cases, the inclusion of PDF evolution improves drastically
the performance of the fit.
The improvement due to this acceleration is quantified in
Table~\ref{tab:FKtimings}.

\begin{table}[t]
\begin{center}
\begin{tabular}{c|c|c|c}
  \hline Observable & \texttt{APPLgrid} & \texttt{FKTable} & optimized \texttt{FKTable} \\
  \hline $W^+$ production &1.03 ms & 0.41 ms (2.5x) & 0.32 ms (3.2x) \\
  Inclusive jet production
  &2.45 ms & 20.1 $\mu$s (120x) & 6.57 $\mu$s (370x) \\ 
  \hline
\end{tabular}
\caption{Comparison of \texttt{APPLgrid} and \texttt{FKTable} convolution
  timings. Results are provided for two different observables: the
  total cross-section for $W^+$ production and for inclusive jet
  production for typical cuts of $p_T$ and rapidity.
In parenthesis we show the relative speed-up compared to the the reference
convolution based on \texttt{APPLgrid}.
In the last column we use SSE acceleration in the convolution
product. \label{tab:FKtimings}}
\end{center}
\end{table}

\begin{figure}
  \centering
  \includegraphics[scale=0.4]{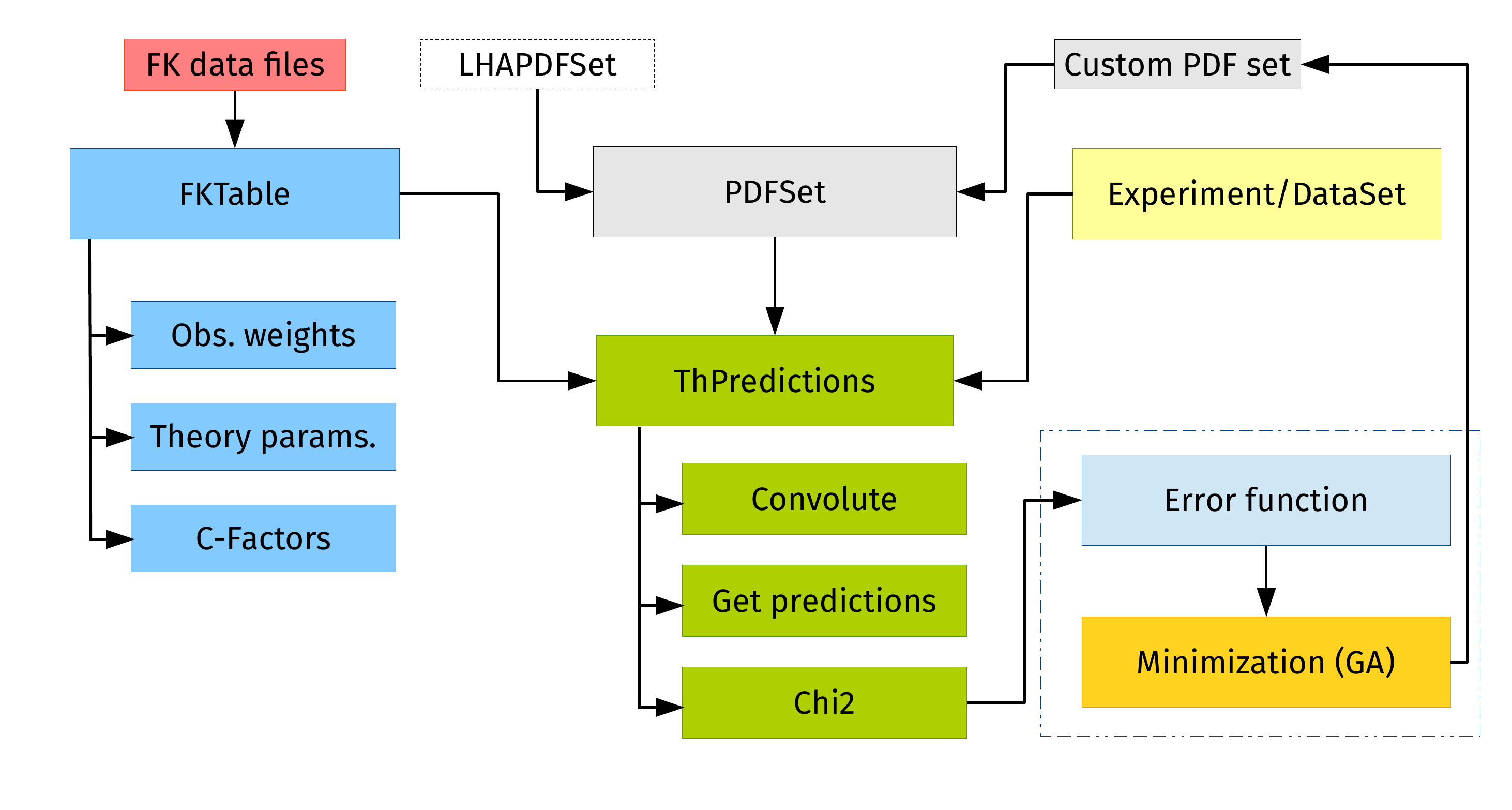}
  \caption{Pictorial representation of the NNPDF theoretical
    predictions framework.}
  \label{fig:fkscheme}
\end{figure}

The code layout for the \texttt{FastKernel} procedure is presented in
Figure~\ref{fig:fkscheme}.
The \texttt{FKTable} class reads the \texttt{FastKernel} objects from
files stored on disk. For each dataset, this class makes available the
convolution kernel, the theoretical setup and the eventual $C$-factors
to other modules of the code.
As we have explained previously, the new fitting code has been
designed with an explicit separation between experiment and theory.
Therefore, the kinematic cuts upon an experimental dataset can now be
performed algorithmically by selecting the points in the
\texttt{CommonData} format which pass the required cuts according to
their bundled kinematic information, and matching with the equivalent
points in the \texttt{FKTable}. This is a considerable improvement
over the earlier regeneration of the precomputed theory tables due to
the monolithic treatment of the experimental data in the
\texttt{Fortran77} code.
Note that this layout allows the introduction of PDF positivity
constrains through the convolution of PDFs with artificial observables
encoded in \texttt{FastKernel} tables, which are tested during the
minimization algorithm and in the case of violation it penalizes the
error function.

The PDF convolution is performed in the \texttt{ThPredictions} class,
which takes as input: a PDF set through the abstract \texttt{PDFSet}
class and a \texttt{FKTable} object, which can be passed automatically
from the \texttt{DataSet} and \texttt{Experiment} classes.
PDFs are accessible through the \texttt{LHAPDFSet} interface, which
calls PDFs from the \texttt{LHAPDF} library, or by any other custom
set obtained by extending the \texttt{PDFSet} class, this is exactly
what the minimization algorithm does.
The \texttt{ThPredictions} object provides methods for the
\texttt{FastKernel} convolution product. This class computes
theoretical predictions but also determines the $\chi^2$ to data when
used in combination with \texttt{Experiment/DataSet}.

Concerning optimizations, in order to ensure a fast and efficient
minimization procedure, the \texttt{FKTable} class has been designed
such that the \texttt{FastKernel} table is stored with the optimal
alignment in machine memory for use with \texttt{SIMD} (Single
Instruction Multiple Data) instructions, which allow for an
acceleration of the observable calculation by performing multiple
numerical operations simultaneously.
The large size of a typical \texttt{FastKernel} product makes the
careful memory alignment of the \texttt{FastKernel} table and PDFs
extremely beneficial.
A number of \texttt{SIMD} instruction sets are available depending on
the individual processor.
By default we use a 16-byte memory alignment for suitability with
Streaming \texttt{SIMD} Extensions (SSE) instructions, although this
can be modified by a parameter to 32-bytes for use with processors
enabled with Advanced Vector Extensions (AVX).
The product itself is performed both with \texttt{SIMD} instructions
and, where available, \texttt{OpenMP} is used to provide acceleration
using multiple CPU cores, parallelizing the computation of predictions
for each experimental data point.
We have also investigated about a further level of improvement of the
\texttt{FastKernel} product by using GPUs, while presenting no
technical objections, has so far not been developed due to scalability
concerns on available computing clusters. Moreover, several
technologies such as \texttt{NVIDIA
  CUDA}\footnote{\url{www.nvidia.com}} or
\texttt{OpenCL}\footnote{\url{https://www.khronos.org/opencl/}} show
optimal performance only on dedicated devices, disfavoring
portability.

The performance improvements are clearly visible when comparing with
the calculation of the hadronic convolution Eq.~(\ref{eq:colFK}) using
the optimized settings with that using non-optimized settings.
To illustrate this point, we compare in Table~\ref{tab:FKtimings} the
timings for a couple of representative LHC observables, for the
convolution performed using \texttt{APPLgrid}, the standard
double-precision version of the {FKTable} implementation, and the
optimized \texttt{FKTable} implementation using the SSE-accelerated
calculation, for two representative observables.
The results shows a massive improvement in speed by precomputing the
PDF evolution in the \texttt{FKTable}, with further improvements
obtained by the careful optimization of the \texttt{FastKernel}
product, and even further gains possible when combined with
\texttt{OpenMP} on a multiprocessor platform, dividing the
computational cost by the total number of available cores.

\subsection{Minimization algorithm}
\label{sec:minim}

The minimization is performed using Genetic Algorithms, which are
especially suitable for dealing with very large parameter space.
Note that the current ANN architecture (2-5-3-1) corresponds to 37
free parameters for each PDF, \textit{i.e.}~a total of 259 free
parameters, to be compared to less than a total of 30 free parameters
for PDF fits based on conventional polynomial functional forms.
Because of the extreme flexibility of the fitting functions and the
large number of parameters, the optimal fit is not necessarily the
absolute minimum of the $\chi^2$ which might correspond to an
`overfit' in which not only the desired best fit is reproduced, but
also statistical fluctuation about it. As a consequence, a stopping
criterion is needed on top of the minimization method. In the next
paragraphs we discuss in turn the GA and the stopping strategies
implemented in NNPDF.

\subsubsection{Genetic Algorithms}
\label{subsec:ga}

In the new framework, we have performed a careful analysis of the
Genetic Algorithm minimization procedure utilized in previous NNPDF
determinations. 
Instead of reproducing the previous methodology, we have introduced
new features only if they resulted in faster fitting.

The GA algorithm implemented here consists of three main steps:
mutation, evaluation and selection. The minimization procedure of each
PDF replica, is completely independent from each other, so the
procedure can be parallelized on multiple machines.
Starting from a large number of mutants, PDF sets are generated based
on a parent set from the previous generation. The goodness of fit to
the data for each mutant is then calculated, with the error function
\begin{equation}
  \chi^{2(k)} = \frac{1}{N_{\textrm{dat}}}
  \sum_{i,j=1}^{N_{\textrm{dat}}} \left( O_{I,i}^{(\textrm{art})(k)} -
    O_{I,i}^{(\textrm{NN})(k)}\right) \left( \textrm{cov}_{t_0} \right)_{i,j}^{-1} \left( O_{J,j}^{(\textrm{art})(k)} - O_{J,j}^{(\textrm{NN})(k)}\right)\,,
\end{equation}
where $O_{I,i}^{(\textrm{NN})(k)}$ is the prediction for replica $k$
of an observable $I$ at a data point $i$ computed with the ANN
parametrization, and $\textrm{cov}_{t_0}$ the covariance matrix based
on the $t_0$ prescription explained in Sect.~\ref{sec:nnpdfdata}.

The best fit mutant is identified and passed on to the next
generation, while the rest are discarded. The algorithm is then
iterated until a set of stopping criteria are satisfied.
The number of mutants tested each generation is now set to 80 for all
generations, removing the two GA `epochs' used in previous
determinations. The choice of this number is arbitrary and depends on
the total number of generations. All mutants are generated from the
single best mutant from the previous generation.

To generate each mutant, the weights of the neural networks from the
parent PDF set are altered by mutations. In fits before NNPDF3.0 the
mutations have consisted of point changes, where individual weights or
thresholds in the networks were mutated at random. However,
investigations of strategies for training neural
networks~\cite{Montana:1989} have found that employing coherent
mutations across the whole network architecture instead leads to
improved fitting performance.
The general principle that explains this is that of changing multiple
weights which are related by the structure of the network, leading to
improvements in both the speed and quality of the training.

In the NNPDF3.0 fits we use a nodal mutation algorithm, which gives
for each node in each network an independent probability of being
mutated.  If a node is selected, its threshold and all of the weights
are mutated according to
\begin{equation}
w \rightarrow w + \frac{\eta r_{\delta}}{N_\textrm{ite}^{r_\textrm{ite}}} \, ,
\end{equation} 
where $\eta$ is the baseline mutation size, $r_{\delta}$ is a uniform
random number between $-1$ and $1$, different for each weight,
$N_\textrm{ite}$ is the number of generations elapsed and
$r_\textrm{ite}$ is a second uniform random number between $0$ and $1$
shared by all of the weights.
An investigation performed on closure test fits in Sect. 4 of
Ref.~\cite{Ball:2014uwa} found that the best value for $\eta$ is 15,
while for the mutation probability the optimal value turns out to be
around 5\%, which corresponds to an average of 3.15 nodal mutations
per mutant PDF set.

As with the removal of the fast- and slow-epochs and their replacement
with a single set of GA parameters, the Targeted Weighted Training
(TWT) procedure adopted in previous fits has also been dropped.
This was originally introduced in order to avoid imbalanced training
between datasets. With the considerably larger dataset of NNPDF3.0
along with numerous methodological improvements, such an imbalance is
no longer observed even in fits without weighted training. Whereas
previously the minimization was initiated with a TWT epoch in which
the fit quality to individual datasets was minimized neglecting their
cross-correlations, in NNPDF3.0 the minimization always includes all
available cross-correlations between experimental datasets.

\begin{figure}
  \centering
  \includegraphics[scale=0.4]{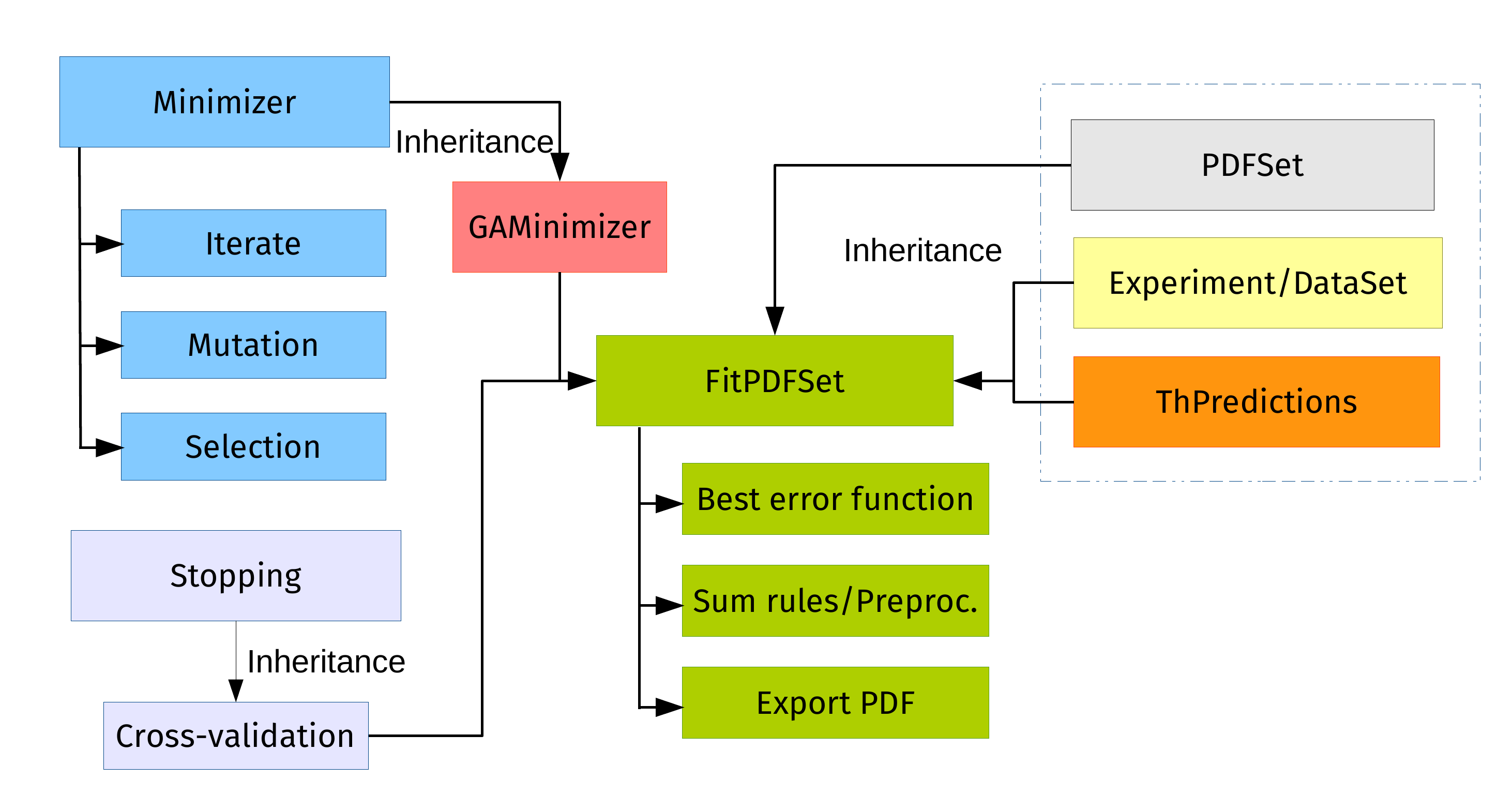}
  \caption{Layout of the NNPDF minimization framework.}
  \label{fig:gascheme}
\end{figure}

\subsubsection{Stopping criterion}
\label{subsec:cv}

The stopping criterion for the GA is the cross-validation method. This
is based on the idea of separating the data in two sets, a training
set, which is fitted, and a validation set, which is not fitted. 
The GA minimizes the $\chi^2$ of the training set, while the $\chi^2$
of the validation set is monitored along the minimization, and the
optimal fit is achieved when the validation $\chi^2$ stops improving.

In PDF fits before NNPDF3.0 this stopping criterion was implemented by
monitoring a moving average of the training and validation $\chi^2$,
and stopping when the validation moving average increased while the
training moving average decreased by an amount which exceeded suitably
chosen threshold values. 
The moving average prevented the fit from stopping due to statistical
fluctuations, but introduced a certain arbitrariness since the value
of these three parameters (the length of the moving average and the
two thresholds) had to be tuned.

In NNPDF3.0 the previous stopping criterion is replaced by the
so-called \emph{look-back} method which stores the PDF parametrization
for the iteration where the fit reaches the absolute minimum of the
validation $\chi^2$ within a given maximum number of generations. This
method reduces the level of arbitrariness introduced in the previous
strategy, however it keeps the total number of iterations for all
replicas.

In Figure~\ref{fig:gascheme} we finalize the description of the new
framework with the minimization layout.
The output set of PDFs is allocated in the \texttt{FitPDFSet} class,
inherited from \texttt{PDFSet}, which drives the minimization and
stores the best mutant and its error function for each iteration of
the GA. This class is also responsible for the computation
of the normalization coefficients and preprocessing of the neural
networks. At the end of the minimization the best PDF parametrization
is exported to file.
From the minimization point of view, we have coded an abstract
\texttt{Minimizer} class with \emph{virtual} methods for the GA
iteration, mutation and selection. This class is extended in
\texttt{GAMinimizer} with the technical choices explained in the
previous section. It contains all elements for a fast computation of
training and validation $\chi^2$ from \texttt{ThPredictions}.
The cross-validation data split is performed at level of
\texttt{Experiment} class, at the beginning of the program, note that
this procedure is parallelized for each artificial replica.
The last point of the code structure is the \texttt{GAMinimizer}
connection to the \texttt{Stopping} class. This class is easily
extended with \textit{i.e.}~\emph{look-back} method.

\section{NNPDF2.3}
\label{sec:nnpdf23}

Now that we have presented the NNPDF methodology through the new code
framework, we conclude this chapter by describing the NNPDF2.3 fitting
configuration in terms of PDF parametrization, minimization setup and
the description of the data included in this fit. Here, we present
this set of PDFs instead of the most recent NNPDF3.0 because QED
correction has been obtained from the baseline NNPDF2.3 set. For a
complete discussion about the phenomenological impact of this set of
PDFs we address the reader to Sect.~\ref{sec:generalfeatures} in
Chap.~\ref{sec:chap1}.

\subsection{Fit configuration}

The PDF parametrization used in the NNPDF2.3 was already shown in
Sect.~\ref{sec:pdfparam}.
In Table~\ref{tab:preproc} the range of the small- and large-$x$
preprocessing exponents used in this fit are presented for each
element of the fitting basis. In the NNPDF2.3 the preprocessing
exponents are the same for both NLO and NNLO determinations.
These ranges have been redetermined self-consistently for different
fits: for example, for fits to reduced datasets, wider ranges are
obtained due to the experimental information being less constraining.

The mutation parameters of the Genetic Algorithm used in NNPDF2.3 are
presented in the left Table~\ref{tab:gapars}: for each PDF basis
element we show the number of mutations $N_{\textrm{mut}}$ and the
respective mutation sizes $\eta$. It interesting to note that this
configuration has changed in NNPDF3.0 by applying a mutation
probability of 5\% per network node, and the mutation size to
$\eta=15$.

In NNPDF2.3 we used the cross-validation method with Targeted Weighted
Training (TWT) for the first $N_{\textrm{gen}}^{\textrm{mut}} =
2500$. In this first phase of the minimization, we use a large number
of mutants $N_{\textrm{mut}}^a = 80$, which is then reduced to
$N_{\textrm{mut}}^b = 30$. 
The dynamic stopping condition, based on the variation of the moving
average of the validation and training $\chi^2$ (see
Sect.~\ref{subsec:cv}), is activated after
$N_{\textrm{gen}}^{\textrm{wt}}=10000$. The moving average criterion
is complemented by a minimum training $E_{\textrm{rm}}^{\textrm{min}}
= 6$. The maximum number of allowed iterations is
$N_{\textrm{gen}}^{\textrm{max}}=50000$. All these parameters are
summarized on the right Table~\ref{tab:gapars}.

\begin{table}
\centering
\begin{tabular}{|c||l|l|}
\hline
 & \multicolumn{2}{c|}{NNPDF2.3 NLO and NNLO}  \\
\cline{2-3}
PDF & [$\alpha_{\textrm{min}},\alpha_{\textrm{max}}$] & [$\beta_{\textrm{min}},\beta_{\textrm{max}}$]  \\
\hline
\hline
$\Sigma$ & [1.05, 1.35] & [2.55, 3.45] \\

$g$ & [1.05, 1.35] & [3.55, 4.45]  \\

$T_3$ & [0.00, 0.50] & [2.55, 3.45] \\

$V$ & [0.00, 0.50] & [2.55, 3.45]  \\

$\Delta_S$ & [-0.95, -0.65] & [12.0, 14.0]  \\

$s^+$ & [1.05, 1.35] & [2.55, 3.45]  \\

$s^-$ & [0.00, 0.50] & [2.55, 3.45]  \\
\hline
\end{tabular}

\caption{The small- and large-$x$ preprocessing exponents in
  Eq.~\ref{eq:preproc} randomly chosen in NNPDF2.3.}
\label{tab:preproc}

\end{table}

\begin{table}
\begin{center}

  \begin{tabular}{|c||c|c|}
\hline
    \multicolumn{3}{|c|}{NNPDF2.3}    \\ \hline
    \multicolumn{3}{|c|}{Single Parameter Mutation}    \\
    \hline 
PDF &   $N_\textrm{mut}$ &  $\eta$  \\
    \hline
\hline 
$\Sigma$    & 2 & 10, 1 \\
$g$  & 3 & 10, 3, 0.4 \\
$T_3$   & 2 &  1, 0.1 \\
$V$   & 3 &  8, 1, 0.1\\
$\Delta_S$   &3 & 5, 1, 0.1 \\
$s^+$  &  2 & 5, 0.5 \\
$s^-$  &  2 & 1, 0.1\\
\hline 
  \end{tabular}\hskip10pt \begin{tabular}{|c||c|}
    \hline
    \multicolumn{2}{|c|}{NNPDF2.3}    \\ \hline
    \multicolumn{2}{|c|}{Minimization Setup}    \\
    \hline 
    Parameter &  Value  \\
    \hline
    \hline
    $N_{\textrm{gen}}^{\textrm{wt}}$ & 10000 \\
    $N_{\textrm{gen}}^{\textrm{mut}}$ & 2500 \\
    $N_{\textrm{gen}}^{\textrm{max}}$ & 50000 \\
    $E^{\textrm{min}}_{\textrm{tr}}$ & 6 \\
    $N_{\textrm{mut}}^{a}$ & 80 \\
    $N_{\textrm{mut}}^{b}$ & 30 \\
    \hline
  \end{tabular}
  \end{center}
  \caption{The mutation parameters are shown for the NNPDF2.3 determination. In the right table, parameters controlling the maximum fit length, number of mutants, target weighted training settings are shown.}
  \label{tab:gapars}
\end{table}

\subsection{Experimental data}

\label{sec-expdata}

After presenting the main characteristics of the NNPDF2.3 methodology,
we now discuss about the data set used by this fit.
Concerning non-LHC data, the NNPDF2.3 data set includes at NLO and
NNLO:
\begin{itemize}

\item NMC~\cite{Arneodo:1996kd,Arneodo:1996qe}, BCDMS~\cite{bcdms1,bcdms2}
  and SLAC~\cite{Whitlow:1991uw} deep-inelastic scattering (DIS) fixed
  target data;
  
\item the combined HERA-I DIS data set~\cite{H1:2009wt}, HERA
  $F_L$~\cite{h1fl} and $F_2^c$ structure function
  data~\cite{Breitweg:1999ad,Chekanov:2003rb,Chekanov:2008yd,Chekanov:2009kj,Adloff:2001zj,
    Collaboration:2009jy,H1F2c10:2009ut}, ZEUS HERA-II DIS
  cross-sections~\cite{Chekanov:2009gm,Chekanov:2008aa}, CHORUS
  inclusive neutrino DIS~\cite{Onengut:2005kv}, and NuTeV dimuon
  production data~\cite{Goncharov:2001qe,MasonPhD};

\item fixed-target E605~\cite{Moreno:1990sf} and
  E866~\cite{Webb:2003ps,Webb:2003bj,Towell:2001nh} Drell-Yan
  production data;
  
\item CDF W asymmetry~\cite{Aaltonen:2009ta} and
CDF~\cite{Aaltonen:2010zza} and D0~\cite{Abazov:2007jy} Z rapidity
distributions;

\item CDF~\cite{Aaltonen:2008eq} and D0~\cite{D0:2008hua} Run-II
  one-jet inclusive cross-sections. 

\end{itemize}

The kinematical cuts of DIS data are the usual $Q^2_{\textrm{min}} =
3$ GeV$^2$ and $W^2_{\textrm{min}} = 12.5$ GeV$^2$. We included also
all currently available LHC data for which the experimental covariance
matrix has been provided:
\begin{itemize}

\item the ATLAS W and Z lepton rapidity distributions from the 2010
  data set~\cite{Aad:2011dm};

\item the CMS W electron asymmetry from the 2011 data
  set~\cite{Chatrchyan:2012xt};

\item the LHCb W lepton rapidity distributions from the 2010 data
  set~\cite{Aaij:2012vn};

\item the ATLAS inclusive jet cross-sections from the 2010
  run with $R=0.4$~\cite{Aad:2011fc}.

\end{itemize}

More recent measurements from the 2011 and 2012 runs, which are very
relevant for PDF fits, like the CMS and LHCb low mass Drell-Yan
differential distributions~\cite{CMSdy,LHCb-CONF-2012-013} and the
inclusive jets and dijets from ATLAS and
CMS~\cite{Aad:2013lpa,Chatrchyan:2012bja} have been included in the
NNPDF3.0 release.

\begin{figure}
\begin{center}
\includegraphics[scale=0.5]{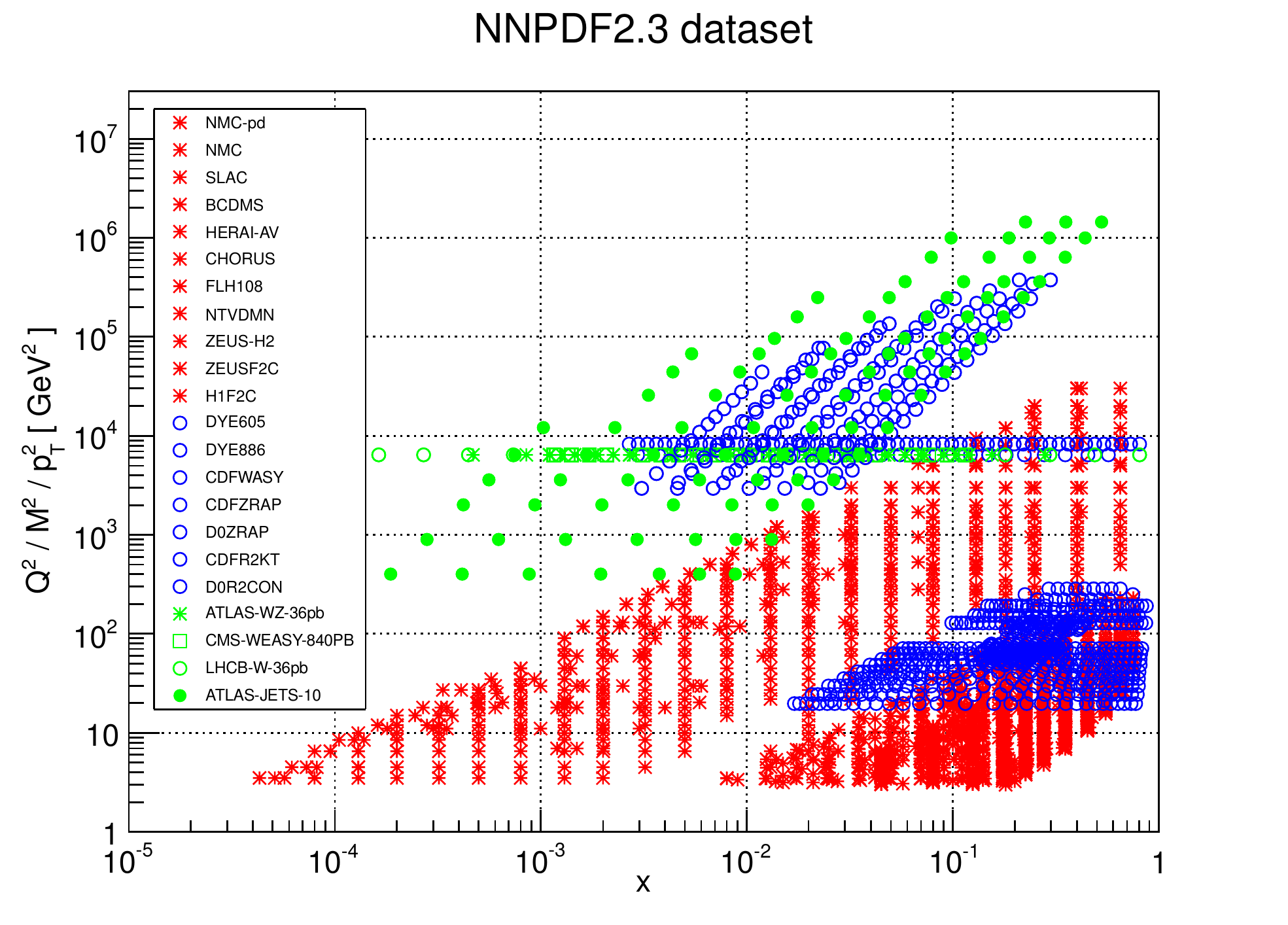}
\caption{The kinematical coverage of the experimental data
used in the NNPDF2.3 PDF determination.
\label{fig:dataplottot}}
\end{center}
\end{figure}

The kinematical coverage of the LHC data sets included in the NNPDF2.3
analysis with the corresponding average experimental uncertainties for
each data set are summarized in
Tab.~\ref{tab:exp-sets-errors}.\footnote{For jets, we plot only the
  $x$ value of the parton with smallest $x$, given by $x=2\frac{p_T
  }{\sqrt{s}}e^{-|\eta|}$}.  A scatter plot of the
kinematical plane for all experimental data from NNPDF2.3 is shown in
Fig.~\ref{fig:dataplottot}.  The LHC electroweak data span a larger
range in Bjorken-$x$ than the Tevatron data thanks to the extended
rapidity coverage (up to $\eta=4.5$), while the inclusive jets span a
much wider kinematical range both in $x$ and $Q^2$ than the one
accessible at the Tevatron.
In Tab.~\ref{tab:sets-numpts} we also give the total number of data
points used for PDF fitting, both for the NLO and the NNLO global
sets.

\begin{table}
 \footnotesize
 \centering
 \begin{tabular}{|c||c|c|c||c|c|c|}
\hline
 Data Set & Ref.  & $N_\textrm{dat}$ &  
$\left[\eta_\textrm{min},\eta_\textrm{max}\right]$ &
$ \sigma_\textrm{stat}$ (\%) &
   $\sigma_\textrm{sys}$  (\%) & $\sigma_\textrm{norm}$  (\%)
 \\ \hline
\hline
CMS $We^-$ asy. $840~\textrm{pb}^{-1}$       &   \cite{Chatrchyan:2012xt}  & 11 & $\left[ 0,2.4\right]$&   2.1  & 4.7  
      & 0   \\  
\hline
ATLAS $\textrm{W}^+$ $36~\textrm{pb}^{-1}$      & \cite{Aad:2011dm} & 11 & $\left[ 0,2.4\right]$ & 1.4 
 & 1.3    & 3.4      \\  
ATLAS $\textrm{W}^-$ $36~\textrm{pb}^{-1}$      & \cite{Aad:2011dm} & 11 & $\left[ 0,2.4\right]$  & 1.6  & 1.4 & 3.4       \\  
ATLAS Z $36~\textrm{pb}^{-1}$       & \cite{Aad:2011dm}  & 8 & $\left[ 0,3.2\right]$ & 2.8
  &  2.4   & 3.4      \\  
\hline
LHCb $\textrm{W}^+$ $36~\textrm{pb}^{-1}$      & \cite{Aaij:2012vn} & 5  & $\left[ 2,4.5\right]$ & 4.7  &   11.1  &  3.4   \\  
LHCb $\textrm{W}^-$ $36~\textrm{pb}^{-1}$      & \cite{Aaij:2012vn} & 5  & $\left[ 2,4.5\right]$ & 3.4   &   7.8  &  3.4   \\  
LHCb Z $36~\textrm{pb}^{-1}$      & \cite{Aaij:2012vn} & 5  & $\left[ 2,4.5\right]$ &  24  &   4.7  &  3.4   \\  
\hline
\hline
ATLAS Incl. Jets 36 pb$^{-1}$       & \cite{Aad:2011fc}  & 90 &
$\left[ 0,4.5\right]$ & 10.2  & 23.4   & 3.4  \\  
\hline        
 \end{tabular}
\caption{\label{tab:exp-sets-errors} The number of data points,
kinematical coverage and average
statistical, systematic and normalization percentage uncertainties for
each of the experimental LHC data sets considered for
the NNPDF2.3 analysis.
}
 \end{table}

\begin{table}                                                
 \footnotesize
 \centering
 \begin{tabular}{|c|cc|}
\hline
    Fit & NLO  &   NNLO    \\
\hline
    NNPDF2.3 noLHC & 3341 &   3360 \\
    NNPDF2.3 Collider only &1217  &   1236 \\
    NNPDF2.3 & 3487 &  3506\\
\hline 
\end{tabular}
\caption{\label{tab:sets-numpts} Total 
number of data points for the various
global sets used for PDF fitting.
}
 \end{table}

 The theoretical predictions for LHC electroweak boson production have
 been computed at NLO with
 \texttt{MCFM}~\cite{MCFMurl,Campbell:2004ch} interfaced with the
 \texttt{APPLgrid} library for fast NLO
 calculations~\cite{Carli:2010rw}.  The NNLO predictions are obtained
 by means of local $C$-factors. These have been computed using the
 \texttt{DYNNLO} code~\cite{Catani:2010en}.
 The kinematical cuts applied to the calculation of the NLO cross
 sections are now discussed in turn. For the ATLAS data, these are the
 following:
\begin{itemize}
\item cuts for the W lepton rapidity  distributions
\begin{equation*}
\centering
p_T^{l} \ge 20~ \textrm{GeV},\qquad  p_T^{\nu} \ge 25~ \textrm{GeV}, \qquad m_T<40~ \textrm{GeV} ,\qquad |\eta_l| \le 2.5;
\end{equation*}
\item cuts for the Z rapidity distribution
\begin{align*}
\centering
& p_T^{l} \ge 20~ \textrm{GeV},\qquad 66~\textrm{GeV} \le m_{l^+l^-} \le 116~\textrm{GeV} ,\qquad \eta_{l^+,l^-} \le 4.9.
\end{align*}
\end{itemize}

In fact, ATLAS measures separately the rapidity distributions in both
the electron and muon channels, and then combines them into a common
data set. The above kinematical cuts correspond to the combination of
electrons and muons, but differ from the cuts applied in individual
leptonic channels. For Z rapidity distributions we have explicitly
verified that results are unchanged if the cut on the rapidity of the
leptons from the Z decay is removed.

For the CMS W electron asymmetry, the only cut is $p_T^{e} \ge 35~
\textrm{GeV}$, with the same binning in electron rapidity as in
Ref.~\cite{Chatrchyan:2012xt}. Finally, for the LHCb we have included
in our determination only the $W$ data because at that time the $Z$
data was being reanalyzed. The kinematical cuts for LHCb are:
\begin{itemize} 
\item cuts for the W muon rapidity distributions
  \begin{equation*} 
    \centering 
    p_T^{\mu} \ge 20~ \textrm{GeV},\qquad 2.0 \le \eta^{\mu} \le 4.5; 
  \end{equation*} 
\end{itemize}

For all three data sets, we performed extensive cross-checks at NLO
using two different codes, \texttt{DYNNLO} and \texttt{MCFM}: we
checked that, once common settings are adopted, the results of the
\texttt{MCFM} and \texttt{DYNNLO} runs agree to better than 1\% for
all the data bins. In the particular case of the ATLAS W and Z
distributions, we also found good agreement with the \texttt{APPLgrid}
tables used in the recent \texttt{HERAfitter} analysis of ATLAS
data~\cite{Aad:2012sb}.
        
Concerning jet data, we have included the measurements from the
Tevatron experiments, which are important for constraining the gluon
PDF, together with the extended kinematics coverage provided by the
LHC jet data. From the 2010 $36 \textrm{pb}^{-1}$ data set inclusive
jet and dijet production have been measured by
CMS~\cite{CMS:2011mea,Chatrchyan:2011qta} and
ATLAS~\cite{Aad:2011fc}, however only ATLAS give the full
experimental covariance matrix. The covariance matrix is particularly
important for these data because they are highly correlated.

The theoretical calculation of NLO jet production cross sections in hadron
collisions can be carried out by exclusive parton level Monte Carlo
codes such as \texttt{NLOjet++} \cite{Nagy:2003tz} and
\texttt{EKS}-\texttt{MEKS} \cite{Ellis:1990ek,Gao:2012he}. These MC
codes provide NLO predictions which are consistently included in a
global PDF analysis using the fast NLO interfaces implemented in
\texttt{FastNLO} \cite{Kluge:2006xs,Wobisch:2011ij} or
\texttt{APPLgrid}~\cite{Carli:2010rw}.

The full NNLO corrections to the inclusive jet production were unknown
at that time. Only recently, results about the exact gluons-only
channel have been published in
Refs.~\cite{Currie:2013dwa,Ridder:2013mf}, but the full channel
prediction is still missing.
At that time only the threshold corrections to the inclusive jet $p_T$
distribution were available \cite{Kidonakis:2000gi}, thus the
inclusion of jet data into an NNLO analysis is necessarily
approximate. On the other hand, in NNPDF3.0 these threshold
corrections have been replaced by the improved predictions based on
threshold resummation published in Ref.~\cite{deFlorian:2013qia},
after applying a rejection criterion~\cite{Carrazza:2014hra} of
kinematical regions based on the difference to the exact gluons-only
channel prediction.

We compute inclusive jet cross-sections using \texttt{NLOjet++}
interfaced to \texttt{APPLgrid}. The jet reconstruction parameters are
identical to those used in the experimental
analysis~\cite{Aad:2010ad}.  The NLO calculation uses the anti-$k_T$
algorithm~\cite{Cacciari:2008gp}, and the factorization and
renormalization scales are set to be $p_T^\textrm{max}$, the
transverse momentum of the hardest jet in each event. We choose to
include in the analysis the data with $R=0.4$. These data are less
sensitive to nonperturbative corrections from the underlying event and
pileup as compared to the $R=0.6$
data~\cite{Dasgupta:2007wa,Cacciari:2008gd}, and though they are a bit
more sensitive to hadronization effects, all in all the
nonperturbative parton to hadron correction factors are smaller for
$R=0.4$ than for $R=0.6$. We have checked that the results are
essentially unchanged, both in terms of impact on PDFs and at the
level of the $\chi^2$ description if the $R=0.6$ data is used instead
of the $R=0.4$ data.

On top of the 86 sources of fully correlated systematic errors, the
ATLAS jet spectra have an additional source of uncertainty due to the
theoretical uncertainty in the computation of the hadron to parton
nonperturbative correction factors. We take these nonperturbative
corrections and their associated uncertainties from the ATLAS
analysis, where they are obtained from the variations of different
leading order Monte Carlo programs. It is clear from
Ref.~\cite{Aad:2010ad} that for a given Monte Carlo model the
nonperturbative correction is strongly correlated between data bins,
and thus conservatively we treat it as an additional source of fully
correlated systematic uncertainty, to be added to the covariance
matrix.

Because NNLO corrections to jet cross-sections are not available,
hadron collider jet data can only be included in a NNLO fit within
some approximations. Here, the NNLO theoretical predictions for CDF
and D0 inclusive jet data are obtained using the approximate NNLO
matrix element obtained from threshold resummation
\cite{Kidonakis:2000gi} as implemented in the \texttt{FastNLO}
framework \cite{Kluge:2006xs,Wobisch:2011ij}. For ATLAS data the
threshold approximation is expected to be worse because of the higher
centre-of-mass energy, and thus we simply used the NLO matrix element
with NNLO PDFs and $\alpha_s$. It was checked in
Ref.~\cite{Ball:2011uy} that the difference between fits with
approximate NNLO jet matrix elements, and fits with purely NLO matrix
elements is significantly smaller than the difference between fits
with and without jet data.
These choices have been updated in the NNPDF3.0 determination by
including consistently threshold resummation
$C$-factors~\cite{deFlorian:2013qia} only for data bins where the
exact gluons-only predictions are close to the
approximation~\cite{Carrazza:2014hra} and excluding all the other
bins.

\chapter{The photon PDF determination}
\label{sec:chap4}

In this chapter we present the determination of the NNPDF2.3QED set of
PDFs. This is the first NNPDF set with QED corrections.
As we have already mentioned at the beginning of this thesis, thanks
to the need of precise phenomenology at the
LHC~\cite{Forte:2010dt,Forte:2013wc,Campbell:2013qaa}, PDFs are determined using the
NNLO order in QCD. However, at this level of accuracy, also LO QED
corrections ($\mathcal{O}(\alpha)$) become relevant.
Some examples about the impact of QED and EW corrections to various
hadron collider processes have been studied in detail,
\textit{i.e.}~the inclusive $W$ and $Z$
production~\cite{Baur:1998kt,Zykunov:2001mn,Dittmaier:2001ay,Baur:2001ze,Baur:2004ig,Arbuzov:2007db,Arbuzov:2005dd,Brensing:2007qm,Balossini:2009sa,CarloniCalame:2007cd,Dittmaier:2009cr},
$W$ and $Z$ boson production in association with
jets~\cite{Denner:2009gj,Denner:2011vu,Denner:2012ts}, dijet
production~\cite{Moretti:2005ut,Dittmaier:2012kx} and top quark pair
production~\cite{Bernreuther:2005is,Kuhn:2005it,Hollik:2007sw,Hollik:2011ps,Kuhn:2013zoa}.

As we have seen in Chapter~\ref{sec:chap2}, the first step to obtain a
set of PDF with QED corrections consist in the implementation of such
corrections to PDF evolution, together with the addition of a new
parton: the photon PDF.
Before the determination of NNPDF2.3QED, we find in literature only
one PDF set with QED corrections: the MRST2004QED
set~\cite{Martin:2004dh}. In this pioneering work, the photon PDF was
determined based on a model inspired by photon radiation off
constituent quarks (though consistency with some HERA data was checked
a posteriori), and therefore not provided with a PDF uncertainty.

The aim of this chapter is to show how we construct a PDF set
including QED corrections, with a photon PDF parametrized in the same
way as all the other PDFs, and determined from a fit to
hard-scattering experimental data using the NNPDF methodology.
The goal is to construct a PDF set where
\begin{itemize}
\item QCD corrections are included up to NLO or NNLO;
\item QED corrections are included to LO;
\item the photon PDF is obtained from a fit to deep-inelastic
  scattering (DIS) and Drell-Yan (both low mass, on-shell $W$ and $Z$
  production, and high mass) data;
\item all other PDFs are constrained by the same data included in the
  NNPDF2.3 PDF determination~\cite{Ball:2012cx}, see Chapter~\ref{sec:chap3}.
\end{itemize}
We will consider negligible the impact of the lepton PDF, as well as
weak contributions to evolution equations~\cite{Ciafaloni:2001mu,
  Ciafaloni:2005fm}.

In principle, this goal could be achieved by simply performing a
global fit including QED and QCD corrections both to perturbative
evolution and to hard matrix elements, and with data which constrain
the photon PDF. In practice, this would require the availability of a
fast interface, like \texttt{APPLgrid}~\cite{Carli:2010rw} or
\texttt{FastNLO}~\cite{Wobisch:2011ij}, to codes which include QED
corrections to processes which are sensitive to the photon PDF, such
as single or double gauge boson production. Because such interfaces
are not available, we adopt instead a reweighting procedure, which
turns out to be sufficiently accurate to accommodate all relevant
existing data.

In Figure~\ref{flowchart} we summarize the steps for the construction
of this special set of PDFs:
\begin{enumerate}

\item In the first step, we construct a set of PDFs (NNPDF2.3QED
  DIS-only), including a photon PDF, by performing a fit to DIS data
  only, based on the same DIS data used for NNPDF2.3 (see
  Sect.~\ref{sec-expdata} in Chap.~\ref{sec:chap3}), and using either
  NLO or NNLO QCD and LO QED theory. To leading order in QED, the
  photon PDF only contributes to DIS through perturbative evolution
  (just like the gluon PDF to leading order in QCD). Therefore, the
  photon PDF is only weakly constrained by DIS data, and thus the
  photon PDF in the NNPDF2.3QED DIS-only set is affected by large
  uncertainties. The result is a pair of PDF sets: NNPDF2.3QED
  DIS-only, NLO or NNLO, according to how QCD evolution has been
  treated.

\item Then, each replica of the photon PDF from the NNPDF2.3QED
  DIS-only set is combined with a random PDF replica of a set of the
  default NNPDF2.3 PDFs, fitted to the global dataset. This works
  because of the small correlation between the photon PDF and other
  PDFs, as we shall explicitly check. Also, the violation of the
  momentum sum rule that this procedure entails is not larger than the
  uncertainty on the momentum sum rule in the global QCD fit. The
  procedure is performed using NLO or NNLO NNPDF2.3 PDFs, for three
  values of $\alpha_s(M_z)=0.117,\>0.118,\>0.119$.  The photon PDF
  determined in the NNPDF2.3QED DIS-only fit is in fact almost
  independent of the value of $\alpha_s$ within this range.  This
  leads to several sets of PDF replicas, which we call NNPDF2.3QED
  prior, at the scale $Q_0^2$.

\item At this stage, we evolve the NNPDF2.3QED prior set to all $Q^2$
  using combined QCD$\otimes$QED evolution equations, to LO in QED and
  either to NLO or NNLO in QCD and with the appropriate value of
  $\alpha_s$, using the strategy explained in details in
  Chap.~\ref{sec:chap2} with the \texttt{APFEL} implementation.

\item The LHC $W$ and $Z/\gamma^*$ production data are now included in
  the fit by Bayesian reweighting~\cite{Ball:2010gb} of the
  NNPDF2.3QED prior PDF set.  

\item Finally, the set of reweighted replicas is then
  unweighted~\cite{Ball:2011gg} in order to obtain a standard set of
  100 replicas of our final NNPDF2.3QED set.

\end{enumerate}

\tikzstyle{block} = [rectangle, draw, fill=green!20, 
    text width=26em, text centered, rounded corners, minimum height=4em]
\tikzstyle{line} = [draw, -latex']
\begin{figure}
  \begin{center}
    \begin{tikzpicture}[node distance = 2.4cm, auto]
      \node [block] (f1) {Perform a fit to DIS data with QED
        corrections:\\
        NNPDF2.3QED DIS-only, $N_\textrm{rep}$ = 500};
      \node [block, below of=f1] (f2) {Construct NNPDF2.3QED prior at $Q_0^2$: \\
        (a) Quark and gluon PDFs from NNPDF2.3 global \\
        (b) Photon PDFs from NNPDF2.3 DIS-only };
      \node [block,below of=f2] (f3) {Evolve NNPDF2.3QED prior to all $Q^2$, \\
        with QCD+QED DGLAP equations};
      \node [block,below of=f3] (f4) {Compute predictions for LHC $W,Z/\gamma^{*}$ production;\\
        reweight NNPDF2.3QED prior }; \node [block,below of=f4] (f5)
      {Unweight the reweighted
        PDF set to get the final NNPDF2.3QED\\
        set of $N_\textrm{rep}$ = 100 replicas}; \path [line] (f1) --
      (f2); \path [line] (f2) -- (f3); \path [line] (f3) -- (f4);
      \path [line] (f4) -- (f5);
    \end{tikzpicture}
  \end{center}
  \caption{\label{flowchart}Flow-chart for the construction of the
    NNPDF2.3QED set. }
\end{figure}
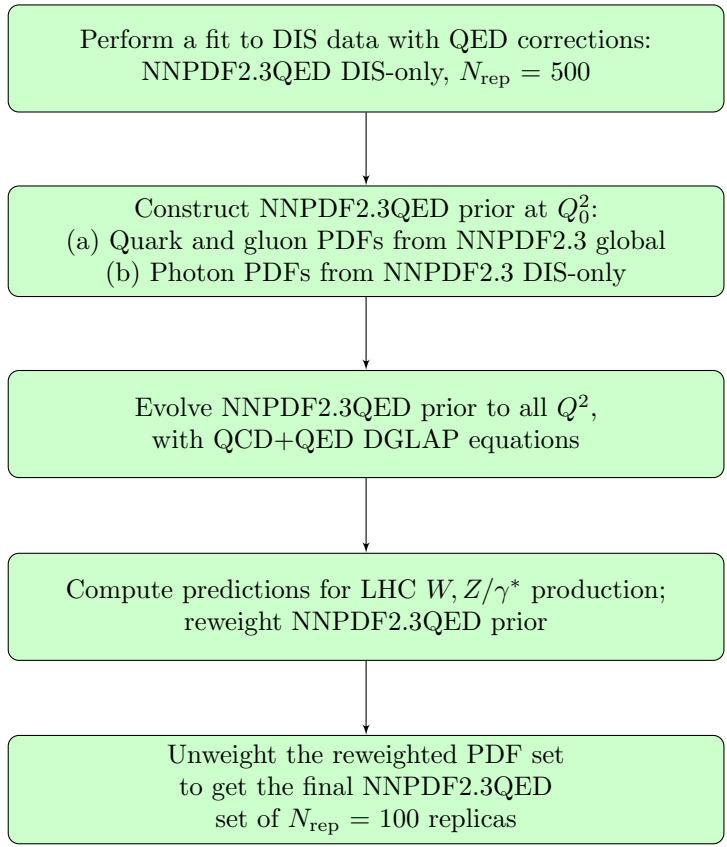

As we will see, the photon PDF in NNPDF2.3QED turns out to be in good
agreement with that from the MRST2004QED set at medium large
$x\gtrsim0.03$, while for smaller $x$ values it is substantially
smaller (by about a factor three for $x\sim10^{-3}$), though
everywhere affected by sizable uncertainties, typically of order 50\%.

This chapter is organized as follows. In Sect.~\ref{sec:disfit} we
also discuss the first step of our procedure, namely, the
determination of NNPDF2.3QED DIS-only PDF set. The subsequent steps,
namely the construction of the NNPDF2.3QED prior set, and its
reweighting and unweighting leading to the final NNPDF2.3QED set are
presented in Sect.~\ref{sec:lhcwz}. Finally, phenomenological
investigations of this set of PDFs are presented in
Chap.~\ref{sec:chap5}.

\section{Deep-inelastic scattering with QED corrections}
\label{sec:disfit}

\begin{figure}
  \begin{centering}
    \includegraphics[scale=0.45]{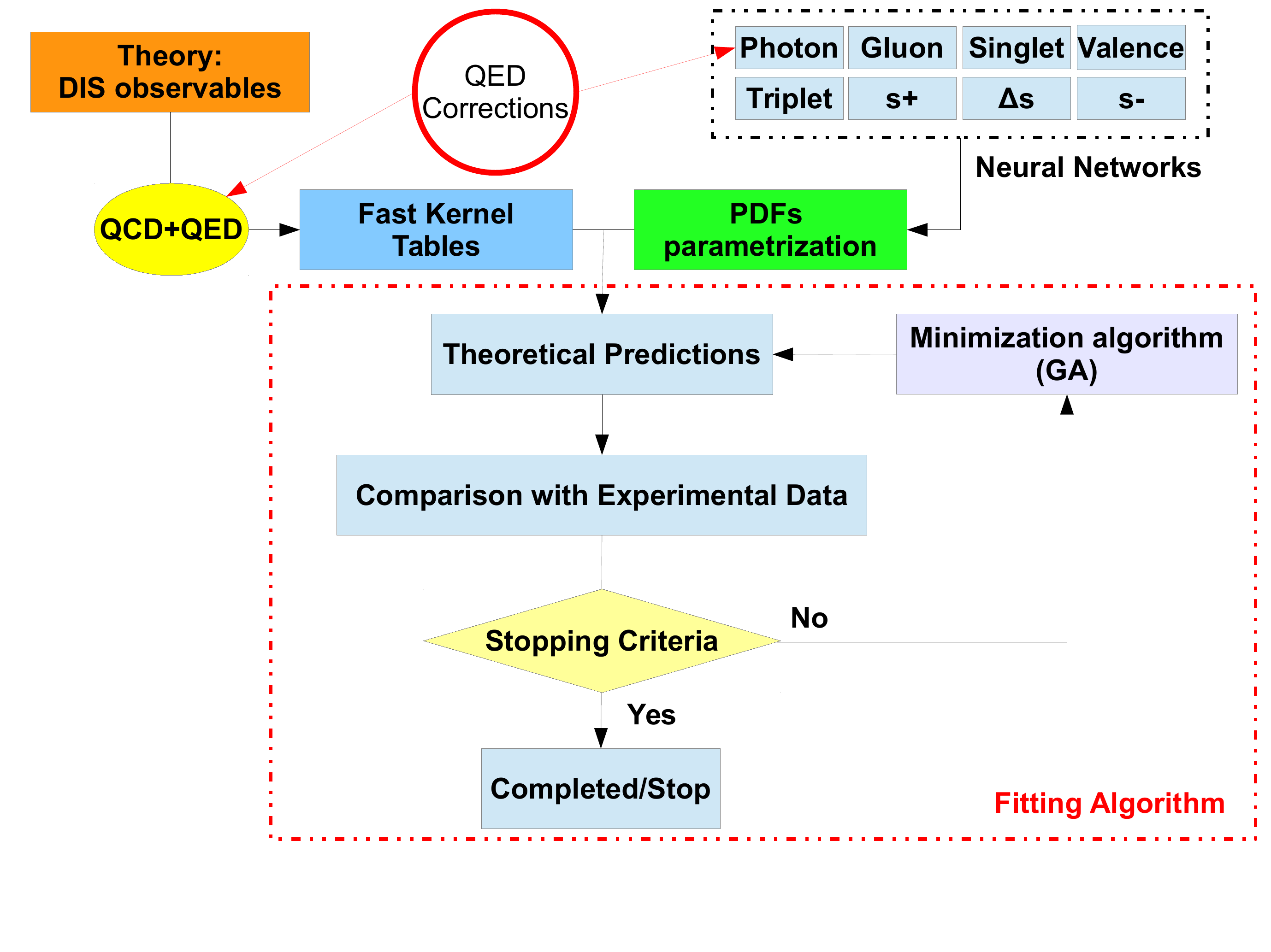}
    \par\end{centering}
  \caption{Graphical representation of the fitting strategy.}
  \label{fig:fitscheme}
\end{figure}

\subsection{Fitting PDFs with QED corrections}
\label{sec:disqed}

Let us now proceed with a first determination of the photon PDF from a
fit to deep-inelastic data. We want to include QED corrections to DIS
at LO, i.e., more accurately, the leading log level. This means that
the splitting functions are computed to $\mathcal{O}(\alpha)$, while
all partonic cross-sections (coefficient functions) are determined to
lowest order in $\alpha$. Because the photon is electrically neutral,
the photon deep-inelastic coefficient function only starts at
$\mathcal{O}(\alpha^2)$, while quark coefficient functions start at
$\mathcal{O}(\alpha)$. This means that at LO the photon coefficient
function vanishes, and the photon only contributes to DIS through its
mixing with quarks due to perturbative evolution. This is fully
analogous to the role of the gluon in the standard LO QCD description
of DIS: the gluon coefficient function only starts at
$\mathcal{O}(\alpha_s)$ while the quark coefficient function starts at
$\mathcal{O}(1)$, so at LO the gluon only contributes to deep-inelastic
scattering through its mixing with quarks upon perturbative evolution.

An important issue when including QED corrections is the choice of the
factorization scheme in the subtraction of QED collinear
singularities~\cite{Diener:2005me,Dittmaier:2009cr}. Different
factorization schemes differ by next-to-leading log terms.  Because
our treatment of QED evolution is at the leading log level, our
results do not depend on the choice of factorization scheme. This
means that if our photon PDF is used in conjunction with a
next-to-leading log computation of QED cross-sections, the latter can
be taken in any (reasonable) factorization scheme. The difference in
results found when changing the QED factorization scheme should be
considered to be part of the theoretical uncertainty.  However, in
practice, in some schemes the perturbative expansion may show faster
convergence (so, for example, next-to-leading log results are closer
to leading-log ones in some schemes than others).  We will indeed see
in the next section that when DIS data are combined with Drell-Yan
data it is advantageous to use the DIS factorization scheme, which is
defined by requiring that the deep-inelastic structure function $F_2$
is given to all orders by its leading-order
expression~\cite{Diener:2005me,Dittmaier:2009cr}.

The starting point of our fit to DIS data including QED corrections is
the NNPDF2.3 PDF determination, in terms of experimental data, theory
settings and methodology.  We will perform fits at NLO and NNLO in
QCD, for three different values of $\alpha_s\left(
  M_Z\right)=0.117,\>0.118$ and 0.119, all with LO QED evolution.
Unless otherwise stated, in the following all results, tables, and
plots will use the $\alpha_s=0.119$ PDF sets.

We add to the NNPDF default set of seven independent PDF combinations
a new, independently parametrized PDF for the photon, in a completely
analogous way to all other PDFs (see Sect.~\ref{sec:pdfparam}), with a
small modification related to positivity to be discussed below:
\begin{eqnarray}
\gamma(x,Q_0^2) = \left( 1-x\right)^{m_{\gamma}}x^{-n_{\gamma}}
 \textrm{NN}_{\gamma}(x),
 \label{eq:photon}
\end{eqnarray}
where $\textrm{NN}_{\gamma}(x)$ is a multi-layer feed-forward neural
network with 2-5-3-1 architecture, with a total of 37 parameters to be
determined by experimental data, and the prefactor is a preprocessing
function used to speed up minimization, and on which the final result
should not depend.  The preprocessing function is parametrized by the
exponents $m_\gamma$ and $n_\gamma$, whose values are chosen at random
for each replica, with uniform distribution in the range
 \begin{eqnarray}\label{prerange}
 1 \le m_{\gamma} \le 20,  \qquad
 -1.5 \le n_{\gamma} \le 1.5.  
 \end{eqnarray}
 We have explicitly checked that the results are independent 
 on the preprocessing range, by computing for each replica the
 effective small- and large-$x$ exponents~\cite{Ball:2013lla}, defined as 
\begin{eqnarray}
n_\gamma[ \gamma(x,Q^2)]=\frac{\ln  \gamma(x,Q^2)}{\ln 1/x}\mbox{ ,}
\qquad m_\gamma[  \gamma(x,Q^2)]=\frac{ \ln  \gamma(x,Q^2) }{\ln(1-x)}\mbox{ ,}
\label{eq:exp}
\end{eqnarray}
and verifying that the range of the effective exponents at small- and
large-$x$ respectively is well within the range of variation of the
preprocessing exponents, thus showing that the small- and large-$x$
behaviour of the best-fit PDFs is not constrained by the choice of
preprocessing but rather determined by experimental data.

A graphical representation of the strategy described above is shown in
Figure~\ref{fig:fitscheme}. The DIS predictions and the combined
QCD$\otimes$QED evolution are encoded in \texttt{FastKernel}
tables. The photon PDF parametrization is added to the other flavors
of the NNPDF2.3 basis. The convolution between both elements produce
theoretical prediction which are compared to experimental data using
the standard NNPDF minimization strategy, presented in details in
Chap.~\ref{sec:chap3}.
 
Parton distributions must satisfy positivity conditions which follow
from the requirement that, even though PDFs are not directly
physically observable, they must lead to positive-definite physical
cross-sections~\cite{Altarelli:1998gn}. Leading-order PDFs are
directly observable, and thus they must be positive-definite: indeed,
they admit a probabilistic interpretation. Because we treat QED
effects at LO, the photon PDF must be positive definite. This is
achieved, as in the construction of the NNPDF2.1 LO PDF
sets~\cite{Ball:2011uy}, by squaring the output of the neuron in the
last (linear) layer of the neural network $\textrm{NN}_{\gamma}(x)$,
so that $\textrm{NN}_{\gamma}(x)$ is a positive semi-definite
function.
  
Once QED evolution is switched on, isospin is no longer a good
symmetry, and thus it can no longer be used to relate the PDFs of the
proton and neutron. Because deuteron deep-inelastic scattering data
are used in the fit, in principle this requires an independent
parametrization for proton and neutron PDFs. Experimental data for the
neutron PDFs would then no longer provide a useful constraint, and in
particular they would no longer constrain the isospin triplet
PDF. Whereas future PDF fits including substantially more LHC data
might allow for an accurate PDF determination without using deuteron
data, this does not seem to be possible at present.

There are two separate issues here: one, is the amount of isospin
violation in the quark and gluon PDFs, and the second is the amount of
isospin violation in the photon PDF. At the scale at which PDFs are
parametrized, which is of the order of the nucleon mass, we expect
isospin violating effects in the quark and gluon PDFs to be of the
same order as that displayed in baryon spectroscopy, which is at the
per mille level, much below the current PDF uncertainties (isospin
violations of this order have been predicted, among others, on the
basis of bag model estimates~\cite{Londergan:2003pq}). The second is
the amount of isospin violation in the photon distribution itself:
this could be somewhat larger (perhaps at the percent level), however
any reasonable amount of isospin violation in the photon is way below
the uncertainty on the photon PDF. Therefore, we will assume that no
isospin violation is present at the initial scale.

Of course, even with isospin conserving PDFs at the starting scale,
isospin violation is then generated by QED evolution: this is
consistently accounted for when solving the evolution equations, by
determining separate solutions for the proton and neutron so that at
any scale $Q\not =Q_0$, $u^p(x,Q^2)\not=d^n(x,Q^2)$ and
$d^p(x,Q^2)\not=u^n(x,Q^2)$. Because of the larger electric charge of
the up quark, the dynamically generated photon PDF ends up being
larger for the proton than it is for the neutron.
 
In Ref.~\cite{Martin:2004dh} isospin violation was parametrized on the
basis of model assumptions.  We will compare our results for isospin
violation to those of this reference in Sect.~\ref{sec:nnpdf23qed}
below: we will see that while indeed the amount of isospin violation
in the photon PDF from that reference is somewhat larger than our own,
it is much smaller than the relevant uncertainty.

\subsection{The photon PDF from DIS data}
\label{sec:disfitgam}

\begin{figure}
  \begin{centering}
    \includegraphics[scale=0.45]{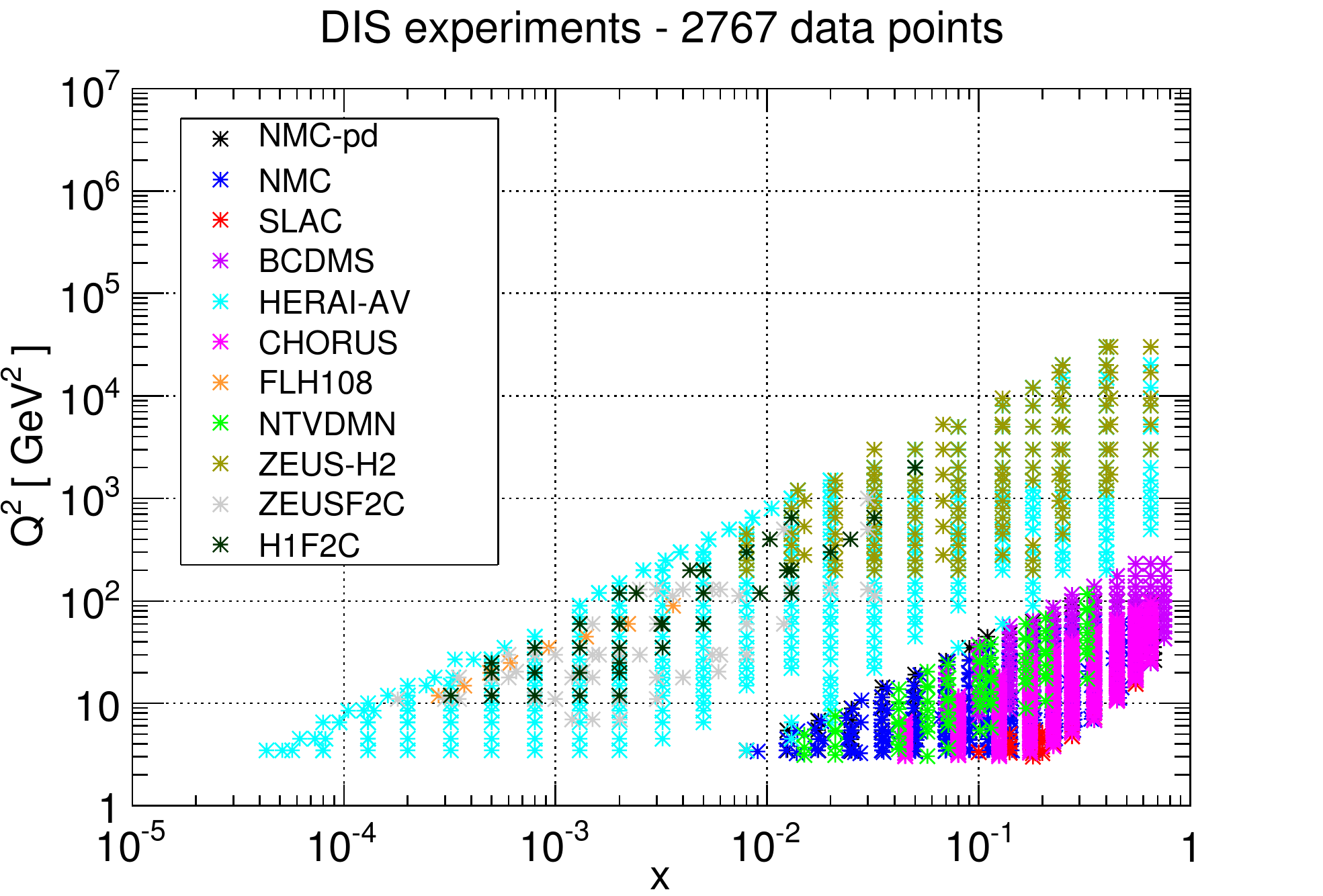}
    \par\end{centering}
  \caption{Kinematic coverage of the experimental DIS data used in the
    determination of the photon PDF.}
  \label{fig:diskin}
\end{figure}

\begin{table}
\centering
\small
\begin{tabular}{c||c|c||c|c}
\hline 
& \multicolumn{2}{c||}{\textbf{NLO}} & \multicolumn{2}{|c}{\textbf{NNLO}}  \\
\hline 
Experiment  & QCD & QCD+QED  & QCD  & QCD+QED  \\
\hline 
\hline 
Total  &  1.10 &  1.10  & 1.10 & 1.10 \\  
 \hline 
NMC-pd              & 0.88  & 0.87      & 0.88   &0.88     \\
 NMC                 & 1.68  &  1.70  & 1.67  &1.69    \\
SLAC                &  1.36   & 1.40   & 1.08 &  1.10     \\
BCDMS               &  1.17  &  1.16  &  1.24 & 1.23     \\
CHORUS              &  1.01  &  1.01  &  0.98  & 0.99    \\
NTVDMN              &  0.54   & 0.54   & 0.56   & 0.54    \\
HERAI-AV            & 1.01    & 1.01   & 1.04  & 1.03     \\
FLH108              &  1.34  &  1.34  &  1.25  &1.24    \\
ZEUS-H2             &  1.26   & 1.25   & 1.24  & 1.25    \\
ZEUS $F_2^c$        &  0.75   &  0.75  & 0.76  &0.78    \\
H1 $F_2^c$          &  1.55  &  1.50  &  1.41  &1.39    \\
 \hline
\hline
\end{tabular}
\caption{\label{tab:chi2} The $\chi^2$ values
per data point for individual experiments computed in the
  NNPDF2.3 DIS-only NLO and NNLO PDF sets, in the QCD-only fits
compared to the results with combined QCD$\otimes$QED evolution.
 All  $\chi^2$ values have been
obtained using  $N_{\textrm{rep}}$=100 replicas with
  $\alpha_s(M_{\textrm{Z}})=0.119$.
  Normalization uncertainties have been included using 
  the experimental definition of the covariance matrix,
  see App. A of Ref.~\cite{Ball:2012wy}, while in the actual
  fitting the $t_0$ definition was used~\cite{Ball:2009qv}. 
}
\end{table}

We have performed two fits at NLO and NNLO to DIS data only, with the
same settings used for NNPDF2.3, but with QED corrections in the PDF
evolution now included, as discussed in Chap.~\ref{sec:chap2}. The
kinematic coverage of experimental DIS data used in this fit is
presented in Figure~\ref{fig:diskin}.

The $\chi^2$ for the fit to the total dataset and the individual DIS
experiments are shown in Table~\ref{tab:chi2}, with and without QED
corrections, and with QCD corrections included either at NLO or at
NNLO. The $\chi^2$ listed in the table use the so-called experimental
definition of the $\chi^2$, in which normalization uncertainties are
included in the covariance matrix: this definition is most suitable
for benchmarking purposes, as it is independent of the fit results,
but it is unsuitable for minimization as it would lead to biased fit
results.  It is clear that there is essentially no difference in fit
quality between the QCD and QED$\otimes$QCD fits.  Indeed, a direct
comparison of the PDFs obtained in the pairs of fits with and without
QED corrections show that they differ very little.
\begin{figure}[ht]
  \begin{centering}
    \includegraphics[scale=0.68]{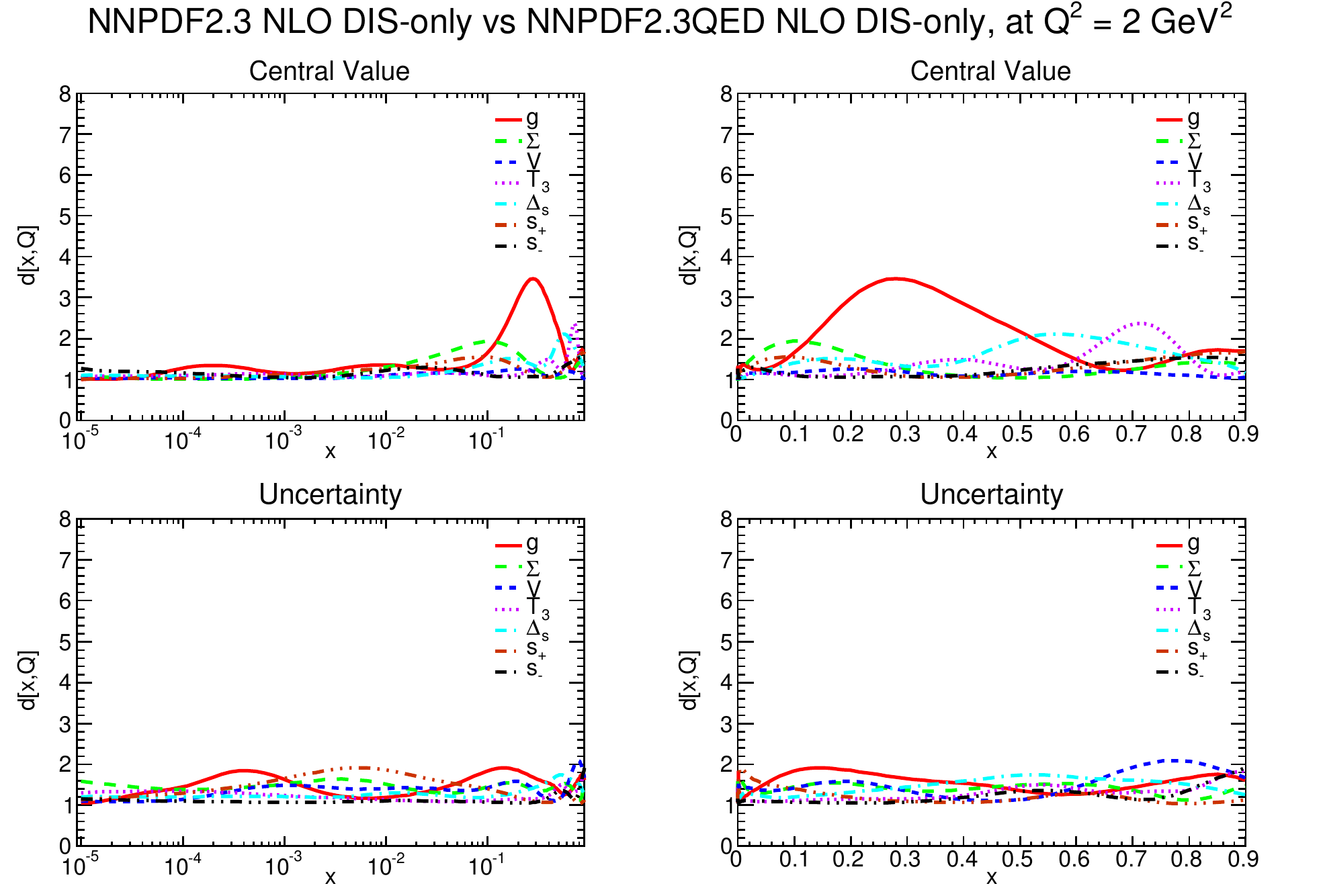}
    \par\end{centering}
  \caption{\label{fig:distances-nlo} Distances
    between PDFs in the NNPDF2.3 NLO DIS-only fit and the fit
    including QED corrections, at the input scale of $Q_0^2$=2
    GeV$^2$. Distances between central values (top) and uncertainties
    (bottom) are shown, on a logarithmic (left) or linear (right)
    scale in $x$.}
\end{figure}

\begin{figure}[ht]
  \begin{centering}
    \includegraphics[scale=0.34]{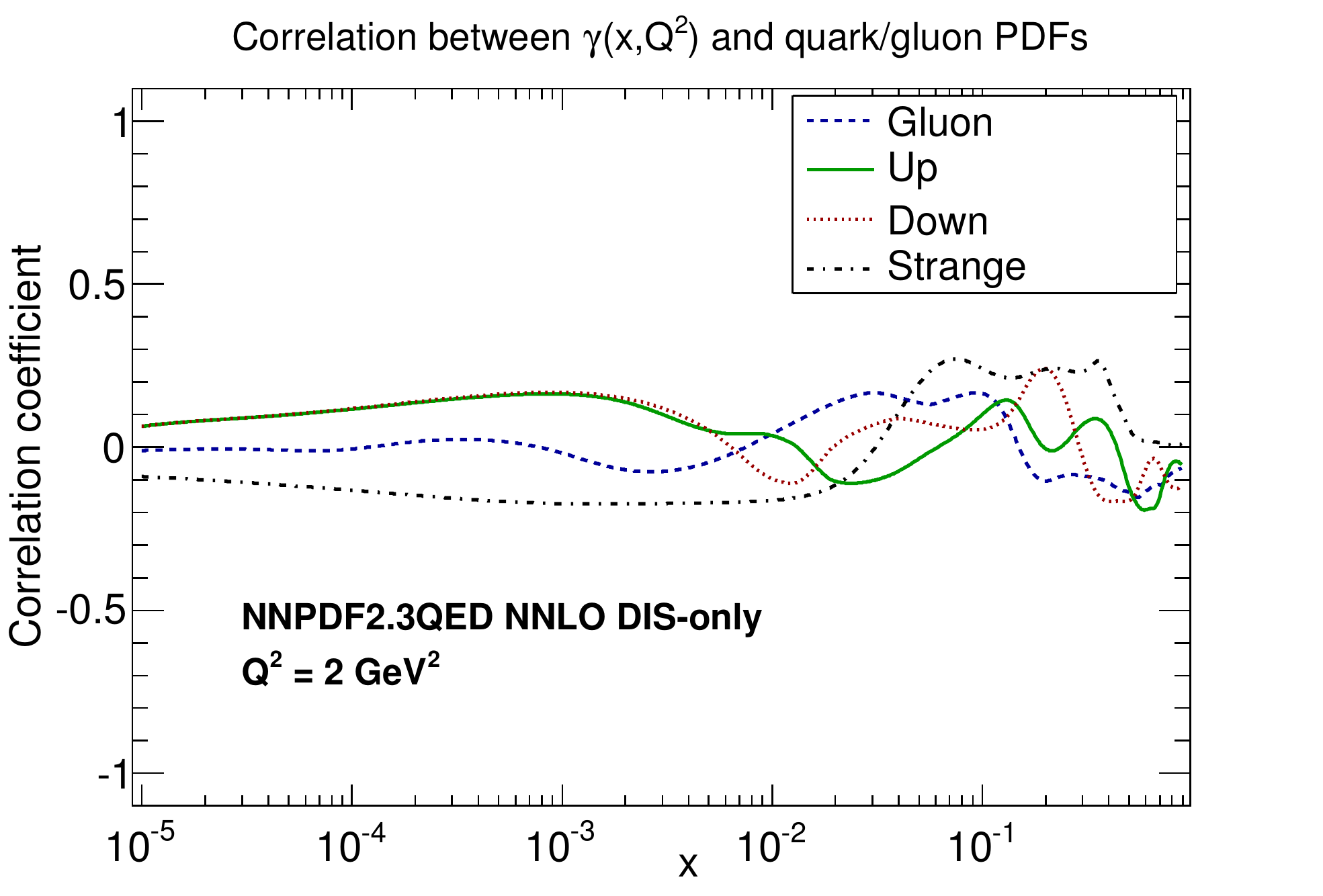}\includegraphics[scale=0.34]{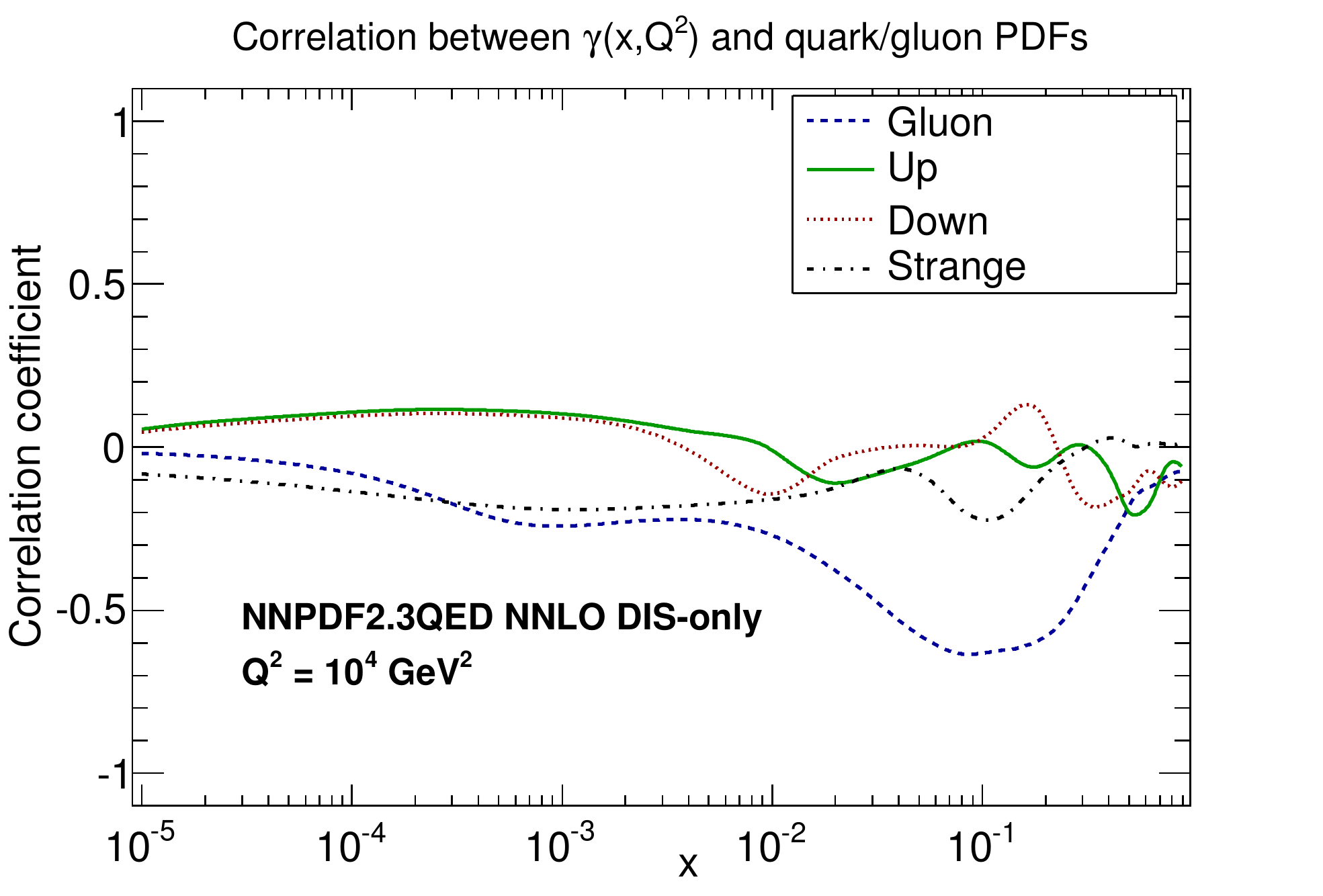}
    \par\end{centering}
  \caption{\label{fig:gammacorr} Correlation between the photon
    and other PDFs in the NNPDF2.3QED NLO DIS-only fit, shown as a
    function of $x$ at the input scale  $Q_0^2$=2
    GeV$^2$ (left) and at $Q^2=10^4$~GeV$^2$.}
\end{figure}
\begin{figure}[ht]
  \begin{centering}
    \includegraphics[scale=0.34]{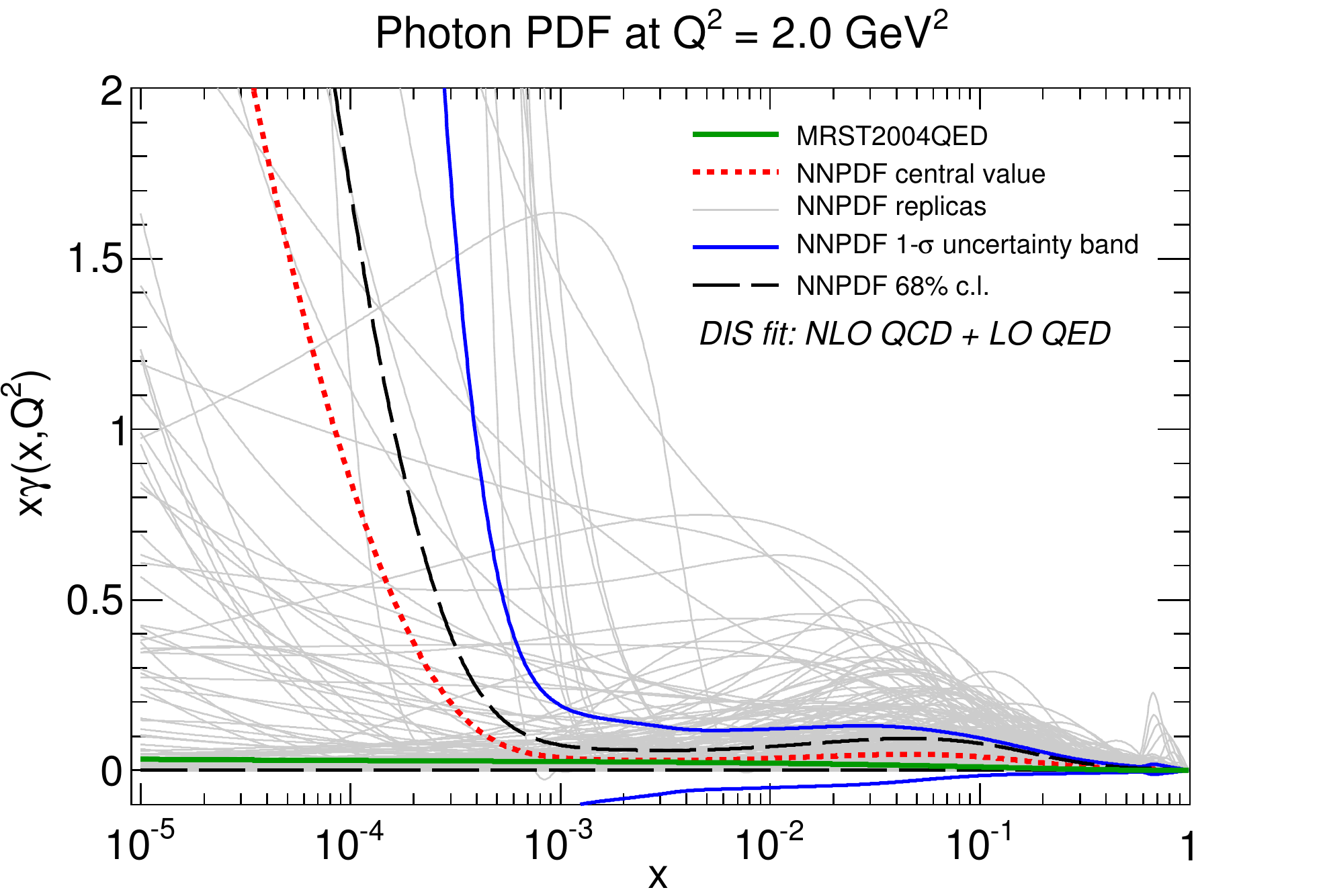}\includegraphics[scale=0.34]{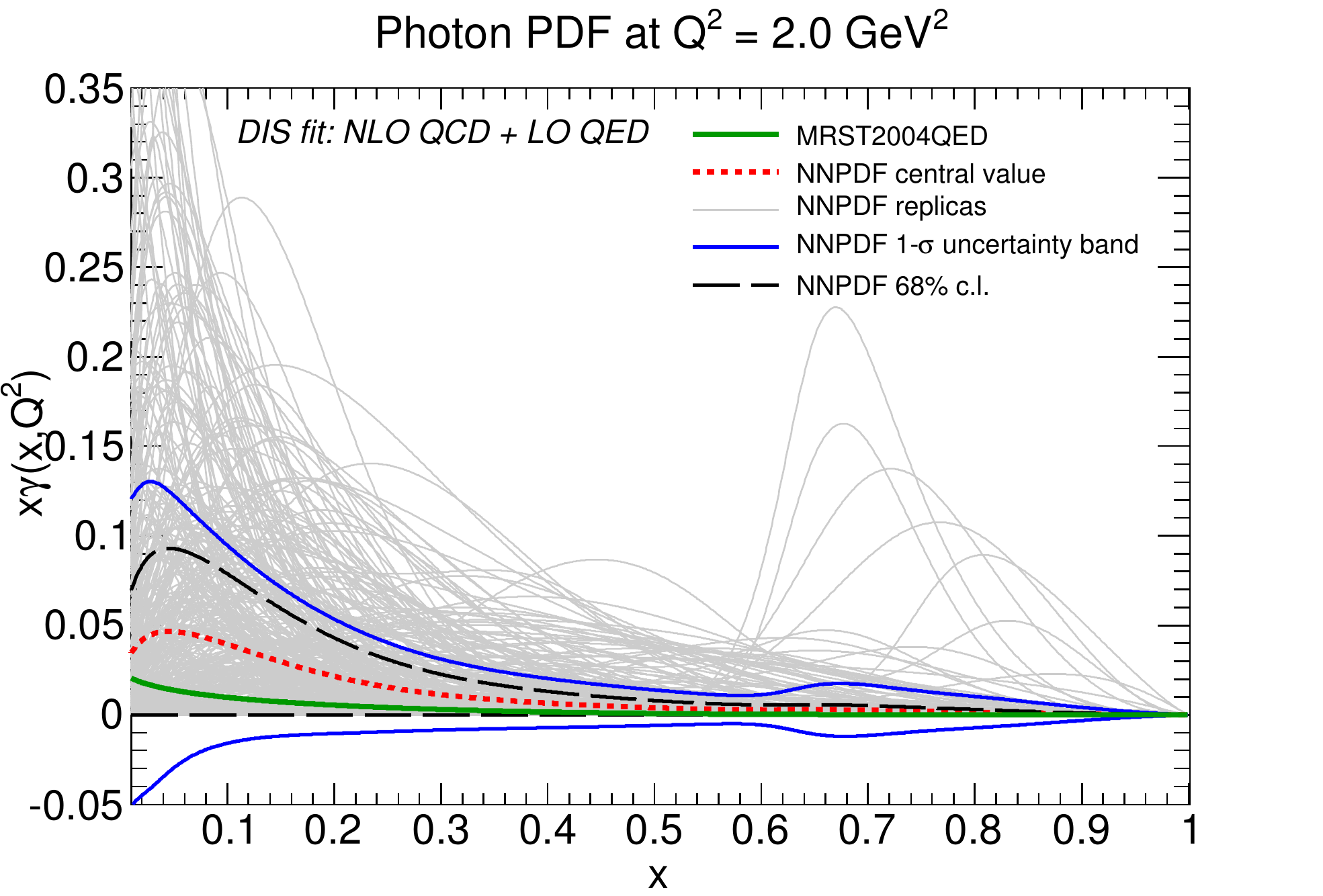}
    \par\end{centering}
  \caption{\label{fig:xpht-nlo} The photon PDF determined from the
    NNPDF2.3QED NLO DIS-only fit, in a linear (left plot) and
    logarithmic (right plot) scales, $N_{\text{rep}}=500$.  We show
    the central value (mean), the individual replicas and the PDF
    uncertainty band defined as a one $\sigma$ sigma interval and as a
    symmetric 68\% confidence level centered at the mean. The
    MRST2004QED photon PDF is also shown. }
\end{figure}
In order to assess this difference quantitatively, in
Figure~\ref{fig:distances-nlo} we plot the distance between central
values and uncertainties of individual combinations of PDFs in the NLO
QCD fit before and after the inclusion of QED corrections. We refer to
Appendix~\ref{app:distances} for the definition of distance.  Recall
that for a set of $N_\textrm{rep}$ PDF replicas, $d\sim1$ corresponds
to PDFs extracted from the same underlying distribution,
\textit{i.e.}~to statistically equivalent PDF sets, while $d\sim
\sqrt{N_{\textrm{rep}}}$ (so $d \sim 10$ in our case) corresponds to
PDFs extracted from distributions whose means or central values differ
by one $\sigma$.  The distances are shown in
Figure~\ref{fig:distances-nlo} for the NLO fit: it is clear that all
PDFs but the gluon from the sets with and without QED corrections are
statistically equivalent, while the gluon shows a change in the
valence region of less than half $\sigma$.  These results are
unchanged when QCD is treated at NNLO order.

The fact that the inclusion of a photon PDF has a negligible impact on
other PDFs can be also seen by determining the correlation between the
photon and other PDFs. Results are shown in
Figure~\ref{fig:gammacorr}. The correlation is negligible at the input
scale, meaning that the particular shape of the photon in each replica
has essentially no effect on the other PDFs of that replica. In
particular, this correlation is much smaller than that which arises at
a higher scale (also shown in Figure~\ref{fig:gammacorr}), due to the
mixing of PDFs with the photon induced by PDF evolution.

Hence, at the initial scale $Q_0^2=2$~GeV$^2$ the sets with and
without QED corrections differ mainly because of the presence of a
photon PDF in the latter. The photon PDF determined in the NLO fit is
shown in Figure~\ref{fig:xpht-nlo} at $Q_0^2=2$~GeV$^2$: the
individual replicas, the mean value, the one-$\sigma$ range and the
68\% confidence interval are all shown. The MRST2004QED photon PDF is
also shown.  It is clear that positivity imposes a strong constraint
on the photon PDF, which is only very loosely constrained by DIS
data. As a consequence, the probability distribution of replicas is
very asymmetric: some replicas may have large positive values of
$\gamma(x,Q^2)$, but positivity always ensures that no replica goes
below zero. It follows that the usual gaussian assumptions cannot be
made, and in particular there is a certain latitude in how to define
the uncertainty. Here and in the remainder of this paper we will
always define central values as the mean of the distribution, and
uncertainties as symmetric 68\% confidence levels centered at the
mean, namely, as the symmetric interval centered at the mean such that
68\% of the replicas falls within it.  All uncertainty bands will be
determined in this way, unless otherwise stated. Because of the
accumulation of replicas just above zero, the lower edge of the
uncertainty band on the photon PDF at the initial scale turns out to
be very close to zero.  Again, results are essentially unchanged when
the fit is done using NNLO QCD theory.

\begin{figure}[ht]
\begin{centering}
\includegraphics[scale=0.34]{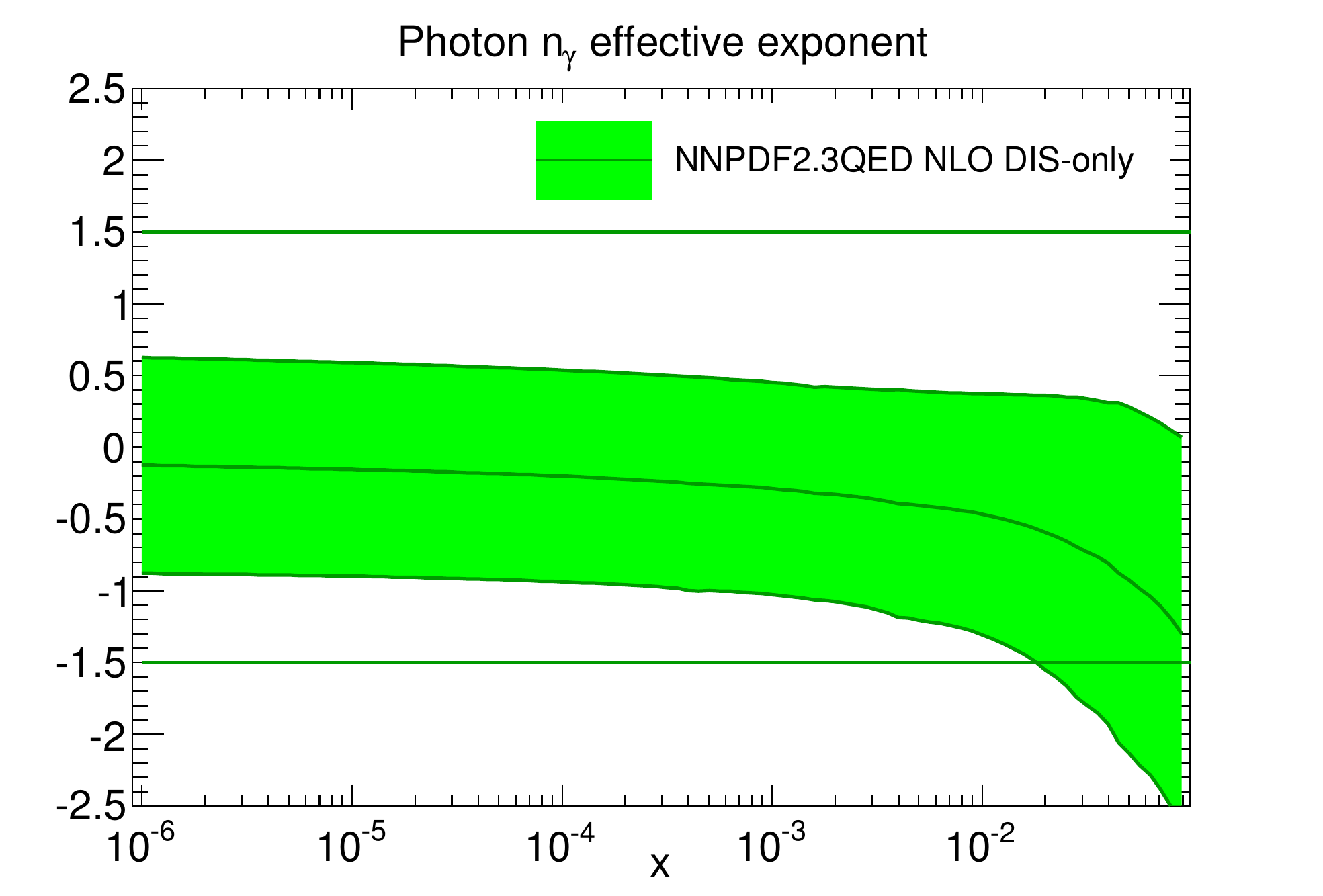}\includegraphics[scale=0.34]{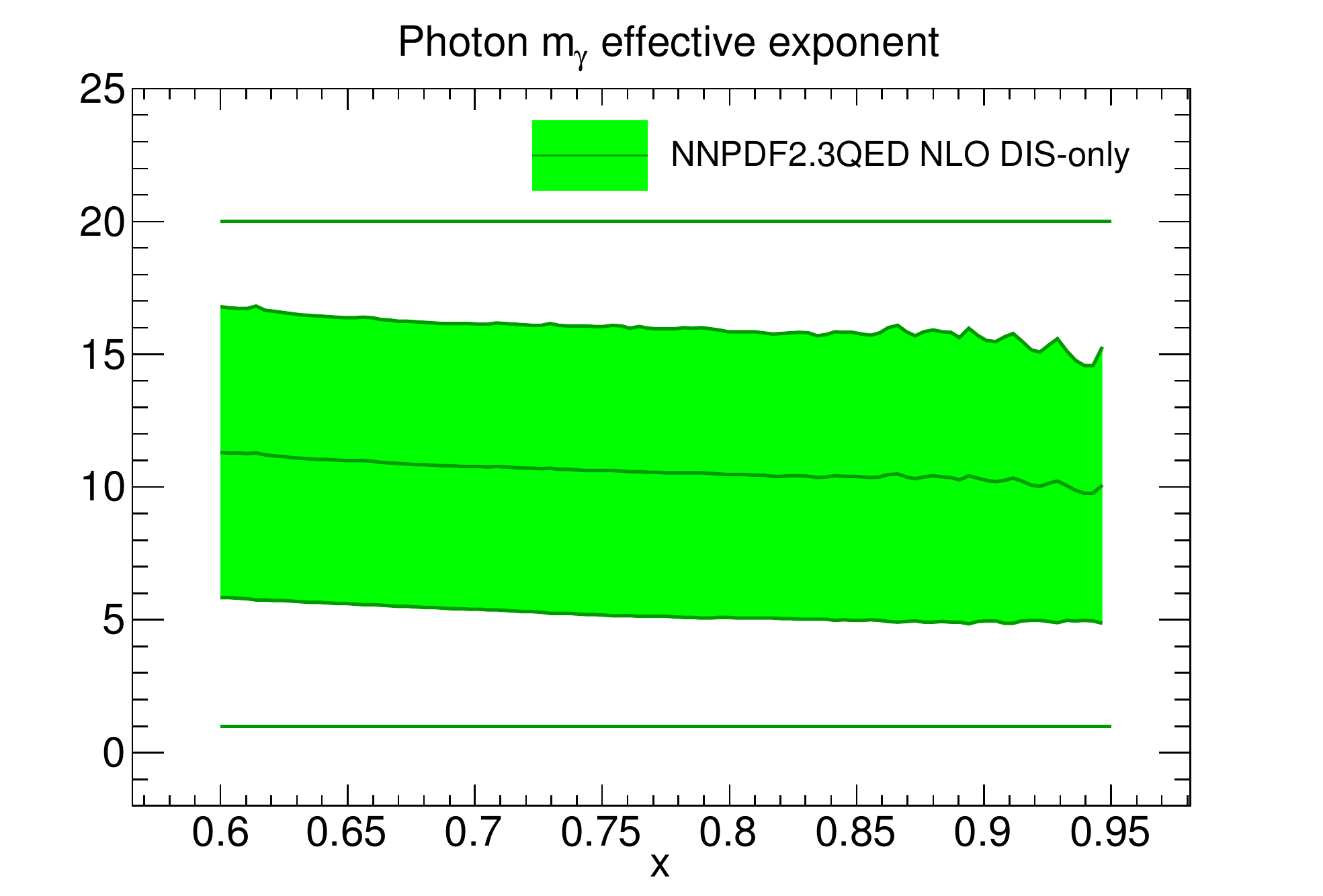}
\par\end{centering}
\caption{\label{fig:preproc} One-$\sigma$ range for the effective
  exponents Eq.~(\ref{eq:exp}) for the photon PDF, compared to the
  range of variation of the preprocessing exponents
  Eq.~(\ref{prerange}) (shown as horizontal lines).}
\end{figure}

As discussed in Sect.~\ref{sec:disqed}, we have determined the
effective exponents Eq.~(\ref{eq:exp}) for the photon PDF, and
compared them to the range of variation of the preprocessing exponents
Eq.~(\ref{prerange}). Given the very loose constraints that the data
impose on the photon PDF, it is especially important to make sure that
preprocessing imposes no bias. The comparison is shown in
Figure~\ref{fig:preproc}: it is clear that the effective exponents are
well within the range chosen for the preprocessing exponents, so that
no bias is being introduced.

The photon PDF at the initial scale shown in Figure~\ref{fig:xpht-nlo}
is essentially compatible with zero, and it remains small even at the
top of its uncertainty band; it is consistent with the MRST2004QED
photon PDF within its large uncertainty band.

\begin{figure}
  \begin{centering}
    \includegraphics[scale=0.37]{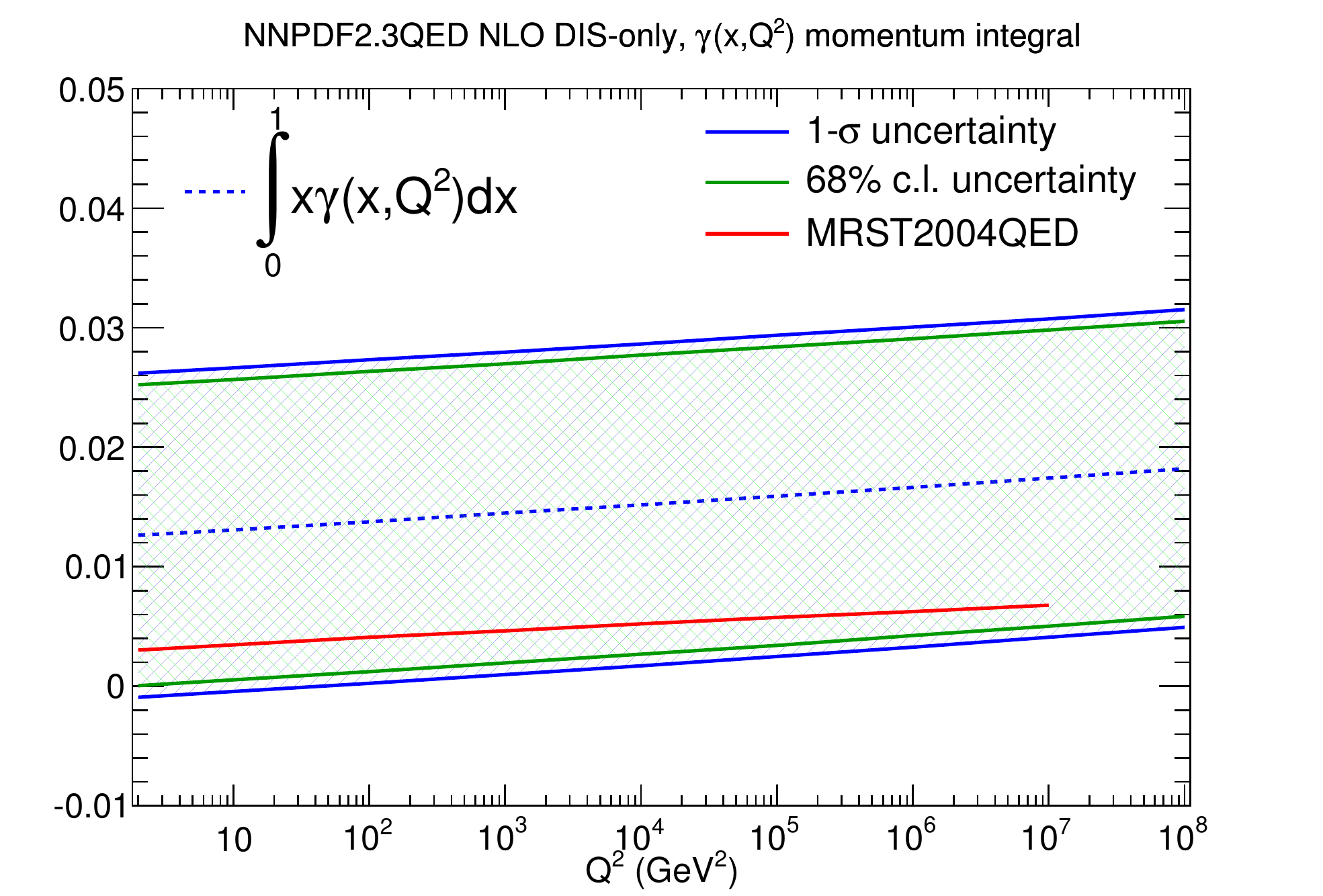}
    \par\end{centering}
  \caption{\label{fig:msr-gamma} The momentum fraction carried
   by the photon PDF in the NLO fit as a function of scale. The
   MRST2004QED result is also shown.}
\end{figure}

The momentum fraction carried by the photon is accordingly small: it
is shown as a function of scale in Figure~\ref{fig:msr-gamma} for the
NLO fit; results at NNLO are very similar.
At the input scale $Q_0^2=2$~GeV$^2$ we find
\begin{eqnarray}
\int_0^1 x \gamma \left(x,Q^{2}_0\right)= \left( 1.26 \pm 1.26 \right)\,\% \,  \, ,\label{eq:msr}
\end{eqnarray}
The symmetric 68\% confidence level uncertainty of Eq.~(\ref{eq:msr})
turns out to be quite close to the standard deviation
$\sigma=1.36\%$. Hence, even at the top of its uncertainty range the
photon momentum fraction hardly exceeds 2\%, and it is compatible with
zero to one $\sigma$. The momentum fraction carried by the the
MRST2004QED photon (also shown in Figure~\ref{fig:msr-gamma}) is well
below 1\%, and thus compatible with our own within uncertainties

\section{The photon PDF from $W$ and $Z$  production at the LHC}
\label{sec:lhcwz}

\begin{figure}
\begin{centering}
\includegraphics[scale=0.65]{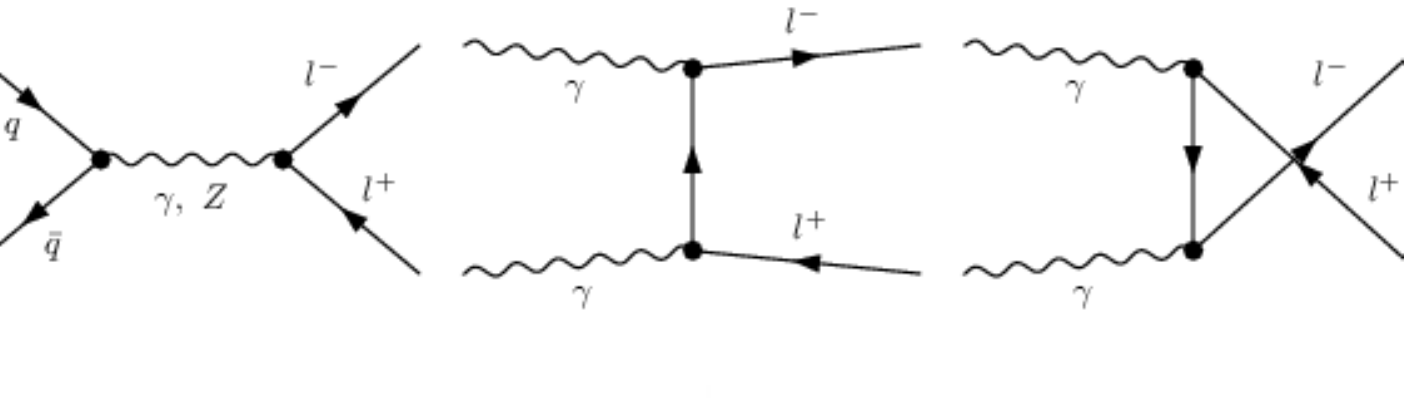}
\par\end{centering}
\caption{\label{fig:lhczborn} Feynman diagrams for the  Born-level partonic
  subprocesses which contribute to the production of dilepton pairs in
  hadronic collisions.
}
\end{figure}

As we have seen in the last section the photon PDF $\gamma(x,Q^2)$
determined from a fit to DIS data is affected by large uncertainties.
This suggests that its impact on predictions for hadron collider
processes to which the photon PDF contributes already at leading order
could be substantial, and thus, conversely, that data on such
processes might provide further constraints. In this section we use
the simplest of such processes, namely, electroweak gauge boson
production, to constrain the photon PDF.

At hadron colliders, the dilepton production process receives
contributions at Born level both from quark-initiated neutral current
$Z/\gamma^*$ exchange and from photon-initiated diagrams, see
Figure~\ref{fig:lhczborn}, and thus the contributions from
$\gamma(x,Q^2)$ must be included even in a pure leading-order
treatment of QED effects.  Photon-initiated contributions to dilepton
production at hadron colliders were recently emphasized in
Ref.~\cite{Dittmaier:2009cr}, where $\mathcal{O}(\alpha)$ radiative
corrections to this
process~\cite{Baur:1998kt,Dittmaier:2001ay,Baur:2001ze,Baur:2004ig,Arbuzov:2007db,Arbuzov:2005dd,Brensing:2007qm,Balossini:2009sa,CarloniCalame:2007cd,Dittmaier:2009cr}
were reassessed, and also kinematic cuts to enhance the sensitivity to
$\gamma(x,Q^2)$ were suggested.

\begin{figure}
\begin{centering}
\includegraphics[trim=0cm 0.5cm 0cm 0cm, clip=true, scale=0.80]{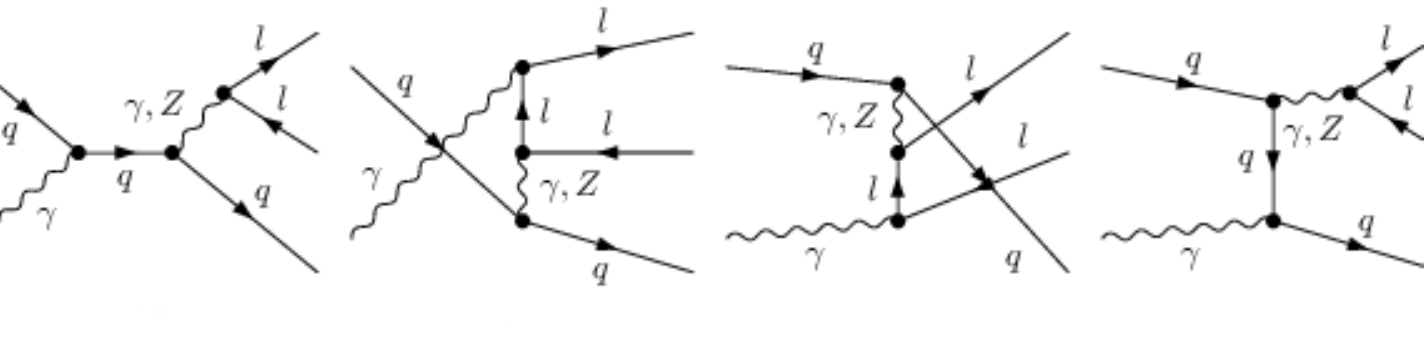}\\
\includegraphics[scale=0.60]{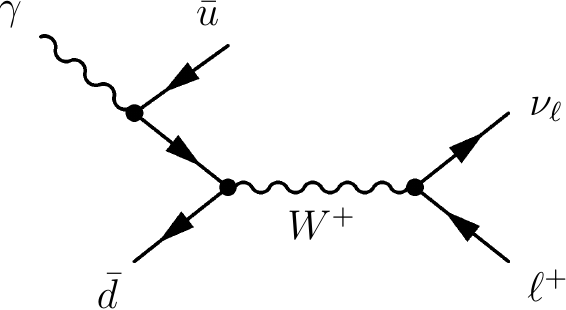}\qquad
\includegraphics[scale=0.60]{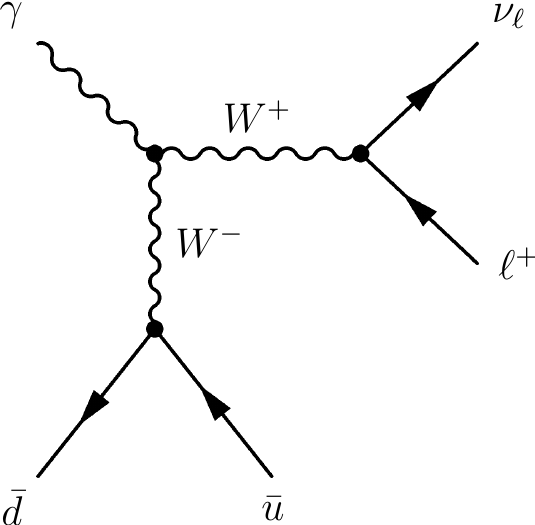}
\par\end{centering}
\caption{\label{fig:lhcwz2} Some Feynman diagrams for
  $\mathcal{O}(\alpha)$ photon-initiated partonic subprocesses which
  contribute to neutral current (top row) and charged current (bottom
  row) dilepton pair production in hadronic collisions.}
\end{figure}

Beyond the Born approximation, radiative corrections to the
neutral-current process, as well as the charged-current process, which
starts at $\mathcal{O}\left( \alpha\right)$ (see
Figure~\ref{fig:lhcwz2} for some representative Feynman diagrams) may
be comparable in size to the Born level contribution, because the
suppression due to the extra power of $\alpha$ might be compensated by
the enhancement arising from the larger size of the quark-photon
parton luminosity in comparison to the photon-photon
luminosity. However, a full inclusion of $\mathcal{O}(\alpha)$
corrections would require solving evolution equations to NLO in the
QED and mixed QED$\otimes$QCD terms, so it is beyond the scope of this
work; we will nevertheless discuss an approximate inclusion of such
corrections which, while not allowing us to claim more than LO
accuracy in QED, should ensure that NLO QED corrections are not
unnaturally large.

We use neutral and charged-current Drell-Yan production data from the
LHC to further constrain the photon PDF, thereby arriving at our final
NNPDF2.3QED PDF sets.
This is obtained by combining the photon PDF from NNPDF2.3 DIS-only
set discussed in the previous section with the standard NNPDF2.3 PDF
set, and then using gauge boson production data to reweight the
result. We discuss first this two-step fitting procedure, and then the
ensuing NNPDF2.3QED PDF set and its features.

\subsection{The prior NNPDF2.3QED and its reweighting}
\label{sec:combination}

As a first step towards the determination of a PDF set with inclusion
of QED corrections, we use the photon PDF determined in the previous
section from a fit to DIS data in conjunction with PDFs which retain
all the information provided by the full NNPDF2.3 data set, which, on
top of DIS, includes Drell-Yan and jet production data from the
Tevatron and the LHC, as we have explained in details in
Sect.~\ref{sec-expdata}.

We have seen in the previous section that all PDFs determined
including QED corrections are statistically equivalent to their
standard counterparts determined when QED corrections are not
included, with the only exception of the gluon, which undergoes a
change by less than half $\sigma$ in a limited kinematic
region. Furthermore, the photon in each PDF replica is essentially
uncorrelated to the shape of other PDFs which are input to
perturbative evolution, the only significant correlation being due to
the mixing induced by the evolution itself.
We can therefore simply combine the photon PDF obtained from the DIS
fit of the previous section with the standard NNPDF2.3 PDFs at the
starting scale $Q_0^2 = 2$ GeV$^2$.
This procedure implies a certain loss of accuracy, which in particular
appears as a violation of the momentum sum rule of the order of the
momentum fraction carried by the photon at the initial scale
Eq.~(\ref{eq:msr}), namely of order 1\%. This is the accuracy to which
the momentum sum rule would be verified if it were not imposed as a
constraint in the fit~\cite{Ball:2011uy}.

The information contained in LHC Drell-Yan production data is included
in the fit through the Bayesian reweighting method presented in
Ref.~\cite{Ball:2011gg,Ball:2010gb} and summarized in
Appendix~\ref{app:rw}.  This method allows for the inclusion of new
data without having to perform a full refit, by using Bayes' theorem
to modify the prior probability distribution of PDF replicas in order
to account for the information contained in the new data. The ensuing
replica set contains an amount of information, and thus allows for the
computation of observables with an accuracy, that corresponds to an
effective number of replicas $N_\textrm{eff}$, which may be determined
from the Shannon entropy of the reweighted set.

\begin{table}
\small
\centering
\begin{tabular}{c|c|c|c|c|c}
\hline
Dataset  &  Observable & Ref.  &  $N_\textrm{dat} $ &$\left[ \eta_\textrm{min}, \eta_\textrm{max}\right]$  &  
$\left[ M^\textrm{min}_\textrm{ll}, M^\textrm{max}_\textrm{ll}\right]$\\
\hline
\hline
LHCb  $\gamma^*/Z$ Low Mass &  $d\sigma(Z)/dM_{ll}$   &  ~\cite{LHCb-CONF-2012-013}  & 9  & [2,4.5]  & 
[5,120] GeV
  \\
ATLAS $W,Z$  &  $d\sigma(W^{\pm},Z)/d\eta$  &   ~\cite{Aad:2011dm} & 30  & [-2.5,2.5]  & 
[60,120] GeV
  \\
 ATLAS $\gamma^*/Z$ High Mass  &  $d\sigma(Z)/dM_{ll}$  &   ~\cite{Aad:2013iua} & 13  & [-2.5,2.5]  & 
[116,1500] GeV
  \\
 \hline
\end{tabular}
\caption{Kinematical coverage of the three LHC datasets used to determinethe photon PDF.  \label{tab:expdata}
}
\end{table}

This new data only constrains significantly the photon PDF, hence we
need to guarantee that good accuracy is obtained by starting with a
large number of photon replicas. The initial prior set is thus
obtained combining 500 photon PDF replicas with a standard set of 100
NNPDF2.3 replicas. In practice, this is done by simply producing five
copies of the NNPDF2.3 100 replica set, and combining each of them at
random with one of the 500 photon PDF replicas obtained from the QED
fit to DIS data discussed in the previous section. The procedure is
performed at NLO and NNLO, in each case combining the photon PDF from
the combined QED$\otimes$QCD fit to DIS data with the other PDFs from
the corresponding standard NNPDF2.3 set.  Furthermore, the procedure
is repeated for three different values of
$\alpha_s=0.117,\>0.118,\>0.119$. We find no dependence of the photon
PDF on the value of $\alpha_s$, though there are minor differences
between the photon determined using NLO or NNLO QCD theory in the DIS
fit.
\begin{figure}[ht]
\begin{centering}
\includegraphics[scale=0.34]{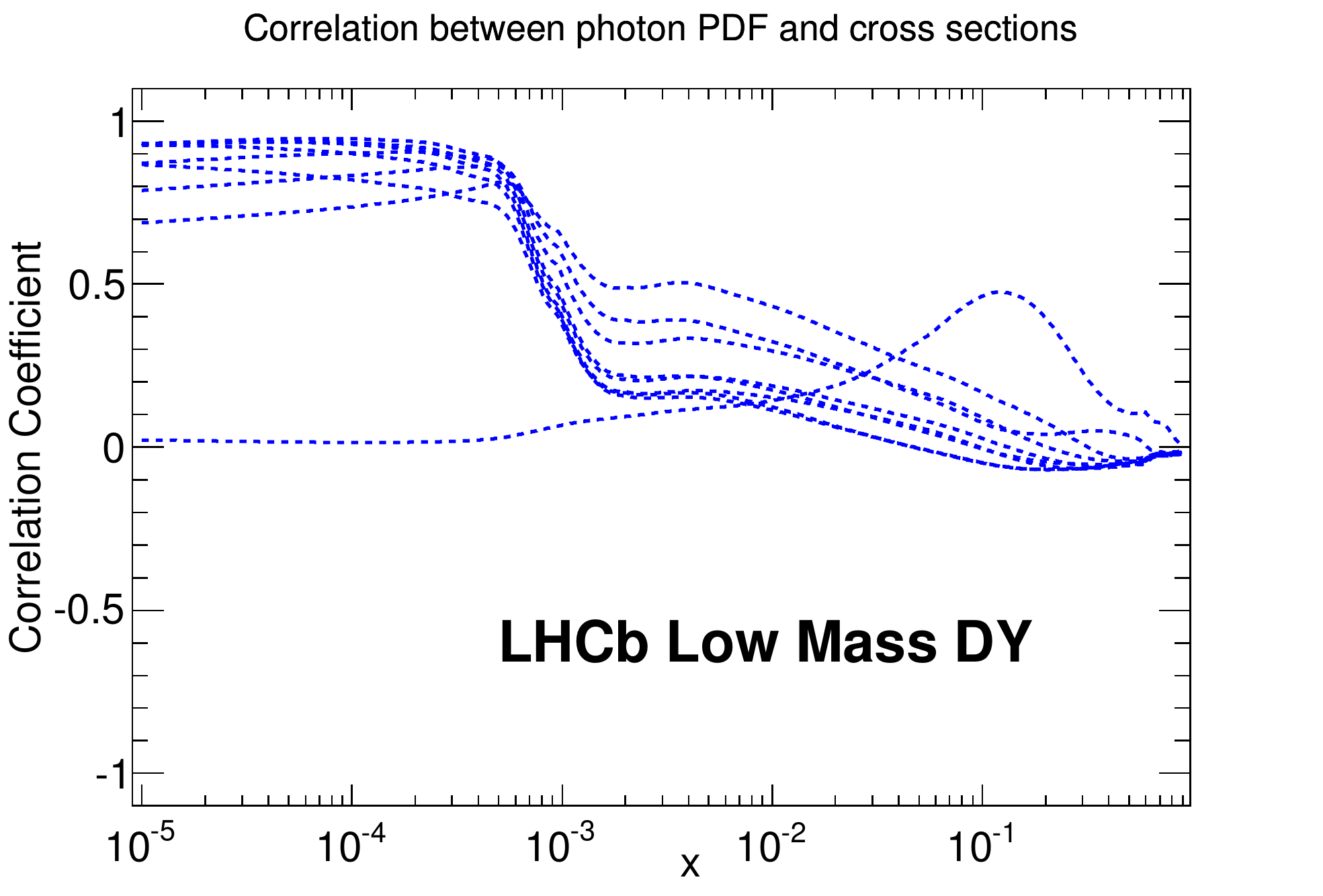}\includegraphics[scale=0.34]{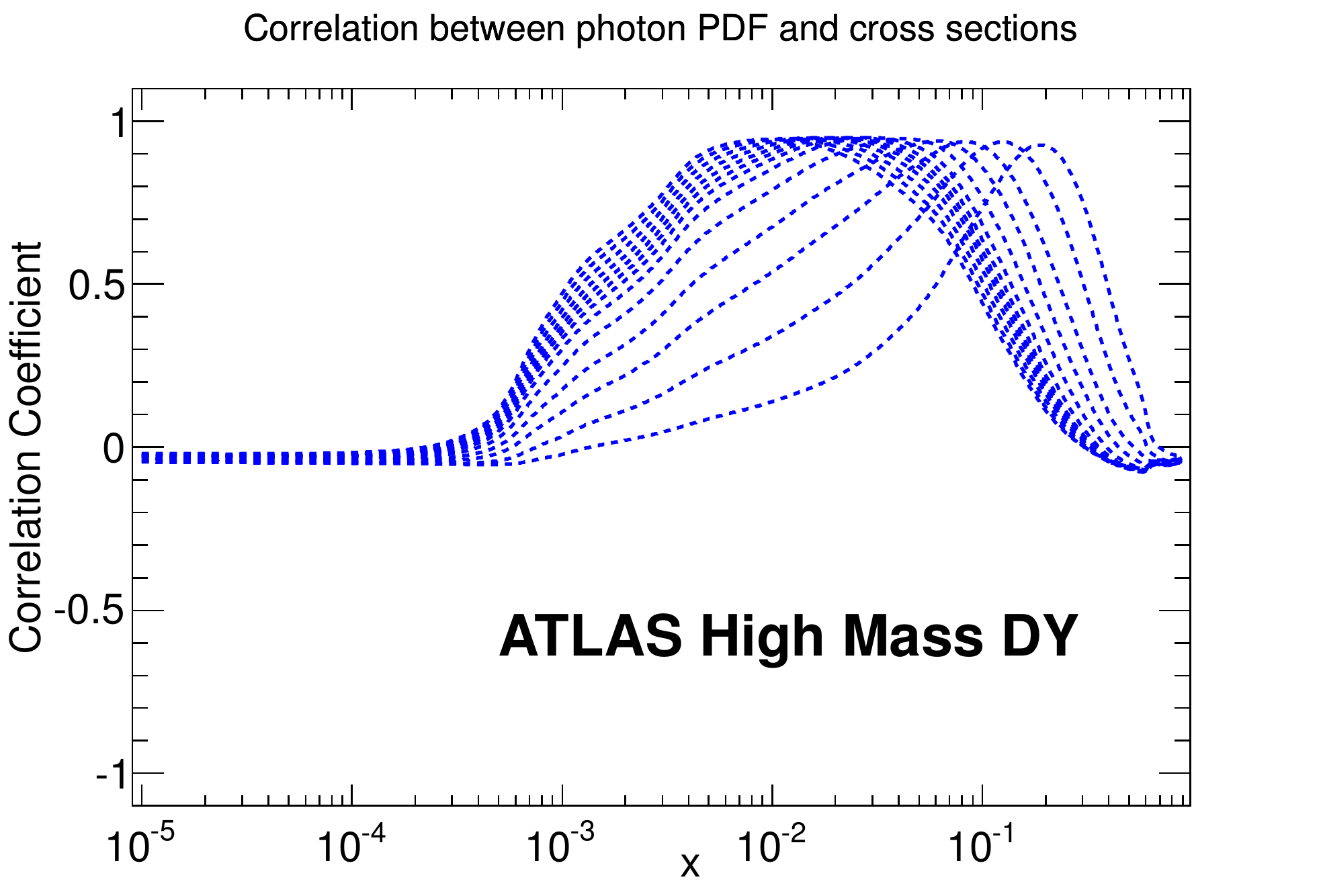} \\
\includegraphics[scale=0.34]{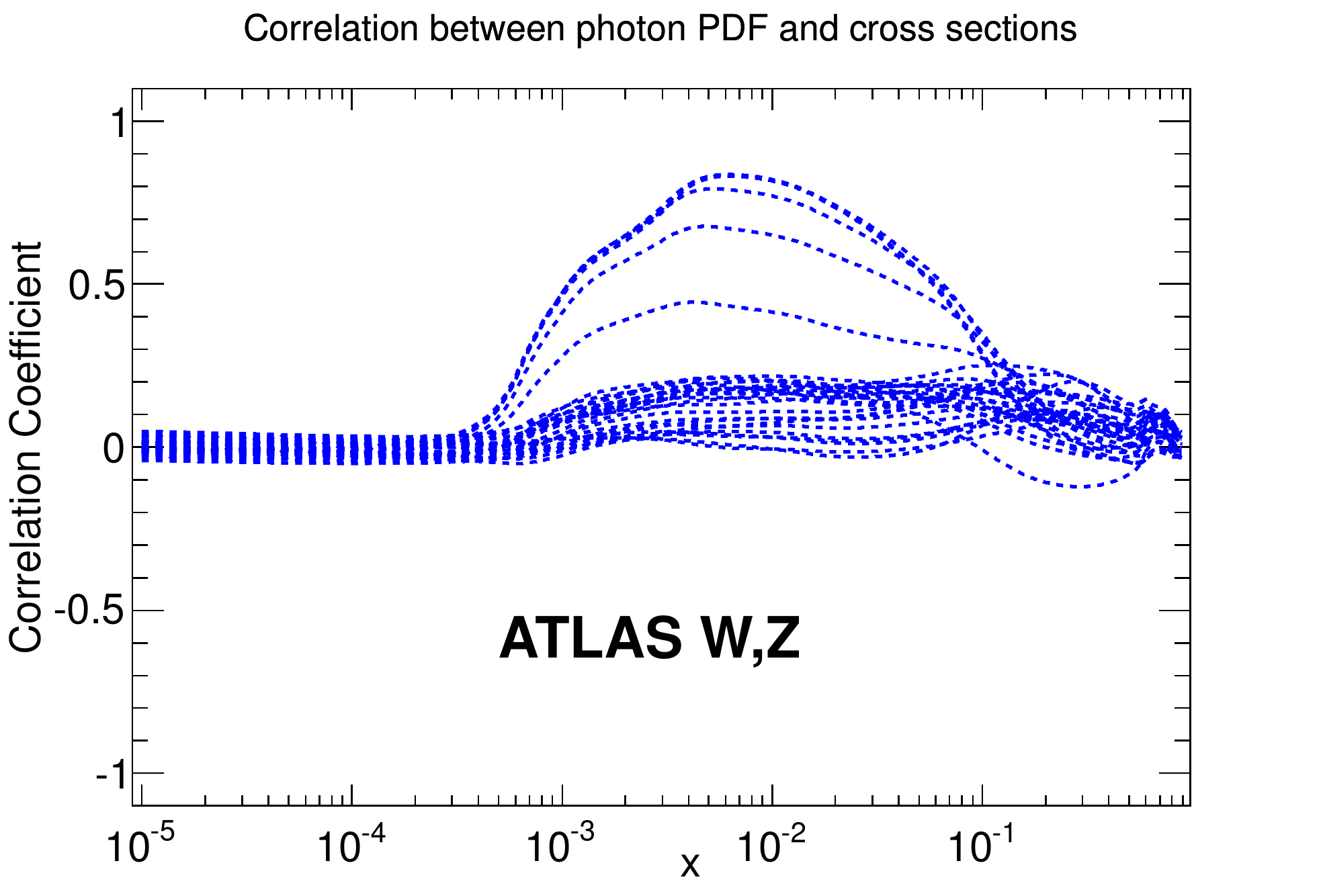}
\par\end{centering}
\caption{\label{fig:correlations} Correlation between the photon PDF
  and the LHC data of Tab.~\ref{tab:expdata}, shown as function of $x$
  for $Q^2=10^4$~GeV$^2$. Each curve corresponds to an individual data
  bin.}
\end{figure}

In each case, the set of $N_\textrm{rep}=500$ replicas is then evolved
to all scales using combined QED$\otimes$QCD evolution. Note that this
in particular implies that no further violation of the momentum sum
rule is introduced on top of that which was present at the initial
scale, up to approximations introduced when solving the evolution
equations.

In this work, the reweighting is performed using the following LHC
datasets:
\begin{itemize}
\item LHCb low-mass $Z/\gamma^*$ Drell-Yan production from the 2010 run~\cite{LHCb-CONF-2012-013}
\item ATLAS inclusive  $W$ and $Z$ production data from the 2010 run~\cite{Aad:2011dm}
\item ATLAS high-mass $Z/\gamma^*$ Drell-Yan production from the 2011 run~\cite{Aad:2013iua},
\end{itemize}
whose kinematic coverage is summarized in Table~\ref{tab:expdata}.
Using data with three different mass ranges for the dilepton pairs,
below, at, and above the $W$ and $Z$ mass, guarantees that both the
low $x$ (from low mass) and high $x$ (from high mass) regions are
covered.

\begin{figure}[t]
  \begin{centering}
    \includegraphics[scale=0.34]{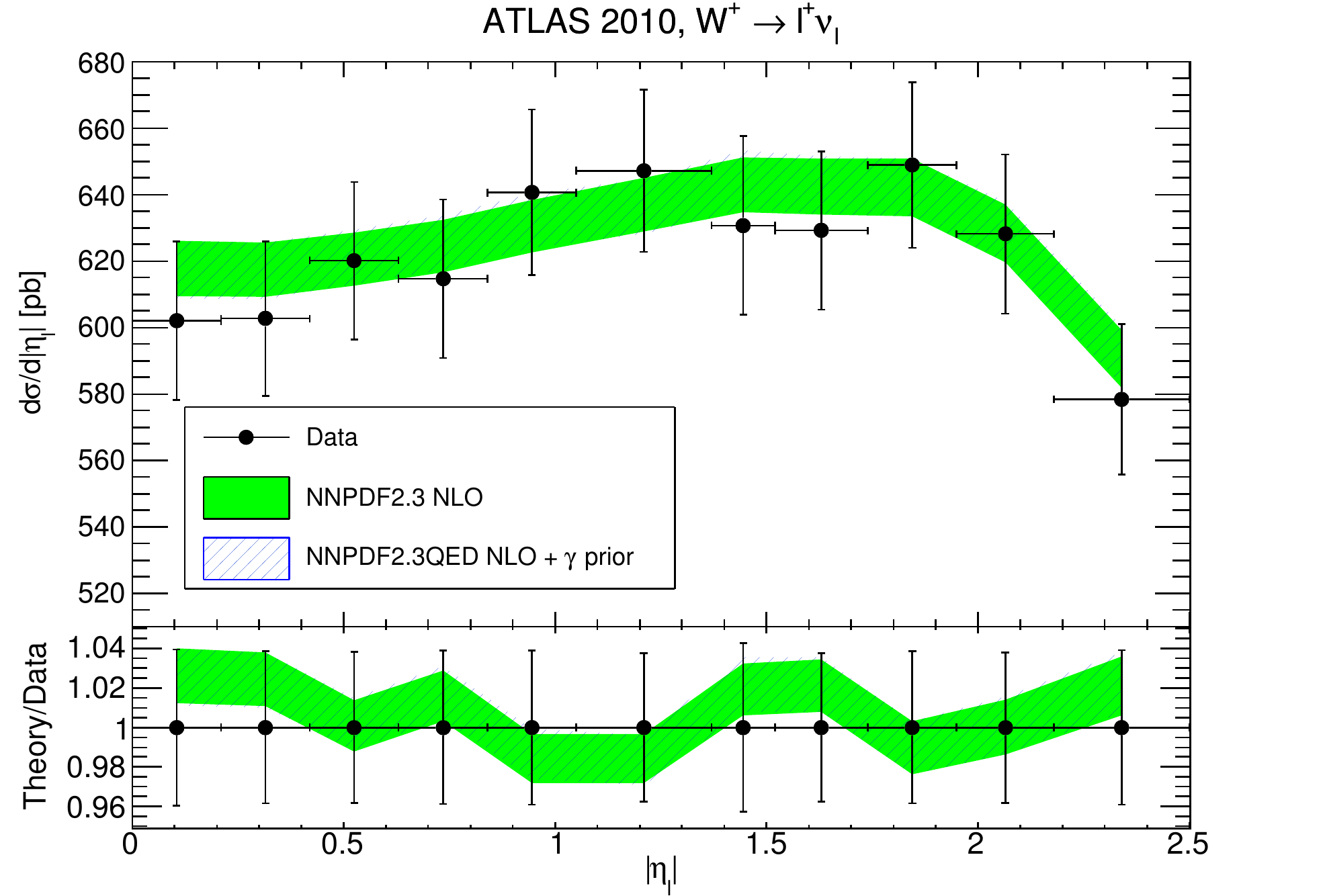}\includegraphics[scale=0.34]{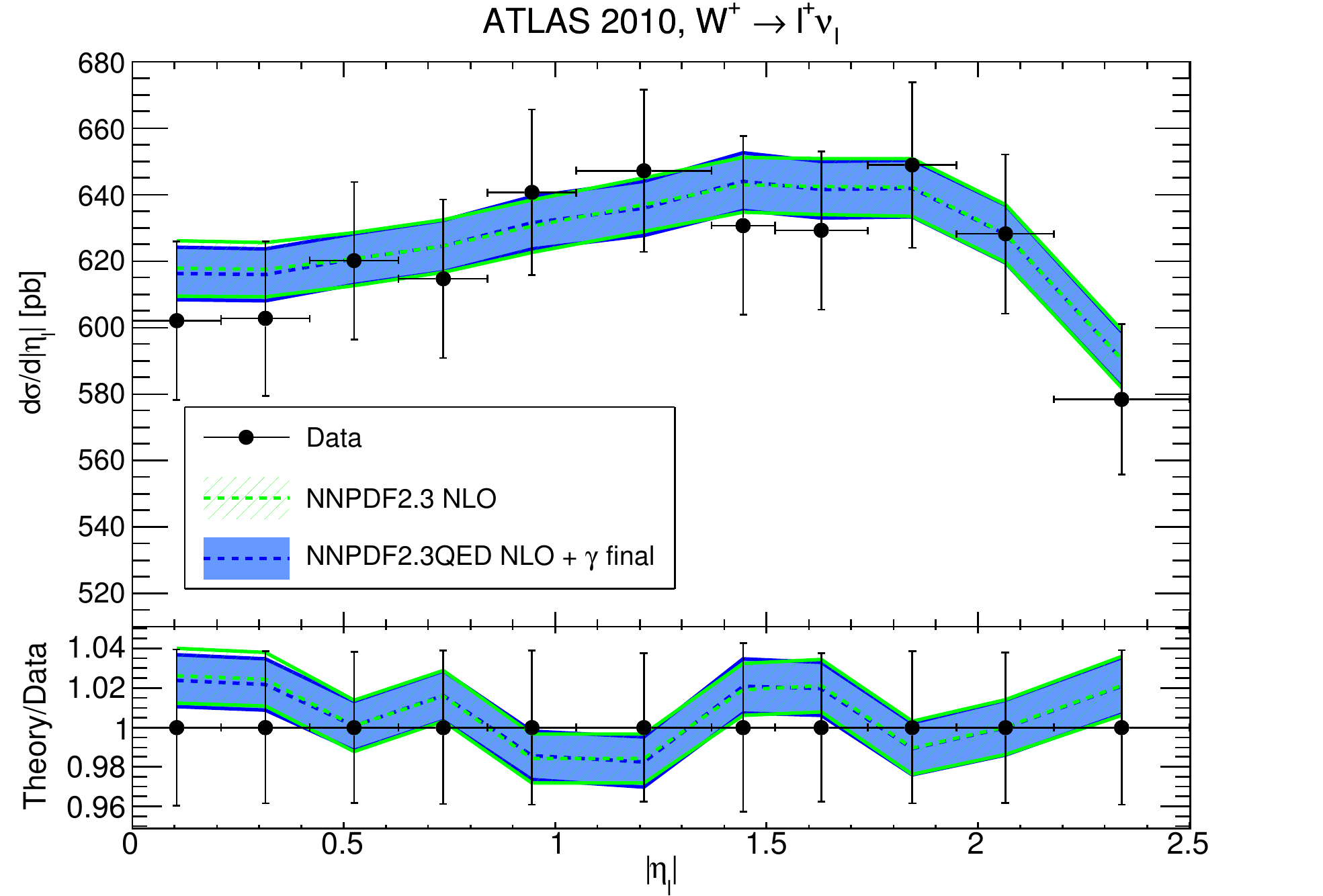}\\
    \includegraphics[scale=0.34]{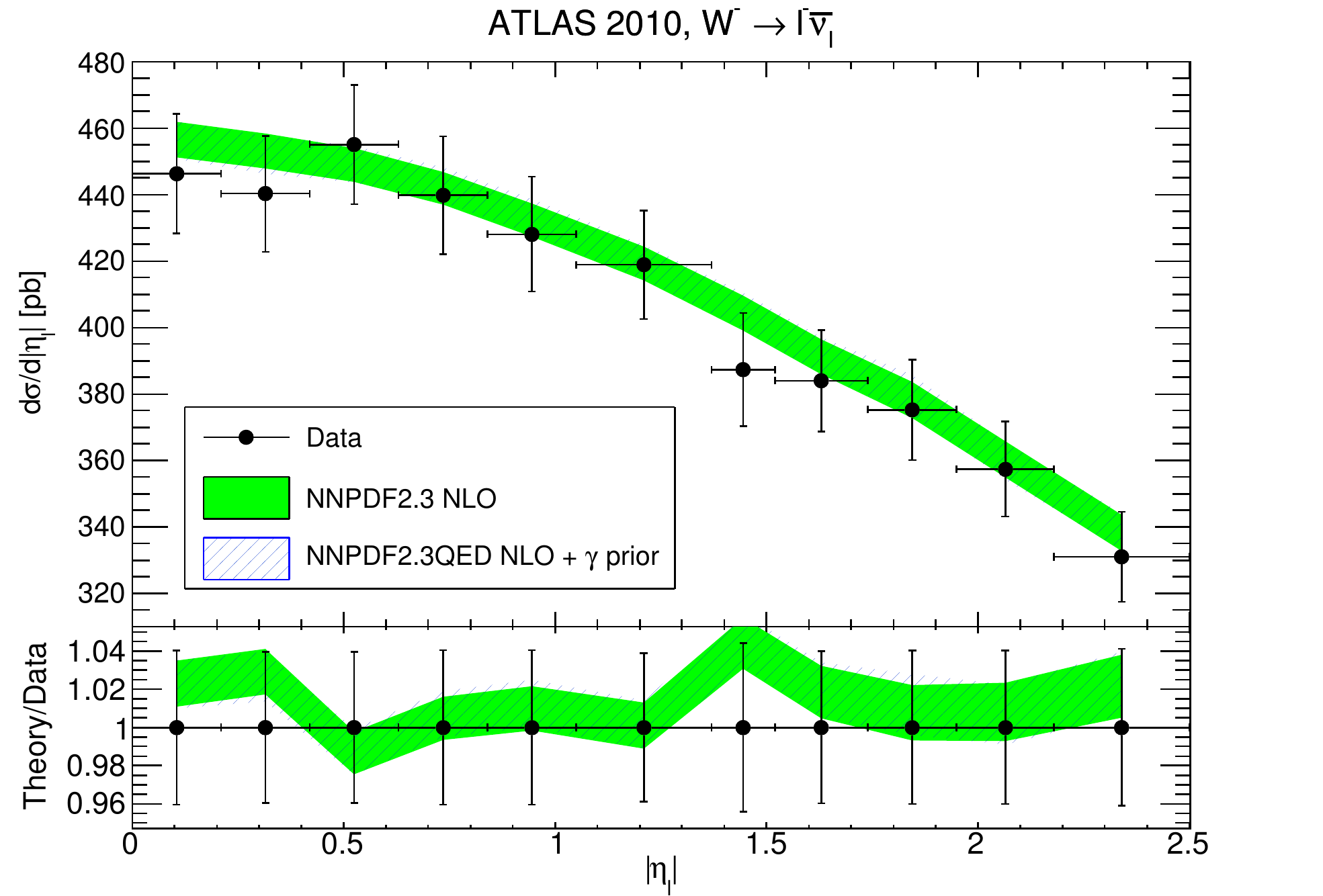}\includegraphics[scale=0.34]{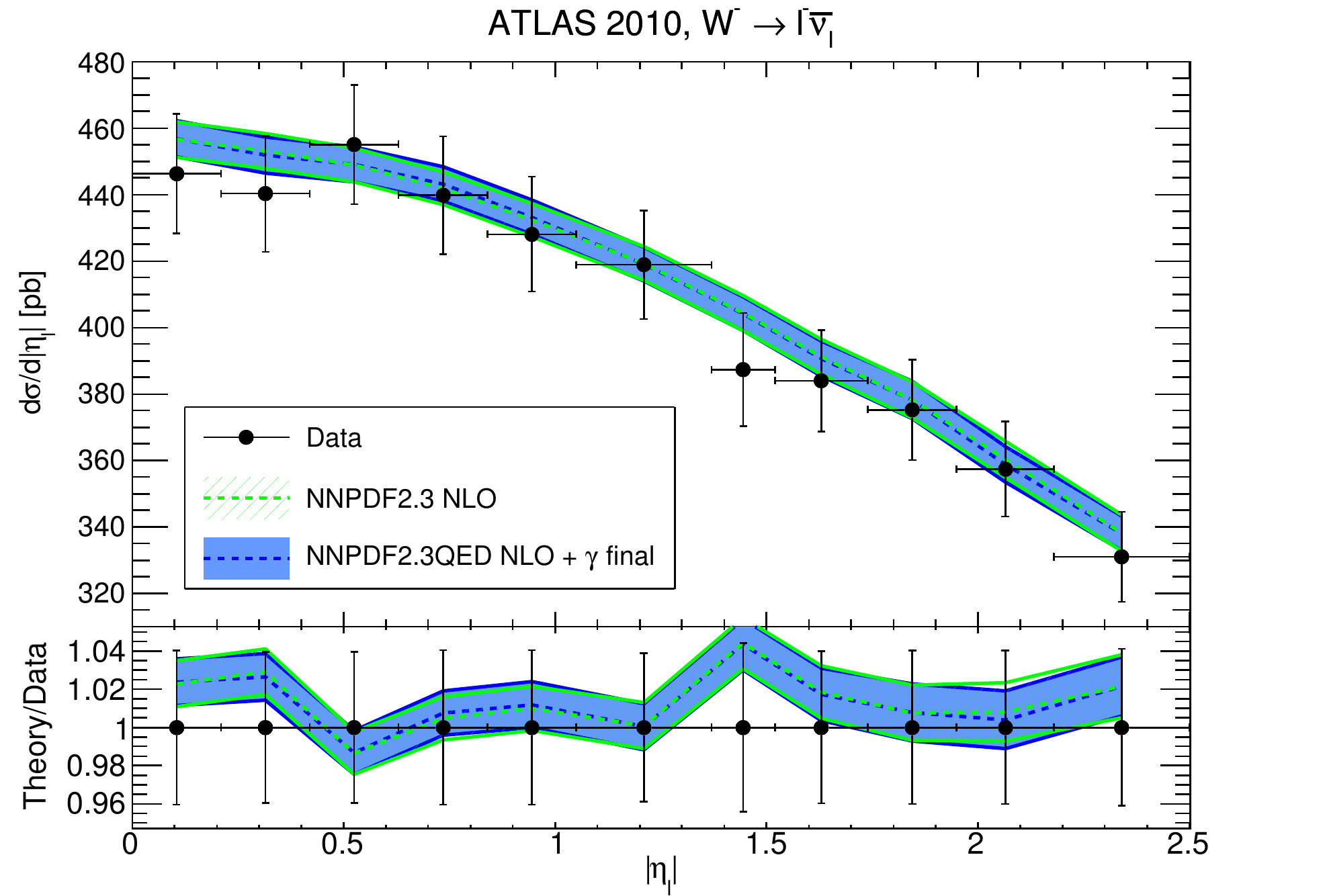}
    \par\end{centering}
  \caption{\label{fig:rwlhc} Comparison of the ATLAS $W$ production
    data with NLO theoretical predictions obtained using PDFs before
    (left) and after (right) reweighting with the data of
    Tab.~\ref{tab:expdata}. In all plots we also show for comparison
    results obtained using the default NNPDF2.3 PDF set, with all QED
    corrections switched off. From top to bottom: $W^+$ and
    $W^-$. Error bands on the theoretical prediction correspond to one
    $\sigma$ uncertainties. Experimental error bars give the total
    combined statistical and systematic uncertainty.  }
\end{figure}

For all the ATLAS data the experimental covariance matrix is
available, hence the $\chi^2$ may be computed fully accounting for
correlated systematics. However, this is not the case for LHCb at that
time: hence, the low-mass data are treated adding statistical and
systematic errors in quadrature, and only including normalization
errors in the covariance matrix. We have checked that if reweighting
is performed using the diagonal covariance matrix, statistically
indistinguishable results are obtained. This means that within the
large uncertainty of the photon PDF, and due to the small impact of
QED corrections on the quark and gluon PDFs, the lack of information
on correlations for the LHCb experiment is immaterial. However, this
implies that $\chi^2$ values quoted for LHCb should only be taken as
indicative.
Unfortunately, at that time the CMS off-peak Drell-Yan
data~\cite{CMSdy} was not publicly available, and thus could not be
used in the present analysis.

 \begin{figure}[ht]
\begin{centering}
\includegraphics[scale=0.34]{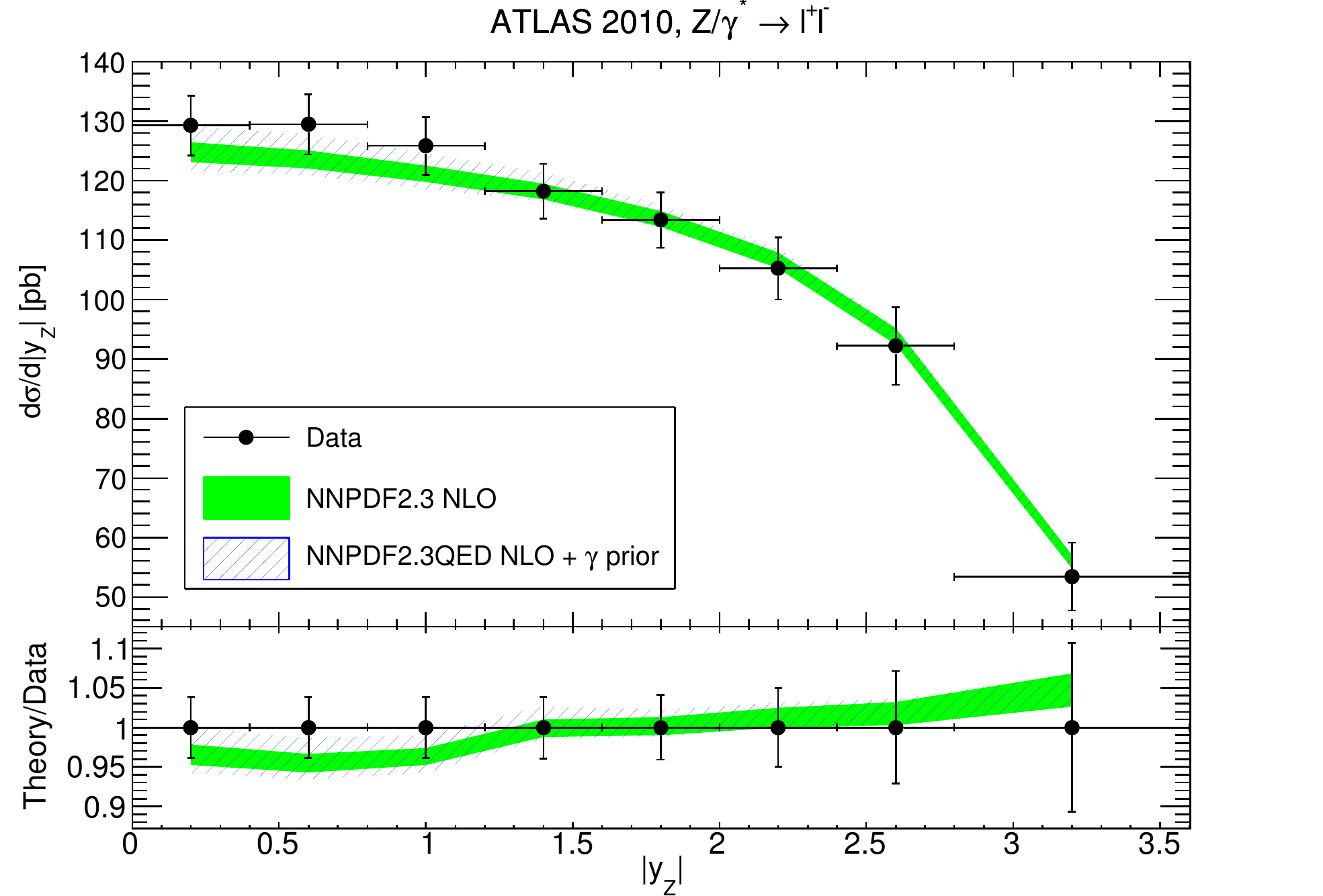}\includegraphics[scale=0.34]{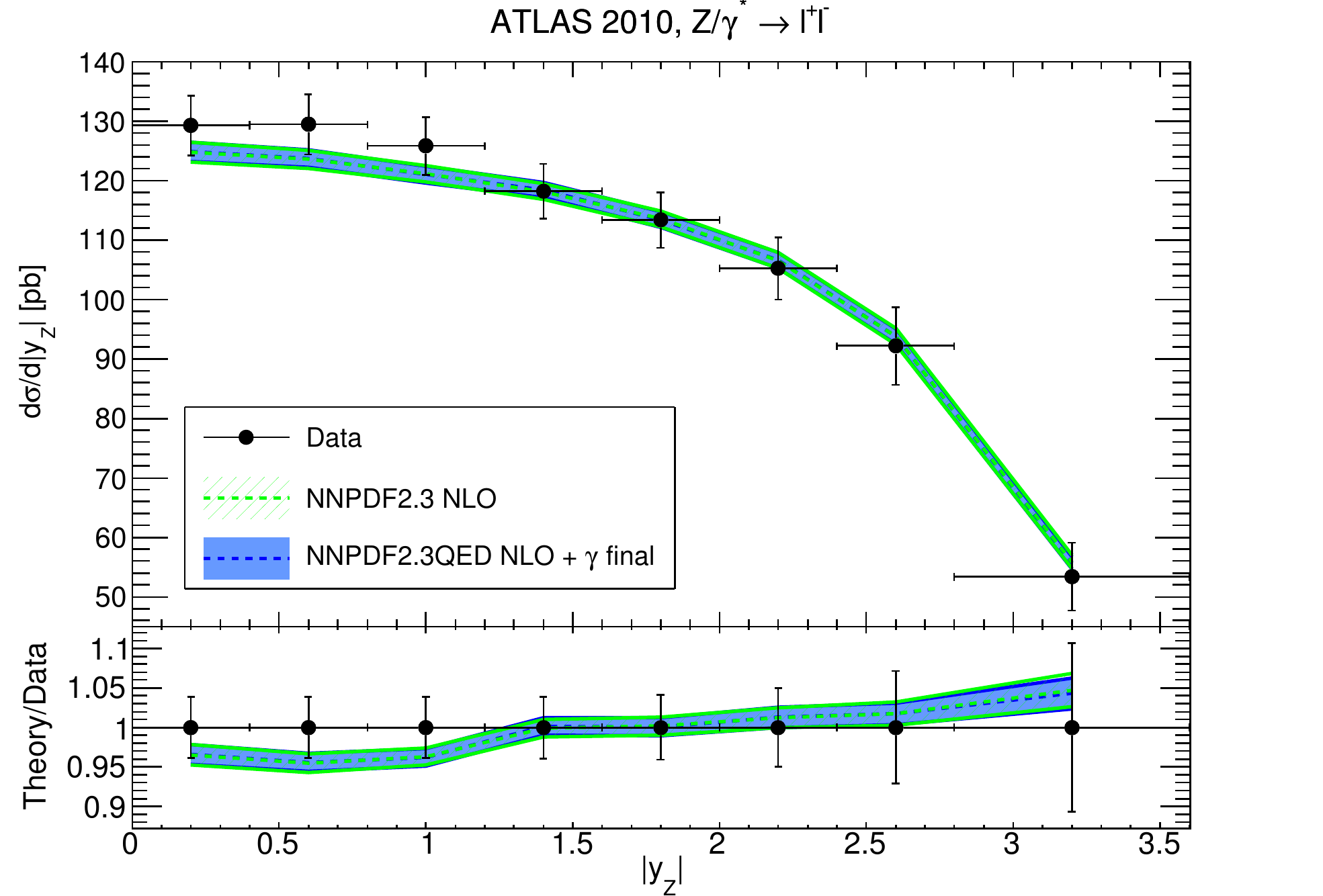}
\par\end{centering}
\caption{\label{fig:rwlhca} Same as Fig.~\ref{fig:rwlhc}, but for the
  neutral current data.  }
\end{figure}

The range of $x$ for the photon PDF which is affected by each of the
datasets of Table~\ref{tab:expdata} can be determined quantitatively
by computing the correlation coefficient (see~\cite{Demartin:2010er}
and Sect. 4.2 of Ref.~\cite{Alekhin:2011sk}) between a given
observable and the PDFs. The correlation coefficients computed using
the NNPDF2.3QED NLO prior set are shown in
Figure~\ref{fig:correlations} for each bin in the experiments in
Table~\ref{tab:expdata}. It is clear that the LHC data guarantee a
good kinematic coverage for all $10^{-5}\lesssim x\lesssim 0.1$.
The correlation is weaker for real $W$ and $Z$ production data, where
the $s$-channel quark contribution dominates as the propagator goes on
shell.
The high-mass (low-mass) Drell-Yan data is thus essential to pin down
$\gamma(x,Q^2)$ at large (small) Bjorken-$x$, where uncertainties are
the largest. A preliminary determination of the photon
distribution~\cite{Carrazza:2013bra}, which did not use the LHCb data,
had significantly larger uncertainties at small $x$, consistently with
the expectations based on the correlation plot of
Figure~\ref{fig:correlations}.

 \begin{figure}[ht]
\begin{centering}
\includegraphics[scale=0.34]{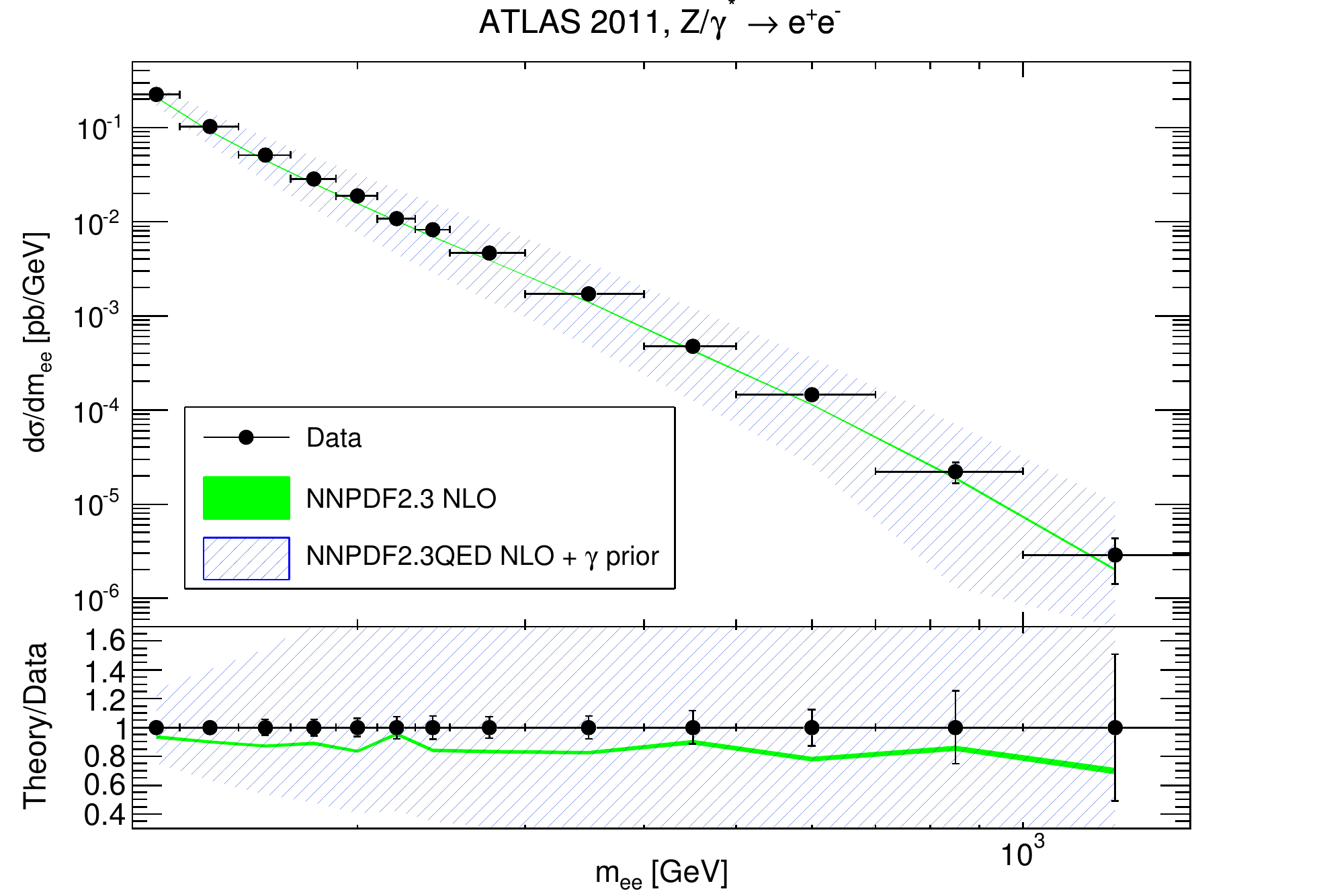}\includegraphics[scale=0.34]{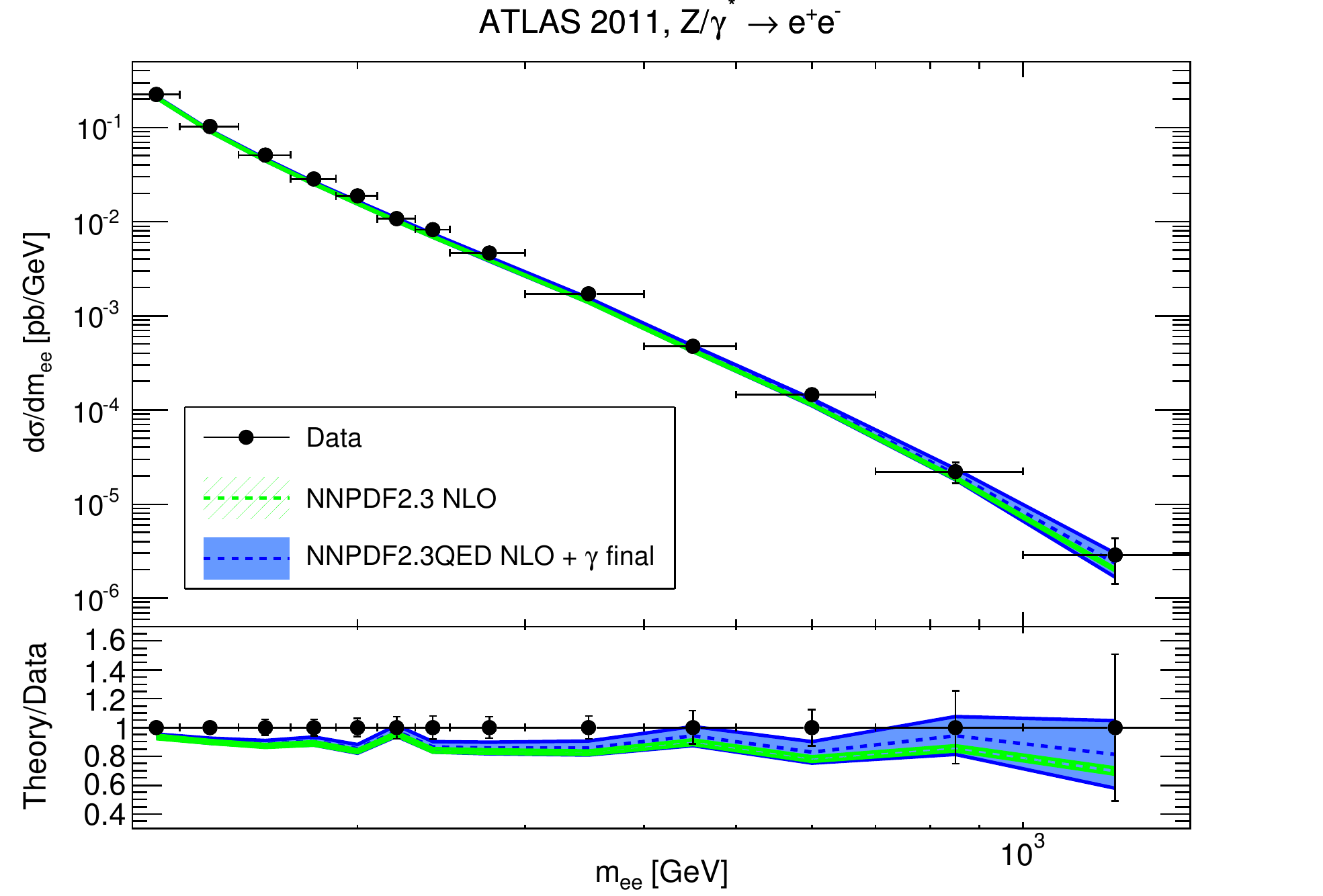}
\par\end{centering}
\caption{\label{fig:rwlhc1} Same as Fig.~\ref{fig:rwlhc}, but for the
  ATLAS high-mass neutral-current data.}
\end{figure}

Theoretical predictions for the datasets in Table~\ref{tab:expdata}
have been computed at NLO and NNLO in QCD using
\texttt{DYNNLO}~\cite{Catani:2010en}, supplemented with Born-level and
$\mathcal{O}\left( \alpha \right)$ QED corrections using
\texttt{HORACE}~\cite{Balossini:2009sa,CarloniCalame:2007cd}. Results
from \texttt{DYNNLO} and \texttt{HORACE} have been combined
additively, avoiding double counting, in order to obtain a consistent
combined QCD$\otimes$QED theory prediction. The additive combination
of QED and QCD corrections avoids introducing
$\mathcal{O}(\alpha\alpha_s)$ terms, which are beyond the accuracy of
our calculation.
In the \texttt{DYNNLO} calculation, the renormalization and
factorization scale have been set to the invariant mass of the
dilepton pair in each bin. The \texttt{HORACE} default settings, with
the renormalization and factorization set to the mass of the gauge
boson, have been used for the ATLAS high-mass data, but we have also
checked that for this data the choice is immaterial, in that the LO
results obtained using \texttt{DYNNLO} and \texttt{HORACE} with the
respective scale settings agree with each other.

For the LHCb low-mass data we have used a modified version of
\texttt{HORACE} in which the scale choice is the same as in
\texttt{DYNNLO}, since for these low scale data the choice of
renormalization and factorization scale does make a significant
difference. Note that the smallest mass values reached by these data
correspond to momentum fractions $x\sim 10^{-3}$ in the central
rapidity regions, for which, at the scale of the data, fixed order
(unresummed) results are expected to be adequate (see
Ref.~\cite{Ball:2007ra}, in particular Figure~1). Indeed we shall see
that our results are perturbatively stable in that the photon PDF at
NLO and NNLO is very similar for all $x$ (see
Figs.\ref{fig:photonrw}-\ref{fig:photonrwnnlo} below).

The same selection and kinematical cuts as in the corresponding
experimental analysis has been adopted: in particular, the same
requirements concerning lepton-photon final state recombination and
the treatment of final state QED radiation have been implemented in
the \texttt{HORACE} computations.

It should be noticed that, whereas the LHCb and ATLAS high-mass data
are only being included now in the fit, the $W$ and $Z$ production
data were already included in the original NNPDF2.3 PDF determination
(where they turned out to have a moderate impact). Therefore, in
principle a modified version of NNPDF2.3 in which these data are
removed from the fit should have been used as a prior. In practice,
however, this would make very little difference. We have verified that
the inclusion of QED evolution affects minimally the prediction for
this data, where differences are at the same level of the Monte Carlo
integration uncertainty, recalling (see Figure~\ref{fig:correlations})
that the main impact of this data is in the $x\sim0.01$ region. This
means that the contributions to this process in the reweighting and in
the original NNPDF2.3 fit in practice only differ because of the
inclusion of the photon contribution. Furthermore, we have explicitly
verified that if the ATLAS $W$ and $Z$ production data are excluded
from the fit, the photon is systematically modified by a small but
non-negligible amount (less then half $\sigma$ at most) in the region
$x\sim 10^{-3}$ where these data are expected to carry information
(see Figure~\ref{fig:correlations}), while all other PDFs are
essentially unaffected.

Whereas our computation is only accurate to leading order in QED, we
did include $\mathcal{O}(\alpha)$ corrections to the electroweak gauge
boson production process through \texttt{HORACE}, with the aim of
avoiding unnaturally large NLO QED corrections. This raises several
issues which we now discuss in turn.
 \begin{figure}[t]
\begin{centering}
  \includegraphics[scale=0.34]{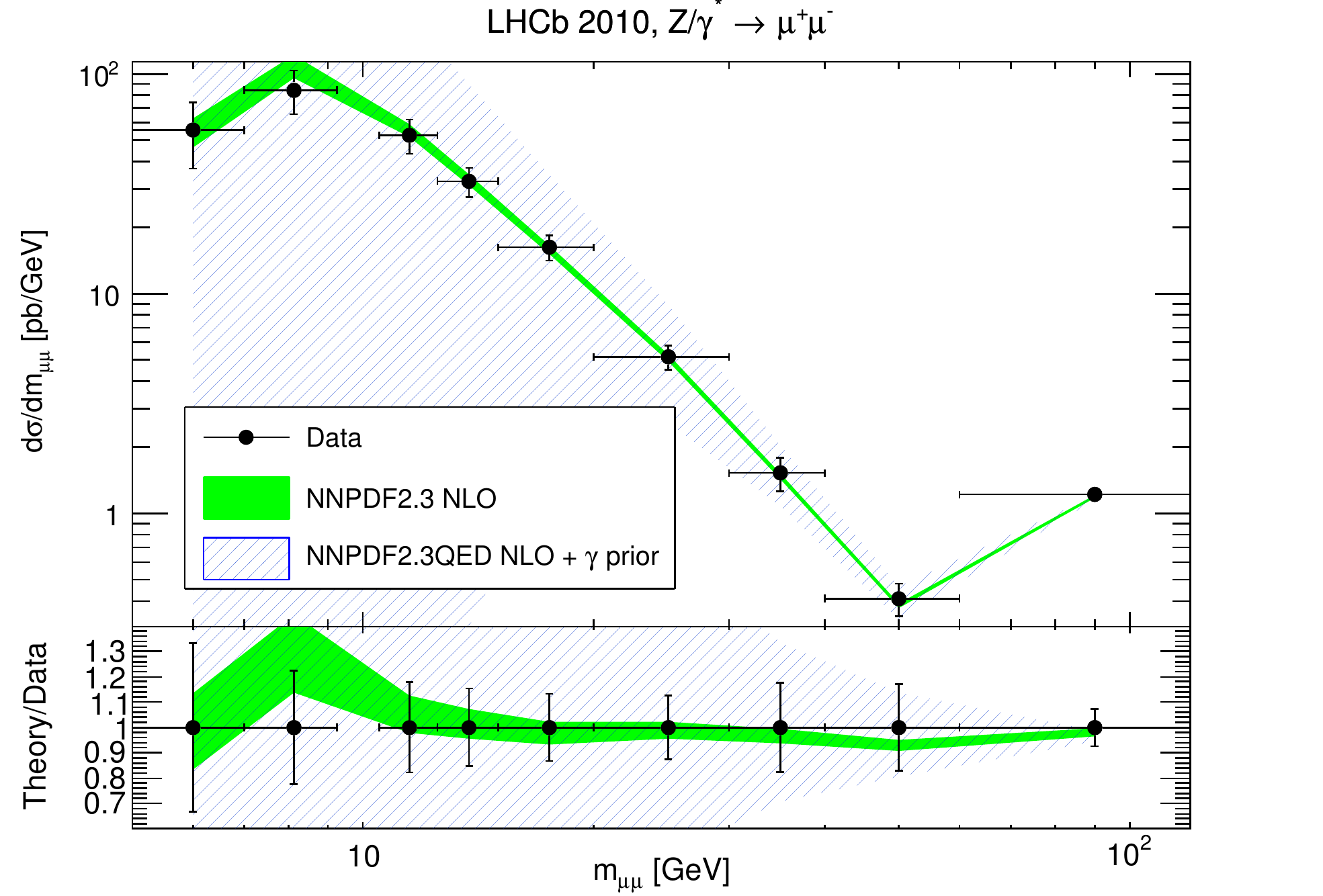}\includegraphics[scale=0.34]{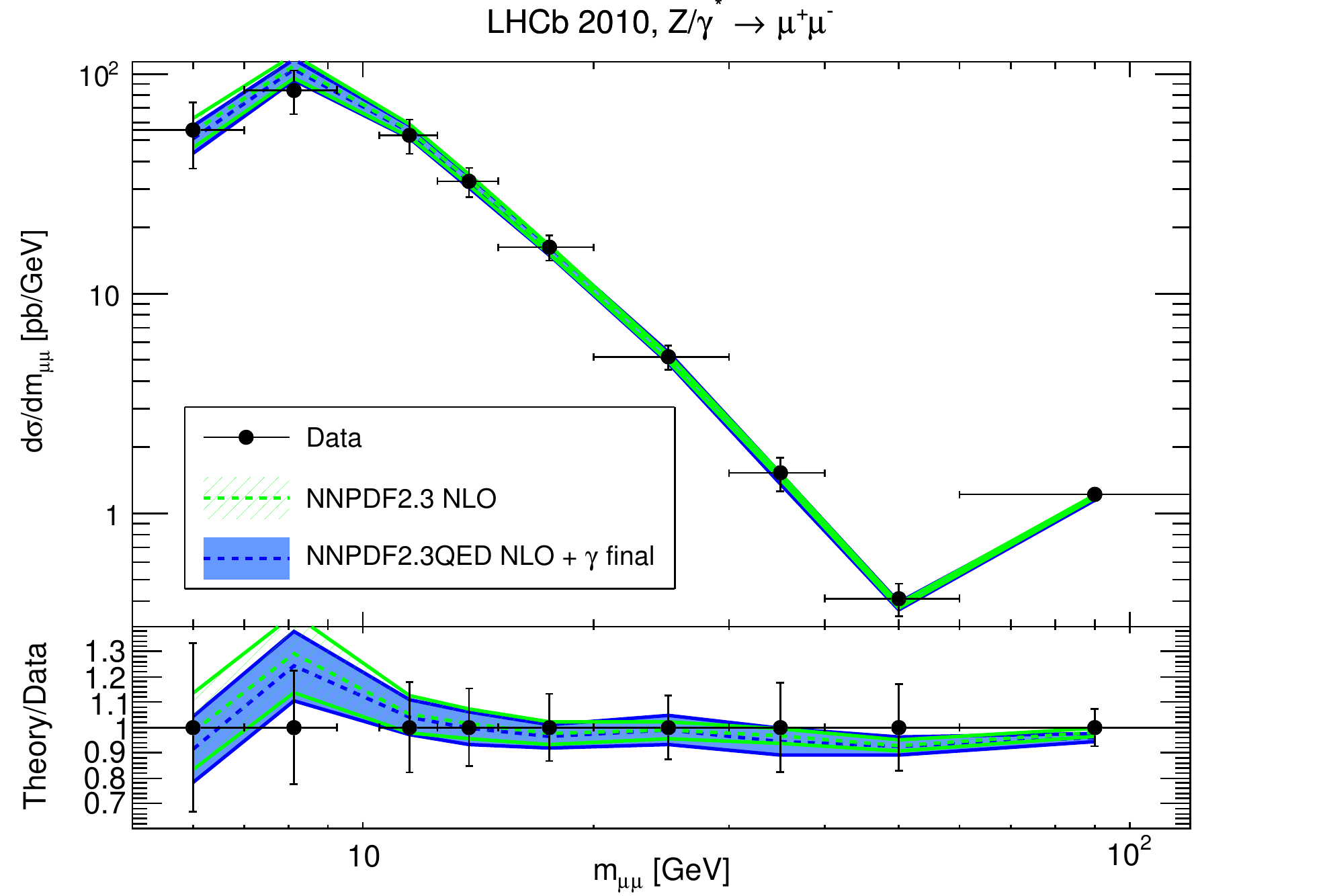}
\par\end{centering}
\caption{\label{fig:rwlhc2} Same as Fig.~\ref{fig:rwlhc}, but for the
  LHCb low-mass neutral current data.}
\end{figure}

As pointed out in Refs.~\cite{Diener:2005me,Dittmaier:2009cr}, usage
of the leading-order expressions in QED for the DIS coefficient
functions can be viewed as the choice of the DIS factorization scheme,
in which deep-inelastic coefficient functions are taken to coincide to
all orders with their leading-order expression, with higher order
corrections factorized into the PDFs. Therefore, use of the DIS scheme
for the QED corrections to the Drell-Yan process ensures that
predictions for Drell-Yan obtained with PDFs determined using DIS data
and LO QED are actually accurate up to NLO, modulo any NLO corrections
from QED evolution. Therefore, we have used the DIS-scheme expressions
for NLO corrections to Drell-Yan as implemented in \texttt{HORACE}.
Of course, in practice, there will be NLO QED evolution effects, even
though there is a certain overlap between the kinematic region of the
HERA DIS data and that of the LHC Drell-Yan data, so we cannot claim
NLO QED accuracy. However we expect this procedure to lead to greater
stability of our results upon the inclusion of NLO QED corrections.

Radiative corrections related to final-state QED radiation have
already been subtracted from the ATLAS data, but not from the LHCb
data. Therefore, for ATLAS we have only included photon-induced
processes in the \texttt{HORACE} runs, while for LHCb we have also
included explicit $\mathcal{O}(\alpha)$ contributions from final-state
QED radiation. Electroweak corrections, which are not subtracted from
any of the data and which are not included in our calculation, could
be potentially relevant in the high-mass
region~\cite{Dittmaier:2009cr}. However, in practice they are always
much smaller than the statistical uncertainty on the ATLAS data.

Finally, to NLO in QED the scheme used in defining electroweak
couplings should be specified. The \texttt{DYNNLO} code uses the
so-called $G_\mu$ scheme for the electroweak couplings, while
\texttt{HORACE} also uses the $G_\mu$ scheme for charged-current
production, but the improved Born approximation (IBA) for
neutral-current production. We have verified the differences in
predictions between the two scheme are negligible in comparison to the
statistical uncertainties of the Monte Carlo integrations, more
details about the IBA scheme will be presented in
Chap.~\ref{sec:chap5}.
 \begin{figure}
\begin{centering}
\includegraphics[scale=0.34]{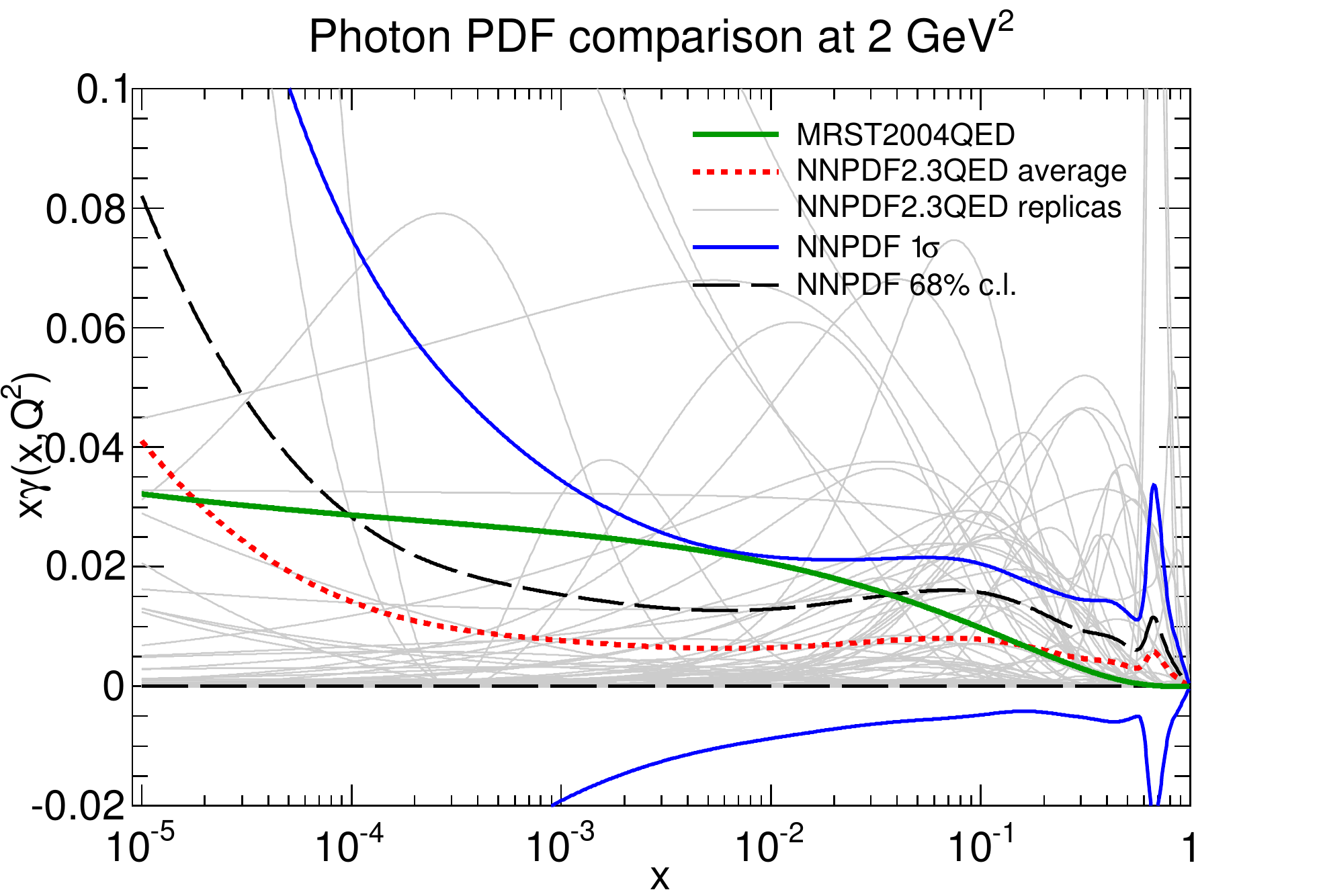}\includegraphics[scale=0.34]{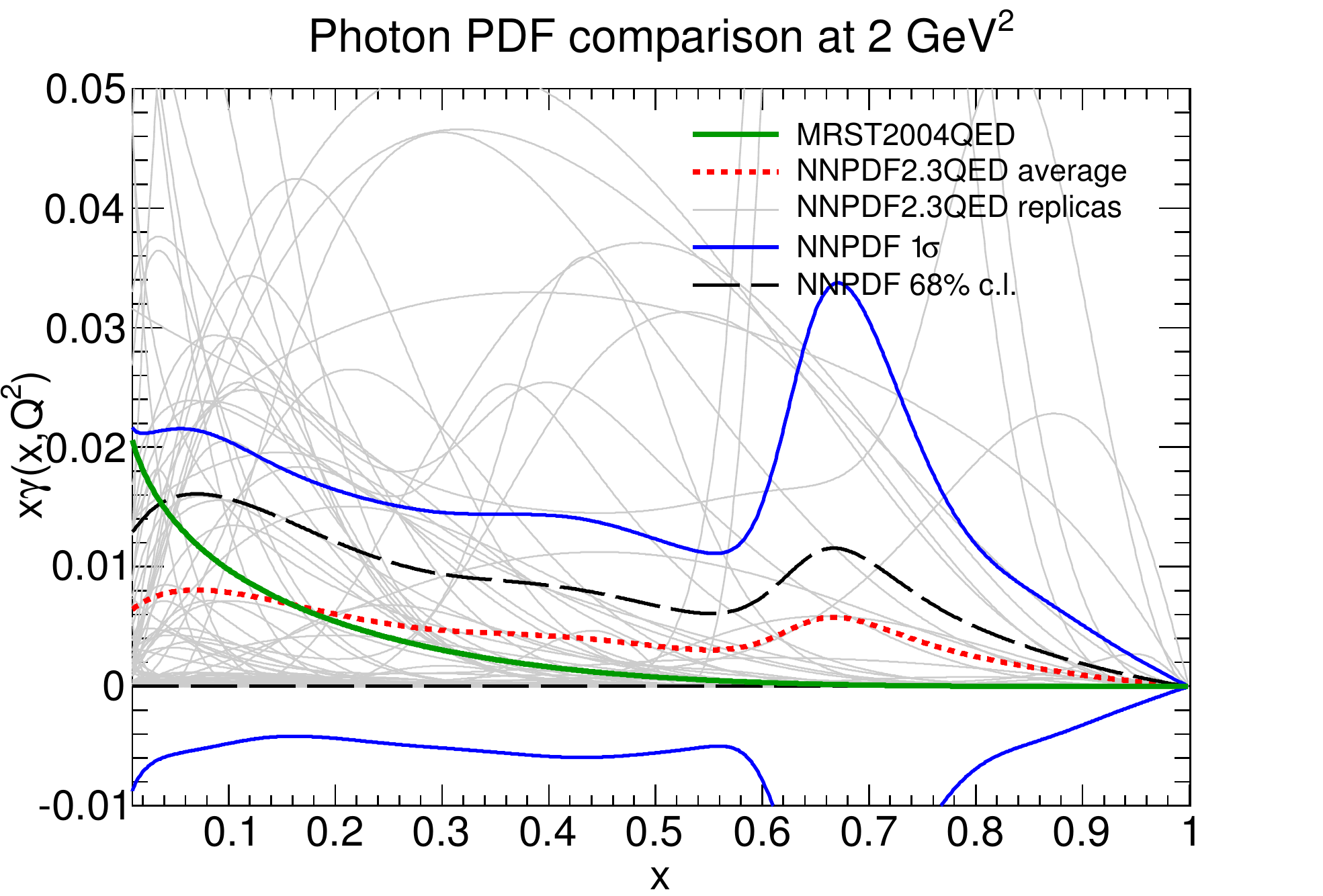}\\
\includegraphics[scale=0.34]{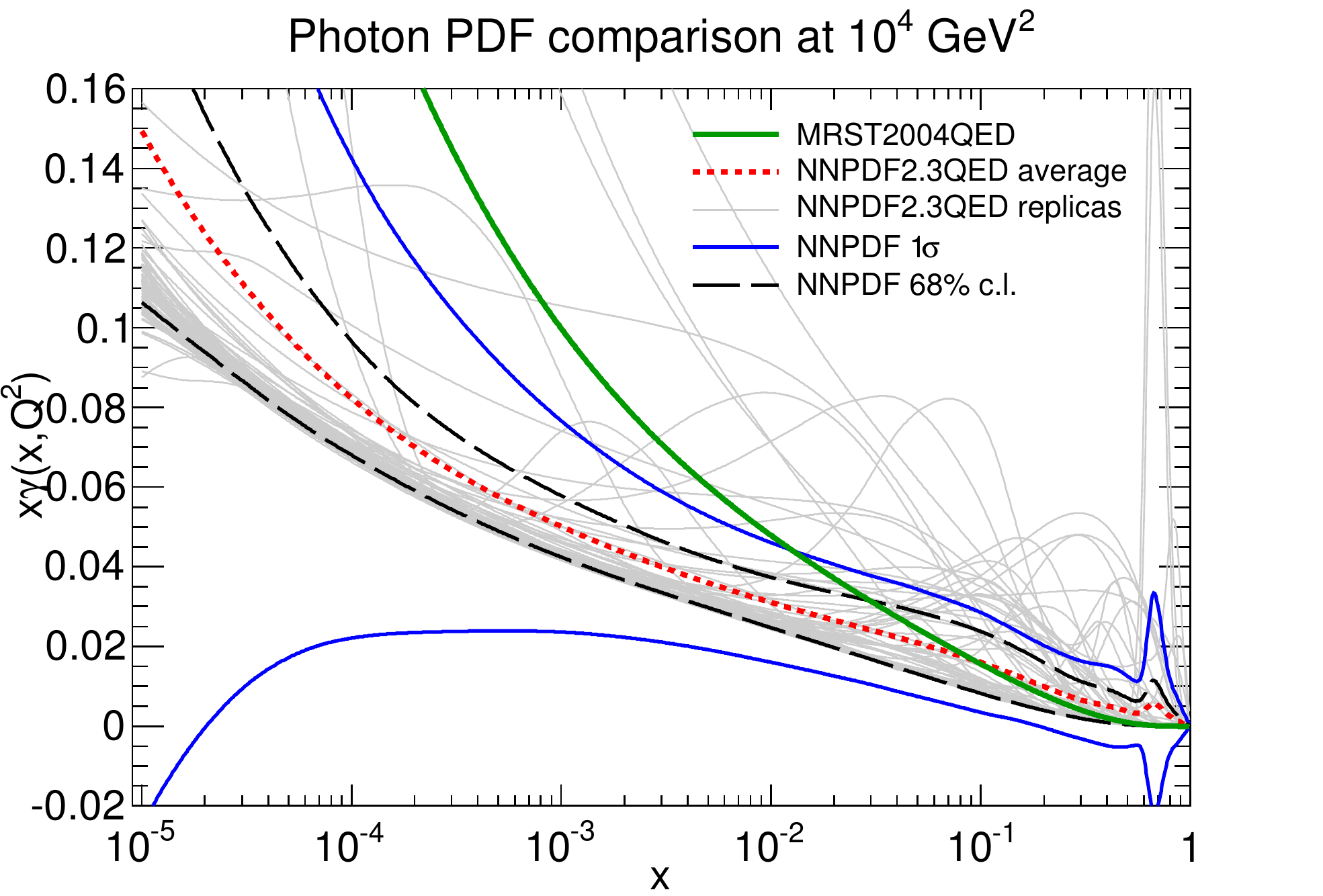}\includegraphics[scale=0.34]{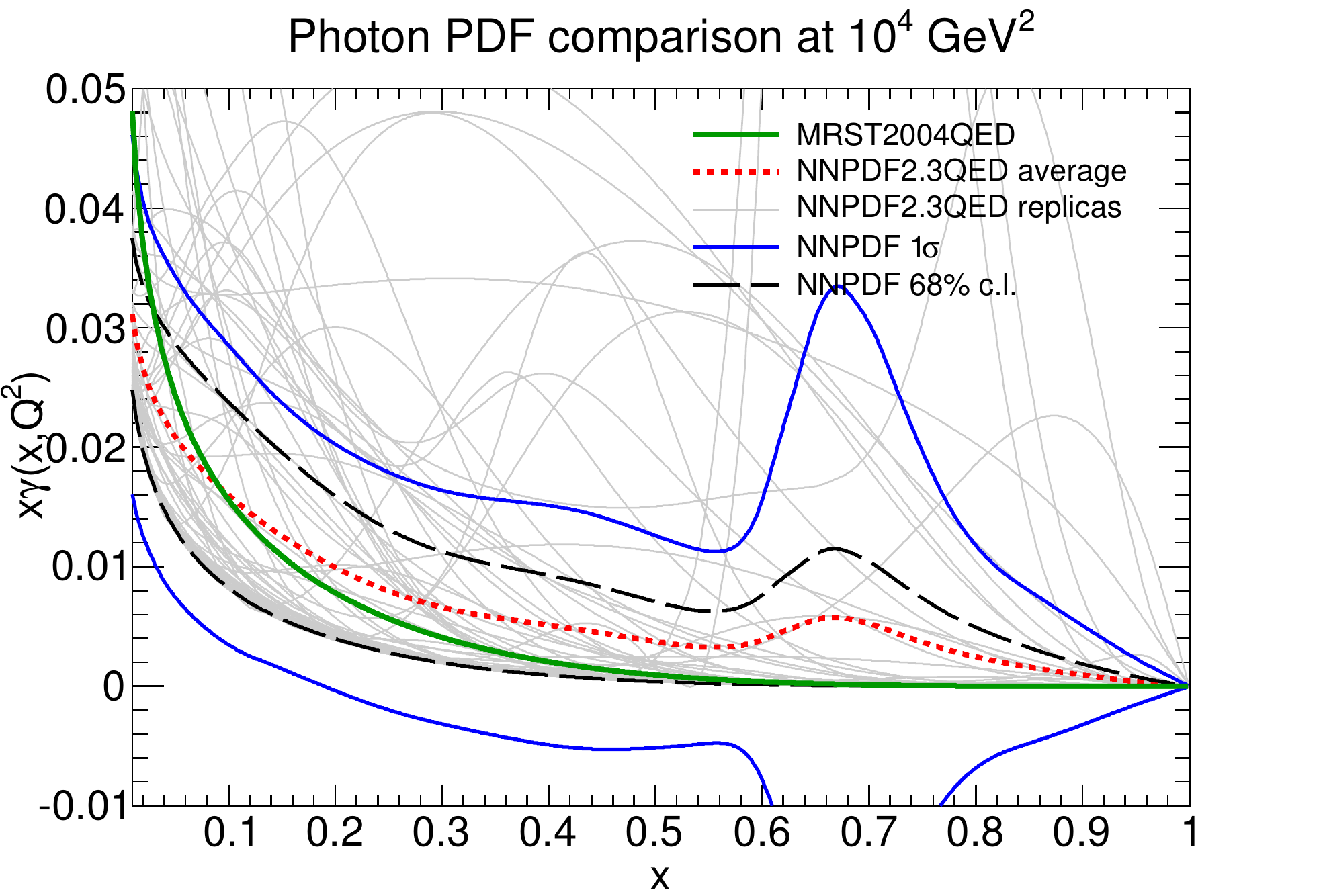}
\par\end{centering}
\caption{\label{fig:photonrw} The NNPDF2.3QED NLO photon PDF at
  $Q^2=2$ GeV$^2$ and $Q^{2}=10^{4}$ GeV$^{2}$ plotted vs. $x$ on a
  log (left) or linear (right) scale. The 100 replicas are shown,
  along with the mean, the one-$\sigma$, and the 68\% confidence level
  ranges. The MRST2004QED photon PDF is also shown for comparison.  }
\end{figure}

 \begin{figure}[ht]
\begin{centering}
\includegraphics[scale=0.34]{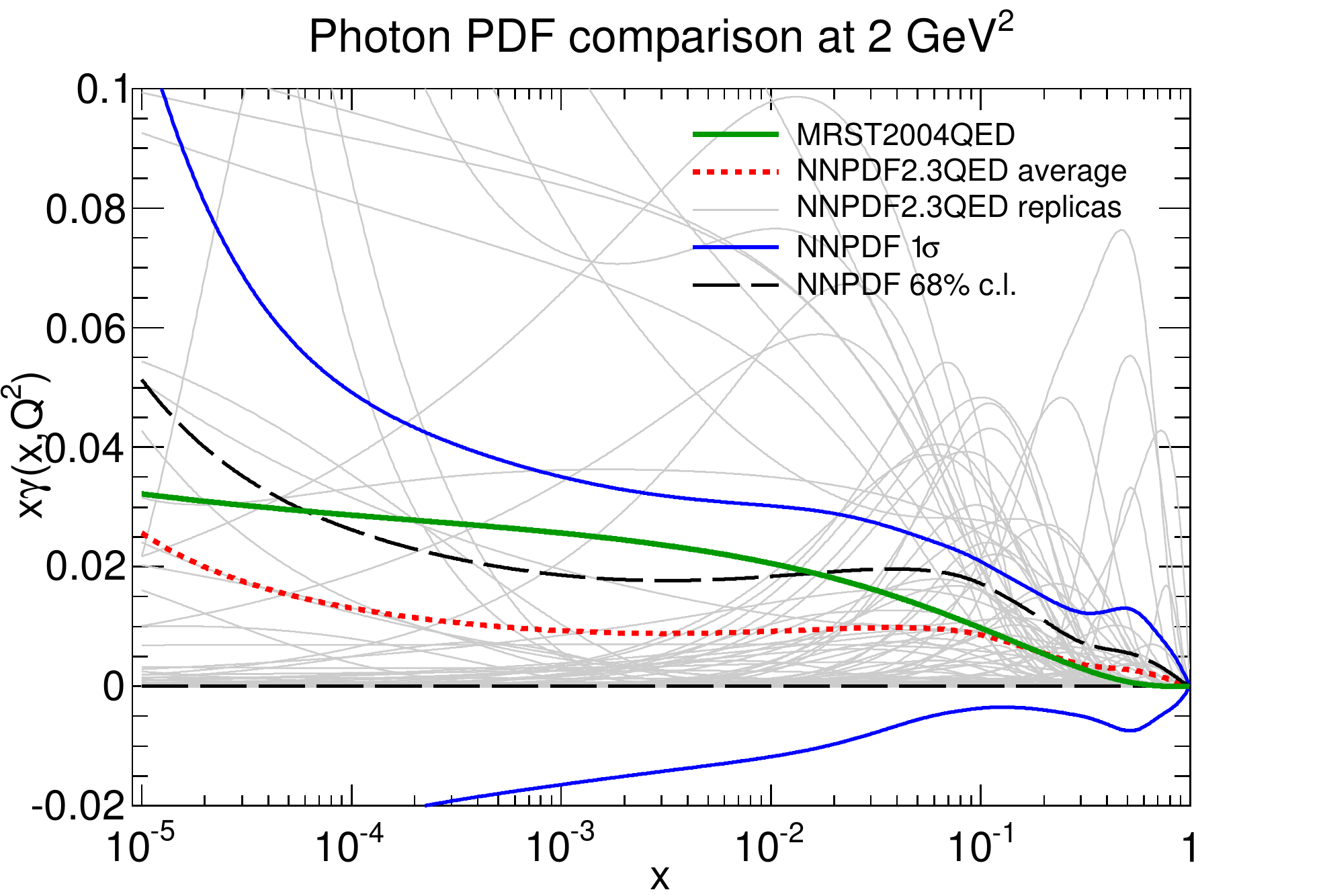}\includegraphics[scale=0.34]{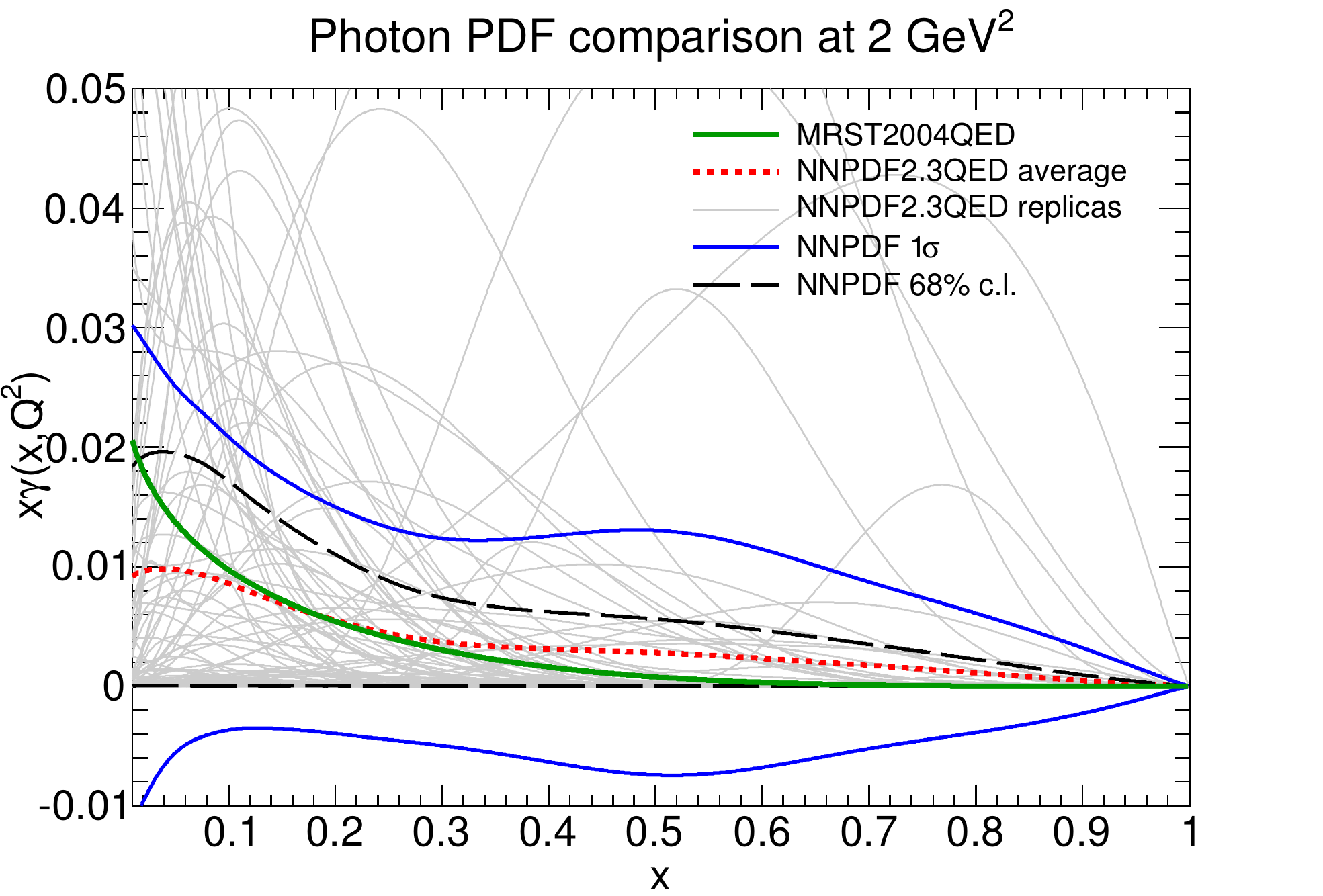}\\
\includegraphics[scale=0.34]{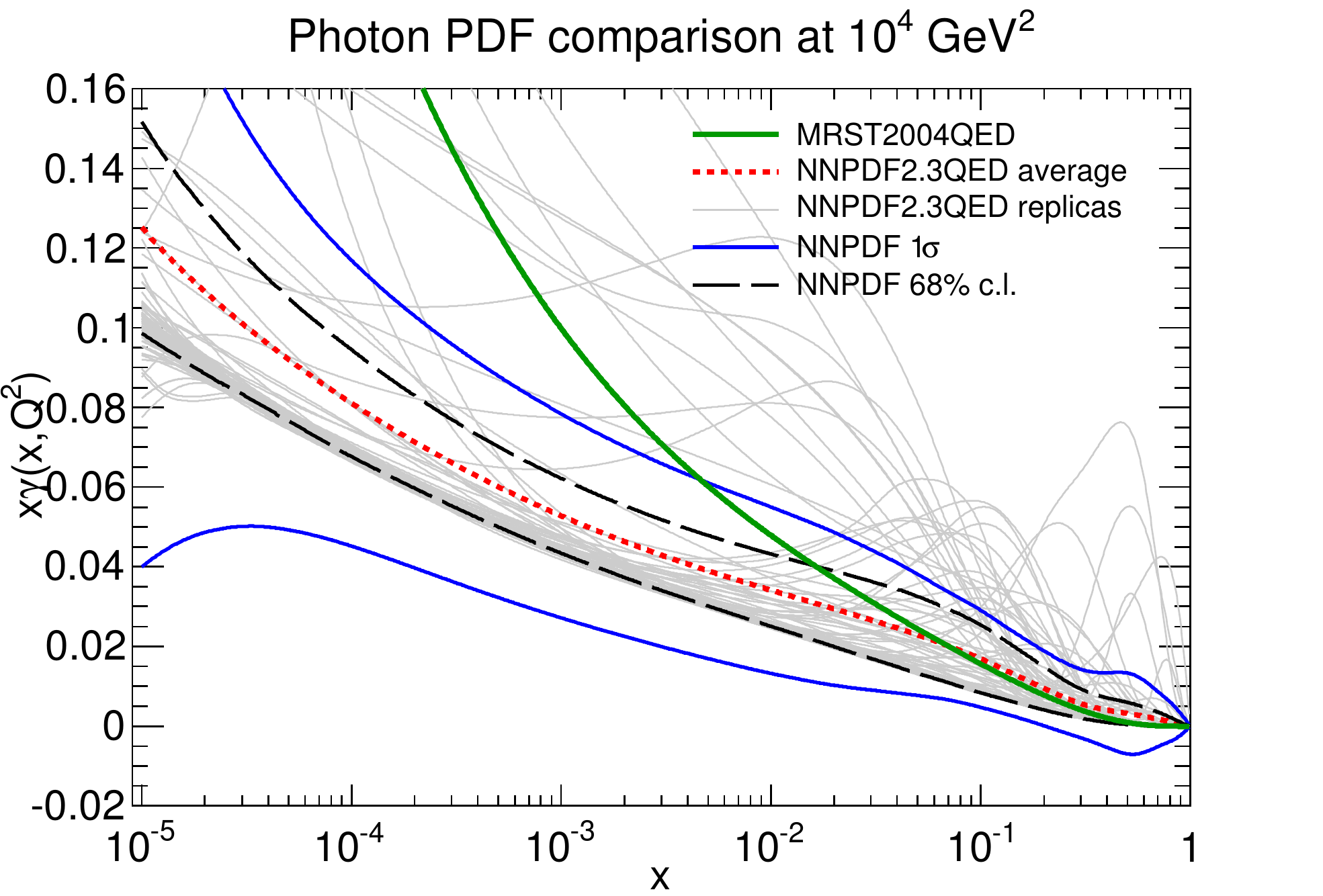}\includegraphics[scale=0.34]{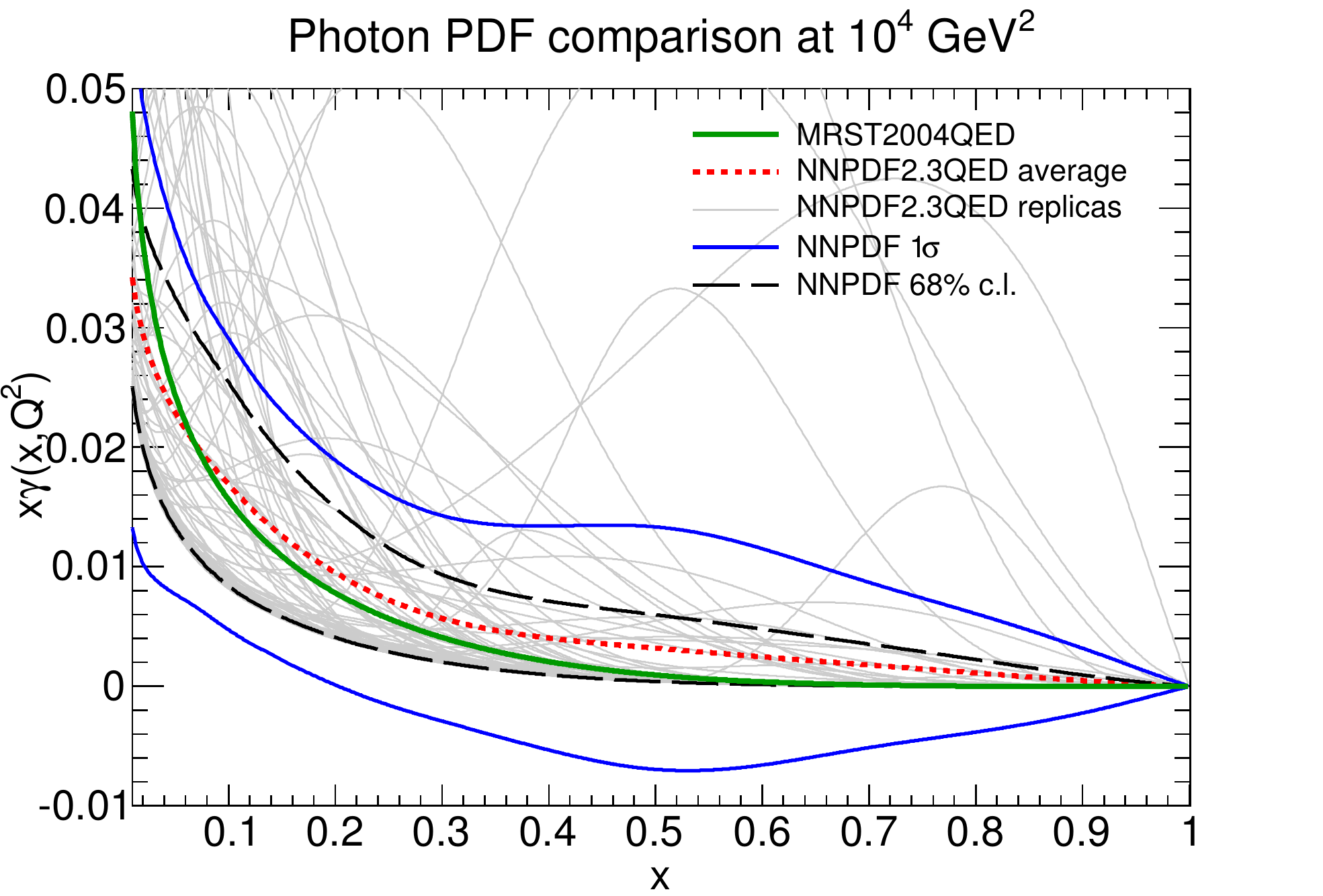}
\par\end{centering}
\caption{\label{fig:photonrwnnlo} Same as~\ref{fig:photonrw} for the
  NNPDF2.3QED NNLO PDF set.  }
\end{figure}

\begin{table}
\small
\centering
\begin{tabular}{c||c||c|c|c}
\hline
\multicolumn{5}{c}{NLO} \\
\hline
&  LHCtot  & ATLAS $W,Z$  & ATLAS high mass DY  & LHCb low-mass DY \\ 
\hline
\hline
$\chi^2_\textrm{in}$  & 2.02 & 1.20 & 3.78 & 2.20 \\
$\chi^2_\textrm{rw}$  &  1.00 &  1.15 & 1.01 & 0.29\\
\hline
$N_\textrm{eff}$  & 287 & 364 & 326 & 267 \\
$\left\langle \alpha \right\rangle$  & 1.41 & 1.24 & 1.53 & 0.89\\
\hline
\end{tabular}
\\ \vspace{0.5cm}
\begin{tabular}{c||c||c|c|c}
\hline
\multicolumn{5}{c}{NNLO} \\
\hline
&  LHCtot  & ATLAS $W,Z$  & ATLAS high mass DY  & LHCb low-mass DY \\ 
\hline
\hline
$\chi^2_\textrm{in}$  &  2.01 & 1.37  &  3.44 & 2.06 \\
$\chi^2_\textrm{rw}$  &  1.08  &  1.21 & 1.00 & 0.66 \\
\hline
$N_\textrm{eff}$  & 197 & 297 & 330 &  363 \\
$\left\langle \alpha \right\rangle$  & 1.48 & 1.33 & 1.52 & 1.20 \\
\hline
\end{tabular}
\caption{Reweighting parameters in the construction of the
  final NNPDF2.3 sets. All $\chi^2$ values are defined as in Tab.~\ref{tab:chi2}.\label{tab:rwnlo}}
\end{table}

\subsection{The NNPDF2.3QED set}
\label{sec:nnpdf23qed}

The NNPDF2.3QED PDF set is obtained by performing a reweighting of the
prior $N_\textrm{rep}=500$ replica set with the data of
Table~\ref{tab:expdata}. The procedure is performed at NLO and NNLO in
QCD, with three different values of $\alpha_s$ in each case. The
theoretical prediction used for reweighting is computed as discussed
in the previous section, and the $\chi^2$ used for reweighting is then
determined from its comparison to the data, using the fully correlated
systematics for the two ATLAS experiments, for which the covariance
matrix is available, but adding statistical and systematic errors in
quadrature for LHCb, for which information on correlations is not
available.  The ensuing weighted set of replicas is then
unweighted~\cite{Ball:2011gg} to obtain a standard set of
$N_\textrm{rep}=100$ replicas.

The parameters of the reweighting are collected in
Table~\ref{tab:rwnlo}: we show the $\chi^2$ (divided by the number of
data points) for the data of Table~\ref{tab:expdata} before and after
reweighting, the effective number of replicas after reweighting, and
the mean value of $\alpha$, the parameter which measures the
consistency of the data which are used for reweighting with those
included in the prior set, by providing the factor by which the
uncertainty on the new data must be rescaled in order of the two sets
to be consistent (so $\alpha\sim 1$ means consistent data). Values are
given for reweighting performed using each individual dataset, and the
three datasets combined. All $\chi^2$ values are computed using the
experimental definition of the covariance matrix as in
Table~\ref{tab:chi2}; the same form of the covariance matrix has also
been used for reweighting for simplicity, as this choice is immaterial
as discussed above.

\begin{figure}[ht]
\begin{centering}
\includegraphics[scale=0.68]{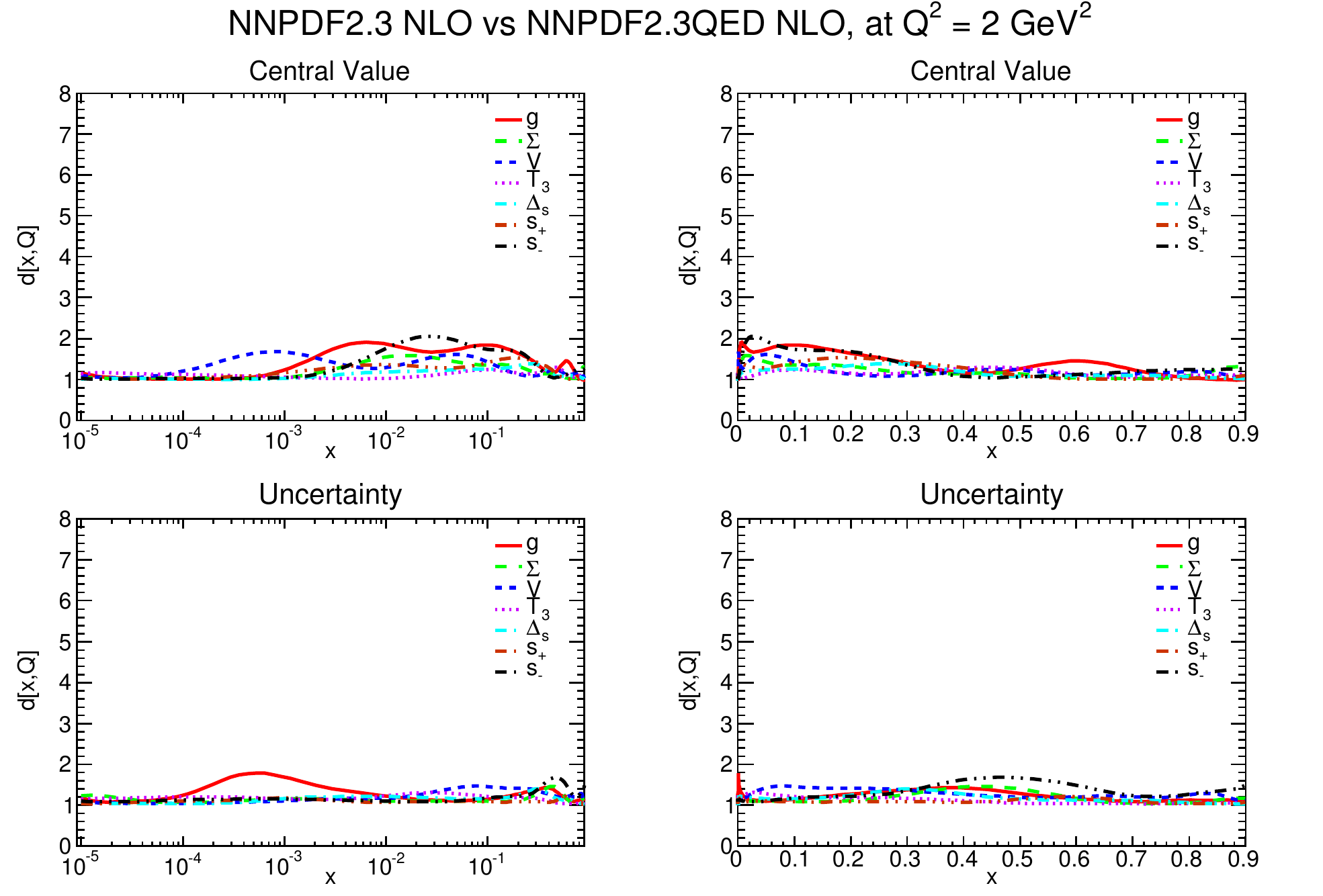}
\par\end{centering}
\caption{\label{fig:distances2} Distances between PDFs in the NNPDF2.3
  and the NNPDF2.3QED NLO sets, at the input scale of
  $Q_0^2$=2~GeV$^2$. Distances between central values (top) and
  uncertainties (bottom) are shown, on a logarithmic (left) and linear
  (right) scale in $x$.  }
\end{figure}

\begin{figure}[ht]
\begin{centering}
\includegraphics[scale=0.68]{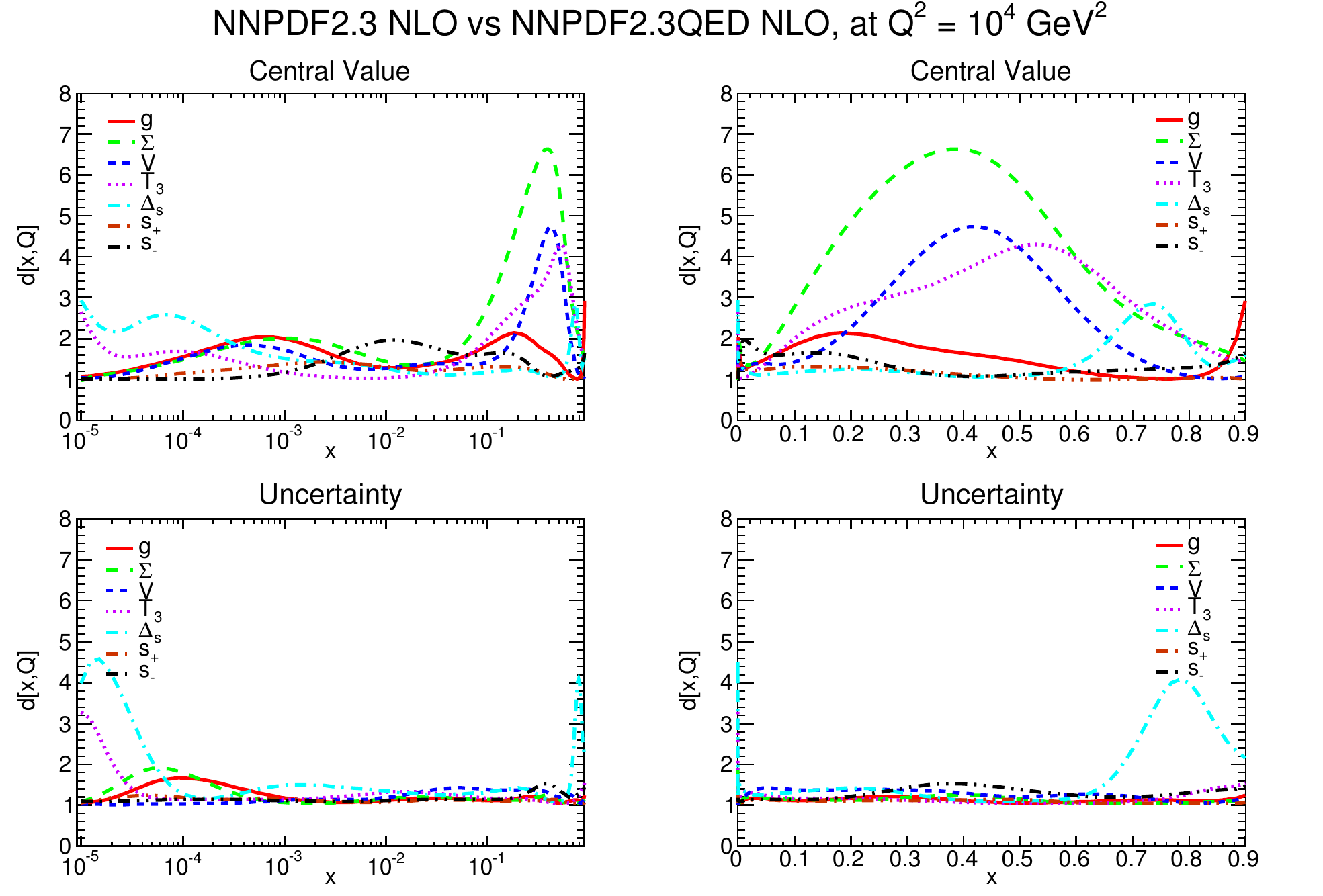}
\par\end{centering}
\caption{\label{fig:distances2bis} Same as Fig.~\ref{fig:distances2}
  but now computed at $Q^2=10^{4}$~GeV$^2$.  }
\end{figure}

In all cases the final effective number replicas turns out to be
$N_\textrm{eff}>100$, thereby guaranteeing the accuracy of the final
unweighted set. All sets show good compatibility with the prior
datasets.  The final $\chi^2$ values show that the reweighted set
provides an essentially perfect fit to the data; the low values for
LHCb are a consequence of the fact that for this experiment the
correlated systematics is not available so statistical and systematic
errors are added in quadrature.  Before reweighting the $\chi^2$ of
individual replicas shows wide fluctuations: indeed, its average and
variance over the starting replica sample are given by $\langle
\chi^2\rangle=25.6 \pm 164.4$. After reweighting the value becomes
$\langle \chi^2\rangle= 1.117\pm 0.098$, thus showing that the
$\chi^2$ of indvidual replicas has become on average almost as good as
that of the central reweighted prediction.

\begin{table}[ht]
\begin{center}
{\footnotesize
\centering
\begin{tabular}{|c|c|c|c|}
\hline
     & NNPDF2.3QED NLO & NNPDF2.3QED NNLO & MRST2004QED \\ 
\hline
\hline $\gamma$; $Q^2=2$~GeV$^2$  &
 $\left( 0.42 \pm 0.42\right)$\% & $\left( 0.34 \pm 0.34 \right)$\%  & $0.30$\% \\  [2ex] 
\hline $\gamma$; $Q^2=10^4$~GeV$^2$  & $\left( 0.68 \pm 0.42\right)$\% &
$\left( 0.61 \pm 0.34\right)$\% & $0.52$\% \\   [2ex] 
\hline\hline
 total; $Q^2=2$~GeV$^2$
  & $\left( 100.43 \pm 0.44\right)$\%  &  $\left( 100.32 \pm 0.34 \right)$\% & $99.95$\%\\   [2ex] 
\hline  total; $Q^2=10^4$~GeV$^2$ & $\left( 100.38 \pm 0.43\right) $\% & 
$\left( 100.29 \pm 0.36 \right) \%$ & $99.92$\% \\   [2ex] 
\hline
\end{tabular}}
\caption{\label{tab:msr} Momentum fractions (in percentage) carried by
  the photon PDF (upper two rows) and by the sum of all partons in the
  proton (lower two rows) in the NNPDF2.3QED NLO, NNLO and MRST2004QED
  PDF sets at two different scales}
\end{center}
\end{table}

\begin{figure}[ht]
\begin{centering}
\includegraphics[scale=0.34]{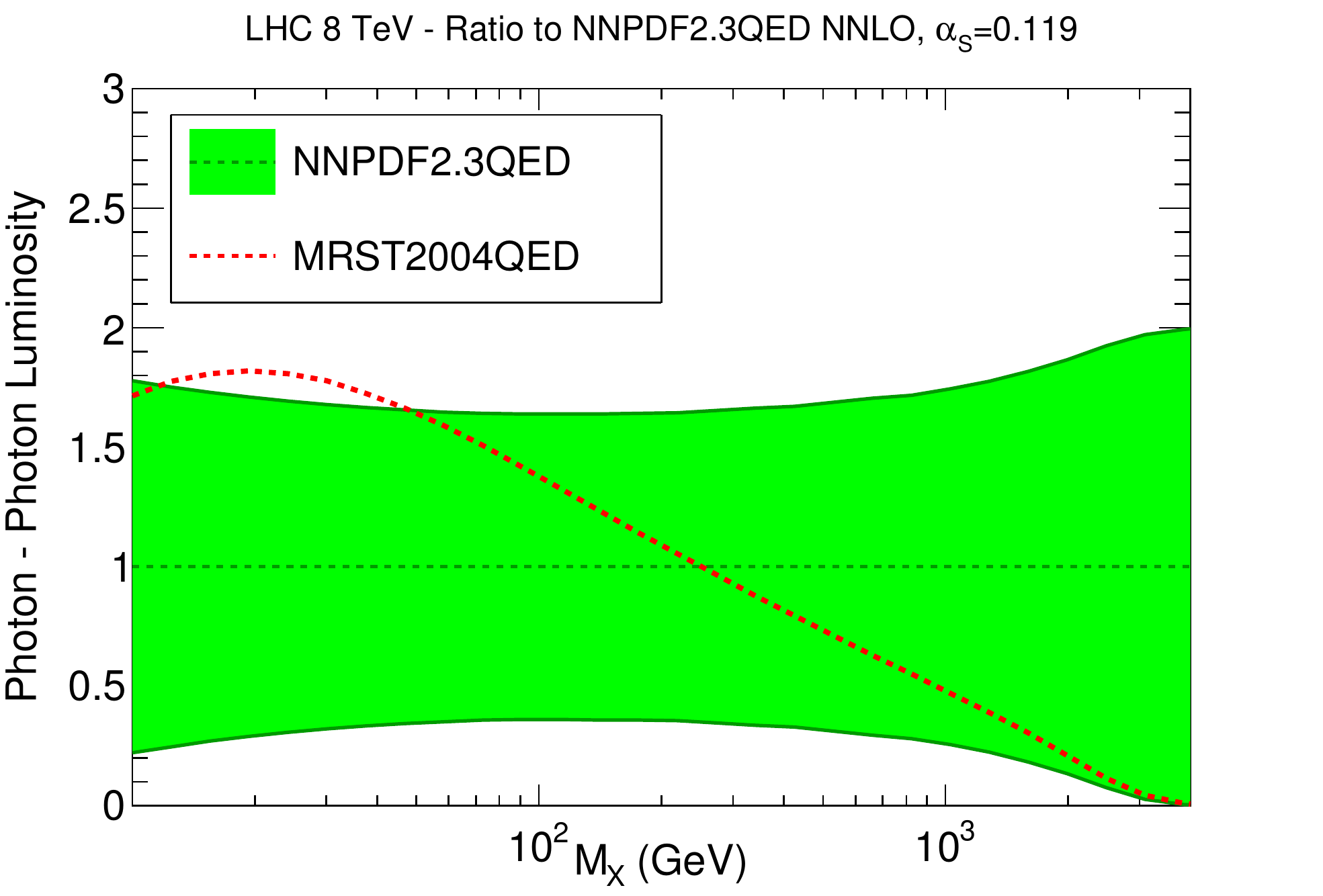}\includegraphics[scale=0.34]{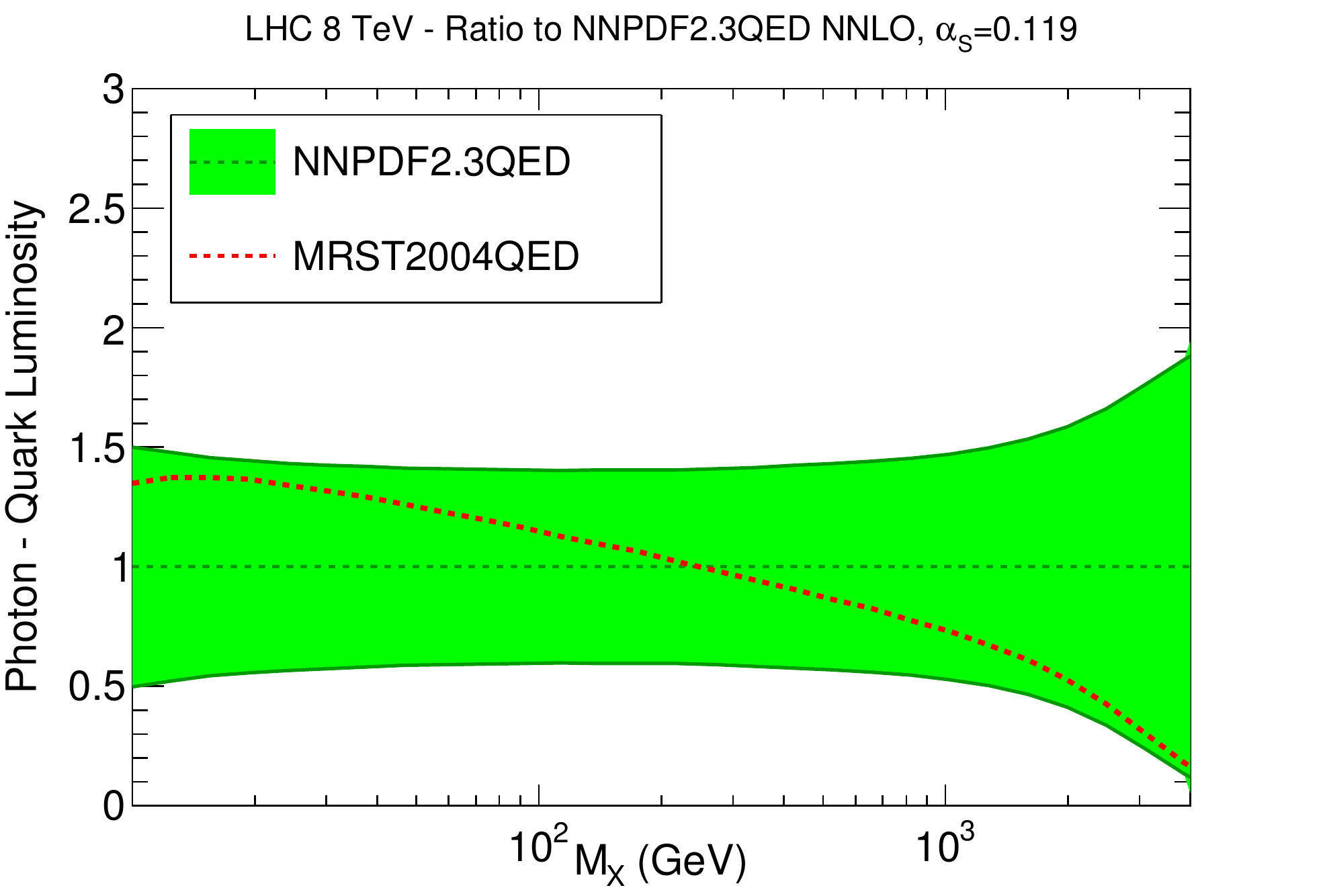}
\par\end{centering}
\caption{\label{fig:lumimrst}The photon-photon $\gamma \gamma$ (left)
  and photon-quark $\gamma q$ (right) parton luminosities at the
  LHC~8~TeV computed using MRST2004QED PDFs, shown as a ratio to the
  NNPDF2.3QED result. The 68\% confidence level on the latter is also
  shown.}
\end{figure}

A first assessment of the impact of the photon-induced corrections and
their effect on the photon PDF can be obtained by comparing the data
to the theoretical prediction obtained using pure QCD theory and the
default NNPDF2.3 set, QCD$\otimes$QED with the prior photon PDF, and
QED$\otimes$QCD with the final NNPDF2.3QED set. The comparison is
shown in Figs.~\ref{fig:rwlhc}-\ref{fig:rwlhc2} for the NLO sets (the
NNLO results are very similar): in the left plots we show the QED+QCD
prediction obtained using the prior PDF set, and in the right plots
the prediction obtained using the final reweighted sets, compared in
both cases to the pure QCD prediction obtained using \texttt{DYNNLO}
and the NNPDF2.3 set. At the $W,Z$ peak, the impact of QED corrections
is quite small, though, in the case of neutral current production, to
which the photon-photon process contributes at Born level, when the
prior photon PDF is used one can see the widening of the uncertainty
band due to the large uncertainty of the photon PDF of
Figure~\ref{fig:xpht-nlo}. At low or high mass, as one moves away from
the peak, the large uncertainty on the prior photon PDF induces an
increasingly large uncertainty on the theoretical prediction,
substantially larger than the data uncertainty. This means that these
data do constrain the photon PDF and indeed after reweighting the
uncertainty is substantially reduced.

 \begin{figure}[ht]
\begin{centering}
\includegraphics[scale=0.34]{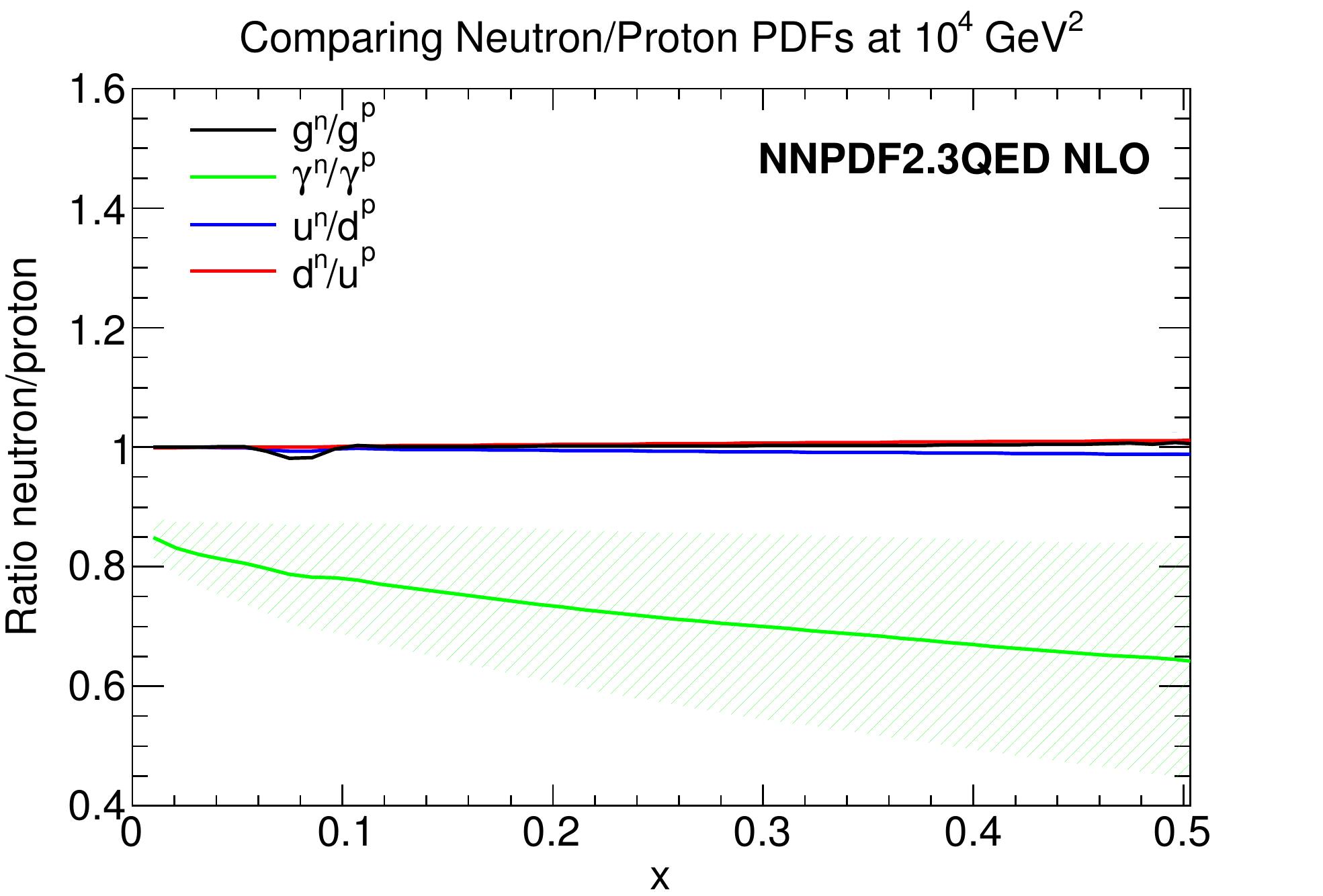}\includegraphics[scale=0.34]{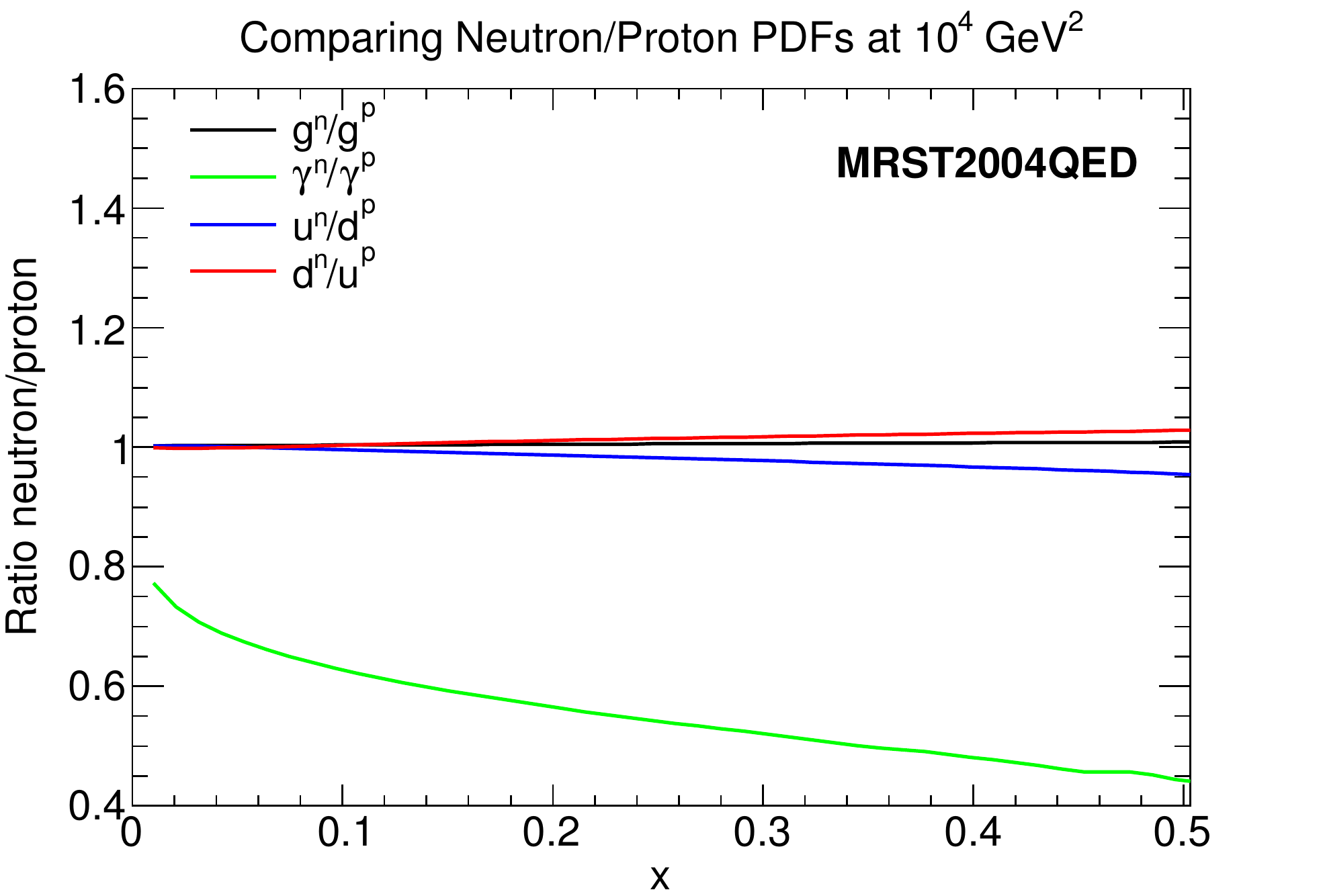}
\par\end{centering}
\caption{\label{fig:isospinLHC} The ratio of the neutron to the proton
  PDFs in the NNPDF2.3QED NLO set at $Q^2=10^4$ GeV$^2$ (left) and
  MRST2004QED set (right).  Results for the photon, gluon, up and down
  quark are shown.  Error bands correspond to one-$\sigma$
  uncertainties.
}
\end{figure}

The final NNPDF2.3QED photon PDF obtained in the NLO and NNLO fits is
respectively shown at $Q^2_0=2$ GeV$^2$ in Figure~\ref{fig:photonrw}
and Figure~\ref{fig:photonrwnnlo}.  We display individual replicas,
the central (mean) photon, and the one-$\sigma$ and 68\% confidence
level ranges, as well as the MRST2004QED result.  The improvement in
accuracy in comparison to the prior PDF of Figure~\ref{fig:xpht-nlo}
is apparent, especially at small and at large $x$.  Note also that,
especially at large $x$, where the experimental information remains
scarce (recall Figure~\ref{fig:correlations}), the positivity bound
still plays an important role in constraining the photon PDF.  Indeed,
at the starting scale $Q_0$ the lower edge of the uncertainty band
(determined as discussed in Sect.~\ref{sec:disfit}) is again very
close to the positivity constraint, and consequently, even after
having used the LHC data, the probability distribution of the photon
PDF is significantly asymmetric, departing substantially from
Gaussian.  This should be kept in mind in phenomenological
applications, in particular when computing uncertainties.

In Table~\ref{tab:msr} we show the momentum fraction carried by the
photon PDF in NNPDF2.3QED at NLO and NNLO, both at a low and high
scale: it is about half of a percent, compatible with zero within
uncertainties, and mildly dependent on scale. The MRST2004QED values,
also shown, are consistent within uncertainties. Note that the
standard deviation would be almost twice the 68\% confidence level
interval given in the table.  We also give the total momentum, which
deviates from unity because of the slightly inconsistent procedure
that we have followed in constructing the prior set, by combining the
photon from a fit to DIS data with the other PDFs from the global
NNPDF2.3 fit as discussed in Sect.~\ref{sec:combination} above. We
also see that the total momentum fraction is not quite scale
independent, because of the approximation introduced when neglecting
terms of $\mathcal{O}(\alpha\alpha_s)$ in the solution of the combined
QED$\otimes$QCD evolution equations. Both effects are well below the
1\% level.

All other PDFs at the initial scale $Q_0$ are left unaffected by the
reweighting.  This can be seen by computing the distances between PDFs
in the starting NNPDF2.3 set and in the final NNPDF2.3QED set; they
are displayed in Figure~\ref{fig:distances2}, at the scale
$Q^2_0=2$~GeV$^2$ at which PDFs are parametrized: it is apparent that
the distances are compatible with statistically equivalent PDFs.  It
is interesting to repeat the same comparison at $Q^2=10^4$~GeV$^2$
(Figure~\ref{fig:distances2bis}): in this case, statistically
significant differences start appearing, as a consequence of the fact
that the statistically equivalent starting PDFs in the two sets are
then evolved respectively with and without QED corrections. However,
the differences are below the one-$\sigma$ level (and concentrated at
large $x$), consistent with the conclusion that the new data are
compatible with those used for the determination of the NNPDF2.3 PDF
set.

In Figs.~\ref{fig:photonrw}-\ref{fig:photonrwnnlo} the photon PDF from
the MRST2004QED set is also shown for comparison. The MRST2004QED
photon PDF is based on a model; an alternative (not publicly
available) version of it, in which consitituent rather than current
quark masses are used as model parameters, has been
used~\cite{Aad:2011dm} to estimate the model uncertainty, though
consitituent masses are considered to be less appropriate by the
authors of Ref.~\cite{Martin:2004dh}. The MRST2004QED photon turns out
to be in good agreement with the central NNPDF2.3QED prediction at
medium and large $x$, but at small $x\lesssim0.03$ it grows more
quickly, and for $x\le 10^{-2}$ it is larger and well outside the
NNPDF2.3QED uncertainty band.

It is also interesting to compare the NNPDF2.3QED and MRST2004QED sets
at the level of the parton luminosities which enter the computation of
hadronic processes.  This comparison is shown in
Figure~\ref{fig:lumimrst}. The two luminosities are in good agreement
for invariant masses of the final state $M_X\sim100$~GeV, but the
agreement is less good for higher or lower final-state masses, with
the MRST2004QED rather smaller at high mass and larger at low mass,
where, for $M_X\sim20$~GeV it is outside the NNPDF2.3QED uncertainty
band.  As we will see in the next section, these differences translate
into differences in the predictions for electroweak processes at the
LHC.

So far, we have shown results for the PDFs of the proton. Note,
however that, as discussed in Sect.~\ref{sec:disfit}, even though we
assume that isospin holds at the scale at which PDFs are parametrized,
QED corrections to perturbative evolution introduce a violation of the
isospin symmetry at all other scales.  Therefore, we provide
independent NNPDF2.3QED PDF sets for proton and neutron. The size of
isospin violation is expected to be comparable to the QED corrections
themselves, so very small for quark and gluon distributions but more
significant for the photon PDF. The expectation is borne out by
Figure~\ref{fig:isospinLHC} where the ratio of the neutron to the
proton PDF at $Q^2=10^4$ GeV$^2$ in NNPDF2.3QED NLO is compared to
that in MRST2004QED set. The comparison shows that while the amount of
isospin violation in the MRST2004QED photon PDF, which had a built-in
model of non-perturbative isospin violation, is somewhat larger than
our own, especially at large $x$, the difference is within the PDF
uncertainty, as anticipated in Sect.~\ref{sec:disqed}. The amount of
isospin violation on quark and gluon PDFs is extremely small, on the
scale of PDF uncertainties, both for MRST2004QED and NNPDF2.3QED. The
same conclusions hold if the NNLO set is used.

\chapter{Phenomenological implications of the photon PDF}
\label{sec:chap5}

\label{sec:searches}

In this chapter we investigate some examples of the use of the
NNPDF2.3QED PDF set.
We analyze several processes which are sensitive to photon-initiated
contributions.
In particular, we will start with the discuss of direct photon
production at HERA, and then we show results about searches for new
massive electroweak gauge bosons and $W$ pair production at small
$p_T$ and large invariant mass, at LHC energies.
After presenting the phenomenological impact for these processes, we
then show details about availability of these sets of PDFs in Monte
Carlo event generators.
Finally, we conclude this chapter with a first determination of lepton
PDFs using the \texttt{APFEL} evolution and sets with photon PDFs.

\section{Photon-induced processes}

\subsection{Direct photon production at HERA}
\label{sec:directphotons}

 \begin{figure}[t]
\begin{centering}
\includegraphics[scale=0.34]{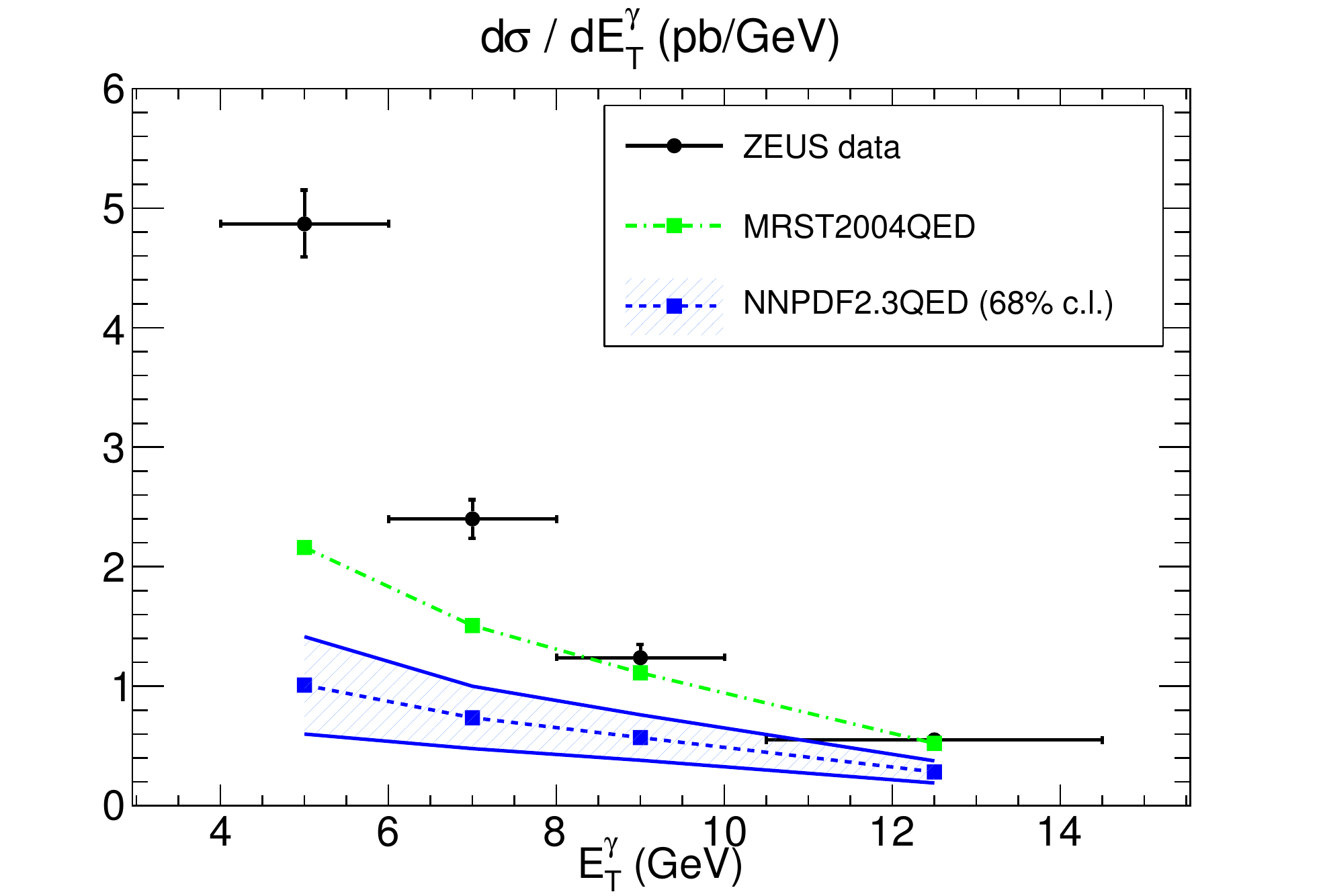}\includegraphics[scale=0.34]{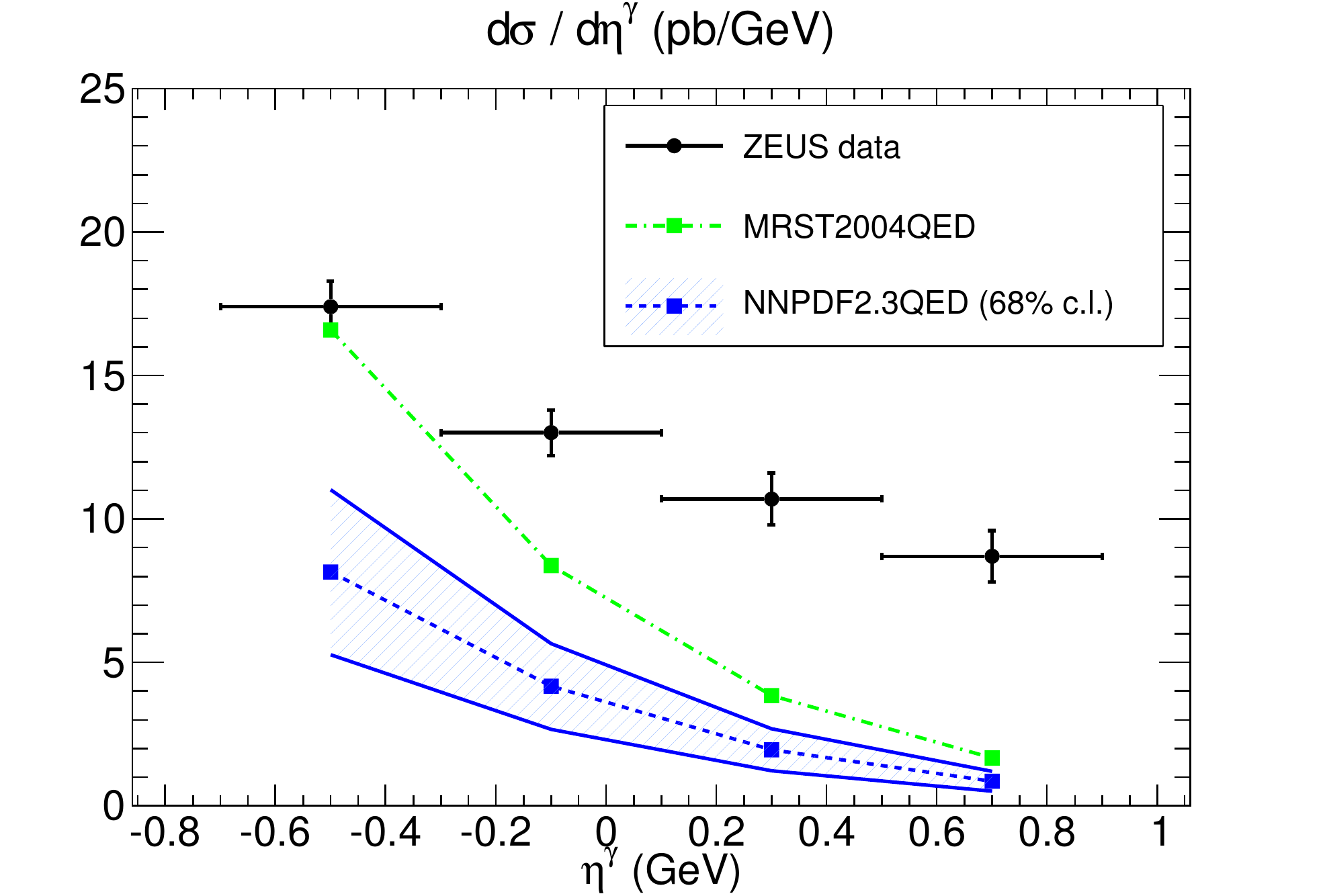}
\par\end{centering}
\caption{\label{fig:photondis} Comparison between the ZEUS
  data~\cite{Chekanov:2009dq} for the photon transverse energy (left)
  and rapidity (right) distributions in deep-inelastic isolated photon
  production and the leading log theoretical prediction obtained
  using NNPDF2.3QED and MRST2004QED PDFs.}
\end{figure}

Deep-inelastic isolated photon production provides a direct handle on
the photon parton distribution of the proton, through Compton
scattering of the incoming electron off the photon component of the
proton~\cite{DeRujula:1998yq}. At the leading log level, this
$\mathcal{O}(\alpha^2)$ partonic subprocess is the only
contribution. In practice, however, the $\mathcal{O}(\alpha^3)$
quark-induced contributions~\cite{GehrmannDeRidder:2006wz} may be
comparable (as for the Drell-Yan process discussed in
Sect.~\ref{sec:lhcwz}) because of the larger size of the quark
distribution.  In Ref.~\cite{Martin:2004dh}, the total cross-section
for this process computed at the leading log level using MRST2004QED
PDFs was shown to be in reasonable agreement with HERA integrated
cross-sections for prompt photon production
data~\cite{Chekanov:2004wr}.

However, more recent HERA data~\cite{Chekanov:2009dq} for the rapidity
and transverse energy distribution of the photon do not agree well
with either the fixed order~\cite{GehrmannDeRidder:2006wz} or the
leading log~\cite{DeRujula:1998yq,Martin:2004dh} results for all
values of the kinematics, suggesting that a calculation matching the
leading-log resummation to the fixed order result would be necessary
in order to obtain good agreement. In the absence of such a
calculation, we did not use these data for the determination of the
photon PDF.

Theoretical predictions obtained using the leading log
calculation~\cite{Martin:2004dh} and the NNPDF2.3QED or MRST2004QED
PDF sets are compared in Fig.~\ref{fig:photondis} to the ZEUS data of
Ref.~\cite{Chekanov:2009dq}.  These predictions have been obtained
using the code of Ref.~\cite{Martin:2004dh}.  The selection cuts are
the same as in~\cite{Chekanov:2009dq}, namely
\begin{eqnarray} 10 \le Q^2 \le 300~\textrm{GeV}^2 \, , \quad 4 \le
  E_T^{\gamma} \le 15~\textrm{GeV} \, , \quad -0.7 \le \eta^{\gamma}
  \le 0.9 \, .
\end{eqnarray}

The fact that the prediction is in better agreement with the data at
large $E_T$ is consistent with the expectation that the leading log
approximation which is being used is more reliable in this region.
However, as already mentioned, a fully matched calculation would be
needed in order to consistently combine the leading log and fixed
order results.

 \begin{figure}[ht]
\begin{centering}
\includegraphics[scale=0.34]{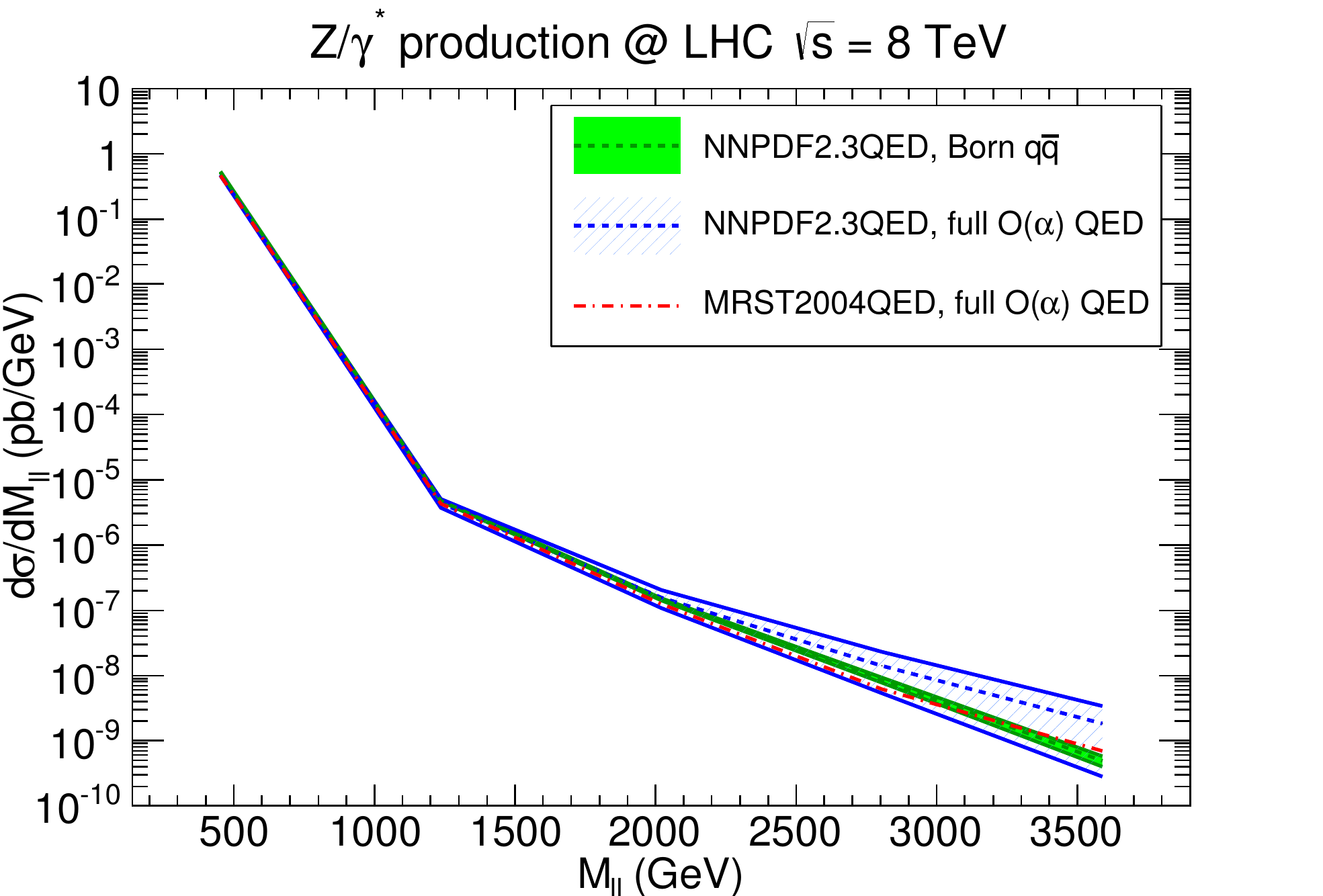}\includegraphics[scale=0.34]{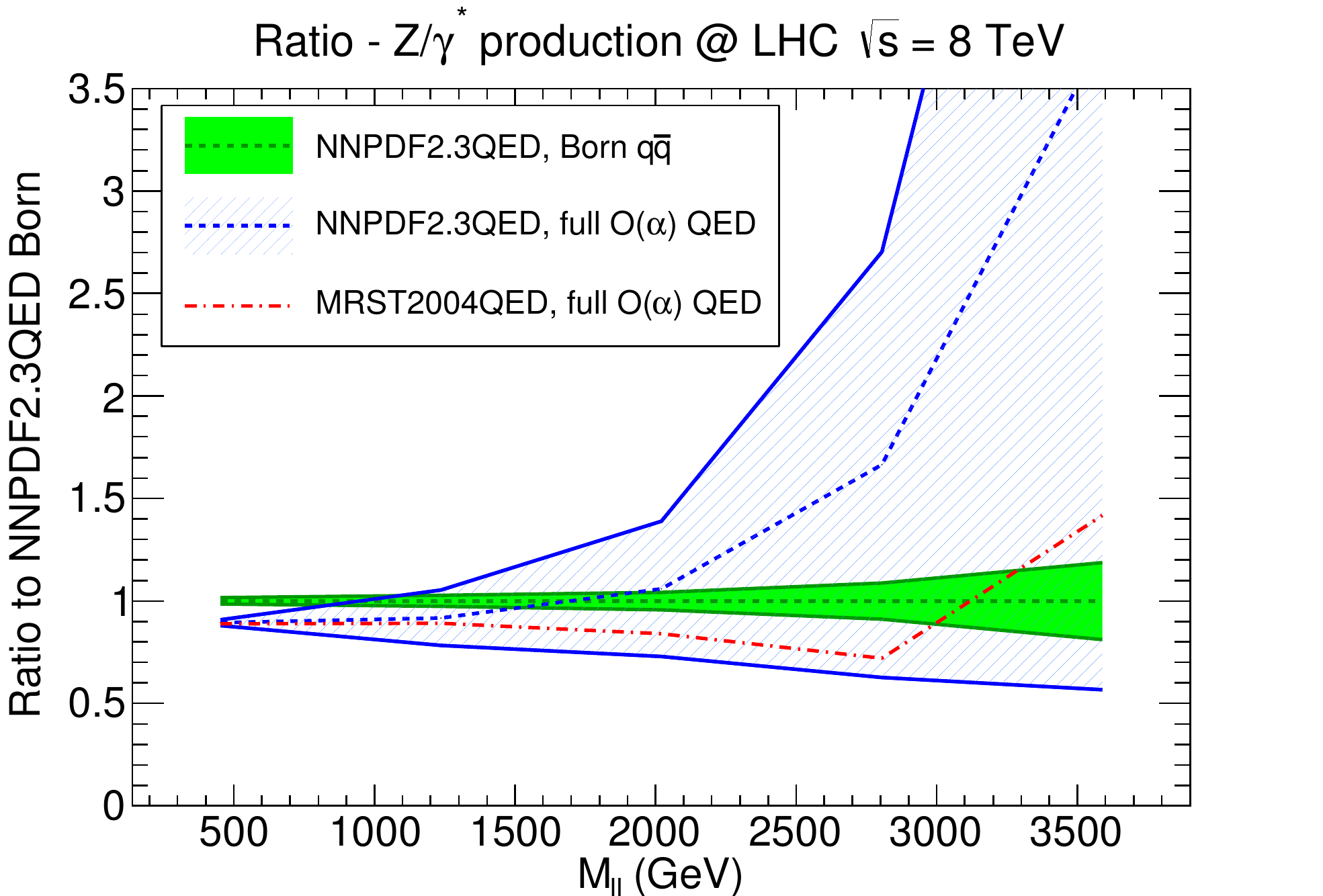}\\
\includegraphics[scale=0.34]{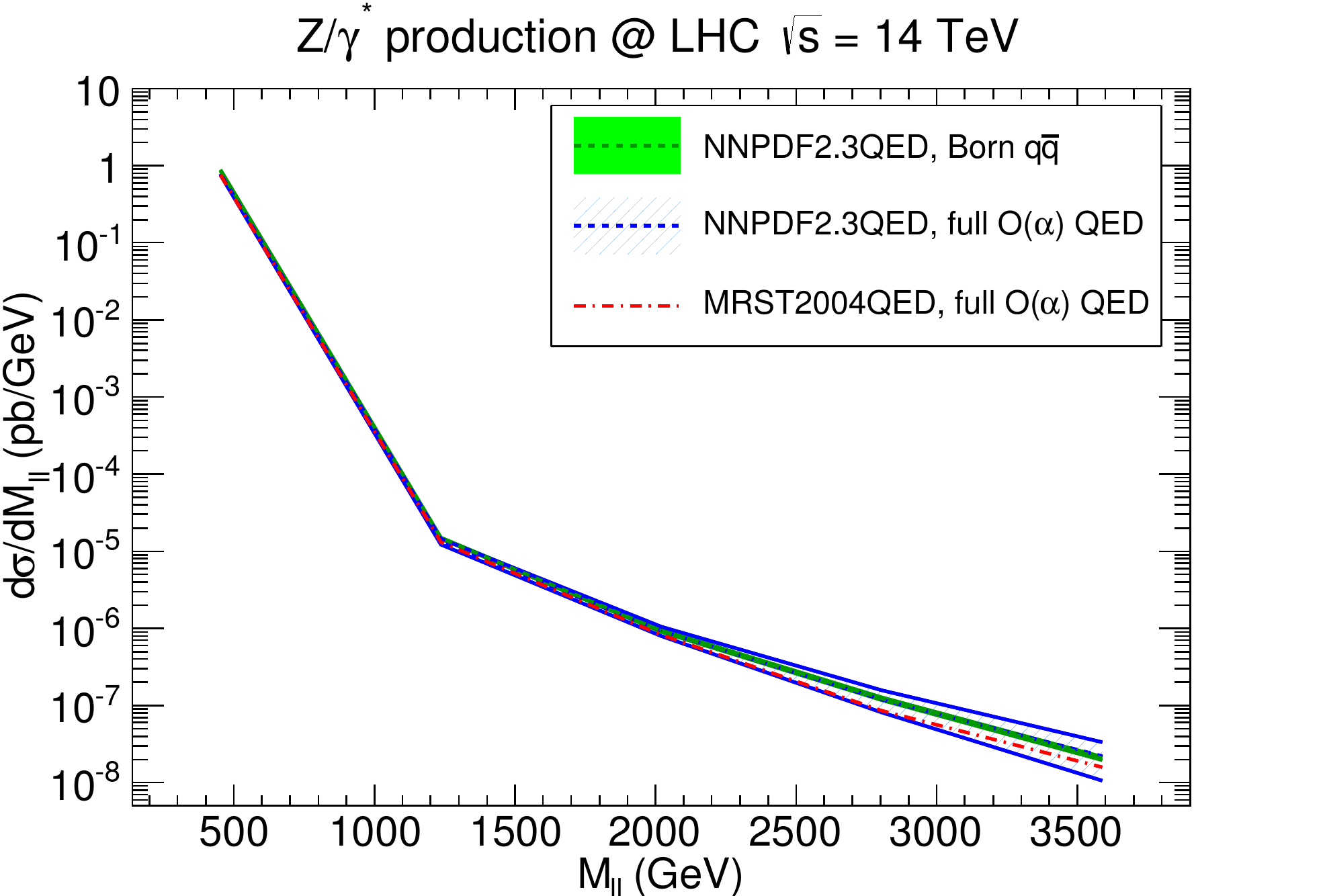}\includegraphics[scale=0.34]{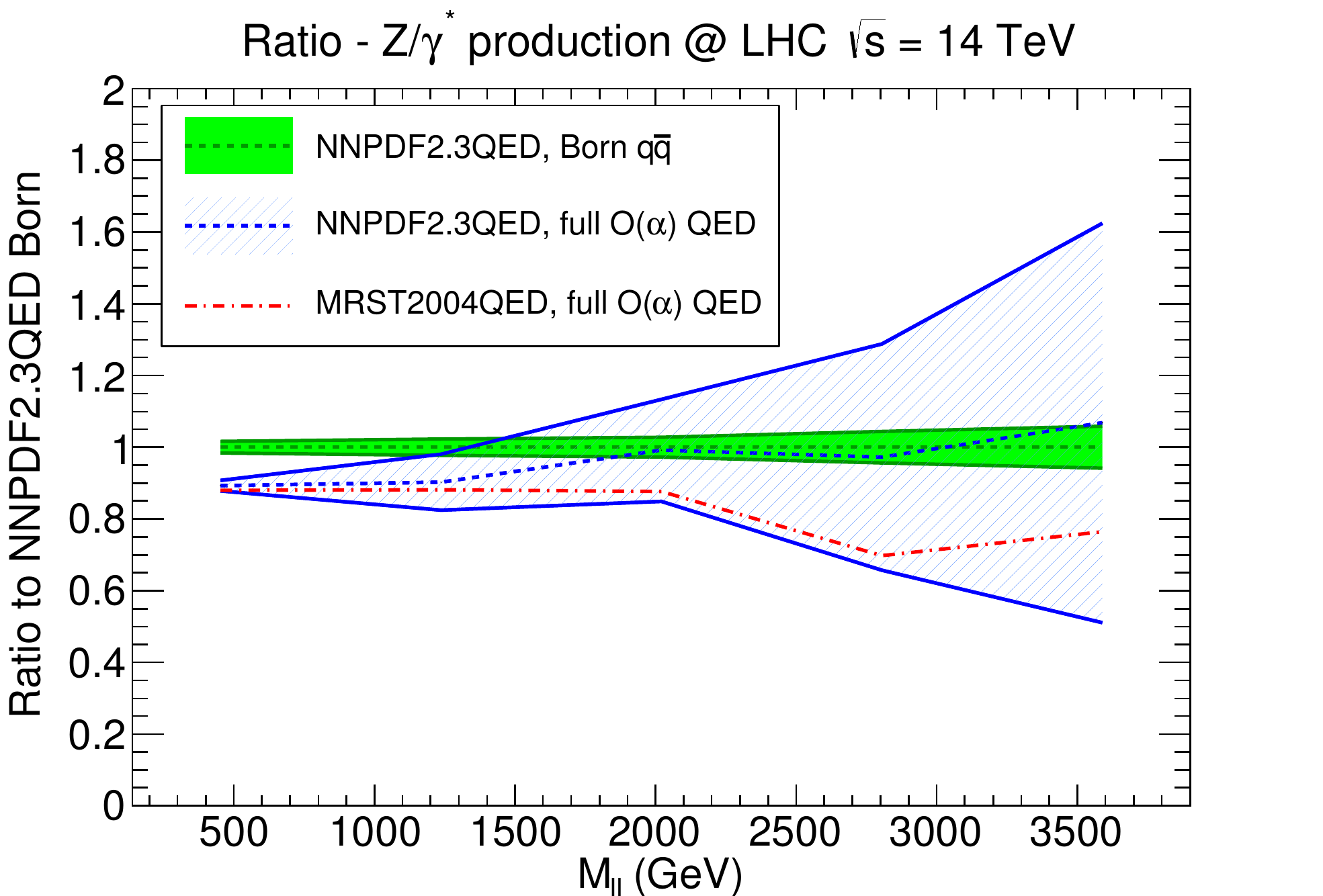}
\par\end{centering}
\caption{\label{fig:Zproduction} Neutral current Drell-Yan
  production at the LHC as a
  function of the invariant mass of the dilepton pair using
  NNPDF2.3QED and MRST2004QED PDFs.
  Theoretical predictions for the Born $q\bar{q}$ and the full
  $\mathcal{O}\left( \alpha\right)$ process (including photon-induced
  contributions) at the LHC 8 TeV (top) and LHC 14 TeV (bottom), are
  shown both on an absolute scale (left) or as a ratio to the central
  value of the Born $q\bar{q}$ cross-section from NNPDF2.3QED.}
\end{figure}

\subsection{Searches for new massive electroweak gauge bosons}
\label{sec:wzprimesearch}

Heavy electroweak gauge bosons, denoted generically by $W'$ and $Z'$,
have been actively searched at the LHC (see
e.g.~\cite{Chatrchyan:2012meb,Chatrchyan:2011dx,Chatrchyan:2011wq,Collaboration:2011dca}),
with current limits for $M_{V'}$ between 1 and 2 TeV depending on the
model assumptions.  The main background for such searches is the
off-resonance production of $W$ and $Z$ bosons respectively.  At such
large invariant masses of the dilepton pair, photon-induced
contributions, of the type shown in
Figs.~\ref{fig:lhczborn}--\ref{fig:lhcwz2}, are potentially large.

\begin{figure}[ht]
\begin{centering}
\includegraphics[scale=0.34]{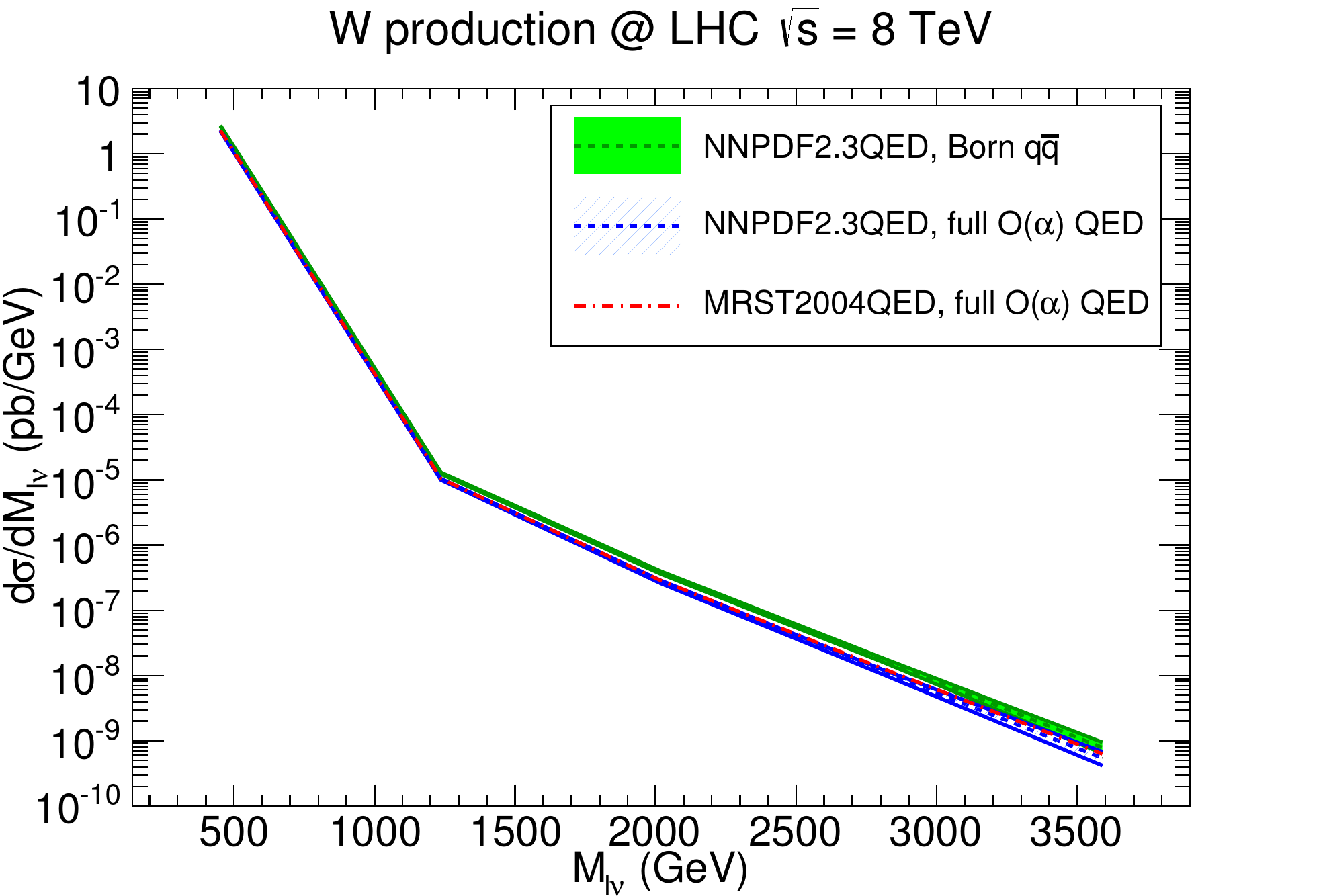}\includegraphics[scale=0.34]{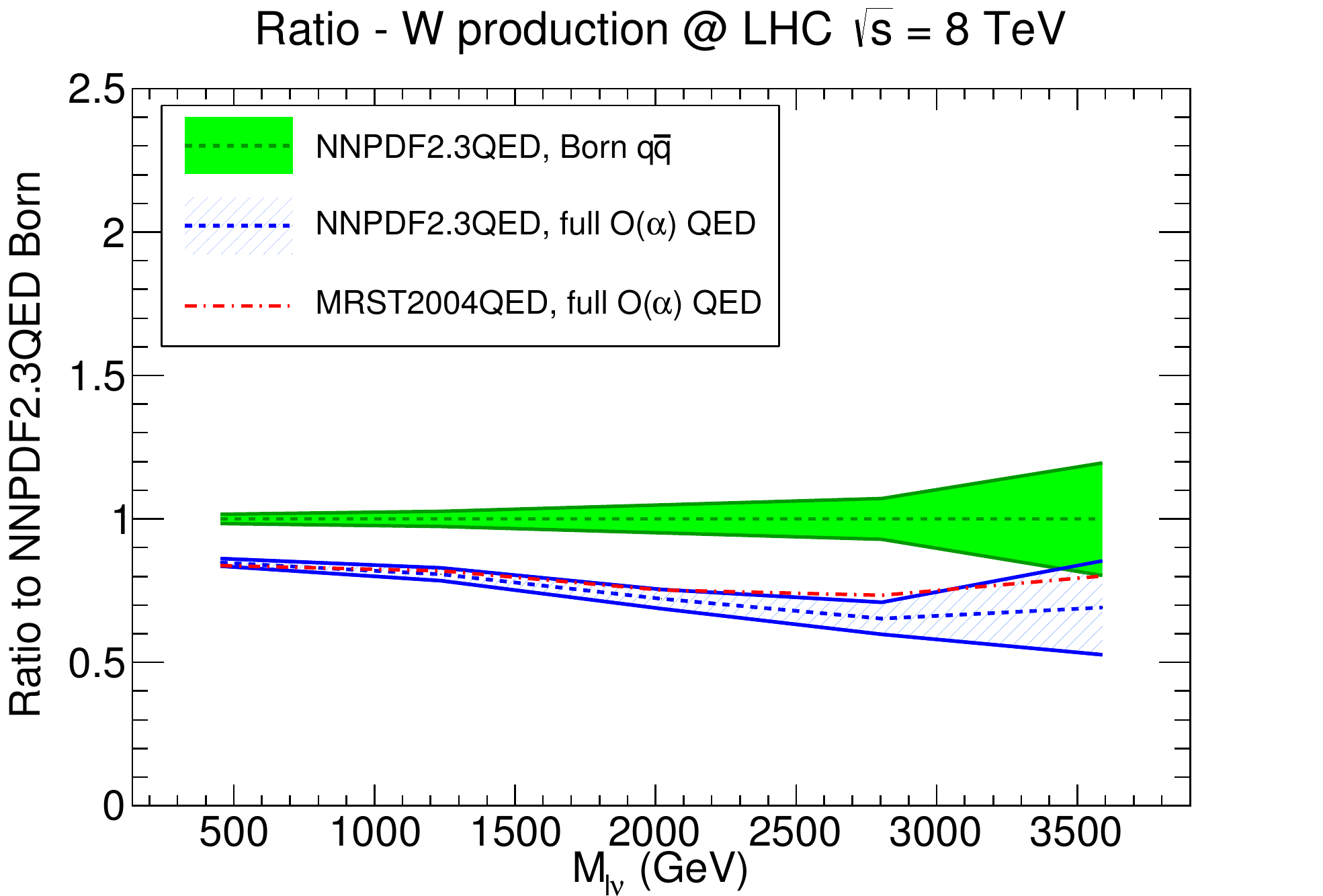}\\
\includegraphics[scale=0.34]{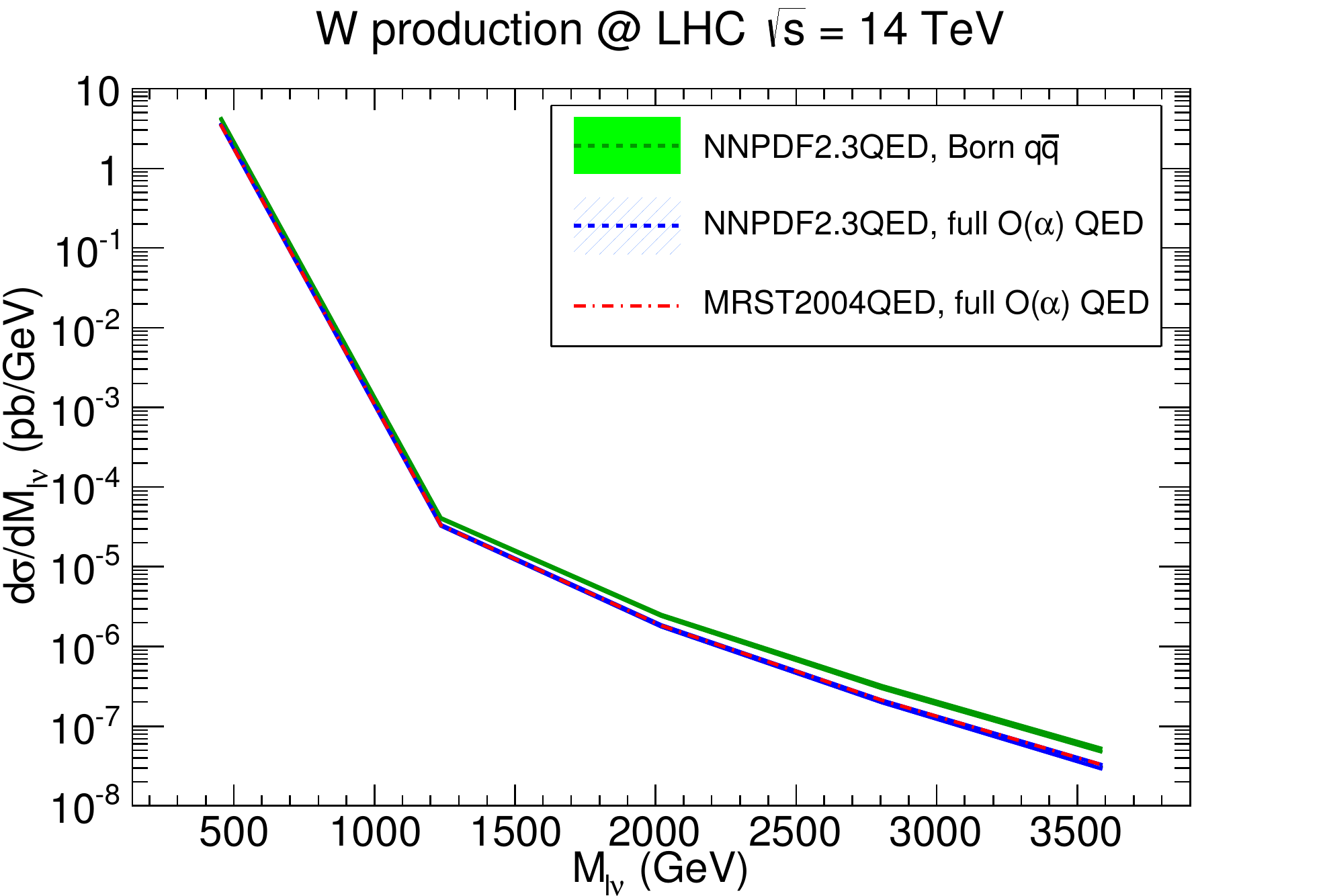}\includegraphics[scale=0.34]{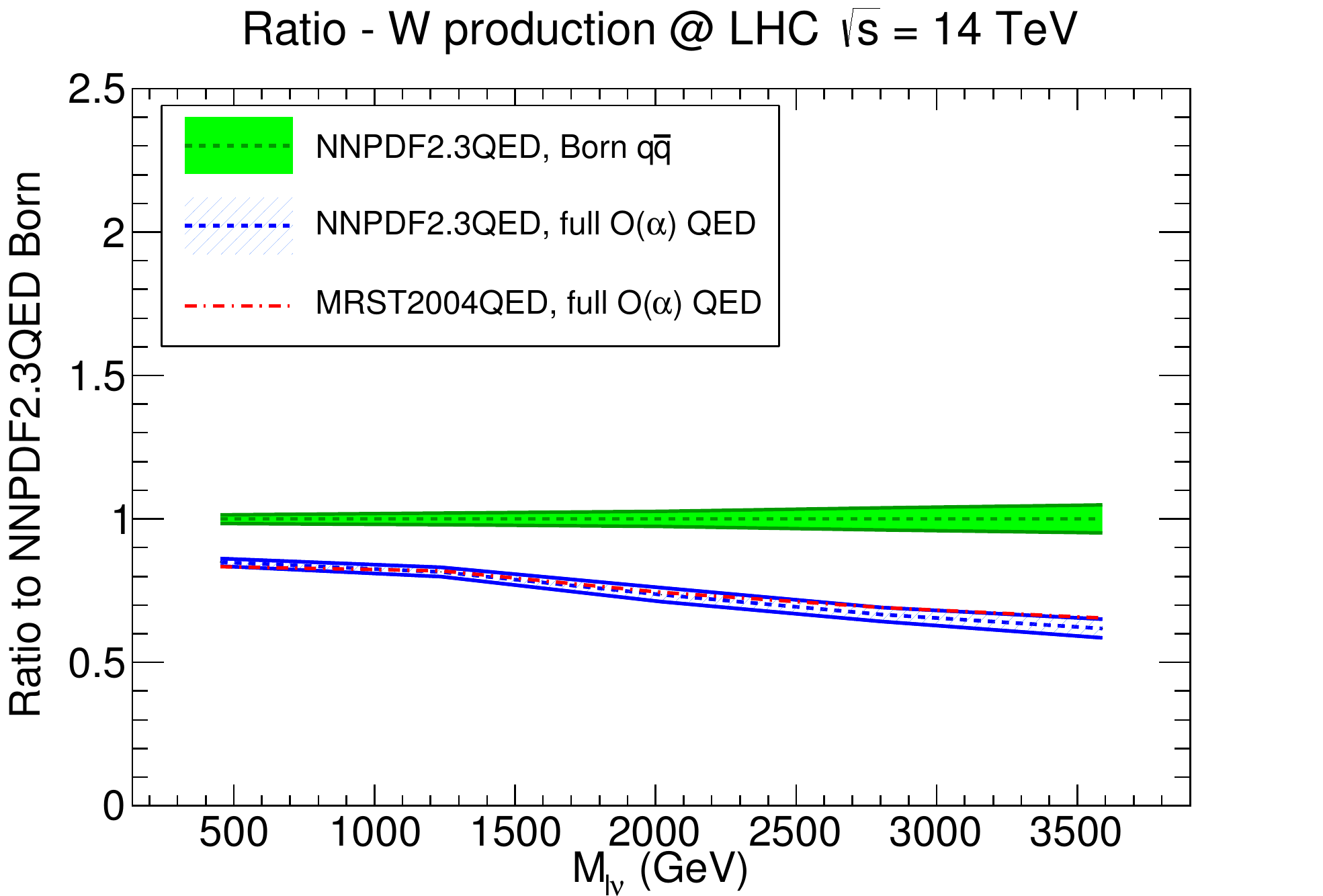}
\par\end{centering}
\caption{\label{fig:Wproduction} Same as Fig.~\ref{fig:Zproduction}
  but for high-mass charged-current production.}
\end{figure}

We have thus computed the theoretical predictions for high mass
off-shell $W$ and $Z$ production using NNPDF2.3QED. We have calculated
separately the $q\bar{q}$ initiated Born contributions, the Born term
supplemented by photon-initiated processes, and the full set of
$\mathcal{O}\left( \alpha \right)$ QED corrections, all determined
with \texttt{HORACE} (hence using LO QCD theory) and the various
electroweak scheme choices discussed in Sect.~\ref{sec:nnpdf23qed}.
We have used the following kinematical cuts, roughly corresponding to
those used in the ATLAS and CMS searches
\begin{eqnarray}
 p_t^l \ge 25~\textrm{GeV} \, , \quad
| \eta^{\gamma} |\le 2.4 \, ,
\end{eqnarray}
and we have generated enough statistics to properly populate the
highest mass bins and reduce the impact of statistical fluctuations.
Results are displayed in Fig.~\ref{fig:Zproduction}, for the
neutral-current and in Fig.~\ref{fig:Wproduction} for charged-current
dilepton production respectively. %
They are provided for LHC~8 TeV and LHC~14 TeV, shown both in an
absolute scale and as a ratio to the central value of the Born
$q\bar{q}$ cross-section from NNPDF2.3QED, using the NLO set.

The contribution from the photon-induced diagrams is generally not
negligible. Especially in the neutral current case, in which the
photon-induced contribution starts at Born level, the uncertainty
induced by the QED corrections in the large invariant mass region is
substantial, because the LHC data we used to constrain the photon PDF
(recall in particular Tab.~\ref{tab:expdata} and
Fig.~\ref{fig:correlations}) have little effect there: the uncertainty
is of order 20\% for $M_{ll} \sim$ 1 TeV at LHC 8 TeV, and it reaches
the 50\% level for $M_{ll} \sim$ 2 TeV.
Of course, for a given value of $M_{ll}$, the photon-induced
uncertainties decrease when going to 14 TeV, since smaller values of
$x$ are probed, closer to the region of the data used for the current
PDF determination.

Currently, the uncertainty on QED corrections is typically estimated
by varying the photon PDF between its MRST2004QED value and zero. Our
results suggest that this might underestimate the size of the
photon-induced contribution; it certainly does underestimate the
uncertainty related to our current knowledge of it. This follows
directly from the behavior of the luminosities of
Fig.~\ref{fig:lumimrst}.  In order to obtain more reliable exclusion
limits for $Z'$ and $W'$ at the LHC, a more accurate determination of
the photon PDF at large $x$ might be necessary. This could come from
the inclusion in the global PDF fit of new observables that are
particularly sensitive to the photon PDF, such as $W$ pair production,
as we now discuss.

\subsection{$W$ pair production at the LHC}

\begin{figure}
  \centering
  \includegraphics[scale=0.4]{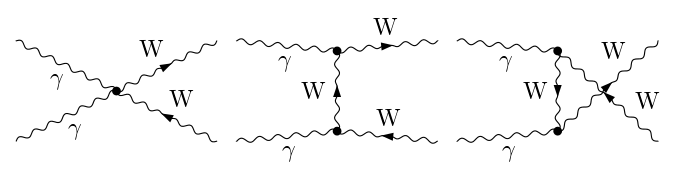}
  \caption{\label{WWphotons} Tree-level diagrams for the LO processes
    $\gamma\gamma \to \mathrm{W}^-\mathrm{W}^+$, from
    Ref.~\cite{Bierweiler:2012kw}.}
\end{figure}

The production of pairs of electroweak gauge bosons is important,
specifically for the determination of triple and quartic gauge boson
couplings~\cite{Chatrchyan:2013oev,Chatrchyan:2011tz,ATLAS:2012mec},
and it is a significant background to
searches~\cite{Chatrchyan:2012ypy,Chatrchyan:2012rva,Chatrchyan:2012kk,Aad:2013wxa,Aad:2012nev}
since several extensions to the Standard Model including warped extra
dimensions~\cite{Randall:1999vf} and dynamical electroweak
symmetry-breaking models~\cite{Andersen:2011yj,Eichten:2007sx} predict
the existence of heavy resonances decaying to pairs of electroweak
gauge bosons.

We consider now specifically the production of $W$ boson pairs for
large values of the invariant mass $M_{WW}$ and moderate values of the
transverse momentum $p_{T,W}$.  Photon-induced contributions to this
process start at Born level (see Fig.~\ref{WWphotons}), and their
contribution can be substantial, in particular at large values of
$M_{WW}$.  NLO QCD corrections, as well as the formally NNLO but
numerically significant gluon-gluon initiated contributions, are
known, and available in public codes such as
\texttt{MCFM}~\cite{Campbell:2011bn}.
Fixed-order electroweak corrections to $W$ pair production are also known~\cite{Bierweiler:2012kw},
as well as the resummation of large Sudakov electroweak logarithms
at NNLL accuracy~\cite{Kuhn:2011mh}; a recent review of theoretical
calculations is in  Ref.~\cite{Baglio:2013toa}.
%
 \begin{figure}
\begin{centering}
\includegraphics[scale=0.34]{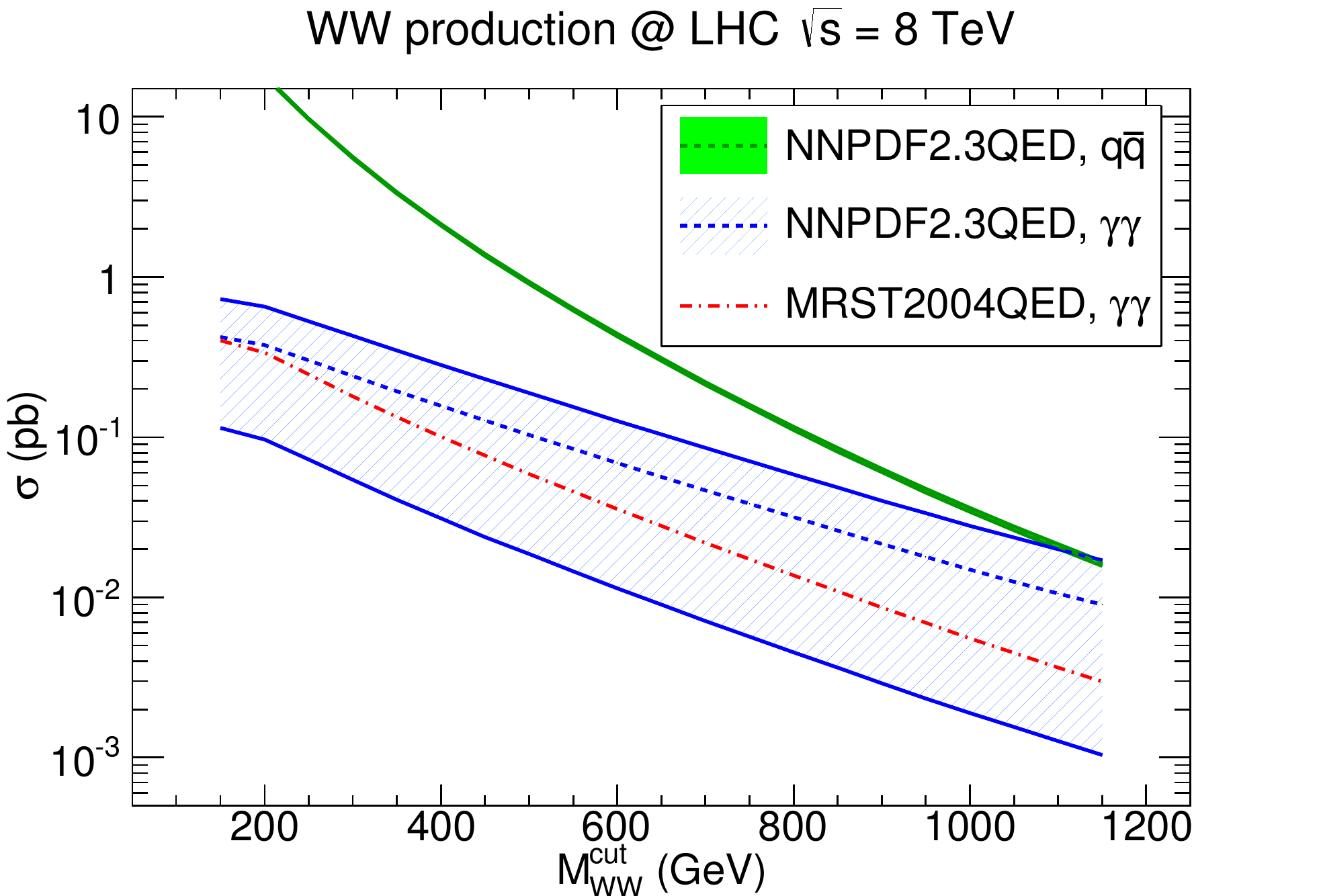}\includegraphics[scale=0.34]{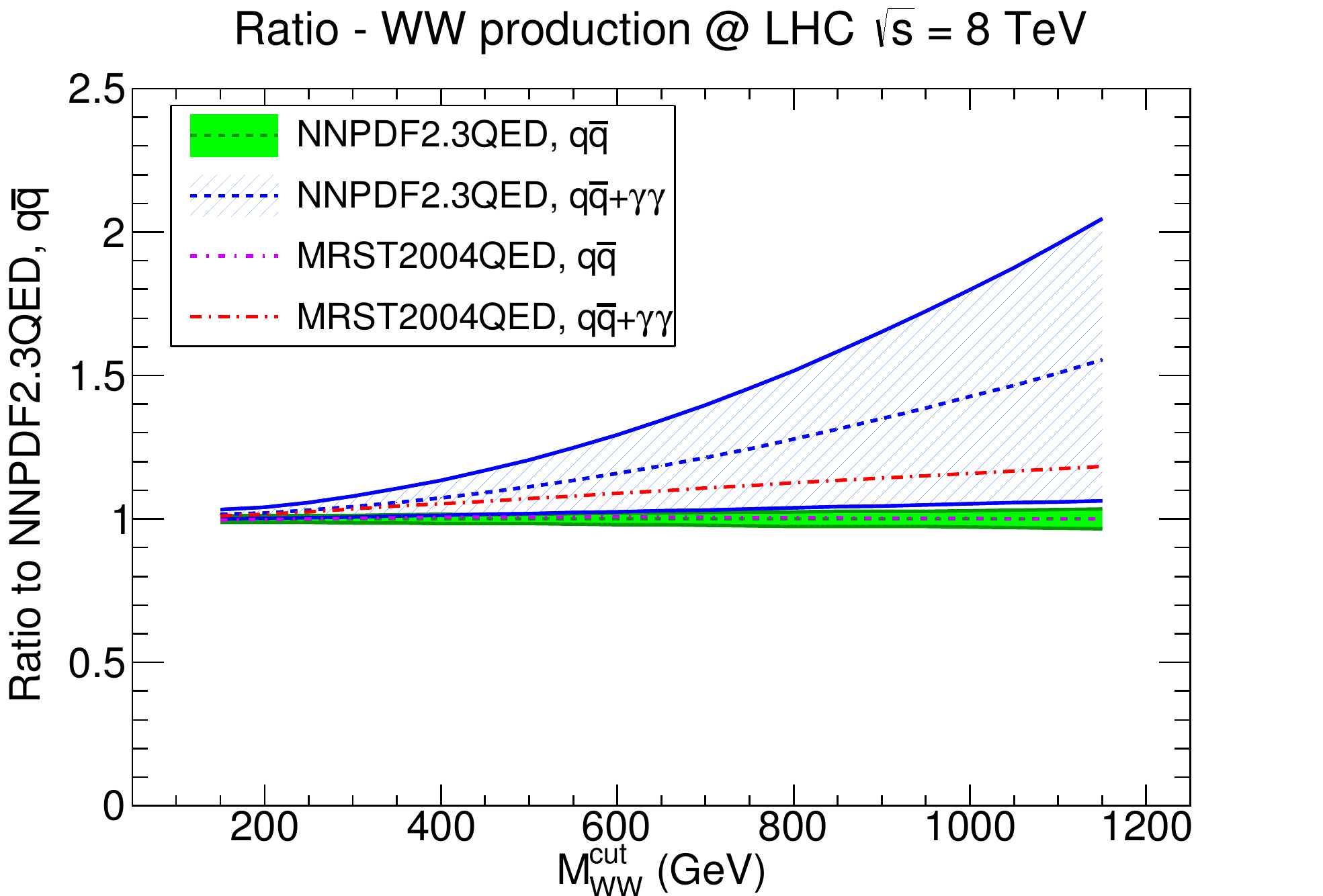}\\
\includegraphics[scale=0.34]{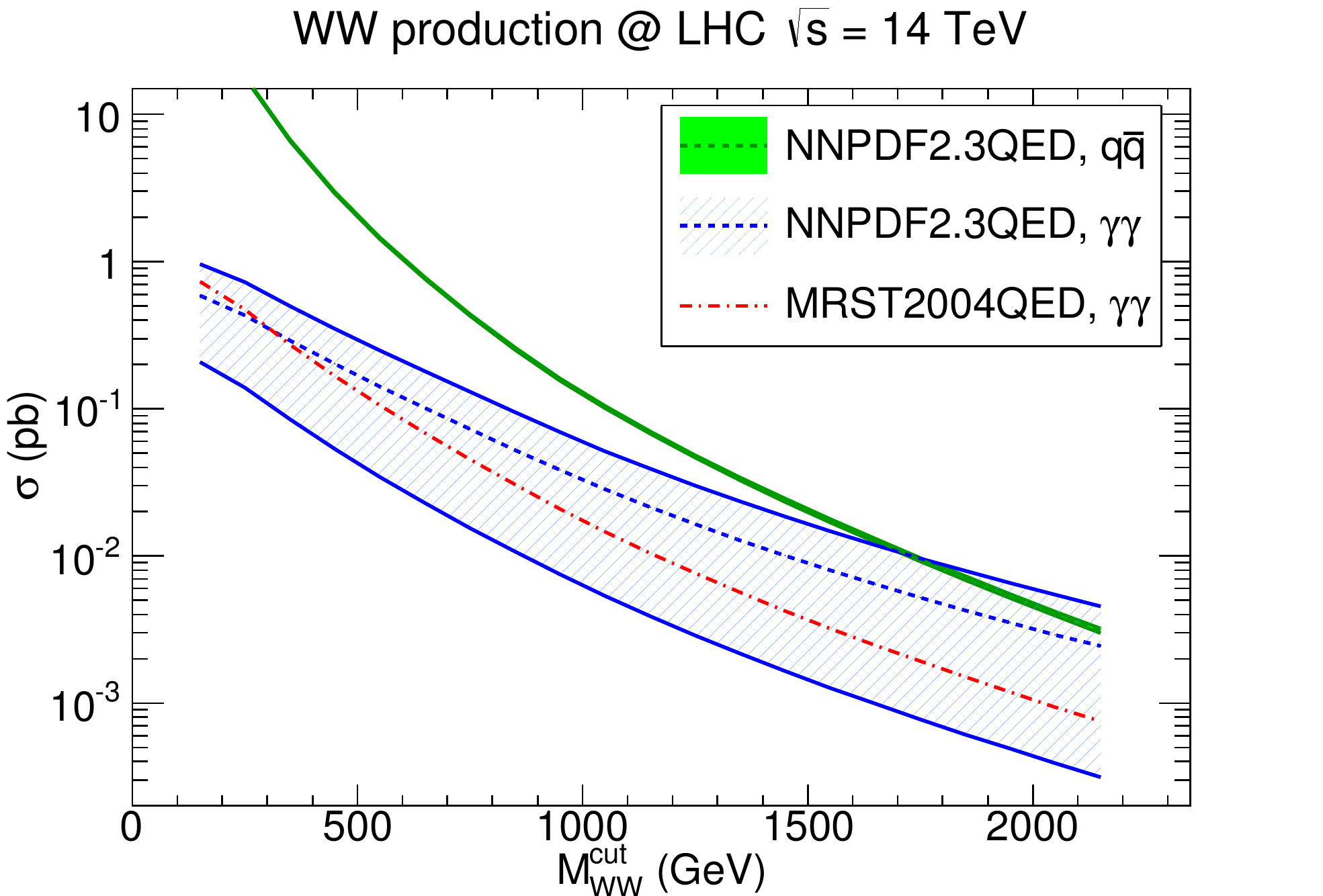}\includegraphics[scale=0.34]{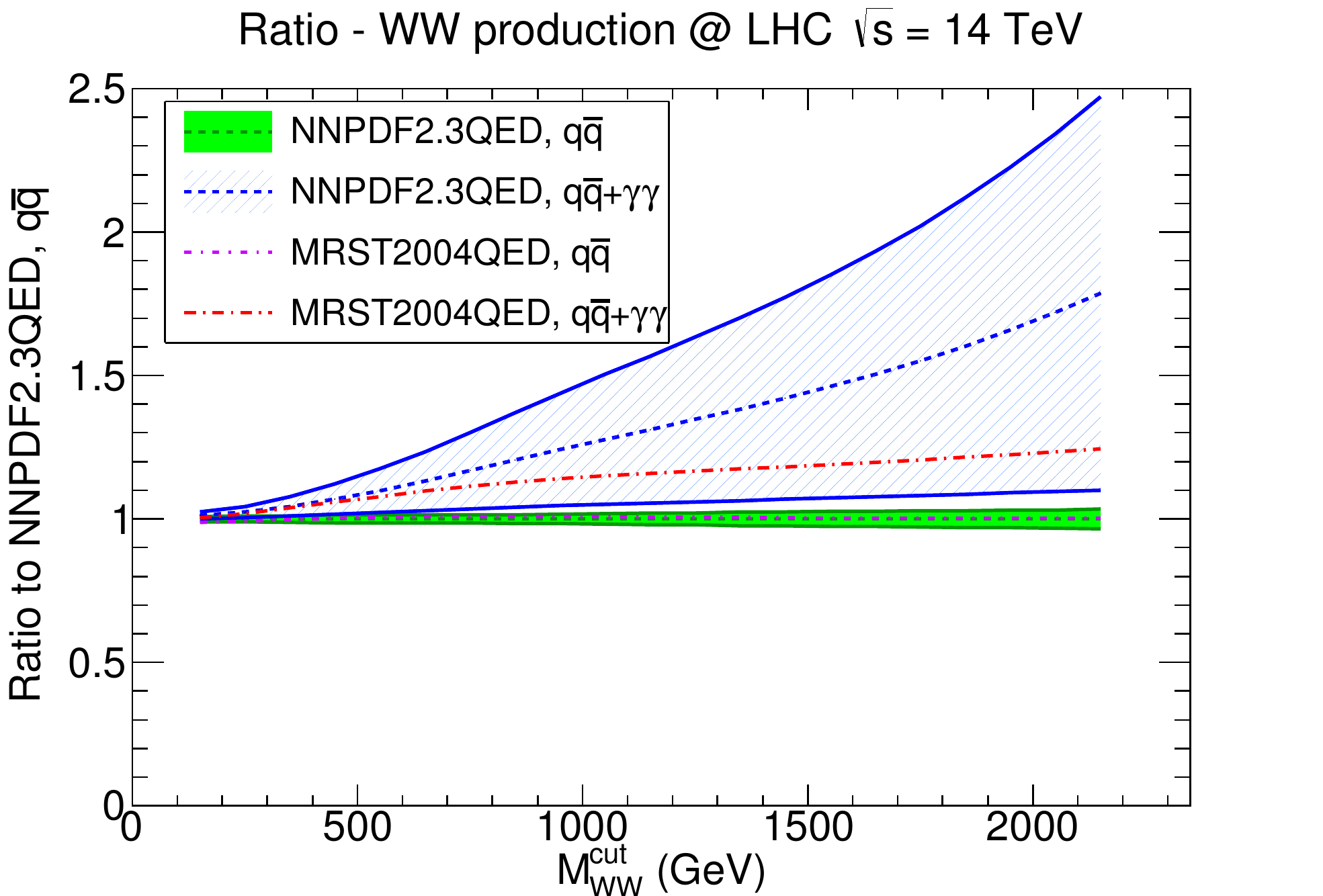}
\par\end{centering}
\caption{\label{fig:WWproduction} 
Photon-induced and quark-induced Born-level contributions to the
production of a $W$ pair with mass $M_{WW}>M_{WW}^\textrm{cut}$ plotted
as a function of  $M_{WW}^\textrm{cut}$ 
at the LHC 8 TeV (top) and LHC 14 TeV (bottom), computed
with the code of Ref.~\cite{Bierweiler:2012kw} and NNPDF2.3QED  NLO and
MRST2004QED PDFs.}
\end{figure}

To estimate the impact of photon-induced contributions to $WW$
production, predictions have been computed with either MRST2004QED or
NNPDF2.3QED NLO PDFs. They have been provided by the authors of
Ref.~\cite{Bierweiler:2012kw} using the code and settings of
Ref.~\cite{Bierweiler:2012kw}.  In particular, the kinematical cuts in
the transverse momentum and rapidity of the $W$ bosons are
\begin{eqnarray}
\label{eq:WWcuts}
p_{T,W} \ge 15~\textrm{GeV} \, ,\quad |y_{W}|\le 2.5 \, .
\end{eqnarray}
%
 \begin{figure}[ht]
\begin{centering}
\includegraphics[scale=0.34]{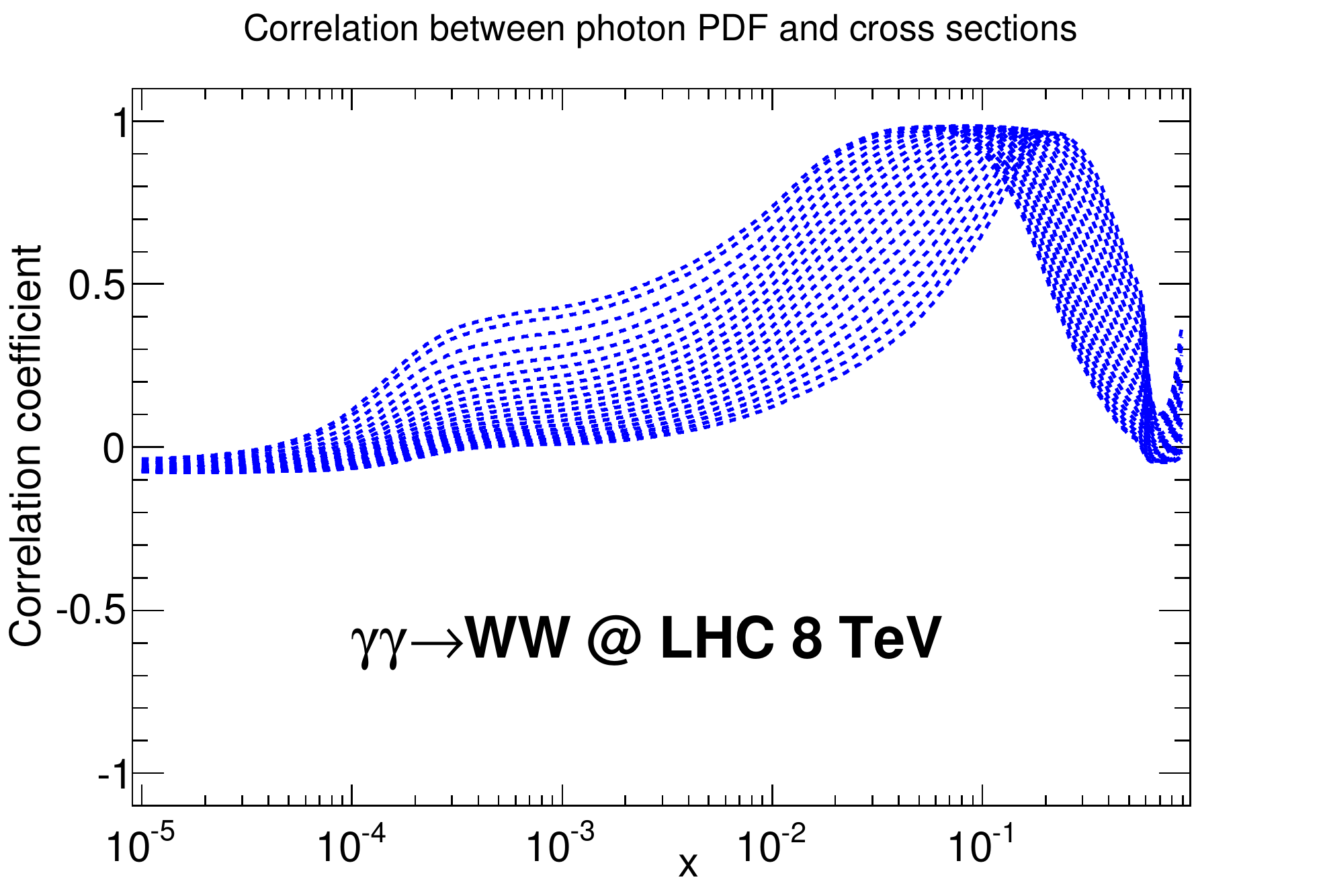}\includegraphics[scale=0.34]{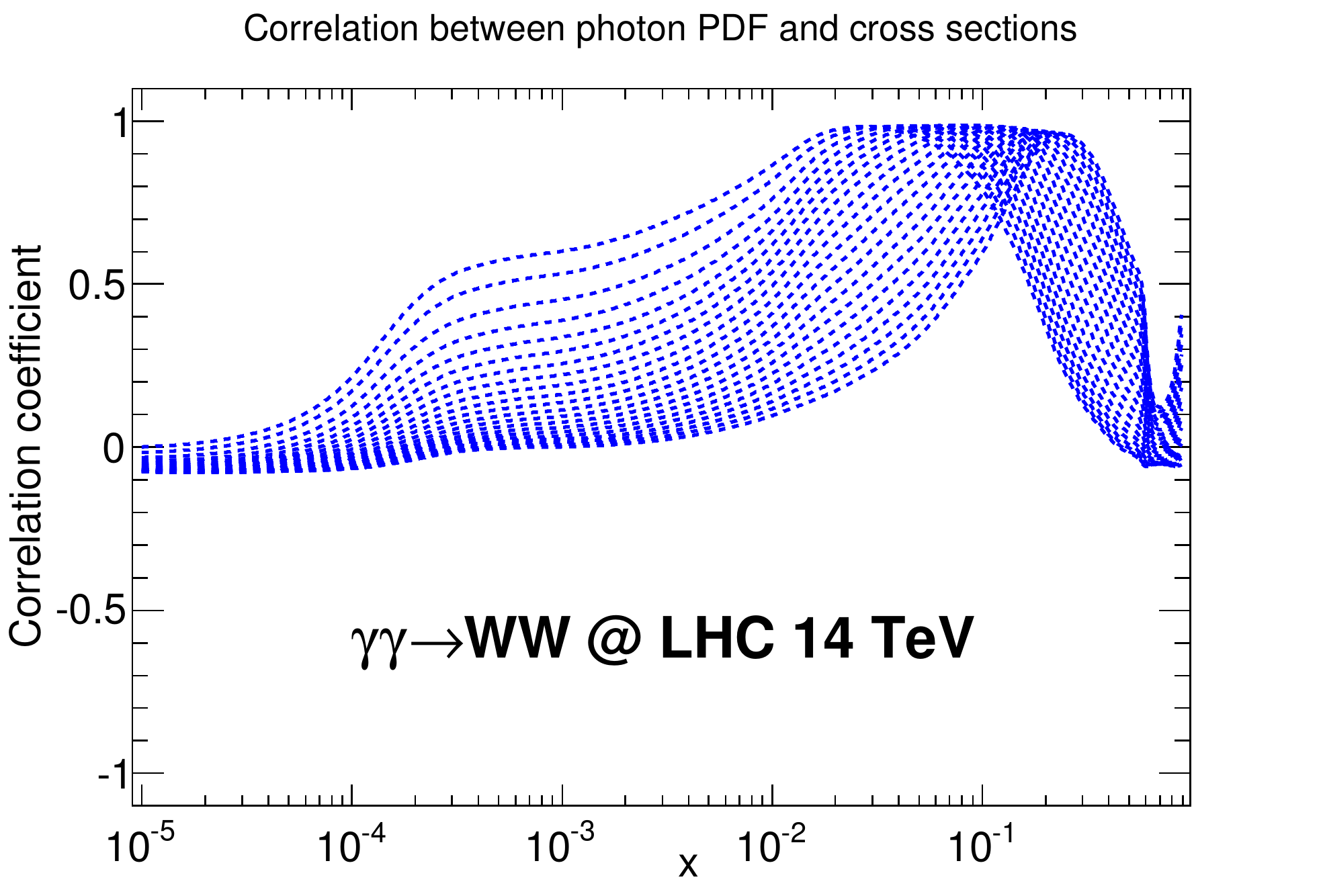}
\par\end{centering}
\caption{\label{fig:WWproductionc} Correlations between the $W$ pair
  production cross-section of Fig.~\ref{fig:WWproduction} and the
  photon PDF from the NNPDF2.3QED NLO set for $Q=10^4$~GeV$^2$.
  Each curve corresponds to one of 40 equally
  spaced bins in which the $M_{WW}^\textrm{cut}$ range of
  Fig.~\ref{fig:WWproduction} has been subdivided.
}
\end{figure}

In Fig.~\ref{fig:WWproduction} the cross-section for production of a
$W$ pair of mass $M_{WW}>M_{WW}^\textrm{cut}$ is displayed as a
function of $M_{WW}^\textrm{cut}$, at the LHC 8 and 14 TeV. The Born
$q\bar{q}$ and $\gamma\gamma$ initiated contributions are shown
(computed using LO QCD), while we refer to
Ref.~\cite{Bierweiler:2012kw} for the full
$\mathcal{O}\left(\alpha\right)$ electroweak corrections, which depend
only weakly on the photon PDF.  It is clear that for large enough
values of the mass of the pair the photon-induced contribution becomes
increasingly important. Again, the relative size of the results
obtained using NNPDF2.3QED or MRST2004QED PDFs can be inferred from
the behavior of the luminosities shown in Fig.~\ref{fig:lumimrst}.

As in the case of Fig.~\ref{fig:Zproduction}, the large uncertainties
found for large values of $M_{WW}^\textrm{cut}$ reflect the lack of
knowledge on the photon PDF at large $x\gtrsim0.1$.  Indeed, in
Fig.~\ref{fig:WWproductionc} we display the correlation between the
cross-section of Fig.~\ref{fig:WWproduction} and the photon PDF at
$Q^2=10^4$~GeV$^2$ as a function of $x$, obtained subdividing the
range of $M_{WW}^\textrm{cut}$ of Fig.~\ref{fig:WWproduction} into 40
bins of equal width, and then computing the correlation for each
bin. It is clear that this process is sensitive to the photon PDF at
large $x$, where the data of Tab.~\ref{tab:expdata} provide little or
no constraint (recall Fig.~\ref{fig:correlations}). Hence, predictions
for $W$ pair production obtained using MRST2004QED or NNPDF2.3QED
should be taken with care: NNPDF2.3QED provides a more conservative
estimate of the uncertainties involved, but perhaps overestimates the
range of reasonable photon PDF shapes.  However, future measurements
of this process could be used to pin down the photon PDF at large $x$,
and thus in turn improve the accuracy of the prediction for very high
mass Drell-Yan production discussed in Sect.~\ref{sec:wzprimesearch}
and Fig.~\ref{fig:Zproduction}, and conversely. Of course, in using
either, or both of these channels for new physics searches, care
should be taken that the sought-for new physics effects are not being
hidden in the PDFs themselves, which could be done by introducing
suitable kinematic cuts.

\subsection{Disentangling electroweak effects in $Z$-boson production }

In this section we estimate and compare to the PDF uncertainties the
contributions to the invariant mass of the Drell-Yan $Z$-boson
production due to electroweak corrections and the photon-induced
channel, by considering the low-mass region, which is below the $Z$
peak resonance and the high-mass tail.

In contrast to what was shown in Ref.~\cite{Boughezal:2013cwa} where
predictions were computed with \texttt{FEWZ}, here we propose to
combine two distinct parton level public codes:
\texttt{DYNNLO}~\cite{Catani:2007vq} for the NLO QCD prediction and
\texttt{HORACE}~\cite{CarloniCalame:2007cd} which provides the exact
$\mathcal{O}(\alpha)$ electroweak radiative correction together with
the photon-induced channel for the $Z$ production.  The motivation for
this combination is the interest to measure the difference between
predictions with electroweak effects at NLO/NNLO QCD accuracy computed
in the improved Born approximation (IBA) instead of using electroweak
correction computed by \texttt{FEWZ} in the $G_{\mu}$ scheme. The main
difference between these choices is that effective couplings in the
IBA reabsorb higher-order electroweak corrections and therefore it
provides predictions in better agreement with experimental data.

Computations are performed exclusively with the NNPDF2.3QED set of
PDFs with $\alpha_s=0.119$, instead of using the respective LO and
NNLO sets because here we will focus only on the NLO QCD accuracy and
that is why we use a NLO set.

In the next sections, we first show the differences at Born level
between the improved Born approximation (IBA), available in
\texttt{HORACE} by default, and the $G_{\mu}$ scheme in
\texttt{DYNNLO}, then, we proceed with the construction of the full
prediction.

\subsubsection{Comparing the improved Born approximation (IBA) with the
  $G_{\mu}$ scheme}

\begin{figure}
  \begin{centering}
    \includegraphics[scale=0.34]{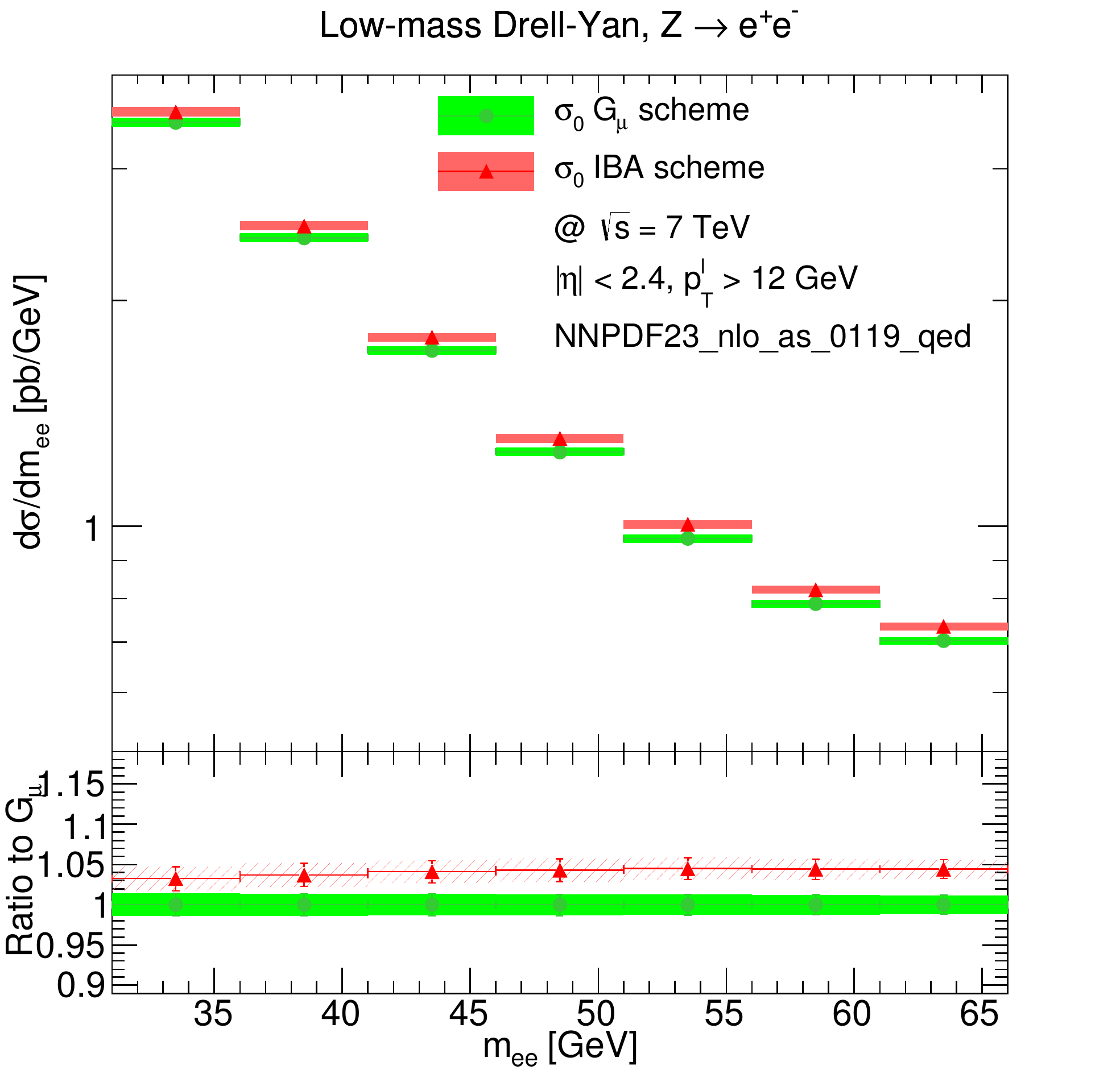}\includegraphics[scale=0.34]{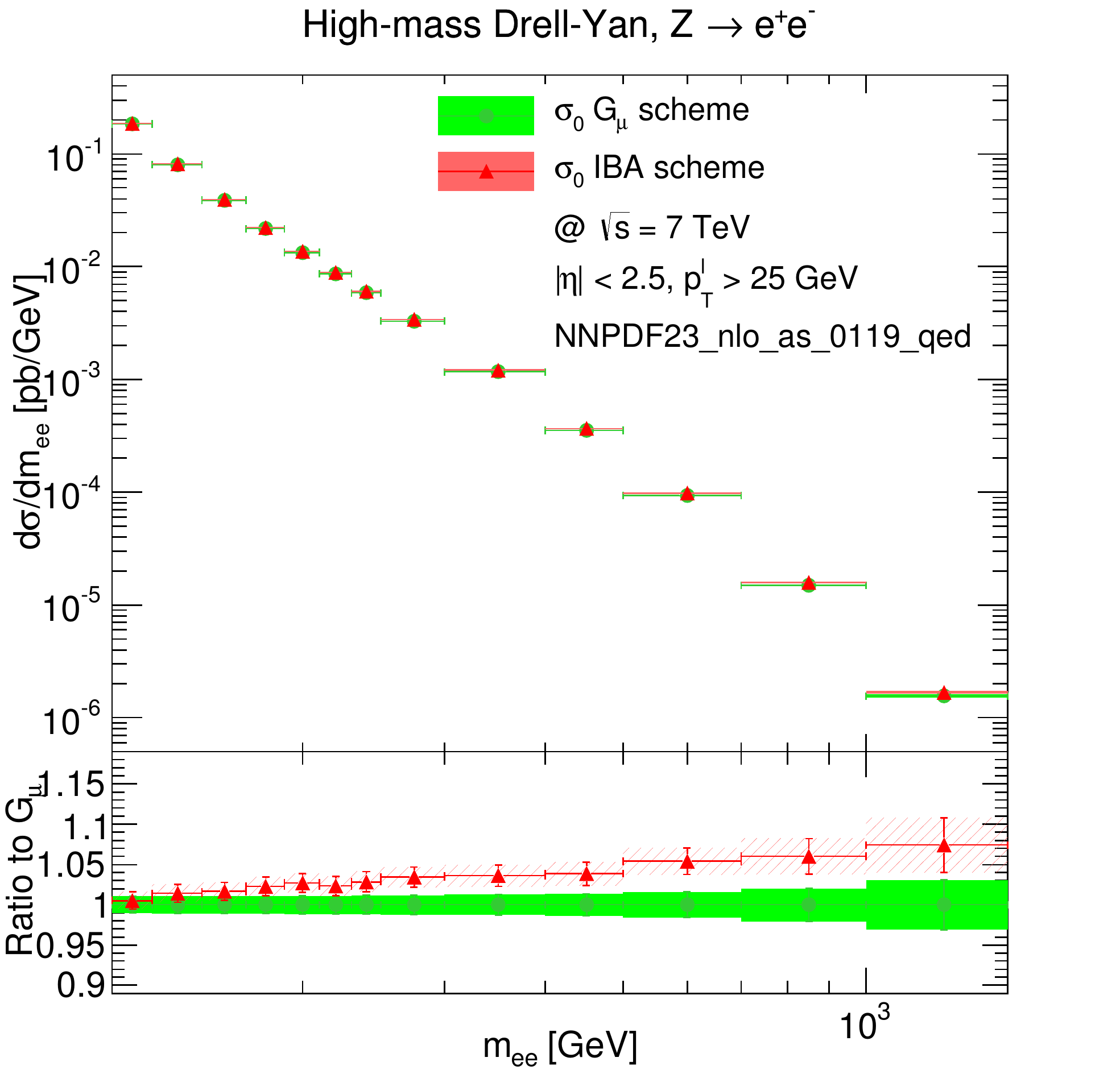}
    \par\end{centering}
    \caption{\label{fig:iba} Born level predictions and respective
      ratios for low- (left) and high-mass (right) Drell-Yan, $Z
      \rightarrow e^{+}e^{-}$ production, using the IBA and the
      $G_{\mu}$ scheme. At low-mass there is a constant gap of 3-4\%
      for all bins, while at high-mass, predictions increase
      progressively with the invariant mass, producing discrepancies
      of 7-8\% in the last bin.}
\end{figure}

In order to obtain realistic results, which are ready for comparisons
with real data, we have selected the kinematic range and cuts inspired
by recent measurements performed by the ATLAS experiment for low- and
high-mass Drell-Yan differential cross-section at $\sqrt{s}=7$ TeV
\cite{Aad:2013iua,Aad:2014qja}.

Figure~\ref{fig:iba} shows the predicted distribution at Born level
using the IBA (\texttt{HORACE}) and the $G_{\mu}$ scheme
(\texttt{DYNNLO}) at low (left plot) and high (right plot) invariant
mass regions, for the Drell-Yan process: $Z \rightarrow
e^{+}e^{-}$. Here, the goal is to measure the numerical differences
due to the choice of these methodologies.

For all distributions, the Monte Carlo uncertainty is below the
percent level.  The uncertainties shown in the figure have been
calculated as the one-$\sigma$ interval obtained after averaging over
the 100 replicas provided by this set.

In the low-mass region, we have applied kinematic cuts to the lepton
pair imposing: $p_{T}^{l} > 12$ GeV and $|\eta^{l}| < 2.4$ as in
ATLAS~\cite{Aad:2014qja}. In this region we observe an almost flat gap
of 3-4\% between the IBA and $G_{\mu}$ predictions, however in the bin
$m_{ee}=51-56$ GeV the difference is slightly higher.

On the other hand, in the high-mass region we have applied the
following kinematic cuts: $p_{T}^{l} > 25$ GeV and $|\eta^{l}| < 2.5$
as in Ref.~\cite{Aad:2013iua}. We observe a progressive increase of
the central value prediction as a function of the invariant mass,
reaching a maximum of 7-8\% at the highest bin in $m_{ee}$. This
suggests that the running of $\alpha(Q^{2})$ in the IBA can play a
crucial role when determining with accuracy the predictions in such
region.

It is important to highlight that in both cases, PDF uncertainties are
smaller than the observed differences induced by the choice of the
scheme. These results are fully consistent with the IBA implementation
discussed in Ref.~\cite{CarloniCalame:2007cd}. In the sequel we are
interested in combining electroweak effects with higher order QCD
corrections in the IBA and then compare these results to pure QCD
$G_{\mu}$ predictions.

\subsubsection{Disentangling electroweak effects}

\begin{figure}
  \begin{centering}
    \includegraphics[scale=0.34]{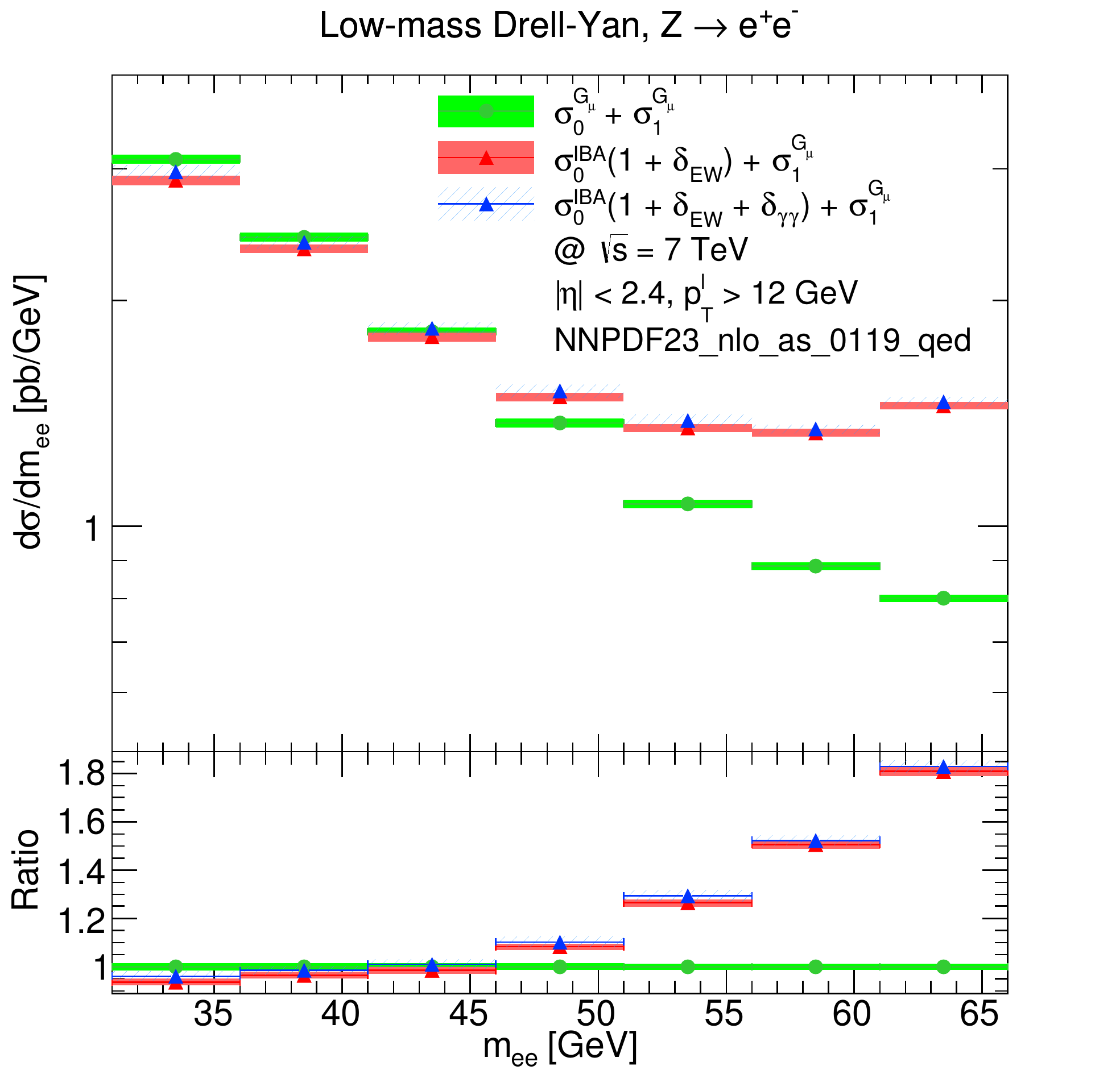}\includegraphics[scale=0.34]{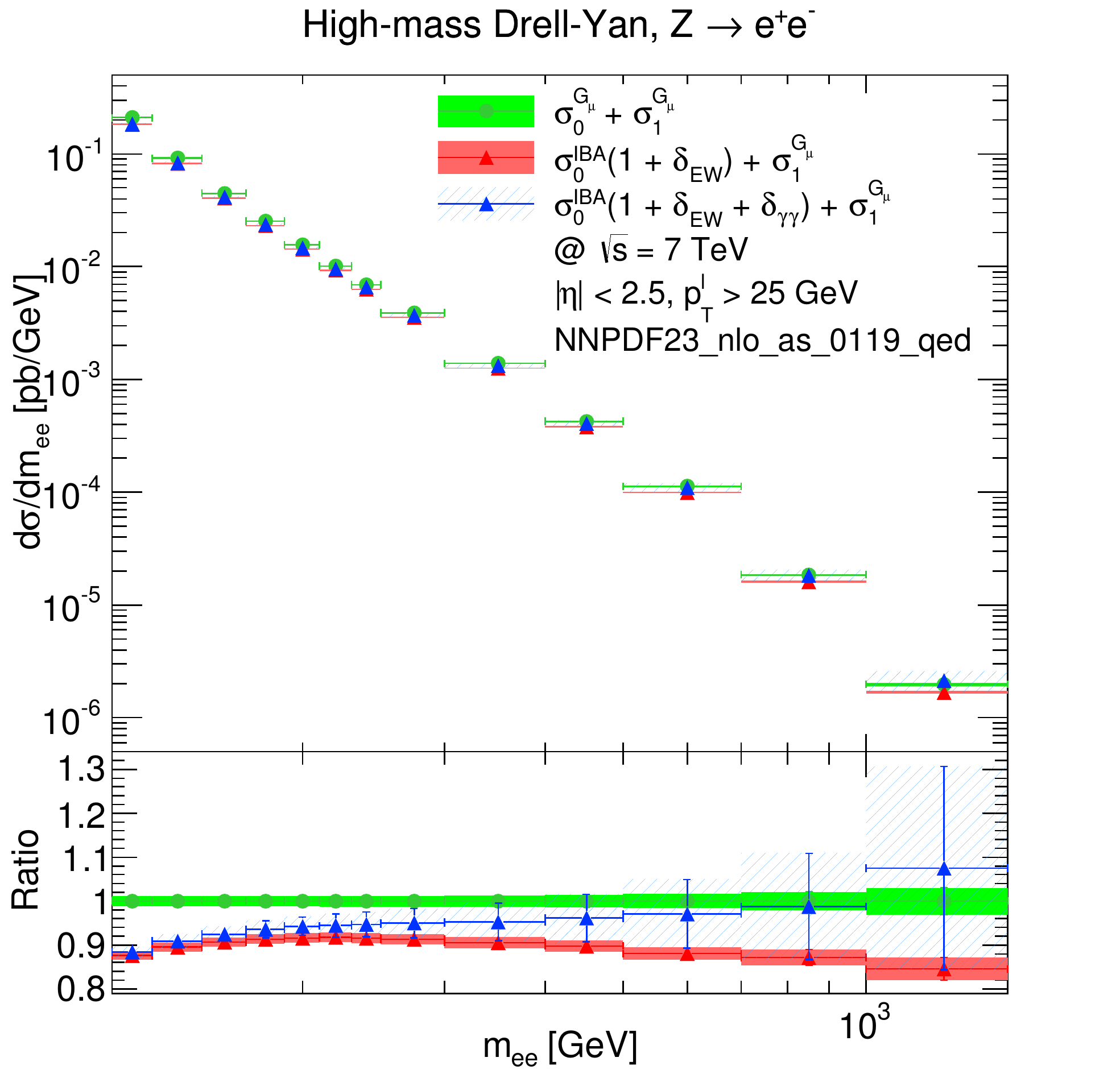}
    \par\end{centering}
    \caption{\label{fig:full} Comparison of predictions and respective
      ratios for low- (left) and high-mass (right) Drell-Yan, $Z
      \rightarrow e^{+}e^{-}$ production. We compare the NLO QCD
      prediction provided by \texttt{ DYNNLO} (green distribution) with:
      the combined prediction with $\delta_{\textrm{EW}}$ (red
      distribution) and with the $\delta_{\textrm{EW}} +
      \delta_{\gamma\gamma}$ (blue distribution).}
\end{figure}

At this point, we are interested in building a prediction based on IBA
which includes NLO QCD with $\mathcal{O}(\alpha)$ correction and the
photon-induced channel. We propose to extract the NLO correction from
\texttt{ DYNNLO} by removing its Born level, which contains the direct and
strong dependence on the $G_{\mu}$ scheme, and combine the result with
the \texttt{ HORACE} prediction. Schematically this can be achieved by
defining the quantities:
\begin{eqnarray}
  \sigma_{\textrm{\texttt{ DYNNLO}}} = \sigma^{G_{\mu}}_{0} +
  \sigma^{G_{\mu}}_{1},\\ \sigma_{\textrm{\texttt{HORACE}}} = \sigma^{\textrm{IBA}}_{0}
  (1+\delta_{\textrm{EW}}+\delta_{\gamma\gamma}),
\end{eqnarray} 
where $\sigma^{\textrm{IBA}}_{0}$ and $\sigma^{G_{\mu}}_{0}$ are the Born
levels presented in Figure~\ref{fig:iba}, $\sigma_{1}^{G_{\mu}}$ the
NLO QCD, $\delta_{\textrm{EW}}$ the $\mathcal{O}(\alpha)$ electroweak
correction and $\delta_{\gamma\gamma}$ the photon-induced
contribution.

The combination is then constructed in the following way:
\begin{eqnarray}
  \sigma_{\textrm{Total}} & = &
  \sigma_{\textrm{\texttt{ DYNNLO}}} + \sigma_{\textrm{\texttt{ HORACE}}} - \sigma^{G_{\mu}}_{0}
  \\ & = & \sigma^{\textrm{IBA}}_{0} + \sigma^{\textrm{IBA}}_{0}
  \delta_{\textrm{EW}}+\sigma^{\textrm{IBA}}_{0} \delta_{\gamma\gamma} +
  \sigma^{G_{\mu}}_{1}.\label{eq:1}
\end{eqnarray} 
where we remove the \texttt{DYNNLO} Born level while we include the
NLO QCD correction in the final prediction.

We are aware that using this methodology we improve the combination
but we do not remove entirely the pure $G_{\mu}$ dependence at
higher orders, however this is the best combination we can propose
without applying technical modifications to both codes.

In Figure~\ref{fig:full} we compare $\sigma_{\textrm{\texttt{
      DYNNLO}}}$ with $\sigma_{\textrm{Total}}$, the combination
presented in Eq.~\ref{eq:1}, with and without the
$\delta_{\gamma\gamma}$ term. For all distributions we compute the
one-$\sigma$ uncertainty except when including the photon-induced
channel where we have used the 68\% c.l. as in
Ref.~\cite{Ball:2013hta}.

In the low-mass region the inclusion of $\mathcal{O}(\alpha)$
electroweak corrections has a strong impact on the last four bins,
where differences can reach $\sim80\%$ in comparison to the pure NLO
QCD $G_{\mu}$ prediction, while the same correction for the high-mass
distribution shows a moderate impact which is below $\sim20\%$ for the
highest invariant mass bin. This behavior is expected and derives from
the shape of the $Z$-boson invariant mass: bins located in a region
lower than the $Z$ peak resonance undergoes large positive corrections
while at high invariant mass we observe a change of sign of such
corrections. It is important to highlight that modern data provided by
the LHC experiments are already corrected by final-state photon
radiation which carries a dominant fraction of the electroweak effects
shown in Figure~\ref{fig:full}.

The photon-induced contribution has a moderate impact in the low-mass
region while for high-mass it is dominant: this behavior is expected
and due to the presence of the $Z$ peak resonance where the
photon-induced channel is negligible.

Also from these plots of Figure~\ref{fig:full}, it is important to
emphasize again that modern PDF sets, as the NNPDF2.3QED, have
uncertainties which are accurate enough to appreciate the differences
due to scheme choices and electroweak effects, including the new
photon PDF, which shows a stable behavior of uncertainties in all
invariant mass regions except at very high-mass bins where
uncertainties grow, reaching more than $\sim20\%$. This situation will
be improved in future by including more relevant and precise data to
constrain the photon PDF.

\section{Photon PDF in Monte Carlo event generators}

\begin{figure}
  \centering
  \includegraphics[scale=0.34]{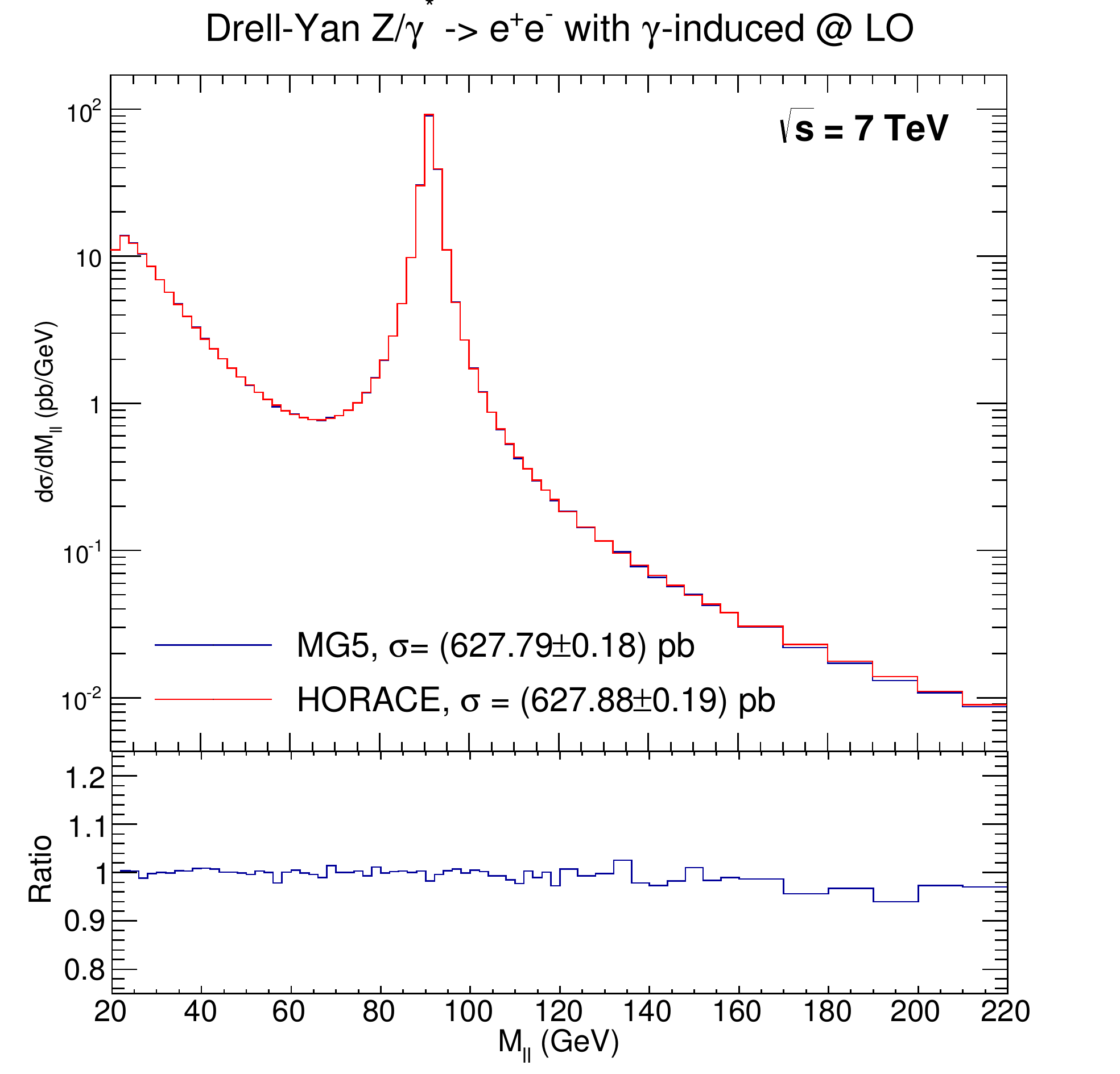}\includegraphics[scale=0.34]{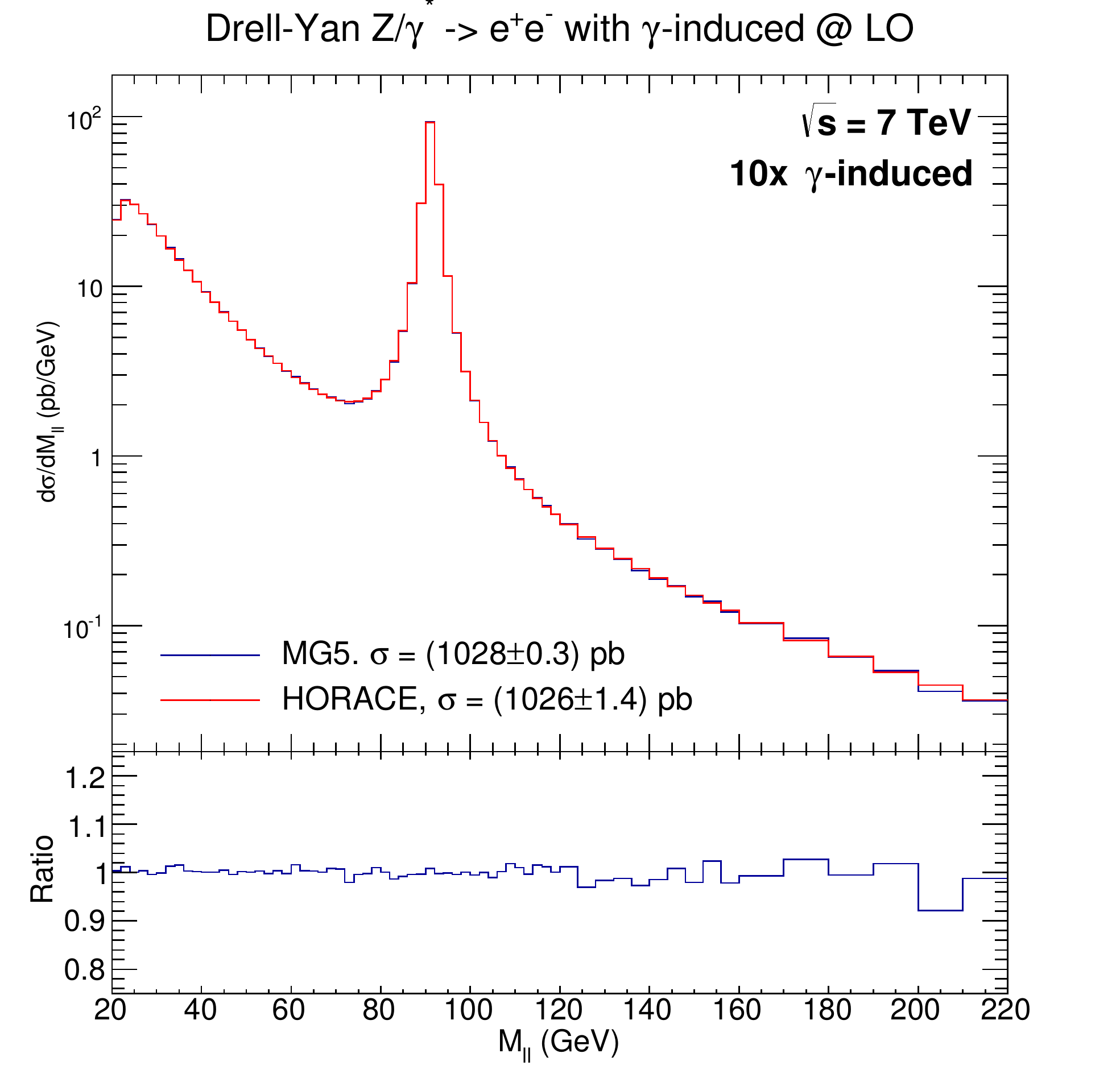}
  \caption{Comparison of \texttt{HORACE} and the
    \texttt{MadGraph5\_aMC@NLO} implementation of the invariant mass
    for Drell-Yan $Z/\gamma^*$ production at leading order with
    photon-induced contributions. The left plots shows predictions
    with the NNPDF2.3QED set of PDFs at $\sqrt{s}=7$ TeV. On the right
    plot the photon PDF contribution is multiplied by a factor 10, in
    order to enlarge the photon-induced contributions and emphasize
    the good level of agreement.}
  \label{fig:amc@nlo}
\end{figure}

After the release of the NNPDF2.3QED set of PDFs, several Monte Carlo
event generators have implemented the possibility to activate
photon-induced channels when computing predictions.
We have released a fast standalone public code for the manipulation of
these sets of PDFs independently of the \texttt{LHAPDF}
library\footnote{The public code is available at
  \url{https://github.com/scarrazza/nnpdfdriver}}. This code written
in \texttt{C++} and \texttt{Fortran77} has been adapted and implemented
in the core of the following Monte Carlo event generators:
\texttt{PYTHIA8}~\cite{Skands:2014pea},
\texttt{MadGraph5\_aMC@NLO}~\cite{Alwall:2014hca} and
\texttt{SHERPA}~\cite{Kallweit:2014xda}.

As an example, the NNPDF2.3QED set of PDFs can be used since
\texttt{PYTHIA8.1}, where, in this release, presented in
Ref.~\cite{Skands:2014pea}, we determine the updated fragmentation
parameters with this new set of PDFs. We use minimum-bias, Drell-Yan,
and underlying-event data from the LHC to constrain the
initial-state-radiation and multi-parton-interaction parameters,
combined with data from SPS and the Tevatron to constrain the energy
scaling. Several distributions show significant improvements with
respect to the current defaults, for both $ee$ and $pp$ collisions,
though we emphasize that interesting discrepancies remain in
particular for strange particles and baryons.

Another example of implementation of this set of PDFs in a MC event
generator is displayed in Figure~\ref{fig:amc@nlo}, where we show an
example of a benchmarking comparison between \texttt{HORACE} and
\texttt{MadGraph5\_aMC@NLO}. In this figure, we compute the LO
invariant mass of the $Z/\gamma^*$ Drell-Yan production at $\sqrt{s} =
7$ TeV.
On the left plot, we estimate the distribution using the NNPDF2.3QED
set of PDFs including photon-induced channels. The level of agreement
is good in the peak and in the tails regions.
However as the photon-induced contributions are small, at least two
orders of magnitude smaller than the total inclusive cross-section, on
the right plot we show the same process but now computed with the
photon PDF from NNPDF2.3QED multiplied by a factor 10.
This plot shows that the agreement is still good, confirming that the
implementation in \texttt{MadGraph5\_aMC@NLO} is correct.

This is an extremely important result, because it opens the possibilty
of implementing a fast NLO interface for computations including
electroweak corrections, and thus the photon-induced contributions
through the \texttt{aMCfast}~\cite{Bertone:2014zva} code.
With this code, we will be able to generate sets of
\texttt{APPLgrid}~\cite{Carli:2010rw} tables with weights associated
to the photon contribution, and consequently, enabling the possibility
to perform new fits of PDFs with QED corrections, improving the
determination and uncertainties of the photon PDF by including more
data in the fit and avoiding the reweighting strategy explained in
Chap.~\ref{sec:chap4}.

\section{Lepton PDFs}

In the previous chapters, we have always neglected the PDFs of charged
leptons, supposing that with the current methodology their
determination is practically impossible from a fit to the available
experimental data. In fact, the PDFs associated to $e^\pm$, $\mu^\pm$
and $\tau^\pm$ are expected to be much smaller than the photon PDF.
However, when computing electroweak corrections to some
hadron-collider processes, such as Drell-Yan, the presence of lepton
PDFs requires the inclusion of new lepton-initiated channels which
might have a non-negligible impact.

Currently, from literature we observe that only the photon content of
the proton has been determined based either on model
assumptions~\cite{Martin:2004dh}, the MRST2004QED set, or on a fit to
data~\cite{Ball:2013hta}, the NNPDF2.3QED, but no attempt to estimated
the lepton PDFs has ever been tried.

Therefore, for the conclusion of this chapter, we propose to give an
estimate on the leptonic content of the proton.
This will be achieved in the following steps: 
\begin{itemize}

\item the implementation of the lepton PDF DGLAP evolution at LO in
  QED in the so called VFN scheme in
  \texttt{APFEL}~\cite{Bertone:2013vaa};

\item the determination of a guess for the lepton PDFs at the initial
  scale $Q_0$, based on the assumption that leptons are generated by photon
  splitting.

\end{itemize}

In the next sections, we discuss the details of both steps.

\subsection{DGLAP Equation in the Presence of Photon and Leptons}
\label{sec:DGLAPwithLeptons}

Following the methodology presented in Chap.~\ref{sec:chap2}, we
extend the DGLAP equations to include the evolution of photon and
lepton PDFs at LO in QED.
At LO in QED, leptons couple directly only to the photon. 
However, since the photon couples to quarks and so, indirectly to
gluon, the lepton PDFs evolution depend on the evolution of all the
other partons. 
Following the notation of Sect.~\ref{sec:apfelevolution}, where QCD
and QED evolutions are treated independently, the inclusion of leptons
does not imply any change to the QCD evolution, while QED evolution
equations are modified with the inclusion of the leptonic terms in the
photon evolution, together with the addition of lepton equations,
namely
\begin{equation}\label{QED_DGLAP}
\begin{array}{rcl}
  \displaystyle \nu^{2}\frac{\partial \gamma}{\partial \nu^{2}}
  &=& \displaystyle \frac{\alpha(\nu)}{4\pi} \left[\left(\sum_{i}N_c e_{i}^{2}\right)
    P_{\gamma\gamma}^{(0)}\otimes\gamma+\sum_ie_{i}^{2}P_{\gamma
  q}^{(0)}\otimes (q_{i}+\bar{q}_{i}) +\sum_j P_{\gamma
  \ell}^{(0)}\otimes (\ell_{j}+\bar{\ell}_{j})
\right]\,,\\
\\
\displaystyle \nu^{2}\frac{\partial \ell_{j}}{\partial
   \nu^{2}} &=&\displaystyle \frac{\alpha(\nu)}{4\pi}
 \left[ P_{\ell\gamma}^{(0)}\otimes \gamma+
 P_{\ell\ell}^{(0)}\otimes \ell_j\right]\,,\\
\\
\displaystyle \nu^{2}\frac{\partial \bar{\ell}_{j}}{\partial
   \nu^{2}} &=&\displaystyle \frac{\alpha(\nu)}{4\pi}
 \left[ P_{\ell\gamma}^{(0)}\otimes \gamma+
 P_{\ell\ell}^{(0)}\otimes \bar{\ell}_j\right]\,,\\
\end{array}
\end{equation}
where $\gamma$, $q_i$, $\bar{q}_i$, $\ell_j$ and $\bar{\ell}_j$ are
respectively the PDFs of the photon, the $i$-th quark, the $i$-th
antiquark, $j$-th lepton and $j$-th anti-lepton, $e_i$ the $i$-th
quark electric charge, $N_c=3$ the number of colors and $\alpha$
the running fine structure constant. Note also that the indices $i$
and $j$ in the first line of eq.~(\ref{QED_DGLAP}) run over the $n_f$
and $n_\ell$ number of active quarks and leptons at the scale $\nu$,
respectively. 
Note that the leading-order QED splitting functions satisfy the
following identities: $P^{(0)}_{q\gamma}=P^{(0)}_{\ell\gamma}$,
$P^{(0)}_{\gamma q}=P^{(0)}_{\gamma\ell}$ and
$P^{(0)}_{qq}=P^{(0)}_{\ell\ell}$.

Combining the system of differential equations in
eq.~(\ref{QED_DGLAP}) with the pure-QCD DGLAP equations that govern
the evolution of gluon and quarks, we obtain the full QCD$\otimes$QED
evolution in the presence of photon and leptons. We have implemented
the solution of this system of differential equations in
\texttt{APFEL} version 2.4.0.

Here, the fine-structure constant $\alpha$ runs with the
renormalization scale that we take to be equal to the factorization
scale $\mu_F$. Consistently with evolution of PDFs, we consider the
leading order running by solving the RG equation
\begin{equation}
\nu^2 \frac{d\alpha}{d\nu^2} = \beta_{\textrm{QED}}^{(0)} \alpha^2(\nu)
\end{equation}
where
\begin{equation}
\beta_{\textrm{QED}}^{(0)} = \frac{8}{12\pi}\left(N_c\sum_{i=1}^{n_f}e_i^2+n_\ell\right)\,,
\end{equation}
with $N_c=3$ the number of colors, $e_i^2$ the electric charge of the
$i$-th quark, $n_f$ the number of light quarks and $n_\ell$ the number
of light leptons. Finally, as a boundary condition for the evolution
we take $\alpha^{-1}(m_\tau) = 133.4$ as in Chap.~\ref{sec:chap2}.

\subsection{Modeling the Lepton PDFs}

The following step consists in the determination of the boundary
condition for the initial scale PDFs. In this case, we can use the
NNPDF2.3QED and MRST2004QED sets for the boundary conditions of
quarks, gluons and photon.
However, lepton PDFs cannot be extracted from data by means of a
fit. The main reason for that is the fact that lepton PDFs are
expected to be very small as compared to the quark and gluon PDFs, and
even much smaller than the photon PDF. In particular, assuming a small
intrinsic leptonic component in the proton, the lepton PDFs are
expected to be of the order of $\alpha$ times the photon PDF, where
$\alpha \sim 10^{-2}$ is the fine structure constant. As a
consequence, being the photon already very small as compared to quark
and gluon PDFs, the contribution of leptons is expected to be
extremely small and this clearly makes a reliable determination of the
lepton PDFs from experimental data extremely hard.

As an alternative to the fit, we can try to guess the functional form
of the PDFs of leptons just by assuming that light leptons,
\textit{i.e.}~electrons and muons, are generated by photon
splitting. At leading-logarithmic accuracy we can then guess their
distributions at the initial scale $Q_0$ as:
\begin{equation}\label{eq:ansatz}
  \ell_\beta(x,Q_0) =\overline{\ell}_\beta(x,Q_0) =
  \frac{\alpha(Q_0)}{4\pi} \ln\left(\frac{Q_0^2}{m_\beta^2}\right)
  \int_x^1\frac{dy}{y} P_{\ell\gamma}^{(0)}\left(\frac{x}{y}\right)
  \gamma(y,Q_0)\,,
\end{equation}
with $\beta=e^{\pm},\mu^{\pm}$. For the light lepton masses, we take $m_{e^{\pm}} =
0.510998928$ MeV and $m_{\mu^{\pm}} = 105.6583715$ MeV, as quoted in the
PDG~\cite{Agashe:2014kda}.

As far as the $\tau^{\pm}$ PDFs are concerned, since $m_{\tau^{\pm}} =
1.777$ GeV $\gtrsim Q_0$, we assume that they are dynamically
generated at the threshold according to the usual scheme matching of
the VFN scheme.

\subsection{Preliminary results}

\begin{table}
  \centering
  \footnotesize
\begin{tabular}{|c|l|c|c|c|c|c|}
  \hline 
  ID & PDF Set & Ref. & QCD & QED & Photon PDF & Lepton PDFs\tabularnewline
  \hline 
  \hline 
  A1 & \texttt{apfel\_nn23nlo0118\_lept0} & \cite{Ball:2012cx} & NLO & LO & $\gamma(x,Q_0) = 0$ & Eq.~(\ref{eq:ansatz0}) \tabularnewline
  \hline 
  A2 & \texttt{apfel\_nn23nnlo0118\_lept0} & \cite{Ball:2012cx} & NNLO  & LO  & $\gamma(x,Q_0) = 0$ &
  Eq.~(\ref{eq:ansatz0}) \tabularnewline
  \hline 
  B1 & \texttt{apfel\_nn23qedlo0118\_lept0} & \cite{Carrazza:2013axa} & LO & LO &
  Internal & Eq.~(\ref{eq:ansatz0}) \tabularnewline
  \hline 
  B2 & \texttt{apfel\_nn23qednlo0118\_lept0} & \cite{Ball:2013hta} & NLO & LO &
  Internal & Eq.~(\ref{eq:ansatz0}) \tabularnewline
  \hline 
  B3 & \texttt{apfel\_nn23qednnlo0118\_lept0} & \cite{Ball:2013hta} & NNLO & LO &
  Internal & Eq.~(\ref{eq:ansatz0}) \tabularnewline
  \hline 
  B4 & \texttt{apfel\_mrst04qed\_lept0} & \cite{Martin:2004dh} & NLO & LO & Internal
  & Eq.~(\ref{eq:ansatz0}) \tabularnewline
  \hline 
  C1 &\texttt{apfel\_nn23qedlo0118\_lept} & \cite{Carrazza:2013axa} & LO & LO &
  Internal & Eq.~(\ref{eq:ansatz}) \tabularnewline
  \hline 
  C2 &\texttt{apfel\_nn23qednlo0118\_lept} & \cite{Ball:2013hta} & NLO & LO &
  Internal & Eq.~(\ref{eq:ansatz}) \tabularnewline
  \hline 
  C3 & \texttt{apfel\_nn23qednnlo0118\_lept} & \cite{Ball:2013hta} & NNLO & LO &
  Internal & Eq.~(\ref{eq:ansatz}) \tabularnewline
  \hline 
  C4 & \texttt{apfel\_mrst04qed\_lept} & \cite{Martin:2004dh} & NLO & LO & Internal & Eq.~(\ref{eq:ansatz}) \tabularnewline
  \hline 
\end{tabular}
\caption{Summary of the sets of PDFs generated with \texttt{APFEL}
  with photons and leptons PDFs.}
\label{tab:sets}
\end{table}

In this section we discuss the results of the implementation of the
lepton PDFs evolution in \texttt{APFEL}. The main goal of this work is
to provide an estimate of the lepton PDFs. As discussed in the
previous section, the determination of lepton PDFs from a direct fit
to data is hard to achieve and thus the alternative is that of
modeling initial scale lepton PDFs based on some theoretical
assumption.

The model presented in the previous section is based on the assumption
that lepton pairs are generated from photon splitting at the
respective mass scale. At leading logarithmic accuracy, this results
in the ansatz in eq.~(\ref{eq:ansatz}) for the light lepton
PDFs. However, in order to test how sensitive the results are to the
initial scale distributions, we also consider the zero-lepton ansatz
where the lepton PDFs at the initial scale $Q_0$ are equal to zero,
that is
\begin{equation}\label{eq:ansatz0}
  \ell_\beta(x,Q_0) =\overline{\ell}_\beta(x,Q_0) = 0\,.
\end{equation}

In this context the construction of PDF sets with leptons requires a
pre-existing PDF set to which we add our model for the lepton
distributions. Of course, in order to apply the ansatz in
eq.~(\ref{eq:ansatz}) we need PDF sets that already contain a photon
PDF. Presently, there are only two sets that contain a photon PDF: the
MRST2004QED set~\cite{Martin:2004dh} and the NNPDF2.3QED
family~\cite{Ball:2013hta}, and we will use both of them to generate
lepton PDFs. On the contrary, the ansatz in eq.~(\ref{eq:ansatz0}) can
be applied to any set so that lepton and photon distributions can be
generated from any PDF set just by evolution.

In order to assess the effect of considering lepton PDFs in the DGLAP
evolution, in this work we consider three different initial scale
configurations that are also summarized in Table~\ref{tab:sets}:

\begin{itemize}

\item Sets where both photon and lepton PDFs are set to zero at the
  initial scale $Q_0$ and dynamically generated by DGLAP
  evolution. For this configuration we have constructed the sets A1
  and A2 in Table~\ref{tab:sets} based on NNPDF2.3 NLO and NNLO
  respectively.

\item Sets where the photon distribution is present in the starting
  set but lepton PDFs are set to zero at the initial scale $Q_0$
  ($i.e.$ eq.~(\ref{eq:ansatz0})) and then evolved as discussed in
  Sect.~\ref{sec:DGLAPwithLeptons}. These configurations are based on
  the NNPDF2.3QED and MRST2004QED sets of PDFs and identified by the
  indices B1, B2, B3 and B4 in Table~\ref{tab:sets}.

\item Sets of PDFs extracted from NNPDF2.3QED and MRST2004QED but
  using the ansatz in eq.~(\ref{eq:ansatz}) for the lepton PDFs (sets
  C1, C2, C3 and C4 in Table~\ref{tab:sets}).

\end{itemize}

The evolution of the PDF sets listed above is performed using
\texttt{APFEL} as discussed in Sect.~\ref{sec:DGLAPwithLeptons} and
tabulated in the \texttt{LHAPDF6} format which allows for the
inclusion of lepton PDFs in a straightforward manner. In the following
we will quantify the differences of the different configurations by
looking at PDFs, momentum fractions and luminosities.


\begin{figure}
  \centering
  \includegraphics[scale=0.45]{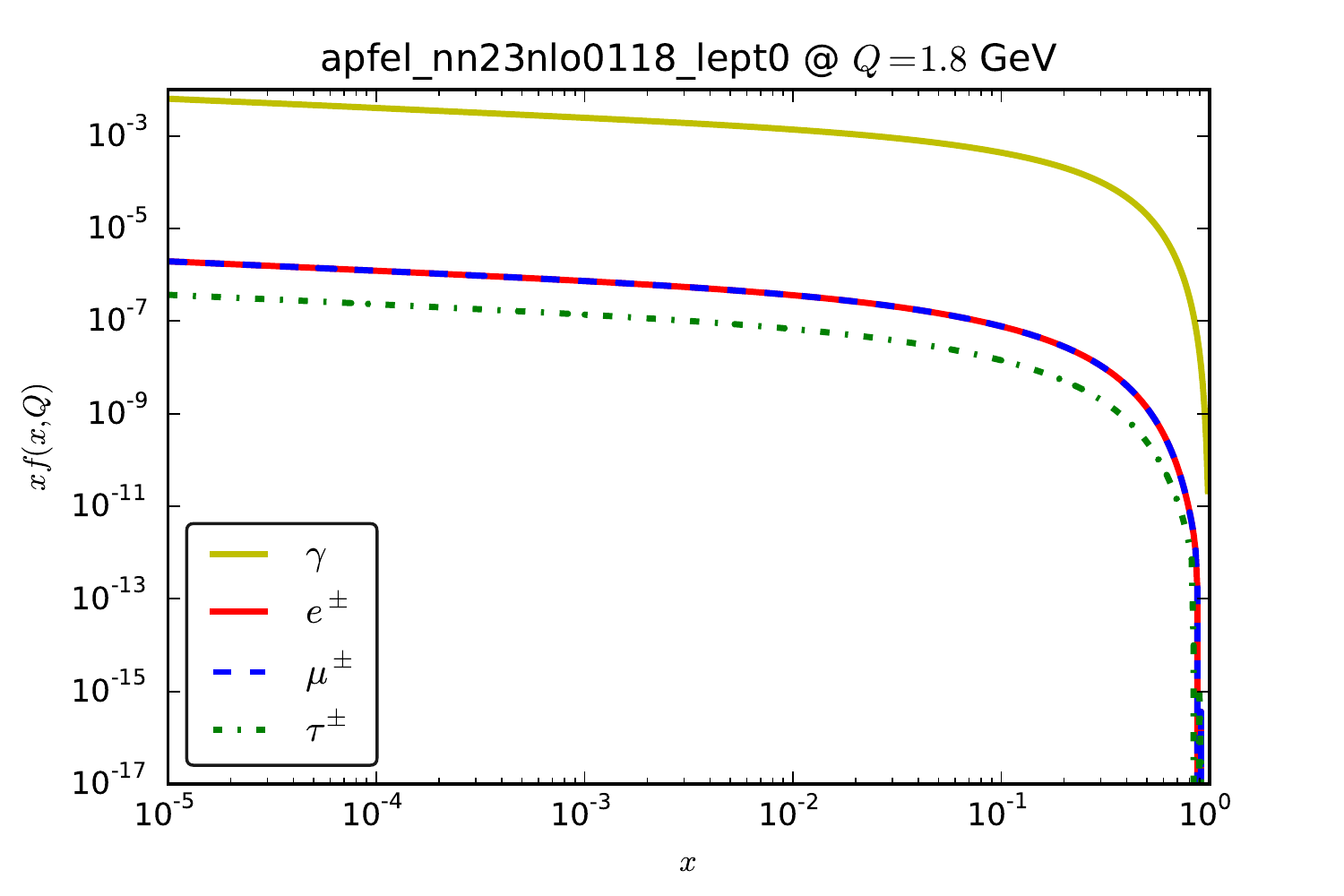}\includegraphics[scale=0.45]{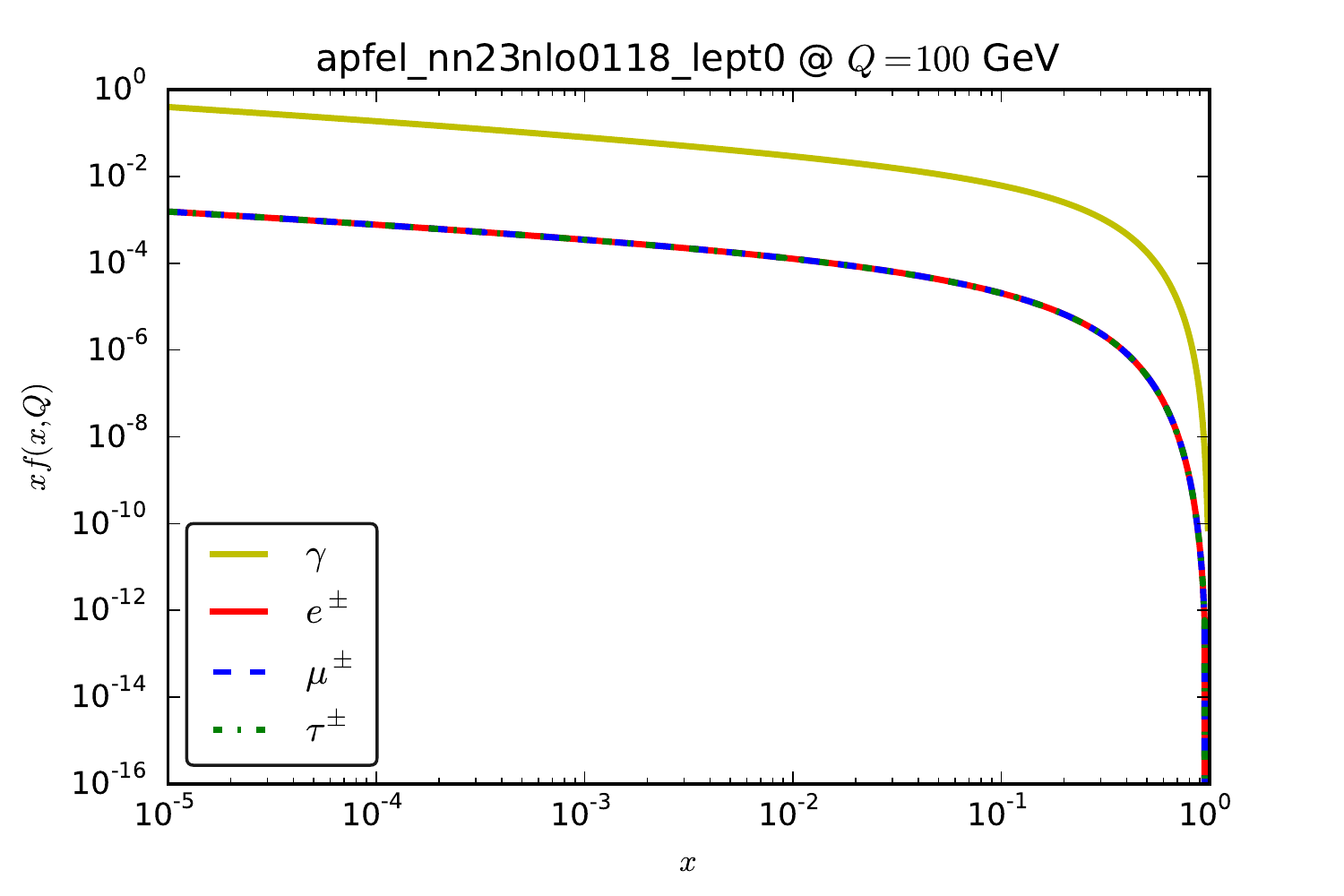}
  \caption{Leptons and photon PDF generated dynamically at NLO in QCD.}
  \label{fig:dyna}
\end{figure}

In Fig.~\ref{fig:dyna} we show the lepton and photon PDF central
values for the A1 configuration. In this configuration photons and
leptons are set to zero at $Q_0=1$ GeV and then dynamically generated
by DGLAP evolution. The left plot shows PDFs at $Q=1.8$ GeV, in this
case electron and muon PDFs are identical (by definition), and the
$\tau$ PDF has just been dynamically generated ($m_\tau = 1.777$
GeV). On the right plot, we display the same comparison but at $Q=100$
GeV, showing that all lepton PDFs are close to each other. Similar
results are obtained also with the NNPDF2.3 NNLO (A2).

\begin{figure}
  \centering
  \includegraphics[scale=0.45]{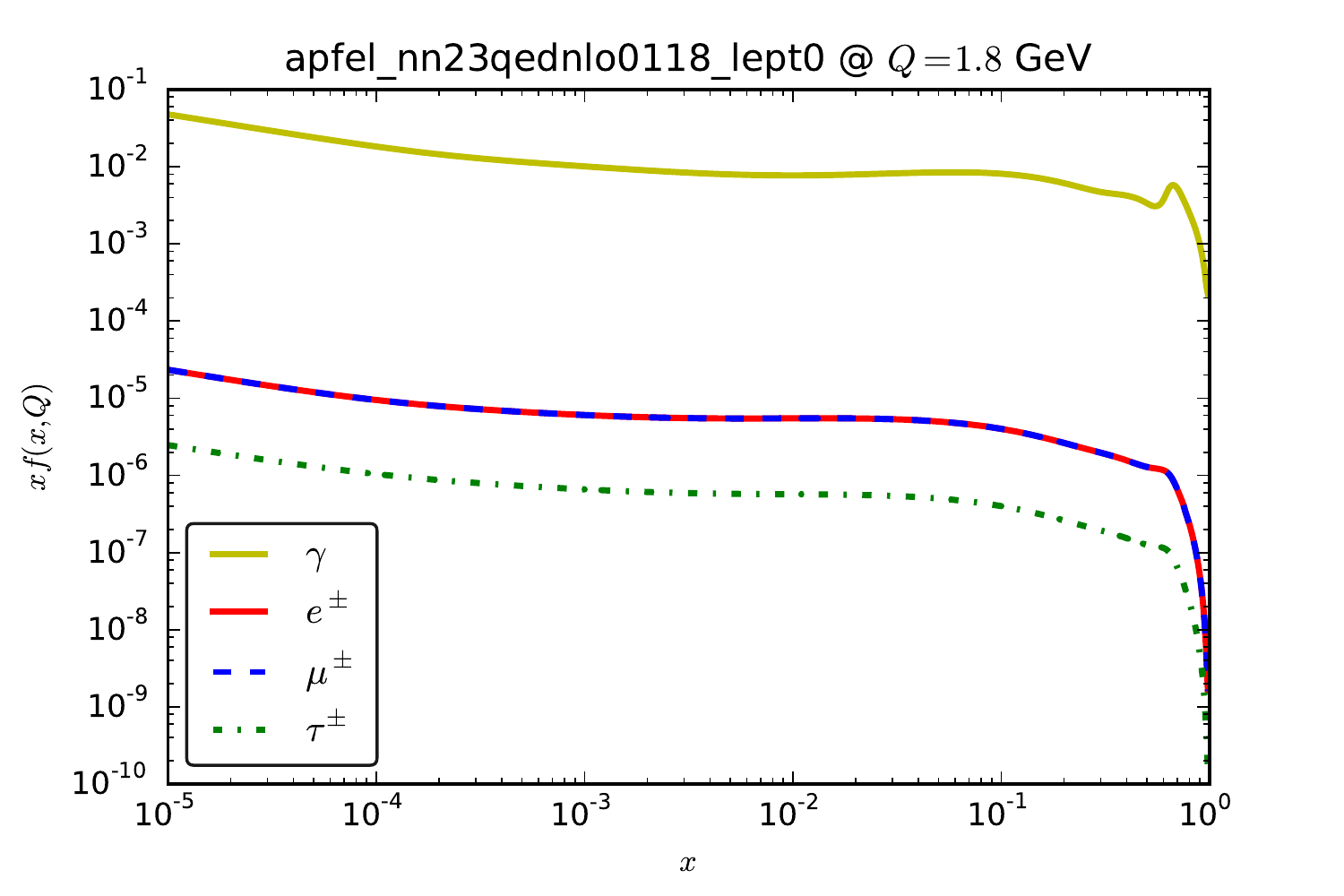}\includegraphics[scale=0.45]{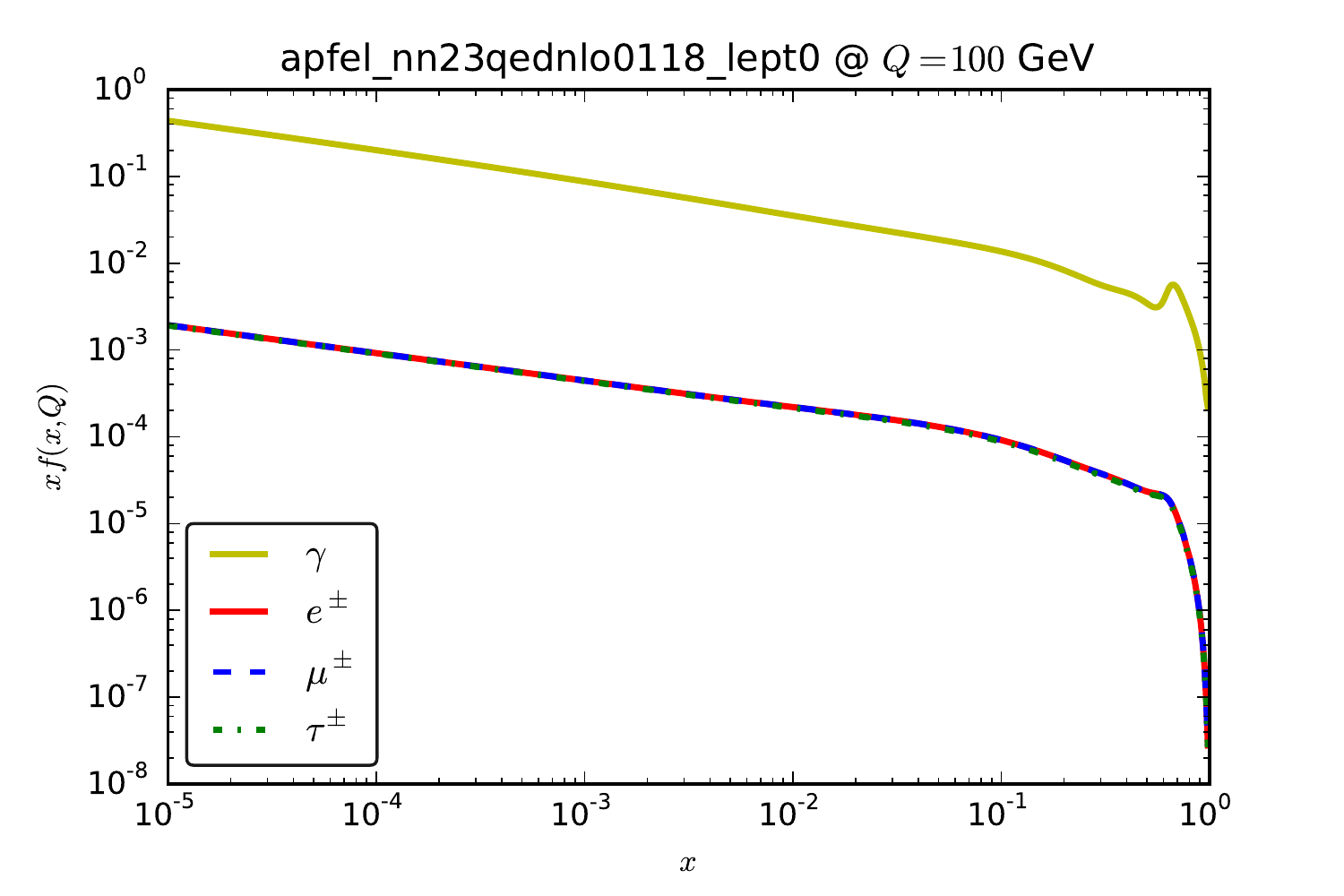}
  \includegraphics[scale=0.45]{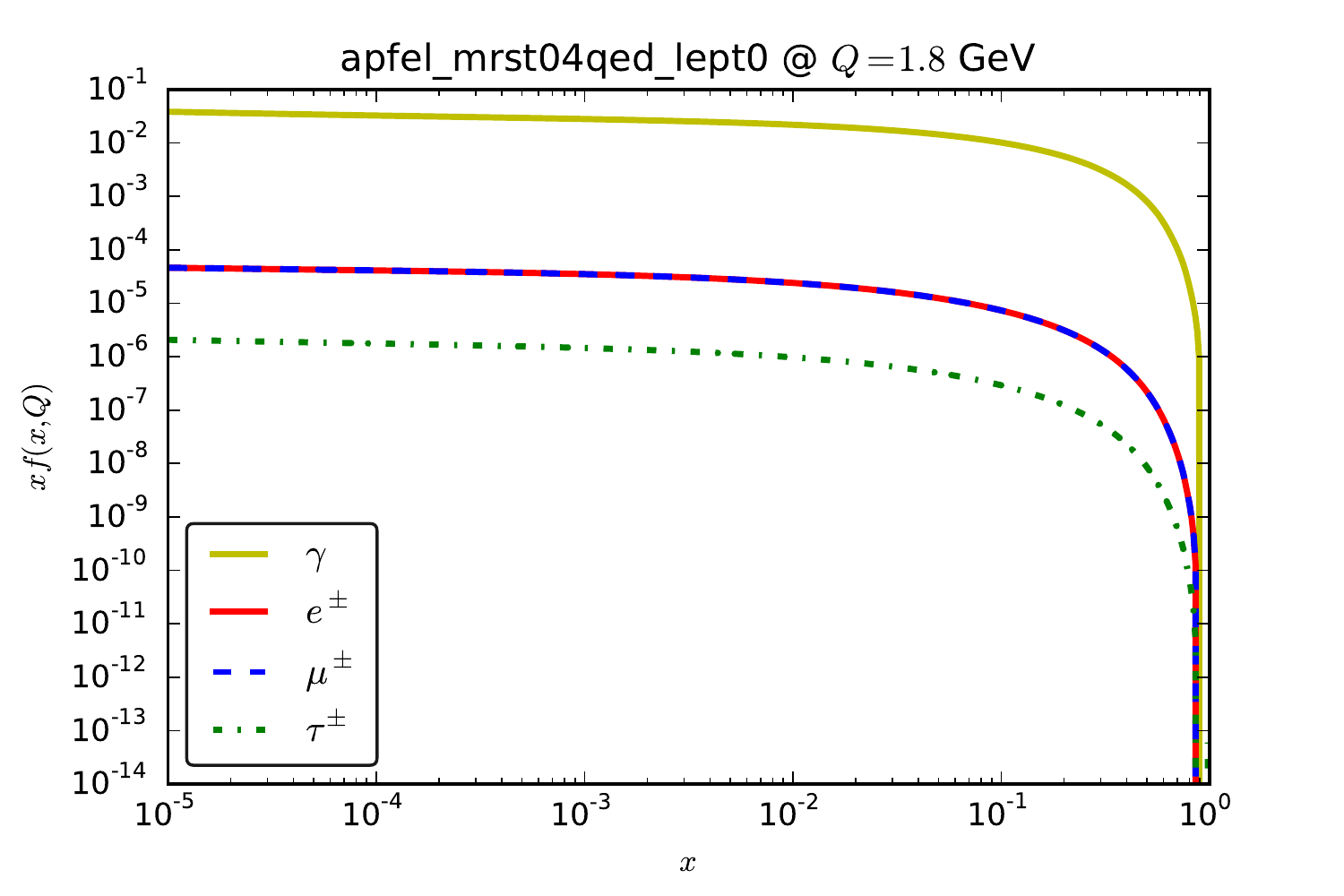}\includegraphics[scale=0.45]{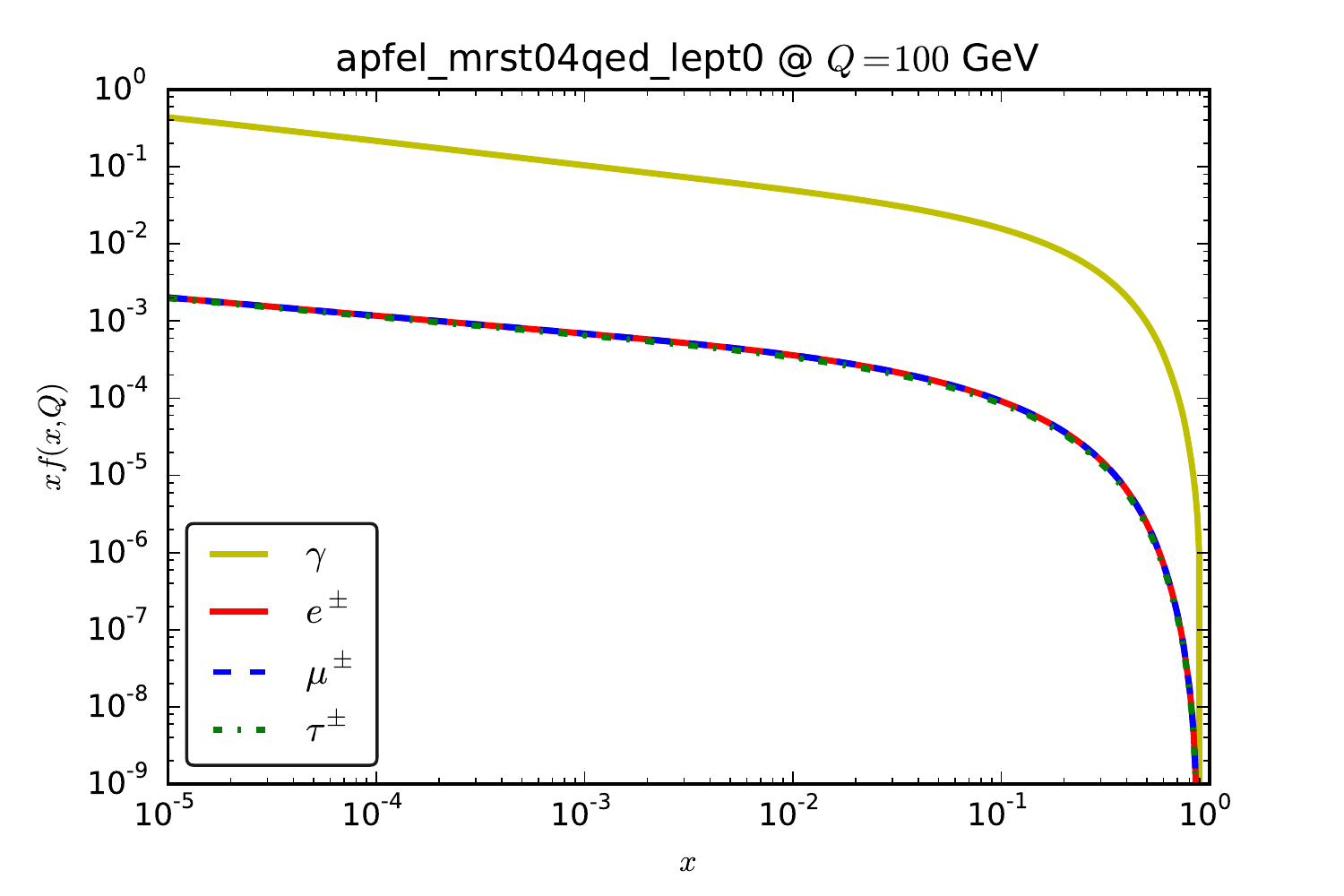}
  \caption{Lepton PDFs generated dynamically from the NNPDF2.3QED NLO
    (top) and MRST2004QED (bottom) photon PDFs.}
  \label{fig:photonfit}
\end{figure}

Configurations B2 and B4 are shown in Fig.~\ref{fig:photonfit}. For
these configurations, the prior set of PDFs contains the photon PDF
while the lepton PDFs are null at the initial scale $Q_0$ and
generated dynamically by DGLAP evolution. A similar behaviors as for
the configuration A is observed with the additional remark that lepton
PDFs present an evident dependence on the shape of the photon
PDF. Again, similar results are obtained for the NNPDF2.3QED NNLO (A2)
set.

\begin{figure}
  \centering
  \includegraphics[scale=0.45]{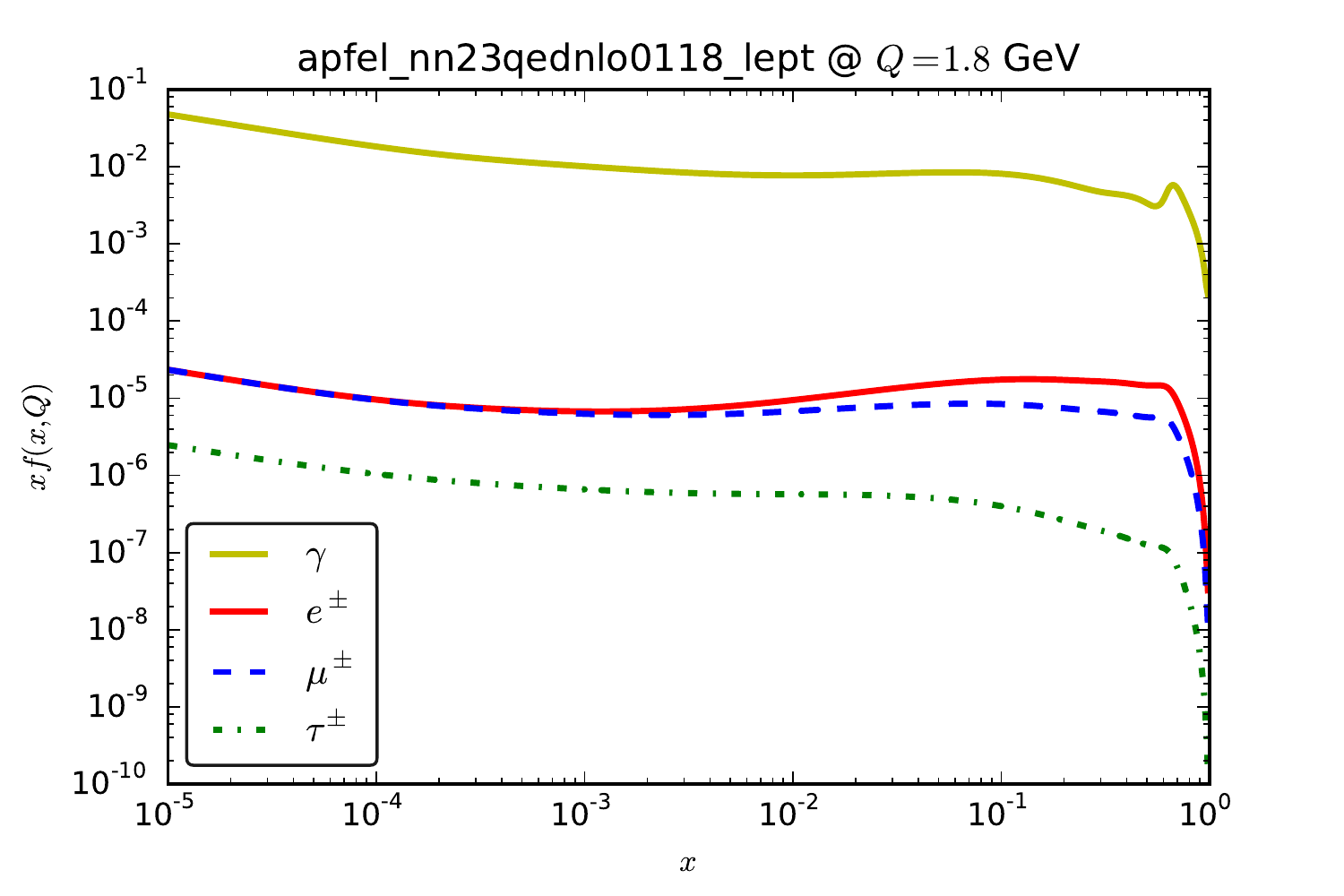}\includegraphics[scale=0.45]{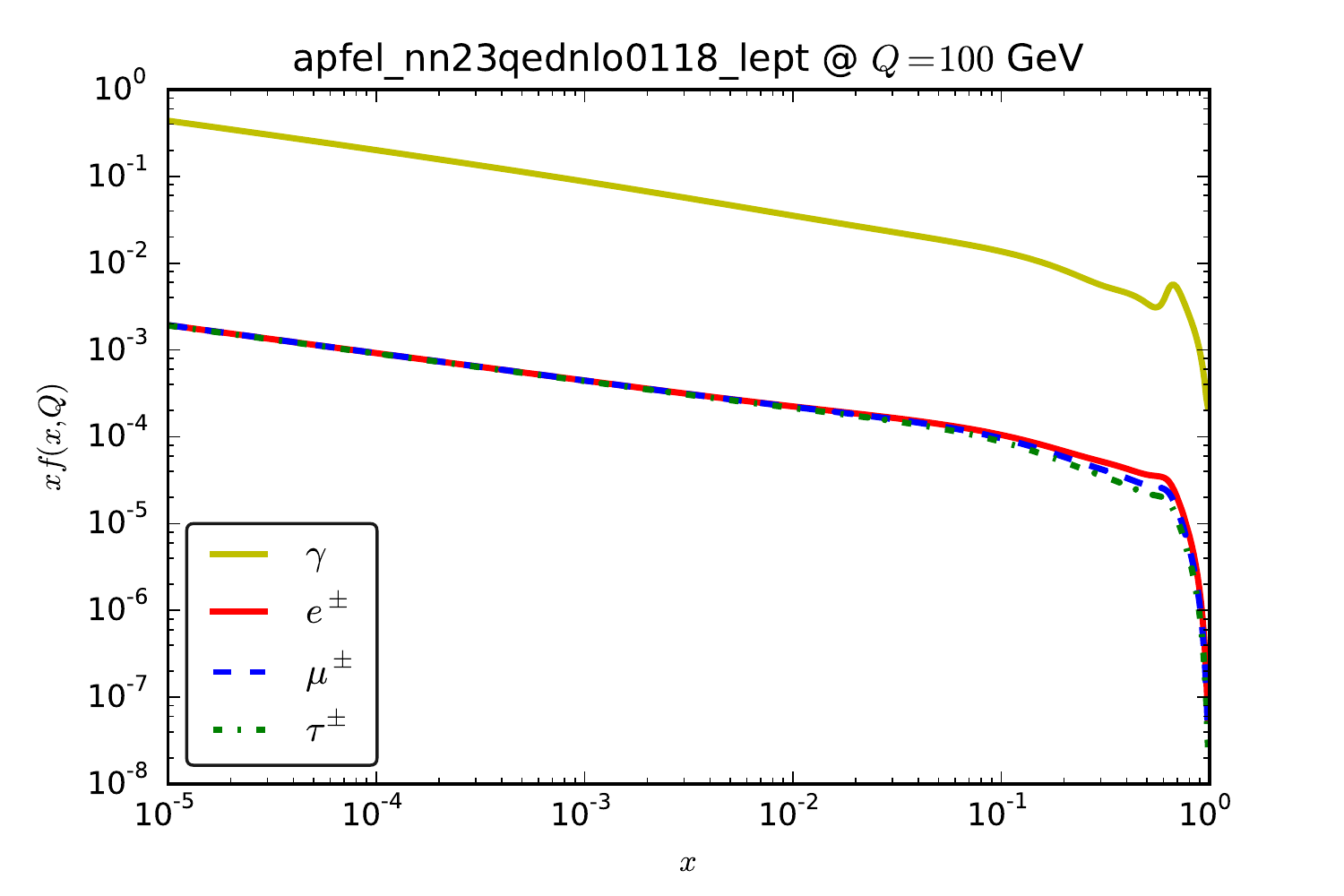}
\includegraphics[scale=0.45]{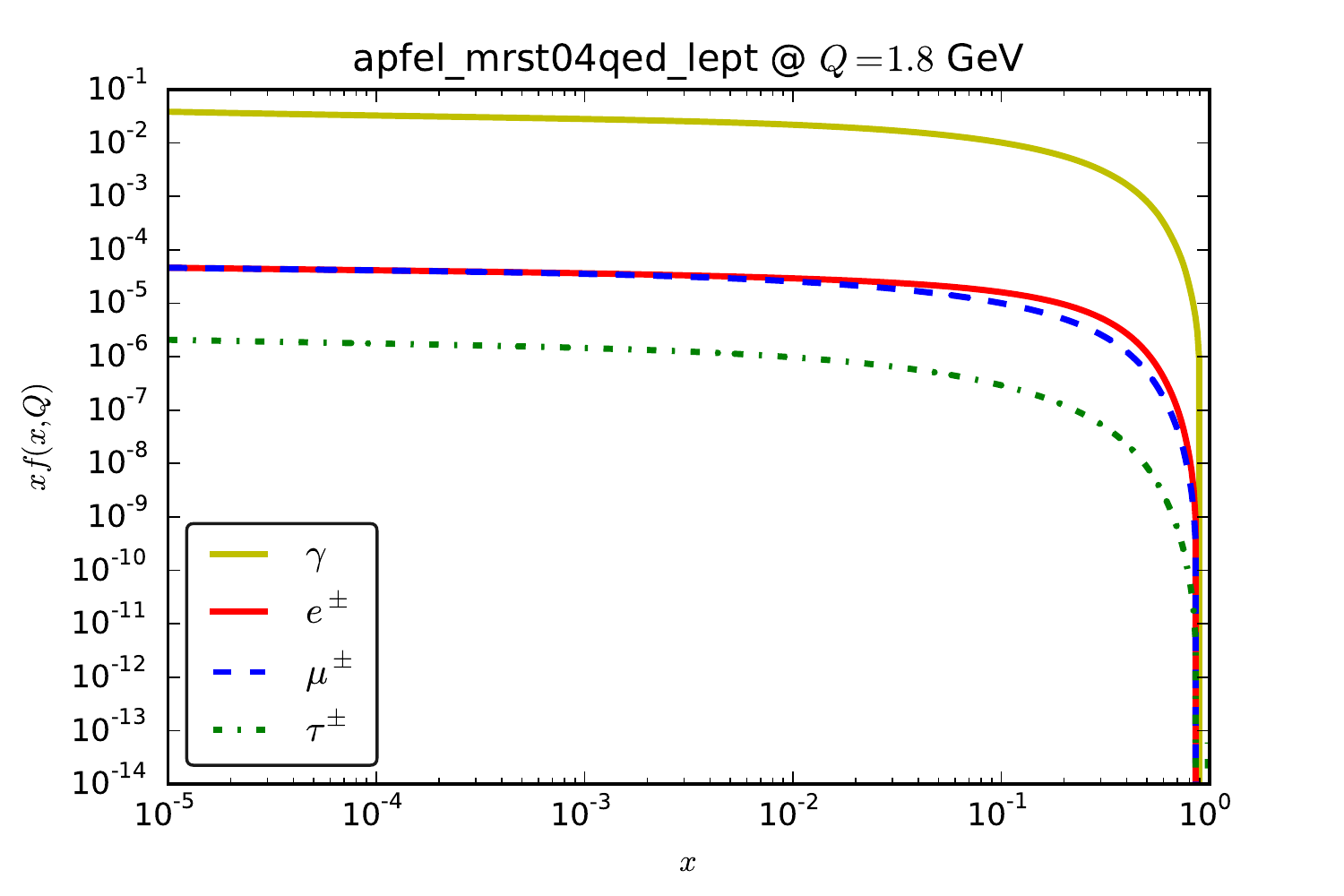}\includegraphics[scale=0.45]{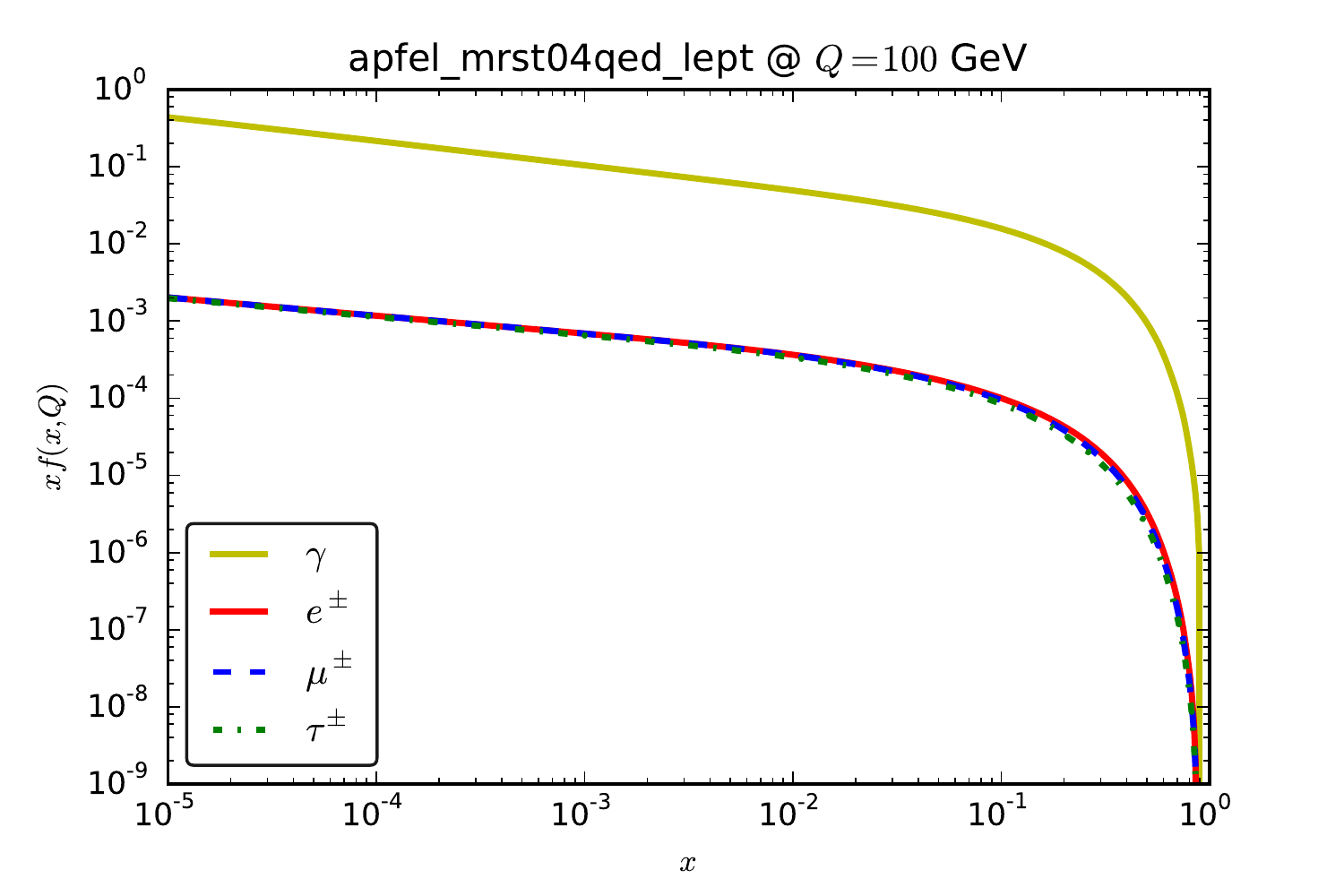}
  \caption{Lepton PDFs based on the ansatz of Eq.~(\ref{eq:ansatz})
    and evolved with the NNPDF2.3QED NLO (top) and MRST2004QED
    (bottom) photon PDFs.}
  \label{fig:leptonevolansatz}
\end{figure}

Now, let us consider the configuration of type C where, starting from
a prior containing a photon PDF, the initial distributions for the leptons
is determined using the ansatz in Eq.~(\ref{eq:ansatz}).
In Fig.~\ref{fig:leptonevolansatz} we show the resulting lepton PDFs
for the configurations C2 (top) and C4 (bottom), at $Q=1.8$ GeV (left)
and $Q=100$ GeV (right). Again, the qualitative behavior is the same
as for the configurations A and B.

\begin{figure}
  \centering
  \includegraphics[scale=0.45]{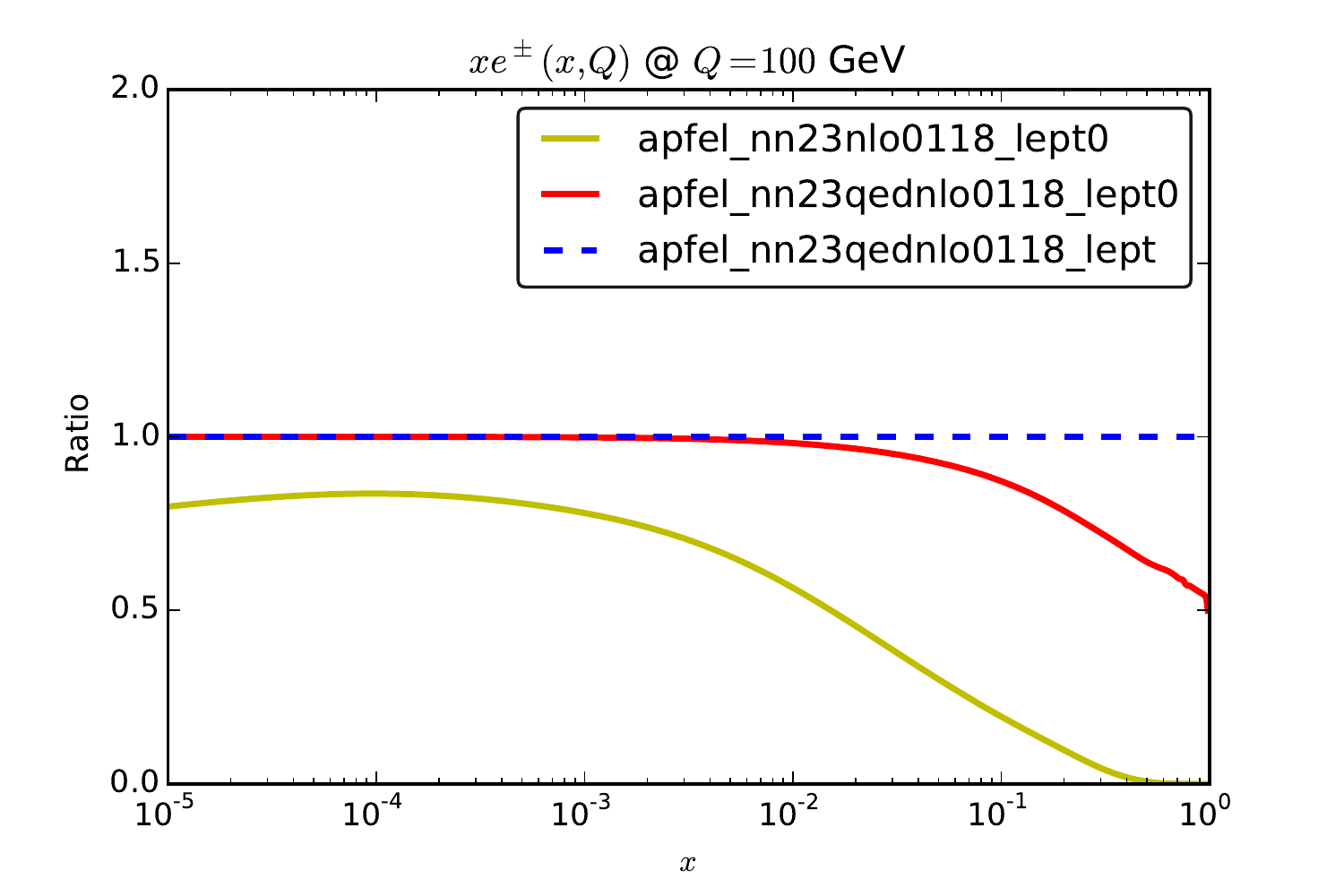}\includegraphics[scale=0.45]{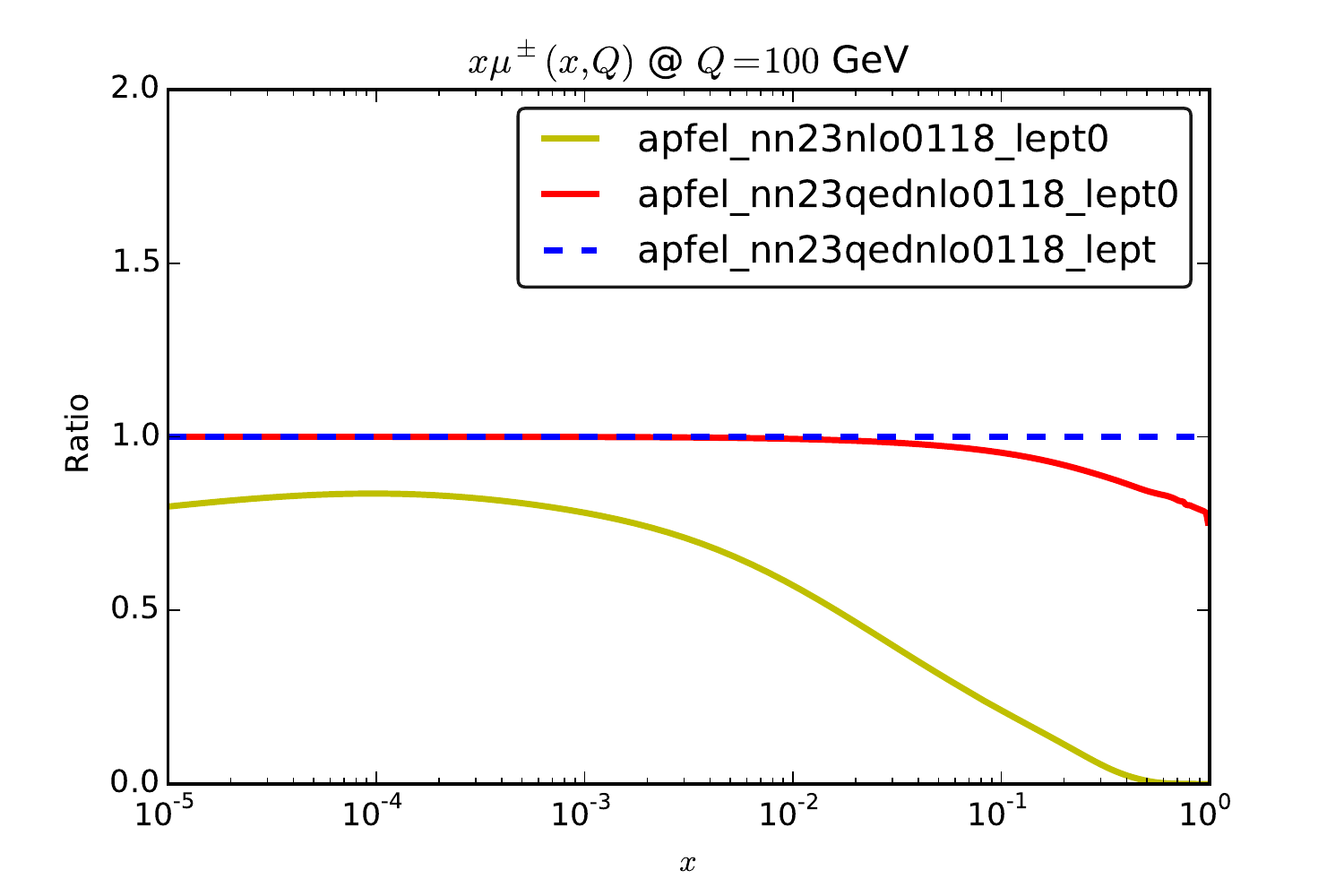}
  \caption{Ratio of electron and muon PDFs for each configuration.}
  \label{fig:leptratios}
\end{figure}

In order to quantify the difference generated by the various initial
conditions on the evolved lepton PDFs, in Fig.~\ref{fig:leptratios} we
show the ratio plots to the configuration C for the light lepton PDFs
produced starting from the NNPDF2.3 sets at NLO at $Q=100$ GeV. For
the electron PDFs (left plot), the ansatz in Eqs.~(\ref{eq:ansatz})
and~(\ref{eq:ansatz0}) applied to a set with a photon PDF lead to
similar results in the small-$x$ region while difference up to 50\%
are observed in the larger-$x$ region. The electron PDFs resulting
from a set without a photon PDF are instead way below all over the $x$
range.  The same behavior is observed also for the muon PDFs (right
plot in Fig.~\ref{fig:leptratios}), with slightly less enhanced
discrepancies as compared to the electrons.

\begin{figure}
  \centering
  \includegraphics[scale=0.45]{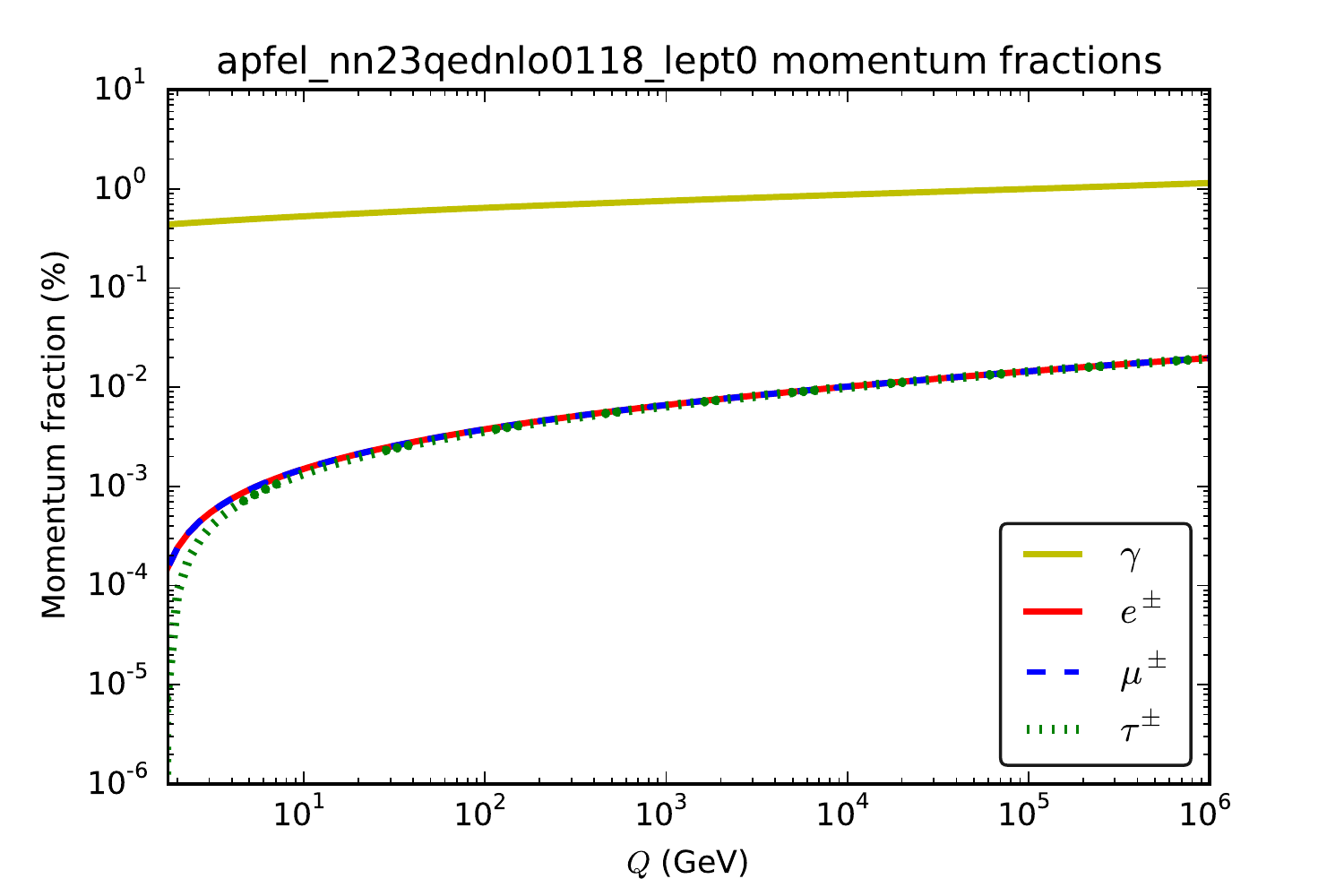}\includegraphics[scale=0.45]{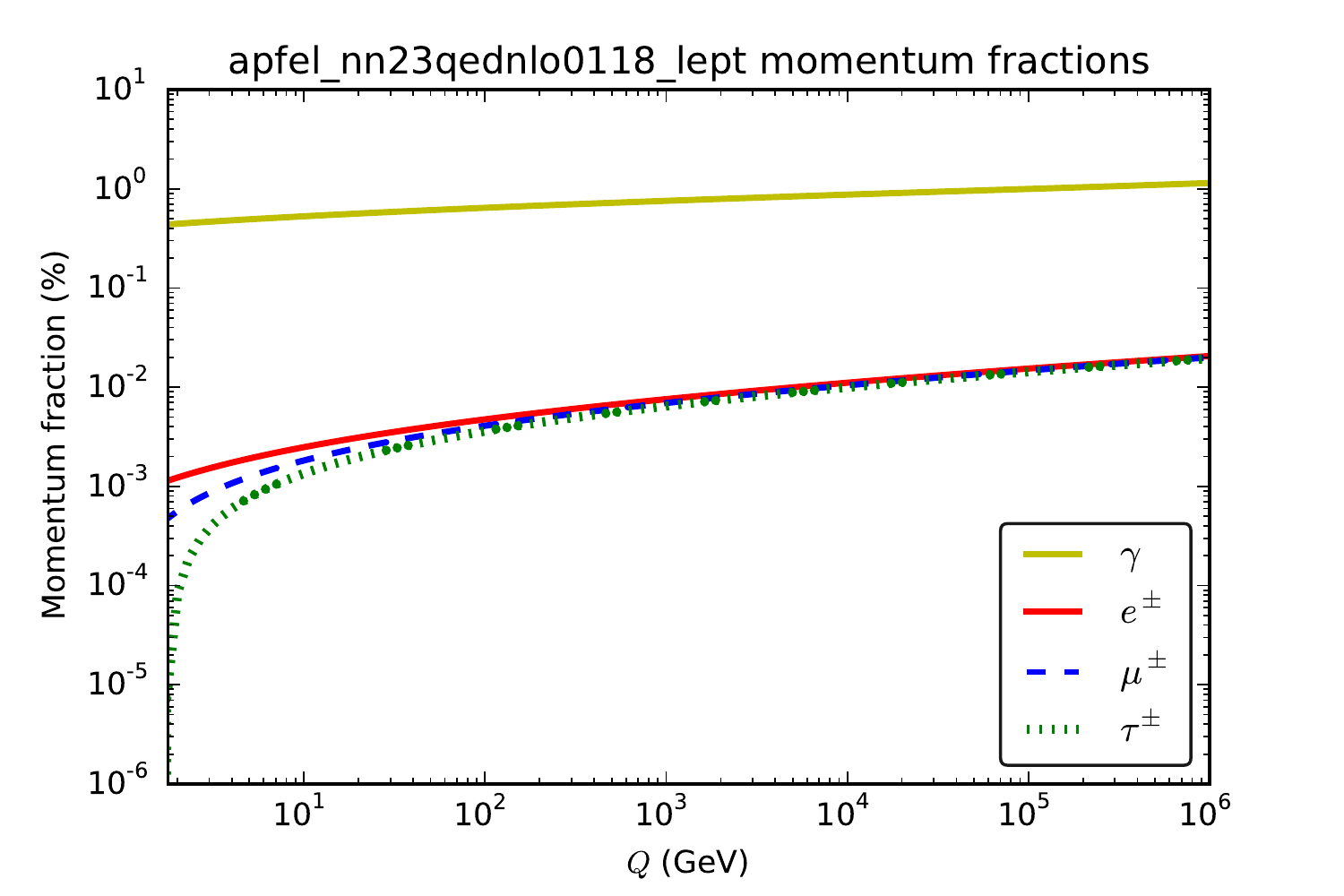}
  \caption{Momentum fractions for the photon and lepton PDFs.}
  \label{fig:msrlept}
\end{figure}

Interesting information about the photon and lepton content of the proton is
provided by the respective momentum fraction defined as:
\begin{equation}
  \textrm{MF}_\gamma(Q) = \int_0^1 dx \, x\gamma(x,Q)\,,\quad
  \textrm{MF}_{\ell^{\pm}}(Q) =  \int_0^1 dx \, x\ell^{\pm}(x,Q)\,.
\end{equation}
In Fig.~\ref{fig:msrlept} we plot the percent momentum fractions as a
function of the energy $Q$ for the configurations B2 (left) and C2
(right). While the photon PDF carries up to around 1\% of the proton
moment fraction of the proton, the lepton PDFs, independently from the
parametrization conditions, carry a much smaller fraction around tow
order of magnitude smaller than that carried by the photon. This is
consistent with the fact that, for both parametrizations, lepton PDFs
are proportional to $\alpha$ times the photon PDF ($\ell \propto
\alpha \times \gamma$). In conclusion, lepton PDFs carry such a small
fraction of the proton momentum that they do not cause a significant
violation of the total momentum sum rule.

\begin{figure}
  \centering
  \includegraphics[scale=0.45]{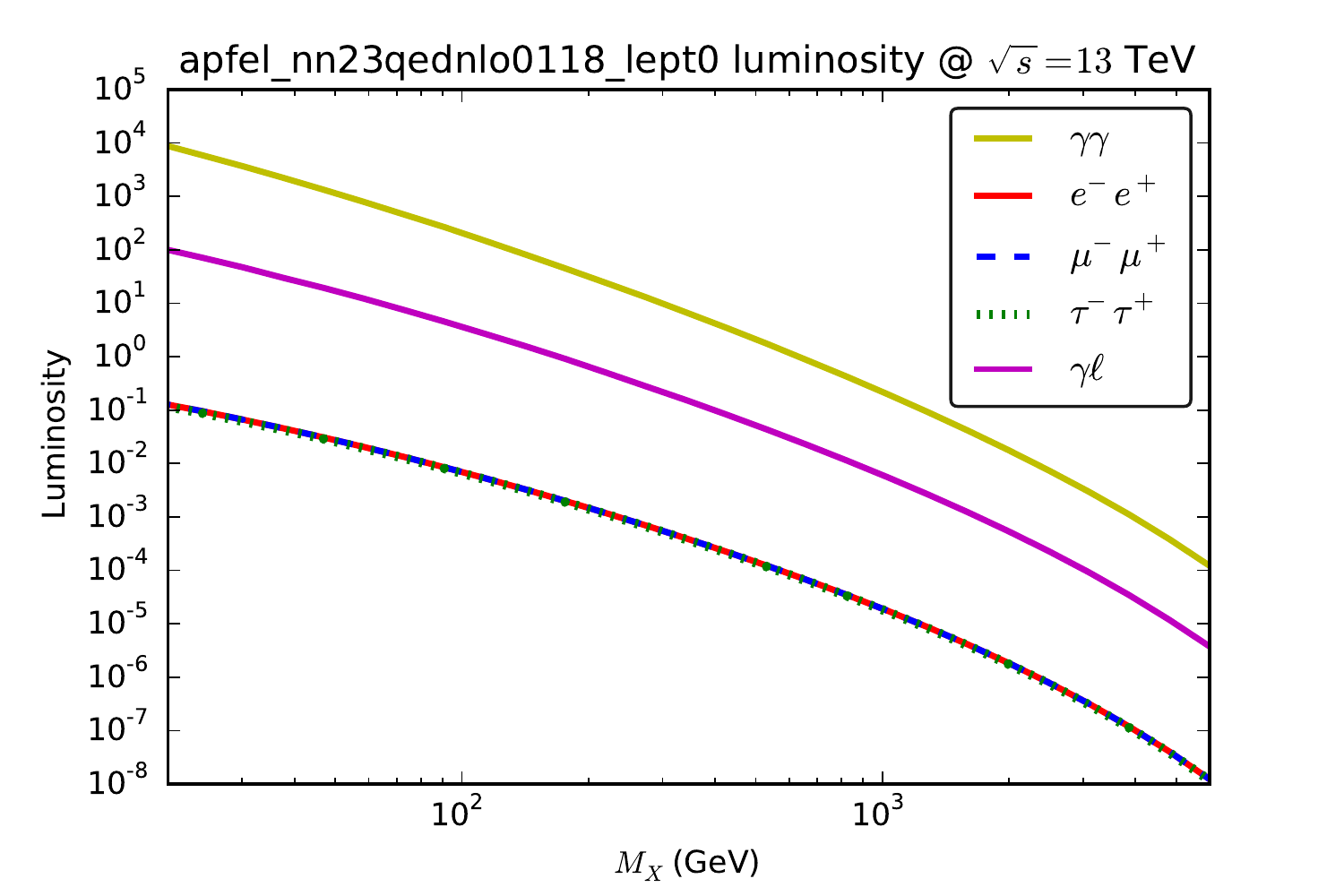}\includegraphics[scale=0.45]{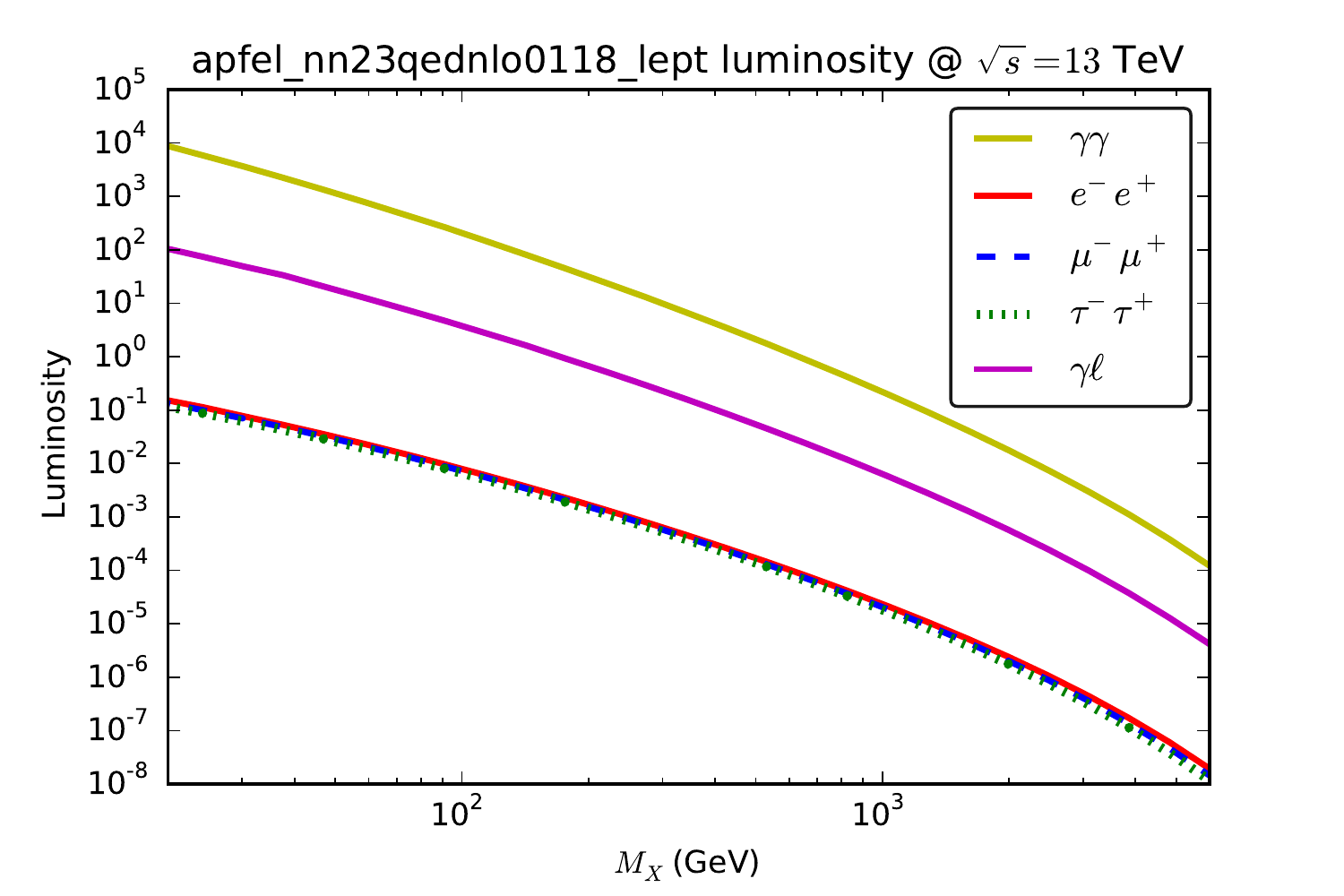}
  \caption{Parton luminosities for photon and lepton PDFs.}
  \label{fig:lumilept}
\end{figure}

In the computation of hadron collider processes, PDFs factorize in the
form of \textit{parton luminosities} as defined in
Eq.~(\ref{eq:lumdef}). Defining
\begin{equation}
  \Phi_{\gamma\ell}\left(M_X\right) =
  \sum_{j=e^\pm,\mu^\pm,\tau^\pm}\Phi_{\gamma j}\left( M_X\right)\,,
\end{equation}
in Fig.~\ref{fig:lumilept} we plot the $\Phi_{\gamma\gamma}$,
$\Phi_{\gamma\ell}$, $\Phi_{e^+e^-}$, $\Phi_{\mu^+\mu^-}$ and
$\Phi_{\tau^+\tau^-}$ parton luminosities as functions of $M_X$ at
$\sqrt{s}=13$ TeV for the sets B2 and C2. The relative size of the
plotted luminosities follows the expected pattern according to which
the $\Phi_{\gamma\ell}$ luminosity is roughly suppressed by one power
of $\alpha$ as compared to $\Phi_{\gamma\gamma}$ while
$\Phi_{\ell^+\ell^-}$, with $\ell=e,\mu,\tau$, is suppressed by two
powers.

\begin{figure}
  \centering
  \includegraphics[scale=0.45]{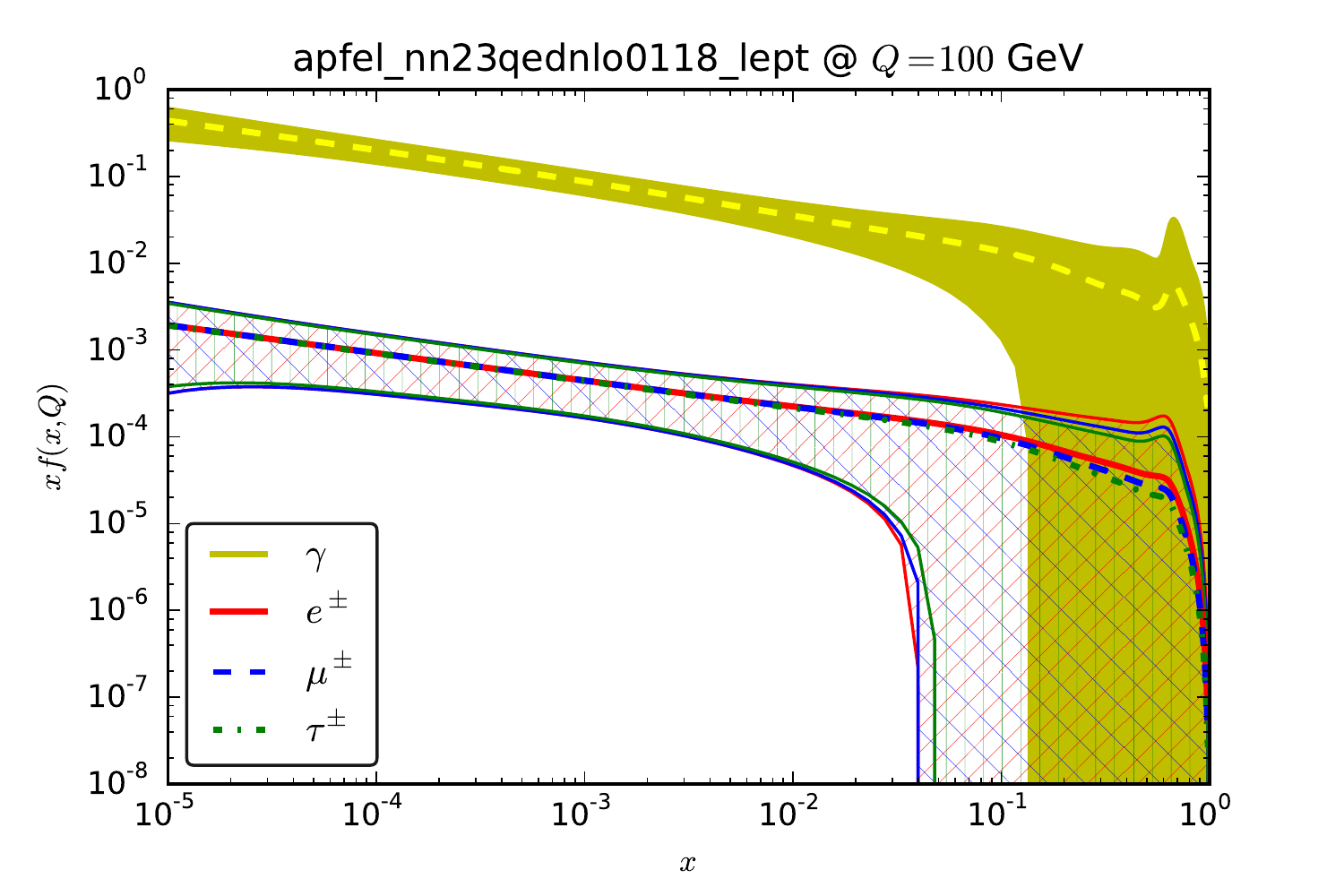}\includegraphics[scale=0.45]{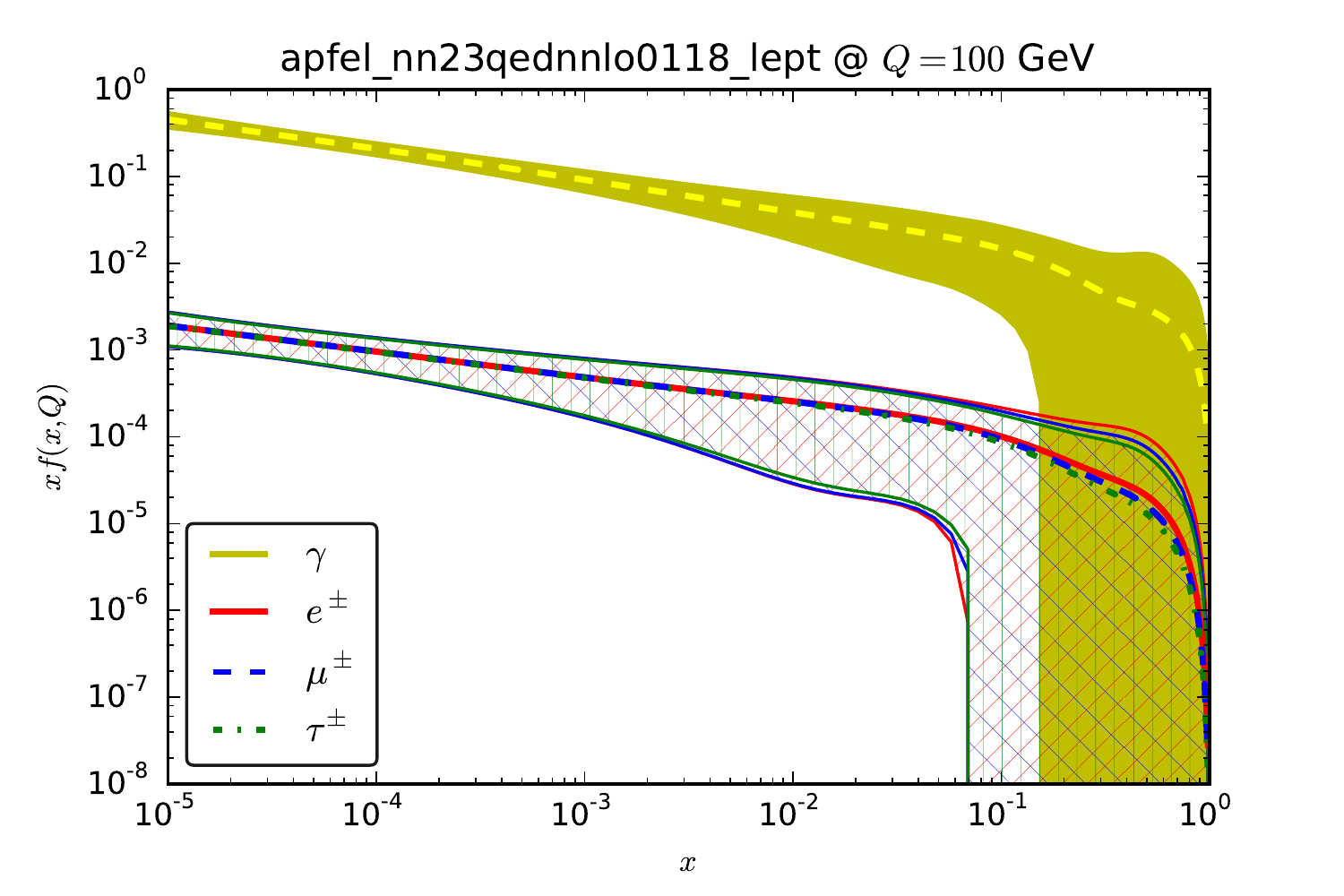}
  \caption{Uncertainties for the lepton PDFs at NLO (left) and NNLO
    (right) in QCD using the NNPDF2.3QED set (C2 and C3). Leptons are
    generated from Eq.~(\ref{eq:ansatz}).}
  \label{fig:leptuncer}
\end{figure}

We now turn to consider the uncertainties of the lepton PDFs. A
realistic estimated of the lepton PDFs requires the estimation of
uncertainty associated to each of them. To this end, using exactly the
same procedure discussed in the previous sections, we generated lepton
PDFs for all replicas of NNPDF2.3QED family sets. This eventually
allowed us to estimate the uncertainty on each lepton PDF.
In Fig.~\ref{fig:leptuncer} we plot the lepton PDFs with the
respective uncertainty for the configurations C2 and C3 at $Q=100$
GeV. Uncertainties are calculated as the one-sigma interval (standard
deviation) from the central value of each PDF flavor. As expected, the
lepton PDF uncertainties follow the pattern of photon PDF
uncertainty. 

\begin{figure}
  \centering
  \includegraphics[scale=0.7]{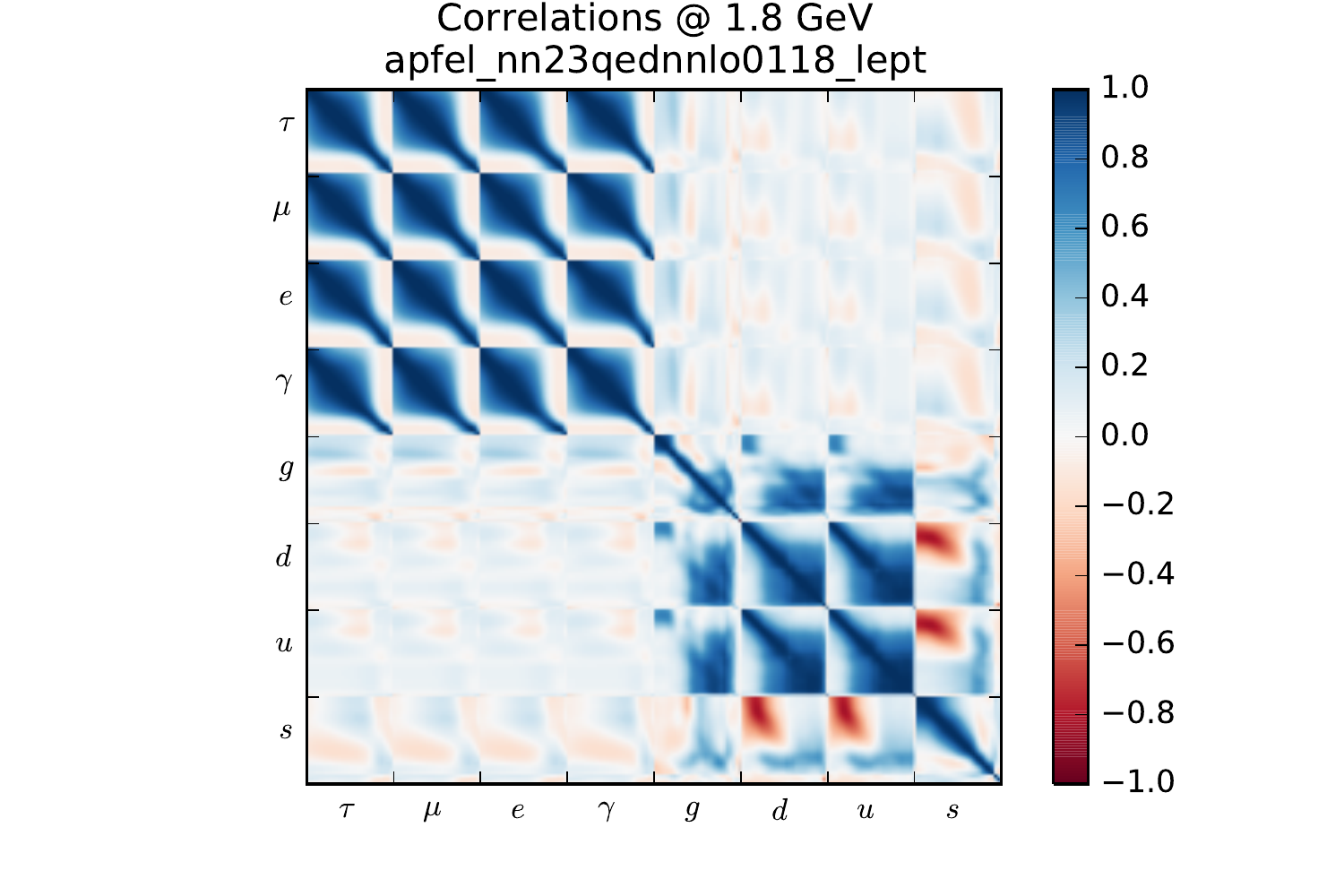}
  \caption{PDF correlation at $Q=1.8$ GeV.}
  \label{fig:pdfcorrelation}
\end{figure}

In Fig.~\ref{fig:pdfcorrelation} we compute the correlation of PDFs
for set C3 at $Q=1.8$ GeV in a grid of $N_x = 50$ points in
$x_1,x_2=[10^{-5},1]$, for the flavors ($\tau,\mu,e,\gamma,g,d,u,s$),
defined as
\begin{equation}
  \small
  \rho_{\alpha\beta}(x_1,x_2,Q) = \frac{N_{\textrm{rep}}}{N_{\textrm{rep}}-1} \left(
    \frac{\left\langle f_\alpha^{(k)}(x_1,Q)f_\beta^{(k)}(x_2,Q)\right\rangle_{\textrm{rep}}- \left\langle
        f_\alpha^{(k)}(x_1,Q)\right\rangle_{\textrm{rep}}\left\langle f_\beta^{(k)}(x_2,Q)\right\rangle_{\textrm{rep}}}{\sigma_\alpha(x_1,Q) \cdot \sigma_\beta(x_2,Q) } \right) \, , 
\end{equation}
where averages are taken over the $k=1,\ldots,N_{\textrm{rep}}$ replicas
and where $\sigma_i(x,Q)$ are the corresponding standard deviations.
Each row of this matrix is expressed in terms of $f_i \cdot N_x +
x_j$, for $j=1,\ldots,N_x$ and $i=\tau,\mu,e,\gamma,g,d,u,s$.

In the first place, we note a clear distinction between the QED (upper
left square region) and the QCD sector (bottom right region). As
expected, there are strong correlations between $(\tau,\mu,e,\gamma)$
due to the fact that leptons are generated by photon splitting. A
similar behavior is also observed for $(g,d,u,s)$. The off-diagonal
elements show that quark and gluon distributions are instead mildly
correlated to lepton and photon PDFs.
Similar results are obtained for the other configurations.

Finally, we remark that we are currently implementing the
lepton-induced channels in the \texttt{MadGraph5\_aMC@NLO} framework
for Drell-Yan production. Preliminary results show that the
contribution of lepton-induced channels is always similar or below to
the photon-induced contribution for lepton pair, dijet and vector
boson pair production at LHC ($\sqrt{s}=$13 TeV) and FCC-hh
($\sqrt{s}=$100 TeV) when applying reasonable kinematic cuts in lepton
$p_T$ and rapidity. The preliminary conclusion is that lepton-induced
processes are rare for the SM predictions tested in this work, and so,
fitting leptons in future sets of PDFs will not lead to significant
results.

We would like to stress that this is the first study which provides a
guess for the lepton PDFs. The sets of PDFs generated with
\texttt{APFEL} are publicly available, so further analysis are
encouraged, in particular on different configurations such as
BSM. Moreover, the current evolution framework opens the possibility
to eventually include the lepton PDFs contributions in PDF fits with
QED correction.

\chapter{Conclusion and outlook}
\label{sec:conclusions}

In this thesis we have presented a first determination of an unbiased
set of PDFs with QED corrections using the NNPDF methodology: the
NNPDF2.3QED set. In this set the photon PDF and its uncertainties are
determined by deep-inelastic scattering and neutral- and
charge-current Drell-Yan production data from the LHC.
We have discussed about the phenomenological impact of the photon PDF,
highlighting the lack of experimental information for large-$x$
region, which induces large uncertainties related to electroweak
corrections in processes which are relevant for new physics searches
at the LHC, such as high mass gauge boson production and double gauge
boson production.

This work has presented a series of important deliverables which have
been developed particularly for the determination of this set of PDFs
with QED corrections. Let us summarize these results:
\begin{itemize}

\item We have implemented \texttt{APFEL}, a new PDF evolution library
  that combines NNLO QCD corrections with LO QED effects in the
  solution of the DGLAP equations. This is the first public evolution
  code that performs the combined QCD$\otimes$QED evolution up to NNLO
  in QCD and LO in QED, both in the FFN and VFN schemes, and using
  either pole or $\overline{\textrm{MS}}$ heavy quark masses. We
  provide two strategies for solving this combined system of evolution
  equations: the \emph{coupled} and the \emph{unified} solutions. We
  have presented a detailed benchmarking exercise between
  \texttt{APFEL} and other public available codes such as:
  \texttt{HOPPET}, \texttt{partonevolution}, \texttt{MRST2004QED} and
  \texttt{QCDNUM}.

\item We have released \texttt{APFEL Web} a new Web-based application,
  born as a spin-off of the \texttt{APFEL} library. \texttt{APFEL Web}
  provides a user-friendly graphical user interface for the
  visualization of PDFs with a wide range of formats: absolute plots,
  ratio plots, compare PDFs from different groups, compare error PDF
  from a single set, plot all PDF flavor combinations at the same
  time, compute parton luminosities and finally compute DIS structure
  functions and \texttt{APPLgrid} observables. Moreover it provides a
  simple interface for the customization of PDF evolution.

\item We have developed a new modern framework for the NNPDF
  methodology. This new framework provides a flexible and fast code
  structure to perform PDF fits, in multiple configurations. This code
  has been used for the determination of this first set of PDFs with
  QED corrections, and it is the production code since NNPDF3.0.

\item We have delivered the NNPDF2.3QED set of PDFs at NLO and NNLO in
  QCD and LO in QED, for $\alpha_s=0.117,0.118,0.119$ values. In this
  set, the photon PDF is parametrized by an artificial neural-network
  and trained with the NNPDF methodology. The photon PDF is extracted
  for the first time from DIS data, and then by reweighting with
  neutral- and charge-current Drell-Yan production data from the
  LHC. The final results provides a first determination of the photon
  PDF and its uncertainties.

\item Several phenomenological implications of this new set of PDFs
  have been studied for: the direct photon production at HERA,
  searches for new massive electroweak gauge bosons and $W$ pair
  production at small $p_T$ and large invariant mass, at LHC
  energies. We have also interfaced this set of PDFs for multiple
  Monte Carlo event generators: \texttt{PYTHIA},
  \texttt{MadGraph5\_aMC@NLO} and \texttt{SHERPA}.

\item We present preliminary results about lepton PDFs. In fact, the
  inclusion of QED corrections requires extending the DGLAP evolution
  equations to include, in the first place, the photon PDF and, for
  consistency, also PDFs for the charged leptons $e^{\pm}$,
  $\mu^{\pm}$ and $\tau^{\pm}$. Here, we have shown how to construct
  those sets with \texttt{APFEL} considering multiple initial
  conditions for photon and lepton PDFs. We discuss about the size of
  these PDFs, its momentum fraction and luminosities. This is the
  first guess for the lepton PDFs, which is useful when considering
  electroweak corrections to some hadron-collider processes.

\end{itemize}

We plan to release in a near future new sets of PDFs with QED
corrections after introducing some technical improvements in the
procedure. First of all, we propose to extend the \texttt{APFEL}
combined QCD$\otimes$QED evolution up to NLO in QED together with the
inclusion of the subleading terms
$\mathcal{O}(\alpha\alpha_s)$. Secondly, we need fast interfaces, such
as \texttt{APPLgrid}, with electroweak corrections including the
photon-initiated channels. Such interface is important because avoids
the reweighting procedure that we have used for this first
determination, moreover it opens the possibility to compute several
photon-induced processes easily. Finally, the last important
ingredient for an improved set of PDFs with QED corrections is the
inclusion of new LHC data in regions where the photon PDF
uncertainties are unconstrained. In particular, based on the results
presented in this work and in the studies performed in
Ref.~\cite{Boughezal:2013cwa}, the most relevant data are: high- and
low-mass Drell-Yan $Z/\gamma^*$ production, diboson pair production at
small-$p_T$ and large invariant mass, dilepton rapidity distributions,
small-$p_T$ distribution for leptons among others.

In conclusion, future sets of PDFs with QED corrections will be
releases after introducing the technological improvements listed in
the last paragraph. Such sets of PDFs will enhance the quality and
reliability of predictions.

\appendix
\chapter{Distance estimators}
\label{app:distances}

The distance estimator assesses the compatibility between two PDFs
sets, and it tests whether two PDF sets are statistically equivalent.

Given a Monte Carlo sample of $N_{\text{rep}}$ replicas representing
the probability distribution of a given PDF set,
$\left\{f^{(k)}\right\}$, the expectation value of the distribution as
a function of $x$ and $Q^2$ is given by
\begin{equation} 
  \bar{f}(x,Q^2)\equiv \left< f (x,Q^2) \right>_{\textrm{rep}} 
= \frac{1}{N_{\text{rep}}}\sum_i^{N_{\text{rep}}} f^{(k)}(x,Q^2)\, ,
\end{equation}
where the index $(k)$ runs over all the replicas in the sample.
The variance of the sample is estimated as
\begin{equation} 
\label{eq:varsample}
  \sigma^2\left[ f (x,Q^2) \right] = \frac{1}{N_{\text{rep}}-1}\sum_i^{N_{\text{rep}}} \left( f^{(k)}(x,Q^2) - \left<f(x,Q^2)\right>_{\textrm{rep}}\right)^2\,.
\end{equation}
The variance of the mean is, in turn, defined in terms of the variance
of the sample by
\begin{equation}
\label{variancemean}
  \sigma^2\left[\left<f(x,Q^2)\right>_{\textrm{rep}}\right] = \frac{1}{N_{\text{rep}}} \sigma^2\left[ f (x,Q^2) \right]\, .
\end{equation}
The variance of the variance itself can be computed using
\begin{equation}
\label{eq:variancevariance}
   \sigma^2\left[\sigma^2\left[f(x,Q^2)\right]\right] = 
   \frac{1}{N_{\text{rep}}} \left[ m_4\left[f(x,Q^2)\right] - \frac{N_{\text{rep}} - 3}{N_{\text{rep}} - 1} 
   \left( \sigma^2\left[f(x,Q^2)\right] \right)^2 \right],
\end{equation}
where $m_4\left[f(x,Q^2)\right]$ denotes the fourth moment of the
probability distribution for $f(x,Q^2)$, namely
\begin{equation}
  m_4\left[f(x,Q^2)\right]=\frac{1}{N_{\mathrm{rep}}} \left[\sum_{k=1}^{N_\mathrm{rep}}\left(f^{(k)}(x,Q^2)- \left< f (x,Q^2) \right>_{\textrm{rep}} \right)^4\right]\,.
\end{equation}

The distance between two sets of PDFs, each characterized by a given
distribution of the Monte Carlo replicas, denoted by
$\left\{f^{(k)}\right\}$ and $\left\{g^{(k)}\right\}$, is defined as
the square root of the square difference of the PDF central values in
units of the uncertainty of the mean:
\begin{equation} 
  d_{\bar{f},\bar{g}}(x,Q^2) = 
  \sqrt{\frac{\left(\bar{f}-\bar{g} \right)^2}
  {\sigma^2 \left[\bar{f}\right] + \sigma^2 \left[\bar{g}\right]}}\,.
  \label{eq:CVdistance}
\end{equation}

In Eq.~(\ref{eq:CVdistance}), the denominator uses the variance of the
mean of the distribution, defined as in Eq.~(\ref{variancemean}).
An analogous distance can be defined for the variances of the two
samples:
\begin{equation} 
   d_{\sigma\left[ f\right],\sigma\left[ g\right]}(x,Q^2) = 
  \sqrt{\frac{\left(\sigma^2\left[ f\right] -
\sigma^2\left[ g\right]\right)^2}
  {\sigma^2 \left[\sigma^2 \left[f\right]\right] + \sigma^2 \left[\sigma^2 \left[g\right]\right]}}\,.
  \label{eq:Vardistance}
\end{equation}
where now in the denominator we have the variance of the variance,
Eq.~(\ref{eq:variancevariance}).

The distances for the central values and for the variances defined in
Eqs.~(\ref{eq:CVdistance}) and~(\ref{eq:Vardistance}) test whether the
underlying distributions from which the two Monte Carlo samples
$\left\{f^{(k)}\right\}$ and $\left\{g^{(k)}\right\}$ are drawn have
respectively the same mean and the same standard deviation.
In particular, it is possible to show that one expects these distances
to fluctuate around $d \sim 1$ if the two samples do indeed come from
the same distribution.
Values of the distances around $d\sim \sqrt{N_\mathrm{rep}}$ indicates
that the central values (the variances) of the two PDF sets differ by
one standard deviation in units of the variance of the distribution
Eq.~(\ref{eq:varsample}) (in units of the variance of the variance
Eq.~(\ref{eq:variancevariance})).

\chapter{Bayesian reweighting}
\label{app:rw}

Let us consider that a set of experimental data is used to construct a
probability distribution for PDFs,
$\mathcal{P}_{\textrm{old}}(f)$. With this probability distribution
any observable can be obtained by performing averages over this
ensemble, equally weighting each PDF.

Suppose that we would like to include new experimental data without
applying the fitting procedure presented in Chap.~\ref{sec:chap3}. The
only option that we have is to extract a new probability distribution
$\mathcal{P}_{\textrm{new}}$ by updating the weights, $w_k$ associated
to each individual PDF $f_k$ of the prior ensemble.

From a practical point of view, the new data is assumed to have
Gaussian errors, so we can write the relative probabilities of the new
data for different choices of PDF in terms of the probability density
of the $\chi$ to the new data
\begin{equation}
  \mathcal{P}(\chi|f) \propto (\chi^2(y,f))^{\frac{1}{2}(n-1)}
  e^{-\frac{1}{2}\chi^2(y,f)}\,,
\end{equation}
where $y=\{y_1,y_2,\ldots,y_n\}$ are the new $n$ experimental data
points and
\begin{equation}
  \chi^2(y,f) = \sum_{i,j=1}^n\frac{(y_i-y_i[f])(y_j-y_j[f])}{\sigma_{ij}}\,,
\end{equation}
where $y_i[f]$ is the value predicted for the data $y_i$ using the PDF
$f$, and $\sigma_{ij}$ is the data uncertainties covariance matrix.

From statistical independece of the old and new data we have
\begin{equation}
  \mathcal{P}_{\textrm{new}}(f) = \mathcal{N}_\chi \mathcal{P}(\chi|f) \mathcal{P}_{\textrm{old}}(f)\,,
\end{equation}
where $\mathcal{N}_\chi$ is a normalization factor, independent of
$f$.
We can show that some observable $\mathcal{O}[f]$ is given in terms of
$N$ reweighted replicas $f_k$.
\begin{equation}
  \langle \mathcal{O}_{\textrm{new}} \rangle = \frac{1}{N} \sum_{k=1}^{N} w_k \mathcal{O}[f_k]\,,
\end{equation}
where
\begin{equation}
  w_k = \mathcal{N}_\chi \mathcal{P}(\chi|f_k) =
  \frac{(\chi^2_k)^{\frac{1}{2}(n-1)}e^{-\frac{1}{2}\chi^2_k}}{\frac{1}{N}\sum_{k=1}^{N}(\chi^2)^{\frac{1}{2}(n-1)}e^{-\frac{1}{2}\chi^2_k}}\,,
\label{eq:wk}
\end{equation}
with $\chi^2 \equiv \chi^2(y,f_k)$.

The definition of weights in Eq.~(\ref{eq:wk}) is used when new
experimental data is included in the fit. We can also quantify the
loss of efficiency by using the Shannon entropy to compute the
effective number of replicas left after reweighting
\begin{equation}
  N_{\textrm{eff}} = \exp \left( \frac{1}{N} \sum_{k=1}^N w_k \ln
      \frac{N}{w_k} \right)\,.
\end{equation}

Finally, after reweighting a set of PDFs it is always possible to
construct an unweighted set where each PDF has equally distributed
weights. More details of the unweighting procedure can be found in
Ref.~\cite{Ball:2011gg}.

\bibliographystyle{JHEP}
{\small\bibliography{thesis}}

\providecommand{\href}[2]{#2}\begingroup\raggedright\begin{thebibliography}{100}

\bibitem{Decamp:1989tu}
\textbf{ALEPH} Collaboration, D.~Decamp et~al., \textit{{Determination of the
  Number of Light Neutrino Species}},  {\em Phys.Lett.} \textbf{B231} (1989)
  519.

\bibitem{Abe:1995hr}
\textbf{CDF} Collaboration, F.~Abe et~al., \textit{{Observation of top quark
  production in $\bar{p}p$ collisions}},  {\em Phys.Rev.Lett.} \textbf{74}
  (1995) 2626--2631,
  [\href{http://arxiv.org/abs/hep-ex/9503002}{\texttt{hep-ex/9503002}}].

\bibitem{Aad:2012tfa}
\textbf{ATLAS} Collaboration, G.~Aad et~al., \textit{{Observation of a new
  particle in the search for the Standard Model Higgs boson with the ATLAS
  detector at the LHC}},  {\em Phys.Lett.} \textbf{B716} (2012) 1--29,
  [\href{http://arxiv.org/abs/1207.7214}{\texttt{arXiv:1207.7214}}].

\bibitem{Chatrchyan:2012ufa}
\textbf{CMS} Collaboration, S.~Chatrchyan et~al., \textit{{Observation of a new
  boson at a mass of 125 GeV with the CMS experiment at the LHC}},  {\em
  Phys.Lett.} \textbf{B716} (2012) 30--61,
  [\href{http://arxiv.org/abs/1207.7235}{\texttt{arXiv:1207.7235}}].

\bibitem{CarloniCalame:2007cd}
C.~Carloni~Calame, G.~Montagna, O.~Nicrosini, and A.~Vicini, \textit{{Precision
  electroweak calculation of the production of a high transverse-momentum
  lepton pair at hadron colliders}},  {\em JHEP} \textbf{0710} (2007) 109,
  [\href{http://arxiv.org/abs/0710.1722}{\texttt{arXiv:0710.1722}}].

\bibitem{Alwall:2014hca}
J.~Alwall, R.~Frederix, S.~Frixione, V.~Hirschi, F.~Maltoni, et~al.,
  \textit{{The automated computation of tree-level and next-to-leading order
  differential cross sections, and their matching to parton shower
  simulations}},  {\em JHEP} \textbf{1407} (2014) 079,
  [\href{http://arxiv.org/abs/1405.0301}{\texttt{arXiv:1405.0301}}].

\bibitem{Li:2012wna}
Y.~Li and F.~Petriello, \textit{{Combining QCD and electroweak corrections to
  dilepton production in FEWZ}},  {\em Phys.Rev.} \textbf{D86} (2012) 094034,
  [\href{http://arxiv.org/abs/1208.5967}{\texttt{arXiv:1208.5967}}].

\bibitem{Ball:2012wy}
R.~D. Ball, S.~Carrazza, L.~Del~Debbio, S.~Forte, J.~Gao, et~al.,
  \textit{{Parton Distribution Benchmarking with LHC Data}},  {\em JHEP}
  \textbf{1304} (2013) 125,
  [\href{http://arxiv.org/abs/1211.5142}{\texttt{arXiv:1211.5142}}].

\bibitem{Bertone:2013vaa}
V.~Bertone, S.~Carrazza, and J.~Rojo, \textit{{APFEL: A PDF Evolution Library
  with QED corrections}},  {\em Comput.Phys.Commun.} \textbf{185} (2014)
  1647--1668,
  [\href{http://arxiv.org/abs/1310.1394}{\texttt{arXiv:1310.1394}}].

\bibitem{Carrazza:2014gfa}
S.~Carrazza, A.~Ferrara, D.~Palazzo, and J.~Rojo, \textit{{APFEL Web: a
  web-based application for the graphical visualization of parton distribution
  functions}},
  \href{http://arxiv.org/abs/1410.5456}{\texttt{arXiv:1410.5456}}.

\bibitem{Ball:2012cx}
R.~D. Ball, V.~Bertone, S.~Carrazza, C.~S. Deans, L.~Del~Debbio, et~al.,
  \textit{{Parton distributions with LHC data}},  {\em Nucl.Phys.}
  \textbf{B867} (2013) 244--289,
  [\href{http://arxiv.org/abs/1207.1303}{\texttt{arXiv:1207.1303}}].

\bibitem{Ball:2014uwa}
\textbf{The NNPDF} Collaboration, R.~D. Ball et~al., \textit{{Parton
  distributions for the LHC Run II}},
  \href{http://arxiv.org/abs/1410.8849}{\texttt{arXiv:1410.8849}}.

\bibitem{Ball:2013lla}
\textbf{The NNPDF} Collaboration, R.~D. Ball et~al., \textit{{Unbiased
  determination of polarized parton distributions and their uncertainties}},
  {\em Nucl.Phys.} \textbf{B874} (2013) 36--84,
  [\href{http://arxiv.org/abs/1303.7236}{\texttt{arXiv:1303.7236}}].

\bibitem{Carrazza:2013wua}
S.~Carrazza, \textit{{Towards an unbiased determination of parton distributions
  with QED corrections}},
  \href{http://arxiv.org/abs/1305.4179}{\texttt{arXiv:1305.4179}}.

\bibitem{Carrazza:2013bra}
S.~Carrazza, \textit{{Towards the determination of the photon parton
  distribution function constrained by LHC data}},
  \href{http://arxiv.org/abs/1307.1131}{\texttt{arXiv:1307.1131}}.

\bibitem{Carrazza:2014cpa}
S.~Carrazza, \textit{{Disentangling electroweak effects in Z-boson
  production}},
  \href{http://arxiv.org/abs/1405.1728}{\texttt{arXiv:1405.1728}}.

\bibitem{Bertone:2015lqa}
V.~Bertone, S.~Carrazza, D.~Pagani, and M.~Zaro, \textit{{On the Impact of
  Lepton PDFs}},
  \href{http://arxiv.org/abs/1508.07002}{\texttt{arXiv:1508.07002}}.

\bibitem{Ellis:1991qj}
R.~K. Ellis, W.~J. Stirling, and B.~R. Webber, {\em QCD and collider physics}.
\newblock {Cambridge University Press}, 1996.

\bibitem{opac-b1131978}
M.~E. Peskin and D.~V. Schroeder, {\em An introduction to quantum field
  theory}.
\newblock Advanced book program. Westview Press Reading (Mass.), Boulder
  (Colo.), 1995.

\bibitem{h1fl}
\textbf{H1} Collaboration, F.~D. Aaron et~al., \textit{{Measurement of the
  Proton Structure Function $F_L$ at Low x}},  {\em Phys. Lett.} \textbf{B665}
  (2008) 139--146,
  [\href{http://arxiv.org/abs/0805.2809}{\texttt{arXiv:0805.2809}}].

\bibitem{ZEUS:2012bx}
\textbf{ZEUS} Collaboration, A.~Cooper~Sarkar, \textit{{Measurement of high-Q2
  neutral current deep inelastic e+p scattering cross sections with a
  longitudinally polarised positron beam at HERA}},
  \href{http://arxiv.org/abs/1208.6138}{\texttt{arXiv:1208.6138}}.

\bibitem{Whitlow:1991uw}
L.~W. Whitlow, E.~M. Riordan, S.~Dasu, S.~Rock, and A.~Bodek, \textit{{Precise
  measurements of the proton and deuteron structure functions from a global
  analysis of the SLAC deep inelastic electron scattering cross-sections}},
  {\em Phys. Lett.} \textbf{B282} (1992) 475--482.

\bibitem{bcdms2}
\textbf{BCDMS} Collaboration, A.~Benvenuti et~al., \textit{{A High Statistics
  Measurement of the Deuteron Structure Functions F2 (X, $Q^2$) and R From Deep
  Inelastic Muon Scattering at High $Q^2$}},  {\em Phys.Lett.} \textbf{B237}
  (1990) 592.

\bibitem{Feynman:1989dd}
R.~Feynman, \textit{{The behavior of hadrond collisions at extreme energies}},
  .

\bibitem{Aad:2013iua}
\textbf{ATLAS} Collaboration, G.~Aad et~al., \textit{{Measurement of the
  high-mass Drell--Yan differential cross-section in pp collisions at
  $\sqrt{s}=7$ TeV with the ATLAS detector}},
  \href{http://arxiv.org/abs/1305.4192}{\texttt{arXiv:1305.4192}}.

\bibitem{Aad:2011dm}
\textbf{ATLAS} Collaboration, G.~Aad et~al., \textit{{Measurement of the
  inclusive $W^{\pm}$ and $Z/\gamma$ cross sections in the electron and muon
  decay channels in $pp$ collisions at $\sqrt{s} = 7$ TeV with the ATLAS
  detector}},  {\em Phys.Rev.} \textbf{D85} (2012) 072004,
  [\href{http://arxiv.org/abs/1109.5141}{\texttt{arXiv:1109.5141}}].

\bibitem{LHCb-CONF-2012-013}
\textit{{Inclusive low mass Drell-Yan production in the forward region at
  $\sqrt{s}$ = 7 TeV}}, . LHCb-CONF-2012-013.

\bibitem{Vogt:2004mw}
A.~Vogt, S.~Moch, and J.~Vermaseren, \textit{{The Three-loop splitting
  functions in QCD: The Singlet case}},  {\em Nucl.Phys.} \textbf{B691} (2004)
  129--181,
  [\href{http://arxiv.org/abs/hep-ph/0404111}{\texttt{hep-ph/0404111}}].

\bibitem{Moch:2004pa}
S.~Moch, J.~Vermaseren, and A.~Vogt, \textit{{The Three loop splitting
  functions in QCD: The Nonsinglet case}},  {\em Nucl.Phys.} \textbf{B688}
  (2004) 101--134,
  [\href{http://arxiv.org/abs/hep-ph/0403192}{\texttt{hep-ph/0403192}}].

\bibitem{Georgi:1951sr}
H.~Georgi and H.~D. Politzer, \textit{{Electroproduction scaling in an
  asymptotically free theory of strong interactions}},  {\em Phys.Rev.}
  \textbf{D9} (1974) 416--420.

\bibitem{Gross:1974cs}
D.~Gross and F.~Wilczek, \textit{{{Asymptotically free gauge theories}}},  {\em
  Phys.Rev.} \textbf{D9} (1974) 980--993.

\bibitem{Alekhin:2012ig}
S.~Alekhin, J.~Blumlein, and S.~Moch, \textit{{Parton Distribution Functions
  and Benchmark Cross Sections at NNLO}},  {\em Phys.Rev.} \textbf{D86} (2012)
  054009, [\href{http://arxiv.org/abs/1202.2281}{\texttt{arXiv:1202.2281}}].

\bibitem{Alekhin:2010sv}
S.~Alekhin and S.~Moch, \textit{{Heavy-quark deep-inelastic scattering with a
  running mass}},  {\em Phys. Lett.} \textbf{B699} (2011) 345--353,
  [\href{http://arxiv.org/abs/1011.5790}{\texttt{arXiv:1011.5790}}].

\bibitem{Nadolsky:2012ia}
P.~Nadolsky, J.~Gao, M.~Guzzi, J.~Huston, H.-L. Lai, et~al., \textit{{Progress
  in CTEQ-TEA PDF analysis}},
  \href{http://arxiv.org/abs/1206.3321}{\texttt{arXiv:1206.3321}}.

\bibitem{Lai:2010vv}
H.-L. Lai et~al., \textit{{New parton distributions for collider physics}},
  {\em Phys. Rev.} \textbf{D82} (2010) 074024,
  [\href{http://arxiv.org/abs/1007.2241}{\texttt{arXiv:1007.2241}}].

\bibitem{Guzzi:2011ew}
M.~Guzzi, P.~M. Nadolsky, H.-L. Lai, and C.-P. Yuan, \textit{{General-Mass
  Treatment for Deep Inelastic Scattering at Two-Loop Accuracy}},  {\em
  Phys.Rev.} \textbf{D86} (2012) 053005,
  [\href{http://arxiv.org/abs/1108.5112}{\texttt{arXiv:1108.5112}}].

\bibitem{Radescu:2010zz}
\textbf{H1 and ZEUS} Collaboration, V.~Radescu, \textit{{Combination and QCD
  analysis of the HERA inclusive cross sections}},  {\em PoS}
  \textbf{ICHEP2010} (2010) 168.

\bibitem{CooperSarkar:2011aa}
\textbf{ZEUS , H1} Collaboration, A.~Cooper-Sarkar, \textit{{PDF Fits at
  HERA}},  {\em PoS} \textbf{EPS-HEP2011} (2011) 320,
  [\href{http://arxiv.org/abs/1112.2107}{\texttt{arXiv:1112.2107}}].

\bibitem{Aaron:2012qi}
\textbf{H1} Collaboration, F.~Aaron et~al., \textit{{Inclusive Deep Inelastic
  Scattering at High $Q^2$ with Longitudinally Polarised Lepton Beams at
  HERA}},  {\em JHEP} \textbf{1209} (2012) 061,
  [\href{http://arxiv.org/abs/1206.7007}{\texttt{arXiv:1206.7007}}].

\bibitem{Martin:2009iq}
A.~D. Martin, W.~J. Stirling, R.~S. Thorne, and G.~Watt, \textit{{Parton
  distributions for the LHC}},  {\em Eur. Phys. J.} \textbf{C63} (2009)
  189--285, [\href{http://arxiv.org/abs/0901.0002}{\texttt{arXiv:0901.0002}}].

\bibitem{Harland-Lang:2014zoa}
L.~Harland-Lang, A.~Martin, P.~Motylinski, and R.~Thorne, \textit{{Parton
  distributions in the LHC era: MMHT 2014 PDFs}},
  \href{http://arxiv.org/abs/1412.3989}{\texttt{arXiv:1412.3989}}.

\bibitem{Carrazza:2015hva}
S.~Carrazza, J.~I. Latorre, J.~Rojo, and G.~Watt, \textit{{A compression
  algorithm for the combination of PDF sets}},
  \href{http://arxiv.org/abs/1504.06469}{\texttt{arXiv:1504.06469}}.

\bibitem{Carrazza:2015aoa}
S.~Carrazza, S.~Forte, Z.~Kassabov, J.~I. Latorre, and J.~Rojo, \textit{{An
  Unbiased Hessian Representation for Monte Carlo PDFs}},
  \href{http://arxiv.org/abs/1505.06736}{\texttt{arXiv:1505.06736}}.

\bibitem{Bonvini:2015ira}
M.~Bonvini, S.~Marzani, J.~Rojo, L.~Rottoli, M.~Ubiali, R.~D. Ball, V.~Bertone,
  S.~Carrazza, and N.~P. Hartland, \textit{{Parton distributions with threshold
  resummation}},
  \href{http://arxiv.org/abs/1507.01006}{\texttt{arXiv:1507.01006}}.

\bibitem{Forte:2010ta}
S.~Forte, E.~Laenen, P.~Nason, and J.~Rojo, \textit{{Heavy quarks in
  deep-inelastic scattering}},  {\em Nucl. Phys.} \textbf{B834} (2010)
  116--162, [\href{http://arxiv.org/abs/1001.2312}{\texttt{arXiv:1001.2312}}].

\bibitem{JimenezDelgado:2009tv}
P.~Jimenez-Delgado and E.~Reya, \textit{{Variable Flavor Number Parton
  Distributions and Weak Gauge and Higgs Boson Production at Hadron Colliders
  at NNLO of QCD}},  {\em Phys. Rev.} \textbf{D80} (2009) 114011,
  [\href{http://arxiv.org/abs/0909.1711}{\texttt{arXiv:0909.1711}}].

\bibitem{Ball:2011mu}
R.~D. Ball et~al., \textit{{Impact of Heavy Quark Masses on Parton
  Distributions and LHC Phenomenology}},  {\em Nucl. Phys.} \textbf{B849}
  (2011) 296--363,
  [\href{http://arxiv.org/abs/1101.1300}{\texttt{arXiv:1101.1300}}].

\bibitem{Ball:2011uy}
\textbf{NNPDF} Collaboration, R.~D. Ball et~al., \textit{{Unbiased global
  determination of parton distributions and their uncertainties at NNLO and at
  LO}},  {\em Nucl.Phys.} \textbf{B855} (2012) 153--221,
  [\href{http://arxiv.org/abs/1107.2652}{\texttt{arXiv:1107.2652}}].

\bibitem{Aad:2012sb}
\textbf{ATLAS} Collaboration, G.~Aad et~al., \textit{{Determination of the
  strange quark density of the proton from ATLAS measurements of the $W \to
  \ell \nu$ and $Z \to \ell\ell$ cross sections}},  {\em Phys.Rev.Lett.}
  \textbf{109} (2012) 012001,
  [\href{http://arxiv.org/abs/1203.4051}{\texttt{arXiv:1203.4051}}].

\bibitem{Campbell:2006wx}
J.~M. Campbell, J.~W. Huston, and W.~J. Stirling, \textit{{Hard interactions of
  quarks and gluons: A primer for LHC physics}},  {\em Rept. Prog. Phys.}
  \textbf{70} (2007) 89,
  [\href{http://arxiv.org/abs/hep-ph/0611148}{\texttt{hep-ph/0611148}}].

\bibitem{Thorne:2012az}
R.~Thorne, \textit{{The Effect of Changes of Variable Flavour Number Scheme on
  PDFs and Predicted Cross Sections}},  {\em Phys. Rev.} \textbf{D86} (2012)
  074017, [\href{http://arxiv.org/abs/1201.6180}{\texttt{arXiv:1201.6180}}].

\bibitem{CooperSarkar:2007ny}
A.~M. Cooper-Sarkar, \textit{{Including heavy flavour production in PDF fits}},
   \href{http://arxiv.org/abs/0709.0191}{\texttt{arXiv:0709.0191}}.

\bibitem{Thorne:2006qt}
R.~Thorne, \textit{{A Variable-flavor number scheme for NNLO}},  {\em
  Phys.Rev.} \textbf{D73} (2006) 054019,
  [\href{http://arxiv.org/abs/hep-ph/0601245}{\texttt{hep-ph/0601245}}].

\bibitem{Thorne:2011kq}
R.~Thorne and G.~Watt, \textit{{PDF dependence of Higgs cross sections at the
  Tevatron and LHC: Response to recent criticism}},  {\em JHEP} \textbf{1108}
  (2011) 100,
  [\href{http://arxiv.org/abs/1106.5789}{\texttt{arXiv:1106.5789}}].

\bibitem{Martin:2003sk}
A.~D. Martin, R.~G. Roberts, W.~J. Stirling, and R.~S. Thorne,
  \textit{{Uncertainties of predictions from parton distributions. II:
  Theoretical errors}},  {\em Eur. Phys. J.} \textbf{C35} (2004) 325--348,
  [\href{http://arxiv.org/abs/hep-ph/0308087}{\texttt{hep-ph/0308087}}].

\bibitem{Ball:2013gsa}
\textbf{The NNPDF} Collaboration, R.~D. Ball et~al., \textit{{Theoretical
  issues in PDF determination and associated uncertainties}},  {\em Phys.Lett.}
  \textbf{B723} (2013) 330--339,
  [\href{http://arxiv.org/abs/1303.1189}{\texttt{arXiv:1303.1189}}].

\bibitem{Anastasiou:2003ds}
C.~Anastasiou, L.~J. Dixon, K.~Melnikov, and F.~Petriello, \textit{{High
  precision QCD at hadron colliders: Electroweak gauge boson rapidity
  distributions at NNLO}},  {\em Phys. Rev.} \textbf{D69} (2004) 094008,
  [\href{http://arxiv.org/abs/hep-ph/0312266}{\texttt{hep-ph/0312266}}].

\bibitem{cmsewk}
\textbf{CMS} Collaboration, \textit{{Inclusive $W/Z$ cross section at 8 TeV}},
  \href{http://arxiv.org/abs/CMS-PAS-SMP-12-011}{\texttt{CMS-PAS-SMP-12-011}}.

\bibitem{Anastasiou:2011pi}
C.~Anastasiou, S.~Buehler, F.~Herzog, and A.~Lazopoulos, \textit{{Total
  cross-section for Higgs boson hadroproduction with anomalous Standard Model
  interactions}},  {\em JHEP} \textbf{1112} (2011) 058,
  [\href{http://arxiv.org/abs/1107.0683}{\texttt{arXiv:1107.0683}}].

\bibitem{Dittmaier:2011ti}
\textbf{LHC Higgs Cross Section Working Group} Collaboration, S.~Dittmaier
  et~al., \textit{{Handbook of LHC Higgs Cross Sections: 1. Inclusive
  Observables}},
  \href{http://arxiv.org/abs/1101.0593}{\texttt{arXiv:1101.0593}}.

\bibitem{Bolzoni:2010xr}
P.~Bolzoni, F.~Maltoni, S.-O. Moch, and M.~Zaro, \textit{{Higgs production via
  vector-boson fusion at NNLO in QCD}},  {\em Phys.Rev.Lett.} \textbf{105}
  (2010) 011801,
  [\href{http://arxiv.org/abs/1003.4451}{\texttt{arXiv:1003.4451}}].

\bibitem{Brein:2003wg}
O.~Brein, A.~Djouadi, and R.~Harlander, \textit{{NNLO QCD corrections to the
  Higgs-strahlung processes a hadron collider}},  {\em Phys.Lett.}
  \textbf{B579} (2004) 149--156,
  [\href{http://arxiv.org/abs/hep-ph/0307206}{\texttt{hep-ph/0307206}}].

\bibitem{Brein:2012ne}
O.~Brein, R.~V. Harlander, and T.~J. Zirke, \textit{{vh@nnlo - Higgs Strahlung
  at hadron colliders}},
  \href{http://arxiv.org/abs/1210.5347}{\texttt{arXiv:1210.5347}}.

\bibitem{Campbell:2002tg}
J.~Campbell and R.~K. Ellis, \textit{{Next-to-leading order corrections to W +
  2jet and Z + 2jet production at hadron colliders}},  {\em Phys. Rev.}
  \textbf{D65} (2002) 113007,
  [\href{http://arxiv.org/abs/hep-ph/0202176}{\texttt{hep-ph/0202176}}].

\bibitem{Czakon:2011xx}
M.~Czakon and A.~Mitov, \textit{{Top++: a program for the calculation of the
  top-pair cross-section at hadron colliders}},
  \href{http://arxiv.org/abs/1112.5675}{\texttt{arXiv:1112.5675}}.

\bibitem{Baernreuther:2012ws}
P.~Baernreuther, M.~Czakon, and A.~Mitov, \textit{{Percent Level Precision
  Physics at the Tevatron: First Genuine NNLO QCD Corrections to $q \bar{q} \to
  t \bar{t} + X$}},  {\em Phys.Rev.Lett.} \textbf{109} (2012) 132001,
  [\href{http://arxiv.org/abs/1204.5201}{\texttt{arXiv:1204.5201}}].

\bibitem{Czakon:2012pz}
M.~Czakon and A.~Mitov, \textit{{NNLO corrections to top pair production at
  hadron colliders: the quark-gluon reaction}},
  \href{http://arxiv.org/abs/1210.6832}{\texttt{arXiv:1210.6832}}.

\bibitem{Czakon:2012zr}
M.~Czakon and A.~Mitov, \textit{{NNLO corrections to top-pair production at
  hadron colliders: the all-fermionic scattering channels}},  {\em JHEP}
  \textbf{1212} (2012) 054,
  [\href{http://arxiv.org/abs/1207.0236}{\texttt{arXiv:1207.0236}}].

\bibitem{Aliev:2010zk}
M.~Aliev et~al., \textit{{-- HATHOR -- HAdronic Top and Heavy quarks crOss
  section calculatoR}},  {\em Comput. Phys. Commun.} \textbf{182} (2011)
  1034--1046,
  [\href{http://arxiv.org/abs/1007.1327}{\texttt{arXiv:1007.1327}}].

\bibitem{Moch:2012mk}
S.~Moch, P.~Uwer, and A.~Vogt, \textit{{On top-pair hadro-production at
  next-to-next-to-leading order}},  {\em Phys.Lett.} \textbf{B714} (2012)
  48--54, [\href{http://arxiv.org/abs/1203.6282}{\texttt{arXiv:1203.6282}}].

\bibitem{Cacciari:2011hy}
M.~Cacciari, M.~Czakon, M.~L. Mangano, A.~Mitov, and P.~Nason,
  \textit{{Top-pair production at hadron colliders with next-to-next-to-leading
  logarithmic soft-gluon resummation}},  {\em Phys.Lett.} \textbf{B710} (2012)
  612--622, [\href{http://arxiv.org/abs/1111.5869}{\texttt{arXiv:1111.5869}}].

\bibitem{cmstop}
\textbf{CMS} Collaboration, \textit{Cross section measurement in the di-lepton
  channel at 8 tev},
  \href{http://arxiv.org/abs/CMS-PAS-TOP-12-007}{\texttt{CMS-PAS-TOP-12-007}}.

\bibitem{cmstopas}
\textbf{CMS} Collaboration, \textit{{First Determination of the Strong Coupling
  Constant from the $t\bar{t}$ Cross Section}},
  \href{http://arxiv.org/abs/CMS-PAS-TOP-12-022}{\texttt{CMS-PAS-TOP-12-022}}.

\bibitem{Forte:2013wc}
S.~Forte and G.~Watt, \textit{{Progress in the Determination of the Partonic
  Structure of the Proton}},
  \href{http://arxiv.org/abs/1301.6754}{\texttt{arXiv:1301.6754}}.

\bibitem{DeRoeck:2011na}
A.~De~Roeck and R.~Thorne, \textit{{Structure Functions}},  {\em
  Prog.Part.Nucl.Phys.} \textbf{66} (2011) 727--781,
  [\href{http://arxiv.org/abs/1103.0555}{\texttt{arXiv:1103.0555}}].

\bibitem{Mishra:2013una}
K.~Mishra, T.~Becher, L.~Barze, M.~Chiesa, S.~Dittmaier, et~al.,
  \textit{{Electroweak Corrections at High Energies}},
  \href{http://arxiv.org/abs/1308.1430}{\texttt{arXiv:1308.1430}}.

\bibitem{Baur:1998kt}
U.~Baur, S.~Keller, and D.~Wackeroth, \textit{{Electroweak radiative
  corrections to $W$ boson production in hadronic collisions}},  {\em
  Phys.Rev.} \textbf{D59} (1999) 013002,
  [\href{http://arxiv.org/abs/hep-ph/9807417}{\texttt{hep-ph/9807417}}].

\bibitem{Zykunov:2001mn}
V.~Zykunov, \textit{{Electroweak corrections to the observables of W boson
  production at RHIC}},  {\em Eur.Phys.J.direct} \textbf{C3} (2001) 9,
  [\href{http://arxiv.org/abs/hep-ph/0107059}{\texttt{hep-ph/0107059}}].

\bibitem{Dittmaier:2001ay}
S.~Dittmaier and .~Kramer, Michael, \textit{{Electroweak radiative corrections
  to W boson production at hadron colliders}},  {\em Phys.Rev.} \textbf{D65}
  (2002) 073007,
  [\href{http://arxiv.org/abs/hep-ph/0109062}{\texttt{hep-ph/0109062}}].

\bibitem{Baur:2001ze}
U.~Baur, O.~Brein, W.~Hollik, C.~Schappacher, and D.~Wackeroth,
  \textit{{Electroweak radiative corrections to neutral current Drell-Yan
  processes at hadron colliders}},  {\em Phys.Rev.} \textbf{D65} (2002) 033007,
  [\href{http://arxiv.org/abs/hep-ph/0108274}{\texttt{hep-ph/0108274}}].

\bibitem{Baur:2004ig}
U.~Baur and D.~Wackeroth, \textit{{Electroweak radiative corrections to $p
  \bar{p} \to W^\pm \to \ell^\pm \nu$ beyond the pole approximation}},  {\em
  Phys.Rev.} \textbf{D70} (2004) 073015,
  [\href{http://arxiv.org/abs/hep-ph/0405191}{\texttt{hep-ph/0405191}}].

\bibitem{Arbuzov:2007db}
A.~Arbuzov, D.~Bardin, S.~Bondarenko, P.~Christova, L.~Kalinovskaya, et~al.,
  \textit{{One-loop corrections to the Drell--Yan process in SANC. (II). The
  Neutral current case}},  {\em Eur.Phys.J.} \textbf{C54} (2008) 451--460,
  [\href{http://arxiv.org/abs/0711.0625}{\texttt{arXiv:0711.0625}}].

\bibitem{Arbuzov:2005dd}
A.~Arbuzov, D.~Bardin, S.~Bondarenko, P.~Christova, L.~Kalinovskaya, et~al.,
  \textit{{One-loop corrections to the Drell-Yan process in SANC. I. The
  Charged current case}},  {\em Eur.Phys.J.} \textbf{C46} (2006) 407--412,
  [\href{http://arxiv.org/abs/hep-ph/0506110}{\texttt{hep-ph/0506110}}].

\bibitem{Brensing:2007qm}
S.~Brensing, S.~Dittmaier, .~Kramer, Michael, and A.~Muck, \textit{{Radiative
  corrections to $W^-$ boson hadroproduction: Higher-order electroweak and
  supersymmetric effects}},  {\em Phys.Rev.} \textbf{D77} (2008) 073006,
  [\href{http://arxiv.org/abs/0710.3309}{\texttt{arXiv:0710.3309}}].

\bibitem{Balossini:2009sa}
G.~Balossini, G.~Montagna, C.~M. Carloni~Calame, M.~Moretti, O.~Nicrosini,
  et~al., \textit{{Combination of electroweak and QCD corrections to single W
  production at the Fermilab Tevatron and the CERN LHC}},  {\em JHEP}
  \textbf{1001} (2010) 013,
  [\href{http://arxiv.org/abs/0907.0276}{\texttt{arXiv:0907.0276}}].

\bibitem{Dittmaier:2009cr}
S.~Dittmaier and M.~Huber, \textit{{Radiative corrections to the
  neutral-current Drell-Yan process in the Standard Model and its minimal
  supersymmetric extension}},  {\em JHEP} \textbf{1001} (2010) 060,
  [\href{http://arxiv.org/abs/0911.2329}{\texttt{arXiv:0911.2329}}].

\bibitem{Denner:2009gj}
A.~Denner, S.~Dittmaier, T.~Kasprzik, and A.~Muck, \textit{{Electroweak
  corrections to W + jet hadroproduction including leptonic W-boson decays}},
  {\em JHEP} \textbf{0908} (2009) 075,
  [\href{http://arxiv.org/abs/0906.1656}{\texttt{arXiv:0906.1656}}].

\bibitem{Denner:2011vu}
A.~Denner, S.~Dittmaier, T.~Kasprzik, and A.~Muck, \textit{{Electroweak
  corrections to dilepton + jet production at hadron colliders}},  {\em JHEP}
  \textbf{1106} (2011) 069,
  [\href{http://arxiv.org/abs/1103.0914}{\texttt{arXiv:1103.0914}}].

\bibitem{Denner:2012ts}
A.~Denner, S.~Dittmaier, T.~Kasprzik, and A.~Muck, \textit{{Electroweak
  corrections to monojet production at the LHC}},  {\em Eur.Phys.J.}
  \textbf{C73} (2013) 2297,
  [\href{http://arxiv.org/abs/1211.5078}{\texttt{arXiv:1211.5078}}].

\bibitem{Baglio:2013toa}
J.~Baglio, L.~D. Ninh, and M.~M. Weber, \textit{{Massive gauge boson pair
  production at the LHC: a next-to-leading order story}},
  \href{http://arxiv.org/abs/1307.4331}{\texttt{arXiv:1307.4331}}.

\bibitem{Bierweiler:2012kw}
A.~Bierweiler, T.~Kasprzik, H.~Kuhn, and S.~Uccirati, \textit{{Electroweak
  corrections to W-boson pair production at the LHC}},  {\em JHEP}
  \textbf{1211} (2012) 093,
  [\href{http://arxiv.org/abs/1208.3147}{\texttt{arXiv:1208.3147}}].

\bibitem{Luszczak:2013ata}
M.~Luszczak and A.~Szczurek, \textit{{Subleading processes in production of
  $W^+ W^-$ pairs in proton-proton collisions}},
  \href{http://arxiv.org/abs/1309.7201}{\texttt{arXiv:1309.7201}}.

\bibitem{Moretti:2005ut}
S.~Moretti, M.~Nolten, and D.~Ross, \textit{{Weak corrections and high E($T$)
  jets at Tevatron}},  {\em Phys.Rev.} \textbf{D74} (2006) 097301,
  [\href{http://arxiv.org/abs/hep-ph/0503152}{\texttt{hep-ph/0503152}}].

\bibitem{Dittmaier:2012kx}
S.~Dittmaier, A.~Huss, and C.~Speckner, \textit{{Weak radiative corrections to
  dijet production at hadron colliders}},  {\em JHEP} \textbf{1211} (2012) 095,
  [\href{http://arxiv.org/abs/1210.0438}{\texttt{arXiv:1210.0438}}].

\bibitem{Bernreuther:2005is}
W.~Bernreuther, M.~Fuecker, and Z.~Si, \textit{{Mixed QCD and weak corrections
  to top quark pair production at hadron colliders}},  {\em Phys.Lett.}
  \textbf{B633} (2006) 54--60,
  [\href{http://arxiv.org/abs/hep-ph/0508091}{\texttt{hep-ph/0508091}}].

\bibitem{Kuhn:2005it}
J.~H. Kuhn, A.~Scharf, and P.~Uwer, \textit{{Electroweak corrections to
  top-quark pair production in quark-antiquark annihilation}},  {\em
  Eur.Phys.J.} \textbf{C45} (2006) 139--150,
  [\href{http://arxiv.org/abs/hep-ph/0508092}{\texttt{hep-ph/0508092}}].

\bibitem{Hollik:2007sw}
W.~Hollik and M.~Kollar, \textit{{NLO QED contributions to top-pair production
  at hadron collider}},  {\em Phys.Rev.} \textbf{D77} (2008) 014008,
  [\href{http://arxiv.org/abs/0708.1697}{\texttt{arXiv:0708.1697}}].

\bibitem{Hollik:2011ps}
W.~Hollik and D.~Pagani, \textit{{The electroweak contribution to the top quark
  forward-backward asymmetry at the Tevatron}},  {\em Phys.Rev.} \textbf{D84}
  (2011) 093003,
  [\href{http://arxiv.org/abs/1107.2606}{\texttt{arXiv:1107.2606}}].

\bibitem{Kuhn:2013zoa}
J.~Kühn, A.~Scharf, and P.~Uwer, \textit{{Weak Interactions in Top-Quark Pair
  Production at Hadron Colliders: An Update}},
  \href{http://arxiv.org/abs/1305.5773}{\texttt{arXiv:1305.5773}}.

\bibitem{DeRujula:1979jj}
A.~De~Rujula, R.~Petronzio, and A.~Savoy-Navarro, \textit{{Radiative
  Corrections to High-Energy Neutrino Scattering}},  {\em Nucl.Phys.}
  \textbf{B154} (1979) 394.

\bibitem{Kripfganz:1988bd}
J.~Kripfganz and H.~Perlt, \textit{{Electroweak radiative corrections and quark
  mass singularities}},  {\em Z.Phys.} \textbf{C41} (1988) 319--321.

\bibitem{Blumlein:1989gk}
J.~Blumlein, \textit{{Leading log radiative corrections to deep inelastic
  neutra and charged current scattering at HERA}},  {\em Z.Phys.} \textbf{C47}
  (1990) 89--94.

\bibitem{Salam:2008qg}
G.~P. Salam and J.~Rojo, \textit{{A Higher Order Perturbative Parton Evolution
  Toolkit (HOPPET)}},  {\em Comput. Phys. Commun.} \textbf{180} (2009)
  120--156, [\href{http://arxiv.org/abs/0804.3755}{\texttt{arXiv:0804.3755}}].

\bibitem{Cafarella:2008du}
A.~Cafarella, C.~Coriano, and M.~Guzzi, \textit{{Precision Studies of the NNLO
  DGLAP Evolution at the LHC with CANDIA}},  {\em Comput.Phys.Commun.}
  \textbf{179} (2008) 665--684,
  [\href{http://arxiv.org/abs/0803.0462}{\texttt{arXiv:0803.0462}}].

\bibitem{Botje:2010ay}
M.~Botje, \textit{{QCDNUM: Fast QCD Evolution and Convolution}},  {\em
  Comput.Phys.Commun.} \textbf{182} (2011) 490--532,
  [\href{http://arxiv.org/abs/1005.1481}{\texttt{arXiv:1005.1481}}].

\bibitem{Ratcliffe:2000kp}
P.~G. Ratcliffe, \textit{{A matrix approach to numerical solution of the DGLAP
  evolution equations}},  {\em Phys.Rev.} \textbf{D63} (2001) 116004,
  [\href{http://arxiv.org/abs/hep-ph/0012376}{\texttt{hep-ph/0012376}}].

\bibitem{Schoeffel:1998tz}
L.~Schoeffel, \textit{{An Elegant and fast method to solve QCD evolution
  equations, application to the determination of the gluon content of the
  pomeron}},  {\em Nucl.Instrum.Meth.} \textbf{A423} (1999) 439--445.

\bibitem{Pascaud:2001bi}
C.~Pascaud and F.~Zomer, \textit{{A Fast and precise method to solve the
  Altarelli-Parisi equations in x space}},
  \href{http://arxiv.org/abs/hep-ph/0104013}{\texttt{hep-ph/0104013}}.

\bibitem{pegasus}
A.~Vogt, \textit{Efficient evolution of unpolarized and polarized parton
  distributions with qcd-pegasus},  {\em Comput. Phys. Commun.} \textbf{170}
  (2005) 65--92,
  [\href{http://arxiv.org/abs/hep-ph/0408244}{\texttt{hep-ph/0408244}}].

\bibitem{Kosower:1997hg}
D.~A. Kosower, \textit{{Evolution of parton distributions}},  {\em Nucl.Phys.}
  \textbf{B506} (1997) 439--467,
  [\href{http://arxiv.org/abs/hep-ph/9706213}{\texttt{hep-ph/9706213}}].

\bibitem{Spiesberger:1994dm}
H.~Spiesberger, \textit{{QED radiative corrections for parton distributions}},
  {\em Phys.Rev.} \textbf{D52} (1995) 4936--4940,
  [\href{http://arxiv.org/abs/hep-ph/9412286}{\texttt{hep-ph/9412286}}].

\bibitem{Roth:2004ti}
M.~Roth and S.~Weinzierl, \textit{{QED corrections to the evolution of parton
  distributions}},  {\em Phys.Lett.} \textbf{B590} (2004) 190--198,
  [\href{http://arxiv.org/abs/hep-ph/0403200}{\texttt{hep-ph/0403200}}].

\bibitem{Martin:2004dh}
A.~D. Martin, R.~G. Roberts, W.~J. Stirling, and R.~S. Thorne, \textit{{Parton
  distributions incorporating QED contributions}},  {\em Eur. Phys. J.}
  \textbf{C39} (2005) 155--161,
  [\href{http://arxiv.org/abs/hep-ph/0411040}{\texttt{hep-ph/0411040}}].

\bibitem{Weinzierl:2002mv}
S.~Weinzierl, \textit{{Fast evolution of parton distributions}},  {\em
  Comput.Phys.Commun.} \textbf{148} (2002) 314--326,
  [\href{http://arxiv.org/abs/hep-ph/0203112}{\texttt{hep-ph/0203112}}].

\bibitem{Buckley:2014ana}
A.~Buckley, J.~Ferrando, S.~Lloyd, K.~Nordstrom, B.~Page, et~al.,
  \textit{{LHAPDF6: parton density access in the LHC precision era}},
  \href{http://arxiv.org/abs/1412.7420}{\texttt{arXiv:1412.7420}}.

\bibitem{DelDebbio:2007ee}
\textbf{The NNPDF collaboration} Collaboration, L.~Del~Debbio, S.~Forte, J.~I.
  Latorre, A.~Piccione, and J.~Rojo, \textit{{Neural network determination of
  parton distributions: The nonsinglet case}},  {\em JHEP} \textbf{03} (2007)
  039, [\href{http://arxiv.org/abs/hep-ph/0701127}{\texttt{hep-ph/0701127}}].

\bibitem{Ball:2008by}
\textbf{The NNPDF} Collaboration, R.~D. Ball et~al., \textit{{A determination
  of parton distributions with faithful uncertainty estimation}},  {\em Nucl.
  Phys.} \textbf{B809} (2009) 1--63,
  [\href{http://arxiv.org/abs/0808.1231}{\texttt{arXiv:0808.1231}}].

\bibitem{ap}
G.~Altarelli and G.~Parisi, \textit{Asymptotic freedom in parton language},
  {\em Nucl. Phys.} \textbf{B126} (1977) 298.

\bibitem{gl}
V.~N. Gribov and L.~N. Lipatov, \textit{Deep inelastic $ep$ scattering in
  perturbation theory},  {\em Sov. J. Nucl. Phys.} \textbf{15} (1972) 438--450.

\bibitem{dok}
Y.~L. Dokshitzer, \textit{Calculation of the structure functions for deep
  inelastic scattering and $e^+e^-$ annihilation by perturbation theory in
  quantum chromodynamics. (in russian)},  {\em Sov. Phys. JETP} \textbf{46}
  (1977) 641--653.

\bibitem{gNLOa}
E.~Floratos, D.~Ross, and C.~Sachrajda {\em Nucl.Phys.} \textbf{B129} (1977)
  66.

\bibitem{gNLOb}
E.~Floratos, D.~Ross, and C.~Sachrajda {\em Nucl.Phys.} \textbf{B152} (1979)
  493.

\bibitem{gNLOc}
A.~Gonzalez-Arroyo, C.~Lopez, and F.~Yndurain {\em Nucl.Phys.} \textbf{B153}
  (1979) 161.

\bibitem{gNLOe}
E.~Floratos, C.~Kounnas, and R.~Lacaze {\em Nucl.Phys.} \textbf{B192} (1981)
  417.

\bibitem{gNLOf}
G.~Curci, W.~Furmanski, and R.~Petronzio {\em Nucl.Phys.} \textbf{B175} (1980)
  27.

\bibitem{gNNLOa}
S.~Moch, J.~Vermaseren, and A.~Vogt {\em Nucl.Phys.} \textbf{B688} (2004) 101.

\bibitem{gNNLOb}
S.~Moch, J.~Vermaseren, and A.~Vogt {\em Phys. Lett.} \textbf{B691} (2004) 129.

\bibitem{Buza:1995ie}
M.~Buza, Y.~Matiounine, J.~Smith, R.~Migneron, and W.~L. van Neerven,
  \textit{{Heavy quark coefficient functions at asymptotic values $Q~2 \gg
  m~2$}},  {\em Nucl. Phys.} \textbf{B472} (1996) 611--658,
  [\href{http://arxiv.org/abs/hep-ph/9601302}{\texttt{hep-ph/9601302}}].

\bibitem{Sadykov:2014aua}
R.~Sadykov, \textit{{Impact of QED radiative corrections on Parton Distribution
  Functions}},
  \href{http://arxiv.org/abs/1401.1133}{\texttt{arXiv:1401.1133}}.

\bibitem{Dittmar:2005ed}
M.~Dittmar et~al., \textit{{Working Group I: Parton distributions: Summary
  report for the HERA LHC Workshop Proceedings}},
  \href{http://arxiv.org/abs/hep-ph/0511119}{\texttt{hep-ph/0511119}}.

\bibitem{Blumlein:1996rp}
J.~Blumlein, S.~Riemersma, M.~Botje, C.~Pascaud, F.~Zomer, et~al., \textit{{A
  Detailed comparison of NLO QCD evolution codes}},
  \href{http://arxiv.org/abs/hep-ph/9609400}{\texttt{hep-ph/9609400}}.

\bibitem{Carli:2010rw}
T.~Carli, D.~Clements, A.~Cooper-Sarkar, C.~Gwenlan, G.~P. Salam, et~al.,
  \textit{{A posteriori inclusion of parton density functions in NLO QCD
  final-state calculations at hadron colliders: The APPLGRID Project}},  {\em
  Eur.Phys.J.} \textbf{C66} (2010) 503--524,
  [\href{http://arxiv.org/abs/0911.2985}{\texttt{arXiv:0911.2985}}].

\bibitem{Agashe:2014kda}
\textbf{Particle Data Group} Collaboration, K.~Olive et~al., \textit{{Review of
  Particle Physics}},  {\em Chin.Phys.} \textbf{C38} (2014) 090001.

\bibitem{Ball:2010de}
\textbf{{The NNPDF collaboration}} Collaboration, R.~D. Ball et~al., \textit{{A
  first unbiased global NLO determination of parton distributions and their
  uncertainties}},  {\em Nucl. Phys.} \textbf{B838} (2010) 136--206,
  [\href{http://arxiv.org/abs/1002.4407}{\texttt{arXiv:1002.4407}}].

\bibitem{Bertone:2014zva}
V.~Bertone, R.~Frederix, S.~Frixione, J.~Rojo, and M.~Sutton, \textit{{aMCfast:
  automation of fast NLO computations for PDF fits}},  {\em JHEP} \textbf{1408}
  (2014) 166,
  [\href{http://arxiv.org/abs/1406.7693}{\texttt{arXiv:1406.7693}}].

\bibitem{Aad:2011fc}
\textbf{ATLAS} Collaboration, G.~Aad et~al., \textit{{Measurement of inclusive
  jet and dijet production in $pp$ collisions at $\sqrt{s} = 7$ TeV using the
  ATLAS detector}},  {\em Phys. Rev.} \textbf{D86} (2012) 014022,
  [\href{http://arxiv.org/abs/1112.6297}{\texttt{arXiv:1112.6297}}].

\bibitem{Bertone:2015cwa}
V.~Bertone, S.~Carrazza, and E.~R. Nocera, \textit{{Reference results for
  time-like evolution up to $ \mathcal{O}\left({\alpha}_s^3\right) $}},  {\em
  JHEP} \textbf{1503} (2015) 046,
  [\href{http://arxiv.org/abs/1501.00494}{\texttt{arXiv:1501.00494}}].

\bibitem{dagos}
G.~D'Agostini, {\em Bayesian reasoning in data analysis: A critical
  introduction}.
\newblock {World Scientific}, 2003.

\bibitem{Kluge:2006xs}
T.~Kluge, K.~Rabbertz, and M.~Wobisch, \textit{{Fast pQCD calculations for PDF
  fits}},  \href{http://arxiv.org/abs/hep-ph/0609285}{\texttt{hep-ph/0609285}}.

\bibitem{Montana:1989}
D.~J. Montana and L.~Davis, \textit{{Training Feedforward Neural Networks Using
  Genetic Algorithms}},  in {\em {Proceedings of the 11th International Joint
  Conference on Artificial Intelligence - Volume 1}}, IJCAI'89, (San Francisco,
  CA, USA), pp.~762--767, Morgan Kaufmann Publishers Inc., 1989.

\bibitem{Arneodo:1996kd}
\textbf{New Muon} Collaboration, M.~Arneodo et~al., \textit{{Accurate
  measurement of $F_2(d)/F_2(p)$ and $R(d)-R(p)$}},  {\em Nucl. Phys.}
  \textbf{B487} (1997) 3--26,
  [\href{http://arxiv.org/abs/hep-ex/9611022}{\texttt{hep-ex/9611022}}].

\bibitem{Arneodo:1996qe}
\textbf{New Muon} Collaboration, M.~Arneodo et~al., \textit{{Measurement of the
  proton and deuteron structure functions, $F_2(p)$ and $F_2(d)$, and of the
  ratio $\sigma(L)/\sigma(T)$}},  {\em Nucl. Phys.} \textbf{B483} (1997) 3--43,
  [\href{http://arxiv.org/abs/hep-ph/9610231}{\texttt{hep-ph/9610231}}].

\bibitem{bcdms1}
\textbf{BCDMS} Collaboration, A.~C. Benvenuti et~al., \textit{{A high
  statistics measurement of the proton structure functions $F_2 (x, Q^2)$ and
  $R$ from deep inelastic muon scattering at high $Q^2$}},  {\em Phys. Lett.}
  \textbf{B223} (1989) 485.

\bibitem{H1:2009wt}
\textbf{H1 and ZEUS} Collaboration, A.~F. et~al., \textit{{Combined Measurement
  and QCD Analysis of the Inclusive ep Scattering Cross Sections at HERA}},
  \href{http://arxiv.org/abs/0911.0884}{\texttt{arXiv:0911.0884}}.

\bibitem{Breitweg:1999ad}
\textbf{ZEUS} Collaboration, J.~Breitweg et~al., \textit{{Measurement of
  $D^{*\pm}$ production and the charm contribution to $F_2$ in deep inelastic
  scattering at HERA}},  {\em Eur. Phys. J.} \textbf{C12} (2000) 35--52,
  [\href{http://arxiv.org/abs/hep-ex/9908012}{\texttt{hep-ex/9908012}}].

\bibitem{Chekanov:2003rb}
\textbf{ZEUS} Collaboration, S.~Chekanov et~al., \textit{{Measurement of
  $D^{*\pm}$ production in deep inelastic $e^\pm p$ scattering at HERA}},  {\em
  Phys. Rev.} \textbf{D69} (2004) 012004,
  [\href{http://arxiv.org/abs/hep-ex/0308068}{\texttt{hep-ex/0308068}}].

\bibitem{Chekanov:2008yd}
\textbf{ZEUS} Collaboration, S.~Chekanov et~al., \textit{{Measurement of
  $D^\pm$ and D0 production in deep inelastic scattering using a lifetime tag
  at HERA}},  {\em Eur. Phys. J.} \textbf{C63} (2009) 171--188,
  [\href{http://arxiv.org/abs/0812.3775}{\texttt{arXiv:0812.3775}}].

\bibitem{Chekanov:2009kj}
\textbf{ZEUS} Collaboration, S.~Chekanov et~al., \textit{{Measurement of charm
  and beauty production in deep inelastic ep scattering from decays into muons
  at HERA}},  {\em Eur. Phys. J.} \textbf{C65} (2010) 65--79,
  [\href{http://arxiv.org/abs/0904.3487}{\texttt{arXiv:0904.3487}}].

\bibitem{Adloff:2001zj}
\textbf{H1} Collaboration, C.~Adloff et~al., \textit{{Measurement of $D^{*\pm}$
  meson production and $F_2(c)$ in deep inelastic scattering at HERA}},  {\em
  Phys. Lett.} \textbf{B528} (2002) 199--214,
  [\href{http://arxiv.org/abs/hep-ex/0108039}{\texttt{hep-ex/0108039}}].

\bibitem{Collaboration:2009jy}
\textbf{H1} Collaboration, F.~D. Aaron et~al., \textit{{Measurement of the D*
  Meson Production Cross Section and $F_2^{ccbar}$, at High $Q^2$, in ep
  Scattering at HERA}},  {\em Phys. Lett.} \textbf{B686} (2010) 91--100,
  [\href{http://arxiv.org/abs/0911.3989}{\texttt{arXiv:0911.3989}}].

\bibitem{H1F2c10:2009ut}
\textbf{H1} Collaboration, F.~D. Aaron et~al., \textit{{Measurement of the
  Charm and Beauty Structure Functions using the H1 Vertex Detector at HERA}},
  {\em Eur. Phys. J.} \textbf{C65} (2010) 89--109,
  [\href{http://arxiv.org/abs/0907.2643}{\texttt{arXiv:0907.2643}}].

\bibitem{Chekanov:2009gm}
\textbf{ZEUS} Collaboration, S.~Chekanov et~al., \textit{{Measurement of
  high-$Q^2$ neutral current deep inelastic $e^- p$ scattering cross sections
  with a longitudinally polarised electron beam at HERA}},  {\em Eur. Phys. J.}
  \textbf{C62} (2009) 625--658,
  [\href{http://arxiv.org/abs/0901.2385}{\texttt{arXiv:0901.2385}}].

\bibitem{Chekanov:2008aa}
\textbf{ZEUS} Collaboration, S.~Chekanov et~al., \textit{Measurement of charged
  current deep inelastic scattering cross sections with a longitudinally
  polarised electron beam at hera},  {\em Eur. Phys. J.} \textbf{C61} (2009)
  223--235, [\href{http://arxiv.org/abs/0812.4620}{\texttt{arXiv:0812.4620}}].

\bibitem{Onengut:2005kv}
\textbf{CHORUS} Collaboration, G.~Onengut et~al., \textit{{Measurement of
  nucleon structure functions in neutrino scattering}},  {\em Phys. Lett.}
  \textbf{B632} (2006) 65--75.

\bibitem{Goncharov:2001qe}
\textbf{NuTeV} Collaboration, M.~Goncharov et~al., \textit{{Precise measurement
  of dimuon production cross-sections in nu/mu Fe and anti-nu/mu Fe deep
  inelastic scattering at the Tevatron}},  {\em Phys. Rev.} \textbf{D64} (2001)
  112006,
  [\href{http://arxiv.org/abs/hep-ex/0102049}{\texttt{hep-ex/0102049}}].

\bibitem{MasonPhD}
D.~A. Mason, \textit{{Measurement of the strange - antistrange asymmetry at NLO
  in QCD from NuTeV dimuon data}}, . FERMILAB-THESIS-2006-01.

\bibitem{Moreno:1990sf}
G.~Moreno et~al., \textit{{Dimuon production in proton - copper collisions at
  $\sqrt{s}$ = 38.8-GeV}},  {\em Phys. Rev.} \textbf{D43} (1991) 2815--2836.

\bibitem{Webb:2003ps}
\textbf{NuSea} Collaboration, J.~C. Webb et~al., \textit{{Absolute Drell-Yan
  dimuon cross sections in 800-GeV/c p p and p d collisions}},
  \href{http://arxiv.org/abs/hep-ex/0302019}{\texttt{hep-ex/0302019}}.

\bibitem{Webb:2003bj}
J.~C. Webb, \textit{{Measurement of continuum dimuon production in 800-GeV/c
  proton nucleon collisions}},
  \href{http://arxiv.org/abs/hep-ex/0301031}{\texttt{hep-ex/0301031}}.

\bibitem{Towell:2001nh}
\textbf{FNAL E866/NuSea} Collaboration, R.~S. Towell et~al., \textit{{Improved
  measurement of the anti-d/anti-u asymmetry in the nucleon sea}},  {\em Phys.
  Rev.} \textbf{D64} (2001) 052002,
  [\href{http://arxiv.org/abs/hep-ex/0103030}{\texttt{hep-ex/0103030}}].

\bibitem{Aaltonen:2009ta}
\textbf{CDF} Collaboration, T.~Aaltonen et~al., \textit{{Direct Measurement of
  the $W$ Production Charge Asymmetry in $p\bar{p}$ Collisions at $\sqrt{s} =
  1.96$ TeV}},  {\em Phys. Rev. Lett.} \textbf{102} (2009) 181801,
  [\href{http://arxiv.org/abs/0901.2169}{\texttt{arXiv:0901.2169}}].

\bibitem{Aaltonen:2010zza}
\textbf{CDF} Collaboration, T.~A. Aaltonen et~al., \textit{{Measurement of
  $d\sigma/dy$ of Drell-Yan $e^+e^-$ pairs in the $Z$ Mass Region from
  $p\bar{p}$ Collisions at $\sqrt{s}=1.96$ TeV}},  {\em Phys. Lett.}
  \textbf{B692} (2010) 232--239,
  [\href{http://arxiv.org/abs/0908.3914}{\texttt{arXiv:0908.3914}}].

\bibitem{Abazov:2007jy}
\textbf{D0} Collaboration, V.~M. Abazov et~al., \textit{{Measurement of the
  shape of the boson rapidity distribution for $p \bar{p} \to Z/gamma^* \to
  e^{+} e^{-}$ + $X$ events produced at $\sqrt{s}$ of 1.96-TeV}},  {\em Phys.
  Rev.} \textbf{D76} (2007) 012003,
  [\href{http://arxiv.org/abs/hep-ex/0702025}{\texttt{hep-ex/0702025}}].

\bibitem{Aaltonen:2008eq}
\textbf{CDF} Collaboration, T.~Aaltonen et~al., \textit{{Measurement of the
  Inclusive Jet Cross Section at the Fermilab Tevatron p-pbar Collider Using a
  Cone-Based Jet Algorithm}},  {\em Phys. Rev.} \textbf{D78} (2008) 052006,
  [\href{http://arxiv.org/abs/0807.2204}{\texttt{arXiv:0807.2204}}].

\bibitem{D0:2008hua}
\textbf{D0} Collaboration, V.~M. Abazov et~al., \textit{{Measurement of the
  inclusive jet cross-section in $p \bar{p}$ collisions at $\sqrt{s}$ =1.96
  TeV}},  {\em Phys. Rev. Lett.} \textbf{101} (2008) 062001,
  [\href{http://arxiv.org/abs/0802.2400}{\texttt{arXiv:0802.2400}}].

\bibitem{Chatrchyan:2012xt}
\textbf{CMS} Collaboration, S.~Chatrchyan et~al., \textit{{Measurement of the
  electron charge asymmetry in inclusive W production in pp collisions at
  $\sqrt{s} = 7$ TeV}},  {\em Phys.Rev.Lett.} \textbf{109} (2012) 111806,
  [\href{http://arxiv.org/abs/1206.2598}{\texttt{arXiv:1206.2598}}].

\bibitem{Aaij:2012vn}
\textbf{LHCb} Collaboration, R.~Aaij et~al., \textit{{Inclusive $W$ and $Z$
  production in the forward region at $\sqrt{s} = 7$ TeV}},
  \href{http://arxiv.org/abs/1204.1620}{\texttt{arXiv:1204.1620}}.

\bibitem{CMSdy}
\textbf{CMS} Collaboration, \textit{{Measurement of the differential and
  double-differential Drell-Yan cross section in proton-proton collisions at 7
  TeV}},
  \href{http://arxiv.org/abs/CMS-PAS-SMP-13-003}{\texttt{CMS-PAS-SMP-13-003}}.

\bibitem{Aad:2013lpa}
\textbf{ATLAS} Collaboration, G.~Aad et~al., \textit{{Measurement of the
  inclusive jet cross section in pp collisions at $\sqrt{s}=2.76$ TeV and
  comparison to the inclusive jet cross section at $\sqrt{s}=7$ TeV using the
  ATLAS detector}},  {\em Eur.Phys.J.} \textbf{C73} (2013) 2509,
  [\href{http://arxiv.org/abs/1304.4739}{\texttt{arXiv:1304.4739}}].

\bibitem{Chatrchyan:2012bja}
\textbf{CMS} Collaboration, S.~Chatrchyan et~al., \textit{{Measurements of
  differential jet cross sections in proton-proton collisions at $\sqrt{s}=7$
  TeV with the CMS detector}},  {\em Phys.Rev.} \textbf{D87} (2013), no.~11
  112002, [\href{http://arxiv.org/abs/1212.6660}{\texttt{arXiv:1212.6660}}].

\bibitem{MCFMurl}
J.~M. Campbell, H.~B. Hartanto, and C.~Williams, \textit{{Next-to-leading order
  predictions for $Z \gamma$+jet and Z $\gamma \gamma$ final states at the
  LHC}},  {\em JHEP} \textbf{1211} (2012) 162,
  [\href{http://arxiv.org/abs/1208.0566}{\texttt{arXiv:1208.0566}}].

\bibitem{Campbell:2004ch}
J.~Campbell, R.~K. Ellis, and F.~Tramontano, \textit{{Single top production and
  decay at next-to-leading order}},  {\em Phys. Rev.} \textbf{D70} (2004)
  094012,
  [\href{http://arxiv.org/abs/hep-ph/0408158}{\texttt{hep-ph/0408158}}].

\bibitem{Catani:2010en}
S.~Catani, G.~Ferrera, and M.~Grazzini, \textit{{W Boson Production at Hadron
  Colliders: The Lepton Charge Asymmetry in NNLO QCD}},  {\em JHEP}
  \textbf{1005} (2010) 006,
  [\href{http://arxiv.org/abs/1002.3115}{\texttt{arXiv:1002.3115}}].

\bibitem{CMS:2011mea}
\textbf{CMS} Collaboration, S.~Chatrchyan et~al., \textit{{Measurement of the
  Inclusive Jet Cross Section in pp Collisions at $\sqrt{s} = 7$ TeV}},  {\em
  Phys.Rev.Lett.} \textbf{107} (2011) 132001,
  [\href{http://arxiv.org/abs/1106.0208}{\texttt{arXiv:1106.0208}}].

\bibitem{Chatrchyan:2011qta}
\textbf{CMS} Collaboration, S.~Chatrchyan et~al., \textit{{Measurement of the
  differential dijet production cross section in proton-proton collisions at
  $\sqrt{s}=7$ TeV}},  {\em Phys.Lett.} \textbf{B700} (2011) 187--206,
  [\href{http://arxiv.org/abs/1104.1693}{\texttt{arXiv:1104.1693}}].

\bibitem{Nagy:2003tz}
Z.~Nagy, \textit{{Next-to-leading order calculation of three-jet observables in
  hadron hadron collision}},  {\em Phys. Rev.} \textbf{D68} (2003) 094002,
  [\href{http://arxiv.org/abs/hep-ph/0307268}{\texttt{hep-ph/0307268}}].

\bibitem{Ellis:1990ek}
S.~D. Ellis, Z.~Kunszt, and D.~E. Soper, \textit{{{The one jet inclusive
  cross-section at order $\alpha_s^{3}$ quarks and gluons}}},  {\em
  Phys.Rev.Lett.} \textbf{64} (1990) 2121.

\bibitem{Gao:2012he}
J.~Gao, Z.~Liang, D.~E. Soper, H.-L. Lai, P.~M. Nadolsky, et~al.,
  \textit{{MEKS: a program for computation of inclusive jet cross sections at
  hadron colliders}},
  \href{http://arxiv.org/abs/1207.0513}{\texttt{arXiv:1207.0513}}.

\bibitem{Wobisch:2011ij}
\textbf{fastNLO} Collaboration, M.~Wobisch, D.~Britzger, T.~Kluge, K.~Rabbertz,
  and F.~Stober, \textit{{Theory-Data Comparisons for Jet Measurements in
  Hadron-Induced Processes}},
  \href{http://arxiv.org/abs/1109.1310}{\texttt{arXiv:1109.1310}}.

\bibitem{Currie:2013dwa}
J.~Currie, A.~Gehrmann-De~Ridder, E.~Glover, and J.~Pires, \textit{{NNLO QCD
  corrections to jet production at hadron colliders from gluon scattering}},
  {\em JHEP} \textbf{1401} (2014) 110,
  [\href{http://arxiv.org/abs/1310.3993}{\texttt{arXiv:1310.3993}}].

\bibitem{Ridder:2013mf}
A.~Gehrmann-De~Ridder, T.~Gehrmann, E.~Glover, and J.~Pires, \textit{{Second
  order QCD corrections to jet production at hadron colliders: the all-gluon
  contribution}},  {\em Phys.Rev.Lett.} \textbf{110} (2013), no.~16 162003,
  [\href{http://arxiv.org/abs/1301.7310}{\texttt{arXiv:1301.7310}}].

\bibitem{Kidonakis:2000gi}
N.~Kidonakis and J.~F. Owens, \textit{{Effects of higher-order threshold
  corrections in high-E(T) jet production}},  {\em Phys. Rev.} \textbf{D63}
  (2001) 054019,
  [\href{http://arxiv.org/abs/hep-ph/0007268}{\texttt{hep-ph/0007268}}].

\bibitem{deFlorian:2013qia}
D.~de~Florian, P.~Hinderer, A.~Mukherjee, F.~Ringer, and W.~Vogelsang,
  \textit{{Approximate next-to-next-to-leading order corrections to hadronic
  jet production}},  {\em Phys.Rev.Lett.} \textbf{112} (2014) 082001,
  [\href{http://arxiv.org/abs/1310.7192}{\texttt{arXiv:1310.7192}}].

\bibitem{Carrazza:2014hra}
S.~Carrazza and J.~Pires, \textit{{Perturbative QCD description of jet data
  from LHC Run-I and Tevatron Run-II}},  {\em JHEP} \textbf{1410} (2014) 145,
  [\href{http://arxiv.org/abs/1407.7031}{\texttt{arXiv:1407.7031}}].

\bibitem{Aad:2010ad}
\textbf{ATLAS} Collaboration, G.~Aad et~al., \textit{{Measurement of inclusive
  jet and dijet cross sections in proton-proton collisions at 7 TeV
  centre-of-mass energy with the ATLAS detector}},  {\em Eur.Phys.J.}
  \textbf{C71} (2011) 1512,
  [\href{http://arxiv.org/abs/1009.5908}{\texttt{arXiv:1009.5908}}].

\bibitem{Cacciari:2008gp}
M.~Cacciari, G.~P. Salam, and G.~Soyez, \textit{{The Anti-k(t) jet clustering
  algorithm}},  {\em JHEP} \textbf{0804} (2008) 063,
  [\href{http://arxiv.org/abs/0802.1189}{\texttt{arXiv:0802.1189}}].

\bibitem{Dasgupta:2007wa}
M.~Dasgupta, L.~Magnea, and G.~P. Salam, \textit{{Non-perturbative QCD effects
  in jets at hadron colliders}},  {\em JHEP} \textbf{02} (2008) 055,
  [\href{http://arxiv.org/abs/0712.3014}{\texttt{arXiv:0712.3014}}].

\bibitem{Cacciari:2008gd}
M.~Cacciari, J.~Rojo, G.~P. Salam, and G.~Soyez, \textit{{Quantifying the
  performance of jet definitions for kinematic reconstruction at the LHC}},
  {\em JHEP} \textbf{12} (2008) 032,
  [\href{http://arxiv.org/abs/0810.1304}{\texttt{arXiv:0810.1304}}].

\bibitem{Forte:2010dt}
S.~Forte, \textit{{Parton distributions at the dawn of the LHC}},  {\em Acta
  Phys.Polon.} \textbf{B41} (2010) 2859--2920,
  [\href{http://arxiv.org/abs/1011.5247}{\texttt{arXiv:1011.5247}}].

\bibitem{Campbell:2013qaa}
J.~M. Campbell et~al., \textit{{Working Group Report: Quantum Chromodynamics}},
   in {\em {Community Summer Study 2013: Snowmass on the Mississippi (CSS2013)
  Minneapolis, MN, USA, July 29-August 6, 2013}}, 2013.
\newblock \href{http://arxiv.org/abs/1310.5189}{\texttt{arXiv:1310.5189}}.

\bibitem{Ciafaloni:2001mu}
M.~Ciafaloni, P.~Ciafaloni, and D.~Comelli, \textit{{Towards collinear
  evolution equations in electroweak theory}},  {\em Phys.Rev.Lett.}
  \textbf{88} (2002) 102001,
  [\href{http://arxiv.org/abs/hep-ph/0111109}{\texttt{hep-ph/0111109}}].

\bibitem{Ciafaloni:2005fm}
P.~Ciafaloni and D.~Comelli, \textit{{Electroweak evolution equations}},  {\em
  JHEP} \textbf{0511} (2005) 022,
  [\href{http://arxiv.org/abs/hep-ph/0505047}{\texttt{hep-ph/0505047}}].

\bibitem{Ball:2010gb}
\textbf{The NNPDF} Collaboration, R.~D. Ball et~al., \textit{{Reweighting
  NNPDFs: the W lepton asymmetry}},  {\em Nucl. Phys.} \textbf{B849} (2011)
  112--143, [\href{http://arxiv.org/abs/1012.0836}{\texttt{arXiv:1012.0836}}].

\bibitem{Ball:2011gg}
R.~D. Ball, V.~Bertone, F.~Cerutti, L.~Del~Debbio, S.~Forte, et~al.,
  \textit{{Reweighting and Unweighting of Parton Distributions and the LHC W
  lepton asymmetry data}},  {\em Nucl.Phys.} \textbf{B855} (2012) 608--638,
  [\href{http://arxiv.org/abs/1108.1758}{\texttt{arXiv:1108.1758}}].

\bibitem{Diener:2005me}
K.-P. Diener, S.~Dittmaier, and W.~Hollik, \textit{{Electroweak higher-order
  effects and theoretical uncertainties in deep-inelastic neutrino
  scattering}},  {\em Phys.Rev.} \textbf{D72} (2005) 093002,
  [\href{http://arxiv.org/abs/hep-ph/0509084}{\texttt{hep-ph/0509084}}].

\bibitem{Altarelli:1998gn}
G.~Altarelli, S.~Forte, and G.~Ridolfi, \textit{{On positivity of parton
  distributions}},  {\em Nucl. Phys.} \textbf{B534} (1998) 277--296,
  [\href{http://arxiv.org/abs/hep-ph/9806345}{\texttt{hep-ph/9806345}}].

\bibitem{Londergan:2003pq}
J.~Londergan and A.~W. Thomas, \textit{{Charge symmetry violating contributions
  to neutrino reactions}},  {\em Phys.Lett.} \textbf{B558} (2003) 132--140,
  [\href{http://arxiv.org/abs/hep-ph/0301147}{\texttt{hep-ph/0301147}}].

\bibitem{Ball:2009qv}
\textbf{The NNPDF collaboration} Collaboration, R.~D. Ball et~al.,
  \textit{{Fitting Parton Distribution Data with Multiplicative Normalization
  Uncertainties}},  {\em JHEP} \textbf{05} (2010) 075,
  [\href{http://arxiv.org/abs/0912.2276}{\texttt{arXiv:0912.2276}}].

\bibitem{Demartin:2010er}
F.~Demartin, S.~Forte, E.~Mariani, J.~Rojo, and A.~Vicini, \textit{{The impact
  of PDF and alphas uncertainties on Higgs Production in gluon fusion at hadron
  colliders}},  {\em Phys. Rev.} \textbf{D82} (2010) 014002,
  [\href{http://arxiv.org/abs/1004.0962}{\texttt{arXiv:1004.0962}}].

\bibitem{Alekhin:2011sk}
S.~Alekhin, S.~Alioli, R.~D. Ball, V.~Bertone, J.~Blumlein, et~al.,
  \textit{{The PDF4LHC Working Group Interim Report}},
  \href{http://arxiv.org/abs/1101.0536}{\texttt{arXiv:1101.0536}}.

\bibitem{Ball:2007ra}
R.~D. Ball, \textit{{Resummation of Hadroproduction Cross-sections at High
  Energy}},  {\em Nucl.Phys.} \textbf{B796} (2008) 137--183,
  [\href{http://arxiv.org/abs/0708.1277}{\texttt{arXiv:0708.1277}}].

\bibitem{Chekanov:2009dq}
\textbf{ZEUS} Collaboration, S.~Chekanov et~al., \textit{{Measurement of
  isolated photon production in deep inelastic ep scattering}},  {\em
  Phys.Lett.} \textbf{B687} (2010) 16--25,
  [\href{http://arxiv.org/abs/0909.4223}{\texttt{arXiv:0909.4223}}].

\bibitem{DeRujula:1998yq}
A.~De~Rujula and W.~Vogelsang, \textit{{On the photon constituency of
  protons}},  {\em Phys.Lett.} \textbf{B451} (1999) 437--444,
  [\href{http://arxiv.org/abs/hep-ph/9812231}{\texttt{hep-ph/9812231}}].

\bibitem{GehrmannDeRidder:2006wz}
A.~Gehrmann-De~Ridder, T.~Gehrmann, and E.~Poulsen, \textit{{Isolated photons
  in deep inelastic scattering}},  {\em Phys.Rev.Lett.} \textbf{96} (2006)
  132002,
  [\href{http://arxiv.org/abs/hep-ph/0601073}{\texttt{hep-ph/0601073}}].

\bibitem{Chekanov:2004wr}
\textbf{ZEUS} Collaboration, S.~Chekanov et~al., \textit{{Observation of
  isolated high E(T) photons in deep inelastic scattering}},  {\em Phys.Lett.}
  \textbf{B595} (2004) 86--100,
  [\href{http://arxiv.org/abs/hep-ex/0402019}{\texttt{hep-ex/0402019}}].

\bibitem{Chatrchyan:2012meb}
\textbf{CMS} Collaboration, S.~Chatrchyan et~al., \textit{{Search for leptonic
  decays of $W$ ' bosons in $pp$ collisions at $\sqrt{s}=7$ TeV}},  {\em JHEP}
  \textbf{1208} (2012) 023,
  [\href{http://arxiv.org/abs/1204.4764}{\texttt{arXiv:1204.4764}}].

\bibitem{Chatrchyan:2011dx}
\textbf{CMS} Collaboration, S.~Chatrchyan et~al., \textit{{Search for a
  $W^\prime$ boson decaying to a muon and a neutrino in $pp$ collisions at
  $\sqrt{s} = 7$ TeV}},  {\em Phys.Lett.} \textbf{B701} (2011) 160--179,
  [\href{http://arxiv.org/abs/1103.0030}{\texttt{arXiv:1103.0030}}].

\bibitem{Chatrchyan:2011wq}
\textbf{CMS} Collaboration, S.~Chatrchyan et~al., \textit{{Search for
  Resonances in the Dilepton Mass Distribution in $pp$ Collisions at $\sqrt{s}
  = 7$ TeV}},  {\em JHEP} \textbf{1105} (2011) 093,
  [\href{http://arxiv.org/abs/1103.0981}{\texttt{arXiv:1103.0981}}].

\bibitem{Collaboration:2011dca}
\textbf{ATLAS} Collaboration, G.~Aad et~al., \textit{{Search for dilepton
  resonances in $pp$ collisions at $\sqrt{s}=7$ TeV with the ATLAS detector}},
  {\em Phys.Rev.Lett.} \textbf{107} (2011) 272002,
  [\href{http://arxiv.org/abs/1108.1582}{\texttt{arXiv:1108.1582}}].

\bibitem{Chatrchyan:2013oev}
\textbf{CMS} Collaboration, S.~Chatrchyan et~al., \textit{{Measurement of
  $W^+W^-$ and $ZZ$ production cross sections in $pp$ collisions at $\sqrt{s} =
  8$ TeV}},  {\em Phys.Lett.} \textbf{B721} (2013) 190--211,
  [\href{http://arxiv.org/abs/1301.4698}{\texttt{arXiv:1301.4698}}].

\bibitem{Chatrchyan:2011tz}
\textbf{CMS} Collaboration, S.~Chatrchyan et~al., \textit{{Measurement of $W^+
  W^-$ Production and Search for the Higgs Boson in $pp$ Collisions at
  $\sqrt{s}=7$ TeV}},  {\em Phys.Lett.} \textbf{B699} (2011) 25--47,
  [\href{http://arxiv.org/abs/1102.5429}{\texttt{arXiv:1102.5429}}].

\bibitem{ATLAS:2012mec}
\textbf{ATLAS} Collaboration, G.~Aad et~al., \textit{{Measurement of $W^+/W^-$
  production in $pp$ collisions at $\sqrt{s}=7$ TeV with the ATLAS detector and
  limits on anomalous $WWZ$ and $WW\gamma$ couplings}},
  \href{http://arxiv.org/abs/1210.2979}{\texttt{arXiv:1210.2979}}.

\bibitem{Chatrchyan:2012ypy}
\textbf{CMS} Collaboration, S.~Chatrchyan et~al., \textit{{Search for heavy
  resonances in the W/Z-tagged dijet mass spectrum in pp collisions at 7 TeV}},
   \href{http://arxiv.org/abs/1212.1910}{\texttt{arXiv:1212.1910}}.

\bibitem{Chatrchyan:2012rva}
\textbf{CMS} Collaboration, S.~Chatrchyan et~al., \textit{{Search for exotic
  resonances decaying into $WZ/ZZ$ in $pp$ collisions at $\sqrt{s}=7$ TeV}},
  {\em JHEP} \textbf{1302} (2013) 036,
  [\href{http://arxiv.org/abs/1211.5779}{\texttt{arXiv:1211.5779}}].

\bibitem{Chatrchyan:2012kk}
\textbf{CMS} Collaboration, S.~Chatrchyan et~al., \textit{{Search for a
  $W^\prime$ or Techni-$\rho$ Decaying into $WZ$ in $pp$ Collisions at
  $\sqrt{s}=7$ TeV}},  {\em Phys.Rev.Lett.} \textbf{109} (2012) 141801,
  [\href{http://arxiv.org/abs/1206.0433}{\texttt{arXiv:1206.0433}}].

\bibitem{Aad:2013wxa}
\textbf{ATLAS} Collaboration, G.~Aad et~al., \textit{{Search for resonant
  diboson production in the $WW/WZ \to l\nu jj$ decay channels with the ATLAS
  detector at $\sqrt{s}$ = 7 TeV}},
  \href{http://arxiv.org/abs/1305.0125}{\texttt{arXiv:1305.0125}}.

\bibitem{Aad:2012nev}
\textbf{ATLAS} Collaboration, G.~Aad et~al., \textit{{Search for new phenomena
  in the $W W$ to $\ell \nu \ell$' $\nu$' final state in $pp$ collisions at
  $\sqrt{s}=7$ TeV with the ATLAS detector}},  {\em Phys.Lett.} \textbf{B718}
  (2013) 860--878,
  [\href{http://arxiv.org/abs/1208.2880}{\texttt{arXiv:1208.2880}}].

\bibitem{Randall:1999vf}
L.~Randall and R.~Sundrum, \textit{{An Alternative to compactification}},  {\em
  Phys.Rev.Lett.} \textbf{83} (1999) 4690--4693,
  [\href{http://arxiv.org/abs/hep-th/9906064}{\texttt{hep-th/9906064}}].

\bibitem{Andersen:2011yj}
J.~Andersen, O.~Antipin, G.~Azuelos, L.~Del~Debbio, E.~Del~Nobile, et~al.,
  \textit{{Discovering Technicolor}},  {\em Eur.Phys.J.Plus} \textbf{126}
  (2011) 81, [\href{http://arxiv.org/abs/1104.1255}{\texttt{arXiv:1104.1255}}].

\bibitem{Eichten:2007sx}
E.~Eichten and K.~Lane, \textit{{Low-scale technicolor at the Tevatron and
  LHC}},  {\em Phys.Lett.} \textbf{B669} (2008) 235--238,
  [\href{http://arxiv.org/abs/0706.2339}{\texttt{arXiv:0706.2339}}].

\bibitem{Campbell:2011bn}
J.~M. Campbell, R.~K. Ellis, and C.~Williams, \textit{{Vector boson pair
  production at the LHC}},  {\em JHEP} \textbf{1107} (2011) 018,
  [\href{http://arxiv.org/abs/1105.0020}{\texttt{arXiv:1105.0020}}].

\bibitem{Kuhn:2011mh}
J.~Kuhn, F.~Metzler, A.~Penin, and S.~Uccirati,
  \textit{{Next-to-Next-to-Leading Electroweak Logarithms for W-Pair Production
  at LHC}},  {\em JHEP} \textbf{1106} (2011) 143,
  [\href{http://arxiv.org/abs/1101.2563}{\texttt{arXiv:1101.2563}}].

\bibitem{Boughezal:2013cwa}
R.~Boughezal, Y.~Li, and F.~Petriello, \textit{{Disentangling radiative
  corrections using the high-mass Drell-Yan process at the LHC}},  {\em
  Phys.Rev.} \textbf{D89} (2014), no.~3 034030,
  [\href{http://arxiv.org/abs/1312.3972}{\texttt{arXiv:1312.3972}}].

\bibitem{Catani:2007vq}
S.~Catani and M.~Grazzini, \textit{{An NNLO subtraction formalism in hadron
  collisions and its application to Higgs boson production at the LHC}},  {\em
  Phys.Rev.Lett.} \textbf{98} (2007) 222002,
  [\href{http://arxiv.org/abs/hep-ph/0703012}{\texttt{hep-ph/0703012}}].

\bibitem{Aad:2014qja}
\textbf{ATLAS} Collaboration, G.~Aad et~al., \textit{{Measurement of the
  low-mass Drell-Yan differential cross section at $\sqrt{s}$ = 7 TeV using the
  ATLAS detector}},  {\em JHEP} \textbf{1406} (2014) 112,
  [\href{http://arxiv.org/abs/1404.1212}{\texttt{arXiv:1404.1212}}].

\bibitem{Ball:2013hta}
\textbf{The NNPDF} Collaboration, R.~D. Ball et~al., \textit{{Parton
  distributions with QED corrections}},
  \href{http://arxiv.org/abs/1308.0598}{\texttt{arXiv:1308.0598}}.

\bibitem{Skands:2014pea}
P.~Skands, S.~Carrazza, and J.~Rojo, \textit{{Tuning PYTHIA 8.1: the Monash
  2013 Tune}},  {\em Eur.Phys.J.} \textbf{C74} (2014), no.~8 3024,
  [\href{http://arxiv.org/abs/1404.5630}{\texttt{arXiv:1404.5630}}].

\bibitem{Kallweit:2014xda}
S.~Kallweit, J.~M. Lindert, P.~Maierhöfer, S.~Pozzorini, and M.~Schönherr,
  \textit{{NLO electroweak automation and precise predictions for W+multijet
  production at the LHC}},
  \href{http://arxiv.org/abs/1412.5157}{\texttt{arXiv:1412.5157}}.

\bibitem{Carrazza:2013axa}
S.~Carrazza, S.~Forte, and J.~Rojo, \textit{{Parton Distributions and Event
  Generators}},
  \href{http://arxiv.org/abs/1311.5887}{\texttt{arXiv:1311.5887}}.

\end{thebibliography}\endgroup

\backmatter

\end{document}